%% file: main.tex
\documentclass[envcountchap,nospthms]{svmono-nospringer}

\def\version{} 

\def\macrosSB{}
\input{macros}

\hypersetup{hypertexnames=TRUE}
\hypersetup{
    urlcolor=black
}

\usepackage{mathptmx}
\usepackage{helvet}
\usepackage{courier}
\usepackage{type1cm}

\usepackage{makeidx}         
\usepackage{multicol}        
\usepackage[bottom]{footmisc}

\DeclareMathAlphabet{\mathcal}{OMS}{cmsy}{m}{n}

\makeindex             

\newtheorem{exercise}[theorem]{Exercise}

\newcommand{\Var}{\mathrm{Var}}
\newcommand{\Cov}{\mathrm{Cov}}

\newcommand{\Kdot}{\dot K}
\newcommand{\Qdot}{\dot Q}
\newcommand{\Vdot}{\dot V}

\newcommand{\VptE}{V_{{\rm pt}}^{(0)}}

\renewcommand{\stab}{{\rm st}}
\newcommand{\DVstab}{\DV^{\stab}}

\newcommand{\CRG}{C_{\rm RG}}

\newcommand{\Aux}{\Zcal}
\newcommand{\aux}{z}
\newcommand{\hAux}{\lambda}
\newcommand{\hAuxg}{\hAux}
\newcommand{\drb}{d} 

\newenvironment{solution}[1]{\bigskip\noindent{\bf Solution to Exercise~\ref{#1}.\label{sol:#1}}}{\bigskip}
\newcommand{\solref}[1]{\hyperref[sol:ex:#1]{[Solution]}}

\newcommand{\Nlast}{\hat{N}}

\newcommand{\Znewlast}{Z_{\hat{N}}}

\newcommand{\Cnewlast}[1]{C_{\hat{N}#1}}

\newcommand{\Vnewlast}{V_{\hat{N}}}

\newcommand{\Unewlast}{U_{\hat{N}}}

\newcommand{\Wnewlast}{W_{\hat{N}}}

\newcommand{\Inewlast}{I_{\hat{N}}}

\newcommand{\Knewlast}{K_{\hat{N}}}

\newcommand{\gnewlast}{g_{\hat{N}}}

\newcommand{\nunewlast}{\nu_{\hat{N}}}

\newcommand{\munewlast}{\mu_{\hat{N}}}

\newcommand{\unewlast}{u_{\hat{N}}}

\newcommand{\zetanewlast}[1]{\zeta_{\hat{N}#1}}

\newcommand{\Auxx}{\Zcal}
\newcommand{\auxx}{z}

\renewcommand{\aux}{y}
\renewcommand{\Aux}{\Ycal}

\usepackage{tikz}
\usetikzlibrary{arrows}
\usetikzlibrary{positioning}

\begin{document}
\author{
  Roland Bauerschmidt,
  David C.\ Brydges, and
  Gordon Slade}
\title{Introduction to a \\ renormalisation group method}
\date{\version}
\maketitle

\frontmatter
\include{preface}
\tableofcontents

\chapter*{Notation}

Throughout this book, we use the following notational conventions.
\begin{itemize}
\item
$x=o (y)$ means that $x/y\rightarrow 0$ as $y\rightarrow y_{0}$,
where $y_{0}$ is supplied by the context.
\item
$x=O (y)$ means that there exist $C,\delta$ such that $|x/y| \le C$ for $|y-y_{0}|<\delta$,
where $y_{0}$ is supplied by the context.
\item
$x=O_{z} (y)$ means that $x=O (y)$ as $y\rightarrow y_{0}$ with $z$ fixed, where $y_{0}$
is supplied by the context.
\item
$A \sim B$ means $A=B(1+o(1))$.
\item
$A \asymp B$ means $C^{-1} A \leq B \leq C A$ for a universal constant $C>0$.
\item
$A \propto B$ means $A = cB$ for some constant $c>0$ (which can depend on parameters).
\item
For $x=(x_i)_{i\in I}$ and $y=(y_i)_{i\in I}$ we write
$(x,y)=\sum_{i\in I} x_iy_i$, where the index set $I$ is supplied by
the context.
\end{itemize}

Some commonly used symbols are listed in the index.

\mainmatter

\part{Spin systems and critical phenomena}
\label{part:intro}
\include{intro}
\include{gauss}
\include{decomp}
\include{hier}

\part{The renormalisation group: Perturbative analysis}
\label{part:hierpt}
\include{hierpt}

\part{The renormalisation group: Nonperturbative analysis}
\label{part:hierK}
\include{Tphi}
\include{hierK}

\part{Self-avoiding walk and supersymmetry}
\label{part:saw}
\include{fermions}

\part{Appendices}
\appendix
\include{euclidean-notes}
\include{solns}

\backmatter
\printindex


\end{document}

%% file: macros.tex
\usepackage{amsfonts}
\usepackage{amsmath,amssymb,amsthm}
\usepackage{appendix}
\usepackage{bbm} 
\usepackage{amsbsy}
\usepackage{enumerate}
\usepackage{cite}

\def\macrosH{}
\def\macrosHarxiv

\InputIfFileExists{./macros_local.tex}{}{}

\ifdefined\macrosPa
  \usepackage[textwidth=465pt,textheight=650pt,centering]{geometry} 
\else\ifdefined\macrosPb
  \usepackage[textwidth=500pt,textheight=650pt,centering]{geometry} 
\fi\fi

\ifdefined\macrosS
  \makeatletter
  \def\boldmath{\pmb}
  
  \makeatother

  \usepackage{mathptmx}
  \DeclareMathAlphabet{\mathcal}{OMS}{cmsy}{m}{n}
\fi

\ifdefined\macrosBirk
\else
\usepackage[dvips]{graphicx}
\fi

\input{def99c} 
\UseSection   
\usepackage[usenames]{color}

\definecolor{bw}{RGB}{240, 120, 0}
\definecolor{at}{rgb}{0.0, 0.5, 0.0} 

\newcommand{\Tay}{{\rm Tay}}

\newcommand{\LT}{{\rm Loc}  }

\newcommand{\DV}{\Dcal}

\renewcommand{\to} {\rightarrow}

\newcommand{\sumtwo}[2]{\sum_{ \mbox{ \scriptsize
    $\begin{array}{c}
                        {#1} \\ {#2}
                        \end{array} $ }
    }
}

\newcommand{\R}{\Rbold}
\newcommand{\Z}{\Zbold}

\newcommand{\N}{\Nbold}
\newcommand{\C}{\mathbb{C}}

\newcommand{\1}{\mathbbm{1}}

\newcommand{\la}{\langle}
\newcommand{\ra}{\rangle}
\newcommand{\supp}{\mathrm{supp}}

\newcommand{\cL}{{\cal L}}

\newcommand{\FDel}{\lambda}

\newcommand{\psib}{\bar\psi}
\newcommand{\ci}{\underline{i}}

\newcommand{\Ex}{\mathbb{E}}

\newcommand{\pair}[1]{\langle #1 \rangle}

\newcommand{\pt}{{\rm pt}}

\newcommand{\Upt}{U_{\rm pt}}
\newcommand{\Vpt}{V_{\rm pt}}

\newcommand{\h}{\mathfrak{h}}
\ifdefined\macrosSB \else
  
  \newcommand{\I}{\mathfrak{I}}
\fi

\newcommand{\Id}{\mathrm{Id}}

\newcommand{\Lcallr}{\stackrel{\leftrightarrow}{\Lcal}}

\newcommand{\gbar}{\bar{g}}

\newcommand{\ggen}{\tilde{g}}

\newcommand{\mgen}{\tilde{m}}
\newcommand{\Iint}{\mathbb{I}}
\newcommand{\Igen}{\tilde{\mathbb{I}}}

\newcommand{\mubar}{\bar{\mu}}

\newcommand{\stab}{\mathrm{stab}}

\newcommand{\domRG}{\mathbb{D}}

\newcommand{\half}{\textstyle{\frac 12}}

\newcommand{\ddp}[2]{\frac{\partial #1}{\partial #2}}

\newcommand{\epdV}{\bar{\epsilon}}

\newcommand{\phib}{\bar\phi}

\newcommand{\sgn}{\mathrm{sgn}}

\newcommand{\Ktilde}{\tilde{K}}

\DeclareMathOperator{\Loc}{Loc} 

\ifdefined\macrosH
  \usepackage{xr-hyper}
  \usepackage{hyperref}
  \hypersetup{hypertexnames=false}
  \hypersetup{colorlinks,citecolor=blue,linkcolor=blue}  

\else\ifdefined\macrosHarxiv
  \usepackage{xr-hyper}
  \usepackage{hyperref}
  \hypersetup{hypertexnames=false, hidelinks}

\else
  \newcommand{\texorpdfstring}[2]{#1}
  \usepackage{xr}
\fi\fi

%% file: def99c.tex
\def\UseSection{
        \numberwithin{equation}{section}
	\theoremstyle{plain}
        \newtheorem{theorem}    {Theorem}[section]
        \DefineTheorems 
}

\def\DefineTheorems{
	
	\newtheorem{lemma}      [theorem] {Lemma}
	
	\newtheorem{prop}       [theorem] {Proposition}
	
	\newtheorem{cor}        [theorem] {Corollary}

	\theoremstyle{definition}
	\newtheorem{defn}       [theorem] {Definition}
	
	\newtheorem{example}       [theorem] {Example}
	\newtheorem{rk} 	[theorem] {Remark}
	\theoremstyle{definition}

}

\newcommand{\bt}   {\begin{theorem}}
\newcommand{\et}   {\end  {theorem}}
\newcommand{\bl}   {\begin{lemma}}
\newcommand{\el}   {\end  {lemma}}
\newcommand{\bp}   {\begin{prop}}
\newcommand{\ep}   {\end  {prop}}
\newcommand{\bc}   {\begin{cor}}
\newcommand{\ec}   {\end  {cor}}
\newcommand{\bd}   {\begin{defn}}
\newcommand{\ed}   {\end  {defn}}

\newcommand{\ba}   {\begin{array}}
\newcommand{\ea}   {\end  {array}}
\newcommand{\be}   {\begin{enumerate}}
\newcommand{\ee}   {\end  {enumerate}}
\newcommand{\bi}   {\begin{itemize}}
\newcommand{\ei}   {\end  {itemize}}

\def\eq#1\en{\begin{equation}#1\end{equation}}  
\def\eqsplit#1\ensplit{
	\begin{equation}\begin{split}#1\end{split}\end{equation}
	}
\def\eqalign#1\enalign{
	\begin{align}#1\end{align}
	}
\def\eqmul#1\enmul{
	\begin{multline}#1\end{multline}
	}
\newcommand{\eqarrstar} {\begin{eqnarray*}} 
\newcommand{\enarrstar} {\end{eqnarray*}} 
\newcommand{\eqarray}   {\begin{eqnarray}} 
\newcommand{\enarray}   {\end{eqnarray}} 
\newcommand{\nnb}	{\nonumber \\} 

\newcommand{\lbeq}[1]  {\label{e:#1}}
\newcommand{\refeq}[1] {\eqref{e:#1}}    
\newcommand{\lbfg}[1]  {\label{fg: #1}}
\newcommand{\reffg}[1] {\ref{fg: #1}}

\makeatletter
\newcommand{\labelcounter}[2]{{%
	\stepcounter{#1}
	\protected@write\@auxout{}%
	{\string\newlabel{#2}{{\csname the#1\endcsname}{\thepage}}}%
	{\ref{#2}}
	}}
\makeatother
\newcommand{\Ebold} {{\mathbb E}}

\newcommand{\Nbold} {{\mathbb N}}

\newcommand{\Rbold} {{\mathbb R}}

\newcommand{\Zbold} {{\mathbb Z}}

\newcommand{\Acal}   {\mathcal{A}} 
\newcommand{\Bcal}   {\mathcal{B}} 
\newcommand{\Ccal}   {\mathcal{C}} 
\newcommand{\Dcal}   {\mathcal{D}} 
 
\newcommand{\Fcal}   {\mathcal{F}}

\newcommand{\Lcal}   {\mathcal{L}} 
 
\newcommand{\Ncal}   {\mathcal{N}} 
 
\newcommand{\Pcal}   {\mathcal{P}}

\newcommand{\Scal}   {\mathcal{S}} 
\newcommand{\Tcal}   {\mathcal{T}} 
\newcommand{\Ucal}   {\mathcal{U}} 
\newcommand{\Vcal}   {\mathcal{V}} 
\newcommand{\Wcal}   {\mathcal{W}} 
\newcommand{\Xcal}   {\mathcal{X}}
\newcommand{\Ycal}   {\mathcal{Y}} 
\newcommand{\Zcal}   {\mathcal{Z}} 

\newcommand{\Ihat} {{\hat{I} }}  

\newcommand{\Khat} {{\hat{K} }}

\newcommand{\Uhat} {{\hat{U} }}
\newcommand{\Vhat} {{\hat{V} }}

\newcommand{\ubar}	{\bar{u}}

\newcommand{\Zd}    {{ {\Zbold}^d }}

\newcommand{\spose}[1] {{\hbox to 0pt{#1\hss}} }
\newcommand{\ltapprox} {\mathrel{\spose{\lower 3pt\hbox{$\mathchar"218$}}
 \raise 2.0pt\hbox{$\mathchar"13C$}}}
\newcommand{\gtapprox} {\mathrel{\spose{\lower 3pt\hbox{$\mathchar"218$}}
 \raise 2.0pt\hbox{$\mathchar"13E$}}}

\newcommand{\dist} {{  \rm dist }}



%% file: preface.tex
\preface

This book provides an introduction to a mathematically rigorous
renormalisation group method which is inspired by Kenneth Wilson's original ideas
from the early 1970s, for which he was awarded the 1982 Nobel Prize in Physics.
The method has been developed and applied over the past ten years in a
series of papers authored by various subsets of the present authors,
along with Martin Lohmann, Alexandre Tomberg and Benjamin Wallace.

We
present the general setting of the problems in critical phenomena that
have been addressed by the method, with focus on the 4-dimensional
$|\varphi|^4$ spin system and the $4$-dimensional continuous-time
weakly self-avoiding walk.  We give a self-contained analysis of the
4-dimensional \emph{hierarchical} $|\varphi|^4$ model, which is
simpler than its Euclidean counterpart but still reveals many of the
ideas and techniques of the renormalisation group method.
We comment on, and give detailed references for,
  the extension of the method to the Euclidean setting
in Appendix~\ref{app:Euclidean}.
The book is intended to be a starting point for a reader who may not have prior
knowledge of the renormalisation group method.

The book originated from lecture notes that were prepared for courses
at several summer schools. Subsequently the lecture notes were
significantly developed and rewritten.  The courses were given at:
\begin{itemize}
\item
the
Summer School in Mathematical Physics, Analysis and Stochastics,
Universit\"at Heidelberg, July 21-26, 2014;
\item
the
MASDOC Summer School on Topics in Renormalisation Group Theory and Regularity
Structures, University of Warwick, May 11-15, 2015;
\item
the Third NIMS Summer School in Probability:  Critical Phenomena, Renormalisation Group, and
Random Interfaces,
National Institute for Mathematical Sciences, Daejeon, June 15-19, 2015;
\item
the Workshop on Renormalization in Statistical Physics and Lattice Field Theories,
Institut Montpelli\'erain Alexander Grothendieck, August 24-28, 2015;
\item
the EMS-IAMP Summer School in Mathematical Physics:
Universality, Scaling Limits and Effective Theories, Rome, July 11-15, 2016;
\item
the
Bilbao Summer School on Probabilistic Approaches in Mathematical Physics,
Basque Center for Applied Mathematics, July 17-22, 2017.
\end{itemize}
We are grateful to Manfred Salmhofer
and Christoph Kopper in Heidelberg; to Stefan Adams in Warwick;
to Kyeong-Hun Kim, Panki Kim and Hyunjae Yoo in Daejeon;
to Damien Calaque and Dominique Manchon in Montpellier;
to Michele Corregi, Alessandro Giuliani, Vieri Mastropietro and Alessandro Pizzo in Rome;
and to Stefan Adams, Jean-Bernard Bru and Walter de Siqueira Pedra in Bilbao;
for organising these events and for
the invitations to lecture.

We are especially grateful to Alexandre Tomberg who gave tutorials for our
courses in Heidelberg and Daejeon, and to Benjamin Wallace who gave tutorials in Bilbao.
Each has contributed in several ways during the early stages of the writing of this book.

This work was supported in part by NSERC of Canada,
by the U.S.\ NSF under agreement DMS-1128155,
and by the Simons Foundation.

\begin{flushright}\noindent
Cambridge, UK \hfill {\it Roland Bauerschmidt}\\
Damariscotta, ME\hfill {\it David C.~Brydges}\\
Vancouver, BC \hfill  {\it Gordon Slade} \\
\end{flushright}

\noindent
June 28, 2019

\bigskip \bigskip

\noindent
Roland Bauerschmidt
\\
Department of Pure Mathematics and Mathematical Statistics
\\
University of Cambridge
\\
Centre for Mathematical Sciences
\\
Wilberforce Road
\\
Cambridge, CB3 0WB, UK
\\
{\tt rb812@cam.ac.uk}

\medskip \noindent
David C.~Brydges
\\
Department of Mathematics
\\
University of British Columbia
\\
Vancouver, BC, Canada V6T 1Z2
\\
{\tt db5d@math.ubc.ca}

\medskip \noindent
Gordon Slade
\\
Department of Mathematics
\\
University of British Columbia
\\
Vancouver, BC, Canada V6T 1Z2
\\
{\tt slade@math.ubc.ca}

%% file: intro.tex
\newcommand{\bbS}{S}
\newcommand{\avg}[1]{\la #1 \ra}
\newcommand{\pa}[1]{\left( #1 \right)}
\newcommand{\He}{\text{Hess}}

\chapter{Spin systems}
\label{ch:intro}

\section{Critical phenomena and the renormalisation group}

The subject of critical phenomena and phase transitions has fascinated
mathematicians for over half a century.  Interest in these topics is
now as great as ever, and models such as percolation, the Ising model,
self-avoiding walk, dimer systems, and others, are prominent in
mathematical physics, in probability theory, and in combinatorics.
The physically relevant and mathematically most interesting aspects of
the subject centre on universal quantities such as critical exponents.
These exponents describe the large-scale behaviour of a system of
strongly dependent random variables as a parameter governing the
strength of dependence, such as temperature, varies near a critical
value at which long-range correlations suddenly appear.  The critical
exponents are independent of many details of how a model is defined,
and for this reason models which are crude in their treatment of local
interactions can nevertheless provide accurate information about the
large-scale behaviour of real physical systems.

An extensive but incomplete mathematical theory of 2-dimensional
critical phenomena has been obtained in recent decades, particularly
with the advent of the Schramm-Loewner Evolution at the turn of the
century.  In high dimensions, namely dimensions $d>4$ for spin systems
and self-avoiding walk, there is a well-developed theory of mean-field
behaviour, based on techniques including reflection positivity,
differential inequalities, and the lace expansion.  The physically
most relevant dimension, $d=3$, has proved intractable to date and
remains an outstanding challenge to mathematicians.

The upper critical dimension, $d=4$, is borderline in the sense that
mean-field theory predicts the correct behaviour in dimensions $d>4$, but not $d<4$, and
typically this borderline behaviour involves logarithmic corrections
to mean-field scaling.  Dimension~4 is also the reference for the
$\epsilon$-expansion, which has provided heuristic results in
dimension $3$ by viewing $d=3$ as $d=4-\epsilon$ with $\epsilon=1$.
This book concerns a method for analysing 4-dimensional critical
phenomena and proving existence of logarithmic corrections to scaling.
The method has also been applied to lower dimensions via a version of
the $\epsilon$-expansion for long-range models.

In the physics literature, critical phenomena are understood via the
renormalisation group method developed by Kenneth G.\ Wilson in the
early 1970s.  Wilson received the 1982 Nobel Prize in Physics for this
development.
Inspiring early references include \cite{WK74,Fish83}.
Although Wilson's renormalisation group method is now
part of the standard toolbox of theoretical physics, there remain
serious challenges to place it on a firm mathematical and non-perturbative
foundation.
This book presents a renormalisation group method, developed by the
authors, which is applicable to the 4-dimensional $n$-component
$|\varphi|^4$ spin system and to the 4-dimensional continuous-time
weakly self-avoiding walk.  The latter is treated rigorously as a
supersymmetric ``$n=0$'' version of the former.  To simplify the
setting, we present the method in the context of the 4-dimensional
$n$-component hierarchical $|\varphi|^4$ model.  Discussion
of the self-avoiding walk is deferred to Chapter~\ref{ch:saw-int-rep}.

Extensions of the methods used in this book can found in
\cite{BBS-phi4-log,BBS-saw4-log,BBS-saw4,ST-phi4,BSTW-clp,Slad17,LSW17,BLS19}
(for $n \ge 0$).
Alternate approaches to the 4-dimensional $|\varphi|^4$ model using
block spin renormalisation can be found in
\cite{GK85,GK86,Hara87,HT87} (for $n=1$), and using phase space
expansion methods in \cite{FMRS87} (for $n=1$).  We make no attempt to
provide a thorough review of the many ways in which renormalisation
group methods have been applied in mathematical physics.  The
low-temperature phase has been studied, e.g., in \cite{Bala95,BO99}.
Renormalisation group methods have recently been applied to gradient
field models in \cite{AKM16}, to the Coulomb gas in
\cite{Falc12,Falc13}, to interacting dimers in \cite{GMT17}, and to
symmetry breaking in low temperature many-boson systems in
\cite{BFKT17}.  The books \cite{Mast08,Riva91,BG95,Salm99} provide
different approaches to the renormalisation group, and \cite{GJ87}
contains useful background.

\index{Universality class}
Two paramount features of critical phenomena are scale invariance and
universality.  The renormalisation group method exploits the scale
invariance to explain universality.  This is done via a multi-scale
analysis, in which a system studied at a particular scale is
represented by an effective Hamiltonian.  Scales are analysed
sequentially, leading to a map that takes the Hamiltonian at one scale
to a Hamiltonian at the next scale.  Advancing the scale gives rise to
a dynamical system defined by this map.  Scale invariance occurs at a
fixed point of the map, and different fixed points correspond to
different universality classes.  The analysis of the dynamical system
at and near the fixed point provides a means to compute universal
quantities such as critical exponents.  In the physics literature, the
analysis is typically performed in a perturbative fashion, without
control of remainder terms.  A mathematically rigorous treatment
requires full control of nonperturbative aspects as well.

This book
presents a self-contained and complete renormalisation group analysis
of the 4-dimensional $n$-component hierarchical $|\varphi|^4$ model.
We have set up the analysis in a fashion parallel to that of its
Euclidean counterpart in \cite{BBS-phi4-log,BBS-saw4-log}; the
Euclidean version involves additional ingredients which make its
analysis more involved.  In Appendix~\ref{app:Euclidean}, we indicate
the main differences and provide references for the Euclidean
analysis.

\bigskip
A \emph{spin system} is a collection of random variables, called
spins, which we denote
$(\varphi_x)_{x\in\Lambda}$ or $(\sigma_x)_{x\in\Lambda}$.
In the examples we discuss, the spins are vectors in $\R^n$.
The spins are indexed by a set
$\Lambda$, which we initially assume to be finite, but large, and
ultimately we are interested in the infinite volume limit $\Lambda \uparrow \Zd$.
The distribution on spin configurations is specified in terms of an energy
$H(\varphi)$ or $H(\sigma)$.    We discuss four examples of spin
systems in this chapter: the Ising model, the mean-field model, the
Gaussian free field, and the $|\varphi|^4$ model.

\section{Ising model}
\label{sec:Ising}

\begin{figure}[h]
\begin{center}
\includegraphics[scale = 0.4]{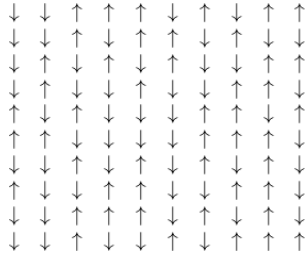}
\end{center}
\caption{
A configuration of the Ising model.}
\label{fig:ising-arrows}
\end{figure}

The prototypical example of a spin system is the Ising model, which is
defined as follows.\index{Ising model} Given a finite box $\Lambda
\subset \Zd$, an Ising configuration is $\sigma =
(\sigma_x)_{x\in\Lambda}$, $\sigma_x \in \{-1,1\}$, as depicted in
Figure~\ref{fig:ising-arrows}.  With $e$ one of the $2d$ unit vectors
in $\Zd$, we define the discrete gradient and Laplacian of a function
$f:\Zd \to \C$ by
\begin{equation}
\lbeq{graddef}
    (\nabla^e f)_x = f_{x+e}-f_x
    ,
    \quad
    (\Delta f)_x
    = - \frac 12 \sum_{e:|e|=1} \nabla^{-e}\nabla^e f_x
    =
    \sum_{e:|e|=1} \nabla^e f_x
    .
\end{equation}
An energy is associated to each configuration $\sigma$ by
\begin{equation}
\label{e:IsingH}
    H_{0,\Lambda}(\sigma)
    =\frac {1}{4} \sum_{e:|e|=1}\sum_{x\in\Lambda} (\nabla^e \sigma)_x^2 ,
\end{equation}
together with a boundary contribution fixing the spins on the outer
boundary of $\Lambda$.  Let $E^{(2)}$ be the set of edges $\{x,y \}$
where $x,y$ are nearest neighbour lattice sites. The energy
\refeq{IsingH} is twice the number of edges in $E^{(2)}$ whose spins
disagree.  Up to an additive constant, it can also be written as
$-\sum_{\{x,y \}\in E^{(2)}} \sigma_x\sigma_{y}$.

\begin{figure}%
\begin{center}
\includegraphics[width = .26 \textwidth]{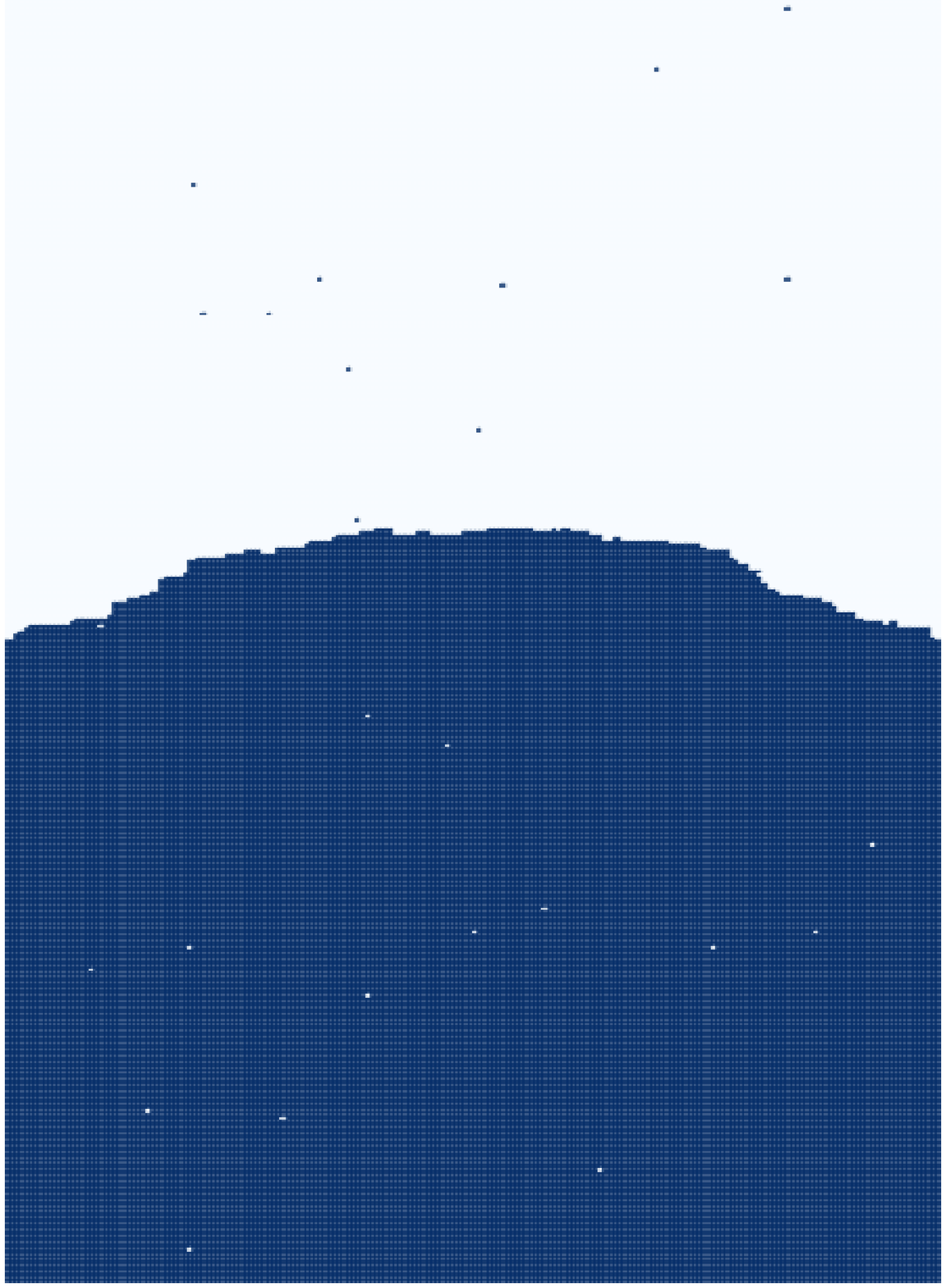}%
\includegraphics[width = .26 \textwidth]{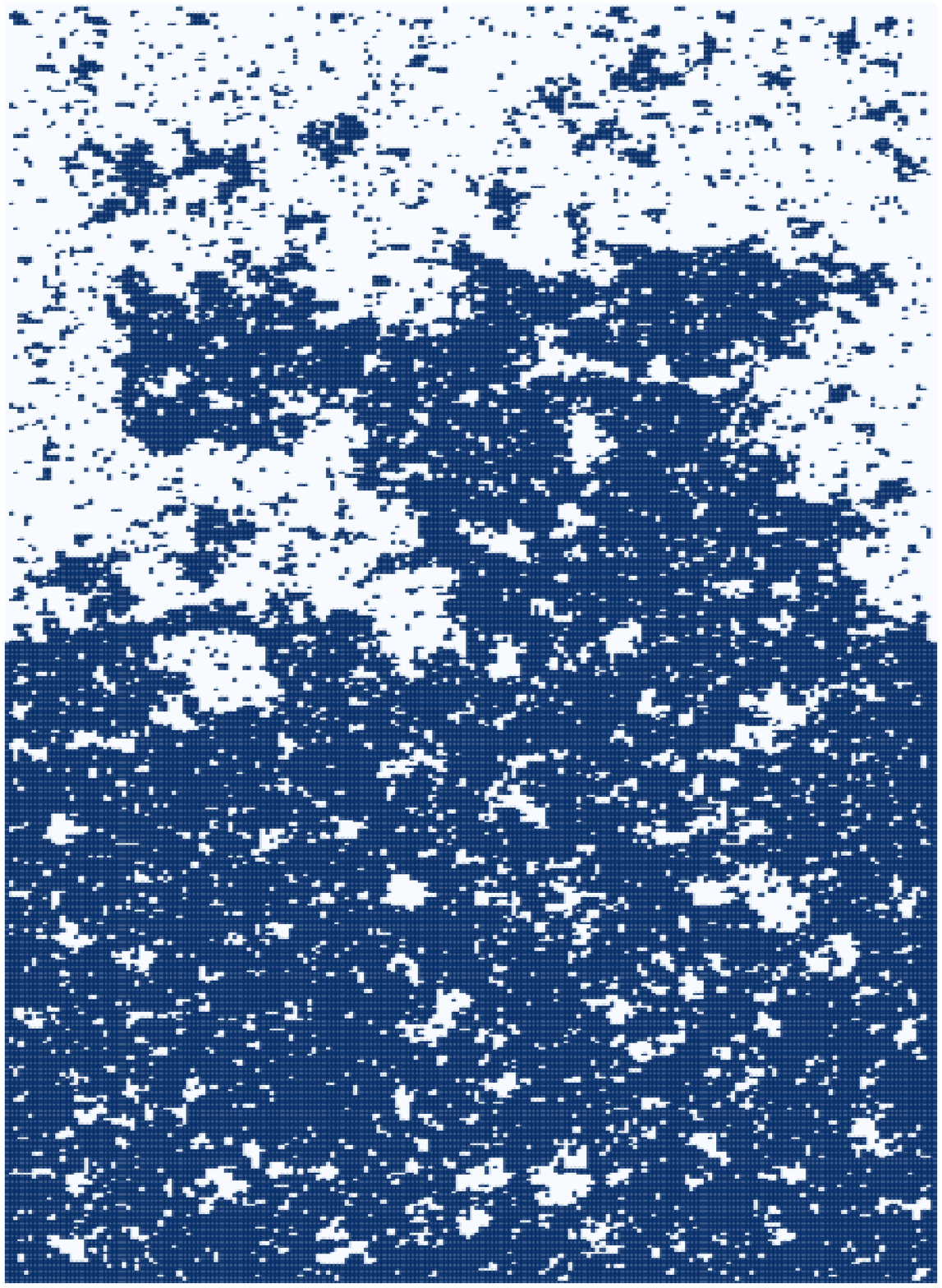}%
\includegraphics[width = .26 \textwidth]{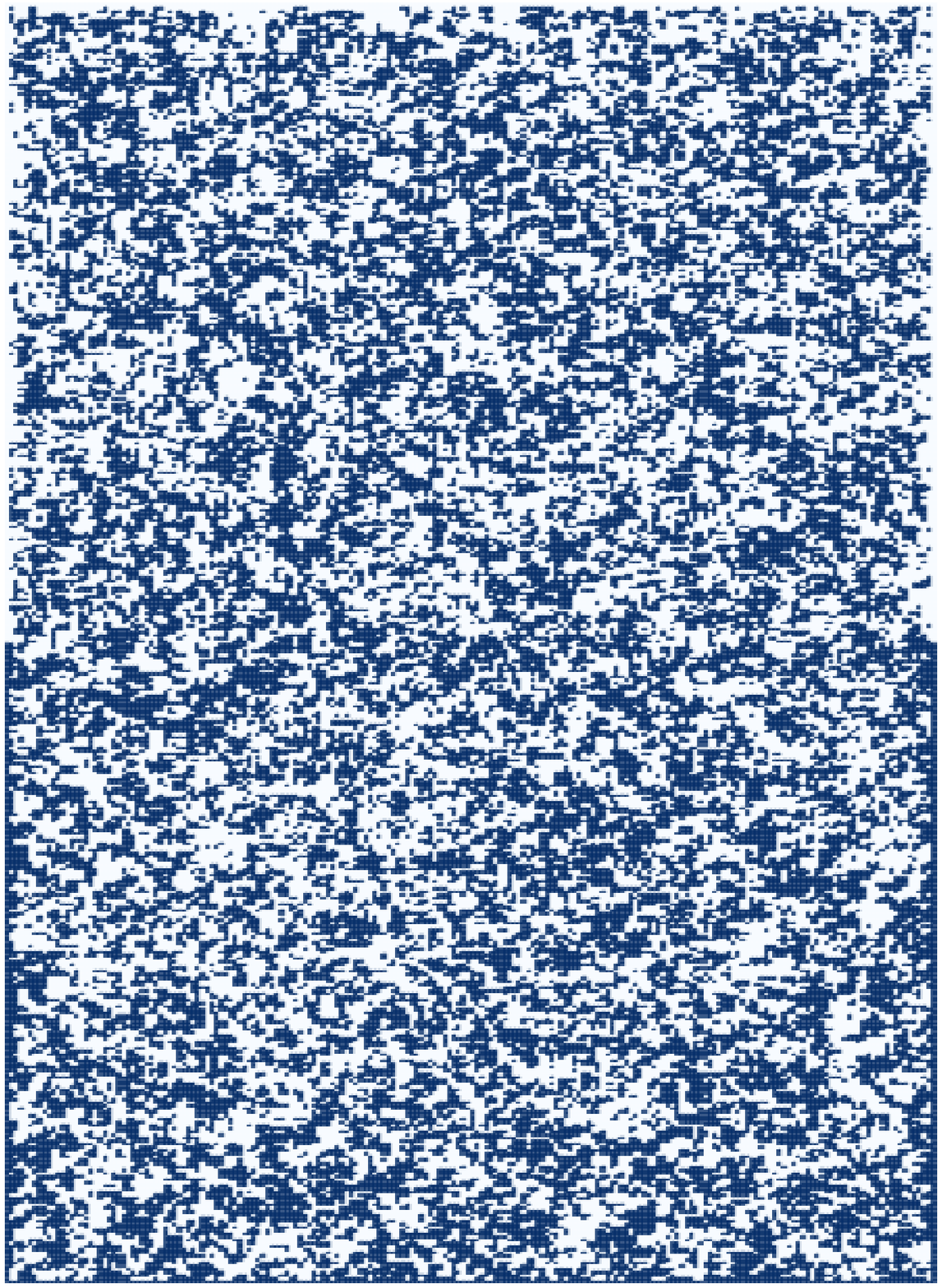}%
\end{center}
\begin{center}
{\scriptsize Low temperature $T<T_c$ \hspace{6mm} Critical temperature $T=T_c$ \hspace{4mm}
High temperature $T>T_c$}
\end{center}
\caption{Typical configurations of the 2-dimensional Ising model,
with boundary spins fixed white for the top half and dark for the
bottom half.
}%
\label{fig:ising-2D}%
\end{figure}

\index{Boltzmann weight}
The probability of a configuration $\sigma$ is given by the
finite-volume Gibbs measure
\begin{equation}
\lbeq{IsingGibbs}
    P_{T,\Lambda}(\sigma)
    \propto
    e^{-H_{0,\Lambda}(\sigma)/T}
    \prod_{x\in \Lambda} (\delta_{\sigma_x,+1} + \delta_{\sigma_x,-1}),
\end{equation}
where $T$ represents \emph{temperature}, and where the constant of
proportionality is such that $P_{T,\Lambda}$ is a probability measure.
The interaction is \emph{ferromagnetic}: configurations with more
neighbouring spins aligned are energetically favourable (lower energy) and have
higher probability.
The configurations with all spins $+1$ or all spins $-1$ have the lowest energy.
For higher energies there is a larger number of
configurations realising that energy, leading to a greater weight---or \emph{entropy}---of these in
the probability measure. The competition of energy and entropy, whose relative weight is controlled
by the temperature, leads to a phase transition at a critical temperature $T_c$.
For $T<T_c$, the dominant mechanism is the minimising of energy,
while for $T>T_c$, it is the effect of entropy that dominates.
Typical configurations look dramatically
different depending on whether $T$ is below, at, or above the critical
temperature $T_c$; see Figure~\ref{fig:ising-2D}.

\index{External field}
\index{Magnetic field}
To model the effect of an \emph{external magnetic field} $h \in \R$,
the Hamiltonian becomes
\begin{equation}
    H_{h,\Lambda}(\sigma)
    = H_{0,\Lambda}(\sigma) -h\sum_{x\in\Lambda}\sigma_x
    = \frac 14 \sum_{e}\sum_{x\in\Lambda} (\nabla^e \sigma)_x^2
    -h\sum_{x\in\Lambda}\sigma_x.
\end{equation}
Associated to this Hamiltonian, there is again a finite-volume Gibbs
measure with $H_{0,\Lambda}$ replaced by $H_{h,\Lambda}$ in
\refeq{IsingGibbs}.
\index{Infinite-volume limit}
\index{Gibbs measure}
The infinite-volume \emph{Gibbs measure} $P_{h,T}$ is defined to be
the limit of the measures $P_{h,T,\Lambda}$ as $\Lambda \uparrow \Zd$.
There is work to do to show existence of the limit, which may depend
on boundary conditions and fail to be unique.  Expectation with
respect to $P_{h,T}$ is denoted $\langle \cdot \rangle_{h,T}$.  See,
e.g., \cite{FV17,Geor11,Simo93} for details about Gibbs measures.

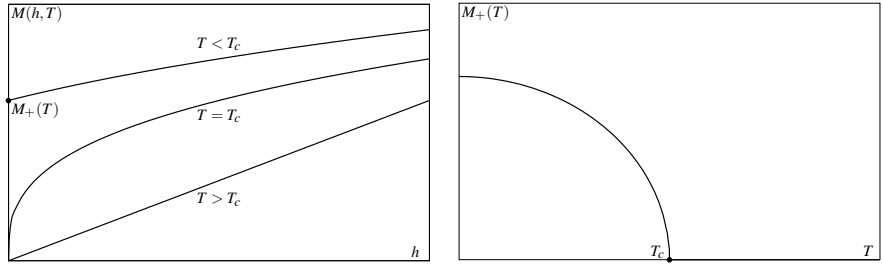
\begin{figure}[h]
\begin{center}
\input{magnetisation.pspdftex}
\hspace{2mm}
\input{spontmagnetisation.pspdftex}
\end{center}
\caption{
Critical behaviour of the magnetisation.}
\label{fig:mag}
\end{figure}

\index{Two-point function}
\index{Correlation length}
\index{Susceptibility}
\index{Magnetisation}
\index{Spontaneous magnetisation}
The \emph{magnetisation} is defined by $M(h,T) = \langle
\sigma_0\rangle_{h,T}$, and the \emph{spontaneous magnetisation} is
$M_+(T) = \lim_{h \downarrow 0} M(h,T)$.  The phase transition for the
Ising model is illustrated in Figure~\ref{fig:mag}.  Above the
critical temperature $T_c$, the spontaneous magnetisation is zero,
whereas below $T_c$ it is positive.  The slope of the magnetisation
$M(h,T)$ at $h=0$ is called the magnetic \emph{susceptibility}; it
diverges as $T \downarrow T_c$.  More precisely, for $T \ge T_c$, we
define:
\begin{align}
\label{e:Ising2ptfcn}
\text{two-point function:}
 &\quad \tau_{0x}(T) = \langle \sigma_0\sigma_x \rangle_{0,T},
 \\
\label{e:Isingcorrlength}
\text{correlation length:}
    & \quad
 \xi(T)^{-1} = -\lim_{n \to \infty} n^{-1} \log \tau_{0,ne_1}(T)
 ,
\\
\label{e:Isingsuscept}
\text{susceptibility:}
& \quad
\chi(T) = \sum_{x\in\Zd} \tau_{0x}(T)
  = \frac{\partial}{\partial h}M(h,T) \Big|_{h=0}.
\end{align}
In \eqref{e:Isingcorrlength}, $e_1=(1,0,\ldots,0)$ is a unit vector in $\Z^d$.
The most subtle and interesting behaviour occurs at and near the phase
transition, where the spins develop strong and non-trivial
correlations.  The scaling of these can be described in terms of
various critical exponents, as follows:
\index{Critical exponent}
\begin{align}
    \chi(T) &\sim A_1 (T-T_c)^{-\gamma}  &(T \downarrow T_c),
    \\
    \xi(T) &\sim A_2 (T-T_c)^{-\nu}  &(T \downarrow T_c),
    \\
    \tau_{0x}(T_c)  &\sim A_3 |x|^{-(d-2+\eta)} &(|x| \to\infty),
    \\
    M(h,T_c)  &\sim A_4 h^{1/\delta} &(h \downarrow 0),
    \\
\lbeq{beta-ce}
    M_+(T)  &\sim A_5 (T_c-T)^{\beta} &(T \uparrow T_c).
\end{align}
\index{Scaling relation}%
\index{Fisher's relation}%
The critical exponents are conjectured to obey certain \emph{scaling
relations}, an example of which is \emph{Fisher's relation} $\gamma =
(2-\eta)\nu$.  The critical exponents are predicted to be universal.
This means that they should depend primarily on the dimension $d$ and
not on fine details of how the model is formulated.  For example, the
exponents are predicted to be the same on the square or triangular or
hexagonal lattices for $d=2$.  The main mathematical problem for the
Ising model, and for spin systems more generally, is to provide
rigorous proof of the existence and universality of the critical
exponents.  The following is an informal summary of what has been
achieved so far.

There has been great success for the case of $d=2$.  For the square
lattice $\Z^2$, it has been proved that the critical temperature is
given by $T_c^{-1}= \frac 12 \log (1+\sqrt{2})$, and that the critical
exponents $\gamma,\beta,\delta,\eta,\nu$ exist and take the values
$\gamma = \frac74$,
$\beta = \frac 18$,
$\delta = 15$,
$\eta = \frac 14$,
$\nu=1$.
In addition, the law of the interface curve in the middle picture in
Figure~\ref{fig:ising-2D} is the Schramm--Loewner Evolution ${\rm
SLE}_3$.  References for these theorems include
\cite{Onsa44,CDHKS14,CGN14,BD-C12}.

In dimensions $d>4$, also much is known.  The critical exponents
$\gamma,\beta,\delta,\eta$ exist and take the values $\gamma=1$,
$\beta = \frac 12$, $\delta =3$, $\eta = 0$.  These exponents have the
same values as for the Ising model defined on the complete graph,
which is called the \emph{Curie--Weiss} or \emph{mean-field} Ising
model.  Precise statements and proofs of these facts can be found in
\cite{Aize82,Froh82,AF86,Saka07}. We discuss the mean-field Ising
model in more detail in Section~\ref{sec:mf}.

\index{Logarithmic corrections}%
Logarithmic corrections to mean-field behaviour are predicted for
$d=4$ \cite{LK69,WR73,BGZ73}, and it is known that there cannot be
corrections which are larger than logarithmic \cite{AG83,AF86}.  It
remains an open problem to prove the precise behaviour for $d=4$, and
in this book we address some closely related problems concerning the
$|\varphi|^4$ model.  For the hierarchical Ising model in dimension 4,
a rigorous renormalisation group analysis is presented in
\cite{HHW01}.

Only recently has it been proved that the spontaneous magnetisation
vanishes at the critical temperature for $\Z^3$ \cite{ADS13}.  It
remains a major open problem to prove the existence of critical
exponents for $d=3$.  In the physics literature, the conformal
bootstrap has been used to compute exponents to high accuracy
\cite{EPPRSV14}.

\section{Spin systems and universality}
\label{sec:universality}
\index{Universality}

The Ising model is only one example of a large class of spin systems.
A general class of $O(n)$-symmetric ferromagnetic spin models can be
defined as follows.

Let $\Lambda$ be a finite set, and let $\beta_{xy}=\beta_{yx}$ be
nonnegative spin-spin \emph{coupling constants} indexed by $\Lambda
\times \Lambda$.  A spin configuration consists of a spin $\varphi_x
\in \R^n$ for each $x\in\Lambda$, and can be considered either as a
map $\varphi:\Lambda \to \R^n$ or as an element $\varphi \in
\R^{n\Lambda}$.  The \emph{bulk energy} of the spin configuration
$\varphi$ is
\begin{equation}
  H(\varphi) = \frac14 \sum_{x,y \in\Lambda} \beta_{xy} |\varphi_x-\varphi_y|^2
  + \sum_{x\in \Lambda}h \cdot \varphi_x
  .
\end{equation}
The constant vector $h$ represents an external magnetic field, which
may be zero.
\index{Single-spin distribution}%
For a given reference measure $\mu$ on $\R^n$ called the
\emph{single-spin distribution}, a probability measure on spin
configurations is defined by the expectation
\begin{equation}
  \la F  \ra \propto \int_{\R^{n\Lambda}} F(\varphi) e^{-H(\varphi)} \prod_{x\in\Lambda} \mu(d\varphi_x).
\end{equation}
\index{Ferromagnetic}%
The assumption $\beta_{xy} \geq 0$ is the assumption that the model is
\emph{ferromagnetic}: it encourages spin alignment.
When $\mu$ is absolutely continuous it is usually convenient to
instead take $\mu$ equal to the Lebesgue measure and equivalently add
a potential to the energy, i.e.,
\begin{equation}
  H(\varphi) = \frac14 \sum_{x,y\in \Lambda} \beta_{xy} |\varphi_x-\varphi_y|^2
  + \sum_{x\in \Lambda}h \cdot \varphi_x
  + \sum_{x\in \Lambda} w(\varphi_x).
\end{equation}

\index{Laplacian}
We associate to $\beta$ the Laplacian matrix $\Delta_\beta$, which
acts on scalar fields $f:\Lambda \to \R$ by
\begin{equation}
\lbeq{Lapbeta}
  (\Delta_\beta f)_x = \sum_{y\in\Lambda} \beta_{xy} (f_y-f_x)
  .
\end{equation}
For the case where $\beta_{xy}=\1_{x\sim y}$ is the indicator that $x$
and $y$ are nearest neighbours in $\Zd$, this recovers the standard
Laplacian of \refeq{graddef}.  For vector-valued fields
$f=(f^1,\ldots, f^n)$ the Laplacian acts component-wise, i.e.,
$(\Delta_\beta f)^i = \Delta_\beta f^i$.
Then we can rewrite $H(\varphi)$ as
\begin{equation}
\lbeq{HLap}
  H(\varphi) = \frac12 \sum_{x \in \Lambda} \varphi_x \cdot (-\Delta_\beta) \varphi_y
  + \sum_{x\in \Lambda}h \cdot \varphi_x
  + \sum_{x\in \Lambda} w(\varphi_x).
\end{equation}
Boundary terms can be included in the energy as well.

\smallskip
\index{Mean-field model}
Examples are given by the following choices of $\mu$ and $w$.  Since
$\mu$ and $w$ provide redundant freedom in the specification of the
model, we either specify $\mu$ and then assume that $w=0$, or we
specify $w$ and then assume that $\mu$ is the Lebesgue measure.
\begin{itemize}
\item
  Ising model: $n=1$ and $\mu = \delta_{+1} + \delta_{-1}$. 
\item
  $O(n)$ model: $\mu$ is the uniform measure on $\bbS^{n-1} \subset \R^n$. 
\item
  Gaussian free field (GFF): $w(\varphi_x) = m^2|\varphi_x|^2$ with $m^2 \geq 0$. 
\item
  $|\varphi|^4$ model: $w(\varphi_x) = \frac14 g|\varphi_x|^4+ \frac12 \nu|\varphi_x|^2$
  with $g>0$ and $\nu\in\R$.
\end{itemize}
\index{Ising model}\index{Rotator model}\index{XY model}\index{Heisenberg model}%
The $O(n)$ model is the Ising model when $n=1$, and it is also called
the rotator model for $n=2$, and the classical Heisenberg model for $n=3$.

\index{Mean-field interaction}%
\index{Nearest-neighbour interaction}%
\index{Finite-range interaction}%
\index{Long-range interaction}%
\index{Hierarchical interaction}%
\smallskip
Examples for the choice of interaction $\beta$ are:
\begin{itemize}
\item
  Mean-field interaction: $\beta_{xy} = \beta/|\Lambda|$ for all $x,y\in \Lambda$.
\item
  Nearest-neighbour interaction: $\Lambda \subset \Z^d$ and $\beta_{xy} = \beta \1_{x\sim y}$.
\item
  Finite-range interaction: $\Lambda \subset \Z^d$ and $\beta_{xy} = \beta \1_{|x-y| \leq R}$
  for some $R\ge 1$.
\item
  Long-range interaction: $\Lambda \subset \Z^d$ and $\beta_{xy}
  \asymp
  |x-y|^{-(d+\alpha)}$
  for some $\alpha \in (0,2)$.
\item
  Hierarchical interaction: discussed in detail in Chapter~\ref{ch:hier}.
\end{itemize}

\index{Universality}
In appropriate limits $|\Lambda|\to\infty$, the above models typically
undergo phase transitions as their respective parameters are varied.
As in the example of the Ising model, the critical behaviour can be
described by critical exponents.
The \emph{universality conjecture} for critical phenomena asserts that
the critical behaviour of spin models is the same within very general
symmetry classes.

The symmetry class is determined by the number of components $n$,
corresponding to the symmetry group $O(n)$, and the
class of coupling constants.  For example, in $\Z^d$, the same
critical behaviour is predicted when the spin-spin coupling $\beta$
has any finite range, or bounded variance $\sum_{x\in \Zd} |x|^2
\beta_{0x}$ (in infinite volume), as long as $\mu$ or $w$ has
appropriate regularity and growth properties.  Also, the same critical
behaviour is predicted for the $O(n)$ and $|\varphi|^4$ models.  A
general proof of the universality conjecture is one of the major open
problems of statistical mechanics.

In the remainder of this chapter, we consider three of the above
examples: the mean-field model, the Gaussian free field, and the
$|\varphi|^4$ model.  For both the mean-field model and the Gaussian
free field, a complete analysis can be carried out. We present
specific instructive cases that illustrate the general phenomena.  The
$|\varphi|^4$ model is a \emph{generic} case, on which much of the
remainder of this book is focussed.

\section{Mean-field model}
\label{sec:mf}
\index{Mean-field model}

\subsection{Critical behaviour of the mean-field model}

Let $n \ge 1$ be an integer, and let $\Lambda=\{0,1,\ldots,N-1\}$ be a
finite set.  As mentioned in the previous section, the mean-field
model corresponds to the choice $\beta_{xy}=\beta/N$ for the coupling
constants.  With this choice, the Laplacian of \refeq{Lapbeta} is
given by
\begin{equation}
    \label{e:P-def-intro}
    -\Delta_\beta = \beta P \quad \text{with} \quad P = \Id-Q,
\end{equation}
where $\Id$ denotes the $N\times N$ identity matrix and $Q$ is the
constant matrix with entries $Q_{xy}=N^{-1}$.  Note that $P$ and $Q$
are orthogonal projections with $P+Q=\Id$.  The energy of the
mean-field $O(n)$ model is then given by
\begin{equation}
  H(\sigma) = \frac12
  \sum_{x \in\Lambda}  \sigma_x \cdot (-\Delta_\beta \sigma)_x
  + \sum_{x\in \Lambda}h \cdot \sigma_x
  .
\end{equation}
The finite-volume expectation is defined by
\begin{equation}
\lbeq{MFavg}
  \avg{F}_{\beta,h,N}
  \propto
  \int_{(\bbS^{n-1})^N} F(\sigma) \, e^{-H(\sigma)} \prod_{x\in\Lambda} \mu(d\sigma_x),
\end{equation}
\index{Curie--Weiss model}%
where the single-spin distribution $\mu$ is the uniform measure on the
sphere $\bbS^{n-1} \subset \R^n$.  In particular, for $n=1$, the
sphere $\bbS^{n-1}$ is the set $\{-1,+1\}$ and we have the mean-field
Ising model, or Curie--Weiss model.  In terms of the temperature
variable $T$ used in our discussion of the Ising model in
Section~\ref{sec:Ising}, here $\beta$ is the inverse temperature
$\beta = 1/T$.

The mean-field Ising model is a canonical example which is discussed
in many books on statistical mechanics, including
\cite{Baxt82,Elli85,FV17}. It is important for various reasons: it is
an example where nontrivial critical behaviour can be worked out
exactly and completely including computation of critical exponents,
its critical exponents have been proven to give bounds on the critical
exponents of other models, and its critical exponents are proven or
predicted to give the same values as other models in dimensions $d>4$.

What makes the mean-field model more tractable is its lack of
geometry.
Apart from an unimportant volume-dependent constant
that is independent of the spin configuration, the energy can be
rewritten in terms of the mean spin $\bar{\sigma} = N^{-1}\sum_x
\sigma_x$ as
\begin{equation}
  H(\sigma) = -\frac12 \frac{\beta}{N}
  \sum_{x,y}  \sigma_x\cdot  \sigma_y
  + \sum_{x} h \cdot \sigma_x
  +\text{const}
  = N \left( -\frac12 \beta \bar{\sigma}\cdot \bar\sigma + h \cdot \bar\sigma \right)
    +\text{const}
  .
\end{equation}
Thus $H$ is actually a function only of the mean spin.  This is the
origin of the name ``mean-field'' model.

\index{Susceptibility}
\index{Magnetisation}
The susceptibility and magnetisation are defined by
\begin{align}
  \label{e:MF-M}
  M(\beta,h) &= \lim_{N\to\infty} \avg{\sigma_0}_{\beta,h,N},
  \\
  \label{e:MF-chi}
  \chi(\beta,h)
  &=
  \ddp{M}{h}(\beta,h)
               .
\end{align}
For the results we focus on the Ising case $n=1$, but we present the
set-up for the general $O(n)$ model.  We will prove the following
theorem, which shows that the critical exponents $\gamma,\delta,\beta$
(for the susceptibility, the vanishing of the magnetisation at the
critical point, and the spontaneous magnetisation) take the
\emph{mean-field values} $\gamma=1$, $\delta=3$, $\bar\beta = \frac
12$.  We have written $\bar\beta$ for the critical exponent of the
spontaneous magnetisation rather than $\beta$ as in \refeq{beta-ce},
since here $\beta$ represents the inverse temperature.  The theorem
also shows that the critical value of $\beta$ is $\beta_c=1$.

\index{Critical exponent}
\begin{theorem}
\label{thm:MFexp}
  Let $\beta_c=1$.

\smallskip \noindent (i)
    The spontaneous magnetisation obeys
    \begin{equation}
      M_+(\beta)  \begin{cases}
        >0 & (\beta > \beta_c)\\
        = 0 & (\beta \leq \beta_c),
      \end{cases}
    \end{equation}
    and
    \begin{equation}
      M_+(\beta) \sim (3(\beta-\beta_c))^{1/2} \quad (\beta \downarrow \beta_c).
    \end{equation}

\smallskip \noindent (ii)
    The magnetisation obeys
    \begin{equation}
      M(\beta_c,h) \sim (3h)^{1/3} \quad (h\downarrow 0).
    \end{equation}

\smallskip \noindent (iii)
   The susceptibility is finite for $\beta<\beta_c$ for any $h$, and also for $\beta>\beta_c$ if $h\neq 0$,
   and
   \begin{equation}
     \chi(\beta,0)
     = \frac{1}{\beta_c-\beta} \quad (\beta < \beta_c),
     \qquad
     \chi(\beta,0_{+}) \sim
     \frac{1}{2(\beta-\beta_c)} \quad (\beta \downarrow \beta_c).
   \end{equation}
\end{theorem}

\subsection{Renormalised measure}

We start with the following elementary lemma.

\begin{lemma} \label{lem:MF-decomp}
Let $\Delta_\beta=-\beta P$ be the mean-field Laplacian.  There is a
constant $c>0$ such that
\begin{equation} \label{e:MF-decomp}
  e^{-\frac12(\sigma,-\Delta_\beta\sigma)} = c \int_{\R^n} e^{-\frac{\beta}{2} (\varphi-\sigma,\varphi-\sigma)} \, d\varphi
  \qquad
  (\sigma \in (\R^n)^N),
\end{equation}
where we identify $\varphi \in \R^n$ as a constant vector $(\varphi,
\dots, \varphi) \in (\R^n)^{N}$, and the parentheses denote the inner
product on $(\R^n)^N$.
\end{lemma}

\begin{proof} Let $\bar\sigma = N^{-1}\sum_x \sigma_x$ denote
the average spin.  We can regard both $\bar\sigma$ and $\varphi$ as
constant vectors in $(\R^n)^N$. By the discussion around
\eqref{e:P-def-intro}, $Q\sigma = \bar\sigma $, and $P=\Id - Q$
projects onto the orthogonal complement of the subspace of constant
fields. Therefore,
\begin{align} \label{e:MF-decomp-pf}
  \big(\varphi-\sigma,\varphi-\sigma\big)
  &=
  \big(\varphi-\sigma,Q (\varphi-\sigma)\big) + \big(\varphi-\sigma,P (\varphi-\sigma)\big)
  \nnb
  &=
  N|\varphi-\bar\sigma|^{2} + \big(\sigma,P \sigma\big) .
\end{align}
We take the exponential $\exp(-\frac 12 \beta (\cdot))$ of both sides
and integrate over $\varphi \in \R^n$.  The term involving $(\sigma, P
\sigma)$ factors out of the integral and gives the desired left-hand
side of \refeq{MF-decomp}, and the remaining integral is seen to be
independent of $\sigma$ after making the change of variables $\varphi
\mapsto \varphi +\bar\sigma$.
\end{proof}

The identity \eqref{e:MF-decomp} allows us to decompose the measure of the mean-field model
$\nu$ on $(\bbS^{n-1})^N$ into two measures,
which we call the renormalised measure and the fluctuation measure.

The \emph{renormalised measure} $\nu_r$ is a measure on $\R^n$ defined as follows.
For $\varphi \in \R^n$, we define the \emph{renormalised potential} by
\begin{equation} \label{e:Vr-def}
  V(\varphi)
  =
  -\log \int_{\bbS^{n-1}} e^{-\frac{\beta}{2} (\varphi-\sigma)\cdot (\varphi-\sigma)+h\cdot \sigma} \, \mu(d\sigma).
\end{equation}
The renormalised measure is then defined by the expectation
\begin{equation}
\lbeq{EGN}
  \Ex_{\nu_r}(G) \propto \int_{\R^n} G(\varphi) \, e^{-NV(\varphi)} \, d\varphi.
\end{equation}
The \emph{fluctuation measure} $\mu_{\varphi}$
is a measure on $(\bbS^{n-1})^N$ but of simpler form than the original $O(n)$ measure.
It is a product measure that depends on the renormalised field $\varphi \in \R^n$, and is defined by
\begin{equation} \label{e:muphidef}
  \Ex_{\mu_\varphi}(F)
  =
  \frac{1}{e^{-NV(\varphi)}}
   \int_{(\bbS^{n-1})^N} F(\sigma) \prod_{x\in\Lambda}
  e^{-\frac{\beta}{2} (\varphi-\sigma_x)\cdot (\varphi - \sigma_x)+h\cdot \sigma_x} \, \mu(d\sigma_x)
  .
\end{equation}

\begin{lemma} \label{lem:MFnudecomp}
  The mean-field measure \refeq{MFavg} has the decomposition
  \begin{equation}
  \label{e:MFnudecomp}
    \avg{F}_{\beta,h,N}
    = \Ex_{\nu_r}(\Ex_{\mu_\varphi}(F)) \quad \text{for $F: (\bbS^{n-1})^N \to \R$.}
  \end{equation}
\end{lemma}

\begin{proof}
  The proof is just a matter of substituting in definitions and using \eqref{e:MF-decomp}:
  \begin{align}
    \avg{F}_{\beta,h,N}
    &\propto
      \int_{(\bbS^{n-1})^N} F(\sigma) \, e^{-\frac12 (\sigma,(-\Delta_\beta)\sigma)+(h,\sigma)} \,
      \prod_{x\in\Lambda} \mu(d\sigma_x)
    \nnb
    &\propto
      \int_{\R^n}
      \int_{(\bbS^{n-1})^N}
      F(\sigma)
      \prod_{x\in\Lambda}
      e^{-\frac{\beta}{2} (\varphi-\sigma_x)\cdot (\varphi-\sigma_x)+h\cdot\sigma_x}
      \mu(d\sigma_x) d\varphi
    \nnb
    &=
      \int_{\R^n} e^{-NV(\varphi)}
      \Ex_{\mu_\varphi}(F) d\varphi
      \nnb
    &
    \propto \Ex_{\nu_r}(\Ex_{\mu_\varphi}(F)).
  \end{align}
  Since $\Ex_{\nu}(1) = 1 = \Ex_{\nu_r}(\Ex_{\mu_\varphi}(1))$,
  the proportional relation becomes an identity.
\end{proof}

The above decomposition of the measure into a fluctuation measure and
a renormalised measure can be seen as a toy example of the idea of renormalisation.
This is further discussed in Example~\ref{ex:MMFdecomp}.

\subsection{Magnetisation and susceptibility: Proof of Theorem~\ref{thm:MFexp}}
\label{sec:MFpf}

To compute the magnetisation, we need the observable $F(\sigma) = \sigma_0$.
Let
\begin{equation}
\lbeq{Gmag}
    G(\varphi) = \Ex_{\mu_{\varphi}}(\sigma_0)
    =
    \frac{1}{e^{-V(\varphi)}}
   \int_{\bbS^{n-1}} \sigma_0 \;
  e^{-\frac{\beta}{2} (\varphi-\sigma_0)\cdot (\varphi - \sigma_0)+h\cdot \sigma_0} \, \mu(d\sigma_0)
  .
\end{equation}
Then \eqref{e:MFnudecomp} and \refeq{EGN}
imply that
\begin{equation}
  \avg{\sigma_0}_{\beta,h,N} 
  = \Ex_{\nu_r}(G(\varphi))
  =
  \frac{\int_{\R^n} G(\varphi) e^{-NV(\varphi)}   d\varphi}{\int_{\R^n} e^{-NV(\varphi) d\varphi}}.
\end{equation}
The right-hand side is a finite-dimensional integral, with dimension $n$ independent of the number of
vertices $N$. Therefore Laplace's Principle can be applied to study the limit as $N \to\infty$.
The following exercise is an instance of Laplace's Principle; for much more on this
kind of result see \cite{Wong01}.

\index{Laplace's Principle}
\begin{theorem}
  \label{thm:laplace-principle}
  Let $V:\R^n \to \R$ be continuous with unique global minimum at $\varphi_0 \in \R^n$.
  Assume that $\int_{\R^n} e^{-V} \, d\varphi$ is finite and that $\{\varphi \in \R^n: V(\varphi) \leq V(\varphi_0)+1\}$
  is compact. Then for any bounded continuous function $g: \R^n\to\R$,
  \begin{equation}
    \lim_{N\to\infty} \frac{\int_{\R^n}g(\varphi)  e^{-NV(\varphi)}  \, d\varphi}
    {\int_{\R^n} e^{-NV(\varphi)} \, d\varphi} = g(\varphi_0).
  \end{equation}
\end{theorem}

\begin{exercise}\label{ex:laplace-principle}
Prove Theorem~\ref{thm:laplace-principle}. \solref{laplace-principle}
\end{exercise}

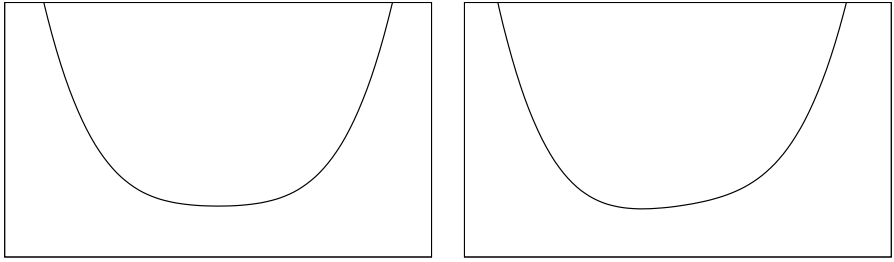
\begin{figure}[t]
  \begin{center}
    \input{mfpot2.pspdftex}
  \end{center}
  \caption{The renormalised potential for $\beta < \beta_c$ with $h=0$ (left) and $h\neq 0$ (right).
  The renormalised potential is convex and the minimum is assumed at a unique point in both cases.}
\label{fig:mean-field-convex}
\end{figure}

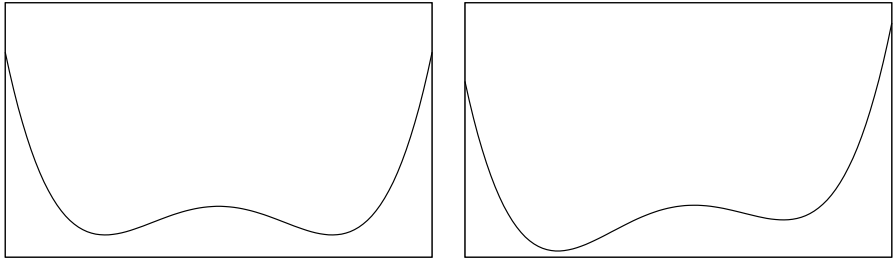
\begin{figure}[t]
  \begin{center}
    \input{mfpot1.pspdftex}
  \end{center}
  \caption{The renormalised potential for $\beta> \beta_c$ with $h=0$ (left) and $h\neq 0$ (right).
    For $h \neq 0$ the minimum is unique, while for $h=0$
    there are two minima for $n=1$ and a set of minima with $O(n)$ symmetry for general $n$.}
\label{fig:mean-field-non-convex}
\end{figure}

Let $G(\varphi) = \Ex_{\mu_\varphi}(\sigma_0)$ be as above.
The critical points $\varphi$ of the renormalised potential $V$ satisfy
\begin{equation} \label{e:self-consistent}
  0 = \nabla V(\varphi)
  = \beta(\varphi-G(\varphi)),
  \quad \text{i.e., $\varphi=G(\varphi)$.}
\end{equation}
The following lemma gives properties of $V$ for the case $n=1$.
See Figure~\ref{fig:mean-field-convex} for part~(ii)
and Figure~\ref{fig:mean-field-non-convex} for part~(iii).

\begin{lemma} \label{lem:V-properties1}
  Let $n=1$ and set $\beta_c=n=1$. Then the renormalised potential $V$ and the function $G$ are
  given by
\begin{equation}
  V(\varphi) = \frac{\beta}{2} \varphi^2 - \log\cosh(\beta\varphi+h) + {\rm const},
  \qquad
  G(\varphi)
  = -\frac{\partial V}{\partial h}
  =\tanh(\beta\varphi+h).
\end{equation}
As a consequence:

\smallskip\noindent
(i)  For $h \neq 0$, $V$ has a unique minimum $\varphi_0(\beta,h)$ with the same sign as $h$.

\smallskip\noindent
(ii)
  For $\beta \leq \beta_c$, $V$ is convex,
  the unique minimum of $V$ tends to $0$ as $h \to 0$,
  and $V''(\varphi) \geq \beta(1-\beta/\beta_c)$ for any $h \in\R$.

\smallskip\noindent
(iii)
  For $\beta > \beta_c$, $V$ is non-convex,
  the minima of $V$ are $\pm r$
  for some $r=r(\beta)>0$ if $h=0$,
  and as $h\downarrow 0$ the unique minimum converges to
  $+r$ or $-r$.

\smallskip\noindent
(iv) The minimum $\varphi_0(\beta,h)$ is differentiable in $h$ whenever $h \neq 0$ or $\beta < \beta_c$.
\end{lemma}

\begin{proof}
  This is a direct computation.
  Note that when $n=1$ the integrals in \refeq{Vr-def} and \refeq{Gmag} are just sums over
  two terms $\sigma=\pm 1$,
  each with measure $\frac 12$.
\end{proof}

\begin{proof}[Proof of Theorem~\ref{thm:MFexp}]
  For $h\neq 0$ or $\beta \leq \beta_c$,
  denote by $\varphi_0(\beta,h)$ the unique minimum of $V$.
  By Theorem~\ref{thm:laplace-principle} and \eqref{e:self-consistent}, the magnetisation is given by
  \begin{equation}
    M(\beta,h)
    = \lim_{N\to\infty} \avg{\sigma_0}_{\beta,h,N}
    =
    \lim_{N\to\infty} \Ex_{\nu_r}(G(\varphi)) = G(\varphi_0(\beta,h))
    = \varphi_0(\beta,h).
  \end{equation}
  The susceptibility is by definition given by
  \begin{equation}
    \chi(\beta,h) = \ddp{M}{h}(\beta,h) = \ddp{\varphi_0}{h}(\beta,h).
  \end{equation}

\noindent
(i)
Lemma~\ref{lem:V-properties1} implies
$\varphi_0(\beta,0_+)=0$ if $\beta \leq \beta_c=1$ and
$\varphi_0(\beta,0_+)>0$ if $\beta >\beta_c$.
Since also $\varphi_0(\beta,0_+) \to 0$ as $\beta\to \beta_c$,
the asymptotics $\tanh(x) = x-\frac13 x^3 + o(x^3)$ imply
  \begin{align}
    \varphi_0(\beta,0_+)
    & = \tanh(\beta \varphi_0(\beta,0_+))
    \nnb  &
    = \beta \varphi_0(\beta,0_+) - \frac13 (\beta \varphi_0(\beta,0_+))^3
    + o(\beta \varphi_0(\beta,0_+))^3,
  \end{align}
  and therefore $\varphi_0 = \varphi_0(\beta,0_+)$ satisfies
  \begin{equation}
    (\beta-1) \varphi_0 = \frac13 (\beta \varphi_0)^3 + o(\beta \varphi_0)^3.
  \end{equation}
  Using $\varphi_0(\beta,0_+) >0$ for $\beta>1$,
  the claim follows by dividing by $\varphi_0/3$ and taking the square root:
  \begin{equation}
    \varphi_0^2 \sim 3 \frac{\beta-1}{\beta^3} \sim 3(\beta-\beta_c) \quad (\beta \downarrow \beta_c).
  \end{equation}

  \smallskip\noindent
  (ii) Similarly, if $\beta=1$ and $h>0$,
  \begin{equation}
    \varphi_0 = \tanh(\varphi_0+h) = \varphi_0+h - \frac13(\varphi_0+h)^{3} + o(\varphi_0+h)^3
  \end{equation}
  implies
  \begin{equation}
    \varphi_0 \sim (3h)^{1/3}  \quad (h\downarrow 0).
  \end{equation}

\smallskip\noindent
(iii) Note that $0 = V'_{\beta,h}(\varphi_0(\beta,h))$ implies
\begin{equation}
  0 = \ddp{^2}{h\partial \varphi} V_{\beta,h}(\varphi_0(\beta,h)) + \ddp{^2}{\varphi^2} V_{\beta,h}(\varphi_0(\beta,h)) \ddp{\varphi_0}{h}(\beta,h).
\end{equation}
Using that
\begin{align}
  \ddp{^2}{h\partial \varphi} V_{\beta,h}(\varphi)
  &=
  - \beta (1-\tanh^2(\beta\varphi+h)),
  \\
  \ddp{^2}{\varphi^2} V_{\beta,h}(\varphi)
  &=
  \beta - \beta^2 (1-\tanh^2(\beta\varphi+h)),
\end{align}
and $\varphi_0 = \tanh(\beta\varphi_0+h)$, therefore
\begin{equation} \label{e:ddp-varphi0}
  \ddp{\varphi_0}{h}(\beta,h)
  = \frac{1}{-\beta + (1-\varphi_0(\beta,h)^2)^{-1}}.
\end{equation}
  This implies
  \begin{equation}
    \chi(\beta,0) = \frac{1}{-\beta+(1-\varphi_0(\beta,0)^2)^{-1}} = \frac{1}{1-\beta} = \frac{1}{\beta_c-\beta} \quad (\beta < \beta_c),
  \end{equation}
  \begin{equation}
    \chi(\beta,0_+) \sim \frac{1}{-\beta+(1-3(\beta-1))^{-1}} \sim \frac{1}{1-\beta+3(\beta-1)} = \frac{1}{2(\beta-\beta_c)} \quad (\beta > \beta_c),
  \end{equation}
  as claimed.
\end{proof}

We conclude this section with two exercises concerning the extension of some
of the above ideas from $n=1$ to $n>1$.

\begin{exercise}\label{ex:V-properties3}
Let $n=3$. Show that
\begin{equation}
  V(\varphi)
  =
  \frac{\beta}{2}|\varphi|^2 - \log\left(\frac{\sinh(|\beta \varphi+h|)}{|\beta\varphi+h|}\right)
  + \frac{\beta}{2},
\end{equation}
where $V (\varphi)$ was defined in \eqref{e:Vr-def}.
\solref{V-properties3}
\end{exercise}

\begin{exercise} \label{ex:V-properties}
  Extend the results of Lemma~\ref{lem:V-properties1} to $n>1$. Let $\beta_c=n$.
  \begin{enumerate}[(i)]
  \item
    For $\beta \leq \beta_c$, the effective potential $V$ is convex and the minimum
    of $V$ tends to $0$ as $h \to 0$.
    Moreover,
    $\He V(\varphi) \geq \beta(1-\beta/\beta_c)$
    for any $h \in \R^n$.
  \item
    For $\beta > \beta_c$, the effective potential $V$ is non-convex.
  \end{enumerate}
  Hint: \cite[Theorem~D.2]{DLS78} is helpful.
  \solref{V-properties}
\end{exercise}

\section{Gaussian free field and simple random walk}

Another fundamental example of a spin system is the Gaussian free field (GFF).
The GFF is a spin system whose distribution is Gaussian.
In this section, we indicate that its critical behaviour can be computed directly,
and establish its
connection to the simple random walk.  We also introduce the bubble diagram, whose
behaviour provides an indication of the special role of dimension $4$.

\subsection{Gaussian free field}
\label{sec:gff}

Let $\Lambda$ be a finite set, and let $\beta = (\beta_{xy})_{x,y\in\Lambda}$ be
non-negative coupling constants with $\beta_{xy} = \beta_{yx}$.  As in \refeq{HLap},
given a spin field $\varphi : \Lambda \to \R^n$, and given $m^2 > 0$,
we define
\begin{align} \label{e:GFF-H-M}
  H(\varphi)
  & = \frac12 (\varphi, (-\Delta_\beta+m^2)\varphi)
  .
\end{align}
We then use $H$ to define a probability measure on field configurations via specification of
the expectation
\begin{equation} \label{e:GFF-measure}
  \avg{F}
  \propto
  \int_{(\R^n)^\Lambda} F(\varphi) \, e^{-H(\varphi)} \, \prod_{x\in\Lambda} d\varphi_x,
\end{equation}
where the integration is with respect to Lebesgue measure on $(\R^n)^\Lambda$.

\index{Gaussian free field}
\index{Free field}
\index{Mass}
\begin{defn}
\label{defn:GFF-1}
  An $n$-component \emph{Gaussian free field} (GFF) with mass $m > 0$ on $\Lambda$ is
  a field distributed according to the above measure.
  An example of particular interest is the case where $\Lambda$ is a finite approximation to $\Z^d$,
and $\beta_{xy} = 1_{x\sim y}$. Then $\Delta_\beta$ is the discrete Laplace operator and we simply
write $\Delta$.
\end{defn}

\begin{exercise} \label{ex:Greenf} \index{Green function}
  Show that
  $(\varphi, -\Delta_{\beta} \varphi) \geq 0$ for all $\varphi \in \R^\Lambda$.
  In particular, $(\varphi, (-\Delta_{\beta} +m^2)\varphi) \geq m^2(\varphi,\varphi) \, > 0$
  for all $\varphi \neq 0$, i.e., $-\Delta_{\beta}+m^2$ is
  strictly positive definite if $m^2>0$ (and thus so is $(-\Delta_{\beta}+m^2)^{-1}$).
  If $\1$ is the constant function on $\Lambda$, defined by $\1_x=1$ for all $x \in \Lambda$,
  then $-\Delta_{\beta} \1 = 0$ and
  \begin{equation}
  \lbeq{freechim2}
    (-\Delta_{\beta}+m^2)^{-1}\1 = m^{-2}\1.
  \end{equation}
  \solref{Greenf}
\end{exercise}

Definition~\ref{defn:GFF-1} can be restated to say that the GFF is defined
as the Gaussian field on $\R^{n\Lambda}$ with mean zero and covariance given by
\begin{equation}
  \avg{\varphi_x^i\varphi_y^j} = \delta_{ij} (-\Delta_\beta+m^2)^{-1}_{xy}.
\end{equation}
For the particular case mentioned in Definition~\ref{defn:GFF-1},
for which the Laplacian is the standard one on a subset $\Lambda \subset \Zd$, we write the covariance as
\begin{equation}
\lbeq{Cxymdef}
    C_{xy;\Lambda}(m^2) = (-\Delta^{(\Lambda)} + m^2)^{-1}_{xy}.
\end{equation}
See Chapter~\ref{ch:gauss} for a detailed introduction to Gaussian fields.
Rather than taking $\Lambda$ as a subset of $\Zd$, we can instead take it to be a
discrete $d$-dimensional torus.  The use of a torus avoids issues concerning boundary
conditions and also preserves translation invariance.
For $m^2>0$ and for all dimensions $d>0$,
it can be proved that in the limit as the period of the torus
goes to infinity, the limit
\begin{equation}
    C_{xy}(m^2) = \lim_{\Lambda \uparrow \Zd} C_{xy;\Lambda}(m^2)
\end{equation}
exists and is given in terms of the Laplacian $\Delta$ on $\ell_2(\Zd)$ by
\begin{equation}
\lbeq{Cxymdef-Zd}
    C_{xy}(m^2) = (-\Delta + m^2)^{-1}_{xy}.
\end{equation}
In addition, for $d>2$ it can be proved that the limit $C_{xy}(0) = \lim_{m^2 \downarrow 0}
C_{xy}(m^2)$ exists.  The restriction to $d>2$ is a reflection of the fact that simple
random walk on $\Zd$ is transient if and only if $d>2$.

\index{Two-point function}
\index{Correlation length}
\index{Susceptibility}
As in the corresponding definitions for the Ising model in
\eqref{e:Ising2ptfcn}--\eqref{e:Isingsuscept}, we define
\begin{align}
\label{e:GFF2ptfcn}
\text{two-point function:}
  &\quad
 \delta_{ij} C_{xy}(m^2),
 \\
\label{e:GFFcorrlength}
\text{correlation length:}
    & \quad
      \xi(m^2)^{-1}
      =
      -\lim_{n \to \infty} n^{-1} \log C_{0,ne_1}(m^2)
 ,
\\
\label{e:GFFsuscept}
\text{susceptibility:}
& \quad
\chi(m^2) = \sum_{x\in\Zd} C_{0x}(m^2).
\end{align}
For the two-point function we allow $m^2 \ge 0$, whereas for the correlation length
and susceptibility we restrict to $m^2>0$.  The susceptibility diverges at the \emph{critical
value} $m^2=0$. The relations
\begin{align}
\lbeq{GFFsusceptm2}
    \chi(m^2) &= m^{-2} & (m^2>0),
    \\
    \xi(m^2)  &\sim m^{-1} &(m^2 \downarrow 0),
    \\
\lbeq{C0xasy}
    C_{0x}(0) & = (-\Delta)^{-1}_{0x} \sim
    c(d) |x|^{-(d-2+\eta)}
    & (|x|\to\infty),
\end{align}
respectively
follow from \refeq{freechim2},
from \cite[Theorem~A.2]{MS93}, and from a standard fact about the lattice
Green function $(-\Delta)^{-1}$ (see, e.g., \cite{Lawl91}).
The above relations show that the critical exponents for the GFF assume the values
\begin{equation}
    \gamma=1,
    \quad
    \nu = \frac 12,
    \quad
    \eta =0.
\end{equation}
\index{Mean-field values}%
\index{Fisher's relation}%
These are conventionally called
mean-field values, although the exponents $\nu$ and $\eta$ involve the
geometry of $\Zd$ and therefore are somewhat unnatural for the mean-field model.
The fact that $\gamma = (2-\eta)\nu$ is an instance of Fisher's relation.

\newcommand{\SRW}{W}

\subsection{Simple random walk}

The GFF is intimately related to the simple random walk.
In this section, we make contact between the two models in the case of $\Z^d$.

Given $d>0$ and $x,y \in \Zd$, an $n$-step walk on $\Zd$ from $x$ to $y$ is a sequence
$\omega = (x=x_0,x_1,\ldots,x_{n-1},x_n=y)$ of neighbouring points ($|x_{i}-x_{i-1}|=1$).
We write $|\omega|=n$ for the length of $\omega$, and write $\Wcal(x,y)$ for the
set of all walks from $x$ to $y$.
Let $V$ be a complex diagonal $\Zd \times \Zd$ matrix
whose elements obey ${\rm Re} v_x \ge c >0$
for some positive $c$.
We define the simple random walk two-point function by
\begin{equation}
    \SRW^{(V)}_{xy} = \sum_{\omega \in \Wcal(x,y)} \prod_{j=0}^{|\omega|} \frac{1}{2d+v_{\omega_j}}.
\end{equation}
The positivity condition on $V$ ensures that the right-hand side converges.
For the special case where $V$ has constant diagonal elements $m^2$, we write
\begin{equation}
\lbeq{Csrwm}
    \SRW^{(m^2)}_{xy}= \sum_{\omega \in \Wcal(x,y)} \prod_{j=0}^{|\omega|} \frac{1}{2d+m^2}.
\end{equation}
The next lemma shows that $W_{xy}$ is related to the covariance of the GFF.

\begin{lemma}
\label{lem:srw}
For $d>0$ and a diagonal matrix $V$ with ${\rm Re} v_x \ge c >0$,
\begin{equation}
\label{e:Csrwrep}
    \SRW^{(V)}_{xy} = (-\Delta + V)^{-1}_{xy}.
\end{equation}
In particular,
\begin{equation}
    \SRW^{(m^2)}_{xy} = C_{xy}(m^2) = (-\Delta + m^2)^{-1}_{xy}.
\end{equation}
\end{lemma}

\begin{proof}
We separate the contribution of the zero-step walk, and for walks taking at least
one step we condition on the first step, to obtain
\begin{equation}
    \SRW^{(V)}_{xy}
    =
    \frac{1}{2d+v_{x}} \delta_{xy} + \frac{1}{2d+v_{x}}\sum_{e:|e|=1} \SRW^V_{x+e,y}.
\end{equation}
We multiply through by $2d+v_{x}$ and rearrange the terms to obtain
\begin{equation}
    (-\Delta \SRW^{(V)})_{xy} + v_x \SRW^{(V)}_{xy} = \delta_{xy},
\end{equation}
which can be restated as $(-\Delta + V) \SRW^{(V)} = I$, and the proof is complete.
\end{proof}

With respect to the uniform measure on $n$-step walks started at $x$, let $p_n(x,y)$
denote the probability that an $n$-step walk started
at $x$ ends at $y$.
Equation~\refeq{Csrwm} can be rewritten as
\begin{equation}
\lbeq{Csrwm2}
    \SRW_{xy}^{(m^2)}= \sum_{n=0}^\infty p_n(x,y)  \frac{(2d)^n}{(2d+m^2)^{n+1}}
    = (-\Delta + m^2)^{-1}_{xy}.
\end{equation}
When $m^2>0$, the sum in \refeq{Csrwm2} is finite in all dimensions.
When $m^2=0$, $\sum_{n=0}^\infty p_n(x,y)$ is the Green function for simple random
walk, which is finite if and only if $d >2$
(see Exercise~\ref{ex:transience}).

The central limit theorem asserts that the distribution of $p_n$ is asymptotically
Gaussian, and the functional central limit theorem
asserts that the scaling limit of simple random walk is Brownian motion.
For random walk,
\emph{universality} is the statement that the critical exponents and limiting distribution
remains
the same, not only for simple random walk, but for any random walk composed
of i.i.d.\ steps $X_i$ having mean zero and finite variance.

\subsection{The bubble diagram}
\label{sec:bubble}

The \emph{bubble diagram} plays a key role in identifying the special role of dimension $4$
in critical phenomena.  It is defined by
\begin{equation}
\lbeq{bubdef}
    B_{m^2}=\sum_{x \in \Zd} \big( C_{0x}(m^2) \big)^2,
\end{equation}
with $C_{0x}(m^2)=(-\Delta+m^2)^{-1}$ as in \refeq{Cxymdef-Zd}.
The Fourier transform is useful for the analysis of the bubble diagram.

The Fourier transform of an absolutely summable function $f:\Zd \to
\C$ is defined by
\begin{align}
    \hat f(k)
    &=
    \sum_{x\in\Zd} f_x \; e^{ ik\cdot x}
    \quad (k \in [-\pi,\pi]^d).
\label{e:FT-def}
\end{align}
The inverse transform is given by
\begin{align}
    f_{x}
    &=
    (2\pi)^{-d} \int_{[-\pi ,\pi]^{d}} \hat f(k) \; e^{ - ik\cdot x}
    \quad (x \in \Z^d) .
\end{align}
With respect to the Fourier transform,
$-\Delta$ acts as a multiplication operator with multiplication by
\begin{equation}
    \label{e:FDel}
    \FDel(k)
    =
    4 \sum_{j=1}^{d} \sin^2 ( k_j/2)
    \qquad
    (k \in [-\pi,\pi]^d)
    .
\end{equation}
This means that
\begin{equation}
    (-\widehat{\Delta f)}(k) = \lambda(k)\hat f(k),
\end{equation}
and hence the Fourier transform of $C_{0x}(m^2)$ is given by
\begin{equation}
\lbeq{Chat}
    \hat C_{m^2}(k) = \frac{1}{\lambda(k)+m^2}.
\end{equation}
Therefore, by Parseval's formula and \refeq{Chat},
\begin{equation}
  \label{e:freebubble}
  B_{m^2}
  =
  \int_{[-\pi,\pi]^d}
  \frac{1}{(\FDel(k) +m^2)^2}
  \frac{dk}{(2\pi)^{d}}
   .
\end{equation}
The logarithmic corrections to scaling for $d=4$ in Theorem~\ref{thm:phi4}
arise via the logarithmic divergence of the $4$-dimensional bubble
diagram.

\begin{exercise} \label{ex:bubble1z} \index{Bubble diagram}
  Show that $B_0<\infty$ if and only if $d>4$, and
  that, as $m^2 \downarrow 0$,
  \begin{equation}
    B_{m^2} \sim
    b_d \times
    \begin{cases}
     m^{-(4-d)} & (d < 4)
    \\
    \log m^{-2} & (d=4),
    \end{cases}
  \end{equation}
with $b_1= \frac 18$, $b_2 = \frac{1}{4\pi}$, $b_3 = \frac{1}{8\pi}$,
$b_4 = \frac{1}{16\pi^2}$.
\solref{bubble1z}
\end{exercise}

The following exercises review the fact that simple random walk
is recurrent in dimensions $d \le 2$ and transient for $d>2$,
and relate the bubble diagram to intersections of random walks.

\begin{exercise} \label{ex:transience}
    (i)
    Let $u$ denote the probability that simple random walk ever returns to the origin. The
    walk is recurrent if $u = 1$ and transient if $u < 1$. Let $N$ denote the random
    number of visits to the origin, including the initial visit at time $0$.
    Show that $EN = (1-u)^{-1}$, so the walk is recurrent if and only if $EN = \infty$.
    \\
    (ii)
    Show that
    \begin{equation}
      EN= \sum_{n= 0}^\infty p_n(0)
      = 2d \int_{[-\pi,\pi]^d}
      \frac{1}{\lambda(k)} \frac{dk}{(2\pi)^d} .
    \end{equation}
    Thus transience is characterised by the integrability of
    $\hat C_{0}(k) =1/\lambda(k)$.
    \\
    (iii)
    Show that simple random walk is recurrent in dimensions $d \leq 2$
    and transient for $d > 2$.
\solref{transience}
\end{exercise}

\begin{exercise} \label{ex:bubble1} \index{Bubble diagram}
  Let $S^1= (S^1_n)_{n\geq 0}$ and $S^2=(S^2_n)_{n\geq 0}$ be two independent
  simple random walks on $\Z^d$
  started at the origin, and let
  \begin{equation}
    I = \sum_{m= 0}^\infty\sum_{n= 0}^\infty \1_{S^1_m = S^2_n}
  \end{equation}
  be the random number of intersections of the two walks. Show that
  \begin{equation}
    E I =  (2d)^2 B_0.
  \end{equation}
  Thus $EI$ is finite if and only if $d > 4$.
  \solref{bubble1}
\end{exercise}

\section{\texorpdfstring{$|\varphi|^4$}{phi4} model}

\subsection{Definition of the \texorpdfstring{$|\varphi|^4$}{phi4} model}
\label{sec:phi4-defn}
\index{Spin model}

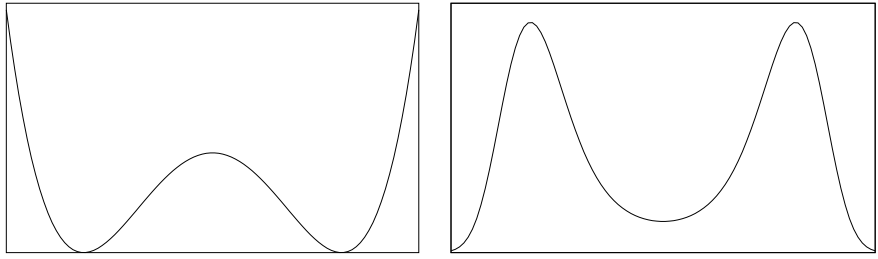
\begin{figure}[h]
\begin{center}
  \input{phi4-doublewell.pspdftex}
\end{center}
\caption{\label{fig:V}
For $n=1$, the density of the single-spin distribution is shown at right,
with its double-well potential at left.}
\end{figure}

As in Section~\ref{sec:universality}, the $n$-component $|\varphi|^4$ model
on a set $\Lambda$ is defined by the expectation
\begin{equation}
\label{e:phi4-exp-new}
  \la F  \ra_{g,\nu,\Lambda}
  =
  \frac{1}{Z_{g,\nu,\Lambda}} \int_{\R^{n\Lambda}} F(\varphi) e^{-H(\varphi)}
   d\varphi
\end{equation}
with
\begin{equation}
\lbeq{H4def}
  H(\varphi) = \frac12 \sum_{x\in \Lambda} \varphi_x \cdot (-\Delta_\beta \varphi)_x
  + \sum_{x\in \Lambda} \left( \frac14 g|\varphi_x|^4+ \frac12 \nu|\varphi_x|^2\right).
\end{equation}
\index{Partition function}%
Here $g>0$, $\nu \in \R$, and $d\varphi = \prod_{x\in \Lambda} d\varphi_x$ is the
Lebesgue measure on $(\R^n)^\Lambda$.
The \emph{partition function}
$Z_{g,\nu,\Lambda}$ is defined by the condition $\langle 1\rangle_{g,\nu,\Lambda} =1$.
An external field $h$ can also be included, but we have omitted it here.
We are primarily concerned here with the nearest-neighbour interaction on a $d$-dimensional
discrete torus, for which
$\Delta_\beta=\Delta$ is the standard Laplacian.
The single-spin distribution is
$e^{-(\frac 14 g |\varphi_x|^4 + \frac 12 \nu |\varphi_x|^2)}d\varphi_x$.
For the case $\nu<0$, which is our principal interest, we have a double-well
potential as depicted for $n=1$ in Figure~\ref{fig:V}.
For $n\geq 2$, it is sometimes called a Mexican hat potential.
\index{Single-spin distribution}

With $\nu=-g \beta$, the single-spin density becomes
proportional to $e^{-\frac{1}{4}g(|\varphi_x|^2- \beta )^2}$.  In the limit
$g\to\infty$, this converges to the $O(n)$ model, whose single-spin distribution is the uniform measure on the surface of
the sphere of radius $\sqrt{\beta}$ in $n$ dimensions.
By rescaling the field by $1/\sqrt{\beta}$, this definition is
equivalent to the more usual one, where spins are on the unit sphere and
an inverse temperature parameter $\beta$ multiplies the spin
coupling term $\varphi\cdot (-\Delta\varphi)$.
Conversely, the
$|\varphi|^4$ model can be realised as a limit of $O(n)$ models
\cite{SG73,DN75}.

\begin{figure}[h]
\begin{center}
\includegraphics[scale = 0.65]{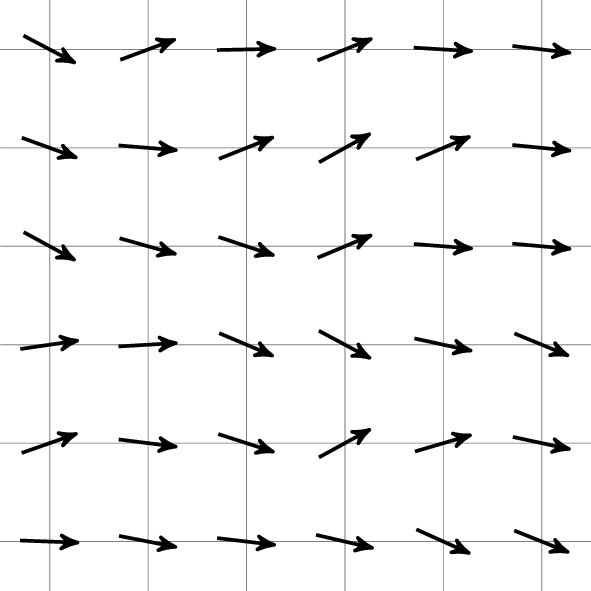} \hspace{20mm}
\includegraphics[scale = 0.65]{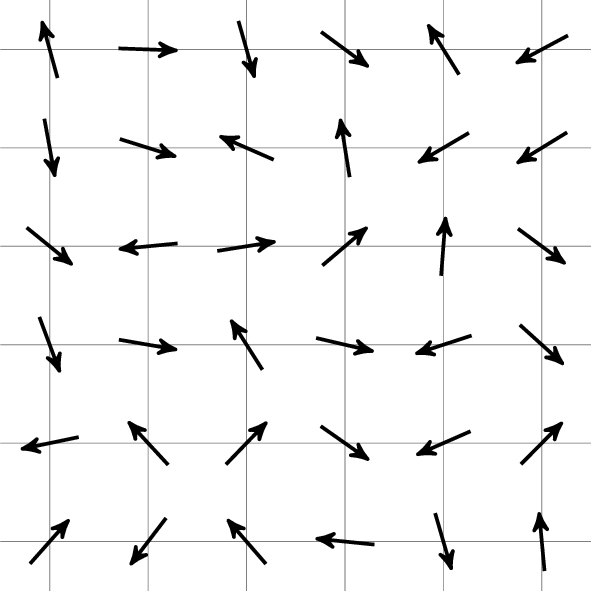}
\end{center}
\caption{
\label{fig:spins}
Typical spin configurations for $\nu<\nu_c$ (spins aligned) and for $\nu>\nu_c$ (spins not aligned).}
\end{figure}

The Ising model Gibbs measure of \refeq{IsingGibbs} is equal to
\begin{equation}
  P_{T,\Lambda}(\sigma)
  \propto
  e^{-\frac{1}{2} \frac{1}{T} \sum_{x\in \Lambda}\sigma_x(-\Delta \sigma)_x}
  \prod_{x\in\Lambda}\frac 12 (\delta_{\sigma_x,1}+\delta_{\sigma_x,-1})
\end{equation}
Let $\varphi_x = T^{-1/2}\sigma_x$. Then
\begin{equation}
  P_{T,\Lambda}(\varphi)
  \propto
  e^{-\frac{1}{2} \sum_{x\in \Lambda}\varphi_x(-\Delta \varphi)_x}
  \prod_{x\in\Lambda}\frac 12 (\delta_{\varphi_x,T^{-1/2}}+\delta_{\varphi_x,-T^{-1/2}}).
\end{equation}
Suppose that we replace the single-spin
distribution $\frac 12 (\delta_{\sigma_x,T^{-1/2}}+\delta_{\sigma_x,-T^{-1/2}})$ by a
smoothed out distribution with two peaks located at $\pm T^{-1/2}$.
It may be expected that, as $T$ is
decreased, such a model will have a phase transition with the same critical exponents
as the Ising model.
This is qualitatively similar to the $|\varphi|^4$ model
with $\nu<0$.
Now $\nu$ plays the role of $T$, and there is again a phase transition
and corresponding critical exponents associated with a (negative)
critical value $\nu_c$ of $\nu$.  Alignment of spins is observed for
$\nu<\nu_c$ but not for $\nu>\nu_c$, as illustrated schematically in Figure~\ref{fig:spins}.

General results on the existence of phase transitions for
multi-component spin systems in dimensions $d \ge 3$ are proved in \cite{FSS76}.
For $d = 2$, the Mermin--Wagner theorem rules out phase transitions for $n \ge 2$.
It is predicted that the $|\varphi|^4$ model is in the same universality
class as the $O(n)$ model, for all $n \ge 1$.  In particular, the critical exponents of
the $n$-component $|\varphi|^4$ are predicted to be the same as those
of the $O(n)$ model.

\index{Pressure}
\index{Susceptibility}
\index{Specific heat}
\index{Correlation length}
We write $\avg{F;G} = \avg{FG} - \avg{F}\avg{G}$ for the covariance of
random variables $F,G$.
Five
quantities of interest are the \emph{pressure}, the \emph{two-point function},
the \emph{susceptibility},
the \emph{correlation length of order} $p>0$, and the \emph{specific heat}.  These are
defined, respectively, as the limits (assuming they exist)
\index{Pressure}%
\index{Two-point function}%
\index{Susceptibility}%
\index{Correlation length of order $p$}%
\index{Specific heat}%
\begin{align}
  \label{e:pressuredef}
    p(g,\nu) &= \lim_{N \to \infty} \frac{1}{|\Lambda_N|} \log Z_{g,\nu,\Lambda_N},
    \\
  \label{e:phi4-2ptfcn}
    \langle \varphi^1_0 \varphi^1_x \rangle_{g,\nu}
    & =
    \lim_{N \to \infty} \avg{\varphi_0^1\varphi_x^1}_{g,\nu,\Lambda_N},
    \\
  \label{e:susceptdef}
    \chi(g,\nu) &= \lim_{N \to \infty} \sum_{x\in \Lambda_N}
    \avg{\varphi_0^1 \varphi_x^1}_{g,\nu,\Lambda_N},
    \\
    \xi_p(g,\nu) &=
    \left( \frac{1}{\chi(g,\nu)}
    \lim_{N \to \infty} \sum_{x\in \Lambda_N}
    |x|^p \avg{\varphi_0^1\varphi_x^1}_{g,\nu,\Lambda_N}\right)^{1/p},
    \\
    c_H(g,\nu) &
    = \frac 14 \lim_{N \to \infty} \sum_{x\in \Lambda_N}
    \avg{|\varphi_0|^2; |\varphi_x|^2}_{g,\nu,\Lambda_N},
\end{align}
for a sequence of boxes $\Lambda_N$ approximating $\Z^d$ as $N\to\infty$.
In making the above definitions,
we used the fact that $\avg{\varphi_x}=0$ for all $x$ due to the $O(n)$ invariance.

In general, the limit defining the pressure has been
proved to exist and to be independent of the boundary conditions for the
$n$-component $|\varphi|^4$ model for any $d>0$, $n \ge 1$, $g>0$ and
$\nu\in \R$ \cite{LP76}.  For $n=1,2$, correlation inequalities
\cite{FFS92} imply that the pressure is convex, and hence also
continuous, in $\nu$, and that for the case of free boundary
conditions the limit defining the susceptibility exists (possibly
infinite) and is monotone non-increasing in $\nu$.  Proofs are lacking
for $n>2$ due to a lack of correlation inequalities in this case (as
discussed, e.g., in \cite{FFS92}), but it is to be expected that these
facts known for $n=1,2$ are true also for $n>2$.

\subsection{Critical exponents of the \texorpdfstring{$|\varphi|^4$}{phi4} model}

\subsubsection*{Dimensions above four}

For $d>4$, the $|\varphi|^4$ model has been proven to exhibit
 mean-field behaviour.  In particular, it is
known \cite{Aize82,Froh82} that for $n=1,2$,
with $\nu = \nu_c + \varepsilon$ and as $\varepsilon \downarrow 0$,
\begin{align}
    \chi(g,\nu)
    \asymp \frac{1}{\varepsilon}
    \quad
    \text{when $d>4$,\; $n=1,2$.}
\end{align}
The proof is based on
correlation inequalities, differential inequalities, and reflection positivity.
\index{Correlation inequalities}
\index{Differential inequalities}
\index{Reflection positivity}
Also, for $n=1,2$, the specific heat does not diverge as $\nu \downarrow \nu_c$
\cite{FFS92,Soka79}.
More recently, the lace expansion has been used to prove
that for $d>4$
and small $g>0$, the critical two-point function has
the Gaussian decay
\begin{equation}
    \langle \varphi^1_0 \varphi^1_x \rangle_{g,\nu_c}
    \sim c \frac{1}{|x|^{d-2}}
    \quad
    \text{as $|x|\to\infty$} ,
\end{equation}
for $n =1$ \cite{Saka15} and for $n=1,2$ \cite{BHH18}.
The above equations are statements that the critical exponents $\gamma,\eta$ take
their mean-field values $\gamma=1$ and $\eta=0$ for $d>4$.

\subsubsection*{Dimension four}

\index{Critical dimension}
For dimension $d=4$, logarithmic corrections to mean-field critical scaling
were predicted in \cite{LK69,BGZ73,WR73}.
In the early 1980s it was established that the deviation from mean-field scaling
is at most logarithmic for $d=4$, for some quantities including the susceptibility
\cite{Aize82,Froh82,AG83}.
A number of rigorous results concerning precise critical behaviour of the 4-dimensional case
were proved during the 1980s using rigorous renormalisation
group methods based on block spins \cite{GK85,GK86,HT87} or phase space expansion \cite{FMRS87}.
The following theorems were proved recently via an approach based on the methods in this book.

\index{Logarithmic corrections}%
\begin{theorem} \cite{BBS-phi4-log}.
\label{thm:phi4}
For $d = 4$, $n\ge 1$, $L$ large, and $g>0$ small, there exists $\nu_c=\nu_c(g,n)<0$ such that,
with $\nu = \nu_c + \varepsilon$ and
as $\varepsilon \downarrow 0$,
\begin{align}
  \label{e:thmphi4-chi}
    \chi(g,\nu) & \sim A_{g,n} \frac{1}{\varepsilon}(\log \varepsilon^{-1})^{(n+2)/(n+8)},
\\
    c_H(g,\nu) & \sim D_{g,n} \times
        \begin{cases}
      (\log \varepsilon^{-1})^{(4-n)/(n+8)} & (n<4)\\
      \log \log \varepsilon^{-1} & (n=4)\\
      1 & (n>4).
    \end{cases}
\end{align}
As $g \downarrow 0$,
$A_{g,n} \sim ((n+8)g/(16\pi^2))^{(n+2)/(n+8)}$, and
 $\nu_c(g,n)\sim -(n+2)gN_4 $ (with $N_4 = (-\Delta)^{-1}_{00}$).
\end{theorem}

\begin{theorem}
\label{thm:BSTW-clp}
\cite{BSTW-clp}.
For $d = 4$, $n\ge 1$, $p>0$, $L$ large, and $g>0$ small (depending on $p,n$),
with $\nu = \nu_c + \varepsilon$ and
as $\varepsilon \downarrow 0$,
\begin{align}
  \label{e:thmphi4-xi}
    \xi_p(g,\nu) & \sim C_{g,n,p} \frac{1}{\varepsilon^{1/2}}(\log \varepsilon^{-1})^{\frac 12 (n+2)/(n+8)}
    .
\end{align}
\end{theorem}

\begin{theorem}
\label{thm:ST-phi4}
\cite{ST-phi4}.
For $d = 4$, $n\ge 1$, $L$ large, and $g>0$ small, as $|x|\to\infty$,
\begin{align}
\label{e:phi42pt}
    \langle \varphi_0^1 \varphi_x^1 \rangle_{g,\nu_c}
    & \sim \frac{A_{g,n}'}{|x|^{2}}
    ,
\\
\label{e:Wasy-1b}
    \langle |\varphi_0|^2 ;  |\varphi_x|^2 \rangle_{g,\nu_c}
    & \sim \frac{nA_{g,n}''}{(\log |x|)^{2(n+2)/(n+8)}}
    \frac{1}{|x|^{4}}
    .
\end{align}
\end{theorem}

Related further results can be found in \cite{BBS-phi4-log,ST-phi4,BSTW-clp}.
In the above theorems, the infinite-volume limits are taken through a sequence
of tori $\Lambda=\Lambda_N= \Z^d/L^N \Z^d$ for sufficiently large $L$,
and it is part of the statements that these limits exist.
In Theorem~\ref{thm:ST-phi4}, the left-hand sides refer to
the limits taken in the order $\lim_{\nu \downarrow
\nu_c}\lim_{N\to\infty}$.

For $n=1$, Theorem~\ref{thm:ST-phi4} was proved thirty years earlier,
in \cite{GK85,GK86}, and the analogue of \refeq{phi42pt} was proved
for a closely related 1-component model in \cite{FMRS87}.  
The logarithmic correction $(\log \varepsilon^{-1})^{1/3}$ in \refeq{thmphi4-chi} was proved in
\cite{HT87}, along with other results including for the correlation
length.

This book describes techniques developed to
prove the above theorems, with focus on the susceptibility.
To keep the focus on the main ideas and avoid further technicalities,
we will prove a statement like \eqref{e:thmphi4-chi} for a
\emph{hierarchical} version of the $|\varphi|^4$ model;  the precise
statement is given in Theorem~\ref{thm:phi4-hier-chi}.

\subsubsection*{Dimensions below four}

\index{$\varepsilon$-expansion}
Dimensions $2<d<4$ are studied in the physics literature using expansions
  in dimension and number of components.
In a seminal paper, Wilson and Fisher initiated the study of dimensions
below 4 by expanding in small positive $\epsilon = 4-d$ \cite{WF72}.
Dimensions above 2 have been studied via expansion in $\epsilon = d-2$,
and it is also common in the literature to expand in $1/n$ for a large number $n$
of field components.

An alternative to expansion in $\epsilon =4-d$ is to consider long-range interactions decaying
with distance $r$ as $r^{-(d+\alpha)}$ with $\alpha \in (0,2)$ \cite{FMN72,SYI72}.  These
models have upper critical dimension $2\alpha$, and the $\epsilon$ expansion can
be carried out in integer dimensions $d=1,2,3$ by choosing $\alpha= \frac 12 (d+\epsilon)$.
Then $2\alpha =d+\epsilon$, so $d$ is slightly below the critical dimension when
$\epsilon$ is small and positive.

Extensions of Theorems~\ref{thm:phi4} and \ref{thm:ST-phi4}
to the long-range setting have been obtained in
\cite{Slad17,LSW17}; see also \cite{BDH98,BMS03,Abde07,ACG13}.  In contrast to
the above theorems, the long-range results involve a \emph{non-Gaussian} renormalisation
group fixed point, with corrections to mean-field scaling that are power law rather than
logarithmic.
An example of a result of this type is the following theorem.
The theorem pertains to the $|\varphi|^4$
model defined with the operator $-\Delta$ in \refeq{H4def} replaced by the fractional power
$(-\Delta)^{\alpha/2}$, with $\alpha =\frac 12 (d+\epsilon)$ for small $\epsilon >0$.
The kernel of this operator decays at large distance as
$-(-\Delta)^{\alpha/2}_{xy} \asymp |x-y|^{-(d+\alpha)}$.

\begin{theorem}
\label{thm:long-range}
\cite{Slad17}.
For $d=1,2,3$, $n \ge 1$, $L$ sufficiently large, and $\epsilon=2\alpha-d>0$ sufficiently small,
there exists $\bar s \asymp \epsilon$ such that, for
$g \in [\frac{63}{64}\bar s, \frac{65}{64}\bar s]$,
there exists $\nu_c=\nu_c(g,n)$ and $C>0$ such that for $\nu=\nu_c+t$ with $t \downarrow 0$,
the susceptibility of the long-range model obeys
\begin{equation}
\lbeq{chigam}
    C^{-1} t^{-(1 +  \frac{n+2}{n+8} \frac{\epsilon}{\alpha} -C\epsilon^2)}
    \le
    \chi (g,\nu ;n)
    \le
    C t^{-(1 +  \frac{n+2}{n+8} \frac{\epsilon}{\alpha} +C\epsilon^2)}.
\end{equation}
This is a statement that the critical exponent $\gamma$ exists to order $\epsilon$, with
\begin{equation}
    \gamma = 1 +  \frac{n+2}{n+8} \frac{\epsilon}{\alpha} +O(\epsilon^2).
\end{equation}
\end{theorem}

\section{Self-avoiding walk}
\label{sec:intro-saw}

\index{Self-avoiding walk}
The self-avoiding walk on $\Zd$ is the uniform probability measure on
the set of $n$-step simple random walk paths on $\Zd$
with no self-intersections.  It is a much studied model of linear polymers \cite{Holl09,Vand98,Genn79}
and is
of independent mathematical interest (see, e.g., \cite{MS93,BDGS12,Hugh95}).
It has long been understood that at a formal (nonrigorous) level, the
critical behaviour of the self-avoiding
walk is predicted from that of the $n$-component $|\varphi|^4$ model by
setting $n=0$.  For example, the asymptotic formula for the susceptibility
of the $4$-dimensional $|\varphi|^4$ model given
by \refeq{thmphi4-chi}, namely
\begin{align}
  \label{e:thmphi4-chi-n}
    \chi(g,\nu) & \sim A_{g,n} \frac{1}{\varepsilon}(\log \varepsilon^{-1})^{(n+2)/(n+8)},
\end{align}
predicts that the susceptibility of the $4$-dimensional self-avoiding walk should obey
\begin{align}
  \label{e:thmphi4-chi-0}
    \chi(g,\nu) & \sim A_{g,0} \frac{1}{\varepsilon}(\log \varepsilon^{-1})^{1/4}.
\end{align}

An advantage of the renormalisation group method presented in this book is that
it applies equally well to a \emph{supersymmetric} version of the $|\varphi|^4$ model
which corresponds exactly (and rigorously) to a model of weakly self-avoiding walk.
In particular,  \refeq{thmphi4-chi-0} can be proved in this setting
\cite{BBS-saw4-log}.
In Chapter~\ref{ch:saw-int-rep}, we define the supersymmetric version of the
$|\varphi|^4$ model and prove
its equivalence to the continuous-time weakly self-avoiding walk.
This provides
a basis for the application of the renormalisation group method.
We also comment in Chapter~\ref{ch:saw-int-rep} on the sense in which the
supersymmetric model corresponds to $n=0$ components.



%% file: magnetisation.pspdftex
\begin{picture}(0,0)%
\includegraphics{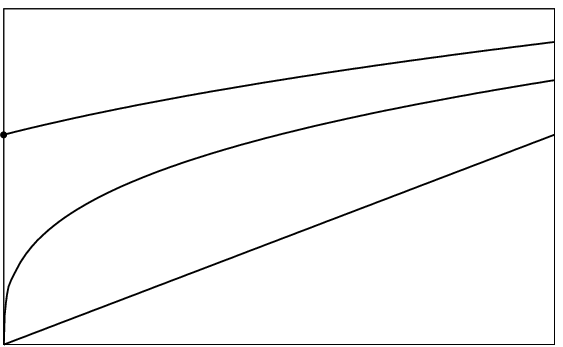}%
\end{picture}%
\setlength{\unitlength}{1973sp}%
\begingroup\makeatletter\ifx\SetFigFont\undefined%
\gdef\SetFigFont#1#2#3#4#5{%
  \reset@font\fontsize{#1}{#2pt}%
  \fontfamily{#3}\fontseries{#4}\fontshape{#5}%
  \selectfont}%
\fi\endgroup%
\begin{picture}(5347,3258)(1650,-3733)
\put(1726,-661){\makebox(0,0)[lb]{\smash{{\SetFigFont{6}{7.2}{\familydefault}{\mddefault}{\updefault}{\color[rgb]{0,0,0}$M(h,T)$}%
}}}}
\put(4051,-2986){\makebox(0,0)[lb]{\smash{{\SetFigFont{6}{7.2}{\familydefault}{\mddefault}{\updefault}{\color[rgb]{0,0,0}$T>T_c$}%
}}}}
\put(4051,-1936){\makebox(0,0)[lb]{\smash{{\SetFigFont{6}{7.2}{\familydefault}{\mddefault}{\updefault}{\color[rgb]{0,0,0}$T=T_c$}%
}}}}
\put(4051,-1036){\makebox(0,0)[lb]{\smash{{\SetFigFont{6}{7.2}{\familydefault}{\mddefault}{\updefault}{\color[rgb]{0,0,0}$T<T_c$}%
}}}}
\put(6751,-3661){\makebox(0,0)[lb]{\smash{{\SetFigFont{6}{7.2}{\familydefault}{\mddefault}{\updefault}{\color[rgb]{0,0,0}$h$}%
}}}}
\put(1726,-1861){\makebox(0,0)[lb]{\smash{{\SetFigFont{6}{7.2}{\familydefault}{\mddefault}{\updefault}{\color[rgb]{0,0,0}$M_+(T)$}%
}}}}
\end{picture}%

%% file: spontmagnetisation.pspdftex
\begin{picture}(0,0)%
\includegraphics{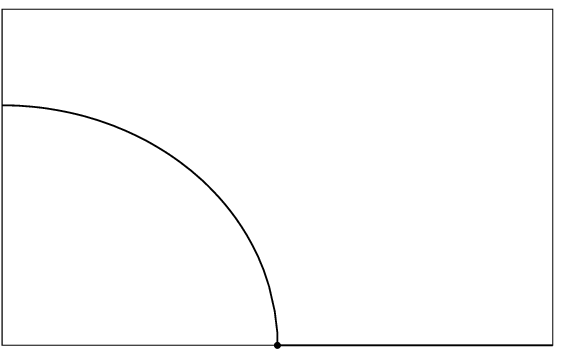}%
\end{picture}%
\setlength{\unitlength}{1973sp}%
\begingroup\makeatletter\ifx\SetFigFont\undefined%
\gdef\SetFigFont#1#2#3#4#5{%
  \reset@font\fontsize{#1}{#2pt}%
  \fontfamily{#3}\fontseries{#4}\fontshape{#5}%
  \selectfont}%
\fi\endgroup%
\begin{picture}(5330,3273)(1666,-3748)
\put(4276,-3661){\makebox(0,0)[rb]{\smash{{\SetFigFont{6}{7.2}{\familydefault}{\mddefault}{\updefault}{\color[rgb]{0,0,0}$T_c$}%
}}}}
\put(6901,-3661){\makebox(0,0)[rb]{\smash{{\SetFigFont{6}{7.2}{\familydefault}{\mddefault}{\updefault}{\color[rgb]{0,0,0}$T$}%
}}}}
\put(1726,-661){\makebox(0,0)[lb]{\smash{{\SetFigFont{6}{7.2}{\rmdefault}{\mddefault}{\updefault}{\color[rgb]{0,0,0}$M_+(T)$}%
}}}}
\end{picture}%

%% file: mfpot2.pspdftex
\begin{picture}(0,0)%
\includegraphics{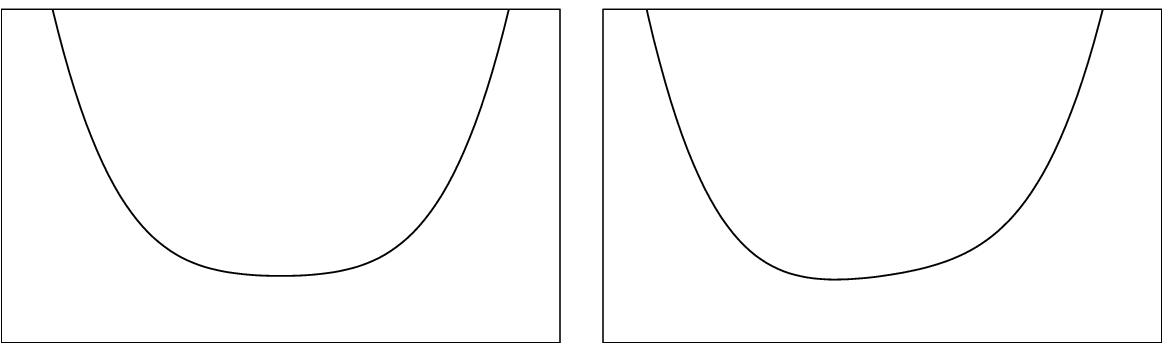}%
\end{picture}%
\setlength{\unitlength}{1973sp}%
\begingroup\makeatletter\ifx\SetFigFont\undefined%
\gdef\SetFigFont#1#2#3#4#5{%
  \reset@font\fontsize{#1}{#2pt}%
  \fontfamily{#3}\fontseries{#4}\fontshape{#5}%
  \selectfont}%
\fi\endgroup%
\begin{picture}(11161,3233)(1076,-3573)
\end{picture}%

%% file: mfpot1.pspdftex
\begin{picture}(0,0)%
\includegraphics{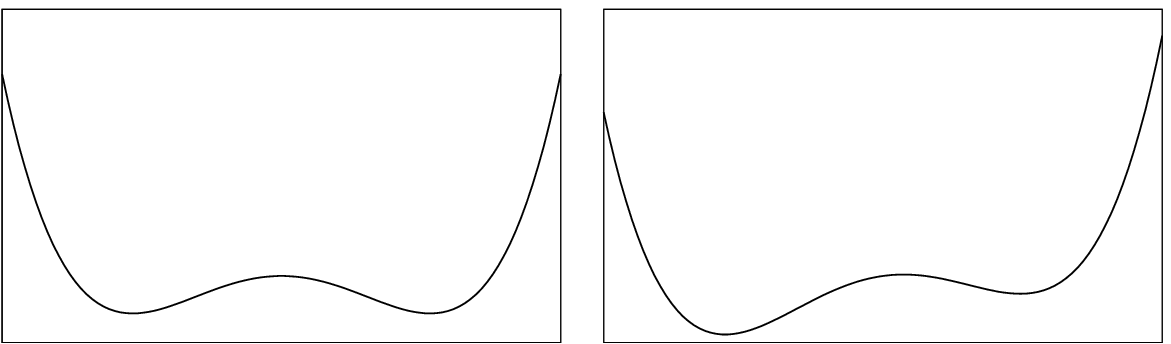}%
\end{picture}%
\setlength{\unitlength}{1973sp}%
\begingroup\makeatletter\ifx\SetFigFont\undefined%
\gdef\SetFigFont#1#2#3#4#5{%
  \reset@font\fontsize{#1}{#2pt}%
  \fontfamily{#3}\fontseries{#4}\fontshape{#5}%
  \selectfont}%
\fi\endgroup%
\begin{picture}(11181,3223)(616,-3423)
\end{picture}%

%% file: phi4-doublewell.pspdftex
\begin{picture}(0,0)%
\includegraphics{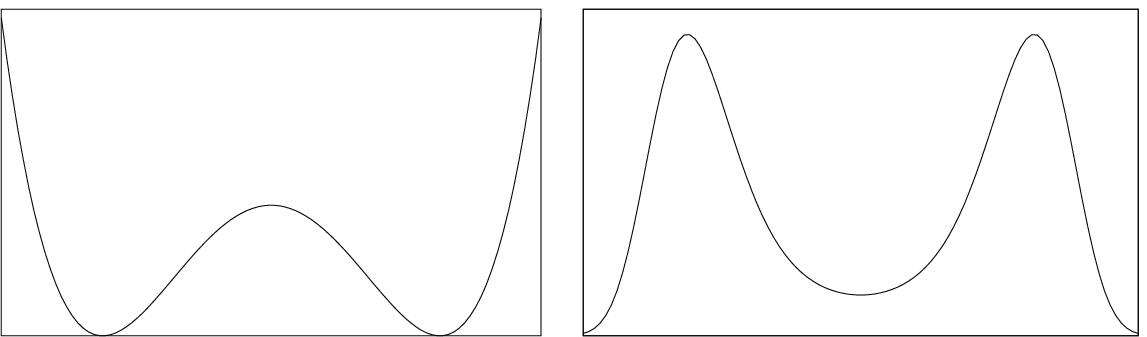}%
\end{picture}%
\setlength{\unitlength}{1934sp}%
\begingroup\makeatletter\ifx\SetFigFont\undefined%
\gdef\SetFigFont#1#2#3#4#5{%
  \reset@font\fontsize{#1}{#2pt}%
  \fontfamily{#3}\fontseries{#4}\fontshape{#5}%
  \selectfont}%
\fi\endgroup%
\begin{picture}(11161,3223)(1601,-3723)
\end{picture}%

%% file: gauss.tex
\chapter{Gaussian fields}
\label{ch:gauss}

In this chapter, we present basic facts about Gaussian integration.
Further material
can be found in many references, e.g., in \cite{Bryd09,Salm99}.

\section{Gaussian integration}

Throughout this chapter,
$X$ is a finite set, we write $\R^X = \{
\varphi : X \to \R \}$, and $(\varphi,\psi) = \sum_{x \in X} \varphi_x
\psi_x$ for $\varphi,\psi \in \R^X$.  We call $\varphi \in \R^X$ a
\emph{field}, and a randomly distributed $\varphi$ is thus a random
field.
We do not make use of any geometric structure of
$X$ here, and only use the fact that $\R^X$ is a finite-dimensional vector
space.

\index{Covariance}
Let $C = (C_{xy})_{x,y \in X}$
denote a symmetric positive semi-definite matrix,
where \emph{positive semi-definite} means that
$(\varphi, C\varphi) \ge 0$
for every  $\varphi \in \R^X$.  If the inequality is
strict for every nonzero $\varphi$, we say that $C$ is positive
definite. This stronger condition implies that the inverse $C^{-1}$
exists.
The following is the higher-dimensional generalisation of the
probability measure $\frac{1}{\sqrt{2\pi}\sigma} e^{-x^2/2\sigma} dx$
of a Gaussian random variable with mean $0$ and variance $\sigma^2$.

\index{Gaussian measure}
\begin{defn}
\label{def:Gm2}
  Let $C$ be positive definite.
  The centred \emph{Gaussian probability measure} $P_C$ on $\R^X$,
  with covariance $C$, is defined by
  \begin{equation}
    P_C(d\varphi) = \det(2\pi C)^{-\frac12} e^{-\frac12
    (\varphi,C^{-1}\varphi)}
    \, d\varphi
    ,
  \end{equation}
  where $d\varphi$ is the Lebesgue measure on $\R^X$.
\end{defn}

To see that $P_C$ really is a probability measure,
it suffices by the spectral theorem to assume that $X=\{1, \dots, n\}$
and that $C$ is diagonal with $C_{ii} = \lambda_i^{-1}$.  In this
case, as required,
\begin{align}
    \int e^{-\frac12 (\varphi,C^{-1}\varphi)} \, d\varphi
    &= \int e^{-\sum_{i=1}^n  \frac{1}{2\lambda_i} \varphi_i^2} \prod_{i=1}^n d\varphi_i
    \nnb
    &= \prod_{i=1}^n \int e^{-\frac1{2 \lambda_i} \varphi_i^2} d\varphi_i
    = \prod_{i=1}^n (2\pi\lambda_i)^{\frac12} = \det(2\pi C)^{\frac12}.
  \lbeq{Gdet}
  \end{align}

In the case that $C$ is positive semi-definite, but not positive
definite, $C$ has a kernel $K$ which is a subspace of $\R^{X}$. We
construct a \emph{degenerate} Gaussian probability measure on $\R^{X}$
as follows.  We set $C'$ equal to the restriction of $C$ to the
orthogonal complement $K^{\perp}$ of $K$ in $\R^{X}$. By the spectral
theorem $K^{\perp}$ is spanned by eigenvectors of $C$ with positive
eigenvalues and therefore is represented by a positive definite matrix
in any orthogonal basis for $K^{\perp}$. We define $P_{C}$ to be the
probability measure on $\R^{X}$ that is supported on $K^{\perp}$ and
which equals the Gaussian measure $P_{C'}$ when restricted to
$K^{\perp}$. To define this construction concretely, we choose an
orthonormal basis of eigenvectors $v_{1},\dots ,v_{n}$ in $\R^{X}$
labelled so that $K$ is spanned by $v_{1},\dots ,v_{k}$ for some $k
\le n$ and define
\begin{equation}
    \label{e:Gm2}
    P_C(d\varphi)
    =
    \det(2\pi C')^{-\frac12}
    e^{-\frac12 (\varphi (t),C'^{-1}\varphi (t))} \,
    \, \prod_{i\le k} \delta (dt_{i})\,
    \prod_{i'=k+1}^{n} dt_{i'} ,
\end{equation}
where $\varphi (t) = \sum_{i=1}^{n}t_{i}v_{i}$. Because of the $\delta$
factors the random variables $(\varphi,v_i)$ with $i \le k$ are
a.s. zero according to this probability law.  Thus it is
straightforward to verify that $C$ continues to be the covariance of
$\varphi$: e.g., $\Var \big((\varphi,v_i)\big) = 0 =
(v_{i},Cv_{i})$ for $i=1,\dots ,k$.

\index{Gaussian measure}
\begin{defn}
\label{def:Gm}
  The centred \emph{Gaussian probability measure} $P_C$ on $\R^X$,
  with covariance $C$, is defined by Definition~\ref{def:Gm2} when $C$
  is positive definite and by \eqref{e:Gm2} if $C$ is positive semi-definite.
  We refer to $\varphi$ with distribution $P_C$ as
  a \emph{Gaussian field with covariance} $C$.
  The \emph{expectation} of a random variable $F: \R^X \to \R$ is
  \index{Expectation}
  \begin{equation} \label{e:Ex}
    \Ex_CF = \int F(\varphi) \; P_C(d\varphi).
  \end{equation}
\end{defn}

\begin{exercise} \label{ex:ibp}
Verify the Gaussian integration by parts identity
\begin{equation}
    \Ex_C(F\varphi_x) = \sum_{y \in X} C_{xy} \Ex_C\left (\ddp{F}{\varphi_y}\right),
\end{equation}
by writing $\Ex_C((C^{-1}\varphi)_xF )$ as a derivative ($C$ is
invertible when restricted to $\varphi$ in the support of $P_{C}$).
\solref{ibp}
\end{exercise}

\begin{example} \label{example:gauss-vect}
The $|\varphi|^4$ model is defined in terms of
vector-valued fields $\varphi = (\varphi_x^i)_{x\in\Lambda,i=1, \dots,n}$.
These are fields $\varphi \in \R^X$ with the special choice
\begin{equation}
  X = n\Lambda = \{(x,i): x\in \Lambda, i=1, \dots, n\}.
\end{equation}
Given a positive semi-definite matrix $C=(C_{xy})_{x,y\in\Lambda}$, we define
an $X\times X$ matrix $(\hat{C}_{(x,i),(y,j)})$ by $\hat{C}_{(x,i),(y,j)}
=\delta_{ij}C_{xy}$.  We refer to the Gaussian
field on $\R^X$ with covariance $\hat{C}$ as the
$n$-component Gaussian field on $\R^\Lambda$ with covariance $C=(C_{xy})_{x,y\in\Lambda}$.
We denote its expectation also by $\Ex_C$.
\end{example}

\begin{defn}
\label{defn:Gconv}
  The \emph{convolution} of $F$ with the Gaussian measure $P_C$ is denoted
  \index{Convolution}
  \begin{equation} \label{e:Extheta}
    \Ex_C\theta F(\varphi) = \int F(\varphi+\zeta) \; P_C(d\zeta) \quad (\varphi \in \R^X),
  \end{equation}
  always assuming the integrals exist.  The above defines
  $\Ex_C\theta$ as a single operation, but we also view it as the
  composition of a map $\theta: F \mapsto F(\cdot + \zeta)$ followed by
  the expectation $\Ex_C$ which integrates with respect to $\zeta$.  The
  map $\theta$ is a homomorphism on the algebra of functions of the
  field $\varphi$.
\end{defn}

The following proposition demonstrates an intimate link between Gaussian integration
and the Laplace operator
\begin{equation}
  \Delta_C = \sum_{x,y \in X} C_{xy} \partial_{\varphi_x} \partial_{\varphi_y}.
\end{equation}
Since we are eventually interested in large $X$ (the vertices of a large graph),
this Laplace operator acts on functions on a high-dimensional space.

\begin{prop} \label{prop:wick} \index{Wick's Lemma} \index{Heat equation}
  For a polynomial $A= A(\varphi)$ in $\varphi$ of degree at most $2p$,
  \begin{equation}
    \Ex_C\theta A = e^{\frac12 \Delta_C} A
    = \left( 1+ \tfrac12 \Delta_C + \cdots + \frac{1}{p! 2^p} \Delta_C^p\right) A.
  \end{equation}
\end{prop}

\begin{proof}
  Set $v(t,\varphi) = \Ex_{tC}\theta A(\varphi)$ and $w(t,\varphi) = e^{\frac12 \Delta_{tC}}A(\varphi)$.
  It can be seen that $v,w$ are both polynomials in $\varphi$ of the same degree as $A$ and that both
  satisfy the heat equation
  \begin{equation}
    \partial_t u = \tfrac12 \Delta_C u, \quad u(0,\varphi) = A(\varphi).
  \end{equation}
  (For $v$, it is convenient to use $v(t,\varphi)
  = \int A(\varphi+\sqrt{t}\psi) \, P_C(d\psi)$
  and Gaussian integration by parts.)
  Since $u=v,w$ are polynomials in $\varphi$, the heat equation is equivalent to a finite-dimensional
  system of linear ODE, with unique solution,
  and we conclude that $v(t,\cdot) = w(t,\cdot)$ for all $t>0$.
\end{proof}

In particular, for a polynomial $A=A(\varphi)$,
\begin{equation}
\lbeq{ECpoly}
  \Ex_CA = \Ex_C\theta A|_{\varphi = 0} = e^{\frac12 \Delta_C} A|_{\varphi=0},
\end{equation}
and thus
\begin{equation} \label{e:Gauss-moments124}
  \Ex_C (\varphi_x) = 0, \quad \Ex_C(\varphi_x\varphi_y) = C_{xy},
  \quad \Ex_C(\varphi_x\varphi_y\varphi_u\varphi_v) = C_{xy}C_{uv} + C_{xu}C_{yv} + C_{xv}C_{yu}.
\end{equation}

\begin{exercise} \label{ex:wickpp}
  By definition, the covariance of random variables $F_1,F_2$ is
  \begin{equation}
   \Cov_C (F_1,F_2)= \Ex_CF_1F_2 - (\Ex_CF_1)(\Ex_CF_2).
  \end{equation}
  By symmetry, $\Cov\big(\varphi_{x}^{p},\varphi_{x'}^{p'}\big)=0$ if
  $p+p'$ is odd.  Show that if $p+p'$ is even then
  $| \Cov(\varphi_{x}^{p},\varphi_{x'}^{p'})|  \leq
  M_{p,p'} \|C\|^{(p+p')/2}$ where $\|C\| = \max_{x} C_{xx}$ and
  $M_{p,p'}$ is a constant depending on $p,p'$.
  \solref{wickpp}
\end{exercise}

Proposition~\ref{prop:wick} is a version of \emph{Wick's Lemma}; it
allows straightforward evaluation of all moments of a Gaussian
measure, in terms only of its covariance. The inverse of this formula
for expectations of polynomials is \emph{Wick ordering}. The Wick
ordering of a polynomial $A$ with respect to a Gaussian measure with
covariance $C$ is commonly denoted by $:\!A\!:_C$.
\begin{defn}
Let $A=A(\varphi)$ be a polynomial. The \emph{Wick ordering} of $A$
with covariance $C$ is
\index{Wick ordering}
\begin{equation}
  {:\!A\!:_C} = e^{-\frac12 \Delta_C}A.
\end{equation}
\end{defn}
Thus, essentially by definition,
\begin{equation}
  \Ex_C\theta {:\!A\!:_C} = A.
\end{equation}
Note that while the heat semigroup $e^{\frac12 \Delta_C}$ is
contractive on suitable function spaces, and can thus be extended to
much more general non-polynomial $A$, Wick ordering can be interpreted
as running the heat equation backwards. For general initial data, this
is problematic, but for nice initial data (and polynomials are
extremely nice) it is perfectly well-defined.  For example, in the
proof of Proposition~\ref{prop:wick}, for polynomials the heat
equation is equivalent to a linear ODE, and any linear ODE can be run
either forward or backward.

\index{Laplace transform} A fundamental property of Gaussian measures
is their characterisation by the Laplace transform, also called the moment
generating function in probability theory.

\begin{prop} \label{prop:Gauss-Laplace}
  A random field $\varphi \in \R^X$
  is Gaussian with covariance $C$ if and only if
  \begin{equation} \label{e:Gauss-Laplace}
    \Ex_C(e^{(f,\varphi)}) = e^{\frac12 (f,Cf)} \quad \text{for all $f \in \R^X$.}
  \end{equation}
\end{prop}
\begin{proof}
Suppose first that $C$ is positive definite.  By completion of the square,
  \begin{equation} \label{e:Gauss-chvar}
    -\tfrac12 (\varphi, C^{-1}\varphi) + (f,\varphi)
    = -\tfrac12 (\varphi - Cf, C^{-1}(\varphi - Cf)) + \tfrac12 (f,Cf).
  \end{equation}
Then \eqref{e:Gauss-Laplace} follows by the change of variables
$\varphi \mapsto \varphi + Cf$, which leaves the Lebesgue measure invariant.
This proves the ``only if'' direction, and the ``if'' direction then follows
from the fact that the Laplace transform characterises probability measures uniquely
\cite[p. 390]{Bill95}.

If $C$ is positive semi-definite but not positive definite, the
Gaussian measure is defined by \refeq{Gm2}.  The restriction $C'$ of
$C$ to the support $K^{\perp}$ of $P_{C}$ is invertible, $C'$ and its
inverse are isomorphisms of $K^\perp$, and $Cf \in K^\perp$.  The
reasoning used for the positive definite case thus applies also here.
\end{proof}

The ``only if'' direction of Proposition~\ref{prop:Gauss-Laplace} has
the following generalisation which we will use later.

\begin{exercise} \label{ex:Gauss-Laplace-Z0}
  For $Z_0 = Z_0(\varphi)$ bounded,
  \begin{equation}
    \Ex_C(e^{(f,\varphi)} Z_0(\varphi))
    = e^{\frac12 (f,Cf)} (\Ex_C\theta Z_0)(Cf) \quad \text{for all $f \in \R^X$.}
  \end{equation}
\solref{Gauss-Laplace-Z0}
\end{exercise}

Proposition~\ref{prop:Gauss-Laplace} also implies the following essential
corollary.

\begin{cor} \label{cor:Gauss-decomp} \index{Convolution}
  Let $\varphi_1$ and $\varphi_2$ be independent Gaussian fields with covariances $C_1$ and $C_2$. Then $\varphi_1+\varphi_2$
  is a Gaussian field with covariance $C_1+C_2$. In terms of convolution,
  \begin{equation}
    \Ex_{C_2}\theta \circ \Ex_{C_1}\theta = \Ex_{C_1+C_2} \theta.
  \end{equation}
\end{cor}

\begin{proof}
  By independence,
  for any $f \in \R^X$,
  \begin{equation}
    \Ex(e^{(f,\varphi_1+\varphi_2)}) = \Ex(e^{(f,\varphi_1)})\Ex(e^{(f,\varphi_2)})
    = e^{\frac12 (f,(C_1+C_2)f)} .
  \end{equation}
  By Proposition~\ref{prop:Gauss-Laplace},
  $\varphi_1+\varphi_2$ is Gaussian with covariance $C_1+C_2$.
\end{proof}

Corollary~\ref{cor:Gauss-decomp} is fundamental for our implementation of the renormalisation
group method,
whose starting point is a decomposition
$C=\sum_{j=1}^N C_j$ of the covariance $C=(-\Delta +m^2)^{-1}$.  This allows
us to rewrite
a Gaussian convolution $\Ex_C\theta Z_0$,
that is difficult to evaluate, as a sequence of convolutions
\begin{equation}
\lbeq{conv-sequence}
  \Ex_C \theta Z_0 = \Ex_{C_N}\theta \circ \cdots \circ \Ex_{C_1}\theta Z_0,
\end{equation}
where each expectation on the right-hand side is more tractable.

\begin{example} \label{ex:MMFdecomp}
Let $\Delta_\beta$ be the mean-field Laplacian matrix \eqref{e:P-def-intro}.
Since $P$ and $Q$ are orthogonal projections with $P+Q=\Id$,
\begin{equation}
  -\Delta_\beta+m^2 = (\beta+m^2)P + m^2 Q.
\end{equation}
For $m^2>0$, it then follows from the spectral theorem that
\begin{equation} \label{e:MMFdecomp}
  (-\Delta_\beta+m^2)^{-1}
  = \frac{1}{\beta+m^2}P + \frac{1}{m^2} Q
  = \frac{1}{\beta+m^2} + \frac{\beta}{m^2(\beta+m^2)} Q
  .
\end{equation}
The left-hand side is the covariance matrix of a Gaussian field
and the two matrices on the right-hand side are each positive definite.
This provides a simple example to which \eqref{e:conv-sequence} can be applied,
with $N=2$.
In fact,
Lemma~\ref{lem:MF-decomp} can be regarded as a limiting case of this fact,
where one of the Gaussian measures becomes degenerate in the limit $m^2 \downarrow 0$.
For Euclidean or hierarchical models, we use the more elaborate covariance
decompositions discussed at length in Chapters~\ref{ch:decomp} and \ref{ch:hier}.
\end{example}

The following exercise establishes properties of the $n$-component Gaussian field
of Example~\ref{example:gauss-vect}.

\begin{exercise} \label{ex:gauss-On}
  Let $C = (C_{xy})_{x,y\in\Lambda}$ be a positive semi-definite matrix on $\R^\Lambda$.
  \\
  (i)
  Verify that the components of the corresponding $n$-comp\-on\-ent Gaussian field are independent and
  identically distributed Gaussian fields on $\Lambda$ with covariance~$C$.
  \\
  (ii)
  Let $T \in O(n)$ act on $\R^{n\Lambda}$ by $(T\varphi)_x = T\varphi_x$ for $x \in \Lambda$,
  and on $F: \R^{n\Lambda}\to \R$ by $TF(\varphi) = F(T\varphi)$.
  We say that $F$ is $O(n)$-\emph{invariant} if $TF = F$ for all $T \in O(n)$.
  Prove that the $n$-component Gaussian field
  is $O(n)$-invariant,
  in the sense that for any bounded measurable $F: \R^{n\Lambda} \to \R$ and $T \in O(n)$,
  \begin{equation}
    \Ex_C(F(\varphi)) = \Ex_C(F(T\varphi)),
    \quad \Ex_C \theta \circ T = T \circ \Ex_C\theta.
  \end{equation}
    In particular, if $F$ is $O(n)$-invariant  then so is $\Ex_C\theta F$,
    and if $F_1,F_2$ are both $O(n)$-invariant then so is $\Cov_C(\theta F_1,\theta F_2)$.
    \solref{gauss-On}
\end{exercise}

A second consequence of
Proposition~\ref{prop:Gauss-Laplace} is the following corollary.

\begin{cor} \label{cor:Gauss-restrict}
  Let $Y \subset X$. The restriction of $P_C$ to $\R^Y$ is the centred Gaussian probability measure
  with covariance $C|_{Y \times Y}$.
\end{cor}

We are ultimately interested in the infinite-volume limit for the $|\varphi|^4$ model.
For this,
we work with finite sets approximating
$\Z^d$, with the aim of obtaining estimates that hold uniformly in the
size of the finite set.  For Gaussian fields, a construction in infinite volume
can be made directly, as a consequence of
Corollary~\ref{cor:Gauss-restrict}.

\begin{exercise} \label{ex:Gauss-infdim} \index{Infinite-volume limit}
  Let $S$ be a possibly infinite set. By definition, an $S \times S$ matrix $C$ is positive definite if $C|_{X\times X}$
  is a positive definite matrix for every \emph{finite} $X \subset S$.
  Let $C$ be positive definite.  Use Corollary~\ref{cor:Gauss-restrict}
  to show that $P_{C|_{X\times X}}$, $(X\subset S \text{ finite})$ forms a consistent family of measures.
  Use the Kolmogorov extension theorem
  (or the nicer Kolmogorov--Nelson extension theorem \cite[Theorem 10.18]{Foll99})
  to conclude that there exists a probability measure $P_C$ on
  $\R^{\Z^d}$ with covariance $C$.
  \solref{Gauss-infdim}
\end{exercise}

\section{Cumulants}

\index{Cumulant}
\begin{defn} \label{Cumulant}
  Let $A_1, \dots, A_n$ be random variables
  (not necessarily Gaussian) such that $\Ex(e^{tA_i}) < \infty$
  for $t$ in some neighbourhood of $t=0$.
  Their \emph{cumulants}, or \emph{truncated expectations}, are defined by
  \begin{equation}
  \lbeq{cumulants}
    \Ex(A_1; \cdots; A_n) =
    \frac{\partial^n}{\partial t_1 \cdots \partial t_n}
    \log \Ex (e^{t_1A_1 + \cdots + t_n A_n}) \Big|_{t_1 = \cdots =t_n = 0}
    .
  \end{equation}
\end{defn}

\index{Truncated expectation} The truncated expectation of a single
random variable is its expectation, and the truncated expectation of a
pair of random variables is their covariance:
\begin{equation}
  \Cov(A_1,A_2) = \Ex(A_1;A_2) = \Ex(A_1A_2) - \Ex(A_1)\Ex(A_2)
  .
\end{equation}
The assumption of exponential moments is not necessary to define
cumulants.  Instead, the logarithm of the expectation on the
right-hand side of \refeq{cumulants} may be regarded as a formal power
series in $t$, upon which the derivative acts.

\begin{exercise} \label{ex:trunc-corr-existence}
Show that the truncated expectations up to order $n$ exist if and only if
the expectations of the product of up to $n$ of the $A_i$ exist,
and that the latter up to order $n$
determine the truncated expectations up to order $n$ and vice-versa.
Hint: Let $I=\{i_{1},\dots ,i_{n} \}$.  A partition $\pi$
of $I$ is a collection of disjoint nonempty subsets of $I$ whose union
is $I$.  Let  $\Pi (I)$ denote the set of all partitions of $I$.
Then if we define $\mu_{I}=\Ex (A_{i_{1}} \cdots A_{i_{n}})$ and
$\kappa_{I} = \Ex (A_{i_{1}};\cdots
;A_{i_{n}})$,
\begin{equation}
    \mu_{I}
    =
    \sum_{\pi \in \Pi (I)}\prod_{J \in \pi} \kappa_{J}
    .
\end{equation}
This system of equations, one for each $I$, uniquely defines
$\kappa_{I}$ for all $I$.
\solref{trunc-corr-existence}
\end{exercise}

The next exercise shows that a collection of random variables is
Gaussian if and only if all higher truncated expectations vanish.

\begin{exercise} \label{ex:Gauss-cum} \label{Laplace transform}
Use Proposition~\ref{prop:Gauss-Laplace} and
Exercise~\ref{ex:trunc-corr-existence} to show that a random field
$\varphi$ on $X$ is a Gaussian field with mean zero and covariance $C$ if
and only if for all $p \in \N$ and $x_1, \dots, x_p \in X$,
\begin{equation}
  \label{e:ex:Gauss-cum}
  \Ex(\varphi_{x_1}; \cdots; \varphi_{x_p})
  =
  \begin{cases} C_{x_1x_2} & (p=2)
      \\
      0 & (p\neq 2).
  \end{cases}
\end{equation}
\solref{Gauss-cum}
\end{exercise}

In the case of Gaussian fields, with $A_i = A_i(\varphi)$, it is
useful to define a convolution version of truncated expectation,
by
\begin{equation} \label{e:Exthetatruncdef}
  \Ex_C (\theta A_1; \cdots; \theta A_n)
  = \frac{\partial^n}{\partial t_1 \cdots \partial t_n} \log \Ex_C \theta (e^{t_1A_1 + \cdots + t_n A_n}) \Big|_{t_1 = \cdots t_n = 0}
  .
\end{equation}
In particular,
\begin{equation} \label{e:ExthetaABdef}
  \Ex_C(\theta A;\theta B)
  = \Cov_C(\theta A,\theta B),
\end{equation}
where, since $\theta (AB)=(\theta A)(\theta B)$,
\begin{equation} \label{e:Covthetadef}
   \Cov_C(\theta A, \theta B)
  = \Ex_C\theta(AB)-(\Ex_C\theta A)(\Ex_C\theta B).
\end{equation}
If $A,B$ are polynomials, then, by Proposition~\ref{prop:wick},
\begin{equation} \label{e:FCwick}
  \Ex_C(\theta A;\theta B)
  = e^{\frac12 \Delta_C}(AB)-(e^{\frac12 \Delta_C}A)(e^{\frac12 \Delta_C}B).
\end{equation}

\begin{exercise} \label{ex:Fexpand}
  For $A,B$ polynomials in $\varphi$, let
  \begin{equation} \label{e:FCex}
    F_C(A,B) = e^{\frac12 \Delta_C}\big((e^{-\frac12 \Delta_C}A)(e^{-\frac12 \Delta_C}B)\big) - AB.
  \end{equation}
  Then $\Ex_C(\theta A;\theta B) = F_C(\Ex_C\theta A, \Ex_C\theta B)$. Show that, if $A,B$ have degree
  at most $p$, then
  \begin{equation}\label{e:FC_derivatives_formula}
    F_C(A,B) = \sum_{n=1}^p \frac{1}{n!} \sum_{x_1,y_1} \cdots \sum_{x_n,y_n} C_{x_1,y_1} \cdots C_{x_n,y_n}
    \ddp{^nA}{\varphi_{x_1}\cdots \partial \varphi_{x_n}}
    \ddp{^nB}{\varphi_{y_1}\cdots \partial \varphi_{y_n}}.
  \end{equation}
  \solref{Fexpand}
\end{exercise}



%% file: decomp.tex
\chapter{Finite-range decomposition}
\label{ch:decomp}

Our implementation of the renormalisation group method relies on the
decomposition of convolution by a Gaussian free field (GFF)
into a sequence of convolutions, as in \refeq{conv-sequence}.  This
requires an appropriate decomposition of the covariance of the
Gaussian field into a sum of simpler covariances.  Such covariance
decompositions, in the context of renormalisation, go back a long way,
early examples can be found in \cite{BCGNOPS78,BCGNOPS80}.

In this chapter, we describe covariance decompositions which have a
\emph{finite-range} property.  This property is an important
ingredient in our renormalisation group method for models defined on
the Euclidean lattice \cite{BS-rg-step}.  We begin in
Section~\ref{sec:frd-pi} by defining the finite-range property and
elaborating on \refeq{conv-sequence} and its role in progressive
integration.  In Section~\ref{sec:frd-continuum}, we motivate the
finite-range decomposition by first discussing it in the much simpler
continuum setting.  In Section~\ref{sec:frd-lattice}, we give a
self-contained presentation of a finite-range decomposition of the
lattice operator $(-\Delta + m^2)^{-1}$ on $\Zd$ following the method
of \cite{Baue13a} (a related method was developed in
\cite{BGM04}).  This easily gives rise to a finite-range decomposition
on the discrete torus, as discussed in Section~\ref{sec:frd-torus}.

After this chapter, we do not return to Euclidean models until
Appendix~\ref{app:Euclidean}, so in a sense this chapter is a cultural
excursion. However, the finite-range decomposition of
Proposition~\ref{prop:decomp} provides a useful motivation for the
hierarchical model that becomes our focus after this chapter.

\section{Progressive integration}
\label{sec:frd-pi}

Recall from \refeq{conv-sequence} that a decomposition
\begin{equation}
  C = C_1+ \cdots + C_N
\end{equation}
of the covariance $C$ provides a way to evaluate a Gaussian
expectation progressively, namely,
\begin{equation}
\lbeq{conv-sequence-bis}
    \Ex_C\theta F = \Ex_{C_{N}}\theta \circ \cdots \circ \Ex_{C_1}\theta F.
\end{equation}
This is the point of departure for the renormalisation group method.
It allows the left-hand side to be evaluated progressively, one $C_j$
at a time.  For this to be useful, the convolutions on the right-hand
side need to be more tractable than the original convolution, and
therefore useful estimates on the $C_j$ are needed.

In this chapter, we explain a method to decompose the covariance
$C=(-\Delta+m^2)^{-1}$ for three
different interpretations of the Laplacian: the continuum operator on
$\R^d$ (with $m^2=0$), the discrete operator on $\Zd$,
and finally the discrete operator on a periodic approximation to
$\Zd$.  In each case, we are interested in decompositions with a
particular finite-range property.

\begin{defn} \index{Finite-range property}
  Let $\zeta$ be a centred Gaussian field on $\Lambda$.
  We say $\zeta$
  is \emph{finite range} with \emph{range} $r$ if
  \begin{equation} \label{e:finrange-def}
    \Ex_C(\zeta_x\zeta_y) = 0\quad \text{if $|x-y|_{1} > r$.}
  \end{equation}
\end{defn}

The following exercise demonstrates that the finite-range property has
an important consequence for independence.

\index{Independence}
\begin{exercise} \label{ex:uncorr-then-indep-Gauss}
  Let $\varphi_x,\varphi_y$ be jointly Gaussian random variables which
  are \emph{uncorrelated}, i.e., $\Ex(\varphi_x\varphi_y) = 0$.  Use
  Proposition~\ref{prop:Gauss-Laplace} to show that $\varphi_x$ and
  $\varphi_y$ are \emph{independent}.  (For general random variables,
  independence is a stronger property than being uncorrelated, but for
  Gaussian random variables the two concepts coincide.)
  \solref{uncorr-then-indep-Gauss}
\end{exercise}

In view of \refeq{conv-sequence-bis}, decomposition of the covariance
$C=(-\Delta+m^2)^{-1}$ as $C=\sum_j C_j$, where the matrices $C_j$ are
symmetric and positive definite, is equivalent to a decomposition of
the GFF $\varphi$ as
\begin{equation} \label{e:varphi-decomp-bis}
  \varphi \stackrel{D}{=} \zeta_1 + \cdots + \zeta_N,
\end{equation}
where the $\zeta_j$ are independent Gaussian fields.
Explicitly, for $C=C_1+C_2$, we have
\begin{align}
    (\Ex_C \theta F)(\varphi') &
     = \Ex_C F(\varphi + \varphi')
     =
     \Ex_{C_2} \Ex_{C_1} F(\zeta_1 + \zeta_2 + \varphi'),
\end{align}
where in the middle the expectation acts on $\varphi$, while on the
right-hand side each expectation with respect to $C_j$ acts on
$\zeta_j$.  The fields $\zeta_{j}$ have the finite-range property
with range $r= \frac12 L^j$
if and only if $C_{j;xy}=0$ for $|x-y|_1 > \frac12 L^j$.

\section{Finite-range decomposition: continuum}
\label{sec:frd-continuum}

\index{Fourier transform}
In this section, we work frequently with the Fourier transform
\begin{equation}
    \label{e:FTRd-def}
    \hat{f}(p) = \int_{\R^d} f(y) e^{-ipy} dy
\end{equation}
of functions $f: \R^d \to \R$ defined on the continuum. The inverse
Fourier transform is
\begin{equation}
     f(x) = \frac{1}{(2\pi)^d}\int_{\R^d} \hat{f}(p) e^{ipx} dp.
\end{equation}

\index{Positive definite}
\begin{defn}
A function $f: \R^d \to \R$ is \emph{positive definite} if it is
continuous and has the property that for every integer $n$ and every
sequence $(x_{1},\dots ,x_{n})$ of points in $\R^{n}$ the $n\times n$
matrix $f (x_{i}-x_{j})$ is positive semi-definite.
\end{defn}

\index{Bochner's theorem}
\begin{exercise} \label{ex:posdef}
For any $h \in C_c(\R^d, \R)$ with $h(-x)=h(x)$, the convolution $h*h$
is positive definite.  More generally,
if $f$ has Fourier transform obeying $\hat f \ge 0$,
then $f$ is positive definite.
(The converse is also true; this is Bochner's theorem
\cite[Theorem~IX.9]{RS75}.) \solref{posdef}
\end{exercise}

\begin{prop} \label{prop:decomp-cont} \index{Covariance decomposition}
  Given $L>1$ and $\alpha > 0$, there exists $u:\R^d \to \R$ which is smooth,
  positive definite, with support in $[-\frac12, \frac12]^d$, such that
  \begin{equation} \label{e:decomp-cont}
    |x|^{-\alpha} = \sum_{j\in\Z} L^{-\alpha j} u(L^{-j}x)
    \qquad (x \neq 0)
    .
  \end{equation}
\end{prop}

For $d \neq 2$, and $\alpha=d-2$, the left-hand side of
\eqref{e:decomp-cont} is a multiple of the Green function of the
Laplace operator $\sum_{i=1}^d \partial_{i}^2$ on $\R^d$.  A similar
representation exists for $d=2$.  The right-hand side of
\refeq{decomp-cont} provides a finite-range decomposition of the Green
function, in the sense that the $j^{\rm th}$ term vanishes if
$|x|_\infty > \frac 12 L^j$. This is an unimportant departure from the
definition in terms of $|x|_1 $ given below
\eqref{e:varphi-decomp-bis}.  The scales $j \le 0 $ which appear in
the sum are absent for a lattice decomposition.  The proof shows that
there is considerable flexibility in the choice of the function $u$.

\begin{proof}[Proof of Proposition~\ref{prop:decomp-cont}] Choose a
function $w \in C_c(\R)$ which is not the zero function.  By the
change of variables $t \mapsto |x| t$,
\begin{equation}
\label{e:xalpha}
  \int_0^\infty t^{-\alpha} w(|x|/t) \frac{dt}{t}
  = c|x|^{-\alpha},
\end{equation}
with $c = \int_0^\infty t^{-\alpha} w(1/t) \frac{dt}{t}$.  After
normalising $w$ by multiplication by a constant so that $c=1$, we
obtain
\begin{equation}
  \label{e:xalpha2}
  |x|^{-\alpha} = \int_0^\infty t^{-\alpha} w(|x|/t) \frac{dt}{t}.
\end{equation}
Now choose $w$ with support in $[-\frac12,\frac12]$ such that
$x\mapsto w(|x|)$ is a smooth, positive definite function on $\R^d$.
By Exercise~\ref{ex:posdef}, a function $w$ with these
properties exists.
Given $L>1$, set
\begin{equation}
  u(x) = \int_{1/L}^1 t^{-\alpha} w(|x|/t) \frac{dt}{t}
  .
\end{equation}
It is not hard to check that this is a positive definite function.  By
change of variables, \eqref{e:decomp-cont} holds, and the proof is
complete.
\end{proof}

A statement analogous to Proposition~\ref{prop:decomp-cont} for the
lattice Green function is more subtle.  The proof for the continuum
exploited in a crucial way two symmetries, homogeneity and rotation
invariance, which are both violated in the discrete case.  To motivate
and prepare for the construction of the finite-range
decomposition for the lattice, we now present another
proof of \eqref{e:decomp-cont}.  As in the previous proof,
it suffices to show that \eqref{e:xalpha2} holds with $w$ a compactly
supported positive definite function.  We will create a radial
function $w$ whose support is a ball of radius $1$ instead of
$\frac12$; this is an unimportant difference.
Our proof exploits a connection with the finite speed of propagation
property of hyperbolic equations that originated in \cite{Baue13a}.

Let $f: \R \to [0,\infty)$ be such that its Fourier transform is
smooth, symmetric, and has support in $[-1,1]$.  We assume that $f$ is
not the zero function.  By multiplication of $f$ by a constant, we can
arrange that
\begin{equation}
    \label{eq:1/lambda-decomp}
    \frac {1} {|k|^{2}}   
    =
    \int_0^\infty t^{2} f(|k| t) \; \frac{dt}{t} 
    \quad\quad
    (k \in \R^d, \, |k|\not = 0) . 
\end{equation}
Indeed, \eqref{eq:1/lambda-decomp} is just \refeq{xalpha} with $w=f$
and $\alpha=2$, after change of variables from $t$ to $1/t$.
For $d>2$, the Green function $|x|^{-(d-2)}$ has Fourier transform
proportional to $1/|k|^{2}$.
By inverting the Fourier transform, we obtain
\begin{equation}
  \label{e:decomp-wave1}
  |x|^{-(d-2)} \;\propto\; \int_0^\infty w(t,x) \; \frac{dt}{t}  
\end{equation}
where
\begin{equation}
  \label{e:wtx}
  w(t,x)
  = (2\pi)^{-d} \int_{\R^d} t^2 f(|k|t) \, e^{ik\cdot x} \, dk.
\end{equation}
Define $w(x) = w(1,x)$. By change of variable, $w(t,x) = t^{-(d-2)}
w(x/t)$ in \eqref{e:decomp-wave1}.  We have achieved a
decomposition like \eqref{e:xalpha2} with $\alpha=d-2$, where $w (t,x)$ has the desired
positive definiteness because $f \ge 0$; it remains to prove that $w (x)$ is supported
in the unit ball.

\index{Wave equation} \index{Finite propagation speed}
By hypothesis the ($1$-dimensional) Fourier transform $\hat f$ is
symmetric and $\supp \hat f \subset [-1,1]$. Therefore
\begin{equation} \label{e:decomp-wave3}
  f(|k|) = (2\pi)^{-1} \int_{-1}^1 \hat f(s)\, \cos(|k|s) \, ds .
\end{equation}
By inserting this into \eqref{e:wtx} and setting $t=1$ we read off the
$d$-dimensional Fourier transform
\begin{equation}
  \label{e:hatw}
  \hat{w}(k)
  =
  (2\pi)^{-1} \int_{-1}^1 \hat f(s)\, \cos(|k|s) \, ds .
\end{equation}
That $w$ has support in the unit ball is a consequence of the finite
propagation speed of the wave equation, as follows.  It suffices to
show, for any smooth function $u_0$ on $\R^d$, that the support of
$w*u_0$ is contained in the $1$-neighbourhood
$\{x \in \R^{d} \mid \dist (x,\supp \,u_{0})\le 1\}$ of $\supp \, u_{0}$, because we
can replace $u_{0}$ by the approximate identity $\epsilon^{-d}u_{0}
(x/\epsilon)$ and let $\epsilon \downarrow 0$.
Let $u(s,x)$ be the solution to the ($d$-dimensional) wave equation
\begin{equation} \label{e:waveequation}
  \ddp{^2u}{s^2}
  =
  \Delta u, \quad u(0,x) = u_0(x), \quad \ddp{}{s} u(0,x) = 0.
\end{equation}
The solution to this equation is
\begin{equation}
   u (s,x)
   =
   (2\pi)^{-d}
   \int_{\R^d} \hat{u}_{0} (k) \cos (s|k|) e^{ik\cdot x} \, dk .
\end{equation}
By combining this with \eqref{e:hatw} we have
\begin{equation}
\lbeq{waveconv}
  w * u_0(x) = (2\pi)^{-1}\int_{-1}^1 \hat f(s) u(s,x) \, ds.
\end{equation}
By the finite propagation speed of the wave equation, the support of
$u(s,\cdot)$ is contained in the $|s|$-neighbourhood of $\supp \,u_0$.
Since the range of the $s$ integral is $s \le 1$ we have proved that
the support of $w*u_0$ is contained in the $1$-neighbourhood of
$\supp \, u_{0}$ as desired.

The formula \eqref{eq:1/lambda-decomp}
generalises to a representation for $|k|^{-\alpha}$ for other values
of $\alpha$ by using a different power of $t$ inside the integral, so
that finite-range decompositions for $|x|^{\alpha -d}$ can also be
constructed by this method. Furthermore, the method applies to the
Green function in dimension $d\leq 2$ with the correct interpretation
of the domain of function on which the Green function acts.

\begin{exercise}
  \label{ex:paley-wiener}
  Use the Schwartz--Paley--Wiener Theorem to deduce from
  \eqref{e:decomp-wave3} that $w$ has support in the unit ball
  without referring to the finite propagation speed of the wave
  equation explicitly. \solref{paley-wiener}
\end{exercise}

\section{Finite-range decomposition: lattice}
\label{sec:frd-lattice}

We present a construction of the finite-range decomposition for the
lattice Green function which is based
on the wave equation perspective
of the continuum decomposition explained in
\eqref{eq:1/lambda-decomp}--\eqref{e:waveconv}.  The wave equation is now
replaced by a discrete wave equation.  For the discrete wave equation,
the Chebyshev polynomials $T_t$ play a role analogous to the functions
$\cos(\sqrt{\cdot}\, t)$ for the continuous wave equation.

\subsection{Statement of the decomposition}

In this section we state a proposition which provides a decomposition
of $(-\Delta_\Zd + m^2)^{-1}$ for all $d>0$ and $m^2>0$.  The
proposition gives the existence and properties of covariances $C_{j}$
on $\Zd$ such that
\begin{equation}
\label{e:ZdCj9}
    (\Delta_\Zd + m^2)^{-1} = \sum_{j=1}^\infty C_j
    ,
\end{equation}
where $C_{j}$ depends on $m^2>0$ and the sum converges in the sense of
quadratic forms, i.e.,
\begin{equation}
    (f, (\Delta_\Zd + m^2)^{-1}f) =
    \sum_{j=1}^\infty (f, C_jf) \quad (f \in \ell^{2}(\Zd)).
\end{equation}
In particular, by polarisation (choose $f=\delta_{x}+\delta_y$ and
$f=\delta_{x}-\delta_y$), it also implies convergence of the matrix
elements $C_{j;xy}$.
The covariances $C_j$ are translation invariant, and have the
\emph{finite-range} property that $C_{j;xy} = 0$ if $|x-y|_1 \geq
\tfrac 12 L^j$.

Finite-difference derivatives are defined as follows.  For $i=1,\dots
,d$ let $e_{i}$ be the unit vector $(0,\dots ,1,0,\dots ,0)$ whose
$i$th component equals $1$, and let $e_{-i}=-e_{i}$ so that, as $i$
ranges over $\{-d,\ldots,-1,1,\ldots, d\}$, $e_{i}$ ranges over the
unit vectors in the lattice $\Z^{d}$. For a function $f:\Z ^{d}
\rightarrow \R$ define $\nabla^{e_{i}}f_{x} = f_{x+e_{i}} -
f_{x}$. For a \index{Multi-index} multi-index $\alpha \in \{-d,\dots
,d \}^{n}$ define
\begin{equation}
  \nabla^\alpha f = \nabla^{e_{\alpha_1}} \cdots \nabla^{e_{\alpha_n}} f.
\end{equation}
For example, for $\alpha = (1,-2)$,
\begin{equation}
  \nabla^\alpha f_{x} =
  (\nabla^{e_{1}}\nabla^{e_{-2}}f)_x
  = (\nabla^{e_{-2}}f)_{x+e_1}-(\nabla^{e_{-2}}f)_x
  = f_{x+e_1-e_2}-f_{x+e_1}-f_{x-e_2}+f_x
  .
\end{equation}

Dependence of $C_j$ on $m^2$ is captured in terms of the parameter
$\vartheta$ defined, for $s,t,m^2\geq 0$ and $j\ge 1$, by
\begin{equation}
\label{e:varthdef}
  \vartheta(t,m^2;s) = \frac{1}{2d+m^2} \left(1+\frac{m^2t^2}{2d+m^2} \right)^{-s},
  \quad
  \vartheta_j(m^2;s) = \vartheta(L^{j},m^2;s).
\end{equation}

\begin{prop} \label{prop:decomp} \index{Covariance decomposition}
  Let $d>0$ and $L>1$.
  For all $m^2>0$
  there exist positive semi-definite matrices $(C_j)_{j \ge 1}$
  such that \refeq{ZdCj9} holds,
  and such that for all $j \ge 1$, 
  \begin{equation} \label{e:C-finrange}
    C_{j;xy} = 0 \quad \text{if $|x-y|_{1} \ge \tfrac12 L^j$}
    \quad \text{(finite-range property).}
  \end{equation}
  The matrix elements
  $C_{j;xy}$ are functions of $x-y$,
  are continuous functions of $m^2$
  and have limits as $m^2 \downarrow 0$.
  Moreover, for all multi-indices $\alpha$ and all $s\geq 0$,
  there are constants $c_{\alpha,s}$ such that,
  for all $m^2\in [0,\infty)$ and  $j \ge 1$,
  \begin{equation} \label{e:C-est}
    |\nabla^{\alpha} C_{j;xy}|
    \leq c_{\alpha,s}
    f_d(L)\vartheta_{j-1}(m^2;s) L^{-(d-2 + |\alpha|_1)(j-1)}
    \quad \text{(scaling estimates),}
  \end{equation}
  with $f_d(L)=1$ for $d>2$, $f_2(L)=\log L$, and $f_d(L)=L^{2-d}$
  for $d<2$. The discrete gradients can act either on $x$ or $y$.
\end{prop}

Estimates on derivatives of $C_j$ with respect to $m^2$ can be found
in \cite{Baue13a}.  We prove Proposition~\ref{prop:decomp} using the
construction of \cite{Baue13a}.  Finite-range decompositions for the
lattice Green function were first constructed in \cite{BGM04}, using a
different method.  Yet another method, which is very general, is used
in \cite{BT06,AKM13,Runa15,Buch16}.  Such decompositions have also
been obtained for fractional powers of the Laplacian
\cite{BGM04,Mitt16,Mitt17,Slad17}.

\subsection{Integral decomposition}

The decomposition we use is structurally similar to that discussed in
connection with the wave equation in Section~\ref{sec:frd-continuum}.
Roughly speaking, the Fourier multiplier $|k|^{2}$ of the continuum
Laplacian is replaced by the Fourier multiplier $\FDel(k) = 4
\sum_{j=1}^{d} \sin^2 ( k_j/2)$ of the discrete Laplacian given in
\eqref{e:FDel}.

Let $f$ be as in \eqref{eq:1/lambda-decomp}.  For $t>0$, we set
  \begin{equation} \label{e:varphi*-defn}
    f^*_t(x)
    = \sum_{n \in \Z} f(xt-2\pi nt)
    \quad (x\in \R)
    .
  \end{equation}
Since $\hat{f}$ is smooth, $f$ decays rapidly and therefore the sum on
the right-hand side is well-defined for $t > 0$. Moreover, $f^*_t \geq
0$ since $f \geq 0$.

\begin{lemma}
  For $x \in \R \setminus 2\pi \Z$,
  \begin{equation} \label{eq:sin2-decomp}
    \frac14
    \sin^{-2}(\frac12 x)
    = \int_0^\infty t^{2} f^*_t(x) \; \frac{dt}{t}
    .
  \end{equation}
\end{lemma}

\begin{proof}
  The left-hand side is a meromorphic function on $\C$ with poles at $2\pi \Z$.
  Its development into partial fractions is (see e.g.\ \cite[p.~204]{Ahlf78})
  \begin{equation}
    \frac14
    \sin^{-2} (\frac12 x) = \sum_{n \in \Z} (x-2\pi n)^{-2}
    \quad (x \in \C \setminus 2\pi\Z).
  \end{equation}
  From \eqref{eq:1/lambda-decomp} with $|k|$ replaced by $x-2\pi n$,
  it follows that
  \begin{equation}
    \frac14
    \sin^{-2} (\frac12 x)
    = \sum_{n \in \Z} \int_0^\infty t^{2}  f\big((x-2\pi n)t\big) \; \frac{dt}{t} .
  \end{equation}
  By hypothesis, $f$ is symmetric, so \eqref{eq:1/lambda-decomp} holds when $|k|$
  in the right-hand side is replaced by the possibly negative $x-2\pi n$.
  The order of the sum and the integral can be exchanged, by non-negativity of the integrand,
  and the proof is complete.
\end{proof}

For $t>0$ and $\zeta \in [0,4]$, we set
\begin{equation} \label{e:Pdef}
  P_t(\zeta)
  =
  f_t^*\left(\arccos\left(1-\frac12  \zeta\right)\right)  .
\end{equation}
Since $f^*_t \geq 0$, also $P_t(\zeta) \geq 0$.

\begin{lemma}
\label{lem:1zeta}
For $\zeta \in (0,4)$,
\begin{equation} \label{e:decomp}
  \frac{1}{\zeta}
  = \int_0^\infty t^2 \, P_t(\zeta) \, \frac{dt}{t}.
\end{equation}
\end{lemma}

\begin{proof}
Let $x=\arccos(1-\half \zeta)$, so that $\zeta = 2(1-\cos x) = 4
\sin^2(\frac{1}{2} x)$.  By \eqref{eq:sin2-decomp} and \refeq{Pdef},
\begin{equation}
    \frac{1}{\zeta} = \int_0^\infty t^2 f_t^*(x) \; \frac{dt}{t}
    =
    \int_0^\infty t^2 \, P_t(\zeta) \, \frac{dt}{t},
\end{equation}
and the proof is complete.
\end{proof}

We wish to apply \eqref{e:decomp} with $\zeta = \lambda(k)+m^2$ for $k \in [-\pi,\pi]^d$,
but for $m^2$ large this choice may not be in $(0,4)$. Therefore let $M^2 = 2d+m^2$ and
set $\zeta = (\lambda(k)+m^2)/M^2$; then $\zeta \in (0,2]$ provided
$m|k| \neq 0$.  By \eqref{e:decomp},
\begin{equation} \label{e:decompintpf}
  \frac{1}{\lambda(k)+m^2} =  \int_0^\infty  \hat w(t,k)\, \frac{dt}{t}
  ,
\end{equation}
with
\begin{equation}
  \label{e:w_hat}
  \hat w(t,k)
  =
  \frac{t^2}{M^2} P_t\left(\frac{1}{M^2} (\lambda(k)+m^2)\right)
  \quad (k \in [-\pi,\pi]^d)
  .
\end{equation}
Here and below, $k \in [-\pi,\pi]^d$ denotes the Fourier variable
of a function defined on the discrete space $\Z^d$ whose points are denoted by $x\in\Z^d$.
We use the same letter $w$ to denote the discrete analogue of
the function \eqref{e:wtx}
(which is on the continuum).
By \refeq{Chat}, inversion of this $d$-dimensional discrete Fourier transform
gives
\begin{equation}
\label{e:wint}
(-\Delta_{\Z^d}+m^2)^{-1}_{0x} = \int_0^\infty w(t,x) \frac{dt}{t},
\end{equation}
where
\begin{equation}
\lbeq{wxspace}
  w(t,x)
  =
  \frac{1}{(2\pi)^d}
  \int_{[-\pi,\pi]^d} \hat w(t,k)  e^{ik\cdot x} dk
  \quad (x\in \Z^d).
\end{equation}

The identity \refeq{wint} is the essential ingredient for the
finite-range decomposition.  We decompose the integral into intervals
$[0,\frac 12 L]$ and $[\frac12 L^{j-1}, \frac12 L^j]$ (for $j \ge 2$),
and define, for $x\in \Z^d$,
\begin{align}
\label{e:C1x}
  C_{1;0x} &= \int_{0}^{\frac12 L} w(t,x) \, \frac{dt}{t},
\\
\label{e:Cjx}
  C_{j;0x} &= \int_{\frac12 L^{j-1}}^{\frac12 L^j} w(t,x) \, \frac{dt}{t} \quad (j \ge 2).
\end{align}
By \eqref{e:decompintpf}, this gives, for $k \in [-\pi,\pi]^d$ and $m^2 \neq 0$,
\begin{equation}
  \frac{1}{\lambda(k)+m^2} =  \sum_{j=1}^\infty \hat C_j(k),
\end{equation}
where $\hat C_j \geq 0$ is the discrete Fourier transform of $C_j$.
Thus, for any $f \in \ell^2(\Z^d)$,
\begin{equation}
\lbeq{qf}
    (f,(-\Delta_{\Z^d} + m^2)^{-1}f) = \sum_{j=1}^\infty (f,C_jf),
\end{equation}
which proves \refeq{ZdCj9}. Furthermore, by \eqref{e:w_hat},
\eqref{e:C1x} and \eqref{e:Cjx}, the inequality $P_t(\zeta) \geq 0$
implies that this decomposition is positive semi-definite.

In Section~\ref{sec:chebyshev}, we will prove that
$P_t (\zeta)$ is a polynomial in $\zeta$ of degree at most
$t$.  This implies the finite-range property  \eqref{e:C-finrange}.
In fact, by \eqref{e:w_hat}, up to a scalar multiple,
$w (t,x-y)$ is the kernel that represents the operator $P_{t}
(M^{-2} (-\Delta + m^{2}))$, which is then a polynomial in
$-\Delta + m^{2}$ of degree at most $t$.
Since $-\Delta_{xy}$ vanishes unless $|x-y|_1 \le 1$, it follows that $w (t,x)=0$
if $|x|_{1}>t$.  By \eqref{e:C1x} and \eqref{e:Cjx}, this gives the
finite-range property \eqref{e:C-finrange}.

The integration domain for the covariance $C_1$ differs from the
domain for $C_j$ with $j \ge 2$.  It is therefore natural to decompose
it as $C_0+C_1'$ with
\begin{align}
\lbeq{C0def}
    C_{0;0x} & = \int_0^1 w(t,x) \, \frac{dt}{t},
    \quad
    C_{1;0x}' = \int_1^{\frac 12 L} w(t,x) \, \frac{dt}{t}.
\end{align}
Then $C_1'$ is of the same form as $C_j$ with $j \ge 2$.  We show in
Section~\ref{sec:decomp-pf} that the integral $C_{0;0x}$ can be
computed exactly:
\begin{equation}\label{e:C_0}
    C_{0;0x} = \frac{1}{2d+m^2}\frac{\hat f(0)}{2\pi}\1_{x=0} .
\end{equation}

In summary, we have proved that there exist positive
semi-definite matrices $(C_j)_{j \ge 1}$ such that \refeq{ZdCj9} holds
as asserted in Proposition~\ref{prop:decomp} and reduced the
finite-range property \eqref{e:C-finrange} to the claim that $P_t (\zeta)$ is
a polynomial in $\zeta$ of degree at most $t$.

\subsection{Chebyshev polynomials}
\label{sec:chebyshev}

We now obtain properties of $P_t$ defined in \refeq{Pdef}.  In
particular, we show that $P_t (\zeta)$ is a polynomial in $\zeta$ of
degree at most $t$.
At the end of the section, we discuss parallels with the finite speed
of propagation argument in Section~\ref{sec:frd-continuum}.
Now it is the discrete wave equation that is relevant,
as is the fact that its fundamental solution can be written in terms of Chebyshev polynomials.

Recall the definition of $P_t(\zeta)$ in \refeq{Pdef}. It involves the
function $f_t^*$. By its definition in \refeq{varphi*-defn}, $f_t^*$
is periodic with period $2\pi$.  By Poisson summation, it can be
written in terms of the continuum Fourier transform of $f$ as
\begin{equation} \label{e:varphi*-defn2}
  f^*_t(x)
  = (2\pi)^{-1} \sum_{p \in \Z} t^{-1}\hat{f}(p/t) \cos(px)
  \quad (x\in \R)
  .
\end{equation}

\begin{exercise}\label{ex:poisson-summation}
Prove \eqref{e:varphi*-defn2}.
\solref{poisson-summation}
\end{exercise}

\index{Chebyshev polynomials}
The Chebyshev polynomials $T_p$ of the first kind are the polynomials
of degree $|p|$ defined by
\begin{equation} \label{eq:Chebyshev}
  T_p(\theta) = \cos(p\arccos(\theta))
  \quad (\theta \in [-1,1],\; p \in \Z).
\end{equation}

\begin{lemma}
\label{lem:Ptpoly} For any $t>0$, when restricted to the interval
$\zeta \in [0,4]$, $P_t(\zeta)$ is a polynomial in $\zeta$, of degree
bounded by $t$.
\end{lemma}

\begin{proof}
  By \eqref{e:Pdef}, \eqref{e:varphi*-defn2}, \eqref{eq:Chebyshev}, and
  $\supp(\hat{f}) \subseteq [-1,1]$,
  \begin{align}
    P_t(\zeta)
    &= \frac{1}{2\pi} \sum_{p\in\Z} t^{-1} \hat{f}(p/t) \cos(p\arccos(1-\frac12 \zeta))
    \nnb
    &= \frac{1}{2\pi} \sum_{p \in \Z\cap[-t,t]} t^{-1} \hat{f}(p/t) T_p(1-\frac12 \zeta).
    \label{e:lem:Ptpoly}
  \end{align}
  This shows that $P_t(\zeta)$ is indeed
  the restriction of a polynomial in $\zeta$ of degree at most $t$ to the
  interval $\zeta \in [0,4]$.
\end{proof}

The following lemma provides an identity and an estimate for the
polynomial $P_t(\zeta)$.  Note that $P_t$ is constant for $t<1$, by
Lemma~\ref{lem:Ptpoly}.

\begin{lemma}
  For any $s \geq 0$, there exists $c_s>0$ such that,
  for $\zeta \in [0,4]$,
  \begin{align} \label{e:Ptle1}
    P_t(\zeta) &= \frac{\hat f(0)}{2\pi t} \quad\quad ( t < 1),
  \\
  \label{e:Pest}
    P_t(\zeta)
    &\le
    c_{s} (1+t^{2}|\zeta|)^{-s} \quad\quad ( t \ge 1).
  \end{align}
\end{lemma}

\begin{proof}
Let $\zeta \in [0,4]$ and set $x = \arccos(1-\frac12\zeta) \in
[0,\pi]$.  By \refeq{Pdef}, $P_t(\zeta) = f_t^*(x)$.

Case $t < 1$. By \eqref{e:varphi*-defn2},
\begin{equation} \label{e:varphi*-defn2-z}
  P_t(\zeta) 
  = \frac{1}{2\pi} \sum_{p \in \Z} t^{-1}\hat{f}(p/t) \cos(px)
  = \frac{1}{2\pi} t^{-1}\hat{f}(0),
\end{equation}
because the sum reduces to the single term $p=0$ by the support
property of $\hat{f}$ which implies $p\le t <1$.  This proves
\eqref{e:Ptle1}.

Case $t \ge 1$.  It suffices to consider integers $s \ge 1$.  Since
$\hat{f}$ is smooth and compactly supported, $f$ decays faster than
any inverse power, i.e., for every $s \ge 1$, $|f (x)| =
O_{s}\left(|x|^{-s} \right)$ as $|x|\rightarrow \infty$.  Therefore,
by \eqref{e:varphi*-defn}, there exist $c_{s}', c_s$ such that for $x
\in [0,\pi]$,
\begin{align}
    |P_t(\zeta)|
    &\le
    c_{s}'\sum_{n \in \Z} (1+t|x-2\pi n|)^{-4s}
    \le
    c_{s}'\sum_{n \in \Z} (1+tx+t\pi |n|)^{-4s}
    \nnb
    &\le
    c_{s}'(1+tx)^{-2s} \sum_{n \in \Z} (1+t\pi |n|)^{-2s}
    \le c_s(1+tx)^{-2s},
\end{align}
since the last sum converges.  Since $\zeta = 4 \sin^{2} (\frac{x}{2})
\le x^{2}$, we have $x \ge \sqrt{\zeta}$.  Therefore,
\begin{align}
    |P_t(\zeta)|
    &
    \le c_s(1+t\sqrt{\zeta})^{- 2s} \le c_s(1+t^2 \zeta)^{-s},
\end{align}
and the proof is complete.
\end{proof}

\index{Wave equation} \index{Finite propagation speed}
Lemma~\ref{lem:Ptpoly} can be understood as a consequence of the
finite propagation speed of the discrete wave equation
\begin{equation} \label{e:discretewave}
  u_{p+1}+u_{p-1}-2u_p = -\zeta u_p,
  \quad u_0 \text{ given},
  \quad u_{1}-u_{-1} = 0,
\end{equation}
which is analogous to \eqref{e:waveequation}, with derivatives in $s$
replaced by discrete derivatives in $p$ and with $-\Delta$ replaced by
$\zeta$.  Its solution is given by
\begin{equation} \label{e:discretewavesoln}
  u_p = T_p(1-\frac12 \zeta) u_0 .
\end{equation}
The Chebyshev polynomials $T_p$ satisfy the recursion relation
$T_{p+1}(\theta)+T_{p-1}(\theta)-2\theta T_p(\theta)=0$ so that
\eqref{e:discretewavesoln} solves \eqref{e:discretewave}.

The equation \eqref{e:lem:Ptpoly} is analogous to
\eqref{e:decomp-wave3} with the continuum wave operator
$\cos(\sqrt{-\Delta}s)$ replaced by the fundamental solution
$T_p(1+\frac12\Delta)$ to the discrete wave equation.

\subsection{Proof of Proposition~\ref{prop:decomp}}
\label{sec:decomp-pf}

To complete the proof of Proposition~\ref{prop:decomp}, the main
remaining step is to obtain estimates on the function $w(t,x)$ defined
in \refeq{wxspace}.  The next lemma provides the required estimates.

\begin{lemma}
\label{lem:wtbds}
Fix any dimension $d>0$.
For any $x \in \Z^d$, any multi-index $\alpha$, and any $s \geq 0$,
there exists $c_{s,\alpha} \geq 0$ such that
  \begin{align}
  \label{e:wtle1}
    w(t,x) &= \frac{t}{2d+m^2} \frac{\hat{f}(0)}{2\pi}  \1_{x=0} \quad (t<1),
  \\
  \label{e:wbd}
    |\nabla^\alpha w(t,x)|
    &\leq  c_{s,\alpha}  \vartheta(t,m^2;s) t^{-(d-2 + |\alpha|_1)}
    \quad (t \ge 1)
    .
  \end{align}
\end{lemma}

\begin{proof}
Case $t < 1$.
By \eqref{e:w_hat} and \eqref{e:Ptle1},
\begin{align}
  w(t,x)
  &=
  (2\pi)^{-d} \int_{[-\pi,\pi]^{d}}
  \frac{t^2}{M^2} P_t\left(\frac{1}{M^2} (\lambda(k)+m^2)\right)\, e^{ik\cdot x}\,dk
  \nnb & =
  \frac{t}{M^2}
  \frac{1}{2\pi} \hat{f}(0)
  \1_{x=0} .
\end{align}
This proves \eqref{e:wtle1}.

Case $t \geq 1$.
Recall from \eqref{e:varthdef} that
\begin{equation}
  \vartheta(t,m^2;s) = \frac{1}{M^2} \left(1+\frac{m^2t^2}{M^2} \right)^{-s}.
\end{equation}
By definition,
\begin{equation}
    \nabla^\alpha w(t,x) =
    \frac{t^2}{M^2}    \frac{1}{(2\pi)^d} \int_{[-\pi,\pi]^d}
      P_t\left(\frac{1}{M^2} (\lambda(k)+m^2)\right)
       \nabla^\alpha e^{ik\cdot x} dk.
\end{equation}
We use $|\nabla^\alpha e^{ik\cdot x}| \le C_{\alpha}|k|^{|\alpha|_1}$,
and apply \refeq{Pest} with $s=s'+s''$ to obtain
\begin{equation}
    \left| P_t\left(\frac{1}{M^2} (\lambda(k)+m^2)\right) \right|
    \le O(1+ t^2\lambda(k)/M^2)^{-s'} (1+tm^2/M^2)^{-s''}.
\end{equation}
  Elementary calculus shows that $\lambda(k) \asymp |k|^2$ for $k \in [-\pi,\pi]^d$.
  Also, with $s'$ chosen larger than $|\alpha_1|+d/2$, we have
  \begin{align}
    \int_{[-\pi,\pi]^d}
    \frac{|k|^{|\alpha|_1}}{(1+ t^2|k|^2/M^2)^{s'}}  dk
    & =
    O\left( \left( M/t \right)^{d+|\alpha|_1} \wedge  \pi^{|\alpha |_{1}}   \right),
  \end{align}
  where the first option on the right-hand side arises from extending the domain
  of integration to $\R^d$ and making the change of variables $k \mapsto (M/t)k$,
  and the second arises by bounding the integrand by $\pi^{|\alpha |_{1}}$.
  With the choice $s''=s+(d+|\alpha|_1)/2$, it follows that
  \begin{align}
    |\nabla^\alpha| w(t,x)|
    &\leq
    O(1+t^2m^2/M^2)^{-s-(d+|\alpha|_1)/2}
    O_{\alpha }\left( (t/M)^{2-d-|\alpha|_1} \wedge (t/M)^2 \right)
      .
  \end{align}
For $m^2 \leq 1$, we have $M^2 \asymp 1$ and \refeq{wbd} follows
immediately by choosing the first option in the minimum on the
right-hand side.  For $m^2 \geq 1$, we have instead $M^2 \asymp m^2$,
and by choosing the second option in the minimum we now obtain
\begin{equation}
  |\nabla^\alpha w(t,x)|
  = O_{\alpha}\left( (1+t^2)^{-s-(d+|\alpha|_1)/2} \frac{t^2}{m^2}\right)
  = O_{\alpha}\left(\frac{1}{m^2} (1+t^2)^{-s}   t^{2-d-|\alpha|_1}  \right)
  .
\end{equation}
This completes the proof for $t \geq 1$.
\end{proof}

\begin{proof}[Proof of Proposition~\ref{prop:decomp}] As in
\refeq{C1x}--\refeq{Cjx}, we define
\begin{align}
\label{e:C1x-4}
  C_{1;0x} &= \int_{0}^{\frac12 L} w(t,x) \, \frac{dt}{t},
\\
\label{e:Cjx-4}
  C_{j;0x} &= \int_{\frac12 L^{j-1}}^{\frac12 L^j} w(t,x) \, \frac{dt}{t} \quad (j \ge 2).
\end{align}
By \eqref{e:decompintpf}, this gives
\begin{equation}
  \frac{1}{\lambda(k)+m^2} =  \sum_{j=1}^\infty \hat C_j(k),
\end{equation}
where $\hat C_j \geq 0$ is the discrete Fourier transform of $C_j$.
Thus, for any $f \in \ell^2(\Z^d)$,
\begin{equation}
    (f,(-\Delta + m^2)^{-1}f) = \sum_{j=1}^\infty (f,C_jf),
\end{equation}
which proves \refeq{ZdCj9}.  Continuity of $C_{j;0x}$ in the mass
$m^2$ can be seen via an application of the dominated convergence
theorem to the integrals \refeq{C1x-4}--\refeq{Cjx-4}.  Since $w(t,x)
= 0$ for $|x|_1 > t$ (as pointed out below \refeq{qf}), $C_j$ has the
finite-range property \eqref{e:C-finrange}.

It remains to prove \refeq{C-est}, which we restate here as
  \begin{equation} \label{e:C-est-pf}
    |\nabla^{\alpha} C_{j;xy}|
    \leq c_{\alpha,s}
    f_d(L)
    \vartheta_{j-1}(m^2;s)
    L^{-(d-2 + |\alpha|_1)(j-1)}
    ,
  \end{equation}
with $f_d(L)=1$ for $d>2$, $f_2(L)=\log L$, and $f_d(L)=L^{2-d}$ for
$d<2$, and with
\begin{align}
\lbeq{vartheta-recall}
    \vartheta_{j-1}(m^2;s)
    =
    \frac{1}{2d+m^2} \left(1+\frac{L^{2(j-1)}m^2}{(2d+m^2)} \right)^{-s}
    .
\end{align}
By \eqref{e:wbd} and the change of variables $\tau=L^{j-1}t$,
\begin{align}
    \Big|\nabla^{\alpha} \int_{\frac 12 L^{j-1}}^{\frac 12 L^j} w (t,x-y) \, \frac{dt}{t}\Big|
    & \leq
    c \int_{\frac 12 L^{j-1}}^{\frac 12 L^j}
    \vartheta(t,m^2;s) t^{-(d-2 + |\alpha|_1)}
    \, \frac{dt}{t}
    \\ \nonumber
    & \le
    c' \vartheta_{j-1}(m^2;s) L^{-(j-1)(d-2 + |\alpha|_1)}
    \int_{\frac 12}^{\frac 12 L}
    \tau^{-(d-1 + |\alpha|_1)}
    \,  d\tau
    ,
\end{align}
where the constants can depend on $\alpha ,s$.
The $\tau$ integral is bounded by $f_d(L)$ in the worst case
$\alpha=0$. By \eqref{e:Cjx-4} the left-hand side equals
$|\nabla^{\alpha}C_{j;xy}|$ for $j \ge 2$, which proves the desired
bound for $j \ge 2$.

For $j=1$ the left-hand side is not equal to $C_{1;xy}$ because the
lower bound on the $t$ integral is $1$ instead of zero.  The above
argument does provide the desired estimate on the contribution to
$C_1$ due to integration over $[1,\frac 12 L]$.  The remaining
contribution to $C_1$ is $C_0$ defined in \refeq{C0def}, i.e.,
\begin{equation}
    C_{0;0x} = \int_0^1 w(t,x) \, \frac{dt}{t}.
\end{equation}
According to \eqref{e:wtle1},
  \begin{align}
  \label{e:wtle1-C0}
    w(t,x) &= \frac{t}{2d+m^2} \frac{\hat{f}(0)}{2\pi}  \1_{x=0} \quad (t<1),
  \end{align}
and therefore, as claimed in \refeq{C_0},
\begin{equation}
    C_{0;0x} =   \frac{1}{2d+m^2} \frac{\hat{f}(0)}{2\pi}  \1_{x=0}.
\end{equation}
This contribution to $C_1$ also obeys \eqref{e:C-est-pf} with $j=1$.  Indeed,
since $\vartheta_0(m^2,s) \ge 2^{-s}(2d+m^2)^{-1}$, we have
\begin{align}
    |\nabla^\alpha C_{0;0x}|
    & \le 2^s \vartheta_0(m^2;s) \frac{\hat{f}(0)}{2\pi}  |\nabla^\alpha \1_{x=0}|.
\end{align}
This completes the proof.
\end{proof}

\section{Finite-range decomposition: torus}
\label{sec:frd-torus}

For $L>1$, $N\ge 1$, $m^2>0$, and $d >0$,
let $\Lambda_N = \Z^d/L^N\Z^d$ be the $d$-dimensional discrete
torus of period $L^N$.
Define
\begin{equation}
\lbeq{341}
     C_{N,j;x,y} = \sum_{z \in \Z^d} C_{j;x,y+zL^N} \quad (j<N).
\end{equation}
We also define
\begin{equation} \label{e:CNNdef}
     C_{N,N;x,y} = \sum_{z \in \Z^d} \sum_{j=N}^\infty C_{j;x,y+zL^N}.
\end{equation}
Since
\begin{equation}
  (-\Delta_\Lambda + m^2)^{-1}_{x,y} = \sum_{z \in \Z^d} (-\Delta_{\Z^d}+m^2)^{-1}_{x,y+zL^N},
\end{equation}
it follows from Proposition~\ref{prop:decomp} that
\begin{equation}
\lbeq{torus-frd}
    (-\Delta_\Lambda + m^2)^{-1}
    = \sum_{j=1}^{N-1} C_{N,j} + C_{N,N}.
\end{equation}
In this finite-range decomposition of the torus covariance, the dependence of $C_{N,j}$ on $N$
is concentrated in the term $C_{N,N}$ in the following sense: by the finite range property
\eqref{e:C-finrange}, for a given $x,y$ and $j<N$, at most one term in the sum over $z$ in \refeq{341} contributes;
another way to say this is that the Gaussian process with covariance $C_{N,j}$ restricted to a subset of
the torus with diameter less than $L^N/2$ is in distribution equal to the Gaussian field
on $\mathbb{Z}^d$ with covariance $C_j$.
Estimates on $C_{N,N}$ can be derived from Proposition~\ref{prop:decomp}.

The following is an immediate consequence of \refeq{torus-frd}.

\begin{cor} \label{cor:varphi-decomp}
  Let $N \ge 1$, and let $\varphi$ be the GFF with mass $m>0$ on $\Lambda_N$.
  There exist independent  Gaussian fields
  $\zeta_j$ $(j=1,\dots, N)$, such that $\zeta_j = (\zeta_{j,x})_{x\in \Lambda_N}$
  are finite range with range $\frac12 L^j$ and
  \begin{equation} \label{e:varphi-decomp}
    \varphi \stackrel{D}{=} \zeta_1 + \cdots + \zeta_N.
  \end{equation}
\end{cor}

\begin{proof}
This follows from \refeq{torus-frd} and
Exercise~\ref{ex:uncorr-then-indep-Gauss}.
\end{proof}



%% file: hier.tex
\chapter{The hierarchical model}
\label{ch:hier}

In Section~\ref{sec:hGFF}, we define a \emph{hierarchical Gaussian
field} as a field that satisfies a strengthened version of the
finite-range decomposition of Chapter~\ref{ch:decomp}.
The hierarchical Gaussian free field (hGFF) is a hierarchical field
that has comparable large distance behaviour to the lattice Gaussian free
field.  We explicitly construct a version of it and verify that it
indeed has the desired properties.  In Section~\ref{sec:hierGFF}, we
define the hierarchical $|\varphi|^4$ model, and in
Theorem~\ref{thm:phi4-hier-chi} state the counterpart of the
asymptotic formula \eqref{e:thmphi4-chi} for the hierarchical model's
susceptibility.  In Section~\ref{sec:Gpert}, we reformulate the
hierarchical $|\varphi|^4$ model as a perturbation of a Gaussian
integral, in preparation for its renormalisation group analysis.

\section{Hierarchical GFF}
\label{sec:hGFF}

\subsection{Hierarchical fields}
\label{sec:hierarchical-fields}

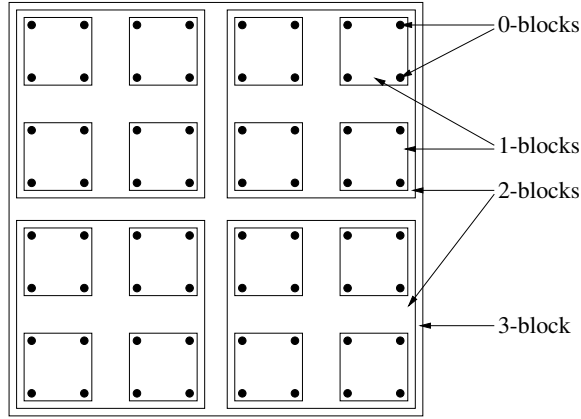
\begin{figure}
\begin{center}
\input{hier1.pspdftex}
\end{center}
\caption{\label{fig:reblock} Blocks in ${\cal B}_j$ for
$j=0,1,2,3$ when $d=2$, $N=3$, $L=2$.}
\label{fig:RG_hierarchy1}
\end{figure}

Periodic boundary conditions are not appropriate for
hierarchical fields.
Throughout our discussion and analysis of the hierarchical field,
$\Lambda_N$ is the hypercube $[0,L^N-1]\times\cdots\times [0,L^N-1]
\subset \Z^d$, with $L>1$ fixed.
As illustrated in Figure~\ref{fig:reblock}, we partition $\Lambda_N$
into disjoint blocks of side length $L^j$, with $0 \leq j \leq N$.

\begin{defn}
\label{def:Bcal}
  For $0\leq j \leq N$, $\Bcal_j$ is the set of disjoint blocks
  $B$ of side length $L^j$ (number of vertices)
  such that $\Lambda_N = \cup_{B \in \Bcal_j} B$.
  An element $B \in \Bcal_j$ is called a \emph{block}, or $j$-\emph{block}.
  We say that two $j$-blocks $B,B'$ \emph{do not touch} if any pair of
  vertices $(x,x') \in B \times B'$ has $|x-x'|_\infty > 1$.
\end{defn}

\index{Block}

The sets $\Bcal_j$ are nested, in the sense that for
every $j$-block $B \in \Bcal_j$ and $k>j$, there is a unique $k$-block
$B'\in \Bcal_k$ such that $B \subset B'$.

By Proposition~\ref{prop:decomp}, the Gaussian fields $\zeta_j$ in the
finite-range decomposition of Corollary~\ref{cor:varphi-decomp} have
the following two properties:
\begin{enumerate}[(i)]
\item
  Given two blocks $B,B' \in \Bcal_j$ that \emph{do not touch},
  $\zeta_j|_B$ and $\zeta_j|_{B'}$ are independent
  identically distributed Gaussian fields.
\item
  Given any block $b \in \Bcal_{j-1}$, the field $\zeta_j|_{b}$ is
  \emph{approximately constant} in the sense that
  the gradient of the covariance obeys an upper bound that is smaller
  by a factor $L^{-(j-1)}$ than the upper bound for the covariance itself.
\end{enumerate}

A \emph{hierarchical} field is a Gaussian field on $\Lambda_N$ with a
decomposition $\varphi=\zeta_1 + \cdots + \zeta_N$ in which the two
properties (i) and (ii) above are replaced by the following stronger
versions (i') and (ii').

\begin{defn}
  \label{def:hierarchical}
  A Gaussian field $\varphi$ on $\Lambda_N$ is \emph{hierarchical} if there
  exist independent Gaussian fields $\zeta_1, \dots, \zeta_N$ on $\Lambda_N$,
  called \emph{the fluctuation fields},
  such that
  \begin{equation}
    \varphi \stackrel{D}{=} \zeta_1 + \dots +\zeta_N,
  \end{equation}
  where the fields $\zeta_j$ obey:
\begin{enumerate}[(i')]
\item
  Given two blocks $B,B' \in \Bcal_j$ that are \emph{not identical},
  $\zeta_j|_B$ and $\zeta_j|_{B'}$ are independent
  identically distributed Gaussian fields.
\item
  Given any block $b \in \Bcal_{j-1}$, the field $\zeta_j|_{b}$ is
  \emph{constant}: $\zeta_{j,x}=\zeta_{j,y}$ almost surely for all $x,y \in b$.
\end{enumerate}
\end{defn}

The replacement of (i--ii) by (i'--ii') is a major technical
simplification for the study of the renormalisation
group.
Condition (ii') means that when $x,y$ are in the same block,
$\zeta_{x}-\zeta_{y}$ has zero variance and therefore the covariance
of $\zeta$ is not positive definite; we have allowed for this in
Definition~\ref{def:Gm}. The condition \eqref{e:hierarchical-C} that
appears below implies other linear combinations also have zero
variance.

\begin{exercise} \label{ex:hier-field-tree-repr}
  The nesting of blocks can be represented as a rooted tree, in which the root is given
  by the unique block $\Lambda_N \in \Bcal_N$, the blocks
  $B \in \Bcal_j$ are the vertices at distance $N-j$ to the root,
  and the children of $B \in \Bcal_{j}$
  are the $b \in \Bcal_{j-1}$ with $b \subset B$.
  Represent the hierarchical field 
  in terms of independent Gaussian variables
  associated to the edges of the tree.
  \solref{hier-field-tree-repr}
\end{exercise}

\index{Block spin}
\begin{rk}
  The \emph{finite-range} decomposition
  of Corollary~\ref{cor:varphi-decomp}
  is a representation of the GFF in which property~(i)
  is as close to its hierarchical version~(i') as possible.
  The price is that property~(ii')
  needs to be weakened to~(ii).
  There is an alternative decomposition of the GFF
  such that property~(i) is replaced by dependence that decays
  exponentially with distance, and property~(ii) holds.
  In this alternate decomposition, known as the \emph{block spin} decomposition,
  $\zeta_{j}$ has the hierarchical features that it
  is a function of $L^{d}$ independent Gaussian fields per block, subject to
  a zero-sum rule as in \eqref{e:hierarchical-C} below.
  The block spin decomposition was used, e.g., in \cite{GK80,GK85,Hara87,HT87}.
\end{rk}

\subsection{Construction of hierarchical GFF}

The hierarchical GFF is defined in terms of the hierarchical
Laplacian, which is itself defined in terms of certain projections.
We start with the projections.

Let $d \geq 2$.  Given a scale $j=0,1,\ldots,N$ and $x \in \Lambda$,
we write $B_x$ for the unique $j$-block that contains $x$.  Then we
define the matrices of symmetric operators $Q_{j}$ and $P_j$, acting
on $\ell^{2} (\Lambda)$, by
\begin{align}
    \label{e:Qj-def}
    Q_{j;xy}
    &=
    \begin{cases}
    L^{-dj} & B_x=B_y \\
    0 & B_x \neq B_y
    \end{cases}
    \quad
    (j=0,1,\ldots,N),
\\
    \label{e:Pj-def}
    P_{j}
    &=
    Q_{j-1} - Q_{j}
    \quad
    (j = 1,\dots ,N).
\end{align}

\begin{lemma}
\label{lem:PQproj} The operators $P_{1},\dots ,P_{N}, Q_{N}$ are
orthogonal projections whose ranges are disjoint and provide a direct
sum decomposition of $\ell^2(\Lambda)$:
\begin{gather}
    \label{e:spectral-projections}
    P_{j}P_{k}
    =
    P_{k}P_{j}
    =
    \begin{cases}
    P_{j}& (j=k)\\
    0   & (j\neq k),
    \end{cases}
    \qquad\quad
    \sum_{j=1}^{N} P_{j} +Q_{N} = \Id .
\end{gather}
\end{lemma}

\begin{proof}
The second equation is an immediate consequence of the definition
\eqref{e:Pj-def} of $P_{j}$, together with the fact that $Q_0=\Id$.
For the other properties, we claim that
\begin{equation}
    \label{e:Q-props}
    Q_{j}Q_{k}
    =
    Q_{j\vee k}
    =
    Q_{k}Q_{j}
 .
\end{equation}
In particular, the case $j=k$ shows that $Q_{j}$ is an orthogonal
projection. To prove \eqref{e:Q-props}, it suffices to consider
$j \le k$.  We use primes to denote blocks in the larger scale
$\Bcal_{k}$, and unprimed blocks are in $\Bcal_{j}$.  Then the $x,y$
matrix element of the product is given by
\begin{align}
    \sum_{z}Q_{j;xz}Q_{k;zy}
    &=
    L^{-d (j+k)}\sum_{z}\1_{B_{x} = B_{z}}\1_{B'_{z} = B'_{y}}
    =
    L^{-d (j+k)}\sum_{z \in B_{x}}\1_{B_{x} \subset B'_{y}}
    \nonumber\\
    &=
    L^{-d k}\1_{B_{x} \subset B'_{y}}
    =
    L^{-d k}\1_{B'_{x} = B'_{y}}
    =
    Q_{k;xy} ,
\end{align}
as claimed.  Thus $\{Q_{j} \}_{j=0,\dots ,N}$ is a sequence of
commuting decreasing projections that starts with $Q_{0} = \Id$.  By
\eqref{e:Q-props} it readily follows that $P_{1},\dots ,P_{N}, Q_{N}$
are orthogonal projections that obey \eqref{e:spectral-projections}.
\end{proof}

The next exercise identifies the subspaces in the direct sum
decomposition given in Lemma~\ref{lem:PQproj}.

\begin{exercise}
\label{ex:PQproj} For $j=0,\ldots,N$, let $X_j$ denote the subspace of
$\ell^2(\Lambda)$ consisting of vectors that are constant on blocks in
$\Bcal_j$, so $X_0 = \ell^2(\Lambda) \supset X_1 \supset \cdots
\supset X_N = {\rm span} (1,\ldots,1)$.  For $j=0,\ldots,N$, show that
the range of the projection $Q_j$ is $X_j$.  For $j=1,\ldots,N$, show
that the range of the projection $P_j$ is the orthogonal complement of
$X_{j}$ in $X_{j-1}$, i.e., the set of vectors constant on
$(j-1)$-blocks whose restriction to any $j$-block has zero sum.
\solref{PQproj}
\end{exercise}

\index{Laplacian}
\index{Hierarchical Laplacian}
\begin{defn} \label{e:defn:hGFF}
  The \emph{hierarchical Laplacian} $\Delta_{H,N}$ is the operator on
  $\ell^{2} (\Lambda)$ given by
\begin{equation}
    \label{e:hierarchical-laplacian}
    - \Delta_{H,N} = \sum_{j=1}^{N}L^{-2 (j-1)}P_{j} .
\end{equation}
\end{defn}

The hierarchical Laplacian generates a certain hierarchical random
walk; this point of view is developed in the next exercise (see also
\cite{BEI92}).  Its decay properties mirror those of the Laplacian on
$\Zd$, and this fact is established in
Exercise~\ref{ex:hier-cov-asym}.

\begin{exercise}
\label{ex:hier-rw} Let $j_x$ be the smallest $j$ such that $0$ and $x$
are in the same $j$-block; we call $j_x$ the \emph{coalescence scale}
for the points $0,x$.  Show that
\begin{align}
\lbeq{DelH0x}
    \Delta_{H,N;0x} & =
    \begin{cases}
    -\frac{1-L^{-d}}{1-L^{-(d+2)}} (1-L^{-(d+2)N})& (x=0)
    \\
    \frac{L^2-1}{1-L^{-(d+2)}} L^{-(d+2)j_x} +
    \frac{1-L^{-d}}{1-L^{-(d+2)}} L^{-(d+2)N}
    & (x\neq 0).
    \end{cases}
\end{align}
In particular, $\Delta_{H,N;00}<0$ and, for $x \neq 0$,
$\Delta_{H,N:0x} >0$.  Show also that $\sum_{x\in \Lambda}
\Delta_{H,N;0,x} = 0$.  This implies that $\Delta_{H,N}$ is the
infinitesimal generator (also called a $Q$-matrix \cite{Norr97}) of a
continuous-time random walk.  What steps does it take?
\solref{hier-rw}
\end{exercise}

Given $m^2>0$, we set
\begin{equation} \label{e:gammadef}
  \gamma_{j}
  = \frac{L^{2(j-1)}}{1+L^{2(j-1)}m^2}
\end{equation}
and for $j=1,\ldots,N$ define matrices
\begin{equation}
  \label{e:Cjhier}
  C_{j;xy}(m^2)
  =
  \gamma_{j}P_{j;xy},
  \qquad
  \Cnewlast{;xy}(m^2)
  =
  \frac{1}{m^2} Q_{N;xy}.
\end{equation}
It follows from Lemma~\ref{lem:PQproj} that
\begin{equation}
\lbeq{CC0}
    C_jC_k = 0 \quad (j \neq k).
\end{equation}
Note that $C_{j;xy} (m^{2})$ is actually well defined for all $m^2 \ge
0$, and is independent of $N$ in the sense that the $C_j$ defined in
terms of any $N \ge j$ are naturally identified.  In contrast,
$\Cnewlast{;xy}(m^2)$ is not defined for $m^2=0$ and does depend on
$N$. In fact, $C_{\hat{N};xy}(m^2)=m^{-2}L^{-dN}$ for all $x,y\in
\Lambda$ because $\Lambda$ is a single block at scale $N$.

The special role of $\Cnewlast{}$ is analogous the the situation for
the Euclidean torus decomposition $(-\Delta_\Lambda + m^2)^{-1} =
\sum_{j=1}^{N-1}C_j + C_{N,N}$ of \refeq{torus-frd}.  There the term
$C_{N,N}$ is special as it is the term that takes the finite-volume
torus into account.  Similarly, in the hierarchical setting we isolate
the finite-volume effect by writing the decomposition in the form
$\sum_{j=1}^N C_j +\Cnewlast{}$, with $\zetanewlast{}$ the field that
takes the finite volume into account.

\index{Zero-sum condition}
\begin{prop}
\label{prop:hGFF}
  For $m^2>0$, $x,y \in \Lambda_{N}$
  and $j=0,\dots ,N-1$,
  \begin{align}\label{}
    \label{e:scaling-estimate-hier}
    C_{j+1;xx} (m^{2})
    &=
    \frac{1}{1+m^2L^{2j}}L^{-(d-2)j} (1-L^{-d}),
    \\
    \label{e:hierarchical-C}
    \sum_x C_{j+1;0x}
    &=
    0 ,
    \\
  \label{e:scaling-estimateN-hier}
    \Cnewlast{;xy}
    (m^{2})
    & = L^{-dN}m^{-2}.
  \end{align}
    The  matrix
  $C = (m^{2}-\Delta_{H,N})^{-1}$ has the decomposition
  \begin{equation}
    \label{e:hierarchical-decomp}
    C
    = C_1 + \cdots C_{N} + \Cnewlast{}.
  \end{equation}
  Let $\zeta_j$ be independent fields, Gaussian with covariance $C_j$.
 Then the field
  $\varphi = \zeta_1 + \dots +\zeta_N + \zetanewlast{}$
  is a hierarchical field
  as in Definition~\ref{def:hierarchical}.
\end{prop}

\begin{proof}
The variance
statement \eqref{e:scaling-estimate-hier} is immediate by setting
$x=y$ in the definition \eqref{e:Pj-def} of $P_{j}$.  The identity
\eqref{e:hierarchical-C} follows from
\eqref{e:Qj-def}--\eqref{e:Pj-def}, since
\begin{equation}
\lbeq{Psum0}
    \sum_{x\in \Lambda} P_{j+1;0x}
    =
    \sum_{x\in \Lambda} Q_{j;0x} - \sum_{x\in \Lambda} Q_{j+1;0x}
    =
    L^{dj}L^{-dj} - L^{d(j+1)}L^{-d(j+1)} = 0.
\end{equation}
Also, \refeq{scaling-estimateN-hier} follows from the
definition \eqref{e:Cjhier} of $\Cnewlast{}$.  

The decomposition statement \eqref{e:hierarchical-decomp} and
$C=(-\Delta_{H,N}+m^2)^{-1}$ follow from the independence of the
fields in the decomposition $\varphi = \zeta_1 + \dots +\zeta_N +
\zetanewlast{}$ and from \eqref{e:spectral-projections} which together
with \eqref{e:hierarchical-laplacian} shows that
$P_{1},..,P_{N},Q_{N}$ are spectral projections for
$-\Delta_{H,N}$. In fact, let $f (t) = (t+m^{2})^{-1}$ and
$\lambda_{j} = L^{- 2 (j-1)}$. By the spectral calculus,
\begin{align}
    (-\Delta_{H,N}+m^{2})^{-1}
    &=
    f (-\Delta_{H,N})
    =
    f \left(\sum_{j=1}^{N} \lambda_{j}P_{j} + 0 \;Q_{N}\right)
    \nonumber\\
    &=
    \sum_{j=1}^{N} f (\lambda_{j})P_{j} + f (0)Q_{N}
    \nonumber\\
    &=
    \sum_{j=1}^{N} C_{j} (m^{2}) + \Cnewlast{} (m^{2}) ,
\end{align}
because $f (\lambda_{j}) = \gamma_{j}$ and $f (0) = m^{-2}$.

The independence required by Definition~\ref{def:hierarchical}(i')
holds by construction, and (ii') follows from the easily checked fact
that $\Var_j(\zeta_{j;x}-\zeta_{j;y})=0$ if $x,y$ both lie in the same
block $b \in \Bcal_{j-1}$.  This completes the proof.
\end{proof}

Although we have the explicit formulas \eqref{e:Cjhier} and
\eqref{e:hierarchical-laplacian} for the covariances $C_{j}$ and for
$\Delta_{H,N}$, for our purposes these explicit formulas are not very
important because almost everything in the following chapters uses
only the properties listed in Proposition~\ref{prop:hGFF} and
Definition~\ref{def:hierarchical}.  However, to be concrete, we call
the particular random field $\varphi$ defined by these explicit
formulas the \emph{hierarchical Gaussian free field (hGFF)}.  The
justification for this terminology is that
\eqref{e:scaling-estimate-hier} has the same scaling as its
counterpart for the Gaussian free field, according to \eqref{e:C-est}.
Similarly, $\Delta_{H,N}$ has properties in common with the standard
lattice Laplacian.  Note that we use $C$ for both the hierarchical and
usual covariances.  It should be clear from context which is intended.

Equation~\eqref{e:hierarchical-C} holds both for block spins and for
the hierarchical model, and this leads to simplifications in
perturbation theory.  However, it does not hold for the Euclidean
model with finite-range decomposition, and perturbation theory is
therefore more involved \cite{BBS-rg-pt}. Not all authors include the
$\Delta_{H,N}$ properties or property \eqref{e:hierarchical-C} when
defining massless hierarchical fields.

\subsection{Properties of hierarchical covariances}

\begin{exercise} \label{ex:hier-freechi}
  Show that $\sum_{x \in \Lambda} C_{0x}(m^2) = m^{-2}$.
  (Cf. Exercise~\ref{ex:Greenf}.)
\solref{hier-freechi}
\end{exercise}

By \refeq{Cjhier}, the hierarchical covariance is given, for $j<N$ and
for $x,y$ in the same $(j+1)$-block, by
\begin{equation}
\lbeq{hiercov}
    C_{j+1;xy}(m^2) =
    \begin{cases}
    L^{-(d-2)j} (1+m^2L^{2j})^{-1} (1-L^{-d}) & (b_x=b_y)
    \\
    -L^{-(d-2)j} (1+m^2L^{2j})^{-1} L^{-d} & (b_x\neq b_y),
    \end{cases}
\end{equation}
where $b_x$ denotes the $j$-block containing $x$; if $x,y$ are not in
the same $(j+1)$-block then $C_{j+1;xy}(m^2)=0$.  We write the
diagonal entry as
\begin{equation}
\lbeq{cjdef}
c_j = C_{j+1;00}(m^2)
= L^{-(d-2)j}(1+m^2L^{2j})^{-1} (1-L^{-d}),
\end{equation}
and for $n\in \N$ define
\begin{equation}
\lbeq{cjndef}
    c_j^{(n)} = \sum_{x\in \Lambda} (C_{j+1;0x}(m^2))^n.
\end{equation}
The fact that $C_{j+1}$ is positive definite and translation invariant
implies that $c_j^{(1)} \ge 0$.  For our specific choice of $C_{j+1}$,
if follows from \refeq{hierarchical-C} (or directly from
\refeq{hiercov}) that $c_j^{(1)}=0$.

\begin{exercise}
\label{ex:cjns}
Use \refeq{hiercov} to show that, for $j<N$,
\begin{align}
    c_{j}^{(2)} 
    &= L^{-(d-4)j} (1+m^2L^{2j})^{-2}(1-L^{-d}),
    \\
    c_{j}^{(3)} 
    & = L^{-(2d-6)j}(1+m^2L^{2j})^{-3} (1-3L^{-d}-2L^{-2d}),
    \\
    c_{j}^{(4)} 
    & =
    L^{-(3d-8)j} (1+m^2L^{2j})^{-4} \big(1 - 4L^{-d} + 6L^{-2d} - 3 L^{-3d}\big)
    .
\end{align}
\solref{cjns}
\end{exercise}

\index{Bubble diagram}
\begin{exercise} \label{ex:bubble} \index{Bubble diagram}
Recall the bubble diagram $B_{m^2} = \sum_{x\in\Z^d}
((-\Delta+m^2)_{0x}^{-1})^2$ defined in \refeq{bubdef}.  The infinite-volume
hierarchical bubble diagram is defined by
\begin{equation}
    B^{H}_{m^2}
    =
    \lim_{N \to \infty}\sum_{x\in \Lambda_N}((-\Delta_{H,N}+m^2)_{0x}^{-1})^2,
\end{equation}
where $\Delta_{H,N}$ is the hierarchical Laplacian on $\Lambda_N$.
Prove that, for $m^2 \ge 0$,
\begin{equation}
    B^{H}_{m^2} = \sum_{j=0}^\infty c_j^{(2)} ,
\end{equation}
with $c^{(2)}_j$ given by \refeq{cjndef}.  In particular,
$B^{H}_{m^2}$ is finite in all dimensions for $m^2>0$, whereas
$B^{H}_0$ is finite if and only if $d>4$.  Prove that, as $m^2
\downarrow 0$,
  \begin{equation}
  \lbeq{hbubble}
    B^{H}_{m^2} \sim
    \begin{cases}
    {\rm const}\; m^{-(4-d)} & (d < 4)
    \\
    \frac{1-L^{-d}}{\log L}\log m^{-1} & (d=4).
    \end{cases}
  \end{equation}
\solref{bubble}
\end{exercise}

The asymptotic behaviour for the hierarchical bubble in
\refeq{hbubble} is analogous to that of Exercise~\ref{ex:bubble1z} for
the bubble diagram of the GFF.  Another correspondence between the
hGFF and the GFF is that in the critical case $m^2=0$ in the
infinite-volume limit, the covariance of the hGFF has the same
large-$|x|$ decay as the GFF.  This is shown in the following
exercise.

\begin{exercise} \label{ex:hier-cov-asym}
(i) Verify that
\begin{equation}
    C(m^2) = \gamma_1Q_0 + \sum_{j=1}^{N-1} (\gamma_{j+1}-\gamma_j)Q_j
    + (m^{-2}-\gamma_N) Q_N.
\end{equation}
(ii)
Using the result of part~(i), prove that
as $|x| \to \infty$ the hierarchical
covariance obeys
\begin{equation} \label{e:GFFdecay}
  \lim_{m^2\downarrow 0} \lim_{N\to\infty} [C_{0x}(m^2) - C_{00}(m^2)]
  \begin{cases}
  \asymp -|x| & (d=1)
  \\
  =  -(1-L^{-2}) \log_L |x|  
  + O(1) & (d = 2)
\end{cases}
\end{equation}
and
\begin{equation} \label{e:GFFdecay2}
  \lim_{m^2\downarrow 0} \lim_{N\to\infty} C_{0x}(m^2)
  \asymp |x|^{-(d-2)} \quad (d>2).
\end{equation}
\solref{hier-cov-asym}
\end{exercise}

On the other hand, the effect of the mass $m>0$ is not as strong for
the hierarchical covariance as it is for the Euclidean one.  The
Euclidean covariance with mass $m$ decays exponentially with rate
$\sim m$ as $m \downarrow 0$, while the hierarchical covariance decays
only polynomially in $m|x|$. This results from the fact that
$-\Delta_H$ is not local; its matrix elements decay only polynomially.

\section{Hierarchical \texorpdfstring{$|\varphi|^{4}$}{phi4} model}
\label{sec:hierGFF}
\index{Hierarchical approximation}
\index{Hierarchical model}

Recall from Section~\ref{sec:phi4-defn}
that the $n$-component $|\varphi|^4$ model
on a set $\Lambda$ is defined by the expectation
\begin{equation}
\label{e:phi4-exp-new-again}
  \la F  \ra_{g,\nu,\Lambda}
  =
  \frac{1}{Z_{g,\nu,\Lambda}} \int_{\R^{n\Lambda}} F(\varphi) e^{-H(\varphi)}
   d\varphi
\end{equation}
with
\begin{equation}
\lbeq{H4def-again}
  H(\varphi) = \frac12 \sum_{x\in \Lambda} \varphi_x \cdot (-\Delta_\beta \varphi)_x
  + \sum_{x\in \Lambda} \left( \frac14 g|\varphi_x|^4+ \frac12 \nu|\varphi_x|^2\right).
\end{equation}
Here $g>0$, $\nu \in \R$, $d\varphi = \prod_{x\in \Lambda} d\varphi_x$ is the
Lebesgue measure on $(\R^n)^\Lambda$,
and $\beta$ is a $\Lambda \times \Lambda$ symmetric matrix with non-negative entries.
The GFF is the degenerate case
$w(\varphi) = \frac12 m^2 |\varphi|^2$.
The commonest short-range
spin-spin interaction is the nearest-neighbour choice
$\Delta_\beta=\Delta_\Lambda$.

Our topic now is the hierarchical $|\varphi|^4$ model,
in which $\Delta_\beta$ is replaced by the hierarchical Laplacian $\Delta_H$
of Section~\ref{sec:hGFF}.  This choice significantly simplifies the
analysis in the renormalisation group approach.  According to
Exercise~\ref{ex:hier-rw}, $-\Delta_H$ is ferromagnetic.  Moreover,
for $x\neq 0$, and in the simplifying case of the limit $N \to
\infty$, $\Delta_{H;0x}$ is proportional to $L^{-(d+2)j_x}$ where
$j_x$ is the coalescence scale.  Therefore $\Delta_{H;0x}$ is bounded
above and below by multiples of $|x-y|^{-d-2}$.  Thus, although the
matrix $\Delta_H$ is long-range, it is almost short-range in the sense
that its variance is only borderline divergent.  Although it does not
respect the symmetries of the Euclidean lattice $\Z^d$, but rather
those of a hierarchical group, it nevertheless shares essential
features of the Euclidean nearest-neighbour model.

We denote expectation in the $n$-component hierarchical $|\varphi|^4$
model by
\begin{equation}
\lbeq{Ex-hier}
    \langle F \rangle_{g,\nu,N}
    =
    Z_{g,\nu,N}^{-1} \int_{\R^{n\Lambda}} F(\varphi)
    e^{-\sum_{x \in \Lambda} \left( \frac12 \varphi_x \cdot (-\Delta_H \varphi)_x
    + \frac 14 g|\varphi_x|^4 + \frac 12 \nu |\varphi_x|^2 \right)} d\varphi
    .
\end{equation}
The finite-volume susceptibility is
\begin{equation}
\lbeq{chiNdef}
        \chi_N(g,\nu)
        =
        \sum_{x\in\Lambda} \la \varphi_0^{1} \varphi_x^{1} \ra_{g,\nu,N},
\end{equation}
and the susceptibility in infinite volume is
\begin{equation} \label{e:suscept-hier-infvol}
    \chi(g,\nu)= \lim_{N\to \infty}\sum_{x\in\Lambda_N} \langle \varphi_0^{1}
    \varphi_x^{1}\rangle_{g,\nu,N}.
\end{equation}
Existence of this limit is part of the statement of the following
theorem.
The theorem provides the hierarchical version of
\eqref{e:thmphi4-chi}.  Its proof occupies the rest of the book.

\begin{theorem} \label{thm:phi4-hier-chi}
Let $d=4$ and $n\ge 1$, let $L>1$ be large, and let $g>0$ be small.
For the hierarchical $|\varphi|^4$ model, there exists
$\nu_c=\nu_c(g,n)<0$ such that, with $\nu = \nu_c + \varepsilon$ and
as $\varepsilon \downarrow 0$,
\begin{equation}
  \label{e:thmphi4-hier-chi}
  \chi(g,\nu) \sim A_{g,n} \frac{1}{\varepsilon}(\log \varepsilon^{-1})^{(n+2)/(n+8)}
  .
\end{equation}
In particular, the limit defining $\chi(g,\nu)$ exists.  Also, as $g
\downarrow 0$,
\begin{equation}
\lbeq{Anuc-hier}
A_{g,n} \sim \left(\frac{(1-L^{-d})(n+8)g}{\log L} \right)^{\frac{n+2}{n+8}},\qquad
\nu_c(g,n)\sim -(n+2)g (-\Delta_H)^{-1}_{00}.
\end{equation}
\end{theorem}

The $L$-dependence present in \refeq{Anuc-hier} is a symptom of the
fact that in our hierarchical model the definition of the model itself
depends on $L$.  This is in contrast to the Euclidean case, where the
corresponding formulas for $A_{g,n}$ and $\nu_c$ are independent of
$L$ in Theorem~\ref{thm:phi4}.

Hierarchical fields were introduced in 1969 by Dyson \cite{Dyso69} for
the study of the 1-dim\-en\-sional Ising model with long-range
spin-spin coupling with decay $r^{-\alpha}$ ($\alpha \in (1,2)$).
Three years later, the hierarchical model was defined independently by
Baker \cite{Bake72}.  In the context of the renormalisation group, the
idea was taken up by Bleher and Sinai, who investigated both the
Gaussian \cite{BS73} and non-Gaussian regimes \cite{BS75}.

Since then, the hierarchical approximation has played an important
role as a test case for the development of renormalisation group
methods.
The hierarchical 1-component $\varphi^4$ model is studied in
\cite{GK82,Wiec97,Wiec98} for $d=4$, and in \cite{KW94,Wiec99} for $d=3$.
An analysis of the hierarchical 4-dimensional Ising model appears in
\cite{HHW01}.  The hierarchical version of the 4-dimensional weakly
self-avoiding walk is analysed in \cite{BEI92,BI03c,BI03d}.  The
$\epsilon$-expansion in the long-range (non-Gaussian) hierarchical
setting is developed in \cite{GK83a,BS75,CE78}.

Hierarchical models are remarkably parallel to Euclidean models, and
our analysis is designed so that the Euclidean proofs closely follow
the hierarchical proofs.  An alternate approach to hierarchical models
is explored in depth in \cite{ACG13}. In \cite{ACG13}, continuum
limits of hierarchical models are defined with $p$-adic numbers
playing the role of $\R^{d}$, and spatially varying coupling constants
are permitted. The search for parallels continues in \cite{Abde18}
where hierarchical conformal invariance is studied.

\section{GFF and \texorpdfstring{$|\varphi|^4$}{phi4} model}
\label{sec:Gpert}

Now we make the connection between the $n$-component hierarchical
$|\varphi|^4$ measure and an $n$-component Gaussian measure.
The exponent
\begin{equation}
\lbeq{Vgnu1}
\tfrac12 \varphi \cdot(-\Delta_{H,N}\varphi) +
\tfrac14 g |\varphi|^4 + \tfrac12 \nu |\varphi|^2
\end{equation}
in \refeq{Ex-hier} has two quadratic terms, so it is tempting to use
these two terms to define a Gaussian measure and write the
$|\varphi|^4$ measure relative to this Gaussian measure.  However, the
corresponding Gaussian measure does not exist when $\nu$ is negative,
and we are interested in the critical value $\nu_c$ which is negative.
Also, the hierarchical Laplacian itself is not positive definite, so
it is not possible to define a Gaussian measure using only the
$\varphi(-\Delta_{H,N}\varphi)$ term, without restriction on
the domain of $\Delta_{H,N}$.

Given a mass parameter $m^2>0$, we define $\nu_0=\nu-m^2$ and
\begin{equation}
  V_{g,\nu_0}(\varphi) = \tfrac14 g |\varphi|^4 + \tfrac12 \nu_0 |\varphi|^2.
\end{equation}
Leaving implicit the volume parameter $N$ on the right-hand side, and
writing $C=(-\Delta_{H,N} + m^2)^{-1}$, we have
\begin{align}
    \pair{F}_{g,\nu,N} & =
    \frac{\Ex_C  F e^{-\sum_{x \in \Lambda} V_{g,\nu_0}(\varphi_x)}}
    {\Ex_C  e^{-V_{g,\nu_0}}}.
\label{e:Fexpectation}
\end{align}

The finite-volume susceptibility corresponds to the choice $F(\varphi)
= \sum_{x\in \Lambda} \varphi_0^{1} \varphi_x^{1}$ on the left-hand
side of \refeq{Fexpectation}.  It can be studied using the Laplace
transform, as in the next exercise.  We define
\begin{equation}\label{e:Z0}
    Z_0(\varphi)
    =
    e^{-V_{g,\nu_0}(\varphi)}
\end{equation}
and, for $f : \R^\Lambda \to \R^{n}$,
\begin{equation}
  \label{e:Sigma}
  \Sigma_N(f)
  =
  \Ex_C(e^{(f,\varphi)} Z_0(\varphi)).
\end{equation}
By Exercise~\ref{ex:Gauss-Laplace-Z0},
\begin{equation}
  \label{e:Sigma1}
  \Sigma_N(f)
  = e^{\frac12 (f,Cf)} (\Ex_C\theta Z_0)(Cf)
  .
\end{equation}
Derivatives of functionals of fields, in the directions of test
functions $h_i$, are defined by
\begin{equation}
\label{e:derivdef}
    D^n F(f;h_1,\ldots,h_n)
    =
    \frac{d^n}{ds_1\cdots ds_n}
    F(f+s_1h_1+\cdots s_nh_n) \Big|_{s_1=\cdots =s_n=0}.
\end{equation}

\begin{exercise} \label{ex:susceptZN-hier}
  For $\nu_0=\nu-m^2$,
  \begin{align}
    \label{e:susceptZN-hier}
    \sum_{x\in\Lambda} \la \varphi_0^{1} \varphi_x^{1} \ra_{g,\nu,N}
    &=
    \frac{1}{|\Lambda|} \frac{D^2\Sigma_N(0;\1,\1)}{\Sigma_N(0)}
    \nonumber\\
    &=
    \frac{1}{m^2} + \frac{1}{m^4 |\Lambda|} \frac{D^2\Znewlast(0;\1,\1)}{\Znewlast(0)}
    ,
  \end{align}
  where $\1$ denotes the constant test function
  $\1_x = (1,0,\ldots,0)$ for all $x \in \Lambda$,
  and where $\Znewlast = \Ex_{C}\theta Z_0$ with $Z_0$ given by \refeq{Z0} and
  the convolution $\Ex_C\theta$ is given by Definition~\ref{defn:Gconv}.
  Hint: use Exercise~\ref{ex:hier-freechi} for the second equality in
  \eqref{e:susceptZN-hier}.
  \solref{susceptZN-hier}
\end{exercise}

Using the renormalisation group method we will compute the
\emph{effective mass} $m^2>0$, as a function of $\nu > \nu_c$, with
the property that the term involving $\Znewlast$ on the right-hand
side of \refeq{susceptZN-hier} goes to zero as $N \to \infty$.  By
Exercise~\ref{ex:hier-freechi}, this expresses the infinite-volume
susceptibility of the interacting model at $\nu$ as the susceptibility
of the free model at $m^{-2}$.

Much of the literature on the triviality (Gaussian nature) of the
4-dimensional $|\varphi|^4$ model has focussed on the
\emph{renormalised coupling constant} $g_{\rm ren}$, e.g.,
\cite{Froh82,AG83}.  This is defined in terms of the \emph{truncated
four-point function} $\bar{u}_4$, which for simplicity we discuss here
for the $1$-component model.  In finite volume, let
\begin{align}
    \bar{u}_{4,N} & = \sum_{x,y,z\in \Lambda}
    \Big(
    \langle \varphi_0\varphi_x\varphi_y\varphi_z \rangle_N
    -
    \langle \varphi_0\varphi_x\rangle_N \langle\varphi_y\varphi_z \rangle_N
    \nnb & \qquad \qquad
    -
    \langle \varphi_0\varphi_y\rangle_N \langle\varphi_x\varphi_z \rangle_N
    -
    \langle \varphi_0\varphi_z\rangle_N \langle\varphi_x\varphi_y \rangle_N
    \Big)
    \nnb
    & =
    \sum_{x,y,z\in \Lambda}
    \langle \varphi_0\varphi_x\varphi_y\varphi_z \rangle_N
    - 3 |\Lambda|\chi_N^{2}.
\end{align}
Then we define $\bar{u}_4 = \lim_{N \to \infty}\bar{u}_{4,N}$
(assuming the limit exists), and set
\index{Renormalised coupling constant}
\begin{equation}
    g_{{\rm ren}} = - \frac 16 \frac{\bar{u}_{4}}{\xi^d \chi^2}
\end{equation}
where $\xi$ is the correlation length.  The $\frac 16$ is simply a
normalisation factor.

\begin{exercise} \label{ex:gren}
(i)
In the setup of Exercise~\ref{ex:susceptZN-hier} with $n=1$, prove that
\begin{equation}
\lbeq{u4N}
    \bar{u}_{4,N}
    =
    \frac{1}{m^8|\Lambda|} \left( \frac{D^4\Znewlast(0;\1,\1,\1,\1)}{\Znewlast(0)}
    -
    3
    \left(\frac{D^2\Znewlast(0;\1,\1)}{\Znewlast(0)} \right)^2
    \right).
\end{equation}
Here $\1_x = 1$ for all $x \in \Lambda$.

\smallskip \noindent
(ii)
As discussed below Exercise~\ref{ex:susceptZN-hier}, we will prove that in infinite volume
the susceptibility is $\chi = m^{-2}$.  As in Theorem~\ref{thm:BSTW-clp}, for $d=4$ we expect
the correlation length to have the same leading asymptotic behaviour as the square root of
the susceptibility.  Thus, for $d=4$, we define
\begin{equation}
\lbeq{grentil}
    \tilde{g}_{{\rm ren}} = - \frac 16 \frac{\bar{u}_{4}}{\chi^4}
    = - \frac 16 m^8 \bar{u}_{4}.
\end{equation}
If $\Znewlast$ is replaced in \refeq{u4N} by $e^{-V_N(\Lambda)}$ with
$V_N(\Lambda) = \sum_{x\in\Lambda}(\frac 14 g_N\varphi_x^4 + \frac 12 \nu_N \varphi_x^2 + u_N)$,
prove that
the right-hand side of \refeq{grentil} then becomes $g_\infty = \lim_{N \to \infty} g_N$
(assuming again that the limit exists).  This explains the name
``renormalised coupling constant.''
\solref{gren}
\end{exercise}



%% file: hier1.pspdftex
\begin{picture}(0,0)%
\includegraphics{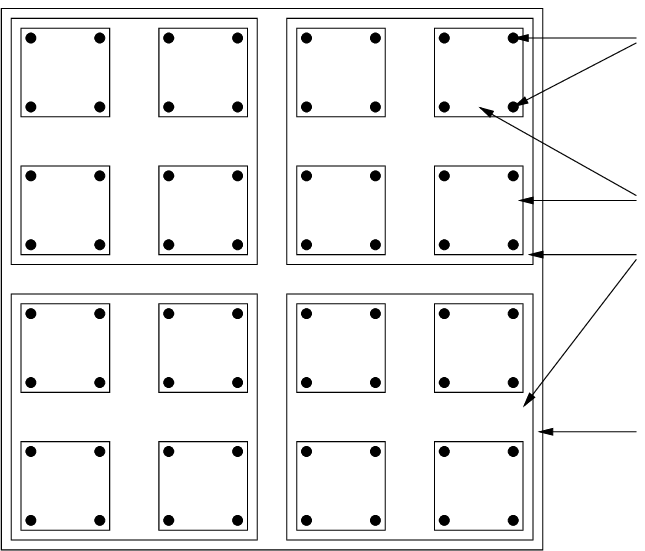}%
\end{picture}%
\setlength{\unitlength}{2072sp}%
\begingroup\makeatletter\ifx\SetFigFont\undefined%
\gdef\SetFigFont#1#2#3#4#5{%
  \reset@font\fontsize{#1}{#2pt}%
  \fontfamily{#3}\fontseries{#4}\fontshape{#5}%
  \selectfont}%
\fi\endgroup%
\begin{picture}(5877,4974)(1249,-5563)
\put(7111,-4561){\makebox(0,0)[lb]{\smash{{\SetFigFont{9}{10.8}{\familydefault}{\mddefault}{\updefault}{\color[rgb]{0,0,0}$3$-block}%
}}}}
\put(7111,-2941){\makebox(0,0)[lb]{\smash{{\SetFigFont{9}{10.8}{\familydefault}{\mddefault}{\updefault}{\color[rgb]{0,0,0}$2$-blocks}%
}}}}
\put(7111,-961){\makebox(0,0)[lb]{\smash{{\SetFigFont{9}{10.8}{\familydefault}{\mddefault}{\updefault}{\color[rgb]{0,0,0}$0$-blocks}%
}}}}
\put(7111,-2401){\makebox(0,0)[lb]{\smash{{\SetFigFont{9}{10.8}{\familydefault}{\mddefault}{\updefault}{\color[rgb]{0,0,0}$1$-blocks}%
}}}}
\end{picture}%

%% file: hierpt.tex
\chapter{The renormalisation group map}
\label{ch:hierpt}

The proof of Theorem~\ref{thm:phi4-hier-chi} uses the renormalisation
group method, and occupies the remainder of the book.  An advantage of
the hierarchical model is that the analysis can be reduced to
individual blocks (recall Definition~\ref{def:Bcal}); this is not the
case in the Euclidean setting.  We explain this reduction in
Section~\ref{sec:hierphi4}.  The renormalisation group map is defined
in Section~\ref{sec:rgmap}.  It involves the notion of flow of
coupling constants $(u_j,g_j,\nu_j)$, as well as the flow of an
infinite-dimensional non-perturbative coordinate $K_j$.  The flow of
coupling constants is given to leading order by perturbation theory,
which is the subject of Section~\ref{sec:Phipt}.

\section{Reduction to block analysis}
\label{sec:hierphi4}

\subsection{Progressive integration}

\index{Finite volume susceptibility}%

Our starting point for the proof
of Theorem~\ref{thm:phi4-hier-chi} is a formula for the finite volume
susceptibility $ \chi_N(g,\nu)$ of \refeq{chiNdef}.  It can be
rewritten, as in Exercise~\ref{ex:susceptZN-hier}, as follows.  Given
$(g_{0},\nu_{0})$ and $m^2>0$, we write
$C=C(m^2)=(-\Delta_{H,N}+m^2)^{-1}$ and
\begin{equation} \label{e:ZNZ0}
  \Znewlast = \Ex_{C}\theta Z_0, \quad Z_0(\varphi) =
  e^{-\sum_{x\in \Lambda}(\frac 14 g_0|\varphi_x|^4 + \frac 12 \nu_0 |\varphi_x|^2)}
  ,
\end{equation}
where the convolution $\Ex_C\theta$ is defined in
Definition~\ref{defn:Gconv}, and we emphasise that it here refers to an
$n$-component Gaussian field as in Example~\ref{example:gauss-vect}.
Then, for any $m^2>0$, and for $g_0=g$ and $\nu_0=\nu-m^2$,
\begin{align}
  \label{e:susceptZN-hier-bis}
  \chi_N(g,\nu)
  &=
  \frac{1}{m^2} + \frac{1}{m^4 |\Lambda|} \frac{D^2\Znewlast(0;\1,\1)}{\Znewlast(0)}
  ,
\end{align}
where $\1$ denotes the constant test function $\1_x = (1,0,\ldots,0)$
for all $x \in \Lambda$.

Thus, to compute the susceptibility, it suffices to understand
$\Znewlast$.  The formula \eqref{e:susceptZN-hier-bis} requires that
$\nu_0=\nu-m^2$, but the right-hand side makes sense as a function of
three independent variables $(m^2,g_0,\nu_0)$, with $m^2> 0$,
$g_0=g>0$, $\nu_0 \in \R$.  Although \eqref{e:susceptZN-hier-bis} no
longer holds without the requirement that $\nu_0=\nu-m^2$, it is
nevertheless useful to analyse $\Znewlast$ as a function of three
independent variables for now, and to restrict $\nu_0$ later.  We will
do so.

The starting point for the renormalisation group is to evaluate
$\Znewlast$ as the last term in a sequence $Z_{0},Z_{1},\dots ,Z_{N},
\Znewlast$ generated by
\begin{equation} \label{e:Zj-def}
  Z_{j+1} = \Ex_{C_{j+1}}\theta Z_j \quad (j<N),
  \qquad \Znewlast = \Ex_{\Cnewlast{}}\theta Z_{N},
\end{equation}
where $C=\sum_{j=1}^{N} C_j+ \Cnewlast{}$ is as in Proposition~\ref{prop:hGFF}.
It follows from the above recursion and
Corollary~\ref{cor:Gauss-decomp} that
\begin{equation}
\label{e:ZNEx}
    \Znewlast = \Ex_{\Cnewlast{}}\theta \circ
    \Ex_{C_{N}}\theta \circ \cdots \circ \Ex_{C_{1}}\theta Z_0 = \Ex_{C(m^2)} \theta Z_0,
\end{equation}
consistent with \eqref{e:ZNZ0}.  The effect of finite volume is
concentrated entirely in the last covariance $\Cnewlast{}$.

The first equation of \refeq{Zj-def} can be rewritten as
\begin{equation}
\lbeq{ZEx}
    Z_{j+1}(\varphi) = \Ex_{C_{j+1}} Z_j( \varphi  + \zeta), 
\end{equation}
where the expectation on the right-hand side integrates with respect
to $\zeta$ leaving $\varphi$ fixed.  By the definition of the
hierarchical GFF in Definition~\ref{def:hierarchical},
\begin{itemize}
\item the restriction of $x \mapsto \zeta_{x}$ to a block $b \in
\Bcal_j$ is constant;
\item the restriction of $x \mapsto \varphi_{x}$ to a block $B \in
\Bcal_{j+1}$ is constant.
\end{itemize}
The fluctuation field $\zeta$ is Gaussian with covariance $C_{j+1}$,
while the block-spin field $\varphi$ is Gaussian with covariance
$C_{j+2}+\cdots+C_N+\Cnewlast{}$.

\index{Scale}
From now on, we often fix a \emph{scale} $j$ and omit it from the notation.
We then write $+$ instead of $j+1$.
In particular, we write $C_+$ for $C_{j+1}$, $\Bcal$ for $\Bcal_j$, and $\Bcal_+$ for $\Bcal_{j+1}$.
We also abbreviate $\Ex_{+} = \Ex_{C_{j+1}}$, and
we typically use $b$ to denote a block at scale $j$
and $B$ to denote a block at scale $j+1$ when blocks at both scales are being used.

\subsection{Polynomials in the hierarchical field}

We use the notation
\begin{equation} \label{e:taudef}
  \tau = \frac12 |\varphi|^2, \quad \tau^2 = \frac14 |\varphi|^4,
\end{equation}
and write, for example, $\tau_{x}^{2} =
\tfrac{1}{4}|\varphi_{x}|^{4}$.

\begin{defn} \label{defn:Vcal}
We set $\Vcal = \R^2$, $\Ucal = \R^3$ and write their elements as $V =
(g,\nu) \in \Vcal$ and $U = (u,V) \in \Ucal$.  We identify $V$ and $U$
with the polynomials $V=g \tau^{2} + \nu \tau$ and $U=g \tau^{2} + \nu
\tau + u$.  Given $V \in \Vcal$ or $U \in \Ucal$ and $X \subset
\Lambda$, we set
\begin{align} \label{e:Vident}
  V(X,\varphi) &= \sum_{x \in X} \left(g \tau_{x}^{2} + \nu \tau_{x} \right),
  \\
  \label{e:Uident}
  U(X,\varphi) &= \sum_{x \in X} \left(g \tau_{x}^{2} + \nu \tau_{x} + u \right) .
\end{align}
Furthermore we set $U_{x} (\varphi) = g \tau_{x}^{2} + \nu \tau_{x} + u$
and similarly for $V_{x} (\varphi)$.
\end{defn}

\begin{exercise} \label{ex:ExUcal-bis}
Show that, for an arbitrary covariance $C$, $\Ex_C \theta$ acts as a
map $\Ucal \to \Ucal$ identified as polynomials in the field by
setting $X=\{x \}$ in \eqref{e:Uident}.  (Recall
Proposition~\ref{prop:wick}). \solref{ExUcal-bis}.
\end{exercise}

Recall that the set $\Bcal_j$ of $j$-blocks is defined in
Definition~\ref{def:Bcal}.  We use multi-index notation:
$\alpha=(\alpha_1, \dots, \alpha_n)$ is a vector of nonnegative
integers, and we write $|\alpha| = \sum_{i=1}^n \alpha_i$, $\alpha! =
\prod_{i=1}^n \alpha_i!$, and $\zeta^\alpha = \prod_{i=1}^n
(\zeta^i)^{\alpha_i}$ for $\zeta \in \R^n$.

\index{Zero-sum condition}
\begin{lemma} \label{lem:covUcal}
If $U,U'\in \Ucal$ and $B \in \Bcal_+$ then
there exist coefficients $p,q,r,s\in \R$, bilinear in $U,U'$, such that
\begin{equation}
    \Cov_+(\theta U_x,\theta U'(B))
    =  p + q\tau_x + r\tau_x^2 + s\tau_x^3 .
\end{equation}
If $c_+^{(1)}=0$ as in \refeq{hierarchical-C} then $s=0$, and hence
$\Cov_+(\theta U_x,\theta U'(B))$ identifies with $(p,r,q) \in
\Ucal$ via \eqref{e:Uident}.
\end{lemma}

\begin{proof}
By Taylor's theorem, $U_x(\varphi+\zeta) = \sum_{|\alpha| \le 4}
\frac{1}{\alpha!}U^{(\alpha)}(\varphi)\zeta_x^\alpha$, so
\begin{equation}
    \Cov_+(\theta U_x,\theta U'(B))
    =
    \sum_{|\alpha|,|\alpha'| \le 4} \frac{1}{\alpha!}\frac{1}{\alpha'!}
    U^{(\alpha)}U'^{(\alpha)}
    \sum_{x'\in B'}
    \Cov_+(\zeta_x^\alpha , \zeta_{x'}^{\alpha'}).
\end{equation}
Terms with $|\alpha|=0$ or $|\alpha'|=0$ vanish since the covariance
does.  The same is true when $|\alpha|+|\alpha'|$ is odd, due to the
$\zeta \mapsto -\zeta$ symmetry.  When $|\alpha|=|\alpha'|=1$, the
covariance vanishes unless $\alpha=\alpha'$, and in this case the sum
over $x'$ is $c_+^{(1)}$.  When $c_+^{(1)}$ is nonzero, the $O(n)$-invariance
of the covariance ensures that the resulting $\varphi$-dependence is of
the form $|\varphi|^6$.

This leaves only terms where $|\alpha|+|\alpha'|\in \{4,6,8\}$.  Such
terms respectively produce contributions which are quartic, quadratic,
and constant in $\varphi$.  The fact that the covariance is
$O(n)$-invariant ensures that the quartic and quadratic terms are
multiples of $|\varphi|^4$ and $|\varphi|^2$, and the proof is
complete.
\end{proof}

\subsection{Functionals of the hierarchical field}

\begin{defn}
\label{def:NcalB} For $B \in \Bcal_j$, let $J(B)$ denote the set of
\emph{constant} maps from $B$ to $\R^n$, and let $j_B:J(B) \to \R^n$
be the map that identifies the constant in the range of a map in
$J(B)$, i.e.,
\begin{equation} \label{e:jBdef}
  j_B(\varphi) = \varphi_x \quad (x\in B,\, \varphi \in J(B)).
\end{equation}
Let $\Ncal(B)$ be the vector space of functions that have the form
$F\circ j_{B}: J(B) \to \R$, where $F:\R^n \to \R$ is a function with
$p_{\Ncal}$ continuous derivatives.  In the proof of
Theorem~\ref{thm:phi4-hier-chi}, we take $p_{\Ncal}=\infty$, though
the proof also works for any finite value $p_{\Ncal} \ge 10$.
\end{defn}

\index{Locality}\index{Spatial homogeneity}\index{$O(n)$-invariance}
\begin{defn}
\label{def:Fcal}
Let $\Fcal = \Fcal_{j} \subset \bigoplus_{B\in\Bcal_j} \Ncal(B)$ be
the vector space of functions $F(B,\varphi)$ that obey the following
properties for all $B \in \Bcal_j$:
\begin{itemize}
\item \emph{locality:} $F(B) \in \Ncal (B)$,
\item
\emph{spatial homogeneity:}
$F(B) = F\circ j_{B}$ where $F:\R^{n}\rightarrow \R$ is the same
for all blocks $B \in \Bcal_j$,
\item \emph{$O(n)$-invariance:}
$F(B,\varphi) = F(B,T\varphi)$ for all $T \in O(n)$,
where $T$ acts on $J(B)$ by $(T\varphi)_x = T\varphi_x$ for $x \in B$.
\end{itemize}
\end{defn}

The property \emph{locality} is already included in the condition that
$\Fcal$ is a subspace of $\bigoplus_B \Ncal(B)$, and is written for
emphasis only.  For $X \subset \Bcal_j$, let $\Bcal_j(X)$ denote the
set of $j$-blocks comprising $X$.  For $F_j\in \bigoplus_{B\in\Bcal_j}
\Ncal(B)$, in particular for $F_j\in \Fcal_j$, we define
\begin{equation}
    \label{e:set-exponent}
    F_j^X = \prod_{B \in \Bcal_j(X)} F_j(B).
\end{equation}

\subsection{Global to local reduction}

Let $V_0=g_0 \tau_x^2 +\nu_0 \tau_x$.
We define $F_0 \in \Fcal_0$ by
\begin{equation}
\lbeq{F0def}
  F_0(\{x\},\varphi) =
  e^{-V_0(\varphi_x)}
  .
\end{equation}
By definition, $Z_0$ of \refeq{ZNZ0} can be written in the
notation \eqref{e:set-exponent} as
\begin{equation} \label{e:Z0prod}
    Z_0 = F_0^\Lambda
\end{equation}
The product $F_0^\Lambda = \prod_{x \in \Lambda} F_0(\{x\})$ in
\eqref{e:Z0prod} is the same as the product over $0$-blocks $B \in
\Bcal_0$, because a $0$-block $B$ equals $\{x\}$ for some lattice
point $x$.  A principal feature of the hierarchical model is the
stability of a product form for $Z_j$ for every $j$, analogous to
\eqref{e:Z0prod}, as in the following lemma.
In its statement, in
accordance with \eqref{e:set-exponent} we write
$F_{k}^{B} = \prod_{b \in \Bcal_k(B)} F_k(b)$ for $B \in \Bcal_{k+1}$, and
$F_j^\Lambda =  \prod_{B \in \Bcal_j} F_j(B)$.

\begin{lemma}
\label{lem:Zprod}
The sequence $F_j$ defined inductively by
\begin{equation}
\label{e:Frec}
  F_{k+1}(B)
  = \Ex_{C_{k+1}} \theta F_{k}^B
  \quad (B \in \Bcal_{k+1})
  ,
\end{equation}
with initial condition \eqref{e:F0def}, defines a sequence $F_j \in
\Fcal_j$ when $F (B) = F (B,\varphi)$ is restricted to the
domain $J (B)$.  Moreover,
\begin{equation} \label{e:Zjprod}
  Z_j = F_j^\Lambda
  .
\end{equation}
\end{lemma}

\begin{proof}
  For $j=0$, the claim \eqref{e:Zjprod} holds by \eqref{e:Z0prod} and $F_{0}
  \in \Fcal_{0}$ as remarked above.
  We apply induction, and assume that \eqref{e:Zjprod} holds for some $j$
  with $F_{j}\in \Fcal_{j}$. In particular, for $B$ in $\Bcal_{j}$,
  $F_{j} (B)$ depends only on $\varphi|_{B}$ and this field is constant on $B$.
  Following the definition of $\Ex_{C_{j+1}}\theta $ we replace $F_{j} (B,\varphi)$
  by $F_{j} (B,\varphi+\zeta )$ where $\zeta$ is Gaussian with covariance $C_{j+1}$
  and the expectation is over $\zeta$. The covariance $C_{j+1}$ is such that
  $\zeta|_B$ and $\zeta|_{B'}$ are independent for distinct blocks
  $B,B' \in \Bcal_{j+1}$. Consequently, by the inductive hypothesis
  \begin{equation}
  \lbeq{ECprod}
    \Ex_{C_{j+1}}\theta Z_j
    = \Ex_{C_{j+1}}\theta \prod_{B \in \Bcal_{j+1}} F_j^B
    = \prod_{B \in \Bcal_{j+1}} \Ex_{C_{j+1}}\theta F_j^B
    = F_{j+1}^\Lambda
  \end{equation}
  as claimed. To prove that $F_{j+1} \in \Fcal_{j+1}$ as in
  Definition~\ref{def:Fcal} we use the inductive
  hypothesis $F_j \in \Fcal_j$, the recursive definition \eqref{e:Frec}
  and that $\varphi_{j+1}$ is constant on $B$.
  These immediately imply that $F_{j+1}$
  satisfies locality and homogeneity.
  By Exercise~\ref{ex:gauss-On},
  $T \circ \Ex_{C_{j+1}}\theta
  = \Ex_{C_{j+1}}\theta\circ T$ for any $T \in O(n)$. This implies that $F_{j+1}$ is $O (n)$-invariant,
  which completes the proof that $F_{j+1} \in \Fcal_{j+1}$, and completes the proof of the lemma.
\end{proof}

According to \refeq{Zjprod} the sequence $Z_j$ is determined by
the sequence $F_j$.
A key point is the simplifying feature that $F_j$ is \emph{local},
i.e., $F(B)$ depends only on $\varphi_x$ for $x \in B$, while $Z_j$ is
\emph{global}, i.e., it depends on $\varphi_x$ for all $x \in \Lambda$.

\begin{figure}
  \centering
  \input{RG0.pspdftex}
  \caption{The map $\Ex_{+}\theta: Z \mapsto Z_+$ is lifted to $F \mapsto F_{+}$.}
  \label{fig:RG0}
\end{figure}
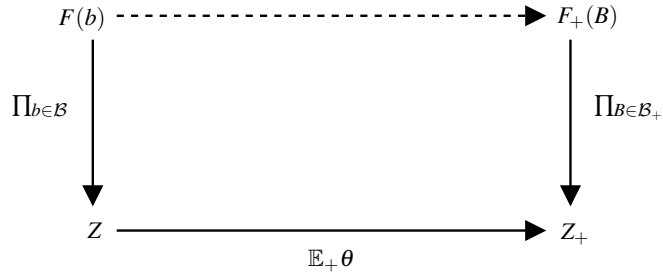

In order to define the renormalisation group map, we make a conceptual
shift in thinking about Lemma~\ref{lem:Zprod}.  Namely, we broaden our
perspective, and no longer consider the input to the expectation
$\Ex_{C_{j+1}}\theta$ as necessarily being determined by a specific
sequence $Z_j$ with initial condition $Z_0$.  Instead, we consider a
generic $F \in \Fcal_j$, define $Z = F^\Lambda$, and assume that $Z$
is integrable.  Then we consider $\Ex_{+}\theta = \Ex_{C_{j+1}}\theta$ as a map acting
on this class of $Z$.  The calculation in \refeq{ECprod} shows that
the map $F \mapsto F_+$ defined by $F_+(B) = \Ex_{+}\theta F^B$
is a lift of the map $Z \mapsto Z_+ = \Ex_{+}Z$.  See
Figure~\ref{fig:RG0}.  As discussed above, this is a global to local
reduction.

\section{The renormalisation group map}
\label{sec:rgmap}

\subsection{Local coordinates}
\label{sec:local-coords}

To describe the map $F \mapsto F_+$, defined by
$F_+(B) = \Ex_{+}\theta F^B$
for integrable $F \in \Fcal$, we introduce coordinates.
Ideally, we would like to replace $F$ by $e^{-U}$ with $U \in \Ucal$.
This is not exactly possible, as we will need more degrees of freedom for
a typical $F$ than just three real parameters $(u,g,\nu)$.  In particular, it is
in general not the case that there exists $U_+ \in \Ucal$ such that
$\Ex_{C_+}\theta e^{-U(B)}$ will be equal to $e^{-U_+(B)}$.
So instead, we make an approximate replacement of $F$ by $e^{-U}$, and keep track of
the error in this replacement.

In detail, given $U = (u,V) \in \Ucal$, we define $I \in \Fcal$ by
\begin{equation}
    I(b) = e^{-V(b)},
\end{equation}
and write $F \in \Fcal$ as
\begin{equation} \label{e:FjIjKj}
  F(b) = e^{-u|b|}(I(b) + K(b))
  ,
\end{equation}
where $K$ is defined so that \eqref{e:FjIjKj} holds:
$
    K(b)
    =
    e^{u|b|}F(b)-I (b)
$.
\index{Coordinates}%
Then \eqref{e:FjIjKj} represents $F$ by \emph{local coordinates}
\begin{equation}
\label{e:uVK}
  (u,V,K) = (U,K),
\end{equation}
with $u \in \R$, $V \in \R^2$, and $K\in \Fcal$.
We can turn this around: given coordinates $(U,K)$, the formula \refeq{FjIjKj} defines $F$.
If we define $Z=F^\Lambda$, then $(U,K)$ also determines $Z$.
Note that $Z_0$ of \refeq{Z0prod} is of this form, with $F=e^{-V}$,
corresponding to $u=0,K=0$.

\begin{figure}[h]
  \centering
  \input{RG1.pspdftex}
  \caption{The map $F \mapsto F_{+}$ is lifted to map $(U,K)\mapsto (U_+,K_{+})$.
    The lift is not unique.
    }
   \label{fig:VK}
\end{figure}
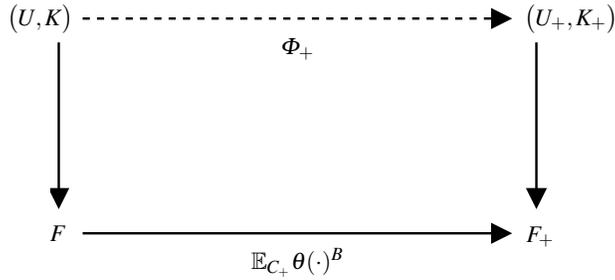

Given \emph{any} $U_+\in \Ucal$, a simple algebraic manipulation shows that
$F_+(B)= \Ex_{C_+}\theta F^B$ can be expressed in the same form
\begin{equation}
\lbeq{FplusB}
  F_+(B)
  = \Ex_{+}\theta F^B
  = e^{-u_+|B|}(I_+(B) + K_+(B))
  ,
\end{equation}
with $I_+\in \Fcal_+$ defined by $I_+(B)=e^{-V_+(B)}$ and with $K_+$ uniquely defined by
\begin{equation}
    \lbeq{K+B-bis}
    K_{+}(B) = e^{(u_{+}-u)|B|}\Ex_{+}\theta (I+K)^B - I_{+}(B) .
\end{equation}
It is straightforward to check that this is the solution that makes
the diagram commutative in Figure~\ref{fig:VK}.
With $Z_+ = \Ex_{+} \theta Z = \Ex_{+} \theta F^\Lambda = F_+^\Lambda$
(we used \refeq{ECprod} for the last equality), we obtain
\index{Factorisation}%
\begin{align}
\lbeq{ZZ+}
    Z_+ = e^{-u_+|\Lambda|}(I_+ + K_+)^\Lambda
    = e^{-u|\Lambda|}\Ex_{+}\theta  (I + K)^\Lambda
    =
    \Ex_{+}\theta Z
    .
\end{align}

To be useful, we will need to make an intelligent choice of $U_+$.
Our choice is made in Section~\ref{sec:rgmapdefn}.  It is designed in such a way
that we will be able to prove that if $K =K_{j}$ is third order
in the coefficients of $U$, then $K_+$ will be third order in the
coefficients of $U_+$ \emph{uniformly in the scale $j$}.  The
coordinate $K$ is thus an error coordinate which gathers third order
errors.  Detailed second-order information is retained in the
polynomials $U$ and $U_+$, and this is the information that is primary
in the computation of critical exponents.

We emphasise that the remainder coordinate $K$ is not written in the
exponent, i.e., we use the form $F(b) = e^{-u|b|}(e^{-V(b)} + K(b))$
instead of $e^{-u|b|-V(b)+K(b)}$.  Since $K(b)$ contains contributions
that are, e.g., degree-6 in the field $\varphi$ and of uncontrolled
sign, it is useful not to exponentiate them.
Note that requiring that $K(b)$ be $O(|\varphi|^6)$ as
  $\varphi \to 0$ would be a natural condition to fix the choice of $V$ and $K$;
  however, we do not impose it and allow $K(b)$ to contain sufficiently small
  contributions of lower order in $\varphi$. This gives a somewhat more flexible
  representation whose generalisation is particularly useful in the Euclidean setting.

\subsection{Localisation}

A crucial idea for the renormalisation group method 
is to choose $V_+$ so that the
    coordinate $K$ in \refeq{uVK} contracts under change in scale.   A
full discussion of this contraction involves the introduction of a
norm to measure the size of $K$.  We defer this to future chapters.
In this section, we restrict attention to the definition of a map
which extracts from a functional of the field, such as $K$, a local
polynomial in the field which represents the parts of the functional
which do \emph{not} contract.  This map is called $\LT$.

The monomials which comprise the range of $\LT$ are those which do not
contract under change of scale, in the following sense.  For dimension
$\drb =4$, by \eqref{e:scaling-estimate-hier} the approximate size
(square root of the variance) of the fluctuation field $|\zeta_{j+1}|$
is $L^{-j}$.  For $V = |\zeta_{j+1}|^{p}$, the size of $V (b)$ is
approximately $L^{4j}L^{-pj}=L^{(4-p)j}$ because this is the number
$|b| = L^{\drb j}$ of fields in $b$ times the approximate size
$L^{-pj}$ of the monomial $|\zeta_{j+1;x}|^{p}$ at a point $x\in b$.
Under this measure of size, the monomial grows exponentially with the
scale if $p<4$, it neither grows nor contracts if $p=4$, and it
contracts with the scale if $p>4$.  This motivates the following
definition.

\begin{defn}
\label{def:relmarirr}
A homogeneous polynomial in $\zeta$ of degree $p$ is
\emph{relevant} if $p < 4$, \emph{marginal} if $p = 4$, and
\emph{irrelevant} if $p > 4$.
\index{Relevant} \index{Marginal} \index{Irrelevant}
\end{defn}

For a functional $F$ of the field, we will use Taylor expansion to define
$\Loc F$ as the projection onto the relevant and marginal monomials of
$F$.  For this, we first develop the theory of Taylor expansion.

Recall that for a sufficiently smooth function $F:\R^{n}\rightarrow
\R$ and a point $\varphi \in \R^{n}$, the $p^{\rm th}$ derivative
$F^{(p)} (\varphi)$ of $F$ at $\varphi$ is the $p$-linear function of
directions $\dot{\varphi}^{p}=(\dot{\varphi}_{1},\dots
,\dot{\varphi}_{p})\in (\R^n)^p$ given by
\begin{equation}
    F^{(p)}(\varphi;\dot{\varphi}^{p})
    =
    \frac{\partial}{\partial t_{1}} \dots
    \frac{\partial}{\partial t_{p}}
    F (\varphi + \textstyle{\sum_{i=1,\dots ,p}} t_{i}\dot{\varphi}_{i}) ,
\end{equation}
with the derivatives evaluated at $t_{1}=\dots =t_{p}=0$.

\begin{defn}\label{def:Loc-hier}
For smooth $F : \R^n \to \R$ and
$k \ge 0$, we define $\Tay_k F$ to be the $k^{\rm th}$-order
Taylor polynomial at $0$, i.e., for $\varphi \in \R^n$,
\begin{equation}
\lbeq{Taydef}
    \Tay_k F(\varphi)
    = \sum_{p=0}^k \frac{1}{p!} F^{(p)}(0;\varphi^{p}),
    \qquad
    \varphi^p = \varphi,\ldots,\varphi .
\end{equation}
For $b\in\Bcal$ and $F (b) = F \circ j_{b} \in \Ncal (b)$, and for a
field $\varphi$ that is constant on $b$, we define
\begin{equation}
    \label{e:Taybdef}
    \Tay_k F(b) =
    (\Tay F) \circ j_{b} .
\end{equation}
We define the $\emph{localisation}$ operator $\LT$ by
\begin{equation}
    \LT F(b) = \Tay_4 F(b).
\end{equation}
\end{defn}

By definition, $\Tay_k$ is a projection.  More generally,
$\Tay_k\Tay_l F(b) = \Tay_{k\wedge l}F(b)$.  We need $\LT=\Tay_4$ only
for $O(n)$-invariant $F(b)$, and in this case it simplifies.

\begin{lemma} \label{lem:Locexample}
Let $b \in \Bcal$ and suppose that $F(b)=F\circ j_b$ is in $\Ncal (b)$
and is $O(n)$-invariant.  Then, for constant $\varphi$ on $b$,
\begin{align}
\lbeq{LTdef}
  \LT \,F(b,\varphi)
  &=
  F(0) + \frac{1}{2!} F^{(2)}(0;e_{1}^{2}) |\varphi|^2
  + \frac{1}{4!} F^{(4)}(0;e_{1}^{4})|\varphi|^4
  ,
\end{align}
where $e_{1}= (1,0,\dots ,0)$ in $\R^{n}$ and $e_{1}^{p} = e_{1},\dots
,e_{1}$ in $\prod_{i=1,\dots ,p}\R^{n}$. In particular, there is a
unique element $U_F$ of $\Ucal$ such that $\Loc F(b)=U_F(b)$, and we
identify $\Loc F$ with this element $U_F$.
\end{lemma}

\begin{proof}
By hypothesis, $F(T\varphi)=F(\varphi)$ for every $T \in O(n)$.  With
the choice $T=-I$, we see that $F^{(\alpha)}(0)=0$ for all odd
$|\alpha|$.  With $T$ chosen so that $T\varphi =
(|\varphi|,0,\ldots,0)$ (a rotation), we obtain \refeq{LTdef}.
\end{proof}

\begin{example}
Consider $1$-component fields $\varphi$ defined on $\R^{\Lambda_{N}}$
which are constant on the block $b \in \Bcal$.
\\
(i) Let
\begin{equation}
    F (b,\varphi) = \sum_{x\in b} \varphi_{x}^{2} .
\end{equation}
This is an element of $\Ncal (b)$ as in Definition~\ref{def:NcalB}. In
particular, $F (b)=F\circ j_{b}$ with $F (u) = |b|u^{2}$, so $\Tay_{4}
F (u) = |b|u^{2}$. Therefore $(\Tay_{4} F) \circ j_{b} (\varphi) =
|b|\big(j_b (\varphi)\big)^{2}$. Equivalently, $(\Tay_{4} F) \circ
j_{b} (\varphi) = \sum_{x\in b}\varphi_{x}^{2}$.  By
Definition~\ref{def:Loc-hier},
\begin{equation}
\lbeq{LTexi}
    \LT \,F (b,\varphi)
    =
    \sum_{x\in b}\varphi_{x}^{2}
    = F(b,\varphi).
\end{equation}
Similarly, $\LT \,F(b,\varphi) = F(b,\varphi)$ if
\begin{equation}
  F(b,\varphi)
  =
  \sum_{x \in b} \left(\frac14 g\varphi_x^4 +
  \frac12 \nu \varphi_x^2 + u \right).
\end{equation}

\smallskip \noindent
(ii) Let
\begin{equation}
    F (b,\varphi) = e^{\sum_{x\in b} \nu \varphi_{x}^{2}} .
\end{equation}
Then $F (b) = F \circ j_{b}$ with $F (u) = e^{|b|\nu u^{2}}$, so
$\Tay_{4} F (u) = 1 + |b|\nu u^{2} +
\tfrac{1}{2}|b|^{2} \nu^2 u^{4}$.  Therefore $(\Tay_{4} F) \circ
j_{b} (\varphi) = 1 + |b|\nu \big(j_b (\varphi)\big)^{2} +
\tfrac{1}{2}|b|^{2} \nu^2
\big(j_{b}(\varphi)\big)^{4}$. Equivalently, $(\Tay_{4} F) \circ j_{b}
(\varphi) = 1 + \sum_{x\in b} \nu \varphi_{x}^{2} +
\tfrac{1}{2}|b|\sum_{x\in b} \nu^2\varphi_{x}^{4}$. By Definition~\ref{def:Loc-hier},
\begin{equation}
\lbeq{LTexii}
    \LT\,
    F (b,\varphi)
    =
    \sum_{x\in b}
    \left(
    \frac{1}{|b|} + \nu  \varphi_{x}^{2}
    + \tfrac{1}{2}|b| \nu^2 \varphi_{x}^{4}
    \right) .
\end{equation}
In \refeq{LTexi} and \refeq{LTexii}, the output of $\LT$ has
been written as a local polynomial in the field, summed over the block
$b$.  For the hierarchical model this could be seen as a redundant
formulation, since the field is constant on $b$ and hence, e.g.,
$\sum_{x\in b}\varphi_x^2 = |b|\varphi^2$.  However, in the Euclidean
model the field is no longer constant on blocks, and \refeq{LTexi} and
\refeq{LTexii} have direct Euclidean counterparts.  This
illustrates the general theme that Euclidean formulas specialised to
the case where fields are constant on blocks reduce to hierarchical
formulas.
\end{example}

Ultimately, the proof that $K_+$ contracts relative to $K$ requires an
estimate on $1-\LT$.  A general version of this crucial estimate is
given in Section~\ref{sec:Taylor}, and its specific application occurs
in Section~\ref{sec:rgest-contraction}.

\subsection{Perturbative map}

In this section, $C_+$ is any covariance with the property that the corresponding
fields are constants on blocks in $b \in \Bcal$. It will be taken to be either $C_{j+1}$
for some $j < N$, or $\Cnewlast{}$ when $j=N$.
We sometimes write $\Ex_+$ in place of $\Ex_{C_+}$.

As discussed in Section~\ref{sec:local-coords}, it is in general not
the case that there exists $U_+\in\Ucal$ such that $\Ex_{+}e^{-U(B)}$
will be equal to $e^{-U_+(B)}$.  The \emph{perturbative map} is a map
$U \mapsto \Upt$ such that, in a sense to be made precise below,
$\Ex_{+}e^{-U(B)}$ is approximately equal to $e^{-\Upt (B)}$.  The map
is defined as follows.

\begin{defn}
\label{def:Phipt}
Recall that $\Ex(\theta A;\theta B)$ is defined in
\eqref{e:ExthetaABdef}; it is the same as the covariance
$\Cov (\theta A, \theta B)$.  Given $U \in \Ucal$, we define
\begin{equation}
    \label{e:Vplus-hier-def}
    \Upt (B)
    =
      \Ex_{C_{+}}\theta U (B)
      - \frac{1}{2} \LT\, \Ex_{C_{+}}\big(\theta U (B);\theta U (B)\big)
    \qquad
    (B \in \Bcal_{j+1})
    .
\end{equation}
Exercise~\ref{ex:ExUcal-bis} shows that
$\Ex_{C_{+}}\theta U (B)$
determines an element of $\Ucal$, and the range of $\LT$ is also
$\Ucal$, so $\Upt(B)$ determines an element of $\Ucal$.  We define the
\emph{perturbative map} $\Phi_\pt: \Ucal \to \Ucal$ by setting
$\Phi_\pt(U)$ to be this element. Then \eqref{e:Vplus-hier-def} can
also be written as
\begin{equation}
\lbeq{Phiptdef}
    \Phi_\pt(U;B) =
    \Ex_{C_{+}}\theta U (B)
    - \frac{1}{2} \LT\, \Ex_{C_{+}} \big(\theta U (B);\theta U (B)\big).
\end{equation}
We also define
\begin{equation}
    \label{e:W-def}
    W_{+}(B)
    =
    \frac{1}{2} (1-\LT) \Ex_{C_{+}}\big(\theta U (B);\theta U (B)\big).
\end{equation}
\end{defn}

By Lemma~\ref{lem:covUcal}, if $c^{(1)}=0$ then the definition of
$\Upt(B)$ is not changed if $\LT$ is removed from the right-hand side
of \refeq{Vplus-hier-def}.  Also by Lemma~\ref{lem:covUcal}, $W(B)$ is
proportional to $\sum_{x\in B} \tau_x^3$, and is in fact zero if
$c^{(1)}=0$.

The polynomial $\Upt\in \Ucal$ can be calculated explicitly, and the
result of this calculation is given in
Proposition~\ref{prop:barflow-a}.  The coefficients of $\Upt$ are
explicit quadratic polynomials in the coefficients of $U$, and the
coefficients of these quadratic polynomials are explicit functions of
the covariance $C_{+}$.

The sense in which the expectation $\Ex_+ \theta e^{- U}$ is
approximately $e^{-\Upt( U)}$ is made precise by Lemma~\ref{lem:K0PT}.
Given $U$, we define
\begin{equation}
\label{e:dUdef}
    \delta U = \theta U - \Upt(U).
\end{equation}
The following lemma illustrates what the definition of $\Upt$
achieves.  It shows that the difference between $\Ex e^{-\theta U}$
and $e^{-\Upt}$ is the sum of three terms.  The term involving $W$ is
second order in $U$ but is zero as long as $c^{(1)}=0$, which does
hold for all scales except the last scale by \refeq{hierarchical-C}.
The term involving $(\LT \Var (\theta U) )^{2}$ is fourth-order in $U$,
and there is a term that is formally third order in $\delta U$,
defined in terms of
\begin{equation}
    \lbeq{A3def}
    A_3(B)
    =
    \frac{1}{2!}\int_0^1 (-\delta U(B))^3 e^{-t\delta U(B)} (1-t)^{2} dt.
\end{equation}
In Section~\ref{sec:S0pf},
we  provide a careful analysis of the term involving $A_3$.

\begin{lemma}
\label{lem:K0PT} For any polynomial $U \in \Ucal$ such that the
expectations exist, and for $B \in \Bcal_+$,
\begin{align}
\lbeq{ExW5}
    \Ex_{+} e^{-\theta U(B)}
    &=
    e^{-\Upt(B)}
    \left(
    1 +  W_+(B) +
    \tfrac{1}{8} \left(\LT \Var (\theta U(B)) \right)^2
    + \Ex_{+} A_3(B)
    \right)
    .
\end{align}
In particular, if the covariance satisfies the zero-sum condition $c^{(1)}=0$, then
\begin{align}
\lbeq{ExW5-zerosum}
    \Ex_{+} e^{-\theta U(B)}
    &=
    e^{-\Upt(B)}
    \left(
    1 +
    \tfrac{1}{8} \left( \Var \big(\theta U (B)\big) \right)^2
    + \Ex_{+} A_3(B)
    \right)
    .
\end{align}
\end{lemma}

\begin{proof}
We drop the block $B$ and subscript $+$ from the notation.
Then we can rewrite the desired formula \refeq{ExW5} as
\begin{align}
\lbeq{ExW5-pf}
    \Ex e^{-\delta U}  &=
    1 +  W +
    \tfrac{1}{8} \left(\LT \Var (\theta U) \right)^2
    + \Ex A_3
    .
\end{align}
By the Taylor remainder formula,
\begin{equation}
    \label{e:edeltaUpt-N}
    e^{-\delta U}
    =
    1 - \delta U + \half (\delta U)^2 + A_3.
\end{equation}
By Definition~\ref{def:Phipt},
\begin{equation}
    \Ex \delta U = \tfrac{1}{2} \LT \Var (\theta U)
    .
\end{equation}
Also, $\Var ( \delta U  ) = \Var(\theta U)$ and $W=\frac 12 (1-\LT) \Var \theta U$, so
\begin{align}
    \Ex
    \left(
    - \delta U + \tfrac{1}{2} (\delta U)^{2}
    \right)
    &=
    -
    \Ex ( \delta U  )
    +
    \tfrac{1}{2} \Var \big( \delta U \big)
    +
    \tfrac{1}{2} \left(\Ex ( \delta U )\right)^{2}
    \nnb & =
    - \tfrac{1}{2} \LT \Var (\theta U) + \tfrac{1}{2}  \Var (\theta U)
    + \tfrac{1}{8} \left(\LT \Var (\theta U) \right)^{2}
    \nnb & =
    W + \tfrac{1}{8} \left(\LT \Var (\theta U) \right)^{2}
    .
\label{e:K0PTpf-N}
\end{align}
By \eqref{e:edeltaUpt-N}, this leads to \refeq{ExW5-pf}.
Finally, \refeq{ExW5-zerosum} then follows immediately since when $c^{(1)}=0$ we
have $W=0$ and $\LT\Var\theta U = \Var\theta U$.
This completes the proof.
\end{proof}

A more naive idea would be to expand $e^{-\theta U}$ into a power
series before computing the expectation $\Ex_+e^{-\theta U}$, but such
expansions can behave badly under the expectation, as illustrated by
the following exercise.

\begin{exercise} \label{ex:phi4integral}
Observe that, for any $g \geq 0$,
\begin{equation}
  \frac{1}{\sqrt{2\pi}}
  \int_{-\infty}^\infty e^{-gx^4} e^{-\frac12 x^2} dx \leq 1,
\end{equation}
whereas, on the other hand, the series
\begin{equation} \label{e:x4series}
  \sum_{n=0}^\infty \frac{1}{n!}
  \frac{1}{\sqrt{2\pi}}
  \int_{-\infty}^\infty (-gx^4)^n e^{-\frac12 x^2} dx
\end{equation}
is not absolutely convergent for any $g\neq 0$.
\solref{phi4integral}
\end{exercise}

\subsection{Definition of the renormalisation group map}
\label{sec:rgmapdefn}

We now have all the ingredients needed to define the renormalisation group map.
We define the map $\Phi_\pt$ with $C_+=C_{j+1}$ as in Proposition~\ref{prop:hGFF}.
As before, the scale $j$ is fixed and omitted from the notation.
We assume here that $j<N$. In particular, this means that
$c_+^{(1)}=0$ holds so that $\Loc$ can be dropped from \eqref{e:Phiptdef}
  and $W_+=0$.

\index{Renormalisation group map}
\begin{defn}
\label{def:RGmap}
For $m^2 \geq 0$, the \emph{renormalisation group map}
\begin{equation}
\lbeq{Phi+def}
    \Phi_+ :(V,K) \mapsto (U_+,K_+) = (u_+,V_+,K_+)
\end{equation}
is defined by
\begin{align}
    \lbeq{U+def}
    U_+ & = \Phi_\pt(V-  \Loc \,(e^V K)),
    \\
\lbeq{K+B}
  K_{+}(B) &= e^{u_{+}|B|} \Ex_{+}\theta (I+K)^B - I_{+}(B)  ,
\end{align}
where $I=e^{-V}$, and $I_+=e^{-V_+}$.  The domain of $\Phi_+$ consists
of those $(V,K)\in \Vcal \times \Fcal$ such that
$\Ex_{C_+}\theta(I(V)+K)^B$ is defined.  We write the components of
$\Phi_+$ as
\begin{equation}
    \label{e:Phi-components}
    \Phi_+
    =
    (\Phi_+^U,\Phi_+^K)
    =
    (\Phi_+^u,\Phi_+^V,\Phi_+^K) = (\Phi_+^u, \Phi_+^{(0)}).
\end{equation}
\end{defn}

Note that in \refeq{Phi+def} the input polynomial is $V \in \Vcal$
rather than $U =u+V \in \Ucal$.  The reason why it is sufficient to
consider the case $u=0$ is discussed in Remark~\ref{rk:u0}.  Note also
that $\Uhat(b) = V(b) - \LT(e^{V(b)}K(b))$ in \refeq{U+def} defines a
$b$-independent element $\Uhat \in \Ucal$, due to the spatial
homogeneity imposed on $K\in\Fcal$ by Definition~\ref{def:Fcal}.  The
formula \refeq{K+B} for $K_+$ is identical to \refeq{K+B-bis}, with
the specific choice \refeq{U+def} for $U_+$, and with $u=0$.

\begin{rk} \label{rk:u0}
The domain of $\Phi_+$ involves $V \in \Vcal$ instead of $U \in
\Ucal$, i.e., has $u=0$, while the output of $\Phi_+$ has a
$u$-component.  This is because the dependence of the expectation on
$u$ is of a trivial nature.  Let $U = (u,g,\nu)=(u,V)$.  Then $\Uhat
= u+ \Vhat$ and $\Upt(U) = u + \Upt(V)$, and thus $U_+(U,K) = u +
U_+(V,K)$ and $K_+(U,K) = e^{u|B|}K_+(V,K)$.  The effect of nonzero
$u$ can thus be incorporated in this manner.  We refer to the
transition from $U=(u,g,\nu)$ to $U_+=(u_+,g_+,\nu_+)$ as the
\emph{flow of coupling constants}.
\end{rk}

\begin{rk}
We emphasise that, while the hierarchical model is originally defined
only for $m^2 >0$, the covariances $C_1, \dots, C_N$ (but not
$\Cnewlast{}$) are well-defined also for $m^2=0$. This allows us to
define the renormalisation group map also for $m^2=0$, as in
Definition~\ref{def:RGmap}.
Furthermore, the maps $\Phi_j$ for $\Lambda_N$ are the same for all $N \ge j$.
\end{rk}

There are two aspects to our choice of $U_+$, which is the basis for
the definition of the renormalisation group map $\Phi_+$, with $K_+$
given by \refeq{K+B-bis}.

\medskip \noindent \emph{Nonperturbative aspect.}  In $e^{-V}+K =
e^{-V} (1+e^{V}K)$, the term $e^{V}K$ can contain relevant and
marginal contributions, so we isolate these as $e^{V}K = \Loc \,e^{V}K
+ (1-\Loc)e^{V}K$.  We wish to absorb $\LT \,e^{V}K$ into $V$, which
is in the exponent, so we approximate $e^{-V} (1 + \LT \, e^{V}K)$ by
$e^{-\Uhat}$ with $\Uhat = V - \LT\, e^{V}K$.  This approximate
transfer of the marginal and relevant terms turns out to be sufficient
since we will impose a hypothesis that the remainder $K$ is higher
order and therefore does not significantly affect the evolution of
$V$.

\medskip \noindent \emph{Perturbative aspect.}  The expectation $\Ex
\theta e^{-\Uhat}$ is approximately $e^{-\Upt(\Uhat)}$, in the sense
that is made precise by Lemma~\ref{lem:K0PT}.  The occurrence of
$\Phi_\pt$ in \refeq{U+def} is for this reason.

\subsection{The last renormalisation group step}

The final renormalisation group step concerns the integration with
covariance $\Cnewlast{} = m^{-2}Q_N$.  This step is special.  There is
no more reblocking because after the integration with covariance $C_N$
only the one block $B=\Lambda_N$ remains.  Also, $\hat{c}_N^{(1)}=
m^{-2} \neq 0$ so $\Wnewlast$ is not zero, where $\Wnewlast$ is given by
\refeq{W-def} to be
\begin{equation}\label{e:Wnewlast-def}
    \Wnewlast(B)
    =
    \frac{1}{2} (1-\LT)
    \Ex_{\Cnewlast{}}\big(\theta V(B);\theta V(B)\big)
    \qquad
    (B = \Lambda ).
\end{equation}

\begin{defn}
\index{Renormalisation group map}
\label{def:laststep} The \emph{final renormalisation group map}
$(V,K)\mapsto (\Unewlast ,\Knewlast)$ is the map from $\Vcal \times
\Fcal$ to $\Ucal \times \Fcal$ defined by
\begin{align}
    \Unewlast & = \Upt(V),
    \\
\lbeq{Klastdef}
    \Knewlast(B) & =
    e^{-\Vnewlast(B)}
    \left(
     \tfrac{1}{8} \left(\LT \Var (\theta U) \right)^2
    + \Ex_{\Cnewlast{}} A_3(B) \right)
    + e^{\unewlast |B|} \Ex_{\Cnewlast{}}\theta K(B) ,
\end{align}
where $B=\Lambda_N$, $\Unewlast=\unewlast + \Vnewlast$ with $\Vnewlast
\in \Vcal$, $\delta U$ and $A_3$ are as in \refeq{dUdef} and
\refeq{A3def}. We assume the expectations in \refeq{Klastdef} exist.
\end{defn}

The formula for $\Unewlast$ in Definition~\ref{def:laststep}
does not have the $\Loc$ term present in \eqref{e:U+def}, because it is not
necessary to remove expanding parts of $K$ when there are no more renormalisation group
steps to cause $K$ to expand.

The following proposition shows that Definition~\ref{def:laststep} and
$\hat c^{(1)}_N \neq 0$ lead to a revised version of the
representation \refeq{ZZ+} where now $\Inewlast$ is given by
$e^{-\Vnewlast}(1+\Wnewlast)$ rather than simply by $e^{-\Vnewlast}$ as in all earlier
renormalisation group steps.

\begin{prop}
\label{prop:Elast}
With $\Unewlast$ and $\Knewlast$ as in Definition~\ref{def:laststep},
with $B=\Lambda_N$, and assuming that the expectations exist,
\begin{equation}
\lbeq{Elast5}
    \Ex_{\Cnewlast{}} \left(e^{-\theta V(B)} +\theta K(B)\right)
    =
    e^{-\unewlast |B|}
    \left(e^{-\Vnewlast(B)}\big(1+\Wnewlast(B)\big) + \Knewlast(B)\right).
\end{equation}
\end{prop}

\begin{proof}
We drop $B$ from the notation.
By Lemma~\ref{lem:K0PT}, the left-hand side of \refeq{Elast5} is equal to
\begin{equation}
    e^{-\Unewlast}
    \left(
    1 +  \Wnewlast  +
    \tfrac{1}{8} \left(\LT \Var (\theta U) \right)^2
    + \Ex_{\Cnewlast{}} A_3
    \right)
    +
    \Ex_{\Cnewlast{}}\theta K.
\end{equation}
After an algebraic reorganisation, this is seen to equal the right-hand
side of \refeq{Elast5}.
\end{proof}

\begin{rk}
\label{rk:WEuclidean}
  As discussed further in Chapter~\ref{app:Euclidean}, for the Euclidean
  finite-range decomposition it is the case that $c^{(1)} \neq 0$ for all covariances.
  This creates a need for a term $W$ in all renormalisation group steps, not just
  in the last step as we have here for the hierarchical model.
\end{rk}

\section{Perturbative flow of coupling constants: the map \texorpdfstring{$\Phi_\pt$}{Phipt}}
\label{sec:Phipt}

In this section, we explicitly compute the map $\Phi_\pt (V) =
\Phi_+^U(V,0)$, which by definition is the map $V \mapsto \Upt$ of
\refeq{Vplus-hier-def}.  Note that $\Phi_\pt$ depends only on $V$, and
not on $K$.  We allow nonzero $u$ in this section, so that, as
discussed in Remark~\ref{rk:u0}, $\Phi_\pt$ acts on $U=u+V \in \Ucal$
rather than on $V\in \Vcal$.  As we show in Chapter~\ref{ch:pt},
$\Phi_\pt$ represents the second-order part of the map $\Phi_{+}$,
whose remaining parts are third-order.  We write the image of $U$
under $\Phi_\pt$ as $( u_\pt,g_\pt,\nu_\pt)$.  Thus our goal is the
calculation of $( u_\pt,g_\pt,\nu_\pt)$ as a function of $(u,g,\nu)$.
This functional dependence of the former on the latter is referred to
as the \emph{perturbative flow of coupling constants}.

\subsection{Statement of the perturbative flow}

The perturbative flow of coupling constants is best expressed in terms
of the rescaled variables:
\begin{equation} \label{e:Greeksrescaled}
  \mu = L^{2j} \nu, \quad \mu_\pt = L^{2(j+1)} \nu_\pt,
  \qquad
  E_{\pt} = L^{d(j+1)} ( u_{\pt}-u) .
\end{equation}
We generally omit the scale index $j$, and regard variables with index
$\pt$ as scale-$(j+1)$ quantities.  The powers of $L$ in
\refeq{Greeksrescaled} correspond to the scaling of the monomials on a
block as discussed above Definition~\ref{def:relmarirr}:
$\nu\varphi^2$ scales like $\nu L^{2j}$, $u\varphi^0$ scales like
$uL^{\drb j}$, and $g$ is unscaled since $\varphi^4$ is marginal.

\index{Perturbation theory}
\begin{prop}
\label{prop:barflow-a}
Let $d=4$, $\gamma = (n+2)/(n+8)$, and suppose that $c^{(1)}=0$.  Then
the map $U \mapsto \Upt$ of \eqref{e:Vplus-hier-def} can be written
as
\begin{align}
  \label{e:gbar}
  g_{\pt} &= g - \beta g^2,
  \\
  \label{e:mubar}
  \mu_\pt &= L^2 \left(\mu (1-\gamma \beta g) + \eta g - \xi g^2 \right)
  ,
  \\
  \label{e:ubar}
  E_\pt   &=    L^d\left(  \kappa_g g + \kappa_\mu \mu
  - \kappa_{g\mu} g\mu
  - \kappa_{gg} g^2 - \kappa_{\mu\mu} \mu^2 \right) ,
\end{align}
where $\beta, \eta, \xi, \kappa_*$
are $j$-dependent constants defined
in \eqref{e:Greekdef2}--\eqref{e:kappadef2} below.
\end{prop}

Ultimately, the coefficient $\gamma$ in \eqref{e:mubar} will become
the exponent of the logarithm in Theorem~\ref{thm:phi4-hier-chi}.  To
define the coefficients that appear in \eqref{e:gbar}--\eqref{e:ubar},
we recall the definitions \refeq{cjdef}--\refeq{cjndef}, namely
\begin{align}
\lbeq{cjdef-bis}
  c_j &=
  L^{-(d-2)j}(1+m^2L^{2j})^{-1} (1-L^{-d}),
\\
\lbeq{cjndef-bis}
    c_j^{(n)}& = \sum_{x\in \Lambda} (C_{j+1;0,x}(m^2))^n.
\end{align}
We define the coefficients
\begin{align}
\label{e:Greekdef}
    \eta_j' &= (n+2)c_j,
    \quad
    \beta_j' = (n+8)c_j^{(2)},
    \quad
    \xi_j' = 2(n+2)c_j^{(3)} + (n+2)^2 c_j c_j^{(2)},
    \\
\lbeq{kappadef0}
    \kappa_{g,j}' & = \tfrac{1}{4}n(n+2) c_j^2,
    \quad
    \kappa_{\nu,j}'   = \tfrac{1}{2} n c_j,
    \quad
    \kappa_{g\nu,j}' = \tfrac{1}{2} n(n+2) c_j c_j^{(2)},
    \\
\lbeq{kappadef}
    \kappa_{gg,j}'  &= \tfrac{1}{4} n (n+2) \left( c_j^{(4)} + (n+2)c_j^2 c_j^{(2)} \right),
    \quad
    \kappa_{\nu\nu,j}'  = \tfrac{1}{4} n c_j^{(2)}.
\end{align}
The primes in the above definitions indicate that they refer to
unscaled variables; these primes are dropped in rescaled versions.
For $d=4$, the rescaled versions are defined by
\begin{align}
\lbeq{Greekdef2}
    \eta_j &= L^{2j}\eta_j',
    \quad
    \beta_j = \beta_j',
    \quad
    \xi_j = L^{2j}\xi_j',
    \\
\lbeq{kappadef1}
    \kappa_{g,j} & = L^{4j} \kappa_{g,j}',
    \quad
    \kappa_{\nu,j}   = L^{2j} \kappa_{\nu,j}',
    \quad
    \kappa_{g\mu,j} = L^{2j} \kappa_{g\nu,j}',
    \\
\lbeq{kappadef2}
    \quad
    \kappa_{gg,j}  &= L^{4j} \kappa_{gg,j}',
    \quad
    \kappa_{\mu\mu,j}  = \kappa_{\nu\nu,j}'.
\end{align}
All the above coefficients depend on the mass $m^2$ occurring in the
covariance.

The coefficient $\beta_j$ is of particular importance.  The use of the Greek letter $\beta$
is not entirely consistent with the term ``beta function'' in physics, which in
\index{Beta function}%
our context would represent the difference between the coupling constant at two
successive scales.  In our formulation, $\beta$ represents the coefficient of $g^2$
in the beta function.

\begin{rk}
A term corresponding to $\kappa_{g\nu}'$ was incorrectly omitted in
\cite[(3.27)--(3.28)]{BBS-phi4-log}.  Its inclusion does not affect
the conclusions of \cite{BBS-phi4-log}.
\end{rk}

\begin{defn}
\label{def:jm} For $m>0$, let $j_m$ be the greatest integer $j$ such
that $L^{j} m \leq 1$, and set $j_m =\infty$ if $m=0$.  We call $j_m$
\index{Mass scale}
the \emph{mass scale}.
\end{defn}

The mass scale is the scale $j$ at which the effect of the mass
becomes important in estimates.  For the mass-dependent factor in
\eqref{e:hiercov}, for $L \ge 2$ we have
\begin{equation}
  \lbeq{varthbd}
  (1+m^2L^{2j})^{-1} \leq L^{-2(j-j_m)_+}
  \le
  4^{-(j-j_m)_+} = \vartheta_j^2 \le \vartheta_j
  ,
\end{equation}
where the equality defines
\index{$\vartheta_j$}
\begin{equation}
\lbeq{varthbdx}
    \vartheta_j =2^{-(j-j_m)_+}.
\end{equation}
The advantage of $\vartheta_j$ over the stronger upper bound
$L^{-2(j-j_m)_+}$ is that $\vartheta_j$ is independent of $L$.  We
often use $\vartheta_j$ as an adequate way to take into account decay
above the mass scale.

\begin{lemma}
\label{lem:Greekbds}
For $d=4$,
\begin{gather} \label{e:betacomputed}
    \eta_j  = \eta_0^{0} (1+m^2L^{2j})^{-1},
    \quad
    \beta_j = \beta_0^{0} (1+m^2L^{2j})^{-2},
        \\
  \xi_j = \xi_{0}^{0} (1+m^2L^{2j})^{-3},
    \quad
    \kappa_{*,j} = O((1+m^2L^{2j})^{-1}),
\end{gather}
where $\beta_{0}^0 = \beta_0(m^2=0)=(n+8)(1-L^{-d})$, and analogously
for $\xi$ and $\eta$.  In particular, each of
$\eta_j,\beta_j,\xi_j,\kappa_{*,j}$ is bounded above by
$O(\vartheta_j)$.
\end{lemma}

\begin{proof}
This follows from the definitions and Exercise~\ref{ex:cjns}.
\end{proof}

An essential property is that $\beta_0^{0} >0$.  Recall from
Exercise~\ref{ex:bubble} that the hierarchical bubble diagram
$B_{m^2}^{H}$ is given by
\begin{equation}
    B_{m^2}^{H} = \sum_{j=1}^\infty c_j^{(2)}.
\end{equation}
Therefore,
\begin{equation}
\lbeq{betasumbub}
    \sum_{j=1}^\infty \beta_j = (n+8)B_{m^2}^{H},
\end{equation}
which is finite for $d=4$ if and only if $m^2>0$, and diverges
logarithmically as $m^2 \downarrow 0$ for $d=4$.

\subsection{Proof of the perturbative flow}

The flow of coupling constants is stated in
Proposition~\ref{prop:barflow-a} in terms of rescaled variables, but
for the proof we find it more convenient to work with the original
variables.  Also, although $c^{(1)}$ of \refeq{cjndef-bis} is equal to
zero by \refeq{hierarchical-C}, the final covariance $\Cnewlast{}$ does
not sum to zero.  We allow for nonzero $c^{(1)}$ in the following
proposition, so that it also handles the case of $\Cnewlast{}$.  For
this, we introduce the two coefficients
\begin{align}
    \label{e:tree-coeffs2}
    s'_{\tau^2,j} & = 4\big(g^2(n+2)c_j + g\nu\big)c_j^{(1)},
    \\
    \label{e:tree-coeffs1}
    s'_{\tau,j} & = \big(g^2(n+2)^2 c_j^2 + 2g\nu(n+2)c_j + \nu^2\big) c_j^{(1)},
\end{align}
which each vanish when $c_j^{(1)}=0$.
Proposition~\ref{prop:barflow-a} is an immediate consequence of
Proposition~\ref{prop:Upt}.

\begin{prop}
\label{prop:Upt} For $U=u+g\tau^2+\nu\tau$, the polynomial $\Upt$
defined in \eqref{e:Vplus-hier-def} has the form
$\Upt = g_\pt \tau^2 + \nu_\pt \tau + u_\pt$, with
\begin{align}
  \label{e:gpt}
  g_\pt &= g - \beta_j' g^2 -s'_{\tau^2,j},
  \\
  \label{e:nupt}
  \nu_\pt &=  \nu(1-\gamma \beta_j' g) + \eta_j' g - \xi_j' g^2 -s'_{\tau,j}
  ,
  \\
  \label{e:upt}
   u_\pt &=   u+ \kappa_{g,j}' g + \kappa_{\nu,j}' \nu
   - \kappa_{g\nu,j}' g\nu
  - \kappa_{gg,j}' g^2 - \kappa_{\nu\nu,j}' \nu^2,
\end{align}
with $\beta_j', \eta_j', \xi_j', \kappa_{*,j}'$ defined
in \eqref{e:Greekdef}--\eqref{e:kappadef}.
Also,
\begin{equation}
\lbeq{Wplus}
    W_+ = -4c^{(1)}g^2 \tau^3.
\end{equation}
In particular, if $c^{(1)}=0$, then $W_+=0$,
$U_\pt$ contains no term proportional to $\tau^3$, and hence $\Upt \in \Ucal$.
\end{prop}

Recall the definition of $U_\pt$ from \eqref{e:Vplus-hier-def} and
recall Exercise~\ref{ex:ExUcal-bis}.
The following lemma computes the terms in $\Upt$ that are linear in $V$.

\begin{lemma}
\label{lem:ECV}
For $U= u+ \tfrac{1}{4}g |\varphi|^4 + \tfrac{1}{2} \nu|\varphi|^2$,
\begin{equation}
    \Ex_{C_{j+1}}\theta U
    =
    \tfrac{1}{4} g|\varphi|^4 + \tfrac{1}{2}(\nu + \eta_j'g)|\varphi|^2
    + (u+ \kappa_{g,j}' g + \kappa_{\nu,j}' \nu).
\end{equation}
\end{lemma}

\begin{proof}
We write $C=C_{j+1}$, and sometimes also omit other labels $j$.
Recall the formula $\Ex_{C}\theta U = e^{\frac12 \Delta_{C}} U$ from
Proposition~\ref{prop:wick}.  Using this, we obtain
\begin{equation}
    \Ex_C \theta U
    =
    U
    +
    \tfrac{1}{2} \Delta_{C} (\tfrac{1}{4}g |\varphi|^4 + \tfrac{1}{2}\nu |\varphi|^2 )
    +
    \tfrac{1}{8} \Delta_{C}^2 \tfrac{1}{4}g |\varphi|^4 .
\end{equation}
The $\eta'$ term in \refeq{nupt} arises from the coefficient of $\frac
12 |\varphi|^2$ in $\frac12 \Delta_{C} \tfrac{1}{4} |\varphi|^4$.  By
definition,
\begin{align}
     \Delta_{C }  |\varphi|^4
     & = c_j \sum_{i=1}^n \frac{\partial^2}{\partial (\varphi^i)^2}
     \left( |\varphi|^2 \right)^2.
\end{align}
Since
\begin{align}
\lbeq{diff4}
    \frac{\partial^2}{\partial (\varphi^i)^2}
     \left(  |\varphi|^2 \right)^2
     & =
     4 \frac{\partial}{\partial \varphi^i} (|\varphi|^2 \varphi^i)
     =
     8(\varphi^i)^2 + 4|\varphi|^2
     ,
\end{align}
this coefficient is $\eta_j'$ given by \eqref{e:Greekdef}, as required.
The constant terms are $\kappa_{\nu}'= \frac 14 \Delta_{C} |\varphi|^2
$ and $\kappa_{g}'= \tfrac{1}{32} \Delta_{C}^2 |\varphi|^4 $.  We
leave the verification of the formulas for $\kappa_{\nu}'
,\kappa_{g}'$ in \refeq{kappadef} to Exercise~\ref{ex:Greeks}.
\end{proof}

\begin{proof}[Proof of Proposition~\ref{prop:Upt}] The definition of
$\Upt$ is given in \refeq{Vplus-hier-def}.  We again write $C=C_{j+1}$
and omit other labels $j$.  The linear terms in
\refeq{gpt}--\refeq{upt} are given by Lemma~\ref{lem:ECV}.  Let $x \in
B\in \Bcal_{j+1}$.  For the quadratic terms, we must compute
\begin{align}
\lbeq{pert1}
    \sum_{y \in B} \Ex_{C}\big(\theta V_x;\theta V_y\big)
    & =
     \sum_{y \in B}
     \Big(
    \tfrac{1}{16}g^2 \Ex_{C}\big(\theta |\varphi_x|^4;\theta |\varphi_y|^4\big)
    +
    \tfrac{1}{4}g \tfrac{1}{2} \nu
    \Ex_{C} \big(\theta |\varphi_x|^2;\theta |\varphi_y|^4\big)
    \nnb & \quad \quad
    +
    \tfrac{1}{4}g \tfrac{1}{2}
    \nu \Ex_{C} \big(\theta |\varphi_x|^4;\theta |\varphi_y|^2\big)
     +
    \tfrac{1}{4}  \nu^2 \Ex_{C}\big(\theta |\varphi_x|^2;\theta |\varphi_y|^2\big)
     \Big).
\end{align}

By Exercise~\ref{ex:Fexpand}, $\Ex_C(\theta P;\theta Q) = F_C(\Ex_C\theta P,
\Ex_C\theta Q)$, and hence it follows from Lemma~\ref{lem:ECV} that
the summand in \refeq{pert1} is equal to
\begin{align}
\lbeq{EFid}
    &
    \tfrac{1}{16}g^2 F_C \big(|\varphi_x|^4+2\eta'|\varphi_x|^2;|\varphi_y|^4+2\eta'|\varphi_y|^2\big)
    \nnb &
    +
    \tfrac{1}{4}g\tfrac{1}{2} \nu F_C \big(|\varphi_x|^2; |\varphi_y|^4+2\eta'|\varphi_y|^2\big)
     +
    \tfrac{1}{4}g\tfrac{1}{2} \nu F_C \big(|\varphi_x|^4+2\eta'|\varphi_x|^2;|\varphi_y|^2\big)
    \nnb &
    +
    \tfrac{1}{4} \nu^2 F_C \big(|\varphi_x|^2;|\varphi_y|^2\big)
    .
\end{align}
The above is equal to
\begin{align}
\lbeq{EFid2}
    &
    \tfrac{1}{16}g^2
     F_C \big(|\varphi_x|^4;|\varphi_y|^4\big)
    \nnb
    &
    +
    \Big(\tfrac{1}{16}g^2 (2\eta')^2 + \tfrac{1}{4}g\nu 2\eta' + \tfrac{1}{4} \nu^2 \Big)
    F_C \big(|\varphi_x|^2;|\varphi_y|^2\big)
    \\ \nonumber
    &
    +
    \Big(\tfrac{1}{16}g^2 2\eta' +  \tfrac{1}{4}g \tfrac{1}{2}\nu
    \Big)
    \Big(F_C \big(|\varphi_x|^2;|\varphi_y|^4\big) + F_C \big(|\varphi_x|^4;|\varphi_y|^2\big)\Big)
     .
\end{align}
This can be evaluated using the formula from
Exercise~\ref{ex:Fexpand}:
\begin{equation}
\lbeq{Fder}
    F_{C} \big(P_x;Q_y\big)
    =
    \sum_{p=1}^4 \frac{1}{p!} C_{x,y}^p \sum_{i_1,\ldots,i_p=1}^n
    \ddp{^p  P_x}{\varphi_{x}^{i_1} \cdots \partial \varphi_x^{i_p}}
    \ddp{^p  Q_y}{\varphi_{y}^{i_1} \cdots \partial \varphi_y^{i_p}}
    .
\end{equation}
In the following, we examine an important sample term, and leave most
details for Exercise~\ref{ex:Greeks}.

Consider the term $F_C \big(|\varphi|^4;|\varphi|^4\big)$.
For $p=2$, four of the eight fields are differentiated and this
produces a $|\varphi|^4$ term.
Calculation as in \refeq{diff4} gives
\begin{align}
    \sum_{i,j=1}^n
    \ddp{^2 |\varphi_x|^4}{\varphi_{x}^{i}  \partial \varphi_x^{j}}
    \ddp{^2 |\varphi_y|^4}{\varphi_{y}^{i}  \partial \varphi_y^{j}}
    & =
    16(n+8)|\varphi|^4,
\end{align}
where the subscript has been dropped on $\varphi$ on the right-hand
side to reflect the fact that the field is constant on $B$.  This
shows that the contribution due to $p=2$ that arises from $- \half
\tfrac{1}{16}g^2 \sum_{y\in B} F_C
\big(|\varphi_x|^4;|\varphi_y|^4\big)$ is
\begin{equation}
    - \Big(\tfrac{1}{2} \tfrac{1}{16} g^2 \tfrac{1}{2!} c_j^{(2)} 16(n+8)  \Big) |\varphi|^4
    = -\beta_j g^2 \tfrac{1}{4}|\varphi|^4,
\end{equation}
which is a term in \refeq{gpt}.  The $p=1$ term gives rise to
$-4c_j^{(1)} g^2 \tau^3$ in $\Upt$.  The $p=1$ term from the third
line of \eqref{e:EFid2} gives rise to $s'_{\tau^2}$.  No other
$|\varphi|^4$ terms can arise from \eqref{e:EFid2}, and the proof of
\refeq{gpt} is complete.  For $p=3$, a contribution to $\xi'$ results,
and for $p=4$, a contribution to $\kappa_{gg}'$ results.  We leave
these, as well as the contributions due to $F_C
\big(|\varphi|^2;|\varphi|^2\big)$ and $F_C
\big(|\varphi|^2;|\varphi|^4\big)$, for Exercise~\ref{ex:Greeks}.
\end{proof}

\begin{exercise}
\label{ex:Greeks} Verify the formulas given for the $\kappa'$
coefficients in \refeq{kappadef}, and the omitted details in
Proposition~\ref{prop:Upt} for the coefficients in \refeq{Greekdef}.
\solref{Greeks}
\end{exercise}

\chapter{Flow equations and main result}
\label{ch:pt}
\index{Flow equations}

In Section~\ref{sec:barflow}, we provide a detailed and elementary analysis of the
perturbative flow of coupling constants,
i.e., of the iteration of the recursion
given by Proposition~\ref{prop:barflow-a}.
We denote this flow by $(\gbar_j,\mubar_j)$.
In particular, we
construct a perturbative critical initial value $\mubar_0$ for which $\mubar_j$ approaches zero
as $j \to \infty$.

In Section~\ref{sec:reduction}, we state extensions of the results of
Section~\ref{sec:barflow} to the nonperturbative setting,
in which the recursion of Proposition~\ref{prop:barflow-a}
  is corrected by higher order terms,
and show that these extensions 
imply the main result Theorem~\ref{thm:phi4-hier-chi}.  The proof of the
nonperturbative versions
is given in Chapters~\ref{ch:pfsus}--\ref{ch:pf-thm:step-mr-K}.

\section{Analysis of perturbative flow}
\label{sec:barflow}

In this section, we study the perturbative flow of coupling constants
$\bar U$, defined as the solution to the recursion $\bar U_{j+1} =
\Phi_{+}^U(\bar U_j,0) = \Phi_\pt(\bar U_j)$.  The analysis of the
susceptibility does not require the sequence $u_j$, so we do not study
$\ubar_j$ here\, though its analysis is analogous.  Moreover, since
$\ubar_{j+1}-\ubar_j$ is a function of $\bar V_j$, $\ubar_j$ can be
computed once $\bar V_j$ is known.  Thus, we are concerned only with
the $\bar V_j$ part of $\bar U_j=(\ubar_j,\bar V_j)$.  We study the
rescaled version $(\gbar_j,\mubar_j)$ of $\bar
V_j=(\gbar_j,\bar\nu_j)$, with $\mubar_j = L^{2j}\bar\nu_j$.

\index{$\gbar$}
\index{$\mubar$}
According to \refeq{gbar}--\refeq{mubar},
\begin{align}
  \label{e:gbar9}
  \gbar_{j+1} &= \gbar_j - \beta_j \gbar_j^2,
  \\
  \label{e:mubar9}
  \mubar_{j+1} &= L^2 \left(\mubar_j(1-\gamma \beta_j \gbar_j) + \eta_j \gbar_j - \xi_j \gbar_j^2 \right)
  .
\end{align}
By Lemma~\ref{lem:Greekbds}, $\beta_j = \beta_0^0 (1+m^2L^{2j})^{-2}$.
In particular, $\beta_j$ is constant when $m^2=0$.
The system of equations \eqref{e:gbar9}--\eqref{e:mubar9} is
\emph{triangular} since the first equation only depends on $\gbar$.
Thus the equations can be solved successively; and they are so simple
that we can calculate anything we want to know.  Triangularity no
longer holds when the effect of $K$ is included, and the analysis of
Chapter~\ref{ch:pfsus} is used to deal with this.

\subsection{Flow of \texorpdfstring{$\gbar$}{gbar}}

The flow of the coupling constant $g_j$ under the renormalisation
group map is fundamental.  This flow adds a higher-order error term to
the perturbative sequence $\gbar_j$.  The analysis of the flow is the
same with or without the error term, so we include the error term from
the outset here.

Thus we generalise \refeq{gbar9} by adding an error term, and for the
moment consider a general sequence of coefficients $a_j$ for the
quadratic term:
\begin{equation}
\lbeq{gae}
    g_{j+1} = g_j - a_j g_j^2 + e_j,
\end{equation}
where we assume
\begin{equation}
    0 \le a  \le a_j \le A , \quad |e_j| \le M_j g_j^3, \quad M_j \le M.
\end{equation}
The recursion for $\gbar_j$ is the case $M=0$ and $a_j=\beta_j$, and
all of our analysis in this section applies also when $M=0$.  The
above recursion appears in many applications and has been studied by
many authors, e.g., \cite[Section~8.5]{Brui81} for the case $a_j=a$
for all $j$.

\begin{exercise}
\label{ex:gsequence} Suppose that $0<a \le A < \infty$.  Prove that if
$g_0>0$ is sufficiently small (depending on $a,A,M$) then $0< \frac 12
g_j< g_{j+1}<g_j$ for all $j \ge 0$.  It follows that the limit
$\lim_{n\to \infty}g_j$ exists and is nonnegative.  Prove that this
limit is zero.  \solref{gsequence}
\end{exercise}

Recall that the mass scale $j_m$ is defined in
Definition~\ref{def:jm}, and that
\begin{equation}
\lbeq{vartheta2}
  \vartheta_j = 2^{-(j-j_m)_+}
\end{equation}
is defined in \refeq{varthbdx}.  In our context, $a_j=\beta_j$ is
independent of $j$ when $m^2=0$, and when $m^2>0$ it begins to decay
exponentially after the mass scale.  This decay, which is an important
feature in our applications, violates the hypothesis $a>0$ in
Exercise~\ref{ex:gsequence} and requires attention.  Its principle
effect is that the flow of $g_j(m^2)$ resembles that of $g_j(0)$ for
scales $j \le j_m$, whereas the flow effectively stops at the mass
scale so that $g_j(m^2)$ resembles $g_{j_m}(0)$ for scales $j > j_m$.

As we show in the next proposition, the solution of the recursion is
essentially the sequence $t_j$ defined by
\begin{equation}
    \label{e:Aj-def}
    A_j = \sum_{i=0}^{j-1} \beta_i, \qquad t_j = \frac{g_0}{1+g_0A_j}.
\end{equation}
In particular, when $m^2=0$,
\begin{equation}
\label{e:At0}
    A_j(0) = \beta_0^0 j, \qquad t_j(0) = \frac{g_0}{1+g_0\beta_0^0 j}.
\end{equation}

\begin{exercise}
\label{ex:Asequence}
For $m^2>0$,
\begin{align}
    A_j(m^2) &= \beta_0^0 (j \wedge j_m) + O(1),
\\
\lbeq{tm0}
    t_j (m^2) &\asymp t_{j\wedge j_m}(0) = \frac{g_0}{1+g_0\beta_0^0 (j\wedge j_m)},
\\ \label{e:tsum1}
    \sum_{l=0}^{j} \vartheta_l t_l
    & \leq O(|\log t_j|).
\end{align}
\solref{Asequence}
\end{exercise}

The following proposition gives the asymptotic behaviour of the
solution to the recursion \refeq{gae} when $a_j=\beta_j$ and
$M_j=M\vartheta_j$.  The leading behaviour is not affected by the
error term $e_j$ in the recursion, as long as $|e_j| \le M\vartheta_j
g_{j}^{3}$.  In particular, $g_j$ and $\gbar_j$ have the same
asymptotic behaviour as $j \to \infty$.

\begin{prop} \label{prop:gjtj}
Let $m^2 \ge 0$ and consider the recursion
\refeq{gae} with $a_j=\beta_j$ and $M_j=M\vartheta_j$.  Let $g_0>0$ be
sufficiently small.

\smallskip \noindent
(i)
As $j \to \infty$,
\begin{equation}
\label{e:gtasy}
    g_j = t_j + O(t_j^2 |\log t_j|),
\end{equation}
with the constant in the error term uniform in $m^2 \ge 0$.  Also,
$g_j = O(g_0)$ and $g_{j+1} \in [\frac 12 g_j, 2 g_j]$.

\smallskip \noindent (ii)
For $m^2=0$, we have $g_j(0) \sim 1/(\beta_0^0 j) \to 0$ as $j \to
\infty$.  For $m^2 >0$, the limit $g_\infty(m^2) = \lim_{j\to\infty}
g_j(m^2) > 0$ exists and obeys $g_\infty(m^2) \sim 1/(\beta_0^0 j_m)$
as $m^2 \downarrow 0$.

\smallskip \noindent (iii)
Suppose that $e_j$ is continuous in $m^2 \ge 0$.  Then $g_\infty(m^2)$
is continuous in $m^2 \ge 0$ and the convergence of $g_j$ to
$g_\infty$ is uniform on compact intervals of $m^2>0$.
\end{prop}

\begin{proof}
(i)
We assume by induction that $g_j \le 2t_j$.  The induction hypothesis
holds for $j=0$ since $g_0=t_0$.  The recursion gives
\begin{align}
    \frac{1}{g_{j+1}}
    & =
    \frac{1}{g_j} \frac{1}{1-a_jg_j + e_j/g_j}
    =
    \frac{1}{g_j}
    +
    a_j + O(a_j+M_j)g_j.
\end{align}
We solve by iteration to get
\begin{align}
\lbeq{gAE-1}
    \frac{1}{g_{j+1}}
    & =
    \frac{1}{g_0} + A_{j+1} + E_{j+1},
\end{align}
with $|E_{j+1}| \le \sum_{i=0}^{j}O(a_i+M_i) g_i$.  By the induction
hypothesis and \refeq{tsum1}, $|E_{j+1}| \le \sum_{i=1}^j
O(\vartheta_i t_i) \le O(|\log t_j|)$.  This gives
\begin{align}
\lbeq{gAE}
    g_{j+1} = \frac{g_0}{1+g_0A_{j+1} + g_0 E_{j+1}} = t_{j+1}(1+O(t_{j+1}E_{j+1})),
\end{align}
which in particular allows the induction to be advanced.
It also proves the desired formula for $g_j$.

Finally, \refeq{tm0} implies that $t_j = O(g_{0})$, and by $g_{j}\le 2
t_{j}$ this proves that $g_j = O(g_0)$. For the proof of $g_{j+1} \in
[\frac 12 g_j, 2 g_j]$ see Exercise~\ref{ex:gsequence}.

\smallskip \noindent (ii)
For $m^2=0$, \refeq{gAE-1} becomes
\begin{equation}
    \frac{1}{g_{j+1}(0)}
    =
    \frac{1}{g_0} + \beta_0^0 j + O(\log j),
\end{equation}
which proves that $g_j(0) \sim 1/(\beta_0^0 j)$.  For $m^2>0$,
\refeq{gAE-1} becomes instead
\begin{equation}
\lbeq{gAE-jm}
    \frac{1}{g_{j+1}(m^2)}
    =
    \frac{1}{g_0} + \beta_0^0 (j \wedge j_m) + O(1) + O(\log (j \wedge j_m)),
\end{equation}
which proves that the limit $g_\infty(m^2)$ exists and is asymptotic
to $1/(\beta_0^0 j_m)$ as $m^2 \downarrow 0$.

\smallskip \noindent (iii)
By definition, $\beta_j(m^2)$ is continuous in $m^2>0$, and $e_j(m^2)$
is continuous by hypothesis.  On a compact subinterval of $m^2\in
(0,\infty)$, both $\beta_j$ and $e_j$ are uniformly bounded by
exponentially decaying sequences.  Consequently the sums $A_{j+1}$ and
$E_{j+1}$ which appear in \refeq{gAE-1} converge uniformly to limits,
and these limits are continuous by dominated convergence.  This proves
the uniform continuity on compact mass subintervals.  The continuity
at $m^2=0$ follows from the fact that the $j\rightarrow \infty$ limit
of the right-hand side of \refeq{gAE-jm} tends to infinity
as $m^2 \downarrow 0$, and hence $\lim_{m^2 \downarrow 0}g_\infty(m^2)
= 0 = g_\infty(0)$.  This completes the proof.
\end{proof}

The next exercise provides an extension of \refeq{tsum1}.

\begin{exercise}
\label{ex:tsequence}
Each of the sequences $g_j,\gbar_j,t_j$ obeys
$g_{j+1} = g_j(1+O(g_0))$, as well as the inequalities
$\vartheta_j(m^2) g_j(m^2) \le O(g_j(0))$ and
  \begin{align} \label{e:gbarsum}
    \sum_{l=j}^{\infty} \vartheta_l g_l^p
    &\leq O(\vartheta_j g_j^{p-1}) \qquad (p>1),
  \\ \label{e:gbarsum1}
    \sum_{l=0}^{j} \vartheta_l g_l
    & \leq O(|\log g_j|).
  \end{align}
(The combination $\vartheta_j \gbar_j$ typically appears in our upper
bounds.)  \solref{tsequence}
\end{exercise}

Given $\mgen^2 \geq 0$, we define the \emph{mass domain}
\begin{equation} \label{e:Ijdef}
   \Iint_j(\mgen^2) = \begin{cases}
   [0,L^{-2j}] & (\mgen^2 = 0)\\
   [\tfrac{1}{2} \mgen^2, 2 \mgen^2] & (\mgen^2 > 0).
 \end{cases}
\end{equation}
The next exercise implies that any of the sequences $g_j,\gbar_j,t_j$
are comparable in value when evaluated at $m^2$ or $\mgen^2$ if $m^2
\in \Iint_j(\mgen^2)$.

\begin{exercise} \label{ex:gbarcomp}
  For $\mgen^2 \geq 0$ and $m^2\in \Iint_j(\mgen^2)$,
  each of the sequences $g_j,\gbar_j,t_j$ obeys
  \begin{equation}\label{e:gbarcomp}
    g_j(m^2) = g_j(\mgen^2) + O(g_j(\mgen^2)^2)
    .
  \end{equation}
  \solref{gbarcomp}
\end{exercise}

\subsection{Perturbative stable manifold}
\label{sec:mubar}

In this section, we obtain a simple 2-dimensional version of an
infinite-dimensional counterpart in the next section.  It is useful
for illustrative purposes, though we do not use the 2-dimensional
version later.  We do however use the following lemma both for the
two-dimensional and infinite-dimensional results.

\begin{lemma}
\label{lem:Piprod} Assume that the sequence $g$ satisfies the
recursion \eqref{e:gae} with $a_j = \beta_j$ and $M_j=M\vartheta_j$,
and with $e_j$ continuous in $m^2 \ge 0$.  For any fixed $\gamma \in
\R$, let
\begin{equation}
    \Pi_{i,j} = \prod_{k=i}^{j} (1 - \gamma \beta_kg_k).
\end{equation}
There exists $c_i=1+O(\vartheta_i\gbar_i)$, which is a continuous
function of $m^2 \ge 0$, such that
\begin{equation}
    \Pi_{i,j}
    =
    \left( \frac{g_{j+1}}{g_i}\right)^\gamma (c_i+O(\vartheta_j\gbar_j)).
\end{equation}
\end{lemma}

\begin{proof}
By Proposition~\ref{prop:gjtj} the sequences $g_j$ and $\gbar_j$ are
comparable; we use $\gbar_j$ for error terms.  Since $(1 - x)^\gamma =
(1 - \gamma x)(1 + O(x^2))$ as $x\rightarrow 0$, there exist $s_k =
O(\vartheta_k^2 \gbar_k^2)$ such that
\begin{align}
    \Pi_{i,j}
    &= \prod_{k=i}^{j} (1 - \beta_k g_k )^\gamma (1 + s_k) .
\end{align}
By \refeq{gae}, and since $g_{j+1} \in [\frac 12 g_j, 2 g_j]$ by
Proposition~\ref{prop:gjtj},
\begin{equation}
    1 - \beta_k g_k
    =
     \frac{g_{k+1} - e_{k}}{g_k}
    =
    \frac{g_{k+1}}{g_k}\left( 1 + O(\vartheta_k \gbar_k^2) \right).
\end{equation}
Therefore, there exist $v_k = O(\vartheta_k \gbar_k^2)$ such that
\begin{align}
    \Pi_{i,j}
    &
    = \prod_{k=i}^{j} \left(\frac{g_{k+1}}{g_{k}}\right)^\gamma (1 + v_k)
    = \left(\frac{g_{j+1}}{g_i}\right)^{\gamma} \prod_{k=i}^{j}(1 + v_k).
\end{align}
Since $\log(1 + x) = O(x)$, the product obeys
\begin{align}
    \prod_{k=i}^{j}(1 + v_k)
    &=
    \exp \left(\sum_{k=i}^{j} O(v_k)\right)
    = \exp \left(O(1) \, \sum_{k=i}^{j} \vartheta_k \gbar_k^2\right).
\end{align}
By \refeq{gbarsum}, the infinite product converges and we can define
\begin{align}
    c_i
    =
    \prod_{k=i}^{\infty}  (1 + v_k)
    =
    1+O(\vartheta_i\gbar_i).
\end{align}
We then obtain the desired formula for $\Pi_{i,j}$ from
\begin{align}
\prod_{k=i}^{j}(1 + v_k)
= c_i  \exp \left(-\sum_{k=j+1}^{\infty} \log (1 + v_k)\right)
= c_i + O(\vartheta_j\gbar_{j+1})
.
\end{align}

Finally, the continuity of $c_i$ in $m^2$ follows from the uniform
upper bound $v_k \le O((\gbar_k(0))^2)$ by
Exercise~\ref{ex:tsequence}, the continuity of the $\beta_k$ and
$r_{g,k}$ and therefore of the $g_k$ and $t_k$, and the dominated
convergence theorem.  This completes the proof.
\end{proof}

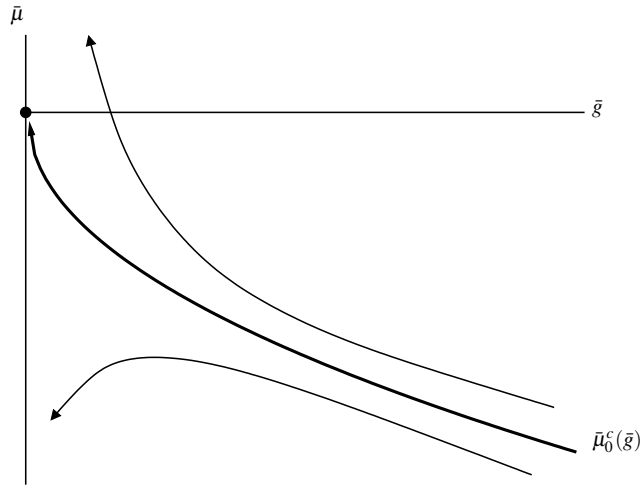
\begin{figure}
  \input{stable2d.pspdftex}
  \caption{Schematic depiction of the stable manifold
  for the perturbative flow $(\gbar_j,\mubar_j)$.}
  \label{fig:stable-manifold-bar}
\end{figure}

The next proposition constructs an initial condition $\mubar_0^c$ for
which the perturbative flow $(\gbar_j,\mubar_j)$ satisfies $\mubar_j
\to 0$.  For $m^2=0$, the set $(g,\mubar_0^c(g))$ plays the role of a
stable manifold for the fixed point $(0,0)$ of the dynamical system
$(\gbar,\mubar) \mapsto (\gbar_+,\mubar_+)$.  A schematic depiction of
the stable manifold is given in Figure~\ref{fig:stable-manifold-bar}.

\begin{prop}
\label{prop:barflow}
  Given $m^2\geq 0$, and given $g_0 >0$ sufficiently small,
  there exists a unique
  $\mubar_0^c = \mubar_0^c(g_0,m^2)$ such that
  if $\bar V_0 = (g_0, \mubar_0^c)$ then
  the solution to the recursion \refeq{gbar9}--\refeq{mubar9}
  satisfies
  \begin{equation}
    \gbar_j \to \gbar_\infty \geq 0, \quad \mubar_j \to 0. 
  \end{equation}
  More precisely, it obeys $\mubar_j = O(\vartheta_j \gbar_j)$, and
  $\gbar_\infty \sim c (\log m^{-2})^{-1}$ as $m^2 \downarrow 0$ (for some $c>0$).
\end{prop}

\index{Infrared asymptotic freedom}
\index{Asymptotic freedom}
In particular, if $m^2=0$, then $\gbar_\infty=0$ and $\bar V_j \to
0$. This is the famous observation of infrared \emph{asymptotic
freedom}, and inspires the prediction that scaling limits of the model
near the critical point are described by the free field.

\index{Stable manifold}

\begin{proof}[Proof of Proposition~\ref{prop:barflow}] Given an
initial condition $\gbar_0$, a sequence $\gbar_j$ is determined by
\refeq{gbar9}, and this sequence obeys the conclusions of
Proposition~\ref{prop:gjtj}.  For the sequence $\mubar_j$, we rewrite
the recursion \refeq{mubar9} backwards as
\begin{equation}
  \mubar_j = (1-\gamma\beta_j\gbar_j)^{-1} (
  L^{-2}\mubar_{j+1} -\eta_j\gbar_j + \xi_j \gbar_j^2 ).
\end{equation}
By iteration, with $\Pi_{i,j}$ given by Lemma~\ref{lem:Piprod}, it
follows that
\begin{equation}
  \mubar_j
  =
  L^{-2(k+1-j)} \Pi_{j,k}^{-1} \mubar_{k+1}
  +
  \sum_{l=j}^k L^{-2(l-j)} \Pi_{j,l}^{-1} (-\eta_l\gbar_l + \xi_l \gbar_l^2).
\end{equation}
Motivated by this, we define
\begin{equation}
  \lbeq{mubarsum}
  \mubar_j
  =
  \sum_{l=j}^\infty L^{-2(l-j)} \Pi_{j,l}^{-1} (-\eta_l\gbar_l + \xi_l \gbar_l^2).
\end{equation}
Since $\gbar_l = O(1)$ and, by Lemma~\ref{lem:Piprod},
$\Pi_{j,l}^{-1}$ is slowly varying compared with $L^{-2(l-j)}$, the
above sum converges, and $\mubar_j = O(\vartheta_j\gbar_j)$.  It is
easy to check that $\mubar_j$ given by \refeq{mubarsum} obeys the
recursion \refeq{mubar9}, and that $\mubar_0$ is the unique initial
value that leads to a zero limit for the sequence.
\end{proof}

Given any initial condition $(\gbar_0,\mubar_0)$, the equations
\refeq{gbar9}--\refeq{mubar9} can be solved by forward iteration.
This defines sequences $\gbar_j,\mubar_j$ for arbitrary initial
conditions, and the sequence $\mubar_j$ can be differentiated with
respect to $\mubar_0$.  This derivative is considered in the next
proposition; its value is independent of the initial condition
$\mubar_0$.

\begin{prop} \label{prop:mubard}
  Given any small $\gbar_0>0$,
  \begin{equation}
  \lbeq{muder}
    \ddp{\mubar_j}{\mubar_0}
    = L^{2j} \left(\frac{\gbar_j}{\gbar_0}\right)^{\gamma}(c+O( \vartheta_j \gbar_j))
    \quad \text{with} \quad c= 1+O(\gbar_0).
  \end{equation}
  Consequently, there exists $c'>0$ such that
  \begin{equation}
  \lbeq{muderlim}
    \lim_{j\to\infty} L^{-2j} \ddp{\mubar_j}{\mubar_0} \sim c' (\log m^{-2})^{\gamma} \quad \text{as $m^2\downarrow 0$.}
  \end{equation}
\end{prop}

\begin{proof}
By the chain rule, $\ddp{\mubar_j}{\mubar_0} = \prod_{k=0}^{j-1}
\ddp{\mubar_{k+1}}{\mubar_{k}}$.  We compute the factors in the
product by differentiating the recursion relation \refeq{mubar9} for
$\mubar$.  Since $\gbar_j$ is independent of $\mubar$, we obtain
\begin{align}
\ddp{\mubar_j}{\mubar_0}
= \prod_{k=0}^{j-1} L^2 (1 - \gamma \beta_k \gbar_k).
\end{align}
Now we apply Lemma~\ref{lem:Piprod} for \refeq{muder}, and
Proposition~\ref{prop:gjtj}(ii) for \refeq{muderlim}.
\end{proof}

\section{Reduction of proof of Theorem~\ref{thm:phi4-hier-chi}}
\label{sec:reduction}

In this section, we prove Theorem~\ref{thm:phi4-hier-chi} subject to
Theorem~\ref{thm:VK} and Proposition~\ref{prop:N}.
Theorem~\ref{thm:VK} is a non-perturbative versions of
Propositions~\ref{prop:barflow}--\ref{prop:mubard}, with no
uncontrolled remainder.  Proposition~\ref{prop:N} is a relatively minor
result which incorporates the effect of the last renormalisation group step,
corresponding to the Gaussian integration with covariance $\Cnewlast{}$.
Their proofs occupy the rest of the book.

Throughout this section, we fix
\begin{equation}
    g_0=g
\end{equation}
and drop $g$ from the notation when its role is insignificant.
Our starting point is \refeq{susceptZN-hier-bis}, which asserts that
for $m^2>0$ and for
\begin{equation}
\lbeq{nu0m2}
    \nu_0=\nu-m^2,
\end{equation}
the susceptibility is given by
\begin{align} \label{e:susceptZN}
  \chi_N(\nu)
  &=
  \frac{1}{m^2} + \frac{1}{m^4 |\Lambda|} \frac{D^2 \Znewlast(0;\1,\1)}{\Znewlast(0)}
  ,
\end{align}
with $\Znewlast = \Ex_C\theta Z_0$ and $Z_0 = e^{-\sum_{x\in
\Lambda}(g_0\tau_x^2 + \nu_0\tau_x)}$.  As discussed below
\refeq{susceptZN-hier-bis}, we can regard the right-hand side as a
function of two independent variables $(m^2,\nu_0)$, without enforcing
\refeq{nu0m2}, even though the equality in \refeq{susceptZN} is
guaranteed only when \refeq{nu0m2} does hold.  We define a function
$\hat\chi_N(m^2,\nu_0)$ by the right-hand side of \eqref{e:susceptZN}
with \emph{independent} variables $(m^2,\nu_0)$.  Thus $\hat\chi_N$ is
a function of two variables (with dependence on $g$ left implicit),
and
\begin{equation}
\lbeq{chichihatN}
    \chi_N(\nu_0+m^2) = \hat\chi_N(m^2,\nu_0).
\end{equation}

To prove Theorem~\ref{thm:phi4-hier-chi}, the general strategy is to
prove that for $m^2 \ge 0$ there is a \emph{critical} initial value
$\nu_0=\nu_0(m^2)$ (depending also on $g_0$ but independent of the
volume parameter $N$) such that starting from the initial condition
$V_0= g_0 \tau^2 + \nu_0 \tau$ and $K_0=0$ it is possible to iterate
the renormalisation group map indefinitely.  This iteration produces a
sequence $(U_j,K_j)=(u_j,V_j,K_j)$ which represents $Z_j$ via
\refeq{ZZ+} as long as $j\le N$.  The sequence
$(U_j,K_j)=(u_j,V_j,K_j)$ is \emph{independent} of $N$ for $j \le N$,
and thus in the limit $N \to \infty$ is a global renormalisation group
trajectory.  For finite $N$, $(U_N,K_N)$ represents $Z_N$.  Finally,
there is the step of \eqref{e:Zj-def} which is the first and only step
where a finite volume system deviates from the global trajectory. This
step maps $Z_{N}$ to $\Znewlast$ with $\Znewlast$ represented by
$(\Unewlast,\Knewlast)=(\unewlast,\Vnewlast,\Knewlast)$ from which
$\hat\chi_N(m^2,\nu_0)$ is computed with \eqref{e:susceptZN}.  The
critical initial value is intimately related to the critical point
$\nu_c$.  The global trajectory has the property that $V_j$ and $K_j$
both go to zero as $j \to \infty$, which is \emph{infrared asymptotic
freedom}.  This property characterises $\nu_0(m^2)$ uniquely.

\begin{figure}
  \input{rgphaseportrait-small.pspdftex}
  \caption{Phase diagram for the dynamical system.}
  \label{fig:phase}
\end{figure}
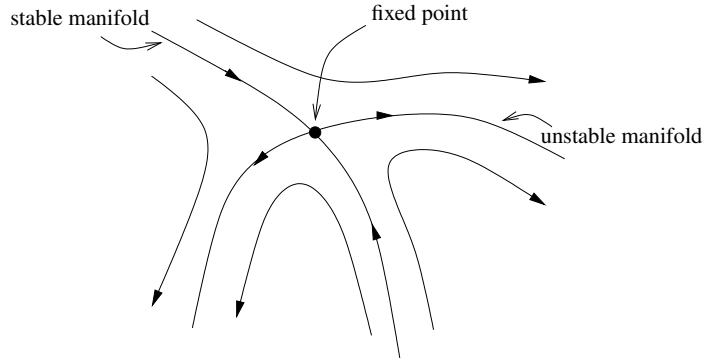

Given $m^2\ge 0$, we can regard $\nu_0$ as a function $\nu_0(g_0)$ of
the initial value $g_0=g$.  The construction of $\nu_0(g_0)$
corresponds schematically to the construction of the stable manifold
depicted in Figure~\ref{fig:stable-manifold-bar} for the perturbative
flow.  However, the dynamical system here is more elaborate than the
perturbative 2-dimensional dynamical system.  Now it is instead
infinite-dimensional due to the presence of the non-perturbative
coordinate $K_j$, and it is also non-autonomous because $K_j$ lies in
different spaces $\Fcal_j$ (which will be equipped with different
norms) for different values of $j$.  The dynamical system is
nonhyperbolic, with expanding coordinate $\mu_j$, contracting
coordinate $K_j$, and with coordinate $g_j$ which is neither
contracting nor expanding.  Its local phase diagram is shown
schematically in Figure~\ref{fig:phase}.  The fixed point is
$(g,\mu,K)=(0,0,0)$.  Given small $g_0$ and $K_0$, the flow of the
dynamical system is towards $(0,0,0)$ when $\nu_0=\nu_0(g_0,K_0)$ is
chosen correctly (we focus on the case $K_0=0$ which is the only case
we need).  This choice defines the stable manifold, which has
co-dimension $1$ corresponding to the variable $\mu_j$.  If $\nu_0$
were chosen off the stable manifold, the flow of $\mu_j$ would explode
exponentially taking the trajectory outside the domain of our RG map.

The sequence $U_j$ is determined recursively from
\begin{equation}
    U_{j+1}(U_j,K_j) = u_j + U_{j+1}(V_j,K_j) = u_j + \Phi_{j+1}^U(V_j,K_j),
\end{equation}
with $U_{j+1}(V_j,K_j) = \Phi_\pt(V_j - \LT (e^{V_j}K_j))$ as in
\refeq{U+def}.  We have already analysed the map $\Phi_\pt$ explicitly
and in detail.  It is defined by \refeq{Phiptdef}, and its explicit
quadratic form is given in Proposition~\ref{prop:barflow-a}.  Thus, to
understand the sequence $U_j$, it suffices to analyse the sequence
\begin{equation}
\label{e:RUdef-5a}
    R_{j+1}^U(V,K) = (r_{g,j},r_{\nu,j},r_{u,j})
\end{equation}
defined by
\begin{equation}
\label{e:RUdef-5}
    R_{j+1}^U(V_j,K_j) = \Phi_{j+1}^U(V_j,K_j) - \Phi_{\pt}(V_j) .
\end{equation}

The following is a non-perturbative version of
Propositions~\ref{prop:barflow}--\ref{prop:mubard}.  Its proof is given in
Section~\ref{sec:pfprops}.

\begin{theorem}
\label{thm:VK}
Fix $L$ sufficiently large and $g_0>0$ sufficiently small.

\smallskip
\noindent
(i)
There exists a continuous function $\nu_0^c(m^2)$ of $m^2\geq 0$
(depending on $g_0$) such that if $\nu_0 = \nu_0^c(m^2)$ then,
for all $j \in \N$,
\begin{equation}
  \label{e:VKN-Rj-bis}
  r_{g,j} = O(\vartheta_j^3g_j^3)   , \quad
  L^{2j} r_{\nu,j} = O(\vartheta_j^3 g_j^3) , \quad
  L^{\drb j} r_{u,j} = O(\vartheta_j^3 g_j^3),
\end{equation}
and
\begin{equation} \label{e:VKN-Kj-bis}
  L^{2j}|\nu_j| = O(\vartheta_j g_j),\qquad
  |K_j(0)| + L^{-2j} |D^2K_j(0;\1,\1)|
  = O(\vartheta_j^3 g_j^3).
\end{equation}

\noindent (ii)
There exists $c=1+O(g_0)$ such that for $m^2 \geq 0$ and $j \in \N$, and
with all derivatives evaluated at $(m^2,\nu_0^c(m^2))$,
\begin{equation} \label{e:nuNp-5}
  \ddp{\mu_j}{\nu_0}
  =
  L^{2j} \left(\frac{g_j}{g_0}\right)^{\gamma}\big(c+O(\vartheta_jg_j)\big),
  \quad
  \ddp{g_j}{\nu_0} = O\left(g_j^{2}  \ddp{\mu_j}{\nu_0} \right),
\end{equation}
\begin{equation} \label{e:gKNp-5}
  L^{-2j} \left|\ddp{}{\nu_0} K_j(0)\right|
  +
  L^{-4j} \left|\ddp{}{\nu_0} D^2 K_j(0; \1,\1)\right|
  = O\left(\vartheta_j^{3} g_j^2\left(\frac{g_j}{g_0}\right)^{\gamma}\right)
  .
\end{equation}
\end{theorem}

From the first bound in \eqref{e:VKN-Rj-bis} and Proposition~\ref{prop:gjtj},
it follows that
\begin{equation} \label{e:ginfty-lim-new}
  \lim_{j \to \infty} g_j(m^2) = g_\infty(m^2) \quad (m^2 \geq 0),
\end{equation}
where $g_\infty(0)=0$,
$g_\infty(m^2)>0$ if $m^2>0$,
and the  limit is uniform on compact intervals of $m^2>0$.
Also by Proposition~\ref{prop:gjtj},
\begin{equation}
\label{e:ginfty-asymp-new}
    g_\infty(m^2) \sim \frac{1}{\beta_0^0 \log_L m^{-2}}  \quad (m^2 \downarrow 0).
\end{equation}

The first $N$ renormalisation group steps
correspond to integration over the
covariances $C_1 + \cdots + C_N$. In finite volume, we are left with
the final covariance $\Cnewlast{}$.
Unlike the other covariances, it has $\sum_y \Cnewlast{;xy} \neq 0$.
The next proposition shows that its contribution is negligible.
The proof, which requires only slight modifications to
the analysis of a typical renormalisation group step, is given in Section~\ref{sec:last-step}.

\begin{prop}
\label{prop:N}
Fix $L$ sufficiently large and $g_0>0$ sufficiently small, and suppose that
$m^2L^{2N} \ge 1$ (i.e., that the last scale $N$ is beyond the mass scale $j_m$).
The expectations in Proposition~\ref{prop:Elast} do exist
at scale $N$ for $(V,K)=(V_N,K_N)$
and \refeq{Elast5} holds, i.e.,
\begin{equation}
\lbeq{ElastN}
    \Ex_{\Cnewlast{}} \left(e^{-\theta V_N(\Lambda)} +K_N(\Lambda)\right)
    =
    e^{-\unewlast |\Lambda|}
    \left(e^{-\Vnewlast(\Lambda)}\big(1+\Wnewlast(\Lambda)\big) + \Knewlast(\Lambda)\right),
\end{equation}
with $\Wnewlast = -\frac 12 c_{\hat{N}}^{(1)}g_N^2 |\varphi|^6$.
The estimates \refeq{VKN-Kj-bis}--\refeq{gKNp-5}  hold with $j$ replaced by
$j=\hat{N}$ on the left-hand
sides and $j=N$ on the right-hand sides, and $\gnewlast = g_N(1+O(\vartheta_Ng_N))$.
\end{prop}

The following corollary shows that the susceptibility
$\chi(\nu_0^c(m^2) + m^2)$ is simply the susceptibility
of the hierarchical GFF with mass $m$ (recall
Exercise~\ref{ex:hier-freechi}).  Thus $\nu_0^c(m^2)$ has the property
that
\begin{equation}
    \label{e:effective}
    \chi(g,\nu_0^c(m^2) + m^2) = \chi(0,m^2) .
\end{equation}
In other words, $m^2$ represents the deviation from the critical value
$\nu_0^c(m^2,g)$ such that the susceptibility of the interacting model
is equal to the susceptibility of the noninteracting model with mass
$m^2$. Physicists refer to $m^{2}$ as an \emph{effective} or
\emph{renormalised mass} and $\nu_0^c(m^2) + m^2$ as a \emph{bare mass}.

\begin{cor} \label{cor:chihatm}
Fix $L$ sufficiently large and $g>0$ sufficiently small, and let $m^2 > 0$
and $\nu_0 = \nu_0^c(m^2)$.
  The limit $\chi(\nu_0^c(m^2)+ m^2)= \lim_{N\to\infty} \chi_N(\nu_0^c(m^2)+ m^2)$ exists,
  uniformly on compact intervals of $m^2 > 0$,
  and
  \begin{equation} \label{e:chihatm}
    \chi(\nu_0^c(m^2) + m^2) = \frac{1}{m^2}.
  \end{equation}
\end{cor}

\begin{proof}
  We will in fact prove the finite volume estimate
  \begin{equation}
  \lbeq{chiNhatm}
    \chi_N(\nu_0^c(m^2)+m^2) = \frac{1}{m^2} \left( 1 + \frac{O(\vartheta_Ng_N)}{m^2L^{2N}}\right),
  \end{equation}
  which implies \refeq{chihatm}.
  By \refeq{susceptZN}, to prove \refeq{chiNhatm}
  it is sufficient to show that, for $\nu_0 = \nu_0^c(m^2)$,
  \begin{equation}
  \label{e:I0}
    \frac{1}{|\Lambda|}\frac{D^2 \Znewlast(0;\1,\1)}{\Znewlast(0)}
    =
    O(L^{-2N}\vartheta_Ng_N).
  \end{equation}
  At scale $N$, $\Lambda$ is a single block,
  so $\Znewlast = e^{-\unewlast|\Lambda|}(\Inewlast+\Knewlast)$,
  where by definition $\Inewlast = e^{-\Vnewlast(\Lambda)}(1+\Wnewlast(\Lambda))$.
  By Proposition~\ref{prop:N},
  $\Wnewlast$ is proportional to $g_N^2 |\varphi|^6$.
  Since $\Vnewlast(\Lambda) = \Wnewlast (\Lambda) = 0$ when $\varphi=0$,
  we have $\Inewlast = 1$ when $\varphi=0$.  Also,
  \begin{equation} \label{e:D2Ilast}
    \frac{1}{|\Lambda|} D^2\Inewlast(\varphi=0;\1,\1)
    = -\nunewlast
    ,
  \end{equation}
  and hence
  \begin{equation}
  \label{e:I0-pf}
    \frac{1}{|\Lambda|}\frac{D^2 \Znewlast(0;\1,\1)}{\Znewlast(0)}
    =
    \frac{-\nunewlast + L^{-\drb N}D^2 \Knewlast(0;\1,\1)}{1+\Knewlast(0)}
    = O(L^{-2N} \vartheta_N g_N),
  \end{equation}
  where the final estimate holds
  by Proposition~\ref{prop:N}
  (with the final scale version of \refeq{VKN-Kj-bis}).
  This proves \refeq{I0}.
  The convergence is uniform in compact intervals of $m^2>0$,
  due to the factor $L^{-2N}$.
  This completes the proof.
\end{proof}

\begin{cor} \label{cor:chihatmp}
Fix $L$ sufficiently large and $g>0$ sufficiently small.
There exists $B = B_{g,n} > 0$ such that
\begin{equation} \label{e:chihatp}
  \left. \ddp{\chi}{\nu}\right|_{\nu=\nu_0^c(m^2)+m^2} \sim - B \frac{1}{m^4(\log m^{-2})^\gamma}
  \quad \text{as $m^2\downarrow 0$}.
\end{equation}
\end{cor}

\begin{proof}
By \refeq{chichihatN}, the finite-volume version $\ddp{}{\nu}\chi_{N}$ of the left-hand side of
\eqref{e:chihatp} is equal to $\ddp{}{\nu_0}\hat\chi_{N}(m^2,\nu_0)$ evaluated at $\nu_0=\nu_0^c(m^2)$.
All $\nu_0$ derivatives in the proof are evaluated at this value, and we denote them by primes.
We compute $\ddp{}{\nu_0}\hat\chi_{N}(m^2,\nu_0)$ by differentiation of
the right-hand side of \refeq{susceptZN}, using
$\Znewlast = e^{-\unewlast|\Lambda|}(\Inewlast + \Knewlast)$ with $\Inewlast$ as in the proof
of Corollary~\ref{cor:chihatm}.
This gives
\begin{align}
    \ddp{\chi_N}{\nu}
    &=
    \frac{1}{m^4 L^{\drb N}}
    \ddp{}{\nu_0}\frac{D^2 \Znewlast(0; \1,\1)}{\Znewlast(0)}
    \nnb &
    =
    \frac{1}{m^4 L^{\drb N}}
    \left(
    \frac{D^2 \Znewlast'(0; \1,\1)}{\Znewlast(0)}
    -
    \frac{\Znewlast'(0) D^2 \Znewlast(0; \1,\1)}{\Znewlast(0)^2}
    \right).
\end{align}
The factor
$e^{-\unewlast|\Lambda|}$ cancels in numerator and denominator of
\refeq{susceptZN}, and in particular need not be differentiated.
We view $\Vnewlast$ and $\Wnewlast$ as functions of $\nu_0$.
As in \eqref{e:D2Ilast},
\begin{equation}
  \Inewlast(\varphi=0) = 1,
  \quad
  \frac{1}{|\Lambda|} D^2 \Inewlast(\varphi=0; \1,\1)
  =
  -\nunewlast,
\end{equation}
and hence
\begin{equation}
  \Inewlast'(\varphi=0) = 0,
  \quad
  \frac{1}{|\Lambda|} D^2\Inewlast'(\varphi=0;\1,\1)
  = - \nunewlast' .  
\end{equation}
With some arguments omitted to simplify the notation, this leads to
\begin{equation}
    \ddp{\chi_N}{\nu}
    =
    \frac{1}{m^4 }
    \left(
    \frac{-\nunewlast' + L^{-\drb N}D^2 \Knewlast'}{1+\Knewlast}
    -
    \frac{\Knewlast' (-\nunewlast + L^{-\drb N}D^2 \Knewlast)}{(1+\Knewlast)^2}
    \right).
\end{equation}
By Proposition~\ref{prop:N}, as $N\to\infty$ the derivative
$\ddp{\chi_N}{\nu}$ has the same limit as $-m^{-4}\nunewlast'$,
and the omitted terms go to zero uniformly on compact
intervals in $m^2>0$ because $\vartheta_N$ does.  Therefore,
by Proposition~\ref{prop:N} with \refeq{nuNp-5}, and by \refeq{ginfty-lim-new},
\begin{equation}
  \lim_{N\to\infty}
  \ddp{\chi_N}{\nu}
  =
  - \frac{c}{m^4} \left(\frac{g_\infty}{g_0} \right)^{\gamma}.
\end{equation}
The limit is again uniform on compact mass intervals, since the same is true of
the limit in \refeq{ginfty-lim-new}.

Since
\begin{equation} \label{e:chihatchi}
    \ddp{}{\nu}\chi_N(\nu)
    =
    \ddp{}{\nu_0}\hat\chi_N (m^2,\nu_0^c(m^2)),
\end{equation}
and since $\nu_0^c$ is a continuous function of $m^2$, the limit
$\lim_{N\to\infty} \ddp{}{\nu}\chi_N(\nu)$
converges uniformly in compact intervals of $\nu$ in the image of $m^2+\nu_0^c(m^2,g_0)$
for $m^2>0$.
Therefore the differentiation and the limit may be interchanged, so that
\begin{equation}
    \ddp{}{\nu}\chi(\nu)
    =
    \lim_{N\to\infty} \ddp{}{\nu}\chi_N(\nu)
    =
    - \frac{c}{m^4}
    \left(\frac{g_\infty}{g_0} \right)^{\gamma} .
\end{equation}
By \refeq{ginfty-asymp-new},
there is a positive constant $B_{g,n}$ such that
\begin{equation}
\lbeq{Bappears}
    c
    \left(\frac{g_\infty}{g_0} \right)^{\gamma}
    \sim
    \frac{B_{g,n}}{(\log m^{-2})^\gamma}
    \quad
    (m^2 \downarrow 0).
\end{equation}
This proves \refeq{chihatp}, and the proof is complete.
\end{proof}

\label{sec:hierchi}
\index{Susceptibility}
\index{Logarithmic corrections}

Finally, we need the next lemma which establishes that
$\nu_0^c(m^2)+m^2$ is an increasing function of small $m^2$.

\begin{lemma}
\label{lem:nu0cont}
Fix $L$ sufficiently large and $g>0$ sufficiently small.
For $\delta>0$ sufficiently small and for $m^2 \in [0,\delta]$,
$\nu_0^c(m^2)+m^2$ is a continuous increasing function of $m^2$.
\end{lemma}

\begin{proof}
Set $\nu^*(m^2) = \nu_0^c(m^2)+m^2$.
The continuity of $\nu^*$ in $m^2\in [0,\delta)$ is immediate from
Theorem~\ref{thm:VK}.
For $m^2>0$, \eqref{e:chihatm} and \eqref{e:chihatp} imply
\begin{align}
\lbeq{chistar}
  \chi(\nu^*(m^2))
  &
    = \frac{1}{m^2} < \infty,
  \\
\lbeq{chiprimestar}
  \ddp{}{\nu} \chi(\nu^*(m^2))
  &
    < 0.
\end{align}
We used the hypothesis $m^{2} \in [0,\delta]$ with $\delta$ small to obtain \eqref{e:chiprimestar}.
Let $I= \{\nu^*(m^2): m^2 \in [0,\delta]\}$.
By continuity of $\nu^*$ in $m^2$, and since continuous
functions map an interval to an interval, $I$ is an interval (which
cannot be a single point due to \refeq{chistar}).  Since $\chi(\nu)$
is decreasing in $\nu\in I$ for small $m^2$ by \refeq{chiprimestar},
and since the composition $\chi(\nu^*(m^2)) = \frac{1}{m^2}$ is
decreasing in $m^2 >0$, it follows that $\nu^*(m^2)$ is increasing in
small $m^2$.
\end{proof}

Now we can complete the proof of Theorem~\ref{thm:phi4-hier-chi},
subject to
Theorem~\ref{thm:VK} and Proposition~\ref{prop:N},
using
Corollaries~\ref{cor:chihatm} and \ref{cor:chihatmp} and Lemma~\ref{lem:nu0cont}.

\begin{proof}[Proof of Theorem~\ref{thm:phi4-hier-chi}]
Define
\begin{equation}
  \nu_c = \nu_0^c(0).
\end{equation}
By Lemma~\ref{lem:nu0cont},
the function $m^2 \mapsto \nu_0^c(m^2)+m^2$ is continuous and increasing
as a function of $m^2\in [0,\delta]$.
It therefore has a continuous inverse.
Its range is a closed interval of the
form $[\nu_c,\nu_c+ \varepsilon']$
for some $\varepsilon'>0$.
The inverse map associates to each $\nu=\nu_c+\varepsilon$,
for $\varepsilon \in [0,\varepsilon']$, a unique $m^2$.
Using this relationship, we see from \eqref{e:chihatp} and \refeq{chihatm} that,
as $m^2 \downarrow 0$ or equivalently $\nu\downarrow \nu_c$,
\begin{equation} \label{e:chipdr}
  \ddp{}{\nu} \chi(g,\nu)
  \sim - B \frac{1}{m^4 (\log m^{-2})^\gamma}
  \sim - B \chi(g,\nu)^2 (\log \chi(g,\nu))^{-\gamma}.
\end{equation}
It is now an exercise in calculus to deduce that, as $\varepsilon \downarrow 0$,
\begin{equation}
\label{e:thmphi4-hier-chi-pf}
  \chi(g,\nu_c+\varepsilon)
  \sim
  \frac{1}{B} \frac{1}{\varepsilon}(\log \varepsilon^{-1})^{\gamma}
  .
\end{equation}
This proves
\eqref{e:thmphi4-hier-chi} with $A=B^{-1}$.

The constant $B$ arises in \refeq{Bappears}, and by Theorem~\ref{thm:VK},
\begin{equation}
    B=(1+O(g))(\log L)^\gamma (g\beta_0^0)^{-\gamma}
\end{equation}
with $\beta_0^0=(n+8)(1-L^{-d})$
given by Lemma~\ref{lem:Greekbds}.  This proves that $A=B^{-1} \sim (g\beta_0^0/(\log L))^\gamma$,
as claimed in \refeq{Anuc-hier}.

It remains to prove the asymptotic formula for the critical point in \refeq{Anuc-hier}.
For the rest of the proof, we set $m^2=0$.
To begin, we note that it follows from
Proposition~\ref{prop:barflow-a}, \refeq{RUdef-5} and \refeq{VKN-Rj-bis} that
$g_j$ and $\mu_j$ are determined recursively from
\begin{align}
  \label{e:grec-bis}
  g_{j+1} &= g_j - \beta_j g_j^2 + O(\vartheta_j^3 g_j^3),
  \\
  \label{e:murec}
  \mu_{j+1} &= L^2 \left(\mu_j(1-\gamma \beta_j g_j) + \eta_j g_j - \xi_j g_j^2 \right)
              + O(\vartheta_j^3g_j^3)
  ,
\end{align}
with initial condition $(g_0,\mu_0) = (g,\nu_c)$.
Just as \refeq{mubarsum} defines a solution to the perturbative flow with zero
final condition, backwards solution of \refeq{murec} gives
\begin{equation} \label{e:mubar0}
  \mu_0
  =
  - \sum_{l=0}^\infty
  L^{-2l} \Pi_{0,l}^{-1}
  (\eta_l g_l + O(\vartheta_l\gbar_l^2)).
\end{equation}
By \refeq{Greekdef}, \refeq{Greekdef2}, and \refeq{cjdef},
\begin{equation}
\lbeq{etadef}
    \eta_l = (n+2)L^{2l}C_{l+1;0,0}(0).
\end{equation}
By Lemma~\ref{lem:Piprod}, we obtain from \eqref{e:mubar0}--\eqref{e:etadef} that
\begin{equation}
  \mu_0 =
  - (n+2) (1+O(g_0))g_0^\gamma \sum_{l=0}^\infty
  \left( C_{l+1;0,0} g_l^{1-\gamma} + O(L^{-2l} \gbar_l^{2-\gamma})\right).
\end{equation}
Since $C(0) = \sum_{l=0}^\infty C_{l+1;0,0}$, this gives
\begin{align} \label{e:mubar0r}
  \mu_0
  &= - (n+2)C(0) g_0 (1+O(g_0))
  - (n+2)(1+O(g_0))g_0^\gamma\sum_{l=0}^\infty C_{l+1;0,0} (g_l^{1-\gamma}-g_0^{1-\gamma})
  \nnb & \qquad\qquad
  + g_0^\gamma \sum_{l=0}^\infty  O(L^{-2l}\gbar_l^{2-\gamma}).
\end{align}
We show that the last two terms are $O(g_0^2)$.
This suffices, as it gives the desired result
\begin{equation}
  \mu_0
  = - (n+2)C(0) g_0 + O(g_0^2).
\end{equation}

The last term in \refeq{mubar0r} is $O(g_0^2)$, since
$g_l = O(g_0)$ by Proposition~\ref{prop:gjtj}(i).
For the more substantial sum in \refeq{mubar0r}, by Taylor's theorem,
and again using the fact $g_l=O(g_0)$, for any $\gamma < 1$  we have
\begin{equation}
    g_l^{1-\gamma}-g_0^{1-\gamma}
    =
    (g_l-g_0) O(g_0^{-\gamma}).
\end{equation}
By the recursion  \eqref{e:grec-bis}  for $g_l$ and $C_{l+1;0,0} = O(L^{-2l})$, this gives
\begin{align}
  g_0^{\gamma}\sum_{l=0}^\infty C_{l+1;0,0} (g_l^{1-\gamma}-g_0^{1-\gamma})
  &=
  O(1)
  \sum_{l=0}^\infty L^{-2l}
  \sum_{k=0}^{l} \vartheta_k g_k^{2}
  = O(g_0^2).
\end{align}
This completes the proof.
\end{proof}

\begin{exercise}
\label{ex:chi-asy}
  Use \eqref{e:chipdr} to prove \eqref{e:thmphi4-hier-chi-pf}.
\solref{chi-asy}
\end{exercise}



%% file: RG0.pspdftex
\begin{picture}(0,0)%
\includegraphics{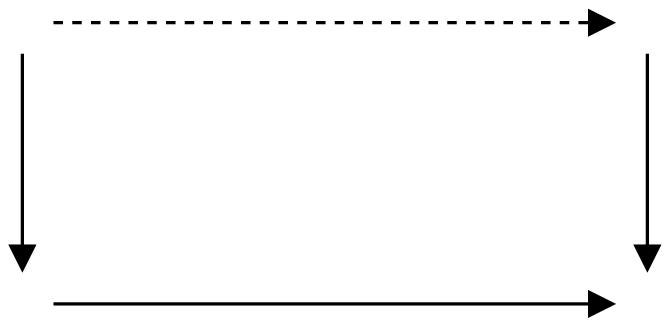}%
\end{picture}%
\setlength{\unitlength}{3947sp}%
\begingroup\makeatletter\ifx\SetFigFont\undefined%
\gdef\SetFigFont#1#2#3#4#5{%
  \reset@font\fontsize{#1}{#2pt}%
  \fontfamily{#3}\fontseries{#4}\fontshape{#5}%
  \selectfont}%
\fi\endgroup%
\begin{picture}(3330,1738)(2386,-6800)
\put(2401,-5761){\makebox(0,0)[rb]{\smash{{\SetFigFont{9}{10.8}{\familydefault}{\mddefault}{\updefault}{\color[rgb]{0,0,0}$\prod_{b\in\Bcal}$}%
}}}}
\put(5701,-5761){\makebox(0,0)[lb]{\smash{{\SetFigFont{9}{10.8}{\familydefault}{\mddefault}{\updefault}{\color[rgb]{0,0,0}$\prod_{B\in\Bcal_{+}}$}%
}}}}
\put(2633,-5217){\makebox(0,0)[rb]{\smash{{\SetFigFont{9}{10.8}{\familydefault}{\mddefault}{\updefault}{\color[rgb]{0,0,0}$F(b)$}%
}}}}
\put(5469,-5209){\makebox(0,0)[lb]{\smash{{\SetFigFont{9}{10.8}{\familydefault}{\mddefault}{\updefault}{\color[rgb]{0,0,0}$F_{+}(B)$}%
}}}}
\put(5492,-6567){\makebox(0,0)[lb]{\smash{{\SetFigFont{9}{10.8}{\familydefault}{\mddefault}{\updefault}{\color[rgb]{0,0,0}$Z_{+}$}%
}}}}
\put(2611,-6561){\makebox(0,0)[rb]{\smash{{\SetFigFont{9}{10.8}{\familydefault}{\mddefault}{\updefault}{\color[rgb]{0,0,0}$Z$}%
}}}}
\put(4051,-6736){\makebox(0,0)[b]{\smash{{\SetFigFont{9}{10.8}{\familydefault}{\mddefault}{\updefault}{\color[rgb]{0,0,0}$\mathbb{E}_{+}\theta$}%
}}}}
\end{picture}%

%% file: RG1.pspdftex
\begin{picture}(0,0)%
\includegraphics{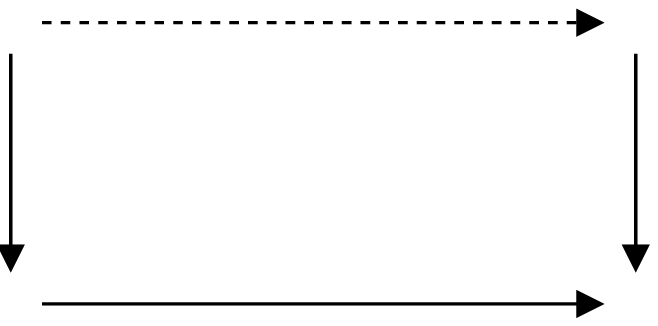}%
\end{picture}%
\setlength{\unitlength}{3947sp}%
\begingroup\makeatletter\ifx\SetFigFont\undefined%
\gdef\SetFigFont#1#2#3#4#5{%
  \reset@font\fontsize{#1}{#2pt}%
  \fontfamily{#3}\fontseries{#4}\fontshape{#5}%
  \selectfont}%
\fi\endgroup%
\begin{picture}(3104,1745)(2499,-6800)
\put(4051,-5386){\makebox(0,0)[b]{\smash{{\SetFigFont{9}{10.8}{\familydefault}{\mddefault}{\updefault}{\color[rgb]{0,0,0}$\Phi_{+}$}%
}}}}
\put(4051,-6736){\makebox(0,0)[b]{\smash{{\SetFigFont{9}{10.8}{\familydefault}{\mddefault}{\updefault}{\color[rgb]{0,0,0}$\mathbb{E}_{C_{+}}\theta (\cdot)^B$}%
}}}}
\put(2640,-5202){\makebox(0,0)[rb]{\smash{{\SetFigFont{9}{10.8}{\familydefault}{\mddefault}{\updefault}{\color[rgb]{0,0,0}$\big(U,K\big)$}%
}}}}
\put(5476,-5202){\makebox(0,0)[lb]{\smash{{\SetFigFont{9}{10.8}{\familydefault}{\mddefault}{\updefault}{\color[rgb]{0,0,0}$\big(U_{+},K_{+}\big)$}%
}}}}
\put(5489,-6566){\makebox(0,0)[lb]{\smash{{\SetFigFont{9}{10.8}{\familydefault}{\mddefault}{\updefault}{\color[rgb]{0,0,0}$F_{+}$}%
}}}}
\put(2600,-6559){\makebox(0,0)[rb]{\smash{{\SetFigFont{9}{10.8}{\familydefault}{\mddefault}{\updefault}{\color[rgb]{0,0,0}$F$}%
}}}}
\end{picture}%

%% file: stable2d.pspdftex
\begin{picture}(0,0)%
\includegraphics{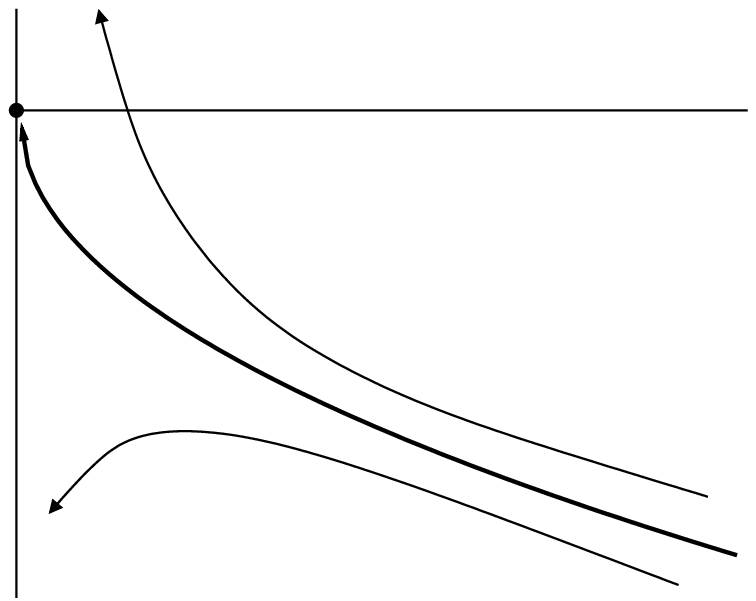}%
\end{picture}%
\setlength{\unitlength}{2565sp}%
\begingroup\makeatletter\ifx\SetFigFont\undefined%
\gdef\SetFigFont#1#2#3#4#5{%
  \reset@font\fontsize{#1}{#2pt}%
  \fontfamily{#3}\fontseries{#4}\fontshape{#5}%
  \selectfont}%
\fi\endgroup%
\begin{picture}(5655,4669)(1486,-7583)
\put(7126,-3961){\makebox(0,0)[lb]{\smash{{\SetFigFont{8}{9.6}{\familydefault}{\mddefault}{\updefault}{\color[rgb]{0,0,0}$\bar g$}%
}}}}
\put(7126,-7186){\makebox(0,0)[lb]{\smash{{\SetFigFont{8}{9.6}{\familydefault}{\mddefault}{\updefault}{\color[rgb]{0,0,0}$\bar\mu_0^c(\bar g)$}%
}}}}
\put(1501,-3061){\makebox(0,0)[lb]{\smash{{\SetFigFont{8}{9.6}{\familydefault}{\mddefault}{\updefault}{\color[rgb]{0,0,0}$\bar\mu$}%
}}}}
\end{picture}%

%% file: rgphaseportrait-small.pspdftex
\begin{picture}(0,0)%
\includegraphics{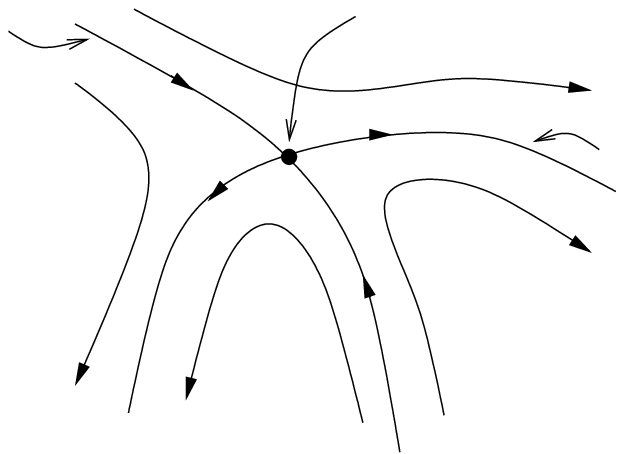}%
\end{picture}%
\setlength{\unitlength}{3108sp}%
\begin{picture}(5767,2826)(5251,-4843)
\put(8146,-2176){\makebox(0,0)[lb]{fixed point}%
}
\put(5266,-2176){\makebox(0,0)[lb]{stable manifold}%
}
\put(9496,-3121){\makebox(0,0)[lb]{unstable manifold}%
}
\end{picture}%

%% file: Tphi.tex
\chapter{The \texorpdfstring{$T_z$}{Tz}-seminorm}
\label{ch:Tphi}

\begin{quote}
``\ldots I went to the hotel, quite tired, and I went
                  to sleep. I dreamed I was in a very long
            corridor,
                  with no ceiling, and nothing in front of me,
            only
                  two very long walls extremely high. Then I
            woke up and
                  understood immediately that I was trapped
            inside a norm!!"\footnote{From an email from Benedetto Scoppola in 2005}

\end{quote}

\bigskip\bigskip \noindent
In order to analyse the renormalisation group map defined in
Definition~\ref{def:RGmap}, we use certain seminorms.  The seminorms
are designed to measure the size of the nonperturbative coordinate
$K_+(B)$ defined in \eqref{e:K+B}, which is a function of the field
$\varphi$ and also of $(V,K)$.  Since $K_+(B)$ is a function of fields
that are constant on blocks $B$ as in Definition~\ref{def:NcalB}, it
is natural (and sufficient) to define the seminorm on functions of the
constant value $\varphi\in \R^n$.  We encode estimates of $K_+(B)$,
and of its derivatives with respect to the three variables
$\varphi,V,K$, in a single seminorm.

In this chapter, we define the seminorm that will be used for this
purpose, the $T_\auxx$-seminorm, and the $T_{\varphi}$-seminorm which
is a special case.  The $T_{\auxx}$-seminorm is defined on functions
of a variable $\auxx$ that lies in a product $\Auxx = \Auxx_{1} \times
\Auxx_{2} \times \Auxx_{3}$ of three normed spaces $\Auxx_{s}$
($s=1,2,3$).  The space $\Auxx_{1} = \R^{n}$ is a space of values of
field configurations $\varphi$. The space $\Auxx_{2} = \Vcal$ is the
space of interactions $V$ as in Definition~\ref{defn:Vcal} and
$\Auxx_{3} = \Ncal(b)$ is the space of $K$ as in
Definition~\ref{def:NcalB}. These choices motivate this chapter,
but the results are valid for arbitrary normed spaces
$\Auxx_{2},\Auxx_{3}$.

\section{Definition of the \texorpdfstring{$T_{\auxx}$}{Tz}-seminorm}
\label{sec:Tphi-props}

Let $X$ be a normed vector space, and let $X^{p}$ denote the Cartesian
product of $p$ copies of $X$.  Given $\dot{x}_i \in X$, we write
$\dot{x}^{p}=\dot{x}_{1},\dots ,\dot{x}_{p}$.  A function $M:X^p\to
\R$ is said to be $p$-\emph{linear} if $M (\dot{x}^{p})$ is linear in
each of its $p$ arguments $\dot{x}_{1},\dots ,\dot{x}_{p}$.  The norm
of $M$ is defined by
\begin{equation}
    \label{e:M-norm}
    \|M\|_{X}
    =
    \sup_{\dot{x}^{p} \in X(1)^{p}} |M(\dot{x}^{p})| ,
\end{equation}
where $X (1)$ is the unit ball in $X$ centred on the origin.

Given a function $F:X \to \R$, the Fr\'echet derivative $F^{(p)} (x)$
of order $p$, when it exists, is a \emph{symmetric} $p$-linear
function of $p$ \emph{directions} $\dot{x}^{p} = \dot{x}_{1},\dots
,\dot{x}_{p}$. It obeys, in particular,
\begin{gather}
    F^{(p)} (x;\dot{x}^{p})
    =
     \left. \frac{\partial^p}{\partial t_{1} \cdots \partial t_{p}}
     \right|_{t_1=\cdots=t_p=0}
     F (x+\sum t_{i}\dot{x}_{i}) ,
     \label{e:frechet-1}
     \\
     \big\|
     F (x+t\dot{x}) -
     \sum_{p'<p}\frac{t^{p'}}{p'!}F^{(p')} \big(x;\dot{x},\dots ,\dot{x}\big)
     \big\|_{X}
     =
     o (t^{p}) ,
     \label{e:frechet-2}
\end{gather}
where $o (t)$ is uniform for $\dot{x} \in X(1)^p$.  For differential
calculus of functions on Banach spaces see \cite{Dieu69}, and for the
more general setting of
normed vector spaces, see \cite[Appendix~D.2]{AKM16}.

\begin{example}
\label{example:Tphi}
Let $M$ be a symmetric $k$-linear function on
$X$, and let $F (x) = M (x,\ldots,x)$, i.e., $F$ is $M$ evaluated on
the diagonal sequence whose $k$-components are all equal to $x$.  Then
the $p$th derivative $F^{(p)} (x)$ at $x$ is zero for $p>k$, and
otherwise is the $p$-linear form $\dot{x}_{1},\dots ,\dot{x}_{p}
\mapsto \frac{k!}{(k-p)!}  M (\dot{x}_{1},\dots
,\dot{x}_{p},x,\ldots,x)$ (there are $k-p$ entries $x$). The
combinatorial factor arises via the symmetry of $M$. By the definition
\eqref{e:M-norm},
\begin{equation}
\lbeq{Tphinorm}
    \|F^{(p)}(x)\|_{X}
    \le
    \tfrac{k!}{(k-p)!}  \|M \|_{X} \|x\|^{k-p} .
\end{equation}
\end{example}

Given normed spaces $\Auxx_{s}$ ($s=1,2,3$), let $\Auxx =
\Auxx_{1}\times\Auxx_{2}\times\Auxx_{3}$, and let $F: \Auxx \to \R$ be
a function on $\Auxx$.  Consider the Fr\'echet derivative of order
$p_{1}$ with respect to $\auxx_{1}$, of order $p_{2}$ with respect to
$\auxx_{2}$, and of order $p_{3}$ with respect to $\auxx_{3}$. Let
$p=p_{1},p_{2},p_{3}$. Then the Fr\'echet derivative
\begin{equation}
    F^{(p)}
    \big(\auxx;\,
    \dot{\auxx}_{1}^{p_{1}};\,
    \dot{\auxx}_{2}^{p_{2}};\,
    \dot{\auxx}_{3}^{p_{3}}
    \big)
\end{equation}
of $F$ at $\auxx \in \Auxx$ is $p_{1}$-linear in
$\dot{\auxx}_{1}^{p_{1}}$, $p_{2}$-linear in
$\dot{\auxx}_{2}^{p_{2}}$, and $p_{3}$-linear in
$\dot{\auxx}_{3}^{p_{3}}$, where
$\dot{\auxx}_{s}^{p_{s}}=\dot{\auxx}_{s,1},\dots
,\dot{\auxx}_{s,p_{s}} \in\Auxx_{s}^{p_{s}}$ for $s=1,2,3$. The norm
of this derivative is, by definition,
\begin{equation}
    \label{e:M-norm2}
    \|F^{(p)} (\auxx)\|_{\Auxx}
    =
    \sup_{\dot{\auxx}_{s}^{p_{s}} \in \Auxx_{s}(1)^{p_{s}},\,s=1,2,3}
    \big|F^{(p)}
    \big(
    \auxx;\,\auxx_{1}^{p_{1}};\,\dot{\auxx}_{2}^{p_{2}};\,\dot{\auxx}_{3}^{p_{3}}
    \big)\big| .
\end{equation}
We use this three-variable formalism in preference to uniting
arguments using a larger normed space, in order to avoid testing
differentiation in unwanted directions such as
$\dot{\varphi}+\dot{V}$.

Let $\h$ be a positive number and let $|\cdot |$ be the Euclidean norm
on $\R^n$. For the remainder of Chapter~\ref{ch:Tphi}, we set
\begin{equation}\label{e:Rn-norm}
        \text{$\Auxx_{1} \; = \; \R_\h^n \;\; = \;\; \R^{n}$ with norm $\h^{-1}|\cdot|$}.
\end{equation}
\index{$\Aux$}%
\index{$\Auxx$}%
Let $\Aux = \Auxx_{2}\times \Auxx_{3}$ so that $\Auxx =
\R_\h^{n}\times \Aux$.  We write $\auxx = (\varphi,\aux)\in \Auxx$.
We use multi-index notation, in which we write $p! = \prod_{s}p_{s}!$,
we write $p \le p'$ to mean that $p_{s}\le p'_{s}$ for each $s$, and
we use multi-binomial coefficients defined by
\begin{equation}
    \label{e:multibinomial}
    \binom{p'}{p}
    =
    \frac{p'!}{p!(p'-p)!} =
    \prod_{s}\frac{p'_{s}!}{p_{s}!(p'_{s}-p_{s})!}.
\end{equation}

\begin{defn}
\label{def:Tnorm} Let $p_\Auxx =
(p_{\Auxx_{1}},p_{\Auxx_{2}},p_{\Auxx_{3}})$ where each
$p_{\Auxx_{s}}$ is a non-negative integer or $\infty$, and define
$p_{\Ncal}=p_{\Auxx_{1}}$ and $p_{\Aux} =
(p_{\Auxx_{2}},p_{\Auxx_{3}})$.  Given a function $F:\Auxx \rightarrow
\R$ with norm-continuous Fr\'echet derivatives of orders up to
$p_{\Auxx}$ we define the $T_\auxx=T_{\auxx} (\h)$-seminorm of $F$ by
\begin{equation}
\lbeq{Tznormdef}
  \|F\|_{T_\auxx}
  =
  \sum_{p\le p_{\Auxx}}  \frac{1}{p!} \|F^{(p)}(\auxx)\|_\Auxx .
\end{equation}
The triangle inequality holds for $\|\cdot\|_{T_\auxx}$ by definition.
The $T_\varphi$-seminorm is defined by the same formula with $p_{\Aux
}=(0,0)$, and is denoted by $\|F\|_{T_{\varphi}} = \|F\|_{T_{\varphi}
(\h)}$.  The $T_\varphi$-seminorm does not examine derivatives with
respect to $\aux$, and is defined on functions $F : \Auxx_1 \to \R$,
or on functions $F : \Auxx \to \R$ with $\auxx=(\varphi,\aux)$ where
$\aux$ is held fixed.
\end{defn}

Later in this chapter and also in subsequent chapters, in a slight
abuse of notation we apply the $T_\auxx$-seminorm to elements of the
space $\Ncal(B)$ of Definition~\ref{def:NcalB}.  An element $F(B) \in
\Ncal(B)$ determines a function $F :\R^n \to \R$ via the relation
$F(B)=F\circ j_B$, and when we write $\|F(B)\|_{T_\auxx}$ we mean
$\|F\|_{T_\auxx}$.

We only need the case $p_{\Auxx}= (\infty ,\infty ,\infty)$, but we
include finite choices to emphasise that there is no need for
analyticity in $\varphi,V,K$ in this chapter.  For $p_{\Auxx} =
(0,0,0)$, the $T_{\auxx}$-seminorm is simply the absolute value of
$F(\auxx)\in\R$.  The name $T_{\auxx}$ refers to the Taylor expansion
at $\auxx$.  Just as the Taylor expansion of the product of two
functions is the product of the Taylor expansions, the seminorm of
Definition~\ref{def:Tnorm} shares with the absolute value the
following \emph{product property}.  A more general product property is
proved in \cite{BS-rg-norm}.

\index{Product property}
\begin{lemma} \label{lem:Tphi-prod}
For $F,G:\Auxx \rightarrow \R$ and $\auxx\in\Auxx$,
\begin{equation}
    \label{e:T-product-prop}
    \|F G\|_{T_{\auxx}}
    \le
    \|F\|_{T_{\auxx}} \|G\|_{T_{\auxx}}
\end{equation}
\end{lemma}

\begin{proof}
It is a consequence of the definition of the norm and the product rule for differentiation
that, for $p \le p_\Auxx$,
\begin{equation}
    \|(FG)^{(p)}(\auxx)\|_{\Auxx}
    \le
    \sum_{p'\le p} \binom{p}{p'} \|F^{(p')}(\auxx)\|_{\Auxx} \|G^{(p-p')}(\auxx)\|_{\Auxx} ,
\end{equation}
where we have used the notation \eqref{e:multibinomial}. Therefore,
\begin{align}
    \|FG\|_{T_{\auxx}}
    & \le
    \sum_{p\le p_{\Auxx}} \frac{1}{p!}
    \sum_{p'\le p} \binom{p}{p'} \|F^{(p')}(\auxx)\|_{\Auxx} \|G^{(p-p')}(\auxx)\|_{\Auxx}
    \nnb
    &=
    \sum_{p' \le p_{\Auxx}}
    \frac{1}{p'!}
    \|F^{(p')}(\auxx)\|_{\Auxx}
    \sum_{p : p' \le p\le p_{\Auxx}}
    \frac{1}{(p-p')!}
    \|G^{(p-p')}(\auxx)\|_{\Auxx}
    \nnb
    &=
    \sum_{p' \le p_{\Auxx}}
    \frac{1}{p'!}
    \|F^{(p')}(\auxx)\|_{\Auxx}
    \sum_{q\le p_{\Auxx}-p'}
    \frac{1}{q!}
    \|G^{(q)}(\auxx)\|_{\Auxx}
    \nnb
    &\le
    \|F\|_{T_{\auxx}}\|G\|_{T_{\auxx}} ,
\end{align}
and the proof is complete.
\end{proof}

The product property simplifies control of smoothness. For example,
\begin{equation}
\label{e:expT}
    \|e^{F}\|_{T_{\auxx}} \le e^{\|F\|_{T_{\auxx}}} .
\end{equation}
This follows by expanding the exponential in a Taylor expansion and
applying the product property term by term.

Given $\h>0$, we define the function
\begin{equation}
    \label{e:P-def-6}
    P_{\h} (t) = 1 + |t|/\h
    \quad
    (t \in \R)
    .
\end{equation}

\begin{exercise}
\label{ex:Tphi} Let $M$ be a symmetric $k$-linear function on
$\R_\h^{n}$ that does not depend on the variables $\aux\in\Aux$.  Let
$F (\varphi) = M (\varphi,\ldots,\varphi)$ denote the result of
evaluating $M$ on the sequence $\varphi,\dots ,\varphi$ with $k$
components. Then
\begin{equation} \label{e:prodTphi}
    \|F\|_{T_{\auxx}}
    \le
    \|M\|_{\Auxx}
    P_{\h}^{k} (\varphi) .
\end{equation}
Combine this with the product property to prove that, for a
nonnegative integer $p$, $\| (\varphi\cdot\varphi)^p \|_{T_{\auxx}}
\le (|\varphi|+\h)^{2p}$. Similarly, for a vector $\zeta \in \R^{n}$,
$\|(\zeta \cdot\varphi) (\varphi\cdot\varphi)^p \|_{T_{\auxx}} \le
|\zeta| (|\varphi|+\h)^{2p+1}$.
\solref{Tphi}
\end{exercise}

Let $F: \Auxx \to \R$ and recall that $\Auxx = \R_\h^n \times \Aux$ with elements denoted $z=(\varphi,y)$.
We define the norm
\begin{equation}
  \label{e:jX-norm-def}
  \|F \|_{T_{\infty,\aux}}
  =
  \sup_{\varphi \in \R^{n}} \|F \|_{T_{\auxx}} ,
\end{equation}
where, in $\auxx = (\varphi,\aux)$, $\aux$ is held fixed.  The
$T_{\auxx}$-seminorm and the $T_\infty$-norm are monotone decreasing
in the norms on $\R_\h^n$ and $\Aux$ and therefore monotone increasing
in $\h$.  The product property for the $T_{\auxx}$-seminorm
immediately implies that the $T_{\infty}$-norm also has the analogous
product property.  When $p_{\Aux}=(0,0)$ the norm
\eqref{e:jX-norm-def} is equivalent to the $\Ccal^{p_{\Ncal}}$ norm,
but is preferable for our purposes because it has the product
property.

The following lemma provides an estimate which compares the norm of a polynomial
in $\varphi$ for two different values of the parameter $\h$ for the norm on
$\Auxx_1 = \R^n_\h$ of \refeq{Rn-norm}, with the norm on $\Aux$ unchanged.

\begin{lemma}
\label{lem:hoverell}
For a function $F:\Auxx \rightarrow \R$, which is polynomial
of degree $k \le p_{\Auxx_1}$ in $\varphi$, and for $\h,\h'>0$,
\begin{align}
  \|F\|_{T_{0,\aux} (\h)}
  &\le
  \left(\frac{\h}{\h'}\vee 1 \right)^{k}
  \|F\|_{T_{0,\aux} (\h')} .
\end{align}
\end{lemma}

\begin{proof}
A unit norm direction in $\Auxx_{1} (\h)$ is a direction in $\Auxx_{1}
(\h')$ with norm $\frac{\h}{\h'}$, whereas, for $s=2,3$, norms of
directions $\dot{y}$ in $\Aux_{s}$ are the same in $\Auxx_{s} (\h)$
and $\Auxx_{s} (\h')$.  Consequently, $\|F^{(p)}((0,\aux)\|_{\Auxx
(\h)} = (\frac{\h}{\h'})^{p_{1}}\|F^{(p)}((0,\aux)\|_{\Auxx (\h')}$.
Therefore, with $\alpha = \frac{\h}{\h'}\vee 1$,
\begin{align}
  \|F\|_{T_{0,\aux} (\h)}
  &=
  \sum_{p\le (k,p_{\Aux})}  \frac{1}{p!} \|F^{(p)}((0,\aux)\|_{\Auxx (\h)}
  \nnb
  &\le
  \alpha^{k}
  \sum_{p\le (k,p_{\Aux})}  \frac{1}{p!} \|F^{(p)}((0,\aux)\|_{\Auxx (\h')}
  =
  \alpha^{k}
  \|F\|_{T_{0,\aux} (\h')} ,
\end{align}
and the proof is complete.
\end{proof}

\section{Control of derivatives}

The following two lemmas indicate how the $T_\auxx$-seminorm provides
estimates on derivatives.

In the statement of the next lemma, for $F: \Auxx \rightarrow \R$, we
define $F^{(p)}(y)$ to be the function $\varphi \mapsto F^{(p)}(\varphi,y)$
with $y$ held fixed.

\begin{lemma}
\label{lem:normderiv-1} For $F: \Auxx \rightarrow \R$, for $p \le
p_\Aux$, and for directions $\dot{\aux}^q =
(\dot{\auxx}_2^{p_2},\dot{\auxx}_3^{p_3})$ which have unit norm in
$\Aux = \Auxx_2 \times \Auxx_3$, for any $(\varphi,y) \in \Auxx$,
\begin{equation}
\label{e:DDFr}
    \|
    F^{(0,p_{2},p_{3})} (y; \dot{\aux}^q)
    \|_{T_{\varphi}}
    \le
    p_{2}!\,p_{3}!\, \|F\|_{T_{\varphi,\aux}} .
\end{equation}
\end{lemma}

\begin{proof}
By Definition~\ref{def:Tnorm}, the $T_{\varphi, \aux}$-seminorm obeys
the inequality
\begin{equation}
    \sum_{p_{1}} \frac{1}{p_{1}!}
    \|F^{(p_{1},q)}(\auxx)\|_\Auxx
    \le
    q!
    \|F\|_{T_{\varphi,\aux}} ,
\end{equation}
where $q=(p_{2},p_{3})$ and $q!=p_{2}!\,p_{3}!$.  This implies
that, for any fixed unit directions $\dot{\aux}^q$,
\begin{equation}
    \sum_{p_{1}} \frac{1}{p_{1}!}
    \sup_{\dot{\varphi}^{p}\in \Auxx_{1} (1)^{p_{1}}}
    \left|F^{(p_{1},q)}(\auxx;\dot{\varphi}^{p_{1}},\dot{\aux}^{q})\right|
    \le
    q!
    \|F\|_{T_{\varphi,\aux}}
    ,
\end{equation}
and this is \eqref{e:DDFr} by the definition of the $T_{\varphi}$
norm and of $F^{(p_1,q)}(y)$.
\end{proof}

\begin{lemma}\label{lem:derivatives2}
Let $F:\Auxx \rightarrow \R$ be polynomial in $\varphi$ of degree $k
\le p_{\Ncal}$.  Then for $r \le k$ and for directions
$\dot{\varphi}^{r}$ which have unit norm in $\R^n$,
\begin{equation}
    \label{e:deriv-Tphi}
    \|F^{(r,0,0)} (\dot{\varphi}^{r})\|_{T_{\varphi,\aux}}
    \le
    2^{k} \frac{r!}{\h^{r}} \|F\|_{T_{\varphi,\aux}}  .
\end{equation}
\end{lemma}

\begin{proof} By Definition~\ref{def:Tnorm}  and the hypotheses,
\begin{equation}
    \|F^{(r,0,0)} (\dot{\varphi}^{r}) \|_{T_{\varphi,\aux}}
    \le
    \sum_{p_{1}\le k-r}
    \frac{1}{p_{1}!}
    \sum_{q}
    \frac{1}{q!}
    \|F^{(p_{1}+r,q)} (\auxx)\|_{\Auxx} \,\h^{-r} .
\end{equation}
The $\h^{-r}$ factor in the right-hand side occurs because the $\Auxx$
norm on the right-hand side is defined in \eqref{e:M-norm2} as a
supremum over directions with unit norm in $\R_{\h}^{n}$ whereas in
the left-hand side we are testing the derivative on directions with
unit norm in $\R^{n}$.  We shift the index by writing $p_{1}'=
p_{1}+r$, and use $\frac{1}{(p'_{1}-r)!} =
\frac{r!}{p'_{1}!}\binom{p'_{1}}{r}$ followed by $\binom{p'_{1}}{r}
\le \sum_{r\le p'_{1}}\binom{p'_{1}}{r} = 2^{p_1'} \le 2^{k}$, and
obtain
\begin{align}
    \|F^{(r,0,0)} (\dot{\varphi}^{r}) \|_{T_{\varphi,\aux}}
    &\le
    \sum_{p'_{1}=r}^{k}
    \frac{1}{(p'_{1}-r)!}
    \sum_{q}
    \frac{1}{q!}
    \|F^{(p'_{1},q)} (\auxx)\|_{\Auxx} \,\h^{-r}
    \nnb
    &\le
    r!\,2^{k}\,
    \sum_{p'_{1},q}
    \frac{1}{p'_{1}!\,q!}
    \|F^{(p'_{1},q)} (\auxx)\|_{\Auxx}  \,\h^{-r} .
\end{align}
The right-hand side is $r!2^{k} \h^{-r}
\|F^{(r,q)}\|_{T_{\varphi,\aux}}$, as desired.
\end{proof}

\section{Expectation and the \texorpdfstring{$T_\auxx$}{Tz}-seminorm}

In \refeq{ZEx}, we encounter an expectation $(\Ex_+ \theta Z)(\varphi)
= \Ex_+ Z(\varphi + \zeta)$, where the integration is with respect to
$\zeta$ with $\varphi$ held fixed.  Similarly, in the definition
\refeq{K+B} of $K_+(B)$ we encounter $(\Ex_+\theta F^B)(\varphi)
=\Ex_+ (\prod_{b\in B}F(b;\varphi + \zeta))$.  The field $\zeta$ is
constant on blocks $b \in \Bcal$, while $\varphi$ is constant on
blocks $B \in \Bcal_{+}$.  In this section, we show in a general
context how such convolution integrals can be estimated using the
$T_\auxx$-seminorm.

Given a block $B$, an $n$-component field $\varphi$ which is constant
on $B$, an $n$-component field $\zeta$ which is constant on blocks $b
\in \Bcal(B)$, and given $F(\cdot,\zeta) \in \Ncal(b)$ with $\zeta$
regarded as fixed, we define $F_\zeta \in \Ncal(b)$ by $F_\zeta
(\varphi) = F (\varphi,\zeta)$.  Similarly, we define $\theta_\zeta F
\in \Ncal(b)$ by $(\theta_\zeta F)(\varphi) =F(\varphi+\zeta)$.
Although $F$ is a function of $(\varphi,\aux) \in \Auxx$, we do not
exhibit the dependence of $F$ on $\aux$ in our notation.  We can take
the $T_{\varphi,\aux}$-seminorm of $F_\zeta$, obtaining $\|F_{\zeta}
\|_{T_{\varphi,\aux}}$ which depends on the variable $\zeta$ that is
held fixed.  Also, with $\varphi$ fixed, we can integrate
$F_\zeta(\varphi)$ with respect to $\zeta$.  These last two facts are
relevant for the interpretation of \eqref{e:ExCthetaTphi2} in the
following proposition.

\begin{prop} \label{prop:ExCthetaTphi}
For $b \in \Bcal $ and $F \in \Ncal (b)$,
\begin{equation} \label{e:ExCthetaTphi}
  \| \theta_{\zeta} F \|_{T_{\varphi,\aux} }
  =
  \|F \|_{T_{\varphi + \zeta,\aux} }.
\end{equation}
For $B\in\Bcal_+$, for $\zeta$ a field which is constant on blocks
$b\in\Bcal(b)$, and for $F=F(\varphi,\zeta)$ with $F(\cdot,\zeta) \in
\Ncal (B)$,
\begin{equation} \label{e:ExCthetaTphi2}
  \|\Ex_CF_\zeta \|_{T_{\varphi,\aux}}
  \leq
  \Ex_C \|F_{\zeta }\|_{T_{\varphi,\aux}}.
\end{equation}
For a family $F (b) \in \Ncal(b)$, where $b$ ranges over $\Bcal(B)$
with $B \in \Bcal_+$, and for $F^B = \prod_{b\in B}F(b)$ as in
\eqref{e:set-exponent},
\begin{equation} \label{e:ExCthetaTphi3}
  \|\Ex_C \theta F^{B}\|_{T_{\varphi,\aux}}
  \leq
  \Ex_C \left(\prod_{b\in\Bcal(B)} \|F(b)\|_{T_{\varphi+\zeta|_b,\aux}} \right).
\end{equation}
\end{prop}

\begin{proof}
The identity \eqref{e:ExCthetaTphi} follows immediately from the
definition of $\theta_{\zeta}F$, by commuting derivatives with the
translation $\varphi \mapsto \varphi + \zeta $.  The inequality
\eqref{e:ExCthetaTphi2} is obtained by commuting derivatives past the
expectation,
\begin{equation}
      \left|
      \ddp{^p}{\varphi^p}
      \Ex_{C} F(\varphi,\zeta )
      \right|
      \le
      \Ex_{C}
      \left|
      \ddp{^p F_{\zeta} (\varphi)}{\varphi^p}
      \right|
\end{equation}
and then the desired result follows from the
Definition~\ref{def:Tnorm} of the $T_{\varphi,\aux}$-seminorm.  For
the inequality \eqref{e:ExCthetaTphi3}, we first use
\eqref{e:ExCthetaTphi2}, then that $\theta$ is a homomorphism,
then the product property of the $T_\auxx$-seminorm, and
finally \eqref{e:ExCthetaTphi}, to obtain
\begin{align}
    \|\Ex \theta_{\zeta} F^{B}\|_{T_{\varphi,\aux}}
    &\le
    \Ex \|\theta_{\zeta} F^{B}\|_{T_{\varphi,\aux}}
    =
    \Ex \|(\theta_{\zeta} F)^{B}\|_{T_{\varphi,\aux}}
    \nnb
    &\le
    \Ex \left(\|\theta_{\zeta} F\|_{T_{\varphi,\aux}}^{B} \right)
    =
    \Ex \left(\prod_{b\in\Bcal(B)} \|F (b)\|_{T_{\varphi+\zeta,\aux}} \right)
    ,
\end{align}
as required.
\end{proof}

\section{Exponentials and the \texorpdfstring{$T_{\auxx}$}{Tz}-seminorm}
\label{sec:norms2}

As a consequence of the product property, the $T_{\auxx}$-norm
interacts well with the exponential function.  The following lemma,
which is based on \cite[Proposition~3.8]{BS-rg-norm}, is an extension
of \eqref{e:expT}.  It improves on \eqref{e:expT} when $F(\auxx)<0$.

\begin{lemma}
\label{lem:eK}
For $F:\Auxx  \rightarrow \R$,
\begin{equation} \label{e:eK}
    \|e^{F  }\|_{T_{\auxx}}
    \le
    e^{ F(\auxx) + \left(\|F  \|_{T_{\auxx}} - |F(\auxx)|\right)}
    .
\end{equation}
\end{lemma}

Lemma~\ref{lem:eK} is an immediate consequence of the following
proposition, which holds in any unital algebra $\Acal$ with seminorm
obeying the product property $\|FG\| \le \|F\|\|G\|$ for all $F,G \in
\Acal$.  To deduce \eqref{e:eK} from Proposition~\ref{prop:eFineq}, we
simply take $r$ to be the value $F(\auxx)$ (not the function) and use
the fact that $\|F-F(\auxx)\|_{T_{\auxx}} =
\|F\|_{T_{\auxx}}-|F(\auxx)|$ by definition of the $T_{\auxx}$-norm.

\begin{prop} \label{prop:eFineq}
Let $\Acal$ be a unital algebra with seminorm obeying the product
property.  For any $F \in \Acal$ and $r \in \R$,
\begin{equation} \label{e:eFineq}
    \|e^{F}\|
    \leq
    e^{r+\|F-r\|}.
\end{equation}
\end{prop}

\begin{proof}
It suffices to show that
\begin{equation}  \label{e:expprod}
    \|e^{F}\|
    \le
    \liminf_{n\to\infty}
    \|1+F/n\|^{n} ,
\end{equation}
since, for any $r \in \R$ and $n\geq|r|$,
\begin{align}
  \|1+F/n\|
  = \|1+r/n+(F-r)/n\|
  &\leq
    1+r/n + \|F-r\|/n,
\end{align}
and hence, by \eqref{e:expprod} and the fact that $e^x =
\lim_{n\to\infty} (1+x/n)^n$,
\begin{equation}
    \|e^{F}\|
    \leq
    \liminf_{n\to\infty}
    (1 + r/n + \|F-r\|/n)^{n}
    =
    e^{r+\|F-r\|}.
\end{equation}

To prove \eqref{e:expprod}, it suffices to restrict to the case $n >
\|F\|$, so that $(1+F/n)^{-1}$ is well-defined by its power series.
We first use the product property to obtain
\begin{align}
\label{e:eFfact}
    \|e^{F}\|
    &=
    \Big\| (e^{F/n})^{n} (1+F/n)^{-n} (1+F/n)^{n}\Big\|
      \nnb
    &\le
    \Big\|e^{F/n} (1+F/n)^{-1}\Big\|^{n}
    \Big\|(1+F/n)\Big\|^{n} .
\end{align}
Let $R_{n}=e^{F/n}-1-F/n$.  By expanding the exponential, we find that
$\|R_{n}\|=O (n^{-2})$. Therefore,
\begin{equation}
    \|e^{F/n} (1+F/n)^{-1}\|
    =
    \| (1+F/n+R_{n}) (1+F/n)^{-1}\|
    =
    1 + O (n^{-2}) .
\end{equation}
Since $(1+O (n^{-2}))^{n} \rightarrow 1$, \eqref{e:expprod} follows
after taking the $\liminf$ in \eqref{e:eFfact}.
\end{proof}

\section{Taylor's theorem and the \texorpdfstring{$T_\auxx$}{Tz}-seminorm}
\label{sec:Taylor}

As in Definition~\ref{def:Loc-hier}, we write $\Tay_k F$ for the
degree-$k$ Taylor polynomial of $F: \Auxx \to \R$ in
$\varphi\in\R_\h^n$, with $\aux\in\Aux$ held fixed, i.e.,
\begin{equation}
    \label{e:TaykF0}
    \Tay_{k} F (\varphi,\aux)
    =
    \sum_{r\le k} \frac{1}{r!}
    F^{(r,0,0)} (0,\aux;\varphi^{r}) .
\end{equation}
The following lemma relates the seminorms of $\Tay_k F$ and $F$.
Given $\h>0$, we write $P_{\h} (t) = 1 + |t|/\h$ for $t \in \R$,
as in \refeq{P-def-6}.

\begin{lemma}
\label{lem:TayTphi}
For $F:\Auxx \rightarrow \R$ and $k \le p_\Ncal$,
\begin{equation} \label{e:TayTphi}
    \|\Tay_k F \|_{T_{\varphi,\aux}}
    \le
    \|F \|_{T_{0,\aux}}
    P_{\h}^{k} (\varphi)
    .
\end{equation}
In particular,
\begin{equation} \label{e:TayT0}
    \|\Tay_k F\|_{T_{0,\aux}}
    \leq \|F\|_{T_{0,\aux}}
    .
\end{equation}
\end{lemma}

\begin{proof}
According to \eqref{e:TaykF0}, $\Tay_{k} F$ is a sum of
terms $M_{r} (\auxx) = F^{(r,0,0)} (0,\aux;\varphi^{r})$ with $r \le k$, where all the
components of the sequence $\varphi^{r}$ are equal to $\varphi$.
We therefore
begin with an estimate for the $T_{\auxx}$-seminorm of $M_{r}$ obtained by
generalising Example~\ref{example:Tphi} to include $\aux$-dependence.
If $p_{1}>r$ then $M_{r}^{(p)}=0$.
For $p=(p_{1},q)$ with $p_{1} \le r$ and $q=(p_{2},p_{3})$,
and for unit norm directions $\dot{\auxx}^{p}$,
\begin{align}
    \big|M_{r}^{(p)} (\auxx; \dot{\auxx}^{p})\big|
    &=
    \frac{r!}{(r-p_{1})!}
    \Big|
    F^{(r,q)} \big(
    0,\aux; \dot{\varphi}^{p_{1}},\varphi^{r-p_{1}};
    \dot{\aux}^{q}
    \big)
    \Big|
    \nnb
    &\le
    \frac{r!}{(r-p_{1})!}
    \big\|F^{(r,q)} (0,\aux)\big\|_{\Auxx}
    \|\varphi\|_{\R^{n}_{\h}}^{r-p_{1}} .
\end{align}
We take the supremum over $\dot{\auxx}^{p}$ and obtain
\begin{equation}
    \big\|M_{r}^{(p)} (\varphi,\aux)\big\|_{\Auxx}
    \le
    \frac{r!}{(r-p_{1})!}
    \big\|F^{(r,q)} (0,\aux)\big\|_{\Auxx}
    \|\varphi\|_{\R^{n}_{\h}}^{r-p_{1}} .
\end{equation}
By dividing by $p! = p_{1}!q!$ and
summing over $p$ with $p_{1}\le r$, and by
Definition~\ref{def:Tnorm} of the $T_{\auxx}$-seminorm, this gives
\begin{align}
    \|M_{r}\|_{T_{\auxx}}
    &\le
    \sum_{p_{1},q}
    \binom{r}{p_{1}}\frac{1}{q!}
    \big\|F^{(r,q)} (0,\aux)\big\|_{\Auxx}
    \|\varphi\|_{\R^{n}_{\h}}^{r-p_{1}}
    \nnb
    &=
    \sum_{q}
    \frac{1}{q!}
    \big\|F^{(r,q)} (0,\aux)\big\|_{\Auxx}
    P_{\h}^{r} (\varphi) ,
\end{align}
where we evaluated the sum over $p_{1}\le r$ by the binomial theorem,
obtaining $(1+\|\varphi\|_{\R^{n}_{\h}})^{r}$ which equals $P_{\h}^r
(\varphi)$ by \eqref{e:Rn-norm} and \eqref{e:P-def-6}. We replace
$P^{r}$ by $P^{k}$, which is larger because $r\le k$, and insert the
resulting bound into the definition \eqref{e:TaykF0} of $\Tay_{k} F$,
to obtain
\begin{equation}
    \|\Tay_{k} F \|_{T_{\varphi,\aux}}
    \le
    \sum_{r\le k} \frac{1}{r!}
    \|M_{r}\|_{T_{\varphi,\aux}}
    \le
    \|F\|_{T_{0,\aux}}
    P_{\h}^{k} (\varphi) .
\end{equation}
This completes the proof.
\end{proof}

\begin{exercise}
\label{ex:Tphi-poly}
Suppose that $F (\auxx)$ is a polynomial in $\varphi$ of degree $k \le
p_{\Ncal}$, with coefficients that are functions of $\aux$. Then
\begin{equation}
\label{e:Tphi-poly-2}
    \|F\|_{T_{\varphi,\aux}}
    \le
    \|F\|_{T_{0,\aux}}
    P_{\h}^{k}(\varphi)
    .
\end{equation}
\solref{Tphi-poly}
\end{exercise}

Lemma~\ref{lem:TayTphi} shows that the Taylor polynomial of $F$ is
effectively bounded in norm by the norm of $F$.  The following lemma
shows how the Taylor remainder $(1-\Tay_k)F$ can be bounded in terms
of the norm of $F$.  In the remainder estimate, an important feature
is that the norm of $(1-\Tay_k)F$ is computed for the field $\varphi
\in \R_{\h_+}^n$ whereas in the norm of $F$ the field lies in
$\R_\h^n$.  In our applications, the change from $\h$ to $\h_+$
corresponds to a change in scale, with small ratio $\h_+/\h$.  The
small factor $(\tfrac{\h_{+}}{\h} )^{k+1}$ present in the upper bound
of \refeq{Taycontraction} is ultimately what leads to the crucial
contraction estimate for the renormalisation group map; see
Proposition~\ref{prop:crucial-0}.  In the lemma, we make explicit the
$\h$ dependence of the norm on $\Auxx$ by writing $T_{\auxx} (\h)$ and
$\|\cdot\|_{\Auxx (\h)}$.

\begin{lemma}   \label{lem:Taycontraction}
For $k<p_\Ncal$, $\h_{+} \le \h$, and $F: \Auxx \rightarrow \R$,
\begin{equation}
  \label{e:Taycontraction}
  \big\|(1-\Tay_k)F\big\|_{T_{\auxx} (\h_{+})}
  \le
  2\left(\tfrac{\h_{+}}{\h} \right)^{k+1}
  P_{\h_{+}}^{k+1}(\varphi)
  \sup_{0 \le t \le 1}
  \|F\|_{T_{\auxx_{t}} (\h)} ,
\end{equation}
with $\auxx = (\varphi,\aux)$ and $\auxx_{t} = (t\varphi,\aux)$.
\end{lemma}

\begin{proof}
We write $R=(1-\Tay_k)F$ and $\bar F = \sup_{0 \le t \le 1}
\|F\|_{T_{\auxx_{t}} (\h)}$, so that our goal becomes
\begin{equation} \label{e:Taycontraction-1}
  \big\|R \big\|_{T_{\auxx} (\h_{+})}
  \le
  2\left(\tfrac{\h_{+}}{\h} \right)^{k+1}
  P_{\h_{+}}^{k+1}(\varphi) \bar F .
\end{equation}
By definition, with $q = ( p_{2},p_{3})$ and $q!=p_{2}!p_{3}!$,
\begin{equation}
\lbeq{Rpsums}
  \big\|R \big\|_{T_{\auxx} (\h_{+})}
  =
  \sum_{p=0}^{k}
  \sum_{q \le p_{\Aux}}
  \tfrac{1}{p!}
  \tfrac{1}{q!}
  \|R^{(p,q)} (\varphi)\|_{\Auxx (\h_{+})}
  +
  \sum_{p=k+1}^{p_{\Ncal}}
  \sum_{q\le p_{\Aux}}
  \tfrac{1}{p!}
  \tfrac{1}{q!}
  \|F^{(p,q)} (\varphi)\|_{\Auxx (\h_{+})},
\end{equation}
where the replacement of $R$ by $F$ in the second sum is justified by
the fact that $p$ $\varphi$-derivatives of $\Tay_k F$ vanish when
$p>k$.  We estimate the two sums on the right-hand side of
\refeq{Rpsums} separately.

For the first sum, we fix $p \le k$ and use the fact that the first
$k-p$ $\varphi$-derivatives of $R$, evaluated at zero field, are equal
to zero.  Let $f (t) = R^{(p,q)}
(\auxx_{t};\dot{\varphi}^{p};\dot{\aux}^{q})$, where
$\dot{\varphi}^{p}; \dot{\aux}^{q}$ are $\Auxx (\h_{+})$ unit norm
directions of differentiation in $\R_{\h_+}^{n}\times \Aux$.  The
Taylor expansion of $f(t)$ at $t=1$ to order $k-p$ about $t=0$
vanishes, so by the integral form of the Taylor remainder, and again
replacing $R$ by $F$ as in the second sum of \refeq{Rpsums}, we obtain
\begin{equation}
  R^{(p,q)} (\auxx;\dot{\varphi}^{p};\dot{\aux}^{q})
  =
  \int_{0}^{1} dt\; \frac{(1-t)^{k-p}}{(k-p)!}
  F^{(k+1,q)} (\auxx_{t};\dot{\varphi}^{p},\varphi^{k+1-p};\dot{\aux}^{q}) ,
\end{equation}
where $\varphi^{k+1-p} = \varphi,\ldots,\varphi \in (\R^n)^{k+1-p}$.
We take the supremum over the directions and apply the
definition \eqref{e:M-norm2} of the $\Auxx$-norm of derivatives.
This yields
\begin{align}
  \|R^{(p,q)} (\auxx)\|_{\Auxx (\h_+)}
  &\le
  \int_{0}^{1} dt\; \frac{(1-t)^{k-p}}{(k-p)!}
  \|F^{(k+1,q)}(\auxx_{t})\|_{\Auxx (\h)}
  (\tfrac{\h_{+}}{\h})^{p} (\tfrac{|\varphi|}{\h})^{k+1-p}.
\end{align}
Since
\begin{equation}
    \sum_{q\le p_{\Aux}}\tfrac{1}{q!}
    \|F^{(k+1,q)}(\auxx_{t})\|_{\Auxx (\h)}
    \le
    (k+1)!\,  \|F\|_{T_{\auxx_{t}} (\h)}
    \le (k+1)!\, \bar F,
\end{equation}
this gives
\begin{align}
\lbeq{Rpsumk}
  \sum_{p=0}^{k}
  \sum_{q\le p_{\Aux}}
  \tfrac{1}{p!}
  \tfrac{1}{q!}
  \|R^{(p,q)} (\varphi)\|_{\Auxx (\h_{+})}
  & \le
  \frac{\bar F}{\h^{k+1}}
  \sum_{p=0}^{k}
  \binom{k+1}{p}
  \h_{+}^{p} |\varphi|^{k+1-p}
  \nnb &
  \le
  (\tfrac{\h_{+}}{\h})^{k+1} P_{\h_+}^{k+1}(\varphi) \bar F  .
\end{align}
In the last step, we extended the sum to $p \le k+1$, applied the
binomial theorem, and used the definition of $P_{\h_+}$ from
\refeq{P-def-6}.

For the second sum in \refeq{Rpsums}, we observe that the
definition~\ref{e:M-norm2} of the $\Auxx$-norm implies that
$\|F^{(p,q)} (\varphi)\|_{\Auxx (\h_{+})} =
(\frac{\h_{+}}{\h})^{p}\|F^{(p,q)} (\varphi)\|_{\Auxx (\h_{+})}$.
Since $\tfrac{\h_+}{\h} \le 1$, we obtain
\begin{align}
  \sum_{p=k+1}^{p_{\Ncal}}
  \sum_{q\le p_{\Aux}}
  \tfrac{1}{p!}
  \tfrac{1}{q!}
  \|F^{(p,q)} (\varphi)\|_{\Auxx (\h_{+})}
  & =
  \sum_{p=k+1}^{p_{\Ncal}}
  \sum_{q\le p_{\Aux}}
  \tfrac{1}{p!}
  \tfrac{1}{q!}
  \|F^{(p,q)} (\varphi)\|_{\Auxx (\h)}
  (\tfrac{\h_+}{\h})^{p}
  \nnb
  &\le
  (\tfrac{\h_+}{\h})^{k+1}\,\| F\|_{T_\auxx(\h)}
  .
\end{align}
Since $P_{\h_+}(\varphi) \ge 1$ and $\| F\|_{T_\auxx(\h)}\le \bar F$,
the above estimate, together with \refeq{Rpsums} and \refeq{Rpsumk},
completes the proof of \refeq{Taycontraction-1}.
\end{proof}

In subsequent chapters, we will make use of two choices of $\h$,
namely $\ell$ and $h$ with $\ell \leq h$.  The following corollary
shows that the $T_{0,\aux}(\ell)$- and $T_{\infty,\aux}(h)$-seminorms
together also control the $T_{\varphi,\aux}(\ell)$-seminorm.

\begin{cor} \label{cor:norm-comparison}
For $k<p_\Ncal$ and $0< \h \le h$,
\begin{equation} \label{e:norm-comparison-simp}
    \|F\|_{T_{\varphi,\aux} (\h)}
    \le
    P_\h^{k+1}
    (\varphi)
    \left(
    \|F\|_{T_{0,\aux} (\h)}
    +
    2\left(\tfrac{\h}{h}\right)^{k+1}
    \|F\|_{T_{\infty,\aux}(h)}
    \right)
\end{equation}
\end{cor}

\begin{proof}
Let $k<p_\Ncal$. With $\Tay_k$ defined by \refeq{TaykF0}, we write
\begin{equation}
  F = \Tay_k F + (1-\Tay_k)F.
\end{equation}
Then \eqref{e:norm-comparison-simp} follows from
Lemmas~\ref{lem:TayTphi} and \ref{lem:Taycontraction}
with $\h=h$ and $\h_+=\h$.
\end{proof}

\section{Polynomial estimates}

In this section, we obtain estimates on the covariance of two
polynomials in the field $\varphi$, and on $\delta U = \theta U -
\Upt(U)$.  Here $\theta U(B)$ is defined by
\begin{equation}
    \theta_\zeta U(B)
    =
    \sum_{x\in B} U(\varphi_x + \zeta_x)
    =
    \sum_{x\in B} U(\varphi + \zeta_x),
\end{equation}
where $\varphi$ is constant on $B\in\Bcal_+$ and $\zeta$ is constant
on smaller blocks $b \in \Bcal$.

\index{$\mathfrak{c}$} Given a covariance $C_+$, we write
\begin{equation}\label{e:cfrak-def}
    \mathfrak{c}_+^2=\max_{x,y\in \Lambda}|C_{+;x,y}|
    = C_{+;x,x}.
\end{equation}
The second equality follows from the fact that since $C_{+;x,y}$ is
positive-semidefinite, $C_{+;x,y}^{2}$ is bounded by
$C_{+;x,x}C_{+;y,y}$, and $C_{+;x,x}=C_{+;y,y}$ for our covariances.

\begin{lemma}
\label{lem:covariance-6} There exists $c>0$ such that for $\h \ge
\mathfrak{c}_+ >0$, for $U=u+V$ and $U'=u'+V'$ polynomials of degree
$4$ with constant parts $u,u'$, and for $x \in B \in \Bcal_+$,
\begin{align}
\label{e:covariance1-6}
    &
    \| \LT \;\Cov_{+} \big(\theta U_x,\theta U' (B)\big)\|_{T_{\varphi,\aux} (\h)}
    \nnb
    &
    \qquad \qquad
    \le
    c
    \left(\tfrac{\mathfrak{c}_+}{\h}\right)^{4}
    \|V_{x}\|_{T_{0,\aux}(\h)} \|V'(B)\|_{T_{0,\aux}(\h)}
    P_{\h}^{4}(\varphi)
    .
\end{align}
\end{lemma}

\begin{proof}
Without loss of generality, we can and do assume that $u=u'=0$ since
constants do not contribute to the covariance.  Recall the multi-index
notation from above Lemma~\ref{lem:covUcal}.  Let $S =
\{(\alpha,\alpha') : |\alpha|\le 4, \, |\alpha'| \le 4,\,
|\alpha|+|\alpha'| \in \{4,6,8\}\}$.  As in the proof of
Lemma~\ref{lem:covUcal},
\begin{equation}
    \LT \;\Cov_{+} \big(\theta V_{x},\theta V'(B)\big) %
    =
    \sum_{(\alpha,\alpha')\in S}
    \frac{1}{\alpha!\alpha'!} V^{(\alpha)} V'^{(\alpha')}
    \sum_{x' \in B}
    \Cov_{+} \big(\zeta_{x}^{\alpha},\zeta_{x'}^{\alpha'} \big) .
\end{equation}
If $U$ is a polynomial in $\varphi$ of degree at most $4$, and if
$|\alpha|=p \le 4$, then it follows from Exercise~\ref{ex:Tphi-poly} and
Lemma~\ref{lem:derivatives2} that
\begin{equation}
\label{e:Ualphabd}
    \| U^{(\alpha)} \|_{T_{\varphi,\aux}}
    \le
    \|U^{(\alpha)}\|_{T_{0,\aux}} P_{\h}^{4-p}(\varphi)
    \le
    O (\h^{-p})
    \|U\|_{T_{0,\aux}} P_{\h}^{4-p}(\varphi) 
    .
\end{equation}
Therefore, with $p=|\alpha |$ and $p'=|\alpha' |$,
\begin{align}
    &
    \|\LT \;\Cov_{+} \big(\theta V_x,
    \theta V' (B)\big)\|_{T_{\varphi,\aux} }
    \nnb & \quad \le
           O(1)
    \|V 
    \|_{T_{0,\aux}}
    \|V'\|_{T_{0,\aux}}
    \sum_{(\alpha,\alpha')\in S}
    \h^{-p-p'}
    P_{\h}^{8-p-p'}(\varphi)
    \sum_{x'\in B}
    \left|
    \Cov_{+} \big(\zeta_{x}^{\alpha},\zeta_{x'}^{\alpha'} \big)
    \right| .
\end{align}
It follows from Exercise~\ref{ex:wickpp} and the definition of
$\mathfrak{c}_+$ that
the covariance is bounded by $O(\mathfrak{c}_+^{p+p'})$.
After inserting this bound, there is no dependence on $x'$ so the
sum over $x'$ becomes a factor $|B|$ which together with
$\|V'\|_{T_{0,\aux}}$ equals $\|V' (B)\|_{T_{0,\aux}}$ because
$\varphi$ is constant on $B$. Also, $P_\h^{8-p-p'}(\varphi) \leq
P_\h^4(\varphi)$, and since $\h \geq \mathfrak{c}_+$, the proof of
\refeq{covariance1-6} is complete since $\sum_{(\alpha,\alpha')\in S}
(\mathfrak{c}_+/\h)^{p+p'}\leq O(\mathfrak{c}_+/\h)^4$.
This completes the proof.
\end{proof}

\begin{exercise}
\label{ex:EthetaV-6}
By adapting the proof of Lemma~\ref{lem:covariance-6},
and with the same hypotheses, show that
\begin{equation}
    \|
    \Ex_{C_{+}}\big(\theta U (B) - U (B)\big)
    \|_{T_{\varphi,\aux} (\h)}
    \le
    c \left(\tfrac{\mathfrak{c}_{+}}{\h}\right)^{2} \|V(B)\|_{T_{0,\aux} (\h)}
    P_{\h}^{2}(\varphi) .
\end{equation}
\solref{EthetaV-6}
\end{exercise}

The next lemma and proposition include as hypothesis $
\|V(B)\|_{T_{0,\aux}(\h)} \leq 1$.  The upper bound $1$
serves merely to avoid introduction of a new constant, and any finite
upper bound would serve the same purpose.  In both the lemmas and
proposition, $U$ can be replaced by $V$ in the statement and proof.
For the lemma this is because if $U=u+V$ then $\Upt(U)=u+\Upt(V)$ so
$u$ cancels in the left-hand side. Likewise, for $\delta U = \theta U
- \Upt(U)$ as in \refeq{dUdef}, we have $\delta U = \theta U - \Upt(U)
= \theta V - \Upt(V) =\delta V$.

\begin{lemma}
\label{lem:Vpt-6} There exists $c>0$ such that, for $\h \ge
\mathfrak{c}_+>0$, and for all $U=u+V$ with $
\|V(B)\|_{T_{0,\aux}(\h)} \leq 1$,
\begin{equation}
    \|\Upt (B) - U(B)\|_{T_{\varphi,\aux}(\h)}
    \le
    c (\tfrac{\mathfrak{c}_+}{\h}) \|V(B)\|_{T_{0,\aux}(\h)}
    P_{\h}^{4}(\varphi) .
\end{equation}
\end{lemma}

\begin{proof}
As discussed above, we can replace $U$ by $V$.  By
Definition~\ref{def:Phipt},
\begin{equation}
    \Upt (B)
    -
    V (B)
    =
    \Ex_{C_{j+1}}\big(\theta V (B)-V (B)\big) -
    \frac{1}{2} \LT \Var_{+}\big(\theta V (B)\big).
\end{equation}
The desired inequality then follows from Exercise~\ref{ex:EthetaV-6}
and Lemma~\ref{lem:covariance-6}, together with $P_\h^2 \le P_\h^4$
and $\frac{\mathfrak{c}_+}{\h} \le 1$.
\end{proof}

\begin{prop}
\label{prop:dVnorm-6} There exists $c>0$ such that, for $\h \ge
\mathfrak{c}_+>0$, and for all $m \geq 1$ and all $U =u+V$ with
$\|V(B)\|_{T_{0,\aux}(\h)} \leq 1$,
\begin{align} \label{e:dVnorm-6}
    \| \delta U (B)_{\zeta }\|_{T_{\varphi,\aux} (\h)}^{m}
    &\le
    c^m O(\tfrac{\mathfrak{c}_{+}}{\h})^m
    \|V(B)\|_{T_{0,\aux} (\h)}^{m}
    P_{\h}^{4m}(\varphi)
    \frac{1}{|B|}\sum_{x\in B}
    P_{\mathfrak{c}_{+}}^{4m}(\zeta_x).
\end{align}
\end{prop}

\begin{proof}
As discussed above, we can replace $U$ by $V$.  Also, it suffices to
prove the case $m=1$, since this case implies the general case by 
Jensen's inequality in the form
$(|B|^{-1}\sum_{x\in B} |a_x|)^m \leq |B|^{-1} \sum_{x\in B} |a_x|^m$.
Let $m=1$.

By the triangle inequality,
\begin{equation}
    \| \delta V(B)_{\zeta }\|_{T_{\varphi,\aux}}
    =
    \| \theta_{\zeta}V(B) - V(B)\|_{T_{\varphi,\aux}} +
    \| V(B) - \Upt (B) \|_{T_{\varphi,\aux}} .
\end{equation}
The second term obeys the desired estimate, by Lemma~\ref{lem:Vpt-6}.

For the first term, it suffices to prove that
\begin{align}
    \| \tfrac{d}{dt} \theta_{t\zeta}V_{x} \|_{T_{\varphi,\aux}}
    &\le
    O(\tfrac{\mathfrak{c}_{+}}{\h})
    \|V\|_{T_{0}}
    P_{\h}^{3}(\varphi)
    P_{\mathfrak{c}}^{4} (\zeta_{x}) ,
\label{new-e:Tphi-zeta2-6}
\end{align}
since integration over $t\in [0,1]$ and summing over $x\in B$
then leads to the desired estimate.  For $j=1,\ldots,n$, let $e_j$
denote the multi-index which has $1$ in $j^{\rm th}$ position and $0$
elsewhere.  We apply the chain rule,
Proposition~\ref{prop:ExCthetaTphi}, and \refeq{Ualphabd} to obtain
\begin{align}
    \| \tfrac{d}{dt} \theta_{t\zeta}V_{x} \|_{T_{\varphi,\aux}}
    &\le
    \sum_{j=1}^n \| \theta_{t\zeta} V^{(e_j)}_{x} \|_{T_{\varphi,\aux}} |\zeta_{x}^j|
    =
    \sum_{j=1}^n\| V^{(e_j)}_{x} \|_{T_{\varphi+t\zeta,\aux}} |\zeta_{x}^j|
    \nonumber\\
    &\le
    O (\h^{-1})
    \|V_{x}\|_{T_{0,\aux}}
    P_{\h}^{3}(\varphi + t \zeta_{x}) |\zeta_{x}| .
\end{align}
For $t \in [0,1]$, $P_{\h} (\varphi+t\zeta)
\le P_{\h}(\varphi)P_{\h} (\zeta) \le
P_{\h}(\varphi)P_{\mathfrak{c}_{+}} (\zeta)$, where we used
$\h\ge \mathfrak{c}_{+}$ in the final inequality. Also,
$|\zeta|\le \mathfrak{c}_{+} P_{\mathfrak{c}_{+}}(\zeta)$.
This gives \eqref{new-e:Tphi-zeta2-6} and completes the proof.
\end{proof}


%% file: hierK.tex
\chapter{Global flow: Proof of Theorem~\ref{thm:phi4-hier-chi}}
\label{ch:pfsus}

The main theorem proved in this book is
Theorem~\ref{thm:phi4-hier-chi}, which provides the asymptotic
behaviour of the susceptibility of the 4-dimensional hierarchical
model.  Chapter~\ref{ch:pt} proves Theorem~\ref{thm:phi4-hier-chi}
subject to Theorem~\ref{thm:VK} and Proposition~\ref{prop:N}.  In this
chapter, we state the two main theorems concerning the renormalisation
group, namely Theorems~\ref{thm:step-mr-K}--\ref{thm:step-mr-R}, and
use these theorems to prove Theorem~\ref{thm:VK}.  This then proves
Theorem~\ref{thm:phi4-hier-chi} subject to
Theorems~\ref{thm:step-mr-K}--\ref{thm:step-mr-R} and
Proposition~\ref{prop:N}.  The proof of Theorem~\ref{thm:step-mr-R} is
given in Chapter~\ref{ch:R+U} and the proofs of
Theorem~\ref{thm:step-mr-K} and Proposition~\ref{prop:N} are given in
Chapter~\ref{ch:pf-thm:step-mr-K}.

We begin in Section~\ref{sec:fields} with a discussion of fields and
domains for the renormalisation group coordinate $V$.  Our choice of
norms for the coordinate $K$ is introduced in Section~\ref{sec:Wcal}.
The main theorems about the renormalisation group map,
Theorems~\ref{thm:step-mr-K}--\ref{thm:step-mr-R}, are stated in
Section~\ref{sec:mr}.  In Section~\ref{sec:BS1}, we apply these main
theorems to construct the critical point and a global renormalisation
group flow started from the critical point.  Finally, in
Section~\ref{sec:pfprops}, we apply the main theorems to prove
Theorem~\ref{thm:VK}.

\section{Fluctuation and block-spin fields}
\label{sec:fields}

\subsection{Hierarchical field}

For the analysis of the renormalisation group map $\Phi_{j+1}$ defined
in Definition~\ref{def:RGmap}, the scale $0 \leq j<N$ is fixed, and we
often drop the subscript $j$ and replace the subscript $j+1$ by
$+$. Thus we write $(V,K) \mapsto (U_+,K_+)$ when discussing the map
$\Phi_{+}$.  All results are uniform in the scale $j$.  We write $B$
for an arbitrary \emph{fixed} block in $\Bcal_{+} (\Lambda)$, whereas
blocks in $\Bcal (B)$ are denoted by $b$.

We recall the decomposition of the covariance $C = C_1 + \cdots +
C_{N-1} + C_{N,N}$ from Proposition~\ref{prop:hGFF}.  For the last
step, we further divide $C_{N,N}= C_N + \Cnewlast{}$.  Given $j$, only
the covariances $C_{j+1}$ and $C_{j+2} + \cdots + C_{N,N}$ are of
importance.

By definition of the hierarchical GFF,
\begin{itemize}
\item the restriction of $x \mapsto \zeta_{x}$ to a block $b \in
\Bcal$ is constant;
\item the restriction of $x \mapsto \varphi_{x}$ to a block $B \in
\Bcal_{+}$ is constant.
\end{itemize}
When attention is on fields $\zeta_{x}$ with $x$ restricted to a
specified scale $j$ block then we often omit $x$ and write $\zeta$
instead.  For the same reason we write $\varphi$ instead of
$\varphi_{x}$ when $x\in B$.

The analysis of $\Phi_+$ relies on perturbation theory and Taylor
approximation in powers of the field $\varphi$ about $\varphi = 0$.
These are only good approximations when fields are small.  Large
fields are handled by non-perturbative estimates which show that large
fields are unlikely.  Implementing this apparently simple idea leads
to notorious complications in rigorous renormalisation group analysis
that are collectively known as the \emph{large-field problem}.  The
subsequent chapters provide a way to solve the large-field problem in
the hierarchical setting, where the difficulties are fewer than in the
Euclidean setting.

Two mechanisms suppress large fields, one for the fluctuation field
and one for the block-spin field.
\begin{itemize}
\item The fluctuation field suppression comes from the low probability
that a Gaussian field is much larger than its standard deviation.
\item The block-spin suppression comes from the factor
$e^{-g\tau^{2}}$ in $e^{-V}$.  This is more subtle because it is a non-Gaussian effect.
\end{itemize}

\subsection{Fluctuation field}
\index{Fluctuation field}

By \eqref{e:scaling-estimate-hier}, for $j+1 \leq N$, the variance of
the fluctuation field $\zeta_{x} = \zeta_{j+1,x}$ at any point $x$ is
\begin{equation}
  C_{j+1;x,x}(m^2) = (1+m^2L^{2j})^{-1} L^{-(\drb-2)j}(1-L^{-d}).
\end{equation}
Given $m \ge 0$, the mass scale $j_m$ is defined in
Definition~\ref{def:jm}.  As in \eqref{e:varthbdx}, the exponential
decay beyond the mass scale due to the factor $(1+m^2L^{2j})^{-1}$ is
encoded by the larger sequence $\vartheta_j^2$, with
\begin{equation}
\lbeq{varthdef2}
    \vartheta_j = 2^{-(j-j_m)_+}.
\end{equation}

\index{$\ell$}
We fix an $L$-dependent constant
\begin{equation}
\lbeq{ell0def}
    \ell_0 = L^{1+\drb /2},
\end{equation}
and define the \emph{fluctuation-field scale}
\begin{equation}
\label{e:ell-def}
  \ell
  =
  \ell_{j}
  =
  \ell_0 L^{-j(\drb -2)/2}.
\end{equation}
Then $C_{+;x,x}$ is bounded by $\vartheta^2\ell_0^{-2} \ell^2$.
Therefore, with $\mathfrak{c}_+$ given by \eqref{e:cfrak-def}, a
typical fluctuation field has size on the order of
\begin{equation}
    \lbeq{Cellbd}
    \mathfrak{c}_{+}
    = C_{+;x,x}^{1/2}
    \le
    \vartheta \ell_0^{-1} \ell
    \le
    \vartheta_+ \ell_{+},
\end{equation}
where we used \refeq{ell0def} and $L\ge 2$ for the last inequality.

We use $\mathfrak{c}_+$ to control the covariance when it is important
to know its decay as a function of the mass.  The parameter $\ell_j$
is an upper bound for the covariance which is independent of the mass
and which we use in the definition of norms.

\subsection{Block-spin field}
\index{Block spin field}

\index{$h$}
\index{$k_0$}
For the block-spin field, we fix strictly positive parameters
$\ggen = \ggen_j$ and $\ggen_+ = \ggen_{j+1}$ obeying
\begin{equation}
    \label{e:gtilde}
    \ggen_{+}
    \in
    [\tfrac{1}{2}\ggen, 2\ggen] ,
\end{equation}
and also fix a small constant $k_0>0$ whose value is determined in
Proposition~\ref{prop:stability}.  Then we define the
\emph{large-field scale}
\begin{equation}
    \label{e:h-def}
    h =
    h_{j} =
    k_0
    (L^{\drb j} \ggen_j)^{-1/4}
    .
\end{equation}
The definition of $h_j$ is arranged so that if $g \asymp \ggen_j$ and
$|\varphi_x| \asymp h_j$, then $\sum_{x\in b} g |\varphi_x|^4
\asymp k_0^{4}$ is positive uniformly in all parameters
$j,L,g$.  In other words, the exponential decay due to $e^{-\frac 14 g
\sum_{x\in b} |\varphi_x|^4}$ becomes significant once $|\varphi|$
exceeds the large-field scale $h_j$, provided that the coupling
constant $g$ is close to its reference value $\ggen$. The latter
condition is encoded by the \emph{stability domain} for the coupling
constants, defined by
\begin{align}
    \DVstab_{j}
    &= \Big\{(g,\nu,u) :
    k_0 \ggen_{j} < g < k_0^{-1} \ggen_{j}, \;
    |\nu| <
    \ggen_{j} h_j^2, \;
    |u| <
    \ggen_{j} h_j^4
    \Big\}.
\label{e:DVstab}
\end{align}
Indeed, the domain $\DVstab$ is defined to make the following estimate
work.

\begin{exercise}\label{ex:stability}
Show that if $U \in \DVstab$ then
\begin{equation}
    U (\varphi)
    \ge
    \tfrac{1}{8} k_0  \ggen |\varphi|^{4}
    -
    \tfrac{3}{2}k_0^3 L^{- \drb j}
    ,
\end{equation}
and hence, if $k_0$ is chosen small enough that $e^{\tfrac{3}{2}k_0^3}
\le 2$, then
\begin{equation}
    \label{e:stability}
    e^{-   U(\varphi)}
    \le
    \left( e^{\tfrac{3}{2}k_0^3}
    e^{-\tfrac{1}{8} k_0^{5}  |\varphi/h_j|^{4}}
    \right)^{L^{-\drb j}}
    \le
    \left( 2
    e^{-\tfrac{1}{8} k_0^{5}  |\varphi/h_j|^{4}}
    \right)^{L^{-\drb j}} .
\end{equation}
\solref{stability}
\end{exercise}

\section{Main estimate on renormalisation group map}

\subsection{Domain for \texorpdfstring{$V$}{V}}

In addition to the stability domain, which ensures the stability
estimate \eqref{e:stability}, we define a smaller domain $\DV$ which
puts constraints on the coupling constants.  These constraints ensure that
the non-perturbative flow remains close to the perturbative flow
defined by the map $\Phi_\pt$ of \refeq{Phiptdef}.  Thus we estimate
$\Phi_+(V,K)$ for $V$ in the domain in $\Vcal$ defined by
\begin{align}
\label{e:DVdef}
    \DV_{j}
    &= \{(g,\nu) :
    2k_0 \ggen_{j} < g < (2k_0)^{-1} \ggen_{j}, \;
    |\nu| < (2k_0)^{-1} \ggen_{j} L^{- (\drb -2)j}
    \} .
\end{align}

The following lemma shows that $\DV$ is contained in both $\DVstab$
and $\DVstab_+$.

\begin{lemma}
\label{lem:DVnest}
For $d=4$, and for each scale $j$,
\begin{equation}
\label{e:DVnest}
    \DV_j \subset
    \DVstab_j \cap \DVstab_{j+1} .
  \end{equation}
\end{lemma}

\begin{proof}
To see that $\DV \subset \DV^\stab$, we examine the three coupling
constants one by one.  The inclusion for $g$ is immediate from the
definition of the domains.  For $\nu$, the $\DV$ condition is $|\nu |
< (2k_0)^{-1}\ggen L^{-2j}$. For $\ggen$ small depending on $k_{0}$
this implies $|\nu | < k_{0}^{-1/2}\ggen^{1/2} L^{-2j}$ which is the
desired $\DVstab$ condition.  For $u$, since $u=0$ for elements of
$\DV_{j}$, there is nothing to check.  The inclusion $\DV \subset
\DV^\stab_+$ follows similarly, using \eqref{e:gtilde}.  For $\nu$ we
must take $\ggen$ small depending on $L$.
\end{proof}

\subsection{Norms}
\label{sec:Wcal}
\index{$\Wcal$-norm}
\index{Norm}

Our estimates on the renormalisation group map are expressed in terms
of certain norms.  These norms are constructed from the
$T_\varphi(\h)$-seminorm of Definition~\ref{def:Tnorm}.  At present,
we do not use any auxiliary space $\Aux$; that will become advantageous in
Chapters~\ref{ch:R+U}--\ref{ch:pf-thm:step-mr-K}.  To obtain estimates that are useful for both the
fluctuation-field scale $\ell$ and the large-field scale $h$, we use
the two choices $\h=\ell$ and $\h=h$.  We fix the parameter $p_\Ncal$,
which guarantees sufficient smoothness in $\varphi$ in
Definition~\ref{def:Tnorm}, with
\begin{equation}
  \label{e:pN-choice}
  p_\Ncal \geq 10, \quad \text{where $p_\Ncal=\infty$ is permitted.}
\end{equation}
The choice $p_\Ncal=\infty$ provides analyticity, whereas the choice
of finite $p_\Ncal$ shows that analyticity is not required for the
method to apply.  For the fluctuation-field scale $\h=\ell$, we usually set
$\varphi$ equal to $0$ for all estimates, i.e., we use the
$T_{0}(\ell)$-seminorm. The following exercise, in particular
\eqref{e:T0-equivalent}, shows that the $T_{0} (\ell)$ norm of the
polynomial $V (b)$ is $O_{L} (\max\{g,|\mu|\})$, where $\mu=L^{2j}\nu$ was defined 
in \eqref{e:Greeksrescaled}.

\begin{exercise}
\label{ex:Ucal-ident} For $U = \frac 14 g |\varphi|^4 + \frac 12 \nu
|\varphi|^2 + u$ and $\h>0$ prove that
\begin{equation}
    \|U_x\|_{T_{0} (\h)}
    =
    \tfrac{1}{4} |g | \h^{4} +
    \tfrac{1}{2} |\nu | \h^{2} +
    |u| .
\end{equation}
Therefore, by \eqref{e:ell-def}, for $b\in \Bcal$,
\begin{equation}\label{e:T0-equivalent}
    \|V(b)\|_{T_0(\ell)}
    =
    \tfrac{1}{4} \ell_0^{4} |g | L^{-(\drb -4)j} +
    \tfrac{1}{2} \ell_0^{2} |\mu |  .
\end{equation}
In particular, by \eqref{e:DVdef} and \eqref{e:h-def}, for $V \in \DV$,
\begin{equation}\lbeq{VinDV}
    \|V(b)\|_{T_0(\ell)}
    \le
    \frac{\ell_0^4}{2k_0}
    \ggen,
    \quad
    \|V(b)\|_{T_0(h)}
    \le
    \frac{k_0^3}{2}
    .
\end{equation}
Hint: find the norm of each monomial separately.
\solref{Ucal-ident}
\end{exercise}

On the other hand, estimates in the large field scale $\h=h$ will be
uniform in $\varphi$, i.e., we use the $T_{\infty}(h)$-norm of
\eqref{e:jX-norm-def}.

The input bounds on $K$ we require for $\Phi_+$ are:
\begin{align}
\lbeq{Kinput1}
    \|K(b)\|_{T_{0}(\ell)} & \le O( \vartheta^{3} \ggen^3),
  \\
\lbeq{Kinput2}
    \|K(b)\|_{T_{\infty}(h)} & \le O(\vartheta^{3} \ggen^{3/4}).
\end{align}
A hint of the choice of powers of $\ggen$ in the above two right-hand
sides can be gleaned from the intuition that $K$ captures higher-order
corrections to second-order perturbation theory, and is dominated by
contributions containing a third power of $\delta V = \theta V -
\Upt(V)$.  According to Proposition~\ref{prop:dVnorm-6}, the
$T_\varphi(\h)$-seminorm of $(\delta V(b))^3$ has an upper bound that
includes a factor $[(\mathfrak{c}_+/\h) \|V(b)\|_{T_0(\h)}]^3$.  By
\eqref{e:VinDV}, and by the fact that $\mathfrak{c}_+\le \ell$ by
\refeq{Cellbd}, this factor is order $\ggen^3$ for $\h=\ell$ and is
order $\ggen^{3/4}$ for $\h=h$.

Recall the definition of the vector space $\Fcal$ in Definition~\ref{def:Fcal}.
We create a norm on $\Fcal$
that combines \refeq{Kinput1}--\refeq{Kinput2} into a
single estimate, namely (for any $b \in \Bcal$)
\begin{equation}
\label{e:Wcal}
  \|K\|_{\Wcal}
  =\|K(b)\|_{T_{0}(\ell)}
    +
    \ggen^{9/4}
    \|K(b)\|_{T_{\infty}(h)}
  .
\end{equation}
Then the statement that $\|K\|_{\Wcal} \le M\vartheta^{3}\ggen^3$
implies the two estimates $\|K (b)\|_{T_{0}(\ell)} \le M
\vartheta^{3}\ggen^3$ and $\|K (b\|_{T_{\infty}(h)} \le
M\vartheta^{3}\ggen^{3/4}$, as in \refeq{Kinput1}--\refeq{Kinput2}.

The $\Wcal$-norm does not obey the product property, whereas the
$T_{\varphi}$-seminorms do.  For this reason, our procedure is to
first obtain estimates for $T_{0}(\ell)$ and $T_{\varphi}(h)$, and to
then combine them into an estimate for the $\Wcal$-norm.  The next
lemma shows that the $\Wcal$-norm also controls the
$T_{\varphi}(\ell)$-norm for nonzero $\varphi$.

\begin{lemma}\label{lem:norm-comparison}
For $K \in \Fcal$ and $b \in \Bcal$,
\begin{equation} \label{e:norm-comparison}
    \|K(b)\|_{T_{\varphi}(\ell)} \le
    P_\ell^{10}(\varphi)
    \|K\|_{\Wcal}.
\end{equation}
\end{lemma}

\begin{proof}
The inequality is an immediate consequence of
Corollary~\ref{cor:norm-comparison} with $k=9<p_\Ncal$ by
\eqref{e:pN-choice}, since $(2\tfrac{\ell}{h})^{10}$ is $o
(\ggen^{9/4})$.
\end{proof}

\subsection{Main result}
\label{sec:mr}

The renormalisation group map $\Phi_+$ depends on the mass $m^2 \geq
0$.  Our estimates for $\Phi_+$ would be most easily stated for a
fixed $m^2$.  However, to prove the continuity of the critical point
as a function of the mass as in Theorem~\ref{thm:VK}, we regard
$\Phi_+$ as a function jointly in $(V,K,m^2)$.  Because $\Phi_+$ has
strong dependence on $m^2$, this requires care to obtain estimates
that are uniform in $(V,K,m^2)$.  To achieve this, we fix $\mgen^2 \ge
0$ and employ the mass domain $\Iint_j(\mgen^2)$ defined in
\eqref{e:Ijdef}, and regard the renormalisation group map as a
function of $m^2 \in \Iint_+(\mgen^2)$. Then $m^2$ is essentially
fixed to $\mgen^2$, but can still be varied.  We also define the
sequence
\begin{equation} \label{e:varthetadef}
  \tilde \vartheta_j = 2^{-(j-j_{\mgen})_+}.
\end{equation}
By our assumption that $m^2 \geq \frac12 \mgen^2$, we have
$\vartheta_j \leq 2\tilde \vartheta_j$.

\index{Domain}
\index{$\domRG$}
Given $\CRG>0$, for $\ggen > 0$ and $\mgen \geq 0$,
the domain of $\Phi_+$ is defined to be the set of $(V,K)$ in
\begin{equation}
  \label{e:Phi_+-domain}
  \domRG = \DV \times
  \{K \in \Fcal : \|K\|_\Wcal < \CRG\tilde \vartheta^{3}\ggen^3\}
  ,
\end{equation}
where the norm of $V\in\DV$ is $\|V(b)\|_{T_0(\ell)}$.
\index{$\Vcal(\ell)$}
We will write $\Vcal(\ell)$ to denote the vector space $\Vcal$ with norm
$\|V (b)\|_{T_{0} (\ell)}$.
Note that the norm on $\Wcal$ and
the domain $\domRG$ are defined in terms of both $\ggen>0$ and
$\mgen^2 \geq 0$ (through $\tilde \vartheta$).
We sometimes emphasise
this dependence by writing
\begin{equation}
    \domRG = \domRG(\mgen^2,\ggen^2).
\end{equation}
We always assume that $\ggen$ and $\ggen_+$ have bounded ratios as in
\refeq{gtilde}.

The following theorem is proved in Chapter~\ref{ch:pf-thm:step-mr-K}.
Besides providing estimates on $\Phi_+^K$, the theorem also specifies
the constant $\CRG$ occuring in $\domRG$.  The first case in
\refeq{DVKbd} shows that $\|K\|_\Wcal \le \CRG \tilde
\vartheta^3\ggen^3$ implies that $\|K_+\|_{\Wcal_+} \le \CRG \tilde
\vartheta_+^3\ggen_+^3$.  This shows that $K$ does not expand as the
scale is advanced.  The proof of this crucial fact is based on the
third case in \refeq{DVKbd}, which shows that
the $K$-derivative of the map taking $K$ to $K_+$ can be made as small
as desired by a choice of sufficiently large $L$, so the map
$\Phi_+^K$ is contractive.

To formulate bounds on derivatives, we consider maps $F:
\Vcal(\ell) \times \Wcal \to \Xcal$ taking values in a normed space
$\Xcal$, where here $\Xcal$ is $\Ucal_+(\ell_+)$ or $\Wcal_+$.  For
$y=(V,K) \in \Vcal(\ell)\times \Wcal$, the derivative
$D^{p_2}D^{p_3}F(V,K)$ at $(V,K)$ is a multilinear map
$\Vcal(\ell)^{p_2} \times \Wcal^{p_3} \to X$.  We write
\begin{equation}
  \label{e:deriv-norm}
  \|D_V^{p_{2}}D_K^{p_{3}} F(V,K)\|_{\Vcal(\ell) \times \Wcal \to \Xcal}
\end{equation}
for the norm of this multilinear map.
In the next theorem, and for other statements that are uniform in all $(V,K)$ considered,
  we typically omit the argument $(V,K)$ from  the notation.

\index{Renormalisation group step}
\begin{theorem}
\label{thm:step-mr-K}
  Let $\mgen^2 \ge 0$, let $L$ be sufficiently large,
  let $\ggen$ be sufficiently small (depending on $L$), and let $p,q\in \N_0$.
  Let $0\le j<N$.
  There exist $L$-dependent $\CRG,M_{p,q} >0$
  and $\kappa =O(L^{-2})$  such that the map
  \begin{equation}
  \lbeq{Kplusmap}
    \Phi^K_+:\domRG  \times \Iint_+  \to \Wcal_{+}
  \end{equation}
  satisfies the estimates
  \begin{align}
\lbeq{DVKbd}
    \|D_{V}^pD_{K}^{q}\Phi^K_+\|_{\Vcal(\ell) \times \Wcal \to \Wcal_+}
    &\le
    \begin{cases}
    \CRG
    \tilde \vartheta_+^{3}
    \ggen_+^{3}
    &
    (p = 0, \, q=0 )
    \\
    M_{p,0}
    \tilde \vartheta_+^{3}
    \ggen_+^{3-p}
    &
    (p > 0, \, q=0 )
    \\
    \rlap{$\kappa$}\hspace{3.3cm}
    & (p=0,\, q=1)
    \\
    M_{p,q}
    \ggen_+^{-p-\frac{9}{4}(q-1)}
    &
    (p \ge 0,\,  q \ge 1)
    .
    \end{cases}
  \end{align}
In addition, $\Phi^K_+$ and all Fr\'echet derivatives $D_V^pD_K^q
\Phi^K_+$ are jointly continuous in all arguments $V,K, \dot{V},
\dot{K}$, as well as in $m^2 \in \Iint_+$.
\end{theorem}

The map $\Phi_\pt^U(V)$ is the same as $\Phi_+^U(V,0)$, and it has
been analysed explicitly in Section~\ref{sec:Phipt}.  Thus, to
complete the understanding of the map $\Phi_+^U$, we recall from
\eqref{e:RUdef-5} the definition
\begin{equation}
\label{e:RUdef-1}
    R_+^U(V,K) = \Phi_+^U(V,K) - \Phi_+^U(V,0) .
\end{equation}
By definition, $R_+^U(V,K)$ is an element of $\Ucal$.
Similar to the notation $\Vcal(\ell)$ we introduced
for the vector space $\Vcal$ with norm
$\|V (b)\|_{T_{0} (\ell)}$, we write $\Ucal_+ (\ell_{+})$ for the vector space $\Ucal$ with norm
$\|U (B)\|_{T_{0} (\ell_{+})}$ for $B\in \Bcal_+$.
\index{$\Ucal_+(\ell_{+})$}
The following theorem is proved in Chapter~\ref{ch:R+U}.

\begin{theorem}
\label{thm:step-mr-R}
  Let $\mgen^2 \ge 0$,
  let $\ggen$ be sufficiently small (depending on $L$), and let $p,q\in \N_0$.
  Let $0\le j<N$.
  There exists an $L$-dependent constant $M_{p,q}>0$
  such that the map
  \begin{equation}
  \lbeq{Rplusmap}
    R^U_+:\domRG   \times \Iint_+  \to \Ucal_+(\ell_+) 
  \end{equation}
  satisfies the estimates
  \begin{align}
    \label{e:Rmain-g}
    \|D_V^p D_K^q R^U_+\|_{\Vcal(\ell) \times \Wcal \to \Ucal_+(\ell_+)}
    & \le
    \begin{cases}
     M_{p,0}
    \tilde \vartheta_+^{3} \ggen_+^{3} & (p\ge 0,\, q=0)\\
    M_{p,q} & (p\ge 0,\, q = 1,2)\\
    \rlap{$0$}\hspace{1cm}  & (p\ge 0,\, q \ge  3).
    \end{cases}
  \end{align}
  In addition,
  $R_+^U$ and all Fr\'echet derivatives $D_V^pD_K^q R_+^U$
  are jointly continuous in all arguments $V,K, \dot{V}, \dot{K}$,
  as well as in $m^2 \in \Iint_+$.
\end{theorem}

As in \eqref{e:RUdef-5a}, we write the components of $R^U_+$ as
$(r_{g,j},r_{\nu,j},r_{u,j})$.  By \refeq{RUdef-1}, these components
give the remainder terms of the renormalisation
group flow relative to the
perturbative flow of Proposition~\ref{prop:barflow-a}.
By combining \eqref{e:T0-equivalent} at scale $j+1$ and \eqref{e:Rmain-g}, we see that
\begin{equation}
  \label{e:thm:step-mr-R:r-bounds}
  r_{g,j} = O_{L}(\vartheta_j^3g_j^3)   , \quad
  L^{2j} r_{\nu,j} = O_{L}(\vartheta_j^3 g_j^3) .
\end{equation}

\section{Construction of critical point}
\label{sec:BS1}

In this section, for $m^2 \ge 0$ we construct a critical value
$\nu_0^c(m^2)$ such that the renormalisation group flow exists for all
scales, and prove that $\nu_0^c(m^2)$ is continuous in $m^2$.  To do
so, we apply Theorems~\ref{thm:step-mr-K}--\ref{thm:step-mr-R}.

The $u$-component of $\Phi_+$ does not play any role and we therefore
write $\Phi_+ = (\Phi_+^V,\Phi_+^K)$.
Given $m^2 \geq 0$ and an integer $k \geq 0$, the sequence
$(V_j,K_j)_{j \leq k}$ is a \emph{flow} (up to scale $k$) of the
renormalisation group map $\Phi$
if, for all $j<k$, $(V_j,K_j)$ is in the domain $\domRG_j(m^2,g_0)$ and
\begin{equation} \label{e:flowdef}
  (V_{j+1},K_{j+1})
  = \Phi_{j+1}(V_j,K_j; m^2).
\end{equation}

\index{$\ggen$}
To apply
Theorems~\ref{thm:step-mr-K}--\ref{thm:step-mr-R}, we need to fix the
sequence $\ggen$.  Given $\mgen^2 \ge 0$, we define $\ggen_j =
\gbar_j(\mgen^2)$ with $\gbar_{j}$ given by
\eqref{e:gbar9}.  This choice obeys the condition \eqref{e:gtilde} by
Proposition~\ref{prop:gjtj}(i).  Thus we now have three similar sequences:
$g$, $\gbar$, and $\ggen$.  The sequence $g$ is the coupling
constant in the true renormalisation group flow and is given in terms
of a complicated equation involving the nonperturbative coordinate
$K_j$ and also $\mu_j$. On the other hand, the sequence $\gbar$ is
explicit in terms of the parameters $(g_0,m^2)$ and a simple quadratic
recursion.  The reason for introducing $\ggen$ in addition to $\gbar$
is that $\ggen$ does not depend on $m^2$.  However, for $m^2 \in
\Iint_{ +}(\mgen^2)$, the sequences $\gbar$ and $\ggen$ are
comparable, and under the condition that $V$ is in the domain $\DV$,
all three of $g,\gbar,\ggen$ are comparable.
In particular, we will use without further comment that for $V$
in $\DV$, error estimates $O(\tilde \vartheta_j^3\ggen_j^3)$ as in the
statement of Theorem~\ref{thm:step-mr-R} are equivalent to
$O(\vartheta_j^3 g_j^3)$, and similarly with other powers.

Given $c_0\in (0,(2k_0)^{-1})$, we define the intervals
\begin{align}
\lbeq{Gintdef}
    G_j &= G_j(m^2) = [-\tfrac{1}{2}\ggen_j(m^2), 2 \ggen_j(m^2)],
\\
\lbeq{Jintdef}
    J_j
    &=
    J_j(m^2;c_0)
    =
    [-c_0 \vartheta_j(m^2) \ggen_j(m^2), c_0 \vartheta_j(m^2) \ggen_j(m^2)].
\end{align}
By the definition of $\DV_j$ in \refeq{DVdef}, we have $G_j \times
L^{-2j}J_j \subset \Dcal_j$. The set $G_j \times J_j$ is
a domain for $(g_{j},\mu_{j})$ as opposed to $(g_{j},\nu_{j})$, hence
the factor $L^{-2j}$.

For the statement of the next proposition, we fix $c_\eta>0$ such that
the coefficient $\eta$ of \refeq{Greekdef2} obeys $|\eta_j| \le
c_\eta\vartheta_j$.  By \refeq{cjdef}, we can choose $c_\eta=n+2$.
For the following, we assume that $k_0$ is small enough to ensure that
$12c_\eta \le (2k_0)^{-1}$, i.e.,
\begin{equation}
\label{e:k0ceta}
    k_0 \le \frac{1}{24 (n+2)}.
\end{equation}
This insures that the choice $c_0=12c_\eta$ obeys the requirement
$c_0\in (0,(2k_0)^{-1})$; this choice occurs in the proof of
Proposition~\ref{prop:BScont}.  As always, we also assume that $k_0$
is small enough to satisfy the restriction imposed by
Proposition~\ref{prop:stability}.

The next proposition characterises the critical value $\nu_0^c$.  The
essence of the proof of its part~(i) is known as the
\emph{Bleher--Sinai argument} \cite{BS73}.

\index{Bleher--Sinai argument}
\index{Stable manifold}
\index{Critical point}
\begin{prop} \label{prop:BS}
Let $K_0=0$ and fix $g_0>0$ sufficiently small.  Let $m^2 \ge 0$.

\smallskip \noindent
(i)
There exists $\nu_0^c(m^2)$ such that the flow $(V_j,K_j)$ of
$\Phi$ with initial condition $g_0$ and
$\nu_0=\nu_0^c(m^2)$ exists for all $j \in \N$ and is such that
$(g_j,\mu_j) \in G_j(m^2) \times J_j(m^2;4c_\eta)$
and $\|K_j\|_{\Wcal_j} \leq \CRG\tilde \vartheta_{j}^{3}\ggen_{j}^3$
for all $j$.  In fact, $g_j = \ggen_j + O(\ggen_j^2 |\log
\ggen_j|)$.

\smallskip \noindent
(ii)
Let $c_0 \in [4c_\eta,(2k_0)^{-1})$.  The value of $\nu_0^c(m^2)$ is
unique in the sense that if a global flow exists started from some
$\nu_0$ and if this flow obeys $\mu_j \in J_j(m^2;c_0)$ for all $j$,
then $\nu_0=\nu_0^c(m^2)$.
\end{prop}

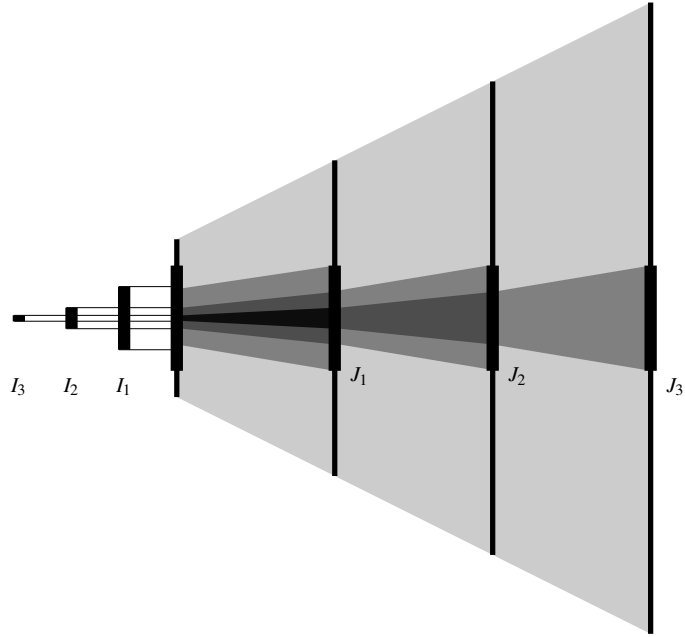
\begin{figure}
  \begin{center}
    \input{BS.pspdftex}
  \end{center}
  \caption{Illustration of the Bleher--Sinai argument.
    The intervals $J_j$ are $I_j$ are indicated by vertical bars,
    where the scale of the figure is chosen such that the intervals $J_j$ have the same length.
    Since the map $\mu_j \mapsto \mu_{j+1}$ expands, the condition that the image of $\mu_0$
    is contained in $J_j$ after $j$ iterations of the map, for all $j$,
    determines $\mu_0$ uniquely. \label{fig:BleherSinai}
  }
\end{figure}

\begin{proof}
(i)
Let $K_0=0$, and fix $g_0>0$, $m^2 \geq 0$, and $\ggen = \ggen(m^2)$.
Throughout the proof, we fix any $c_0\in [4c_\eta,(2k_0)^{-1})$ and
set $J_j=J_j(m^2;c_0)$, and we drop $m^2$ from the notation.

We apply induction in $k$. The induction hypothesis is that there is
a closed interval $I_k$ such that for $\mu_0 \in I_k$, the flow
$(V_j,K_j)_{j\leq k}$ exists, $\mu_k \in J_k$, and every $\mu_k \in
J_k$ has some preimage $\mu_0\in I_k$.  For $k=0$, the inductive
hypothesis clearly holds.
The intervals are illustrated in Figure~\ref{fig:BleherSinai}.

To advance the induction: the inductive hypothesis implies that
$(V_k,K_k) \in \domRG_k$. Then Theorem~\ref{thm:step-mr-K} implies
that $K_{k+1}$ is defined and satisfies $\|K_{k+1}\|_{\Wcal_{k+1}} \le
\CRG\tilde \vartheta_{k+1}^{3}\ggen_{k+1}^3$.  By
Theorem~\ref{thm:step-mr-R}, in particular
\eqref{e:thm:step-mr-R:r-bounds}, $V_{k+1}$ is given by
Proposition~\ref{prop:barflow-a} with corrections from $R^{U}$ which
are $O (\tilde\vartheta^{3}\tilde{g}_{j}^{3})$.  Therefore the flow
exists up to scale $k+1$ and
\begin{align}
\label{e:BSg}
  g_{j+1} &= g_j - \beta_j g_j^2 + O(\vartheta_j^3 g_j^3),
  \\
\label{e:BSmu}
  \mu_{j+1} &= L^2 (\mu_j  + e_j).
\end{align}
Using Proposition~\ref{prop:gjtj}(i) and \refeq{BSg}, followed by
\refeq{gbarcomp}, we obtain that $g_j =
\gbar_j(1+O(\gbar_j|\log\gbar_j|)) =
\ggen_j(1+O(\ggen_j|\log\ggen_j|)$.  In particular, this shows that
$g_j \in G_j$.  By \refeq{mubar}, the term $e_j$ in \eqref{e:BSmu}
satisfies
\begin{equation}
  \lbeq{cetabd}
  |e_j|
  \le |\eta_j| g_j +O(\vartheta_jg_j^2)
  \le  c_\eta \vartheta_j g_j + O(\vartheta_jg_j^2)
  \le 2c_\eta \vartheta_j   \ggen_j.
\end{equation}
By~\refeq{BSmu} and \refeq{cetabd}, $\mu_{k+1} \geq L^2 (\mu_k -
2c_\eta \vartheta_k\ggen_k) \geq L^2 \frac 12 \mu_k$ if $\mu_k =
c_0\vartheta_k \ggen_k$ is the largest point in $J_k$, and $\mu_{k+1}
\le L^2(\mu_k + 2c_\eta \vartheta_k \ggen_k) \leq L^2 \frac 12 \mu_k$
if $\mu_k=-c_0\vartheta_k\ggen_k$ is the smallest point in $J_k$.
Together with the continuity of the map $\mu_0 \mapsto \mu_{k+1}$, it
follows that the set of $\mu_{k+1}$'s produced from $\mu_0 \in I_k$
includes the interval $\frac 12 L^2 J_k$, which strictly includes
$J_{k+1}$ for large $L^2$.  Thus we can define a new interval
$I_{k+1}\subset I_k$ as the inverse image of $J_{k+1}$ under the map
$\mu_0 \mapsto \mu_{k+1}$, and it has the required properties.  This
advances the induction.  By construction, $\mu_j \in J_j$.

Finally, since either $\vartheta_j$ or $\ggen_j$ decreases to $0$, the
sequences $\vartheta_j \ggen_j$ tends to $0$ for any $m^2 \geq 0$, and
the intersection $\cap_{j \ge 1}I_{j}$ must consist of a single point.
We choose $\nu_0^c(m^2)$ to be that point.

\smallskip \noindent (ii) The proof is a corollary of the proof of
part~(i).  We fix $m^2 \ge 0$ and drop it from the notation.

As shown in the proof of part~(i), the existence of a
global flow obeying $\mu_j \in J_j(c_0)$ characterises $\nu_0 =
\nu_0^c$ uniquely, although in principle it could be the case that
$\nu_0$ depends on $c_0$, i.e., $\nu_0=\nu_0^c(c_0)$.  To see that it
does not depend on $c_0$, note that the flow started from
$\nu_0(4c_\eta)$ is such that $\mu_j \in J_j(4c_\eta)$ for all $j$.
Since $J_j(4c_\eta) \subset J_j(c_0)$ when $c_0 \ge 4c_\eta$, the flow
started from $\nu_0^c(4c_\eta)$ is such that $\mu_j \in J_j(c_0)$ as
well.  Since the proof of part~(i) shows that only
$\nu_0^c(c_0)$ has this last property, we conclude that $\nu_0^c(c_0)
= \nu_0^c(4c_\eta)$, and the proof is complete.
\end{proof}

Let $\nu_0^c(m^2)$ be the initial condition uniquely defined by
Proposition~\ref{prop:BS}.  To prepare for a proof of the continuity
of $\nu_0^c(m^2)$ in $m^2$, we first prove the following
lemma. Let $j^*(g_0,\nu_0,m^2)$ be the largest integer $j^*$
such that the flow $(V_j,K_j)_{j\leq j^*}$ of $\Phi$ with initial
condition $V_0=(g_0,\nu_0)$, $K_0=0$ exists. Recall that the term ``flow''
includes the condition $(V_j,K_j) \in \domRG_j(m^2,g_0)$ for all
$j\leq j^*$.

\begin{lemma} \label{lem:Vjcont}
  Let $(V_j,K_j)$ be the flow with initial condition $V_0=(g_0,\nu_0)$ and $K_0=0$.
  Given $(\ggen_0, \tilde \nu_0, \mgen^2)$
  and $j<j^*(\ggen_0, \tilde \nu_0, \mgen^2)$,
  the map $(g_0,\nu_0,m^2) \mapsto V_j$
  is continuous in a neighbourhood of $(\ggen_0, \tilde \nu_0, \mgen^2)$.
\end{lemma}

\begin{proof}
  By Theorems~\ref{thm:step-mr-K}--\ref{thm:step-mr-R},
  the map $\Phi_{j+1}$ is jointly continuous in $(V,K,m^2)$
  as a map $\domRG_j(\mgen^2,\ggen_j) \times \Iint_j(\mgen^2)
  \subset \Vcal \times \Wcal_j(\ggen_j) \times \Igen_j(\mgen^2)
  \to \Vcal \times \Wcal_{j+1}(\ggen_{j+1})$.
  The claim then follows from the continuity of the projection
  $(V,K,m^2) \mapsto V$.
\end{proof}

\begin{prop} \label{prop:BScont}
    The function $m^2 \mapsto \nu_0^c(m^2)$ is continuous in $m^2 \geq 0$,
    including right-continuity at $m^2=0$.
    For the corresponding flow,
    the function $m^2 \mapsto V_j$ is continuous in $m^2 \geq 0$, for each $j \in \N$.
\end{prop}

\begin{proof}
Limit points of the set $\{\nu_0^c(m^2): m^2 \geq 0\}$ exist because
the set is bounded.  Suppose that $m^2 \to \mgen^2$, and let $\tilde
\nu_0$ be any limit point of $\nu_0^c(m^2)$ as $m^2\to \mgen^2$.  It
suffices to show $\tilde \nu_0=\nu^c_0(\mgen^2)$.  For this, consider
the flow $(\tilde V_j,\tilde K_j)$ with mass $\mgen$ and initial
condition $(g_0,\tilde \nu_0)$.  By Proposition~\ref{prop:BS}(ii), the
continuity of $\nu_0^c$ would follow from the condition that $\tilde
V_j \in \DV_j(\mgen^2)$ and $\tilde \mu_j \in J_j(\mgen^2;12c_\eta)$
for all $j \in \N$.  Then the continuity of $V_j$ would follow from
the continuity of $V_j$ at $\mgen^2$.

To verify the above condition, we use the fact that for any given $k$,
the endpoints of $J_k(m^2)$ and those of the intervals defining the
domain $\DV_k(m^2)$ can jump at most by a multiplicative factor $3$
when $m^2$ is varied.  More precisely, we write $\vartheta^+(\mgen^2)
= \limsup_{m^2 \to \mgen^2} \vartheta(m^2)$ and $\vartheta^-(\mgen^2)
= \liminf_{m^2 \to \mgen^2} \vartheta(m^2)$.  Then, for every $k$,
$\frac12 \vartheta_k^-(\mgen^2) \leq \vartheta_k(\mgen^2) \leq
2\vartheta_k^+(\mgen^2)$.  Since the endpoints of $J_k$ and $\DV_k$
are defined in terms of $\vartheta_k$, and of $\ggen_k$ which jumps at
most by a factor $1+O(g_0)$, we conclude that the endpoints of $J_k$
and $\DV_k$ can jump at most by a factor $2+O(g_0) < 3$.

By Lemma~\ref{lem:Vjcont}, $(g_k,\mu_k)$ is continuous in
$(\nu_0,m^2)$ in a neighbourhood of $(\tilde \nu_0,\mgen^2)$.  Since
$m^2 \to \mgen^2$ and $\nu_0^c(m^2)\to\tilde \nu_0$, therefore
$g_j(\nu_0^c(m^2),m^2) \to g_j(\tilde \nu_0, \mgen^2)$ and the
sequence $\mu_j$ with mass $m$ is continuous as $m^2\to \mgen^2$.
Since $V_k(m^2) \in G_k(m^2) \times J_k(m^2;4c_\eta)$ for any $m^2
\geq 0$, and using the above bound on the jumps of the endpoints of
the intervals in $\DV_k$, we see that $V_k \in \DV_k(\mgen^2)$ for all
$k \leq j$, when $m^2 \to \mgen^2$.
Moreover, by Proposition~\ref{prop:BS}, $\mu_j(\mu_0(m^2),m^2) \in
J_j(m^2;4c_\eta)$ for any $m^2$, and thus $\mu_j(\tilde \mu_0,
\mgen^2) \in J_j(\mgen^2; 12 c_\eta)$.  As discussed above, this
completes the proof.
\end{proof}

\section{Proof of Theorem~\ref{thm:VK}}
\label{sec:pfprops}

We now restate and prove Theorem~\ref{thm:VK}, and thereby complete
the proof of Theorem~\ref{thm:phi4-hier-chi} subject to
Theorems~\ref{thm:step-mr-K}--\ref{thm:step-mr-R} and
Proposition~\ref{prop:N}.

\begin{theorem}
\label{thm:VK-8}
Fix $L$ sufficiently large and $g_0>0$ sufficiently small.

\smallskip \noindent
(i) There exists a continuous function $\nu_0^c(m^2)$ of $m^2\geq 0$
(depending on $g_0$) such that if $\nu_0 = \nu_0^c(m^2)$ then, for all
$j \in \N$,
\begin{equation}
  \label{e:VKN-Rj-bis-8}
  r_{g,j} = O(\vartheta_j^3g_j^3)   , \quad
  L^{2j} r_{\nu,j} = O(\vartheta_j^3 g_j^3) , \quad
  L^{\drb j} r_{u,j} = O(\vartheta_j^3 g_j^3),
\end{equation}
\begin{equation} \label{e:VKN-Kj-bis-8}
  L^{2j}|\nu_j| = O(\vartheta_j g_j),\qquad
  |K_j(0)| + L^{-2j} |D^2K_j(0;\1,\1)|
  = O(\vartheta_j^3 g_j^3).
\end{equation}

\noindent (ii)
There exists $c=1+O(g_0)$ such that for $m^2 \geq 0$ and $j \in \N$,
with all derivatives evaluated at $(m^2,\nu_0^c(m^2))$,
\begin{equation} \label{e:nuNp-5-8}
  \ddp{\mu_j}{\nu_0}
  =
  L^{2j} \left(\frac{g_j}{g_0}\right)^{\gamma}\big(c+O(\vartheta_jg_j)\big),
  \quad
  \ddp{g_j}{\nu_0} = O\left(g_j^{2}  \ddp{\mu_j}{\nu_0}   \right),
\end{equation}
\begin{equation} \label{e:gKNp-5-8}
  L^{-2j} \left|\ddp{}{\nu_0} K_j(0)\right|
  +
  L^{-4j} \left|\ddp{}{\nu_0} D^2 K_j(0; \1,\1)\right|
  = O\left(\vartheta_j^3  g_j^2\left(\frac{g_j}{g_0}\right)^{\gamma}\right)
  .
\end{equation}
\end{theorem}

In the following, we fix $g_0>0$ small and drop it from the notation
and discussion. Also, the dependence of $\Wcal_j$ on $(\tilde g_0,\tilde m^2)$ is left implicit.
We use primes to denote derivatives with respect to $\nu_0 =\mu_0$.
Let $V_j = V_j(m^2)$ be the infinite sequence given by
Proposition~\ref{prop:BS} with initial condition
$(g_0,\nu_0^c(m^{2}))$.  Let $K_j = K_j(m^2)$ be the
corresponding sequence given by Proposition~\ref{prop:BS}.  Let
$(V_j',K_j')$ denote the sequence of derivatives along this solution,
with respect to the initial condition $\mu_0$, and with the derivative
$K_j'$ taken in the space $\Wcal_j$.  The following lemma isolates a
continuity property of $V_j'$.

\begin{lemma} \label{lem:Vjprimecont}
  The function $m^2 \mapsto V_j'(m^2)$ is continuous in $m^2 \geq 0$.
\end{lemma}

\begin{proof}
    By definition, $V_j'(m^2)$ is the derivative of $V_j$ with respect to
    the initial condition  $\nu_0$, evaluated at $(g_0,\nu_0^c(m^2),m^2)$.
    By Theorem~\ref{thm:step-mr-K}--\ref{thm:step-mr-R} and the chain rule,
    $V_j'$ is continuous in $(\nu_0,m^2)$.
    By Proposition~\ref{prop:BScont}, the function $\nu_0^c(m^2)$ is continuous,
    so it follows that $V'(m^2)$ is continuous in $m^2$.
\end{proof}

\begin{proof}[Proof of Theorem~\ref{thm:VK-8}]
  (i)
  The function $\nu_0^c(m^2)$ is given by Proposition~\ref{prop:BS}.
  Proposition~\ref{prop:BS} also implies that $(V_j,K_j) \in \domRG_j(m^2)$,
  since $G_j\times J_j \subset \DV_j$ as noted below \refeq{Jintdef}.
  Theorem~\ref{thm:step-mr-R} (in particular \eqref{e:thm:step-mr-R:r-bounds})
  yields \eqref{e:VKN-Rj-bis-8}, and $V_j \in \DV_j$ immediately
  gives the bound on $\nu_j$ in \refeq{VKN-Kj-bis-8}.

  For the bound on $K_j$, we use the fact that the $\Wcal$-norm dominates the $T_0$-seminorm
  by \refeq{Wcal}.  Since the $T_0$-seminorm dominates the absolute value
  and $(V_j,K_j) \in \domRG_j(m^2)$, we have
  in particular that $|K_j(0)| \le \|K_j\|_{\Wcal_j} \le O(\vartheta_j^3 g_j^3)$.
  For the $\varphi$-derivative $D^2K_j$
  in \refeq{VKN-Kj-bis-8}, we use the fact
  that by definition of the $T_\varphi$-seminorm, the derivative in the direction of
  a test function $f$ obeys
\begin{equation}
  |D^2F(0; f, f)|
  \leq 2 \|F\|_{T_{0,j}} \|f\|_{\Phi_j}^2.
\end{equation}
The norm of the constant test function $\1 \in \Phi_j$ is
\begin{equation}
\lbeq{1norm}
  \|\1\|_{\Phi_j} = \ell_j^{-1} \sup_x |\1_x| = \ell_j^{-1}
  = O(L^{j(\drb-2)/2})
  .
\end{equation}
Therefore,
\begin{equation}
\label{e:WNbd}
  L^{-2j}
  \left| D^2 K_j(0; \1,\1)\right|
  \leq
  2\|K_j\|_{T_{0,j}}
  \le
  O(\vartheta_j^3   g_j^2).
\end{equation}
This completes the proof of
\eqref{e:VKN-Rj-bis-8}--\eqref{e:VKN-Kj-bis-8}.

\smallskip \noindent (ii)
We prove that there exists a continuous function $c:[0,\infty) \to
\R$, which satisfies $c(m^2)=1+O(g_0)$, such that for all $j \in
\N_0$:
\begin{equation} \label{e:mugzprime}
  \mu_j'
  =
  L^{2j}
  \left(\frac{g_j}{g_0}\right)^\gamma(c(m^2) + O(\vartheta_j g_j))
  ,
  \quad
  g_j' = O\left(\mu_j' g_j^{2}  \right),
\end{equation}
and
\begin{equation} \label{e:Kprime}
  \|K_j'\|_{\Wcal_j} = O\left(\vartheta_j^3 \mu_j' g_j^2 \right).
\end{equation}
Then \refeq{nuNp-5-8} follows from \eqref{e:mugzprime}.  Also, as in
the proof of part~(i), \refeq{Kprime} implies that
\begin{equation}
    |K_j'(0)| + L^{-2j}|D^2 K_j'(0;\1,\1)| \le O\left(\vartheta_j^3 \mu_j' g_j^2 \right),
\end{equation}
and then \eqref{e:gKNp-5-8} follows from \refeq{nuNp-5-8}.  It remains
to prove \eqref{e:mugzprime}--\eqref{e:Kprime}.

Recall the definition of $\Pi_{i,j}$ from Lemma~\ref{lem:Piprod}.  We
define $\Pi_j^* = \Pi_j(m^2 )$ by
\begin{equation}
  \Pi_j^* = L^{2j}\Pi_{0,j-1}
  = L^{2j}\prod_{l=0}^{j-1}(1-\gamma \beta_l g_l)
  .
\end{equation}
By Lemma~\ref{lem:Piprod}, and by the continuity of $V_j$ in $m^2$
provided by Proposition~\ref{prop:BScont}, there is a continuous
function $\Gamma_\infty (m^2) = O(g_0)$ such that
\begin{equation}  \label{e:prodid}
  \Pi_j^* =
  L^{2j}
  \left( \frac{g_{j}}{g_{0}} \right)^\gamma
  \big(1+\Gamma_\infty(m^2)+O(\vartheta_jg_j)\big)
  .
\end{equation}
We also define $\Sigma_j = \Sigma_j(m^2)$ by
\begin{equation}
  \label{e:muPiSig}
  \mu_j' = \Pi_j^* (1+ \Sigma_j) \quad (j \geq 0), \qquad \Sigma_{-1} = 0
  .
\end{equation}

We will use induction on $j$, where $j<N$. The inductive assumption is
that there exist $M_1 \gg M_2 \gg 1$ such that for $k \le j$,
\begin{gather}
  \label{e:induct1a}
  |\Sigma_{k}-\Sigma_{k-1}| \leq O(M_1+M_2) \vartheta_k g_k^2,
  \quad
  |g_k'| \leq M_1 \vartheta_k^3 \Pi_k^* g_k^2 ,
  \\
  \label{e:induct1b}
  \|K_k'\|_{\Wcal_k} \leq M_2 \vartheta_k^3 \Pi_k^* g_k^2.
\end{gather}
Since $(g_0',\mu_0',K_0') = (0,1,0)$, $\Sigma_{0}=0$ and so the
inductive assumption \eqref{e:induct1a}--\eqref{e:induct1b} is true
for $j=0$.
By summing the first inequality in
\eqref{e:induct1a}--\eqref{e:induct1b} over $k\le j$ and by applying
\eqref{e:prodid}--\eqref{e:muPiSig} we conclude that, if $L \gg 1$ and
if $g_0$ is sufficiently small, then
\begin{equation}
\lbeq{inductionhelpers}
  |\mu_j'| \le 2 \Pi_j^*
  ,\quad
  \vartheta_j^3 \Pi_{j}^*g_j^2 \leq \half \vartheta_{j+1}^3 \Pi_{j+1}^* g_{j+1}^2
  ,
\end{equation}
where we used \refeq{gbarsum} for the first inequality.

As a first step in advancing the induction, we differentiate
\begin{align}
\lbeq{fvflow}
    (V_{j+1},K_{j+1})
    &
    =
    \big(\Phi_\pt^V(V_j) + R_{j+1}^V(V_j,K_j),
    \Phi_{j+1}^K(V_j,K_j)
    \big),
\end{align}
where $R_{j+1}^V$ denotes the $V$-component of $R_{j+1}^U$.  By the
chain rule, with $F = R_{j+1}^V$ or $F=\Phi_{j+1}^K$,
\begin{equation} \label{e:chain}
  F'(V_j,K_j) = D_{V}F(V_j,K_j)V_j' + D_KF(V_j,K_j)K_j'.
\end{equation}
The $g_j'$ estimate in \eqref{e:induct1a} and the $\mu_j'$ estimate in
\eqref{e:inductionhelpers} bound the coefficients of $V_j'$. By
combining this bound on the coefficients with \eqref{e:T0-equivalent}
we obtain $\|V_j' (b)\|_{T_{0} (\ell_{j})} = O_{L} (M_1 +2)\Pi_j^*$.
We also have the estimate for $K_j'$ in \eqref{e:induct1b}. By
applying Theorems~\ref{thm:step-mr-K}--\ref{thm:step-mr-R}, we obtain
\begin{align}
  \|D_VF(V_j,K_j)V_j'\| &\leq O(\vartheta_j^3 g_j^2)(M_1 g_j^2+2)\Pi_j^* \leq O(\vartheta_j^3 \Pi_j^*g_j^2),
  \\
  \|D_KR_{j+1}^{V}(V_j,K_j)K_j'\| &\leq O(M_2)\vartheta_j^3 \Pi_j^*g_j^2,
  \\
  \|D_KK_{j+1}(V_j,K_j)K_j'\| &\leq M_2\vartheta_j^3 \Pi_j^*g_j^2,
\end{align}
with the norms dictated by
Theorems~\ref{thm:step-mr-K}--\ref{thm:step-mr-R} on the left-hand
sides.  For example, if $F=R_{j+1}^V$ (which is in $\Ucal_{j+1}
(\ell_{j+1})$) the norm is $\|R_{j+1}^V (B)\|_{T_{0} (\ell_{j+1})}$, and if
$F=\Phi_{j+1}^K$ the norm is $\Wcal_{j+1}$. This implies, for $M_2 \gg
1$,
\begin{equation} \label{e:phiprime}
  \|(R_{j+1}^{V})'(V_j,K_j)\| \leq O(M_2) \vartheta_j^3 \Pi_j^*g_j^2,
  \quad
  \|K_{j+1}'(V_j,K_j)\| \leq 2M_2\vartheta_j^3 \Pi_j^*g_j^2.
\end{equation}
With the second inequality of \refeq{inductionhelpers}, this advances
the induction for $K'$.

For $\mu'$, the induction is advanced using the recursion \eqref{e:fvflow}
with \eqref{e:mubar}, \eqref{e:phiprime}, together with the estimates
\begin{equation}
  \eta_j g_j',\; \xi_j (g_j^2)' = O(M_1 \vartheta_j^3 \Pi_j^* g_j^2)
\end{equation}
which follow from \eqref{e:induct1a}.
We obtain
\begin{align} \label{e:muchO}
  \mu_{j+1}'
  &=
  L^2 \mu_j'(1-\gamma\beta_j g_j)
  + O\left((M_1+M_2)\vartheta_j^3 \Pi_j^*  g_j^2 \right)
  \nnb
  &
  = \Pi_{j+1}^*(1+\Sigma_j) + O\left((M_1+M_2)\vartheta_{j+1}^3 \Pi_{j+1}^*  g_{j+1}^2 \right)
  .
\end{align}
This advances the induction for $\mu'$, namely the first estimate of
\eqref{e:induct1a}.

The advancement of the induction for $ g$ is similar,
as follows.
We use
the recursion relation \eqref{e:fvflow} with \eqref{e:gbar}
and \eqref{e:phiprime}, and choose $M_1 \gg M_2$ to obtain
\begin{align}
  |g_{j+1}'|
  &\leq (M_1(1+O(g_j))+O(M_2))\vartheta_j^3 \Pi_j^* g_j^2
  \nnb
  &\le 2 M_1 \vartheta_j^3 \Pi_j^* g_j^2
  \leq M_1 \vartheta_{j+1}^3 \Pi_{j+1}^* g_{j+1}^2.
\lbeq{gchO}
\end{align}
This advances the induction for $g'$.

By Lemma~\ref{lem:Vjprimecont}, $V_j'$ is continuous in $m^2$.
Since $\Pi_j^*$ is continuous,
it follows that $\Sigma_j(m^2)$ is continuous in $m^2$,
for each $j \in \N_0$.
Since $\sum_{j=1}^\infty\vartheta_j g_j^2 =O(g_0)$ by \refeq{gbarsum},
it follows from \eqref{e:induct1a} that the limit $\Sigma_\infty =
\lim_{j \to \infty}\Sigma_j =\sum_{j=1}^\infty (\Sigma_j -
\Sigma_{j-1})$ exists with $\Sigma_\infty = O(g_0)$.
Continuity of $\Sigma_\infty$ in $m^2$ follows from the dominated
convergence theorem, with Proposition~\ref{prop:barflow}.
Also, again by \refeq{gbarsum},
\begin{equation}
  \label{e:sumid}
  \Sigma_\infty-\Sigma_j =
  O\left(\sum_{k=j+1}^\infty \vartheta_k g_j^2 \right) = O(\vartheta_j g_j)
  .
\end{equation}
From \eqref{e:prodid}--\eqref{e:muPiSig} and \eqref{e:sumid}, we
obtain the equation for $\mu_j'$ in \eqref{e:mugzprime}, with $c(m^2)
= (1 + \Sigma_\infty(m^2))(1+\Gamma_\infty(m^2))$.  This $c(m^2)$ is
indeed continuous, since $\Sigma_\infty$ and $\Gamma_\infty$ are.
With this, \eqref{e:induct1a}--\eqref{e:induct1b} implies the last
inequality in \eqref{e:mugzprime} and \eqref{e:Kprime}, and the proof
is complete.
\end{proof}

\chapter{Nonperturbative contribution to \texorpdfstring{$\Phi_+^U$}{Phi+U}:
Proof of Theorem~\ref{thm:step-mr-R}}
\label{ch:R+U}

In this chapter, we prove the estimates of
Theorem~\ref{thm:step-mr-R}, which for convenience we restate below as
Theorem~\ref{thm:Rplus}.  The continuity statement of
Theorem~\ref{thm:step-mr-R} is deferred to Section~\ref{sec:masscont}.

The proof of Theorem~\ref{thm:Rplus} makes use of certain norm
estimates on polynomials in the field, which are developed in
Section~\ref{sec:poly-norms}.
A more comprehensive set of estimates on polynomials is needed
for Chapter~\ref{ch:pf-thm:step-mr-K}, and we present these estimates
also in  Section~\ref{sec:poly-norms}.

\section{The polynomial \texorpdfstring{$R_{+}^{U}$}{R+U}}

The non-perturbative contribution $R_{+}^U$ to $\Phi_+^U$ is defined
in \eqref{e:RUdef-5} as
\begin{equation}
\label{e:norm-RUdef}
    R_+^U(V,K) = \Phi_+^U(V,K) - \Phi_+^U(V,0) .
\end{equation}
By definition, $R_+^U(V,K)$ is an element of $\Ucal$.  
To this element of $\mathcal{U}$, we associate an element of
     $\mathcal{N}(B)$ by summation over points in $B$ as in Definition~\ref{defn:Vcal}. 
We denote this
element by $R_+^U(B)$ and then omit the argument $(V,K)$.

The map $\Phi_+^U$ is defined in Definition~\ref{def:RGmap} as
\begin{equation}
\label{e:norm-U+def}
    \Phi_+^U(V,K) = \Phi_\pt (V-  Q),
\end{equation}
with $Q \in \Ucal$ equal to the $b$-independent polynomial defined by
\index{$Q$}%
\begin{equation}
    \label{e:Q-def}
    Q(b)
    =
    \LT (e^{V(b)}K(b))
\end{equation}
for $b \in \Bcal$.  The map $\Phi_\pt$ is given in \refeq{Phiptdef} as
\begin{equation}
\lbeq{Phiptdef-bis}
    \Phi_\pt(U;B) =
    \Ex_{+}\theta U (B)
    - \tfrac{1}{2} \Ex_{+}\big(\theta U (B);\theta U (B)\big).
\end{equation}
Recall that, by definition of the renormalisation group map
  in Definition~\ref{def:RGmap}, the expectation $\Ex_+$ here is
  with respect to a covariance obeying the zero-sum condition $c^{(1)}=0$.
  This allows us to drop the operator $\Loc$ from the definition of $\Phi_\pt$
  in \refeq{Phiptdef-bis},
  as remarked under \eqref{e:Phiptdef}.

After simplification, the above formulas lead to
\begin{equation}
\lbeq{RUrewrite}
    R_+^U(B)
    =
    - \Ex_{C_{+}}\theta Q (B) + \Cov_+ (\theta (V(B)-\half   Q(B)),\theta Q(B))  .
\end{equation}
According to Exercise~\ref{ex:ExUcal-bis} and Lemma~\ref{lem:covUcal},
the right-hand side is indeed an element of $\Ucal$.  By definition,
$Q$ is linear in $K$.  Thus $R_+^U$ is a quadratic function of $K$.
The explicit form of \refeq{RUrewrite} makes the analysis of $R_+^U$
relatively easy.

In this chapter, we prove the estimates of
Theorem~\ref{thm:step-mr-R}, which we restate here as follows.
Although the domain of $R_+^U$ in \refeq{Rplusmap-bis} is stated in
terms of $\domRG$, in which $K$ is measured with the $\Wcal$-norm, in
the proof we actually only use the weaker hypothesis that
$\|K(b)\|_{T_{0}(\ell)} \le \CRG \tilde\vartheta^{3}\ggen^3$.

\begin{theorem}
\label{thm:Rplus}
  Let $\mgen^2 \ge 0$,
  let $\ggen$ be sufficiently small (depending on $L$), and let $p,q\in \N_0$.
  Let $0\le j<N$.
  There exists an $L$-dependent constant $M_{p,q}>0$ such that the map
  \begin{equation}
  \lbeq{Rplusmap-bis}
    \R^U_+:\domRG  \times \Iint_+  \to \Ucal_+
  \end{equation}
  satisfies the estimates
  \begin{align}
    \label{e:Rmain-g-bis}
    \|D_V^p D_K^q R^U_+\|_{\Vcal(\ell) \times \Wcal \to \Ucal_+(\ell_+)}
    & \le
    \begin{cases}
    M_{p,0}
    \tilde \vartheta_+^{3} \ggen_+^{3} & (p\ge 0,\, q=0)\\
    M_{p,q}  & (p\ge 0,\, q = 1,2)\\
    \rlap{$0$}\hspace{1cm}  & (p\ge 0,\, q \ge  3).
    \end{cases}
  \end{align}
\end{theorem}

\section{The standard and extended norms}
\label{sec:extendednorm}

\subsection{Utility of the extended norm}

\index{Standard norm}
\index{Extended norm}
\index{Norm}
Theorems~\ref{thm:step-mr-K}--\ref{thm:step-mr-R} (and so
Theorem~\ref{thm:Rplus}) are expressed in terms of the \emph{standard}
norms.  For $V$, this is the norm $\|V (b)\|_{T_{0} (\ell)}$
on the space $\Vcal (\ell)$.
For $K$, these are the seminorms $T_\infty(h)$ and $T_{0}(\ell)$ and
their combination as the $\Wcal$-norm, on the space $\Fcal$.
Functions of $(V,K)$ such as $\Phi_+$ are functions on $\Vcal \times \Fcal$.
The standard norm is good for estimates that hold for fixed $(V,K)$.
The dependence of a function $F(V,K)$ on $(V,K)$ is controlled by
derivatives with respect to $(V,K)$ in these norms.  For a given
function $F$, these can in principle be computed using the usual rules
of calculus.  In practice, for complicated functions $F$ such as
$\Phi_+$, this can become unwieldy.

To handle derivatives systematically, we use the \emph{extended} norm
which encodes not only point-wise dependence of a function $F(V,K)$ on
$(V,K)$ but also their derivatives. Thus the extended norm of $F(V,K)$
is not a norm for $(V,K)$ fixed, but requires an infinitesimal
neighbourhood of some reference point $(V,K)$. It is in fact an
instance of the $T_z$-norm, with $\Ycal$ chosen to be a subspace of
$\Vcal \times \Fcal$.

We denote the coordinate maps for $(V,K)$ using stars, i.e., $V^*:
\Ycal \to \Vcal$ and $K^* :\Ycal \to \Fcal$ are defined by
\begin{equation}
  V^*(V,K) = V, \quad K^*(V,K) = K.
\end{equation}
For the coordinate maps, by Definition~\ref{def:Tnorm} with $\aux =
(V,K)$, the extended and standard norms are related by
\begin{align}
  \label{e:V*V-general}
  \|V^*(b)\|_{T_{\varphi,y}} = \|V^*(b)\|_{T_{\varphi,V,K}}
  =
  \|V(b)\|_{T_\varphi} +
  \sup_{(\Vdot,\Kdot)} \frac{\|\Vdot(b)\|_{T_\varphi}}{\|(\Vdot,\Kdot)\|_{\Ycal}},
  \\
  \label{e:K*K-general}
  \|K^*(b)\|_{T_{\varphi,y}}
  =
  \|K^*(b)\|_{T_{\varphi,V,K}}
  =
  \|K(b)\|_{T_\varphi} +
  \sup_{(\Vdot,\Kdot)} \frac{\|\Kdot(b)\|_{T_\varphi}}{\|(\Vdot,\Kdot)\|_{\Ycal}}
  ,
\end{align}
where $b$ is an arbitrary block in $\Bcal_{j} (B)$.
In particular, the standard norm can be recovered from the extended
norm with the limiting choice $\|y\|_\Ycal = \|(V,K)\|_\Ycal =
\infty$.  The standard norm of $V$ or $K$ only differs from the
extended norm of $V^*$ or $K^*$ at $(V,K)$ by an additive
constant. Thus, for the coordinate maps, the additional information
encoded by the extended norm is trivial.  However, for maps that are
non-linear in $(V,K)$, the extended norm is a significant help because
it can often be bounded in the same way as the standard norm, yet a
bound of the extended norm of a function of $(V,K)$ yields also bounds
on the derivatives of this function by Lemma~\ref{lem:normderiv-1}
(see also the special case Lemma~\ref{lem:normderiv} below).

\subsection{Choice of the space \texorpdfstring{$\Aux$}{Y}}
\label{sec:Aux}

We now specify the space $\Aux$ used to define the extended norm.
We set $p_\Aux =\infty$ and fix nonnegative parameters $\hAuxg
= (\hAuxg_{V},\hAuxg_{K})$, which for the moment are arbitrary.
Recall that $\Vcal=\R^2$ is defined in Definition~\ref{defn:Vcal}.
We define $\Aux \subset \Vcal \times \Fcal$ to be the space of $y=(V,K)$
with finite norm
\begin{equation}
    \label{e:Aux-norm-def}
    \|\aux\|_\Aux
    =
    \max
    \left\{
    \frac{\|V(b)\|_{T_{0}(\ell)}}{\hAuxg_V},
    \frac{\|K\|_{\Wcal}}{\hAuxg_K}
    \right\}
    ,
\end{equation}
where $b$ is an arbitrary block in $\Bcal (B)$.
We also define
\begin{equation}
    X = \R_\h^n,
    \quad
    \Auxx
    =
    X \times \Aux \subset \R_\h^n \times (\Vcal \times \Fcal).
\end{equation}
The case $\hAux_V=\hAux_K=0$, which superficially appears to prescribe
division by zero in \eqref{e:Aux-norm-def}, is equivalent to taking
$p_\Aux =0$ in \refeq{Tznormdef}, or equivalently to a norm that does
not measure the size of derivatives with respect to $(V,K)$; in this
case the norm on $\Aux$ is not used and there is no division by zero.

\index{Extended norm}
In the same spirit as in the extension of the $T_\varphi$-seminorms to the
$T_{\varphi,y}$-seminorms, we extend the definition \refeq{Wcal} of the
$\Wcal$-norm to incorporate the parameter $\hAux$, by defining the
extended $\Wcal_\aux(\hAux)$ norm of a function $F(V,K)$ to be
\begin{equation}
\lbeq{Wcalgrand}
    \|F\|_{\Wcal_\aux(\hAux)}
    =
    \|F(b) \|_{T_{0,\aux}(\ell,\hAux)} + \ggen^{9/4}\|F(b) \|_{T_{\infty,\aux}(h,\hAux)}.
\end{equation}

The following lemma is a special case of Lemma~\ref{lem:normderiv-1}.

\begin{lemma}  \label{lem:normderiv}
The following hold for any $y=(V,K) \in \Vcal \times \Wcal$.

\smallskip\noindent
\label{lem:normderiv-2}
(i) For $F: \Vcal(\ell) \times \Wcal \rightarrow \Ucal_+(\ell_+)$,
\begin{equation}
\lbeq{normderivT0}
    \|D_V^{p_{2}}D_K^{p_{3}} F(V,K)\|_{\Vcal(\ell) \times \Wcal \to \Ucal_+(\ell_+)}
    \le
    \frac{p_{2}!p_{3}!}{\hAux_V^{p_{2}}\hAux_K^{p_{3}}} \|F  
    \|_{T_{0,\aux(\ell_+,\hAux)}}.
\end{equation}

\smallskip\noindent
(ii) For $F: \Vcal(\ell) \times \Wcal \rightarrow \Wcal_+$,
\begin{equation}
\lbeq{normderiv}
    \|D_V^{p_{2}}D_K^{p_{3}} F(V,K)\|_{\Vcal(\ell) \times \Wcal \to \Wcal_+}
    \le
    \frac{p_{2}!p_{3}!}{\hAux_V^{p_{2}}\hAux_K^{p_{3}}} \|F\|_{\Wcal_{y,+}(\hAux)}.
\end{equation}
\end{lemma}

\begin{proof}
In \eqref{e:Aux-norm-def}, the norm $\|y\|_{\Ycal}$ of $y= (V,
K)$ is defined such that the unit ball $\|y\|_{\Ycal} \leq 1$
corresponds to $\|V(b)\|_{T_{0} (\ell)} \leq \hAuxg_{V}$ and
$\|K\|_{\Wcal} \leq \hAuxg_{K}$.  The norm of a multilinear map
defined on $\Vcal(\ell) \times \Wcal$ is, however, defined with
respect to unit directions of $( V, K)$.
By Lemma~\ref{lem:normderiv-1}, therefore
\begin{align}
    \|
    D_{\auxx_{2}}^{p_{2}}D_{\auxx_{3}}^{p_{3}} F(V,K) 
    \|_{\Vcal(\ell) \times \Wcal \rightarrow T_{\varphi} (\h)}
    & =
    \frac{1}{\hAux_V^{p_2}\hAux_K^{p_3}}
    \|
    D_{\auxx_{2}}^{p_{2}}D_{\auxx_{3}}^{p_{3}} F(V,K) 
    \|_{\Aux \rightarrow T_{\varphi} (\h)}
    \nnb & \le
    \frac{p_2!p_3!}{\hAux_V^{p_2}\hAux_K^{p_3}}
    \|F\|_{T_{\varphi,\aux}(\h,\hAux)}. 
\end{align}

In the above inequality we set (i) $\h = \ell_{+}$ and $\varphi = 0$,
and (ii) $\h=h_{+}$. The result is the two estimates:
\begin{align}
    \|
    D_{\auxx_{2}}^{p_{2}}D_{\auxx_{3}}^{p_{3}} F(V,K) 
    \|_{T_{0} (\ell) \times \Wcal \rightarrow T_0 (\ell_{+})}
    &\le
    \frac{p_2!p_3!}{\hAux_V^{p_2}\hAux_K^{p_3}}
    \|F\|_{T_{0,\aux}(\ell_{+},\hAux)} ,  
    \\
    \|
    D_{\auxx_{2}}^{p_{2}}D_{\auxx_{3}}^{p_{3}} F(V,K) 
    \|_{T_{0} (\ell) \times \Wcal \rightarrow T_{\varphi} (h_{+})}
    &\le
    \frac{p_2!p_3!}{\hAux_V^{p_2}\hAux_K^{p_3}}
    \|F\|_{T_{\varphi,\aux}(h_{+},\hAux)} . 
\lbeq{normderpf}
\end{align}
The first estimate is \eqref{e:normderivT0}, since the $T_0(\ell_+)$-seminorm
on a block is the same as the $\Ucal_+(\ell_+)$ norm by definition of the latter.
In the second estimate
we replace $T_{\varphi}$ by $T_{\infty}$ by taking the supremum over
$\varphi$. By the definitions \eqref{e:Wcal} and \eqref{e:Wcalgrand}
of the $\Wcal$- and $\Wcal (\hAux)$-norms, the desired result
\eqref{e:normderiv} is an immediate consequence of multiplying the
second estimate by $\ggen_{+}^{9/4}$ and adding the two estimates.
\end{proof}

\section{Norms of polynomials}
\label{sec:poly-norms}

In this section, we develop a comprehensive set of norm estimates on
polynomials in the field.  For the proof of Theorem~\ref{thm:Rplus},
we only need the $\h=\ell$ case of the bound on $V$ given in
\refeq{V-norm}, and the bound on $Q$ given in \eqref{e:QbdT0aux}.  The
remaining results in this section prepare the ground for our analysis
of $K$ in Chapter~\ref{ch:pf-thm:step-mr-K}.  Recall that the domains
$\DV$ and $\DVstab$ are defined by \eqref{e:DVdef} and
\eqref{e:DVstab}.

We first observe that it follows from Exercise~\ref{ex:Ucal-ident}
that the norm of $U=g\tau^2 + \nu \tau + u$ on a block $b\in\Bcal$ is given by
\begin{align}
   \label{e:delta}
     \|U(b)\|_{T_{0}(\h)}
    &=
     L^{j\drb}\left( \tfrac{1}{4}g\h^4 + \tfrac{1}{2} |\nu|\h^2 + |u| \right).
\end{align}
By the definitions of $\ell$ and $h$ in \eqref{e:ell-def} and \eqref{e:h-def},
\begin{equation}
    \label{e:ell--h}
    L^{j\drb}\ggen h^{4}
    =
    k_{0}^{4} ,
    \qquad
    \tfrac{\ell}{h}
    =
    \tfrac{\ell_{0}}{k_{0}}\ggen^{1/4}L^{-j (\drb-4)/4} ,
    \qquad h^2=k_0^2 \ggen^{-1/2} L^{-j\drb/2} . 
\end{equation}
The constant $k_0$ is small and independent of $L$, whereas $\ell_0$
is large and is equal to $L^{1+\drb/2}$ by \refeq{ell0def}.  By
\refeq{delta} and the first equality of \refeq{ell--h}, for
$V=g\tau^2+\nu\tau$ we have
\begin{equation}
     \label{e:epdV--1}
     \|V(b)\|_{T_{0}(\h)}
     =     \tfrac{1}{4} k_{0}^{4} \tfrac{g}{\ggen} (\tfrac{\h}{h})^4 +
     \tfrac{1}{2} L^{j\drb} |\nu| h^{2} (\tfrac{\h}{h})^2
     .
\end{equation}
The equalities of \refeq{ell--h} are useful for application of
\eqref{e:epdV--1}.  The following lemma uses \refeq{epdV--1} to obtain
estimates.  For our application to dimension $\drb=4$ the factors
$L^{-(\drb -4)j}$ in the following lemma equal one.

\begin{lemma}
\label{lem:V-norm}
Let $\drb\geq 4$, $U=g\tau^2 + \nu \tau + u = V+u$, and $b \in \Bcal$.

\smallskip\noindent
(i) If $V \in \DV$ then
\begin{align}
    \label{e:V-norm}
    \|V(b)\|_{T_{0}(\h)}
    &\le
     \begin{cases}
    O_L(\ggen) L^{-(\drb -4)j}
    &(\h=\ell ) \\
    k_0^{3}  & (\h=h ) .
    \end{cases}
\end{align}
(ii) If $V \in \DVstab$ then
\begin{align}
   \label{e:V-normstab}
    \|V(b)\|_{T_{0}(\h)}
    &\le
    \begin{cases}
    O_L(\ggen^{1/2}) L^{-(\drb -4)j/2}  & (\h=\ell)
    \\
    k_{0}^{3} & (\h=h).
    \end{cases}
\end{align}
\end{lemma}

\begin{proof}
We use \eqref{e:ell--h} without comment in the proof.

Suppose first that $V \in \DV$.
By the definition of $\DV$ in \eqref{e:DVdef},
$g < (2k_0)^{-1} \ggen$ and $|\nu| < (2k_0)^{-1} \ggen L^{- (\drb -2)j}$,
so
\begin{equation}
    \label{e:epdV--2}
    \|V(b)\|_{T_{0}(\h)}
    \le
    \tfrac{1}{8}  k_{0}^{3}
    (\tfrac{\h}{h})^4 +
    \tfrac{1}{4}
    \ggen^{1/2}
    k_{0}L^{-j (\drb-4)/2}
    (\tfrac{\h}{h})^2 .
\end{equation}
For $\h=\ell$, the right-hand side is
$
    \left(
    \tfrac{1}{8}
    \tfrac{\ell_{0}^{4}}{k_{0}} +
    \tfrac{1}{4}
    \tfrac{\ell_{0}^{2}}{k_{0}}
    \right) \ggen L^{-j (\drb-4)}
$.
Since $\ell_{0}\ge 1$, this is less than
$
    k_{0}^{-1}\ell_{0}^{4}     \ggen L^{-j (\drb-4)},
$
and this proves the case $\h=\ell$ of \eqref{e:V-norm}.  For $\h=h$
and $\ggen \le k_{0}^{4}$, the right-hand side of \eqref{e:epdV--2} is
at most $k_0^3$, as desired for the case $\h=h$ of \eqref{e:V-norm}.

Suppose now that $V\in \DVstab$.  By \eqref{e:DVstab}, we have $g \le
k_0^{-1}\ggen$ and $|\nu| \le \ggen h^2$.  With \eqref{e:epdV--1}, we
obtain
\begin{equation}
     \label{e:epdV--3}
     \|V(b)\|_{T_{0}(\h)}
     \le
     \tfrac{1}{4} k_{0}^{3}  (\tfrac{\h}{h})^4 +
     \tfrac{1}{2} k_{0}^{4} (\tfrac{\h}{h})^2 .
\end{equation}
For $\h=\ell$ we use $k_{0}^{3} \le \ell_{0}^{2}$ (since $k_0\le 1 \le
L$) to obtain
\begin{equation}
     \|V(b)\|_{T_{0}(\ell)}
     \le
     \tfrac{1}{4}
     \tfrac{\ell_{0}^{4}}{k_{0}} \ggen L^{-j (\drb -4)} +
     \tfrac{1}{2} k_{0}^{2} \ell_{0}^{2}
     \ggen^{1/2} L^{-j (\drb -4)/2}
     \le
     \tfrac{\ell_{0}^{4}}{k_{0}} \ggen^{1/2} L^{-j (\drb -4)/2},
\end{equation}
as desired for the case $\h=\ell$ of \eqref{e:V-normstab}.  Finally,
for $\h=h$, the right-hand side of \eqref{e:epdV--3} is
$
     \tfrac{1}{4} k_{0}^{3} +
     \tfrac{1}{2} k_{0}^{4} \le k_{0}^{3},
$
which proves the case $\h=h$ of \eqref{e:V-normstab}.
\end{proof}

Next, we obtain bounds in the extended norm.  By
\eqref{e:V*V-general}--\eqref{e:K*K-general},
\begin{align}
    \label{e:Ustar1}
    \|V^*(b)\|_{T_{0,\aux}(\h,\hAux)}
    &=
    \|V(b)\|_{T_{0}(\h)}
    +
    \hAuxg_{V}
    \sup_{\dot V \in \Vcal}
    \frac{\|\dot V(b)\|_{T_{0}(\h)}}{\|\dot V(b)\|_{T_{0}(\ell)}} ,
    \\
    \label{e:Kstar-h-aa}
    \|K^*(b)\|_{T_{\varphi,\aux}(\h,\hAux)}
    &=
    \|K(b)\|_{T_{\varphi}(\h)}
    +
    \hAuxg_{K}
    \sup_{\dot K \in \Fcal}
    \frac{\|\dot K(b)\|_{T_\varphi(\h)}}{\|\dot K\|_{\Wcal}}.
\end{align}
In particular, this shows that the norms of $U^*$ and $K^*$ are
monotone increasing in $\h$. Furthermore, for $V \in \Vcal$,
\index{Extended norm}%
\begin{align}
\label{e:U*norm1}
    \|V^*(b)\|_{T_{0,\aux}(\ell,\hAux)}
    & =
    \|V(b)\|_{T_{0}(\ell)}
    +
    \hAuxg_{V} ,
\\
    \|K^*(b)\|_{T_{0,\aux}(\ell,\hAux)}
    &\le
    \|K(b)\|_{T_{0}(\ell)}
    +
    \hAuxg_{K}
    ,
\label{e:K*norm-ell}
\\
    \|K^*(b) \|_{T_{\varphi,\aux}(h,\hAux)}
    &\le
    \|K(b)\|_{T_{\varphi}(h)}
    +
    \hAuxg_{K} \ggen^{-9/4}
      .
\label{e:Kstar-h-9}
\end{align}
The bound \eqref{e:U*norm1} is obtained by setting $\h = \ell$ in
\eqref{e:Ustar1}.  Likewise, \eqref{e:K*norm-ell}--\eqref{e:Kstar-h-9}
follow from \eqref{e:Kstar-h-aa} and the definition of the
$\Wcal$-norm in \eqref{e:Wcal}.

The above estimates do not yet include a bound on
$\|V^*(b)\|_{T_{0,\aux}(h,\hAux)}$.  The following lemma fills this
gap.  For the proof of Proposition~\ref{prop:stability}, it will be
important that the coefficient $\frac{3}{8}$ on the first right-hand
side of \eqref{e:gnustar} is smaller than $\frac{1}{2}$.

\begin{lemma}
\label{lem:U*norm-h}
Let $\drb =4$. Let $(g,\nu,0) \in \DVstab$, $\hAux_V \le
\ggen$, $b \in \Bcal$, and let $L$ be sufficiently large. Then
\begin{equation}
\label{e:gnustar}
    \|(g\tau^2)^*(b)\|_{T_{0,\aux}(h,\hAux)}
    \le
    \tfrac{3}{8}\tfrac{g}{\ggen} k_0^4,
    \quad
    \|(\nu\tau)^*(b)\|_{T_{0,\aux}(h,\hAux)}
     \le
    k_0^4 .
\end{equation}
In particular, the extended norm of $V= g \tau^2+ \nu \tau$
obeys
\begin{equation}
\label{e:Vstarhbd}
    \|V^*(b)\|_{T_{0,\aux}(h,\hAux)} \le \frac{11}{8} k_0^3.
\end{equation}
\end{lemma}

\begin{proof}
As in \eqref{e:Ustar1},
\begin{align}
  \|(g\tau^2)^*(b)\|_{T_{0,\aux}(h,\hAux)}
    & =
    \|g\tau^{2}(b)\|_{T_{0}(h)}
    +
      \hAuxg_{V}
      \sup_{\dot V \in \Vcal: \dot \nu=\dot u = 0}
      \frac{\|\dot V(b)\|_{T_{0}(h)}}{\|\dot V(b)\|_{T_{0}(\ell)}} .
\end{align}
By \eqref{e:epdV--1} with $\nu =0$ and $\h=h$,
the first term on the right-hand side is $\frac14 k_0^4 g\ggen^{-1}$.
Similarly, by comparing \eqref{e:epdV--1} with $\h = h$ and with $\h =
\ell$, the ratio in the second term is $h^4\ell^{-4} = \ggen^{-1} k_0^4
\ell_0^{-4}$. Therefore,
\begin{align}
    \|(g\tau^2)^*(b)\|_{T_{0,\aux}(h,\hAux)}
    & =
    \left(
    \tfrac{1}{4}\tfrac{g}{\ggen}  + \tfrac{\hAuxg_{V}}{\ggen \ell_0^{4}}
    \right)k_0^4 .
\label{e:U*norm2z}
\end{align}
By hypothesis, $\frac{\hAuxg_{V}}{\ggen \ell_0^{4}} \le \ell_0^{-4}$,
and by the definition of $\ell_0$ in \refeq{ell0def} we have $\ell_{0}
\rightarrow \infty$ as $L\rightarrow \infty$.  Hence, since
$k_0<\frac{g}{\ggen}$ by definition of $\DVstab$ in \refeq{DVstab}, if
$L$ is sufficiently large then $\ell_0^{-4} \le \frac 18 k_0 < \frac
18 \frac{g}{\ggen}$.  This proves the first inequality of
\eqref{e:gnustar}. Similarly, from \eqref{e:epdV--1} and
\eqref{e:ell--h},
\begin{align}
    \|(\nu\tau)^*(b)\|_{T_{0,\aux}(h,\hAux)}
    & =
    \left(
    \tfrac{1}{2}\tfrac{|\nu|L^{\drb j/2}}{\ggen^{1/2}}
    + \tfrac{\hAuxg_{V}}{\ggen^{1/2}\ell_0^{2}}
    \right)k_0^2
    \le
    \left(
    \tfrac{1}{2}k_0^2
    + \tfrac{\ggen^{1/2}}{\ell_0^{2}} \right)k_0^2 \le k_0^4
    .
    \label{e:U*norm3z}
\end{align}
This proves the second inequality of \eqref{e:gnustar}.  Finally,
\eqref{e:Vstarhbd} follows from the triangle inequality and the bound
$g/\ggen \le k_0^{-1}$ (due to \refeq{DVstab}).
\end{proof}

Finally, we obtain estimates for $Q(b)$ of \refeq{Q-def}.  We view $Q$
as a function of $(V,K)$, so we use the extended norm.

\begin{lemma}
For all $(V,K) \in \Vcal \times \Fcal$, and for all $\h>0$,
\begin{equation}
  \label{e:QbdT0aux-P}
  \|Q(b) \|_{T_{\varphi,\aux}(\h,\hAux)}
  \le e^{\|V^*(b)\|_{T_{0,\aux}(\h,\hAux)}} \|K^*\|_{T_{0,\aux}(\h,\hAux)} P_{\h}^4(\varphi)  .
\end{equation}
In particular,
\begin{equation}
  \label{e:QbdT0aux}
  \|Q(b) \|_{T_{0,\aux}(\ell,\hAux)}
  \le
  e^{\|V(b)\|_{T_{0}(\ell)}+\hAux_{V}}
  \left(\|K\|_{T_{0}(\ell)}+\hAux_{K} \right).
\end{equation}
Suppose now that $\hAuxg_V \le \ggen$, $\hAuxg_K \le \ggen$, $\h_+ \ge \ell$,
$V \in \DVstab$, and $\|K(b)\|_{T_{0}(\ell)} \le \ggen$.
Then, for $L$ sufficiently large,
\begin{equation}
    \label{e:eQbdB}
    \|Q(B)\|_{T_{\varphi,\aux} (\h_{+},\hAux)}
    \le
    O(\ell_0^{-4})\left(\frac{\h_{+}}{h_{+}} \right)^{4}P_{\h_{+}}^4(\varphi).
\end{equation}
\end{lemma}

\begin{proof}
Recall that $\LT = \Tay_4$.  By the definition of $Q$ in
\eqref{e:Q-def} and the bound on $\Tay_4$ in \eqref{e:TayT0}, for
$\h>0$,
\begin{equation}
    \|Q(b) \|_{T_{0,\aux}(\h,\hAux)}
    \le
    \|e^{V^{*}(b)} K^{*}(b)\|_{T_{0,\aux}(\h,\hAux)}
    \le
    e^{\|V^*(b)\|_{T_{0,\aux}(\h,\hAux)}} \|K^*\|_{T_{0,\aux}(\h,\hAux)}
    ,
\end{equation}
which, with \eqref{e:Tphi-poly-2}, proves \refeq{QbdT0aux-P}.
By setting $\h = \ell$ and inserting \eqref{e:U*norm1}--\eqref{e:K*norm-ell},
\begin{align}
    \|Q(b) \|_{T_{0,\aux}(\ell,\hAux)}
    &\le
    e^{\|V(b)\|_{T_{0,\aux}(\ell,\hAux)}+\hAux_{V}}
    \left( \|K\|_{T_{0,\aux}(\ell,\hAux)} + \hAux_{K}\right) ,
\end{align}
which proves \refeq{QbdT0aux}.

Next we prove \eqref{e:eQbdB}.
By \eqref{e:Tphi-poly-2}, it suffices to prove the result just for
$\varphi = 0$.  By the hypothesis $\h_{+}\ge \ell$ and
Lemma~\ref{lem:hoverell} with $\h'=\ell$ and $\h=\h_{+}$,
\begin{align}
  \|Q(b) \|_{T_{0,\aux}(\h_{+},\hAux)}
  &\le
    \left(\frac{\h_{+}}{\ell} \right)^{4}
    \|Q(b) \|_{T_{0,\aux}(\ell,\hAux)} .
    \lbeq{QbdP}
\end{align}
We insert \eqref{e:QbdT0aux} in the right-hand side and use
\eqref{e:V-normstab}, which implies  $e^{\|V(b)\|_{T_{0,\aux}(\ell)}} \leq 2$
(for $\ggen$ small).  Using also the hypotheses, we obtain
\begin{equation}
  \label{e:QbdTphiaux}
  \|Q(b)\|_{T_{0,\aux}(\h_{+},\hAuxg)}
  \le
    \left(\frac{\h_{+}}{\ell} \right)^{4}
    2 e^{\hAux_{V}}
    \left(\ggen+\hAux_{K} \right)
  \le
    5\ggen
    \left(\frac{\h_{+}}{\ell} \right)^{4}
    .
\end{equation}
By \eqref{e:ell-def} and \eqref{e:ell--h}, $\frac{\h_{+}}{\ell} =
L^{-1}\frac{\h_{+}}{\ell_{+}} =
L^{-1}\frac{\h_{+}}{h_{+}}\frac{h_{+}}{\ell_{+}} =
L^{-1}\frac{\h_{+}}{h_{+}}\frac{k_{0}}{\ell_{0}}\ggen_{+}^{-1/4}$. Therefore
\begin{equation}
  \|Q(b)\|_{T_{0,\aux} (\h_{+},\hAux)}
  \le
  O (k_{0}^{4})L^{-4}\ell_{0}^{-4}\left(\frac{\h_{+}}{h_{+}} \right)^{4} ,
\end{equation}
where we used \eqref{e:gtilde}.  Since $\|Q(B)\|_{T_{0,\aux}
(\h_{+},\hAux)} = L^{\drb}\|Q(b)\|_{T_{0,\aux} (h_{+},\hAux)}$, this
implies \eqref{e:eQbdB} for $\varphi=0$ as desired.
The proof is complete.
\end{proof}

Next we bound $\Vhat = V-Q$.

\begin{lemma}
\label{lem:moreQbd} Let $\hAuxg_V \le \ggen$ and $\hAuxg_K \le \ggen$.
Let $V\in\DV$ and $\|K(b)\|_{T_{0}(\ell)} \le \ggen$.
Then, for $L$ sufficiently large,
\begin{align}
\label{e:moreQbd:BVhatell}
    \|\Vhat(B)\|_{T_{0,\aux}(\ell_{+},\hAux)}
    &  \le O_{L}(\ggen_{+}),
  \\
\label{e:moreQbd:BVhat}
    \|\Vhat(B)\|_{T_{0,\aux}(h_{+},\hAux)}
    &  \le 1
    .
\end{align}
The same estimates hold when $\Vhat$ is replaced by $V^*$.
\end{lemma}

\begin{proof}
Since we are using the extended norm,
we write $\Vhat=V^* - Q$.  We will bound $V^*$ and $Q$
individually.  The proof for $V^*$ instead of $\Vhat$ is obtained by
forgetting $Q$. We will use \eqref{e:gtilde} without comment to
replace $O (\ggen)$ by $O (\ggen_{+})$.

We begin with \eqref{e:moreQbd:BVhatell}. The
$T_{0,\aux}(\ell_{+},\hAux)$ norm of $V^{*} (B)$ increases when
$\ell_{+}$ is replaced by $\ell$.  Furthermore, the
$T_{0,\aux}(\ell,\hAux)$ norm of $V^{*} (B)$ is $L^{d}$ times the same
norm of $V^{*} (b)$, which, by \eqref{e:U*norm1} and \eqref{e:V-norm},
is $O_{L} (\ggen)$. Therefore
\begin{equation}
    \|\Vhat(B)\|_{T_{0,\aux}(\ell_{+},\hAux)}
    =
    O_{L} (\ggen_+) .
\end{equation}
By \eqref{e:eQbdB} with $\h_{+}=\ell_{+}$ and $\eqref{e:ell--h}$ to
show that $\left(\frac{\ell_{+}}{h_{+}} \right)^{4} = O_{L}
(\ggen_{+})$ we find that $\|Q(B)\|_{T_{\varphi,\aux}
(\ell_{+},\hAux)}$ is $O_{L} (\ggen_+)$. Therefore, by the triangle
inequality,
\begin{align}
\label{e:BVhatbd}
    \|\Vhat(B)\|_{T_{0,\aux}(\ell_{+},\hAux)}
    \le
    \|V^*(B)\|_{T_{0,\aux}(\ell_{+},\hAux)}
    +
    \|Q(B)\|_{T_{0,\aux}(\ell_{+},\hAux)}
    \le
    O_{L} (\ggen_+)
\end{align}
as desired.

For \eqref{e:moreQbd:BVhat},
by the triangle inequality and \eqref{e:Ustar1} followed by
\eqref{e:epdV--1} and \eqref{e:ell--h}, as in \eqref{e:U*norm2z}--\eqref{e:U*norm3z}
we obtain
\begin{align}
    \|V^*(B)\|_{T_{0,\aux}(h_{+},\hAux)}
    & \le
    \left(
    \tfrac{1}{4}|B|\tfrac{g_{+}}{\ggen_{+}}\ggen_{+}h_{+}^{4}  + \hAuxg_{V} \tfrac{|B|h_{+}^{4}}{|b|\ell^{4}}
    \right)
    +
    \left(
    \tfrac{1}{2}|\nu||B|h_{+}^{2}
    + \hAuxg_{V}\tfrac{|B|h_{+}^{2}}{|b|\ell^{2}}
    \right)  \nnb
    & =
    \left(
    \tfrac{1}{4}\tfrac{g_{+}}{\ggen_{+}}  + \tfrac{\hAuxg_{V}}{\ell_{0}^{4}}
    \right)k_0^4
    +
    \left(
    \tfrac{1}{2}\tfrac{|\nu|L^{\drb (j+1) /2}}{\ggen^{1/2}}
    + \tfrac{\ggen^{1/2}}{\ell_0^{2}}
      \right)k_0^2
      .
\end{align}
By inserting the definition \eqref{e:DVdef} of $V \in \DV$,
\eqref{e:gtilde}, and using \eqref{e:eQbdB} with $\h_{+}=h_{+}$
for $Q$, we obtain
\begin{equation}
    \|V^*(B)\|_{T_{0,\aux}(h_{+},\hAux)}
    +\|Q(B)\|_{T_{0,\aux}(h_{+},\hAux)}
    \le
    \tfrac{1}{4}k_{0}^{3}
    +
    O_{L} (\ggen^{1/2})
    +O(\ell_0^{-4})
    \leq 1
\end{equation}
because $\ggen$ and $\ell_0^{-4}$ are small depending on $L$
and $k_0 \leq 1$. This proves \eqref{e:moreQbd:BVhat}.
\end{proof}

The following lemma is a consequence of the previous two.

\begin{lemma}
\label{lem:moreQbd2}
Let $\hAuxg_V \le \ggen$ and $\hAuxg_K \le \ggen$.
Let $V\in\DV$ and
$\|K(b)\|_{T_{0}(\ell)} \le \ggen$.
Then, for $L$ sufficiently large,
\begin{equation}
  \label{e:moreQbd:U+-V}
  \|U_+(B)-V^{*}(B)\|_{T_{\varphi,\aux}(h_+,\hAux)}
  \le
  O(\ell_0^{-4})P_{h_+}^4(\varphi).
\end{equation}
\end{lemma}

\begin{proof}
By \refeq{Tphi-poly-2}, it suffices to consider the case $\varphi=0$.
By definition, $U_+-V = \Upt(\Vhat) - V$ with $\Vhat=V-Q$.  By the triangle inequality,
\begin{equation}
    \|U_+(B)-V^{*}(B)\|_{T_{0,\aux}(h_+,\hAux)}
    \le
    \| \Upt(\Vhat,B) - \Vhat(B)\|_{T_{0,\aux}(h_+,\hAux)} +\| Q\|_{T_{0,\aux}(h_+,\hAux)}.
\end{equation}
We use \refeq{moreQbd:BVhat}
and Lemma~\ref{lem:Vpt-6} for the first term,
and \eqref{e:eQbdB} for the second term.  With the bound on
$\mathfrak{c}_+$ of \refeq{Cellbd}, this leads to
\begin{equation}
    \|U_+(B)-V^{*}(B)\|_{T_{0,\aux}(h_+,\hAux)}
    \le
    O\left(\frac{\mathfrak{c}_+}{h_+}\right)  + O(\ell_0^{-4})
    = O\left( \ggen_+^{1/4} + \ell_0^{-4} \right),
\end{equation}
and the proof is complete since we can choose $\ggen$ small depending
on $L$.
\end{proof}

\section{Proof of Theorem~\ref{thm:Rplus}}
\label{sec:proof:Rplus}

The following lemma is the basis for the proof of
Theorem~\ref{thm:Rplus}.

\begin{lemma}
\label{lem:norm-Rplus} Let $\Aux=\Vcal \times \Fcal$ have the norm
\eqref{e:Aux-norm-def}, with arbitrary $\hAuxg$.  For $(V,K)\in \Vcal
\times \Fcal$, let
$r_{1} = \|V(b)\|_{T_{0}(\ell)}+\hAux_{V}$
and
$r_{2} = \|K\|_{T_{0}(\ell)}+\hAux_{K}$, and assume $r_2 \le
r_1$.  Then
\begin{align}
\label{e:norm-Rmain-general}
    \|R_{+}^U(B)\|_{T_{0,\aux}(\ell_+,\hAux)}
    & =
    O (L^{2d}) \left(e^{2r_1}r_1r_2 \right) .
\end{align}
\end{lemma}

\begin{proof}
By \eqref{e:QbdT0aux}, $\ell_{+} \le \ell$ and monotonicity in $\h$,
\begin{equation}
    \label{e:Qr12}
    \|Q(B) \|_{T_{0,\aux}(\ell_+,\hAux)}  \le L^d e^{r_1} r_2 ,
\end{equation}
where $L^{d}$ is the number of blocks $b$ in $B$.
By Exercise~\ref{ex:EthetaV-6}, and since $\mathfrak{c}_+ \le \ell_+$
by \refeq{Cellbd}, this implies
\begin{align}
\lbeq{EQbd}
    \|
    \Ex_{C_{+}} \theta Q (B)
    \|_{T_{0,\aux} (\ell_{+},\hAux)}
    &\le
    \| Q (B) \|_{T_{0,\aux} (\ell_{+},\hAux)}
    \left(
    1 + c \left(\tfrac{\mathfrak{c}_{+}}{\ell_{+}}\right)^{2}
    \right)
    \le
    O (L^{d}) (e^{r_1} r_2) .
\end{align}
By Lemma~\ref{lem:covariance-6}, by $\mathfrak{c}_+ \le \ell_+$, and by the
hypothesis $r_2 \le r_1$ and \eqref{e:Qr12},
\begin{align}
\lbeq{CovQVbd}
    &
    \| \Cov_{+} \big(\theta (V(B)-\half   Q(B)),\theta Q(B)\big)\|_{T_{0,\aux} (\ell_{+},\hAux)}
    \nnb
    &\le
    c|B|^{2}
    \|V_{x} - \half Q_{x}\|_{T_{0,\aux}(\ell_{+},\hAux)} \|Q_{x}\|_{T_{0,\aux}(\ell_{+},\hAux)}
    \le
    O (L^{2d}) (e^{2r_1} r_{1} r_2)
    .
\end{align}
The desired result then follows by inserting
\refeq{EQbd}--\refeq{CovQVbd} into \refeq{RUrewrite}.
\end{proof}

\begin{proof}[Proof of Theorem~\ref{thm:Rplus}] Since $R_+^U$ is
quadratic in $K$, the case $q \ge 3$ of \refeq{Rmain-g-bis} is
immediate and we need only consider the cases $q=0,1,2$.  Let $p \ge
0$ and $q=0,1,2$.

We will apply \eqref{e:normderivT0} and \eqref{e:norm-Rmain-general}.
Let $r_{1}=\|V(b)\|_{T_{0}(\ell)}+\hAux_{V}$ and
$r_{2}=\|K\|_{T_{0}(\ell)}+\hAux_{K}$.  By \eqref{e:V-norm}, and the
hypothesis on $K$, we have $r_{1} = O_{L} (\ggen) + \hAuxg_V$ and $r_2
= O (\tilde\vartheta^{3/2}\ggen^3) + \hAuxg_{K}$.  We choose $\hAuxg_{V}=1$.
Then the hypothesis $r_2 \le r_1$ of Lemma~\ref{lem:norm-Rplus}
applies as long as $\hAuxg_K \le 1$, which we also assume.  We apply
Lemma~\ref{lem:norm-Rplus} and \eqref{e:normderivT0} and obtain
\begin{align}
    \label{e:norm-derivative-bound2a}
    \|D_V^p D_K^q R_+^{U}(B)\|_{{\Vcal(\ell) \times \Wcal \to \Ucal_+(\ell_+)}}
    &\le
    \hAuxg_{K}^{-q}
    p!q! \|R_{+}^U(B)\|_{T_{0,\aux}(\h_+,\hAux)}
\nnb
    &
    \le
    \hAuxg_{K}^{-q}
    p!q!
    O_{L} \left(\tilde\vartheta^{3}\ggen^3 + \hAuxg_K\right) .
\end{align}
We obtain the case $q=0$ of \eqref{e:Rmain-g-bis} by setting $\hAuxg_K=0$,
and the cases $q=1,2$ by setting $\hAuxg_{K} = 1$.  
This completes the proof.
\end{proof}

The polynomial $R_+^U$ has the relatively simple explicit formula
\refeq{RUrewrite}, so it is also possible to compute and estimate the
derivatives directly using only the $T_{0}(\ell)$-seminorm, and
without introduction of the extended norm.  This is the subject of the
next exercise.  Direct computation and estimation of derivatives of
$K_+$ is less straightforward, and the profit from using the extended
norm is larger.

\begin{exercise}\label{ex:R+U}
Compute the derivatives $D_V^pD_K^q R_+^U(V,K;\dot V^p, \dot K^q)$
explicitly, and use the result to estimate the norms in
\eqref{e:Rmain-g-bis} directly using only the $T_{0}(\ell)$-seminorms.
\solref{R+U}
\end{exercise}

\chapter{Bounds on \texorpdfstring{$\Phi_+^K$}{Phi+K}: Proof of
Theorem~\ref{thm:step-mr-K}}
\label{ch:pf-thm:step-mr-K}

\renewcommand{\VptE}{U_{\rm pt}^{(0)}}

In this chapter, we prove Theorem~\ref{thm:step-mr-K} and, as a
byproduct, also Proposition~\ref{prop:N}.  We also prove the
continuity assertion in Theorem~\ref{thm:step-mr-R}.  This then
completes the proof of Theorem~\ref{thm:phi4-hier-chi}.

\section{Main result}

Our main goal is to prove the estimates of
Theorem~\ref{thm:step-mr-K}, which we restate here as follows.
In Section~\ref{sec:masscont}, we verify the continuity assertions of
Theorem~\ref{thm:step-mr-K}, and also of Theorem~\ref{thm:step-mr-R}.

\index{Renormalisation group step}
\begin{theorem}
\label{thm:step-mr-K-2}
  Let $\mgen^2 \ge 0$, let
  $L$ be sufficiently large,
  let $\ggen$ be sufficiently small (depending on $L$), and let $p,q\in \N_0$.
  Let $0\le j<N$.
  There exist $L$-dependent $\CRG,M_{p,q} >0$
  and $\kappa =O(L^{-2})$  such that  the map
  \begin{equation}
  \lbeq{Kplusmap-2}
    \Phi^K_+:\domRG  \times \Iint_+  \to \Wcal_{+}
  \end{equation}
  satisfies the estimates
  \begin{align}
\lbeq{DVKbd-2}
    \|D_{V}^pD_{K}^{q}\Phi^K_+\|_{\Vcal(\ell) \times \Wcal \to \Wcal_+}
    &\le
    \begin{cases}
    \CRG
    \tilde \vartheta_+^{3}
    \ggen_+^{3}
    &
    (p = 0, \, q=0 )
    \\
    M_{p,0}
    \tilde \vartheta_+^{3}
    \ggen_+^{3-p}
    &
    (p > 0, \, q=0 )
    \\
    \rlap{$\kappa$}\hspace{3.3cm}
    & (p=0,\, q=1)
    \\
    M_{p,q}
    \ggen_+^{-p-\frac{9}{4}(q-1)}
    &
    (p \ge 0,\, q \ge 1)
    .
    \end{cases}
  \end{align}
\end{theorem}

\begin{figure}
  \includegraphics[width=0.4\textwidth]{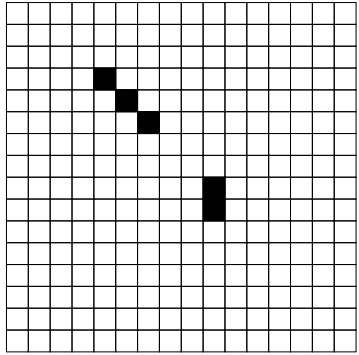}
  \caption{Each block $B$ is divided into $L^\drb$ blocks $b$ at the previous scale.
    The small black blocks represent the region $X$ in \eqref{e:S1def}.\label{fig:Pcal}
  }
\end{figure}

The proof is based on a decomposition of $\Phi_+(V,K)$ as a sum of two
contributions, which are constructed as follows.  Let $K_+ =
\Phi_+^K(V,K)$.  With $Q$ defined in \eqref{e:Q-def}, we let
\begin{equation} \label{e:hatdef}
  \Vhat = V - Q,
  \qquad
  \Khat = e^{-V} - e^{-\Vhat} + K ,
\end{equation}
so that
\begin{equation}
\lbeq{eVKhats}
    e^{-V}+K = e^{-\Vhat}+\Khat.
\end{equation}
By \eqref{e:K+B},
\begin{align}
    K_{+} (B)
    &=
    e^{u_+|B|}
    \left(
    \Ex_+\theta
    \left( e^{-\Vhat} + \Khat  \right)^B
    - e^{-U_+ (B)}
    \right)
    \nnb
    &=
    e^{u_+|B|}
    \left(
    \sum_{X \subset \Bcal (B)}
    \Ex_+\theta
    \left( e^{-\Vhat (B\setminus X)} \Khat^{X}\right)
    - e^{-U_+ (B)}
    \right) .
\lbeq{KofKhat}
\end{align}
We isolate the term with $X= \varnothing$ and thus write $K_+(B)$ as
\begin{equation}
    K_{+}
    =
    S_{0} + S_{1}
    ,
    \label{e:K+setup'}
  \end{equation}
with
\begin{align}
    \label{e:S0def}
    S_{0}
    &=
    e^{u_+|B|}
    \left(
    \Ex_+\theta
    e^{-\Vhat (B)}
    - e^{-U_+ (B)}
    \right)
    ,
    \\
    \label{e:S1def}
    S_{1}
    &=
    e^{u_+|B|}
    \sumtwo{X \subset \Bcal (B)}{|X|\ge 1}
    \Ex_+\theta
    \left(e^{-\Vhat (B\setminus X)} \Khat^{X}\right) .
\end{align}
The region $X$ on the right-hand side of \eqref{e:S1def}
is illustrated in Figure~\ref{fig:Pcal}.

Separate mechanisms are invoked to estimate $S_0$ and $S_1$.  The term
$S_0$ will be shown to be third order in $\ggen_+$ due to the fact
that $U_+=\Phi_\pt(V-Q)$ (recall \refeq{U+def}) has been defined in
such a manner to achieve this.  Indeed, Lemma~\ref{lem:K0PT} implies
(here $W_+=0$) that
\begin{align}
\lbeq{ExW5-10}
    \Ex_{+} e^{-\theta \Vhat(B)} - e^{-U_+(B)}
    &=
    e^{-U_+(B)}
    \left(
    \tfrac{1}{8} \left(\LT \Var (\theta \Vhat) \right)^2
    + \Ex_{+} A_3(B)
    \right)
    ,
\end{align}
where $A_3$ (given by \refeq{A3def} with $U$ replaced by $\Vhat$) is
third order in $\Vhat-U_+$.  The variance term is fourth order in
$\Vhat$.  Using these facts, $S_0$ will be shown to be third order in
$\ggen_+$.

The contribution of $S_1$ to $K_+$ is small because it contains a
factor $\Khat(b)$ for at least one small block $b\in B$.  If $\Khat$
were roughly the same size as $K$, this would provide a good factor of
order $g^3$.  However, there are $L^{\drb}$ small blocks $b\in B$, so
this good factor must be multiplied by $L^{\drb}$.
Theorem~\ref{thm:step-mr-K-2} asserts that the $K$-derivative of the
map $(V,K) \mapsto K_+$ is less than $1$, and the naive argument just
laid out cannot prove this; it is spoiled by the $L^{\drb}$.  Instead,
we make use of the crucial fact that the map $K \to \Khat$ on a single
small block $b$ has derivative of order $L^{-\drb-2}$ because the
terms $e^{-V}-e^{-\Vhat}$ in \refeq{hatdef} effectively cancel the
relevant and marginal parts of $K$, leaving behind irrelevant parts
that scale down.  The good scaling factor $L^{-\drb-2}$ and the bad
entropic factor $L^{\drb}$ combine to give $L^{-2}$.  We can choose
$L$ large enough so that $L^{-2}$ cancels any dimension-dependent (but
$L$-independent) combinatorial factors that arise in the estimates.
Consequently, $S_1$ remains small enough to prove
Theorem~\ref{thm:step-mr-K-2}.

The details for $S_0$ and $S_1$ are presented in the rest of the
chapter.  An important special case is obtained by setting $K=0$ in
\refeq{S0def} to yield $S_{0,0}$ defined by
\begin{equation}
\lbeq{S00def}
    S_{0,0}
    =
    e^{u_{\pt}|B|}
    \left(
    \Ex_+\theta
    e^{-V (B)}
    - e^{-U_\pt (B)}
    \right).
\end{equation}
\index{Perturbative contribution}%
We refer to $S_{0,0}$ as the \emph{perturbative contribution} to
$K_+$, since it is the contribution when $K=0$.  As discussed below
\refeq{uVK}, at the initial scale $0$ we have $K_0=0$, so for the
first application of the renormalisation group map, from scale $0$ to
$1$, $K_1=\Phi_1^K(V_0,K_0)$ is equal to the perturbative
contribution.  This can be considered the genesis of $K$. At
subsequent scales, the previous $K$ creates an additional contribution
to $K_+$.  The following proposition gives an estimate on $S_{0,0}$.

\begin{prop}\label{prop:K0norm-0}
For $L$ sufficiently large and $\ggen$ sufficiently small, there is a
constant $C_{\pt} = C_{\pt}(L)$ such that for all $V \in \DV$ and $B
\in \Bcal_+$,
\begin{equation}
\lbeq{Cpt}
    \|S_{0,0}\|_{\Wcal_+}
    \le
    C_{\pt}
    \vartheta_+^3 \ggen_+^3
    .
\end{equation}
\end{prop}

The subscript in the constant $C_\pt$ in \refeq{Cpt} stands for
``perturbation theory.''  The constant $\CRG$ which appears in the
domain $\domRG$ and in the first estimate of
Theorem~\ref{thm:step-mr-K-2} is defined by
\begin{equation}
\lbeq{CRGdef}
    \CRG = \CRG(L) = 2C_\pt.
\end{equation}

We prove Proposition~\ref{prop:K0norm-0} as a special case of the
following proposition, in which we use the extended norm (see
Section~\ref{sec:extendednorm}).  The definition of the extended norm
requires specification of the parameters $\hAux_V$ and
$\hAux_K$. Throughout this chapter, we always require
\begin{equation}
\label{e:lamstab-V}
    \hAuxg_{V} \le \ggen.   
\end{equation}
For $\hAux_K$, we require either
\begin{equation}
\lbeq{lamstab-K}
    \hAuxg_{K} \le \ggen
    \quad
    \text{or}
    \quad
    \hAuxg_K \le \ggen^{9/4},
\end{equation}
with the choice depending on the estimate being proved.  For
derivatives, we obtain the best estimates by taking the two parameters
to be as large as possible.  Smaller choices are also useful and
permitted, including $\hAux_V=\hAux_K=0$, which gives the best
estimates on the functions themselves.

Proposition~\ref{prop:smoothness} is the main ingredient in the proof
of the derivative estimates in \refeq{DVKbd-2}, but on its own is not
sufficient to prove the cases $(p,q)=(0,0)$ and $(p,q)=(0,1)$ with
constants $\CRG$ and $\kappa$ in their upper bounds.  Those cases are
given separate treatment.  Note that the constant $\bar{C}$ in
\refeq{K+Tz-T2-a} is not the same as $\CRG$ in
Proposition~\ref{prop:smoothness}.  The case $(p,q)=(0,0)$ in
\refeq{DVKbd-2} is proved using the crucial contraction, i.e., the
case $(p,q)=(0,1)$ of \refeq{DVKbd-2}, and is discussed in detail
later in this section.

\begin{prop} \label{prop:smoothness}
Let $L$ be sufficiently large and let $\ggen$ be sufficiently small
depending on $L$.  Let $\CRG$ be given by \refeq{CRGdef}.  Let
$\hAuxg_V \le \ggen$ and $\hAux_K\le \ggen^{9/4}$.  There is a
constant $\bar C = \bar C(L)$ such that if $(V,K,m^2) \in \domRG
\times \Iint_+$ then (recall $y=(V,K)$)
\begin{align}
    \label{e:K+Tz-T2-a}
    \|K_+\|_{\Wcal_{\aux,+}(\hAux)}
    &\le
      \bar{C}
      (\vartheta_+^3 \ggen_+^3  + \hAuxg_K).
\end{align}
\end{prop}

The next two lemmas give estimates on $S_0$ and $S_1$.  The lemmas are
proved in Sections~\ref{sec:S0} and \ref{sec:S1}, respectively.

\begin{lemma} \label{lem:S0}
Let $V \in \DV$ and $\|K(b)\|_{T_{0}(\ell)} \le \ggen$.  If
$\hAuxg_V\le \ggen$ and $\hAux_K \le \ggen$ then
\begin{equation}
\lbeq{S0bd}
    \|S_0\|_{\Wcal_{\aux,+}(\hAux)}
    \le O_L(\vartheta_+^3 \ggen_+^3) .
\end{equation}
\end{lemma}

\begin{lemma} \label{lem:S1}
Let $(V,K) \in \domRG$.  If $\hAuxg_V\le \ggen$ and $\hAux_K \le
\ggen^{9/4}$ then
\begin{align}
  \label{e:S1bd}
  \|S_{1}\|_{\Wcal_{\aux,+}(\hAux)}
  &\le
  O_L (\vartheta_+^3 \ggen_+^3  +\hAux_K).
\end{align}
\end{lemma}

\begin{proof}[Proof of Proposition~\ref{prop:K0norm-0}] We take $K=0$
in \eqref{e:K+setup'} so that $K_+ = S_{0,0}$ with $S_{0,0}$ given by
\refeq{S00def}.  Taking $\hAuxg_V=\hAuxg_K=0$ in Lemma~\ref{lem:S0},
we get 
the estimates $\|S_{0,0}(V)\|_{T_{\varphi}(\ell_+)} \le \bar{C}'\vartheta_+^3
\ggen_+^3$ and $\|S_{0,0}(V)\|_{T_{\varphi}(h_+)} \le \bar{C}'\vartheta_+^3
\ggen_+^{3/4}$.  
The constant $\bar{C}'$ does not depend on $\CRG$,
because $\CRG$ serves only in the definition \eqref{e:Phi_+-domain}
of $\domRG$ to provide a limitation on the size of $K$,
and we have set $K=0$.  Thus we obtain \refeq{Cpt} with $C_\pt =
\bar{C}'$.
\end{proof}

\begin{proof}[Proof of Proposition~\ref{prop:smoothness}] The bound
\refeq{K+Tz-T2-a} is an immediate consequence of
Lemmas~\ref{lem:S0}--\ref{lem:S1}, together with the decomposition
$K_+=S_0+S_1$ of \refeq{K+setup'}.
\end{proof}

\begin{proof}[Proof of \refeq{DVKbd-2} except cases $(p,q)=(0,0)$ and
$(p,q)=(0,1)$] We fix $(V,K,m^2) \in \domRG\times \Iint_+$.  From
\eqref{e:K+Tz-T2-a} and Lemma~\ref{lem:normderiv} we have
\begin{equation}
    \|D_V^pD_K^q K_+\|_{\Vcal(\ell) \times \Wcal \to \Wcal_+}
    \le
    \frac{p!q!}{\hAux_V^p\hAux_K^q} O_L(\vartheta_+^3 \ggen_+^3 + \hAux_K).
\end{equation}

For the case $p\ge 1$ with $q=0$, we set $\hAuxg_K=0$ and take
equality for $\hAuxg_V$ in \refeq{lamstab-V}, to get the desired
result.  For $(p,q)=(0,0)$, the choice $\hAuxg_V=\hAuxg_K=0$ could be used, but
this gives an upper bound with constant $\bar{C}$ rather than the
required $\CRG$; this case is discussed in the next proof.

For the case $p \ge 0$ with $q\ge 1$, we take equality for $\hAuxg_V$
in \refeq{lamstab-V} and set $\hAuxg_K = \ggen^{9/4}$.  This gives the
desired result except for $(p,q)=(0,1)$, where here we see an upper
bound $2\bar{C}(\vartheta^3 \ggen^{3/4} + 1)$ rather than $\kappa
=O(L^{-2})$; the bound with $\kappa$ is the crucial contraction whose
proof is given in Section~\ref{sec:rgest-contraction}.  This completes
the proof.
\end{proof}

\begin{proof}[Proof of case $(p,q)=(0,0)$ in \refeq{DVKbd-2}] This
proof uses the $\Wcal$-norm of \refeq{Wcal}, and not the extended
$\Wcal(\hAux)$-norm of \refeq{Wcalgrand}.

Let $f(t) = \Phi_+^K(V,tK)$.  We apply Taylor's Theorem to $f$, with
integral form of the remainder, and obtain
\begin{align}
\lbeq{K+decomp}
    \Phi_+^K(V,K) & = \Phi_+^K(V,0) + D_{K}\Phi_+^K(V,0;K) + R_+^K(V,K)
\end{align}
with
\begin{equation}
    R_+^K(V,K)
    =
    \int_0^1 (1-t) \frac{d^2}{dt^2} \Phi_+(V,tK) dt
    .
\end{equation}
By the triangle inequality,
\begin{align}
\lbeq{K+decomp-ti}
    \|\Phi_+^K(V,K)\|_{\Wcal_+}
    & \le \|\Phi_+^K(V,0)\|_{\Wcal_+}
    + \|D_{K}\Phi_+^K(V,0;K)\|_{\Wcal_+}
    + \| R_+^K(V,K)\|_{\Wcal_+}
    .
\end{align}

By Proposition~\ref{prop:K0norm-0},
\begin{equation}
    \|\Phi_+^K(V,0)\|_{\Wcal_+} \le C_\pt\vartheta_+^3 \ggen_+^3.
\end{equation}
By the crucial contraction with $\kappa \le c_\kappa L^{-2}$ (the
case $(p,q)=(0,1)$ of \refeq{DVKbd-2} proved in
Section~\ref{sec:rgest-contraction}), and by the assumption that
$\|K\|_\Wcal \le \CRG \vartheta^3 \ggen^3$, we have
\begin{equation}
    \|D_{K}\Phi_+^K(V,0;K)\|_{\Wcal_+}
    \le
    \kappa  \CRG \vartheta^3 \ggen^3
    \le
    c_\kappa L^{-2}2C_\pt \vartheta^3 \ggen^3.
\end{equation}
Since $\vartheta \le 2 \vartheta_+$ and $\ggen \le 2 \ggen_+$, we can
choose $L$ so that $L^2 \ge 2 \cdot 2 \cdot 2^6 c_\kappa$ to conclude
that
\begin{equation}
    \|D_{K}\Phi_+^K(V,0;K)\|_{\Wcal_+}
    \le
    \tfrac{1}{2} C_\pt \vartheta_+^3 \ggen_+^3.
\end{equation}
By the case $(p,q)=(0,2)$ of \refeq{DVKbd-2}, for $\ggen$ chosen
sufficiently small depending on $L$ to ensure that $M_{0,2} (2C_\pt
2^6)^2 \ggen^{3/4} \le \frac 12 C_\pt$, we also have
\begin{align}
    \| R_+^K(V,K)\|_{\Wcal_+}
    &  \le
    M_{0,2} \ggen_+^{-9/4} (\CRG \vartheta^3 \ggen^3)^2  \le \frac 12 C_\pt \vartheta_+^3 \ggen_+^3
    .
\end{align}
This gives the desired result, with $\CRG = 2C_\pt$ as in \refeq{CRGdef}.
\end{proof}

It remains now to prove Lemmas~\ref{lem:S0}--\ref{lem:S1}, as well as
the case $(p,q)=(0,1)$ of \refeq{DVKbd-2}.  We do this in the
remainder of the chapter.

\section{Stability}
\label{sec:stability}
\index{Stability}

This section is concerned with a collection of
estimates which together go by the name of \emph{stability} estimates.
The domain $\DV$ for $V=g\tau^2+\nu\tau$ permits negative values of
the coupling constant $\nu$, as it must in order to approach the
critical value, which is itself negative.  Thus, $V$ can have a double
well shape for $n=1$, and a Mexican-hat shape for $n>1$, so in
$e^{-V(b)}$ there is a growing exponential factor $e^{-\nu\tau(b)}$
which must be compensated, or \emph{stabilised}, by the decaying
factor $e^{-g\tau^2(b)}$.  Moreover, it is not only the value of
$e^{-V(b)}$ itself that must be controlled, but also its derivatives
with respect to the field.  For this, we use the
$T_\varphi(h)$-seminorm with $h=k_0\ggen^{-1/4}L^{-\drb j/4}$ given by
\refeq{h-def}.

Recall the definition of the domain $\DVstab$ in \eqref{e:DVstab}.
The definition guarantees that for $V = (g,\nu,0) \in\DVstab$
the stability estimate \eqref{e:stability} holds.
This is an estimate for $e^{-V(b)}$ pointwise in $\varphi$. In this section, we
extend this estimate to an estimate for $T_{\varphi,y}$-norms and also
consider more general expressions than $e^{-V(b)}$.
The stability domain is useful because,
although it is not the case that $\DV$ is contained in $\DV_+$,
according to Lemma~\ref{lem:DVnest} we do have
$\DV \subset \DVstab \cap \DVstab_{+}$.
Therefore a hypothesis that $V \in \DV$ ensures stability at both scales.
This fact is used, e.g., in the proof of Lemma~\ref{lem:V-Qbd}.

The following proposition is fundamental.  It contains a hypothesis on
the constant $k_{0}$ that appears in the definition \eqref{e:h-def} of
the large-field scale $h_j$ and in the definition \eqref{e:DVstab} of
the stability domain $\DVstab$. Henceforth, we fix $k_{0}$ so that the
conclusions of Proposition~\ref{prop:stability} hold; we also require
that $k_0 \le \frac{1}{24 (n+2)}$ as in \refeq{k0ceta}.  The statement
of the proposition involves the constant $c^{\stab}$ defined by
\begin{equation}\label{e:cstab-def}
    c^{\stab}
    =
    \frac{1}{128}k_{0}^{5} .
\end{equation}
By Exercise~\ref{ex:stability}, $k_0^5$ is the best possible order in
$c^{\stab}$, because the $T_{\varphi,\aux}$ norm dominates the
absolute value.

\begin{prop}
\label{prop:stability} For $k_{0}>0$ sufficiently small, $V=(g,\nu,0)
\in \DVstab$, $\h\le h$, $t \ge 0$, and $\hAux_V \le \ggen$,
\begin{align} \label{e:stabilityb}
    \|e^{-t V^*(b)}\|_{T_{\varphi,\aux} (\h,\hAuxg)}
    \le
    2^{t/8}
    e^{-8t c^{\stab}  | \tfrac{\varphi}{h} |^{4}}.
\end{align}
\end{prop}

\begin{proof}
The $T_{\varphi,\aux}(\h,\hAuxg)$-seminorm is monotone in $\h$, so it
suffices to consider $\h=h$.  By the product property and
Lemma~\ref{lem:eK},
\begin{align}
\label{e:stability-z-GS}
    \|e^{-t V^*(b)}\|_{T_{\varphi,\aux}}
    & \le
    e^{- 2 tg\tau^2 (b)} \;e^{t \|(g\tau^2)^*(b)\|_{T_{\varphi,\aux}} }\;
    e^{t  \|(\nu \tau)^*(b)\|_{T_{\varphi,\aux}}}
    .
\end{align}
We write the total exponent on the right-hand side of
\eqref{e:stability-z-GS} as $tX$, and set $s = |\varphi|/h$.  Then it
suffices to show that
\begin{equation}\label{e:stabilityc}
    X \le \tfrac{1}{8}\log 2 - 8 c^{\stab} s^4 .
\end{equation}

Recall that $\tau = \tfrac{1}{2}|\varphi|^{2}$.  We estimate the
$T_{\varphi,\aux}$-norms in $X$ by $T_{0,\aux}$-norms using
Exercise~\ref{ex:Tphi-poly}, and then apply Lemma~\ref{lem:U*norm-h}
to bound the $T_{0,\aux}$-norms.  The result is
\begin{align}
    X
    & \le
    - \tfrac 24 g s^4 h^4 L^{dj}
    + \tfrac 38 \tfrac{g}{\ggen}k_0^4 (1+s)^{4}
    +
    k_0^4 (1+s)^2
    \nnb & =
    - \tfrac{1}{2} \tfrac{g}{\ggen} k_{0}^{4} s^4
    + \tfrac 38 \tfrac{g}{\ggen}k_0^4 (1+s)^{4}
    +
    k_0^4 (1+s)^2
    ,
\end{align}
where we used the definition \eqref{e:h-def} of $h$ in the first term.
We split $-\frac{1}{2}s^4$ in the first term into $-\tfrac{1}{16}s^4$ and
$-\tfrac{7}{16}s^4$ and obtain
\begin{align}
    X
    &\le
    -\tfrac{1}{16} k_0^4 \tfrac{g}{\ggen} s^4
    +
     k_0^4 \tfrac{g}{\ggen}  \left(
    - \tfrac{7}{16} s^4
    +  \tfrac 38 (1+s)^4
    +  \tfrac{\ggen}{g} (1+s)^2 \right)
    \nnb
    &\le
    -\tfrac{1}{16} k_0^5 s^4
    +
    k_0^3 \max  \Big(
    - \tfrac{7}{16} s^4
    +  \tfrac 38 (1+s)^4
    +  k_{0}^{-1} (1+s)^2 \Big)
    ,
\end{align}
where we used the bounds on $g$ given by $\DVstab$ in
\eqref{e:DVstab}.  The maximum is positive and is $O (k_{0}^{-2})$ as
$k_{0} \downarrow 0$ so $X \le -\tfrac{1}{16} k_0^5 s^4 + O
(k_{0})$, which is the same as $X \le O(k_{0}) - 8 c^{\stab} s^4$
by the definition \eqref{e:cstab-def} of $c^{\stab}$.
Therefore there exists sufficiently small $k_{0}$ such that
\eqref{e:stabilityc} holds.
The proof is complete.
\end{proof}

The next lemma gives an estimate for an extended $T_\varphi$-norm
of $e^{tQ(B)}$.  Since $\varphi$ is a constant field, we have
$Q(b)=L^{-d}Q(B)$, so the choice
$t=L^{-d}$ gives a bound on the norm of $e^{Q(b)}$.  The situation
is similar in subsequent lemmas.

\begin{lemma}
\label{lem:eQbd} Let $\hAuxg_V \le \ggen$ and $\hAuxg_K \le \ggen$.
Let $\h_+ \le h_{+}$ and $t \ge 0$.  Let $V \in \DVstab$ and
$\|K(b)\|_{T_{0}(\ell)} \le \ggen$.  Then, for $L$ sufficiently
large,
\begin{equation}
    \|e^{t Q(B)}\|_{T_{\varphi,\aux} (\h_{+},\hAux)}
    \le
    2^{t/8} e^{\frac 18 t c^{\stab}
    |\frac{\varphi}{h_{+}}|^4}
    .
  \end{equation}
\end{lemma}

\begin{proof}
It is sufficient to set $\h=h_{+}$. By the product property
\begin{equation}
    \|e^{tQ(B)}\|_{T_{\varphi,\aux} (h_{+},\hAux)}
    \le
    e^{t\| Q(B)\|_{T_{\varphi,\aux} (h_{+},\hAux)}} .
\end{equation}
By \eqref{e:eQbdB} and the inequality $P_{h_+}^4(\varphi) \le 2^4(1+
|\frac{\varphi}{h_{+}}|^4)$, the norm in the exponent on the
right-hand side is bounded above by
$O(\ell_0^{-4})(1+|\frac{\varphi}{h_{+}}|^4 )$.  Since $\ell_0 \to
\infty$ as $L \to \infty$ by the definition of $\ell_0$ in
\refeq{ell0def}, the prefactor on the right-hand side of
\eqref{e:eQbdB} can be made as small as we wish, and the desired
result follows.
\end{proof}

For the statement of the remaining results we single out the following
hypotheses:
\begin{equation}
    \label{e:constraints}
    \hAuxg_{V} \le \ggen,
    \qquad
    \hAuxg_K \le \ggen ,
    \qquad
    \|K(b)\|_{T_{0}(\ell)} \le \ggen .
\end{equation}

\begin{lemma}
\label{lem:V-Qbd} Let $\h_+ \le h_{+}$, $t \ge 0$, $0 \le s \le 1$, $V
\in \DV$, and assume \eqref{e:constraints}.  For $L$ sufficiently
large,
\begin{equation} \label{e:stabilityVhat}
    \|e^{-t (V^{*}-s Q)(B)}\|_{T_{\varphi,\aux} (\h_+,\hAuxg)}
    \le
    2^{t/4} e^{- 4tc^{\stab}
    |\frac{\varphi}{h_+}|^4}
    .
\end{equation}
\end{lemma}

\begin{proof}
Since $V \in \DV$, by Lemma~\ref{lem:DVnest}, we have $V \in
\DVstab_{+}$ and we can apply Proposition~\ref{prop:stability} at the
next scale. The claim follows by multiplying the estimates of
Proposition~\ref{prop:stability} at the next scale and
Lemma~\ref{lem:eQbd}.
\end{proof}

\begin{lemma}
\label{lem:Uplusstab} Let $\h_+ \le h_{+}$, $t \ge 0$, $V \in \DV$,
and assume \eqref{e:constraints}.  For $L$ sufficiently large,
\begin{equation}
\label{e:stabilityUplus}
    \|e^{-t U_+(B)}\|_{T_{\varphi,\aux} (\h_+,\hAuxg)}
    \le
    2^{t/2} e^{-2tc^{\stab}
    |\frac{\varphi}{h_{+}}|^4}
    .
\end{equation}
The same estimate holds with $U_+$ replaced by $U_+-u_+$ on the
left-hand side. Furthermore,
$e^{t \|u_+(B)\|_{T_{\varphi,\aux} (\h_+,\hAuxg)}} \le 2^{t/2}$.
\end{lemma}

\begin{proof}
We drop the subscript $T_{\varphi,\aux} (\h_+,\hAuxg)$ from the seminorm in this proof,
and by we monotonicity assume that $\h_+=h_+$.
We write $U_+=[U_+-V^*]+V^*$ and apply the product property to conclude that
\begin{equation}
    \|e^{-t U_+(B)}\|
    \le
    e^{t\|U_+(B)-V^*(B)\|} \|e^{-tV^*(B)}\|.
\end{equation}
We estimate the first factor on the right-hand side using
\eqref{e:moreQbd:U+-V}, and the second factor using
Proposition~\ref{prop:stability} at the next scale.  This gives
\begin{equation}
    \|e^{-t U_+(B)}\|
    \le
    e^{t O(\ell_0^{-4})P_{h_+}^4(\varphi)} 2^{t/8} e^{-8tc^{\stab}|\frac{\varphi}{h_{+}}|^4} .
\end{equation}
To complete the proof, we use $P_{h_+}^4(\varphi) \le 2^4(1+
|\frac{\varphi}{h_{+}}|^4)$ and take $L$ large, using
$\ell_0=L^{1+\drb /2}$ by \refeq{ell0def}.

Since $\|U_+(B)-|B|u_+-V^*(B)\|_{T_{\varphi,\aux}(\h_+,\hAux)} \leq
\|U_+(B)-V^*(B)\|_{T_{\varphi,\aux}(\h_+,\hAux)}$, the same argument
shows that bound holds with $U_+$ replaced by $U_+-u_+$.  Similarly,
for the bound involving $u_+$, we use that
$\|u_+\|_{T_{\varphi,\aux}(\h,\hAux)} = \|u_+\|_{T_{0,\aux}(\h,\hAux)}
\leq \|U_+-V\|_{T_{0,\aux}(\h,\hAux)} \leq O(\ell_0^{-4})$, which
implies the claimed bound on $e^{t \|u_+(B)\|_{T_{\varphi,\aux}
(\h_+,\hAuxg)}}$.
\end{proof}

In the following proposition, the fluctuation field $\zeta$ is
as usual constant on small blocks $b$, and we write its value on
$b$ as $\zeta_b$.
The subscript $\zeta$ notation was introduced above  Proposition~\ref{prop:ExCthetaTphi}.

\begin{prop}
\label{prop:stab1} Let $\h_+ \le h_{+}$, $t \in [0,1]$, $V \in \DV$,
and assume \eqref{e:constraints}.  For $L$ sufficiently large and
$c^{\stab}$ the constant of \refeq{cstab-def},
\begin{equation}
\label{e:stabphit}
    \|(e^{-U_+(B) - t \delta \hat V(B) })_\zeta\|_{T_{\varphi,\aux}(\h_+,\hAuxg)}
    \le
    2 e^{- c^{\stab}   |\varphi/h_{+}|^{4}} + 2
    e^{-c^{\stab} L^{-d}\sum_{b\in \Bcal (B)}|\tfrac{\varphi+\zeta_{b}}{h_{+}} |^{4}}
    ,
\end{equation}
where $\hat V = V-Q$ and $\delta \hat V = \theta \hat V- U_+$.
The same estimate holds with $U_+$ replaced by $U_+-u_+$ on the
left-hand side.
\end{prop}

\begin{proof}
By monotonicity, we may assume that $\h_+=h_+$.
It follows from the definition of $\theta_{\zeta}$ that
\begin{equation}
    (e^{-(1-t)U_+(B) - t \theta \hat V(B) })_\zeta
    =
    e^{-(1-t)U_{+}(B)}
    \prod_{b\in \Bcal(B)}
    (e^{- t \theta \hat V(b) })_\zeta
    .
\end{equation}
By the product property and Lemma~\ref{lem:Uplusstab},
\begin{align}
    &
    \|(e^{-(1-t)U_+(B) - t \theta \hat V(B) })_\zeta\|_{T_{\varphi,\aux} (h_{+},\hAuxg)}\nnb
    & \hspace{10mm} \le
    \Big(
    2^{1/2} e^{- 2c^{\stab} | \tfrac{\varphi }{h_{+}} |^{4}}
    \Big)^{(1-t)}
    \prod_{b\in \Bcal(B)}
    \|(e^{- t \theta \hat V(b) })_\zeta\|_{T_{\varphi,\aux}(h_{+},\hAuxg)}
    .
\end{align}
By Lemma~\ref{lem:V-Qbd}, with $t = L^{-d}$, we estimate each factor
under the product over blocks and multiply the resulting estimate to obtain
\begin{align}
    &
    \|(e^{-(1-t)U_+(B) - t \theta \hat V(B) })_\zeta\|_{T_{\varphi,\aux} (h_{+},\hAuxg)}
    \nnb
    & \hspace{10mm} \le
    \Big(
    2^{1/2} e^{- 2c^{\stab} | \tfrac{\varphi }{h_{+}} |^{4}}
    \Big)^{(1-t)}
    \Big(
    2^{1/4}
    e^{-4c^{\stab}
    L^{-d}\sum_{b \in \Bcal (B)}|\tfrac{\varphi+\zeta_{b}}{h_{+}} |^{4}}
    \Big)^{t}.
\end{align}
Then we apply the arithmetic mean inequality $a^{1-t}b^{t} \le (1-t)a+
t b \le a+b $ to the right-hand side, to obtain the desired
inequality.
\end{proof}

\section{Bound on \texorpdfstring{$S_0$}{S0}: proof of Lemma~\ref{lem:S0}}
\label{sec:S0}

In this section, we prove Lemma~\ref{lem:S0}.  We begin with an estimate
for Gaussian integrals that is useful in the proof of Lemma~\ref{lem:S0} and is
also useful later.

\subsection{Estimation of Gaussian moments}

We exploit the fact that values of $\zeta$ significantly larger than
$\ell_+$ are unlikely, via the existence of high moments implied by
the following lemma for powers of $P_{\ell_+}$ convolved with a
quartic exponential factor.  Recall from \eqref{e:P-def-6} that $P_{\h}
(t) = 1 + |t|/\h$.

\begin{lemma}\label{lem:G}
For $q \ge 0$ there exists $c_2>0$ (depending on $q$)
such that for all $\h_+ \ge \ell_+$ and $0 \le c_1 \le \frac 18
L^{4}$,
\begin{equation}
\label{e:Exphizeta}
  \Ex_+(P_{\ell_+}^q(\zeta_b) e^{-c_1 |\tfrac{\varphi+\zeta_{b}}{\h_{+}} |^{4}})
  \leq c_2 e^{-\frac{c_1}{2}|\tfrac{\varphi}{h_+}|^2}.
\end{equation}
\end{lemma}

\begin{proof}
Throughout the proof, we write $\zeta=\zeta_b$.  For $t,u,v\in \R^n$,
we use the inequalities $|t|^2 \geq |t|-\frac14$, $|u+v| \geq
||u|-|v||$, and $2|u||v| \leq \tfrac{1}{2}|u|^2 + 2|v|^2$, to conclude
that
\begin{align}
    |u+v|^4
    & \geq (|u|-|v|)^2-\frac14
    \nnb &
    = |u|^2 + |v|^2 - 2|u||v| -\frac14
    \geq \tfrac{1}{2} |u|^2 -  |v|^2-\frac14 .
\end{align}
This gives
  \begin{equation}
    e^{-c_1|\tfrac{\varphi+\zeta}{\h_+}|^4}
    \leq e^{\frac{c_1}{4}-\frac{c_1}{2} |\tfrac{\varphi}{\h_+}|^2} e^{c_1|\tfrac{\zeta}{\h_+}|^2}
    .
  \end{equation}
Since $P_{\ell_+}^q(\zeta) = O_{q}(e^{c_1|\frac{\zeta}{\ell_+}|^2})$ and $\h_+ \ge \ell_+$,
\begin{equation}
  \label{e:Exphizeta-2}
  \Ex_+(P_{\ell_+}^q(\zeta) e^{-c_1 |\tfrac{\varphi+\zeta}{\h_{+}} |^{4}})
  \le
  e^{\frac{c_1}{4}-\frac{c_1}{2} |\tfrac{\varphi}{h_+}|^2}\,
  O_{q}(\Ex_+ e^{2c_1|\tfrac{\zeta}{\ell_+}|^2}) .
\end{equation}
To complete the proof it suffices to show that
$\Ex_+ e^{2c_1|\frac{\zeta}{\ell_+}|^2} \le 2^{n/2}$.
As in \refeq{cfrak-def}, we write $\mathfrak{c}_+^2 = C_{+;x,x}$.
The coefficient $\frac{2c_1}{\ell_+^2}$ of $|\zeta|^{2}$ is bounded by
\begin{equation}
    \label{e:Exphizeta-3}
    \frac{2c_1}{\ell_+^2} = \frac{2}{8\mathfrak{c}_+^2} \frac{8c_1 \mathfrak{c}_+^2}{\ell_+^2}
    \le
    \frac{2}{8\mathfrak{c}_+^2}
    \frac{8c_1 L^{2}}{\ell_0^2}
    \le
    \frac{2}{8\mathfrak{c}_+^2}
    ,
\end{equation}
where we used \eqref{e:Cellbd} followed by $\ell_0^2 = L^6$ from
\eqref{e:ell0def} and the $c_{1}$ hypothesis.
Denote by $X$ the first component of
$\mathfrak{c}_+^{-1} \zeta \in \R^n$.  Then $X$ is a standard normal
variable, so
\begin{equation} \label{e:Expnormal}
  \Ex_+
  (
  e^{\tfrac{2}{8} X^{2}}
  )
  =
  \int_{-\infty}^\infty \frac{1}{\sqrt{2\pi}} e^{- \tfrac{1}{2 }z^{2} }
  e^{\tfrac{2}{8}  z^{2} } \, dz =\sqrt{2}.
\end{equation}
The components of $\zeta$ are independent, and hence, by
\eqref{e:Exphizeta-3},
$\Ex_+ e^{2c_1 \ell_+^{-2}|\zeta|^2} \le \Ex (e^{\frac{2}{8} \mathfrak{c}_+^{-2} |\zeta|^2}) = 2^{n/2}$ as desired.
\end{proof}

\subsection{Bound on \texorpdfstring{$S_0$}{S0}}
\label{sec:S0pf}

We begin with preparation and explanation and then
prove Lemma~\ref{lem:S0}.  We recall from the definition of $S_0$ in
\eqref{e:S0def} that
\begin{align}
  S_{0}
  &=
  e^{u_+|B|}
    \left(
    \Ex_+\theta
    e^{-\Vhat (B)}
    - e^{-U_+ (B)}
    \right)
    .
\end{align}
By Lemma~\ref{lem:K0PT} applied with $\Vhat$ in place of $V$,
and using $U_+ = \Upt(\Vhat)$,
\begin{equation}
  \Ex e^{-\theta \Vhat(B)} =
  e^{-U_+(B)}
  \left(
    1 +
    \tfrac{1}{8} \left( \Var_+ \theta \Vhat(B)\right)^2
    + \Ex A_3(B)
  \right)
  ,
\end{equation}
where, with $\delta \Vhat = \theta \Vhat(B) - U_+(B)$,
\begin{equation}
  \label{e:RdV}
  A_3(B)
  = - \frac{1}{2!} (\delta \Vhat)^{3} \int_0^1 e^{-t\delta \Vhat} (1-t)^2 \, dt.
\end{equation}
Thus
\begin{equation}
\lbeq{S0A3}
  S_0 = e^{-U_+(B) +u_+|B|} \left(
    \tfrac{1}{8} \left( \Var_+ \theta \Vhat(B)\right)^2
    + \Ex A_3(B)
  \right)
  .
\end{equation}
Note that there is a cancellation in the exponent on the right-hand side, namely
\begin{equation}
  U_+ -  u_{+}
  = g_+ \tau^2 + \nu_+ \tau.
\end{equation}

In the proof of Lemma~\ref{lem:S0}, the ratio $\mathfrak{c}_+/\h_+$
occurs when estimating expectations in which some fields have been
replaced by their typical values under the fluctuation-field
expectation, which is $\mathfrak{c}_+$, rather than giving them size
$\h_+$ through the norm.  Recall the definitions
$\ell_j= \ell_0 L^{-j}$ and $h_j=k_0\ggen_j^{-1/4} L^{-j}$
from \refeq{ell-def} and \refeq{h-def}, and recall the inequality
$\mathfrak{c}_+ \le \vartheta_j L^{-j}$ from \refeq{Cellbd}.  These
imply that
\begin{equation}
\lbeq{elloverh}
    \frac{\mathfrak{c}_{+}}{\h_+}
    \le
    \begin{cases}
    \vartheta & (\h_+=\ell_+)
    \\
    \vartheta L k_0^{-1} \ggen_+^{1/4} & (\h_+=h_+).
    \end{cases}
\end{equation}

This ratio sometimes occurs multiplied by $\|V(B)\|_{T_{0,\aux}(\h_+)}$;
this factor is the size of the sum of $V$ over a block $B$ if the
field has size $\h_+$.  We define
\begin{equation}
\label{e:epdVdef}
    \epdV = \epdV(\h) =
    \begin{cases}
    \tilde \vartheta \ggen & (\h=\ell)
    \\
    \tilde \vartheta \ggen^{1/4} & (\h=h).
    \end{cases}
\end{equation}
By Lemma~\ref{lem:moreQbd}, if $V \in \DV$ and
$\h_{+} \in \{\ell_{+},h_{+} \}$, then
\begin{align}
  \label{new-e:epdV}
  \frac{\mathfrak{c}_{+}}{\h_+} \|V^*(B)\|_{T_{0,\aux}(\h_+,\hAux)}
  &\le O_L(\epdV)
    \\
  \label{e:epdVhat}
  \frac{\mathfrak{c}_{+}}{\h_+} \|\Vhat(B)\|_{T_{0,\aux}(\h_+,\hAux)}
  &\le O_L(\epdV),
\end{align}
where we also assume $\|K(b))\|_{T_{0}(\ell)} \le \ggen$ for
\eqref{e:epdVhat}.

To prove Lemma~\ref{lem:S0} it suffices to show that,
for $\h_+ \in \{\ell_{+},h_{+}\}$,
\begin{align}
\label{e:S0bd-new}
    \|S_{0}\|_{T_{\varphi,\aux} (\h_+,\hAux)}
    &\le
    O_L(\epdV_+^3(\h))
    P_{\h_+}^{12}(\varphi) e^{-c_L |\varphi/h_+|^2}
    .
\end{align}
The desired inequality \refeq{S0bd} then follows immediately from
\refeq{S0bd-new} and the definition of the $\Wcal_\aux(\lambda)$-norm in
\refeq{Wcalgrand}.

Note that $h_+$ appears in the right-hand side of \refeq{S0bd-new},
regardless of the choice of $\h_+ \le h_+$.  The inequality
\refeq{S0bd-new} reveals why we need $\h_+=h_+$, and why it is not
enough to use $\h_+=\ell_+$.  Indeed, uniformly in $\ggen>0$ small,
the supremum over $\varphi$ of $P_{\h_+}^{p}(\varphi) e^{-c
|\varphi/h_+|^{2}}$ is bounded if $\h_+=h_+$, but diverges as $\ggen
\rightarrow 0$ if $\h_+=\ell_+$.  We control this large field problem
with our choice $\h_+=h_+$.  For $\h_+=\ell_+$ we still use the fact that
trivially $P_{\h_+}^{p}(\varphi) e^{-c |\varphi/h_+|^{2}} = 1$ when
$\varphi=0$.

The following proof relies heavily on our specific choice of the
polynomial $\Upt$.

\begin{proof}[Proof of Lemma~\ref{lem:S0}] As noted above, it suffices
to prove \refeq{S0bd-new}.  For this, by \refeq{S0A3} and
Lemma~\ref{lem:TayTphi}, it is enough to prove that there are
constants $c_L, C_L$ such that, for $\ggen$ sufficiently small, $L$
sufficiently large, $V \in \DV$, $B\in\Bcal_+$, and $\h_+ \in
\{\ell_{+},h_{+}\}$,
\begin{align}
  \label{e:EFTaynorma}
  \|e^{-(U_+-u_+)(B)}\|_{T_{\varphi,\aux}(\h_{+},\hAux)}
   & \le
  2e^{-c |\varphi/h_+|^4}  ,
  \\
    \label{e:EFTaynormb}
  \|\Var_{+}\big(\theta \Vhat(B)\big)\|_{T_{\varphi,\aux}(\h_{+},\hAux)}
  & \le
  C_L
    (\tfrac{\mathfrak{c}_+}{\h_{+}})^{2}\epdV^2 P_{\h_+}^{4}(\varphi)   ,
  \\
\label{e:ERFnorm}
  \|e^{-(U_+-u_+)(B)}
  \Ex_+  A_3\|_{T_{\varphi,\aux}(\h_{+},\hAux)}
  & \le
  C_L
    \epdV^3 P_{\h_+}^{12}(\varphi) e^{-c_{L}|\varphi/h_+|^2} 
    .
\end{align}
The following proof of these estimates shows that they also hold with
$V$ instead of $\Vhat$ and $U_\pt$ instead of $U_+$; in fact, this
replacement simplifies the proof.

The inequality \eqref{e:EFTaynorma} is proved in
Lemma~\ref{lem:Uplusstab}.  For \eqref{e:EFTaynormb},
Lemma~\ref{lem:covariance-6} gives
\begin{align}
    \| \Var_{+} \big(\theta \Vhat (B) \big)\|_{T_{\varphi,\aux}(\h_{+},\hAux)}
    &\le
    c
    (\tfrac{\mathfrak{c}_+}{\h_{+}})^{4}
    |B|^{2}
    \|\Vhat\|_{T_{0,\aux}(\h_{+},\hAux)}^{2}
    P_{\h_{+}}^{4}(\varphi)
    ,
    \label{e:covarianceDV}
\end{align}
where we recall that $\Var_{+} \big(\theta \Vhat (B) \big)$ is a
degree $4$ monomial since $j<N$ and so $\LT$ in
\eqref{e:covariance1-6} is the identity.
Now \refeq{EFTaynormb} follows from \eqref{e:epdVhat}.

It remains to prove \eqref{e:ERFnorm}.  We write $\delta \Vhat$ in
place of $\delta \Vhat (B)$.  Starting with the definition
\eqref{e:RdV} of $A_{3}$, we use $\int_0^1 (1-t)^2dt = \frac 13$,
followed by \refeq{ExCthetaTphi2}, to obtain
\begin{align}
  &\|e^{-(U_+-u_+)(B)}\Ex_{+}A_3\|_{T_{\varphi,\aux}(\h_{+},\hAux)}
  \nnb
  &\le
    \frac{1}{3!} \sup_{t \in [0,1]}
    \|\Ex_+ \left(\delta \Vhat^3 e^{-(U_+-u_+)(B)-t\delta \Vhat} \right) \|_{T_{\varphi,\aux}(\h_{+},\hAux)}
    \nnb
    &\le
    \frac{1}{3!} \sup_{t \in [0,1]}
    \Ex_+ \left(\|(\delta \Vhat)_{\zeta}\|_{T_{\varphi,\aux}(\h_{+},\hAux)}^3
    \|(e^{-(U_+-u_+)(B)-t\delta \Vhat})_{\zeta }\|_{T_{\varphi,\aux}(\h_{+},\hAux)} \right) .
    \label{e:ERFnorm1}
\end{align}
Since $\|\Vhat (B)\|_{T_{0,\aux}(\h_{+},\hAux)} \leq 1$ by
Lemma~\ref{lem:moreQbd},
we can apply Proposition~\ref{prop:dVnorm-6} (with $m=3$).
In the bound of  Lemma~\ref{lem:moreQbd}, the normalised sum over $x$
can be replaced by a normalised sum over $b$ since $\zeta$ is constant on
small blocks.
Together with \eqref{new-e:epdV}, we conclude
that there is a constant $c$ such that
\begin{equation}
    \| \delta \Vhat_{\zeta }\|_{T_{\varphi,\aux}(\h_{+},\hAux)}^{3}
    \le
    c \epdV^{3}
    P_{\h_+}^{12}(\varphi)
    \frac{1}{L^d}\sum_{b\in \Bcal(B)}
    P_{\ell_{+}}^{12}(\zeta_b)
    .
\label{e:ERFnorm2}
\end{equation}
By \refeq{stabphit},
\begin{align}
\label{e:stabphit-bis}
    \|(e^{-(U_+-u_+)(B) - t \delta \Vhat})_\zeta\|_{T_{\varphi,\aux}(\h_{+},\hAux)}
    &\le
    2 e^{- c^{\stab} 
      |\varphi/h_{+}|^{4}} + 2
    e^{- c^{\stab} 
    L^{-d}\sum_{b\in \Bcal (B)}|\tfrac{\varphi+\zeta_{b}}{h_{+}} |^{4}}
    \nnb & \le
    2 e^{- c^{\stab} 
      |\varphi/h_{+}|^{4}} + 2
    e^{- c^{\stab} 
    L^{-d}|\tfrac{\varphi+\zeta_{b}}{h_{+}} |^{4}}
    ,
\end{align}
where in the second line $b$ is an arbitrary block in $\Bcal(B)$.

By combining \eqref{e:ERFnorm1}--\eqref{e:stabphit-bis},
$
\|e^{-(U_+-u_+)(B)}\Ex_{+}A_3\|_{T_{\varphi,\aux}(\h_+,\hAux)}
$
is bounded by
\begin{equation}
    O \big(\epdV^{3}\big)
    P_{\h_+}^{12}(\varphi)
    \frac{1}{L^d}\sum_{b \in \Bcal(B)} 
    \left(
    e^{- c^{\stab}   |\varphi/h_{+}|^{4}}
    \Ex_+ \Big( P_{\ell_{+}}^{12}(\zeta_{b})\Big)
    +
    \Ex_+
    \Big(
    P_{\ell_{+}}^{12}(\zeta_{b})
    e^{- c^{\stab}  L^{-d}|\tfrac{\varphi+\zeta_{b}}{h_{+}} |^{4}}
    \Big)
    \right) .
\end{equation}
By \eqref{e:Exphizeta} with $c_{1}=0$ the first expectation is at most
$c_{2}$; by \eqref{e:Exphizeta} with $c_1 = c^{\stab} L^{-d}$ we also
bound the second expectation. After these bounds there is no longer
any $b$ dependence and the normalised sum drops out. Therefore
\begin{equation}
    \|e^{-(U_+-u_+)(B)}\Ex_{+}A_3\|_{T_{\varphi,\aux}(\h_+,\hAux)}
    \le
    O\big(\epdV^{3}\big)
    P_{\h_+}^{12}(\varphi)
    \left(
    e^{- c^{\stab}
    |\varphi/h_{+}|^{4}}
    +
    e^{-\frac{c^{\stab}
    L^{-d}}{2}|\varphi/h_+|^2}
    \right) .
\end{equation}
This implies \eqref{e:ERFnorm} and completes the proof.
\end{proof}

\section{Bound on \texorpdfstring{$S_1$}{S1}: proof of Lemma~\ref{lem:S1}}
\label{sec:S1}

In this section, we prove Lemma~\ref{lem:S1}.  Recall from
\eqref{e:S1def} that
\begin{align}
    \label{e:S1def-bis}
    S_{1}
    &=
    e^{u_+|B|}
    \sumtwo{X \subset \Bcal (B)}{|X|\ge 1}
    \Ex_+\theta
    \left(e^{-\Vhat (B\setminus X)} \Khat^{X}\right) .
\end{align}
Let $(V,K) \in \domRG$, and recall the hypotheses
that $\hAuxg_V\le \ggen$ and $\hAux_K \le \ggen^{9/4}$.  We define
$\hAuxg_K(\h) \le 1$ by
\begin{equation}
\label{e:hAuxKh}
    \hAuxg_K(\h) =
    \begin{cases}
    \hAuxg_K & (\h=\ell)
    \\
    \hAuxg_K \ggen^{-9/4} & (\h=h).
    \end{cases}
\end{equation}

Recall from \refeq{Wcalgrand} that the extended $\Wcal$-norm is defined to be
$\|K\|_{\Wcal_\aux(\hAux)} =  \|K(b) \|_{T_{0,\aux}(\ell,\hAux)}
+ \ggen^{9/4}\|K(b) \|_{T_{\infty,\aux}(h,\hAux)}$,
where $\|K(b)\|_{T_{\infty}(h)}=\sup_\varphi \|K(b)\|_{T_{\varphi}(h)}$.
To prove Lemma~\ref{lem:S1}, it suffices to prove that, for
$(\varphi,\h_+)$ 
in $T_{\varphi,\aux} (\h_+,\hAux)$
equal to either $(\infty,h_+)$ or $(0,\ell_+)$,
\begin{align}
  \label{e:S1bd-pf}
  \|S_{1}\|_{T_{\varphi,\aux} (\h_+,\hAux)}
  &\le
  O_L (\epdV_+^3(\h) + \hAuxg_{K}(\h))
  ,
\end{align}
since these two bounds then combine to give the desired
estimate $\|S_{1}\|_{\Wcal_{\aux,+}(\hAux)} \le O_L (\vartheta_+^3 \ggen_+^3+ \hAuxg_{K}
)$.  The inequality \refeq{S1bd-pf} is an immediate consequence of
\refeq{S1def-bis}, the triangle inequality, and the following lemma
because there are at most $2^{L^{d}}$ terms in the sum in
\eqref{e:S1def-bis}.

\begin{lemma}
\label{lem:X}
For $(\varphi,\h_+)$ equal to either $(\infty,h_+)$ or $(0,\ell_+)$,
\begin{align}
    \|e^{u_+|B|} \Ex_+\theta
    (e^{-\Vhat (B\setminus X)} \Khat^{X})\|_{T_{\varphi,\aux} (\h_+,\hAux)}
    & \le
    O_L (\epdV_+^3(\h) + \hAuxg_{K}(\h))^{|X|}.
\end{align}
\end{lemma}

To prove Lemma~\ref{lem:X}, we first develop general estimates
relating the norm of an expectation to the expectation of the norm, as
well as estimates on $\Khat$.  Note that it follows exactly as in the
proof of \refeq{norm-comparison} that, for $F \in \Fcal$,
\begin{align}
\lbeq{grand-comparison}
    \|F\|_{T_{\varphi,\aux}(\ell,\hAux)}
    & \le P_\ell^{10}(\varphi)\|F\|_{\Wcal_\aux(\hAux)}.
\end{align}

\begin{lemma}
\label{lem:grand-comparison}
For a family $F (b) \in \Ncal(b)$ where $b\in \Bcal(B)$,
\begin{align}
  \label{e:ExWcal-T0}
  \|\Ex_+\theta F^B\|_{T_{0,\aux}(\ell_+,\hAux)}
  &
    \leq O_L(1) \prod_{b\in \Bcal(B)}\|F(b)\|_{\Wcal_{\aux,+}(\hAux)},
  \\
  \label{e:ExWcal-Tinfty}
  \|\Ex_+\theta F^B\|_{T_{\infty,\aux}(h_+,\hAux)}
  &\leq
    \prod_{b\in \Bcal(B)}\|F(b)\|_{T_{\infty,\aux}(h_+,\hAux)}.
\end{align}
\end{lemma}

\begin{proof}
By \refeq{ExCthetaTphi3},
\begin{equation} \label{e:ExCthetaTphi3-bis}
  \|\Ex_+ \theta F^{B}\|_{T_{\varphi,\aux}(\h_+,\hAux)}
  \leq
  \Ex_+ \left(\prod_{b\in\Bcal(B)} \|F(b)\|_{T_{\varphi+\zeta_b,\aux}(\h_+,\hAux)} \right).
\end{equation}
This immediately implies \refeq{ExWcal-Tinfty}.  With
\refeq{grand-comparison}, it also gives
\begin{equation}
    \|\Ex_+\theta F^B\|_{T_{0,\aux}(\ell_+,\hAux)}
    \le
    O(1) \left( \prod_{b\in \Bcal(B)}\|F(b)\|_{\Wcal_{\aux,+}(\hAux)} \right)
    \Ex_+ \left( \prod_{b\in \Bcal(B)}P_{\ell_+}^{10}(\zeta_b) \right).
\end{equation}
By H\"older's inequality (with any fixed $b$ on the right-hand side),
\begin{align}
    \Ex_+ \left( \prod_{b\in \Bcal(B)}P_{\ell_+}^{10}(\zeta_b) \right)
    & \le
    \Ex_+ P_{\ell_+}^{10L^d}(\zeta_b).
\end{align}
The expectation on the right-hand side is bounded by an $L$-dependent
constant by Lemma~\ref{lem:G}.  This completes the proof.
\end{proof}

\begin{lemma} \label{lem:KA-bound}
For $\hAux_{K} \ge 0$,
\begin{align}
\lbeq{Khath-K}
  \|\Khat (b) \|_{T_{\infty,\aux} (h,\hAux)}
  &\le
    O_{L}\big( \|K^*(b)\|_{T_{\infty,\aux} (h,\hAux)}\big)  ,
  \\
\lbeq{Khatell-K}
    \|\Khat (b)\|_{\Wcal_\aux(\hAux)}
    &\le
    O_{L}\big(\|K^* (b)\|_{\Wcal_{\aux,+}(\hAux)}\big)
  .
\end{align}
In particular, for $(V,K) \in \domRG$,
\begin{align}
\label{e:Khath}
  \|\Khat (b) \|_{T_{\infty,\aux} (h,\hAux)}
  &\le
    O_L\big( \vartheta^{3} \ggen^{3}  + \hAuxg_K  \big) \ggen^{-9/4} ,
  \\
\label{e:Khatell}
    \|\Khat (b)\|_{\Wcal_\aux(\hAux)}
    &\le
    O_L\big( \vartheta^{3}\ggen^3 + \hAuxg_{K}\big)
  .
\end{align}
\end{lemma}

\begin{proof}
We drop the block $b$ from the notation.  By the definition of $\Khat$
in \refeq{hatdef},
\begin{equation}
  \Khat
  =
  K+ e^{-V} - e^{-V+Q}
  =
  K - \int_0^1 Q e^{-V+sQ} \, ds.
\end{equation}
This implies that
\begin{equation}
  \|\Khat\|_{T_{\varphi,\aux}(\h,\hAuxg)}
  \leq
  \|K^*\|_{T_{\varphi,\aux}(\h,\hAuxg)}
  +
  \|Q\|_{T_{\varphi,\aux}(\h,\hAuxg)}
  \sup_{s\in[0,1]} \|e^{-V^*+sQ}\|_{T_{\varphi,\aux}(\h,\hAuxg)}.
\end{equation}
By \refeq{V-norm}, \refeq{U*norm1}, \eqref{e:Vstarhbd}
and \eqref{e:QbdT0aux-P},
\begin{align}
\lbeq{Qagain}
  \|Q\|_{T_{\varphi,\aux}(\h,\hAuxg)}
  \le
  2 P_{\h}^4(\varphi) \|K^*\|_{T_{0,\aux}(\h,\hAuxg)}.
\end{align}
We apply Lemma~\ref{lem:V-Qbd} to bound the exponential, after making
use of the comment above Lemma~\ref{lem:eQbd} which permits it to be applied
on a small block by choosing $t=L^{-d}$.
Consequently, the product
$P_{\h}^4(\varphi)\|e^{-V+sQ}\|_{T_{\varphi,\aux}(\h,\hAuxg)}$ is
bounded by $O(1)$ if $\varphi=0$ and $\h=\ell$, and uniformly in
$\varphi$ by $O (t^{-1}) = O (L^{d})$ if $\h=h$.  Therefore,
\begin{align}
    \|\Khat\|_{T_{\infty,\aux}(h,\hAuxg)} & \le O(L^{d})\|K^*\|_{T_{\infty,\aux}(h,\hAuxg)}
    , \quad
    \|\Khat\|_{T_{0,\aux}(\ell,\hAuxg)} \le O(1)\|K^*\|_{T_{0,\aux}(\ell,\hAuxg)}
    .
\end{align}
This proves \refeq{Khath-K} and \refeq{Khatell-K}.

By \refeq{K*norm-ell}, \refeq{Kstar-h-9} and the definition
\eqref{e:Phi_+-domain} of $\domRG$,
\begin{align}
    \label{e:KA-bound}
    \|K^* \|_{T_{0,\aux} (\ell,\hAux)}
    &\le
    \CRG \vartheta^{3} \ggen^3 + \hAuxg_K,
    \\
    \label{e:KA-bound-2}
    \|K^* \|_{T_{\infty,\aux} (h,\hAux)}
    &\le
    \CRG \vartheta^{3} \ggen^{3/4} + \hAuxg_K \ggen^{-9/4}
    .
\end{align}
This implies \eqref{e:Khath} and \eqref{e:Khatell} and completes the
proof.
\end{proof}

\begin{proof}[Proof of Lemma~\ref{lem:X}]
Let $J=\Ex_+\theta (e^{-\Vhat (B\setminus X)}
\Khat^{X})$.  By the product property followed by
Lemma~\ref{lem:Uplusstab} with $t=1$,
\begin{align}
    \|e^{u_+|B|} J\|_{T_{\varphi,\aux} (\h_+,\hAux)}
    \le
    \|e^{u_+|B|}\|_{T_{\varphi,\aux} (\h_+,\hAux)}
    \|J\|_{T_{\varphi,\aux} (\h_+,\hAux)}
    \le
    2^{1/2}
    \|J\|_{T_{\varphi,\aux} (\h_+,\hAux)} .
\end{align}
Therefore we are reduced to proving that $\|J \|_{T_{\varphi,\aux}
(\h_+,\hAux)} = O_L (\epdV_+^3(\h) + \hAuxg_{K}(\h))^{|X|}$ for
the two cases
$(\varphi,\h_+)=(\infty,h_+)$ and $(\varphi,\h_+)=(0,\ell_+)$.
We write $J = \Ex_+\theta F^B$ with $F$ defined by
\begin{equation}
    F(b) =
    \begin{cases}
    \Khat (b) & (b \in X) ,
    \\
    e^{-\Vhat (b)}  & (b \in B\setminus X) .
    \end{cases}
\end{equation}

Suppose first that
$(\varphi,\h_+)=(\infty,h_+)$.  By \eqref{e:ExWcal-Tinfty},
\begin{equation}
    \label{e:ExWcal-Tinfty2}
    \|J\|_{T_{\infty,\aux}(h_+,\hAux)}
    \le
    \prod_{b \in \Bcal (B)} \|F (b)\|_{T_{\infty,\aux}(h_+,\hAux)}.
\end{equation}
For $b \in \Bcal (X)$ we bound $F (b)$ using \eqref{e:Khath}; for $b \in
\Bcal (B\setminus X)$ we bound $F (b)$ using Lemma~\ref{lem:V-Qbd} with
$t=L^{-d}$ and $s=1$. The result is
\begin{equation}
    \|F(b)\|_{T_{\infty,\aux}(h_+,\hAux)}
    \le
    \begin{cases}
    O_L\big( \vartheta^{3} \ggen^{3}  + \hAuxg_K  \big) \ggen^{-9/4} & (b \in X)
    \\
    2^{L^{-d}/4} & (b \in B\setminus X) .
    \end{cases}
\end{equation}
When $\big( \vartheta^{3} \ggen^{3} + \hAuxg_K \big) \ggen^{-9/4}$ is
rewritten in terms of $\hAuxg_{K}(\h)$ defined in \eqref{e:hAuxKh}
and $\epdV$ defined in \eqref{e:epdVdef}, it becomes $\epdV_+^3(\h) +
\hAuxg_{K}(\h)$. Therefore, from \eqref{e:ExWcal-Tinfty2}, we have
$\|J\|_{T_{\infty,\aux}(h_+,\hAux)} \le O_L (\epdV_+^3(\h) +
\hAuxg_{K}(\h))^{|X|}$ as desired.

Suppose now that $(\varphi,\h_+)=(0,\ell_+)$.  By \eqref{e:ExWcal-T0},
\begin{equation}
    \|J\|_{T_{0,\aux}(\ell_+,\hAux)}
    \leq
    O_L(1) \prod_{b\in \Bcal(B)}\|F(b)\|_{\Wcal_{\aux,+}(\hAux)}.
\end{equation}
For $b \in \Bcal (X)$ we bound $F (b)$ using \eqref{e:Khatell}; for $b
\in \Bcal (B\setminus X)$ we bound $F (b)=e^{-\Vhat (b)}$ using
Lemma~\ref{lem:V-Qbd} with $t=L^{-d}$ and $s=1$. In more detail,
Lemma~\ref{lem:V-Qbd} bounds the $T_{\infty,\aux}(h_+,\hAux)$ norm of
$e^{-\Vhat (b)}$ and this is one of the two terms in the definition
\eqref{e:Wcalgrand} of the $\Wcal_{\aux,+}(\hAux_{+})$-norm. However, the
$T_{\varphi,\aux}(h_+,\hAux)$-seminorm becomes the
$T_{0,\aux}(h_+,\hAux)$-seminorm
by setting $\varphi = 0$, and the $T_{0,\aux}(h_+,\hAux)$-seminorm is
larger than the $T_{0,\aux}(\ell_+,\hAux)$-seminorm because $h_+ \ge
\ell_{+}$. Therefore Lemma~\ref{lem:V-Qbd} also bounds the other term
in the $\Wcal_{\aux,+}(\hAux)$-norm. Thus we have
\begin{align}
    \|F(b)\|_{\Wcal_{\aux,+}(\hAux)}
    & \le
    \begin{cases}
    O_L\big( \vartheta^{3} \ggen^{3}  + \hAuxg_K  \big) \ggen^{-9/4} & (b \in X)
    \\
    2^{L^{-d}/4} (1+\ggen^{9/4}) & (b \in B\setminus X) .
    \end{cases}
\end{align}
By \eqref{e:ExWcal-T0}, this implies $\|J\|_{\Wcal_{\aux,+}(\hAux)} \le O_L
(\epdV_+^3(\h) + \hAuxg_{K}(\h))^{|X|}$ as desired.
\end{proof}

\section{Crucial contraction}
\label{sec:rgest-contraction}

Throughout this section, we work with the $T_\varphi(\h)$-seminorm.
In fact, the analysis presented here also applies for the
$T_{\varphi,\aux}(\h,\hAuxg)$-seminorm, but we do not require the more
detailed information that it encodes.

The crucial contraction is the $(p,q)=(0,1)$ case of \refeq{DVKbd-2},
which asserts that if $(V,K) \in \domRG$ then $\|D_K\Phi_+^K\|\le
\kappa$ with $\kappa=O(L^{-2})$.  This estimate is the key fact used
to prove that $K$ does not grow from one scale to the next as long as
$(V,K)$ remains in the renormalisation group domain $\domRG$.  It
relies heavily on our specific choice in \refeq{U+def} of the
polynomial $U_+ = \Phi_\pt(V - \Loc (e^V K))$ as part of the
definition of the renormalisation group map.  This choice transfers
the growing contributions from $K$ into $V$ where they are dominated
by terms that are quadratic in the coupling constants.

\index{Crucial contraction}
\index{Contraction}
\begin{prop} \label{prop:crucial-0}
Let $L$ be sufficiently large,
and let $\ggen$ be sufficiently small depending on $L$.
For $(V,K) \in \domRG$,
the Fr\'echet derivative of $\Phi_+^K$ as a map from $\Wcal\to \Wcal_+$ at $K=0$
obeys
\begin{equation} \label{e:crucial}
  \|D_K\Phi_+^K(V,0)\|_{\Wcal\to \Wcal_+} \leq \kappa
\end{equation}
with $\kappa = O(L^{-2}) < 1$.
\end{prop}

As in \eqref{e:K+setup'}, we write $K_+ = \Phi_+(V,K)$ and
\begin{equation}
  K_{+} (B)
  =
  S_{0} + S_{1} .
  \label{e:K+setup'-bis}
\end{equation}
The next lemma shows that the $K$-derivative of $S_0$ is negligible.

\begin{lemma}
Under the hypotheses of Proposition~\ref{prop:crucial-0},
\begin{align}
    \|D_K S_0\|_{\Vcal \times \Wcal \to \Wcal_+}
    &\le
    O(\vartheta_+^{3} \ggen_+^{2}) \le O(L^{-2})
    .
\lbeq{DKS0a}
\end{align}
\end{lemma}

\begin{proof}
We apply Lemma~\ref{lem:S0} with $\hAux_V=\hAux_K=\ggen$, and obtain
\begin{equation}
    \|S_0\|_{\Wcal_{\aux,+}(\hAux)} \le O_L(\vartheta_+^3 \ggen_+^3).
\end{equation}
The desired result then follows immediately from
Lemma~\ref{lem:normderiv}.
\end{proof}

Thus the main work in proving the crucial contraction rests with
estimation of the $K$-derivative of $S_1$.  By the definition of $S_1$
in \eqref{e:S1def},
\begin{align}
  S_1
  & =
  e^{u_+|B|}
  \sum_{b \in B}
  \Ex_+\theta
  \big(e^{-\Vhat (B\setminus b)} \Khat(b)\big)
  +
  e^{u_+|B|} \!\!\!
  \sumtwo{X \subset \Bcal (B)}{|X|\ge 2}  \!\!\!
  \Ex_+\theta
  \big(e^{-\Vhat (B\setminus X)} \Khat^{X}\big)
  .
\end{align}
For the first term, we write $ Q(b)= \LT (e^{V(b)}K(b))$ as in
\refeq{Q-def}, and use the definition of $\Khat$ in \eqref{e:hatdef}
to obtain
\begin{align}
    \Khat
    &=
    e^{-V}\left(1 - e^{Q} + e^{V} K \right)
    =
    e^{-V}(1-\Loc) ( e^V K) + A,
    \label{e:Khat2}
\end{align}
with
\begin{equation}
  A(b)
  = e^{-V(b)}(1+Q(b)-e^{Q(b)})
  .
\end{equation}
This gives
\begin{align}
  \sum_{b \in B}
  \Ex_+\theta
  \big(e^{-\Vhat (B\setminus b)} \Khat(b)\big)
  & =
  \Ex_+\theta TK(b) +
  \sum_{b\in B}
  \Ex_+\theta
  \big(e^{-\Vhat (B\setminus b)} A(b)\big)
  ,
\end{align}
with
\begin{equation} \label{e:Tdef}
  TK
  =
  \sum_{b \in \Bcal (B)}
  \left(e^{-V (B)} (1-\Loc) ( e^{V(b)} K(b))\right).
\end{equation}
We write $e^{u_+|B|} = e^{u_\pt|B|} + \delta$ with $\delta =
e^{u_+|B|} - e^{u_\pt|B|}$.  Then the above leads to
\begin{align}
    S_1 & = (e^{u_\pt|B|}+\delta) \Ex_+\theta TK
    + e^{u_+|B|}\sum_{b\in B}
  \Ex_+\theta
  \big(e^{-\Vhat (B\setminus b)} A(b)\big)
  \nnb & \quad +
  e^{u_+|B|} \!\!\!
  \sumtwo{X \subset \Bcal (B)}{|X|\ge 2}  \!\!\!
  \Ex_+\theta
  \big(e^{-\Vhat (B\setminus X)} \Khat^{X}\big).
\lbeq{S1alg}
\end{align}
We will show in the proof of Proposition~\ref{prop:crucial-0} that the
linear term $e^{u_\pt|B|} \Ex_+\theta TK$ on the right-hand side is
the Fr\'echet derivative of $S_1$, and that the other terms are error
terms.

Before doing so, in Lemma~\ref{lem:Tcontract} we obtain an estimate
for the norm of the linear operator $T$.  In $TK(b)$ there are a
dangerous number $|\Bcal(B)|= L^4$ of terms in the sum over $b$. Thus,
naively, the operator norm of $T$ is not obviously small.  On the
other hand, the operator $1-\LT$ has an important contractive
property.  According to Definition~\ref{def:Loc-hier}, $\Loc =
\Tay_4$.  The contractive property of $1-\LT$ is given by
Lemma~\ref{lem:Taycontraction}, which asserts that if $F:\R^n
\rightarrow \R$ is $O(n)$-invariant and if $\h_{+} \le \h$, then
\begin{equation} \label{e:contraction}
  \|(1-\LT)F\|_{T_{\varphi} (\h_{+})}
  \le
  2\left(\frac{\h_{+}}{\h} \right)^{6}
  P_{\h_{+}}^{6}(\varphi)
  \sup_{0 \le t \le 1}
  \|F\|_{T_{t\varphi}(\h)} .
\end{equation}
The hypothesis that $F$ is $O(n)$-invariant has been used here to
replace $\LT=\Tay_4$ by $\Tay_5$, which is possible since $F^{(5)} (0)
= 0$.  (Since we have made the choice $p_\Ncal=\infty$, the hypothesis
concerning $p_\Ncal$ in Lemma~\ref{lem:Taycontraction} is certainly
satisfied.)  We use \refeq{contraction} in Lemma~\ref{lem:Tcontract}
to obtain a factor $(\h_{+}/\h)^{6} = O(L^{-6})$, which more than
compensates for the entropic factor $L^4$, resulting in an estimate of
order $L^{-2}$ for the norm of $T$.

\begin{lemma} \label{lem:Tcontract}
Let $L$ be sufficiently large, and let $\ggen$ be sufficiently small
depending on $L$.  For $V \in \DV$ and $\Kdot \in \Fcal$,
\begin{equation}
  \|T\Kdot\|_{\Wcal_+} \leq O(L^{-2})\|\Kdot\|_{\Wcal}.
\end{equation}
\end{lemma}

\begin{proof}
It suffices to prove that
\begin{align}
  \|T\Kdot\|_{T_\infty(h_+)} &\leq O(L^{-2})\|\Kdot\|_{T_\infty(h)}
  ,
  \\
  \|T\Kdot\|_{T_{0}(\ell_+)} &\leq O(L^{-2})\|\Kdot\|_{T_{0}(\ell)}.
\end{align}

By the definition of $\LT$, and by the $O(n)$ symmetry of $V$ and $K$,
the Taylor expansion of $(1-\Loc)(e^{V(b)}\Kdot(b))$ starts at order
$6$.  Therefore, the same is true for
$e^{-V(b)}(1-\Loc)(e^{V(b)}\Kdot(b))$. Thus $1-\Loc$ acts on it as the
identity, and
\begin{gather}
    e^{-V (B)} (1-\Loc) \big(e^{V(b)} \Kdot (b)\big)
    =
    e^{-V (B\setminus b)}
    e^{-V(b)} (1-\Loc) \big(e^{V(b)} \Kdot (b)\big)
    \nnb
    =
    e^{-V (B\setminus b)}
    (1-\Loc) e^{-V(b)} (1-\Loc) \big(e^{V(b)} \Kdot (b)\big)
    .
\lbeq{LocLoc}
\end{gather}
We insert this equality into the definition \eqref{e:Tdef} of $T\Kdot$
and write the result as $T\Kdot(b) = T_1\Kdot +T_2\Kdot$,
where
\begin{align}
    T_1\Kdot
    &=
    \sum_{b \in \Bcal (B)}
    e^{-V (B\setminus b)} (1-\Loc) \Kdot (b)
    ,
    \\
    T_2\Kdot
    &=
    -
    \sum_{b \in \Bcal (B)}
    e^{-V (B\setminus b)} (1-\Loc) \big( e^{-V(b)} \Qdot (b)\big) ,
\end{align}
with $\Qdot(b) = \Loc \big(e^{V(b)} \Kdot (b)\big)$.  The $T_1$ term
comes from the $1$ and the $T_2$ term from the $-\Loc$ in the inner
$1-\Loc$ on the right-hand side of \refeq{LocLoc}.

Since $\h_{+}/\h = O(L^{-1})$ for both $\h=\ell$ and $\h=h$, it
follows from \refeq{contraction} that
\begin{equation}
    \|T_1 \Kdot\|_{T_{\varphi} (\h_{+})}
    \le
    L^{\drb}
    \|e^{-V (B\setminus b)}\|_{T_{\varphi} (\h_{+})}
    O(L^{-6})
    P_{\h_{+}}^{6}(\varphi)
    \sup_{0 \le t \le 1}
    \|\Kdot (b)\|_{T_{t\varphi} (\h)}.
\end{equation}
For $\h_+=\ell_+$ and $\varphi=0$, this simplifies to
\begin{equation}
    \|T_1 \Kdot\|_{T_{0} (\ell_{+})}
    \le
    O (L^{-2})
    \|\Kdot (b)\|_{T_{0} (\ell)},
\end{equation}
since the norm of $e^{-V (B\setminus b)}$ is $O(1)$ by
Proposition~\ref{prop:stability} at the next scale (with
$t=1-|b|/|B|=1-L^{-d}$).  For $\h_+=h_+$, we again apply
Proposition~\ref{prop:stability} to conclude that
\begin{equation}
    \label{e:large-field-suppression}
    \|e^{-V (B\setminus b)}\|_{T_{\varphi}
    (h_{+})}P_{h_{+}}^6 
    (\varphi)
    =O(1)
\end{equation}
uniformly in $\varphi$.  Therefore,
\begin{equation}
    \|T_1\Kdot \|_{T_{\infty} (h_{+})}
    \le
    O ( L^{-2})
    \|\Kdot (b)\|_{T_{\infty} (h)}
    .
\end{equation}

For $T_2$ the method is the same once we have verified that $e^{-V(b)}
\Qdot (b)$ is bounded in $T_{\infty} (h)$ and in $T_{0}(\ell)$, which
we now do.  By \eqref{e:QbdT0aux} with $\hAux_V=\hAux_K=0$ and
\refeq{V-norm},
\begin{align}
\lbeq{Qhneed}
    \|e^{-V(b)} \Qdot(b)\|_{T_{\varphi} (\h)}
    &\le
    2
     \|e^{-V(b)}\|_{T_{\varphi} (\h)}
    P_{\h}^{4} (\varphi)
    \|\Kdot (b)\|_{T_{0} (\h)}.
\end{align}
By Proposition~\ref{prop:stability}, $\|e^{-V(b)}\|_{T_{\varphi} (\h)}
P_{\h}^{4} (\varphi)$ is bounded by a constant uniformly in $\varphi$
for $\h=h$, and when $\varphi=0$ for $\h=\ell$.  Now the bounds
\begin{align}
    \|T_2\Kdot \|_{T_{0} (\ell_{+})}
    &=
    O (L^{-2}) \|\Kdot\|_{T_{0}(\ell)},
    \\
    \|T_2\Kdot \|_{T_{\infty} (h_{+})}
    &=
    O (L^{-2}) \|\Kdot\|_{T_\infty(h)}
\end{align}
follow as in the analysis of $T_1$.  This completes the proof.
\end{proof}

\begin{proof}[Proof of Proposition~\ref{prop:crucial-0}]
We use the decomposition $\Phi_+^K = S_0 +S_1 $.
By \refeq{DKS0a}, the Fr\'echet derivative of $S_0$ obeys
\begin{equation}
  \|D_K S_0(V,0)\|_{T_{0}(\ell)\times \Wcal \to \Wcal_+}  = O(\ggen^2) =O(L^{-2}).
\end{equation}
Thus, it suffices to
identify $e^{u_\pt |B|} \Ex_+\theta T$ as the Fr\'echet derivative of $S_1$ and to
prove that it is bounded in norm by $O(L^{-2})$.

We begin with the bound.  By Lemma~\ref{lem:Uplusstab}, $e^{u_\pt|B|} \leq 2$.
By Lemma~\ref{lem:Tcontract},
together with \eqref{e:ExWcal-T0}--\eqref{e:ExWcal-Tinfty},
\begin{align}
\lbeq{ETell}
  \|\Ex_+\theta T\Kdot\|_{T_{0}(\ell_+)}
  &
  \leq O(1) \|T\Kdot\|_{\Wcal_+}
  \leq O(L^{-2}) \|\Kdot\|_{\Wcal},
  \\
\lbeq{ETh}
  \|\Ex_+\theta T\Kdot\|_{T_\infty(h_+)}
  &\leq \|T\Kdot\|_{T_{\infty}(h_+)}
  \leq O(L^{-2}) \|\Kdot\|_{T_\infty(h)}.
\end{align}
In particular, we have the desired bound
\begin{equation}
\lbeq{ETbd}
  \|E_+\theta T\Kdot\|_{\Wcal_+} \leq O(L^{-2}) \|\Kdot\|_{\Wcal}.
\end{equation}

It remains to identify $e^{u_\pt |B|} \Ex_+\theta T$ as the Fr\'echet derivative of $S_1$.
For this, it suffices to prove that, for $(V,K) \in \domRG$,
\begin{align}
  \label{e:DKS1-T0}
  \|S_1(V,K) - e^{u_\pt |B|} \Ex_+\theta TK\|_{T_{0}(\ell_+)}
  &= O_{L}(\|K\|_{\Wcal}^2),
  \\
  \label{e:DKS1-Tinfty}
  \|S_1(V,K) - e^{u_\pt |B|} \Ex_+\theta TK\|_{T_\infty(h_+)}
  &= O_{L}(\|K\|_{T_\infty(h)}\|K\|_{\Wcal})
  .
\end{align}
To prove \refeq{DKS1-T0}--\refeq{DKS1-Tinfty}, we will
show that the three terms on the right-hand side of the formula \refeq{S1alg}
for $S_1$ involving
$\delta$, $A$, and $|X|\ge 2$ are bounded by the right-hand sides of
\eqref{e:DKS1-T0}--\eqref{e:DKS1-Tinfty}.

The $\delta$ term is $\delta \Ex_+\theta TK $, with
\begin{align}
    \delta & =
      e^{-u_+|B|} - e^{-u_\pt |B|}
      =
      e^{-u_\pt |B|}O(|u_+ - u_\pt||B|)
      .
\end{align}
By Lemma~\ref{lem:Uplusstab}, the factor $e^{-u_\pt |B|}$ is bounded
by $2$, and the factor $|u_+ - u_\pt||B|$ is bounded by
$O_{L}(\|K\|_{T_{0}(\ell)})$, by Lemma~\ref{lem:norm-Rplus}.
With \refeq{ETell}--\refeq{ETh}, this shows that the $\delta$ term
obeys the required estimate.

The term involving $A$ is $e^{u_+|B|}\sum_{b\in B}\Ex_+\theta
(e^{-\Vhat (B\setminus b)} A(b))$, with $A(b)= e^{-V(b)}(1+Q(b)-e^{Q(b)})$.
By Taylor's formula,
\begin{equation}
  1+Q(b)-e^{Q(b)} = -\int_0^1 (1-s) Q(b)^2 e^{sQ(b)}\, ds
  .
\end{equation}
This gives
\begin{equation}
  \|A(b)\|_{T_\varphi(\h_+)}
  \leq
  \sup_{s\in[0,1]} \|Q(b)\|_{T_\varphi(\h_+)}^2 \|e^{-V(b)+sQ(b)}\|_{T_\varphi(\h_+)}.
\end{equation}
By \eqref{e:QbdT0aux-P} with $\hAux =0$ and \eqref{e:V-norm}
to bound $Q$, and
Lemma~\ref{lem:V-Qbd} with $t=L^{-d}$ to bound the exponential term,
\begin{equation}
  \|A\|_{T_\varphi(\h_+)}
  \leq 2 \| K(b)\|_{T_{0}(\h_+)}^2 P_{\h_+}^8(\varphi) e^{-ct|\varphi/h_+|^4}
  \leq O_{L}\left(\|K(b)\|_{T_{0}(\h_+)}^2\right)
\end{equation}
if $\h_+=h_+$ or $\varphi=0$.  Also, $\|e^{-\Vhat (B\setminus b)}\|_{T_{\varphi}(\h_+)}
\le O(1)$ by Lemma~\ref{lem:V-Qbd}, and $e^{u_+|B|} \le 2$ by Lemma~\ref{lem:Uplusstab}.
Finally, we apply
\eqref{e:ExWcal-T0}--\eqref{e:ExWcal-Tinfty}
to estimate the expectation.

The remaining term in \refeq{S1alg} is
\begin{equation}
    e^{u_+|B|} \!\!\!
  \sumtwo{X \subset \Bcal (B)}{|X|\ge 2}  \!\!\!
  \Ex_+\theta
  \big(e^{-\Vhat (B\setminus X)} \Khat^{X}\big).
\end{equation}
In the proof of Lemma~\ref{lem:X}, an estimate is given for
$T_{0}(\ell_+)$- and $T_\infty(h_+)$-seminorms of the terms in the
above sum.  These estimates show that the norm of the sum is dominated
by the terms with $|X|=2$, and these are respectively
$O(\|K\|_\Wcal^2)$ and $O(\|K\|_{T_\infty(h)}^2)$.  This completes the
proof.
\end{proof}

\section{Continuity in the mass}
\label{sec:masscont}

In this section, we prove the continuity assertions of
Theorems~\ref{thm:step-mr-K}--\ref{thm:step-mr-R}, which we restate as
the following proposition.  With $m^2$ fixed, the continuity in
$(V,K)$ follows from the differentiability in $(V,K)$, so our main
attention is on continuity in the mass parameter $m^2$.

In the proposition, the Fr\'echet derivatives $D_V^p D_K^q R^U_+$ and
$D_V^p D_K^q \Phi^K_+$ are multilinear maps defined on directions
$\dot{V}\in(T_{0}(\ell))^{p}$, $\dot{K}\in\Wcal^{q}$ and taking values
in $T_{0}(\ell_+)$ for $R_+^U$ and in $\Wcal_+$ for $\Phi^K_+$.

\begin{prop}
\label{prop:masscont}
  Let $\mgen^2 \ge 0$,
  let $\ggen$ be sufficiently small (depending on $L$), and let $p,q\in \N_0$.
  Let $0\le j<N$, and let $p,q\ge 0$.  For $\Phi^K_+$, we also assume that $L$ is
  sufficiently large.
  The maps $R^U_+:\domRG  \times \Iint_+  \to \Ucal_+$ and
  $\Phi^K_+:\domRG  \times \Iint_+  \to \Wcal_+$
  and their Fr\'echet derivatives $D_V^p
D_K^q R^U_+$  and $D_V^p
D_K^q \Phi^K_+$  are
jointly continuous in all arguments $V,K, \dot{V}, \dot{K}$, as well
as in $m^2 \in \Iint_+$.
\end{prop}

The proof of Proposition~\ref{prop:masscont} uses the following lemma.
We use the extended norm in the proof as it controls the Fr\'echet
derivatives as in Lemma~\ref{lem:normderiv}.

\begin{lemma} \label{lem:Excont}
    Let $\mgen^2 \geq 0$.
    Let $B \in \Bcal_+$ and suppose that $F: \domRG \times \Iint_+ \to \Ncal(B)$
    obeys $\|F\|_{T_{\varphi,\aux}(\h,\hAux)} \le c_F P^{k}_\h(\varphi)$ for some
    $c_F,k \ge 0$.
    There exists a function $\eta(m^2)$, with $\eta(m^2) \to 0$ as $m^2 \to \mgen^2$, such that
    \begin{equation}
      \|\Ex_{C_+(m^2)}\theta F-\Ex_{C_+(\mgen^2)}\theta F\|_{T_{\varphi,\aux}(\h,\hAux)}
      \le
      \eta(m^2) c_F  P^{ k}_\h(\varphi) .
    \end{equation}
\end{lemma}

\begin{proof}
Let $C=C_+(m^2)$ and $\tilde {C} = C_+(\mgen^2)$.  Note that $C$ and
$\tilde C$ differ only in the multiplicative factor $\gamma_j$ in
\refeq{Cjhier}. Since $j<N$ this factor $\gamma_{j}$ is a
continuous function of $m^{2}$, including at $m^{2}=0$. According to
the interpretation of Gaussian integration with respect to a positive
semi-definite matrix given in \refeq{Gm2}, there is a positive
definite matrix $C'$ and a subspace $Z$ of $\R^{n\Lambda}$ such that
\begin{equation}
    \Ex_C\theta F(\varphi) =  \int_{Z} F(\varphi+\zeta) p_{C'}(\zeta) d \zeta
\end{equation}
with
\begin{equation}
    p_{C'}(\zeta) =  \det(2\pi C')^{-1/2} e^{-\frac12 (\zeta, (C')^{-1} \zeta)}.
\end{equation}
Let $\tilde{C}'$ be the positive definite matrix that similarly
represents $\Ex_{\tilde C}$.

  Since $P_\h(\varphi+\zeta) \leq P_\h(\varphi)P_\h(\zeta)$,
  our assumption on $F$ implies that
  \begin{equation}
    \|\Ex_{C}\theta F - \Ex_{\tilde{C}}\theta F \|_{T_{\varphi,\aux}(\h,\hAux)}
    \leq
    c_F
    P_\h^{k}(\varphi)
     \int_{Z} |p_{C'}(\zeta)- p_{\tilde{C}'}(\zeta)|  P_\h^k(\zeta) \, d\zeta.
  \end{equation}
  We define $\eta(m^2)$ to be the integral in the above right-hand side.
  It goes to zero as $m^2 \to \mgen^2$ by dominated convergence, since
  $\gamma_j$ is continuous.
  This completes the proof.
\end{proof}

\begin{proof}[Proof of Proposition~\ref{prop:masscont}] We write
$K_+=\Phi_+^K(V,K)$ .  By
Theorems~\ref{thm:step-mr-K}--\ref{thm:step-mr-R}, $R_+$ and $K_+$
and derivatives are smooth in $(V,K)$, uniformly in $m^2 \in
\Iint_+$ and in unit directions $\dot V, \dot K$.  To show the desired
joint continuity in $(V,K,\dot V, \dot K, m^2)$, it therefore suffices
to show that $R_+$, $K_+$ and their derivatives are continuous in $m^2
\in \Iint_+$ uniformly in $(V,K) \in \domRG$.  To do so, we will show
that $R_+$ and $K_+$ are continuous in $m^2 \in \Iint_+$, uniformly in
$y=(V,K)$, where we use the $T_{0,y}(\ell_+,\hAux)$-norm for $R_+$ and
the $\Wcal_{\aux,+}(\hAux)$-norm for $K_+$.  We require that $\hAux$
satisfy \eqref{e:lamstab-V}--\eqref{e:lamstab-K}.  The continuity of
the derivatives then follows from Lemma~\ref{lem:normderiv}.  (Note
that although the inverse powers of $\hAux$ in the bounds of
Lemma~\ref{lem:normderiv} may appear dangerous, they do not create
trouble because we are merely proving continuity and make no claim on
the modulus of continuity.)

We begin with $R_+$.  By definition,
\begin{align}
\label{e:RUdef-2}
    R_+  & = \Phi_+^U(V,K) - \Phi_+^U(V,0)
    =
    \Phi_\pt(\Vhat) - \Phi_\pt(V),
\end{align}
where $\Vhat = V - \LT (e^VK)$, and, as in \refeq{Phiptdef},
\begin{equation}
\lbeq{Phiptmasscont}
    \Phi_\pt(V;B) = \Ex_{C_+}\theta V(B) - \frac 12 \Ex_{C_+} (\theta V(B);\theta V(B)).
\end{equation}
The $\LT$ in \refeq{Phiptdef} plays no role here since, with the
hypothesis $j<N$, we have the $c^{(1)}=0$ hypothesis of
Proposition~\ref{prop:Upt}, so $\LT$ is omitted in
\refeq{Phiptmasscont}.  Thus to prove the continuity in $m^2$ of
$R_+$, it suffices to prove the continuity of $\Phi_\pt(\Vhat)$ and of
$\Phi_\pt(V)$. These are entirely analogous and we therefore only
consider $\Phi_\pt(\Vhat)$.
By Lemma~\ref{lem:moreQbd} and \refeq{Tphi-poly-2},
\begin{equation}
    \|\Vhat(B)\|_{T_{\varphi,\aux}(\h_+,\hAux)} \le  P^{4}_\h(\varphi).
\end{equation}
The product property of the norm then implies that the norm of $\Vhat(B)^2$ is
bounded above by $P_\h^8(\varphi)$.
By \refeq{Phiptmasscont},
the continuity of $\Phi_\pt(\Vhat)$ in $m^2$
(in $T_{0,y}(\ell_+,\hAux)$-norm) then follows from Lemma~\ref{lem:Excont}.

Next, we prove the continuity of $K_+$ in $m^2$, with $\Wcal_{\aux,+}(\hAux)$-norm.  By definition,
  \begin{align}
    K_{+}(B)
    &=
    e^{u_+|B|}
    \left(
    \Ex_{C_+(m^2)}\theta
    \left( e^{-V} + K  \right)^B
    - e^{-U_+ (B)}
    \right)
    ,
  \end{align}
where $U_+=\Phi_+^U(V,K)$.
We consider the
two cases in the definition of the $\Wcal_{\aux,+}(\hAux)$-norm separately.
That is, we consider the $T_{\infty,y}(h_+,\hAux)$ norm and the
$T_{0,y}(\ell_+,\hAux)$-seminorm.
Since both norms satisfy the product property,
it suffices to prove the continuity of $e^{u_+|B|}$, $e^{-U_+(B)}$ and of $\Ex_{C_+(m^2)}\theta ( e^{-V} + K )^B$ separately,
in both norms.

We first show that $e^{-U_+(B)}$ is continuous; the continuity of $e^{u_+|B|}$ is analogous
and we do not enter into its details.  We write $U_+=U_+(B,m^2)$ and $\tilde{U}_+=U_+(B,\mgen^2)$.
  By the Fundamental Theorem of Calculus,
  \begin{equation}
    e^{-U_+}-e^{-\tilde U_+}
    =
    \int_0^1 e^{-tU_+-(1-t)\tilde U_+ } (U_+ - \tilde U_+) \, dt.
  \end{equation}
We apply the product property of the norm, and use \eqref{e:stabilityUplus} to bound the
norms of the exponential factors.  This gives
  \begin{align}
    \|e^{-U_+ }-e^{- \tilde U_+ }\|_{T_{\varphi,y}(\h_+,\hAux)}
    & \leq
    \|U_+- \tilde U_+\|_{T_{\varphi,y}(\h_+,\hAux)}
    2^{1/2} e^{-2 c^{\stab} |\varphi_x/h_+|^4}
    \nnb & \le
    \|U_+- \tilde U_+\|_{T_{0,y}(\h_+,\hAux)}P_{\h_+}^4(\varphi)
    2^{1/2} e^{-2 c^{\stab} |\varphi_x/h_+|^4}.
  \end{align}
For the $T_{0,\aux}(\ell_+,\hAux)$-norm, the $\varphi$-dependent factors on the right-hand
side are absent, and the continuity then follows from the fact
shown earlier in the proof that $U_+=\Phi_\pt(\Vhat)$ is continuous in $m^2$ when
considered as a map into $T_{{0,y}}(\ell_+,\hAux)$.
For the $T_{\infty,\aux}(h_+,\hAux)$-norm, we have a uniform bound on
the product
of the exponential and polynomial factors in the last line, and
the norm on the right-hand side goes to zero as $m^2 \to \mgen^2$ as a consequence
of the $\h_+=\ell_+$ case and Lemma~\ref{lem:hoverell}.

Finally, we prove the continuity of $\Ex_{C_+(m^2)}\theta ( e^{-V} + K )^B$ in $m^2$.
Let $G=e^{-V}+K$ and $F=G^B$.
By Lemma~\ref{lem:Excont}, it suffices to prove that there are constants $c_F,k$ such that,
for $h_+=\ell_+$ and $\h_+=h_+$,
\begin{equation}
\lbeq{cFbd}
    \| F\|_{T_{\varphi,\aux}(\h_+,\hAux)} \le c_F P_{\h_+}^k(\varphi).
\end{equation}
By the product property, \refeq{cFbd} will follow once we prove that
there are constants $c_G,m$ such that
\begin{equation}
\lbeq{cGbd}
    \| G\|_{T_{\varphi,\aux}(\h_+,\hAux)} \le c_G P_{\h_+}^m(\varphi).
\end{equation}
By \eqref{e:K*norm-ell}, \eqref{e:Kstar-h-9}, \eqref{e:stabilityVhat},
and the assumption $(V,K) \in \domRG$,
\begin{equation} \label{e:V*K*norm-bis}
    \|e^{-V^{*}(B)}\|_{T_{\infty,\aux} (\h_+,\hAux)} \le 2,
    \quad
    \|K^*\|_{T_{\infty,\aux}(h_+,\hAux)} \leq 2,
    \quad
    \|K^*\|_{T_{\varphi,\aux}(\ell_+,\hAux)} \leq 4 P_\ell(\varphi)^{10},
\end{equation}
where for the last inequality we also used \eqref{e:norm-comparison-simp}.
Therefore, by the triangle inequality,
  \begin{equation}
    \|G\|_{T_{\infty,y}(h_+,\hAux)}  \leq 4, \quad\quad
    \|G\|_{T_{\varphi,y}(\ell_+,\hAux)} \leq 8P_\ell(\varphi)^{10}.
  \end{equation}
This gives \refeq{cGbd} and completes the proof.
\end{proof}

\section{Last renormalisation group step: Proof of Proposition~\ref{prop:N}}
\label{sec:last-step}

In this section, we prove Proposition~\ref{prop:N}, which accounts for
the last renormalisation group step.  This last step is given by the
map defined in Definition~\ref{def:laststep}.  It does not change
scale, and it does not extract the growing contributions from $K$ as
this is unnecessary because the map is not iterated.  Until the last
step, we have relied on the vanishing of $c^{(1)}$, but the last
covariance does not satisfy $c^{(1)}_{\hat N} =0$.  The last step
therefore involves the additional perturbative contribution $\Wnewlast
= -\frac 12 c_{\hat{N}}^{(1)}g_N^2 |\varphi|^6$ (recall
\refeq{Wplus}).

At scale $N$ there is only one block $B = \Lambda \in \Bcal_{N} (\Lambda)$, and
\begin{equation}
\lbeq{ClastQ}
    \Cnewlast{} = m^{-2} Q_N
\end{equation}
with $Q_N$ defined by \refeq{Qj-def}.  Then, by definition,
$c_{\hat{N}}^{(1)} = \sum_{x \in B} Q_{N;0x} = m^{-2}$,
and by \eqref{e:cfrak-def} $\mathfrak{c}_+ = m^{-1}L^{-dN/2}$.

According to Definition~\ref{def:laststep}, the final renormalisation
group map $(V,K)\mapsto (\Unewlast ,\Knewlast)$ is defined by
\begin{align}
\lbeq{Ulast-10}
    \Unewlast  &= \Upt(V)= \gnewlast\tau^2 + \nunewlast \tau + \unewlast,
\\
\lbeq{Klastdef-10}
    \Knewlast(B) & =
    e^{-\Vnewlast(B)}
    \left(
    \tfrac{1}{8} \left(\LT \Var_{\Cnewlast{}} (\theta V) \right)^2
    + \Ex_{\Cnewlast{}} A_3(B) \right)
    + e^{\unewlast |B|} \Ex_{\Cnewlast{}} \theta K(B) .
\end{align}
Here, as in \refeq{dUdef} and \refeq{A3def},
\begin{align}
\label{e:dUdef-10}
    \delta V &= \theta V - \Upt(V),
\\
    \lbeq{A3def-10}
    A_3(B)
    & =
    \frac{1}{2!}\int_0^1 (-\delta V(B))^3 e^{-t\delta V(B)} (1-t)^{2} dt.
\end{align}
By Proposition~\ref{prop:Elast},
provided that the expectations on the right-hand side are well-defined,
\begin{equation}
\lbeq{ElastN-10}
    \Ex_{\Cnewlast{}} \left(e^{-\theta V_N(B)} + \theta  K_N(B)\right)
    =
    e^{-\unewlast |B|}
    \left(e^{-\Vnewlast(B)}\big(1+\Wnewlast(B)\big) + \Knewlast(B)\right) .
\end{equation}

The following proposition is a restatement of Proposition~\ref{prop:N}.

\begin{prop}
\label{prop:N-10}
Fix $L$ sufficiently large and $g_0>0$ sufficiently small, and suppose
that $m^2L^{2N} \ge 1$. Let $(V_N,K_N)\in \domRG_N$. Derivatives with
respect to $\nu_0$ are evaluated at $(m^2,\nu_0^c(m^2))$.

\smallskip \noindent
(i)
The perturbative part of the last map obeys
\begin{equation} \label{e:UN-10}
  \gnewlast = g_N(1+O(\vartheta_Ng_N)), \quad
  L^{2N}|\nunewlast| = O(\vartheta_N g_N), \quad
  \Wnewlast = -\frac 12 c_{\hat{N}}^{(1)}g_N^2 |\varphi|^6,
\end{equation}
\begin{equation} \label{e:nuNp-10}
  \ddp{\nunewlast}{\nu_0}
  =
  \left(\frac{g_N}{g_0}\right)^{\gamma}(c+O(\vartheta_Ng_N)),
  \quad
  \ddp{\gnewlast}{\nu_0}
  = O\left( L^{2N}  g_N^2\left(\frac{g_N}{g_0}\right)^{\gamma} \right),
\end{equation}
with  $c=1+O(g_0)$ from Theorem~\ref{thm:VK}.

\smallskip \noindent
(ii)
At scale $N$,
the expectations on the right-hand side of \refeq{ElastN-10} exist, and
\begin{align} \label{e:VKN-Kj-10}
  |\Knewlast(0)| + L^{-2N} |D^2\Knewlast(0;\1,\1)|
  &= O(\vartheta_N^{3} g_N^{3}),
\\
\label{e:gKNp-10}
  L^{-2N} \left|\ddp{}{\nu_0} \Knewlast(0)\right|
  +
  L^{-4N} \left|\ddp{}{\nu_0} D^2 \Knewlast(0; \1,\1)\right|
  &= O\left(\vartheta_N^3  g_N^2
  \left(\frac{g_N}{g_0}\right)^{\gamma}\right)
  .
\end{align}
\end{prop}

\begin{proof}
We again use the parameter $\epdV_N$ defined in \refeq{epdVdef}, which obeys
\begin{equation}
\lbeq{epdVdef-10}
    \epdV_N = \epdV_N(\h) \asymp
    \begin{cases}
     \vartheta_N g_N & (\h=\ell)
    \\
     \vartheta_N g_N^{1/4} & (\h=h).
    \end{cases}
\end{equation}

\smallskip \noindent
(i)
The formula for $\Wnewlast$ follows from \refeq{Wplus}.

By \refeq{Ulast-10} and \refeq{Phiptdef},
\begin{equation}
\lbeq{Ulastquadratic}
    \Unewlast (B)
    =
    \Ex_{\Cnewlast{}}\theta V_N(B)
    -
    \frac 12 \LT \Var_{\Cnewlast{}} (\theta V_N(B))
    .
\end{equation}

Fix $\hAux_V \le g_N$.  By Lemma~\ref{lem:moreQbd},
$\|V_N(B)\|_{T_{0,\aux}(\h_N,\hAux)}$ is at most $1$ for $\h_N=h_N$
and is at most $O(g_N)$ for $\h_N=\ell_N$.  Since $N>j_m$ by
assumption, we have $(mL^{N})^{-1} \le \vartheta_N$.  Therefore, by
Lemma~\ref{lem:Vpt-6} and
$\mathfrak{c}_+ = m^{-1}L^{- d N/2}$, there exists $c>0$ such that
\begin{align}
    \|\Unewlast (B) - V_N(B)\|_{T_{\varphi,\aux}(\h,\hAux)}
    &\le
    c \tfrac{m^{-1} L^{-d N/2}}{\h_N} \|V_N(B)\|_{T_{0,\aux}(\h,\hAux)}
    P_{\h}^{4}(\varphi)
    \nnb & \le
    O(\epdV_N) P_{\h}^{4}(\varphi).
\end{align}
With $\h_N=\ell_N$ and $\varphi=0$, this implies in particular that
$\munewlast = \mu_N +O(\vartheta_ng_N)$, from which the estimate on
$\nunewlast$ in \refeq{UN-10} holds because $\nu_N$ obeys that
estimate.  For the bound $\gnewlast = g_N(1+O(\vartheta_Ng_N))$, we
observe that $g_N$ is the only contribution to $\gnewlast$ from
$\Ex_{\Cnewlast{}}\theta V_N(B)$, so the difference is contained in
the covariance term in \refeq{Ulastquadratic}, and this term obeys the
quadratic upper bound \refeq{covariance1-6}.  This then gives
$\gnewlast - g_N=O(\vartheta_Ng_N^2))$, and the proof of \refeq{UN-10}
is complete.

The proof of \refeq{nuNp-10} follows as in \refeq{muchO}--\refeq{gchO}
with $j$ replaced by $N$ and $j+1$ replaced by $\Nlast$; in fact it is
easier here because there is no dependence on $K_N$ for
$\gnewlast,\nunewlast$.

\smallskip \noindent
(ii)
Fix $\hAuxg_V \leq g_N$ and let $\hAuxg_K(\h)$ be given by \refeq{hAuxKh}
with $\hAuxg_K \le g_N^{9/4}$.
It suffices to prove that
for $(\varphi,\h_N)$ equal to either $(\infty,h_N)$ or $(0,\ell_N)$,
\begin{equation}
\label{e:Klastbd}
        \|\Knewlast(B)\|_{T_{\varphi,\aux}(\h_N,\hAux)} \leq O( \epdV_N^3 + \hAuxg_K(\h)),
        \quad
        \|\Knewlast'(B)\|_{T_\varphi(\h_N)} \le O(\vartheta_N^3 g_N^2 \mu_N') ,
\end{equation}
where $K'=\ddp{}{\nu_0} \Knewlast$.
Indeed, this is more than is needed, the case $(\varphi,\h)=(0,\ell_N)$ suffices as it implies
\refeq{VKN-Kj-10}--\refeq{gKNp-10} because
$\|\1\|_{\Phi_N(\ell_N)} = O(L^{N})$ 
by \refeq{1norm} (recall \refeq{WNbd}).

According to \refeq{Klastdef-10},
\begin{align}
\lbeq{Klastdef-10a}
    \Knewlast(B) & =
    e^{-\Vnewlast(B)}
    \left(
    \tfrac{1}{8} \left(\LT \Var_{\Cnewlast{}} (\theta V_N) \right)^2
    + \Ex_{\Cnewlast{}} A_3(B) \right)
    + e^{\unewlast |B|} \Ex_{\Cnewlast{}} \theta K_N(B) .
\end{align}
To estimate the terms in \refeq{Klastdef-10a}, we use the bounds,
valid for $(\varphi,\h)$ equal to either $(\infty,h_N)$ or $(0,\ell_N)$,
\begin{align}
\label{e:expUlastbd}
  e^{\|\Unewlast(B)\|_{T_{\varphi,\aux}(\h_N,\hAux)}}
  & \le 2,
\\
\lbeq{Varlast}
    \|\LT\Var_{\Cnewlast{}}(\theta V_N(B))\|_{T_{\varphi,\aux}(\h_N,\hAux)}
    & \le O(\epdV_N^2)P_{\h_N}^4(\varphi)
\\
    \|\Ex_{\Cnewlast{}} A_3(B)\|_{T_{\varphi,\aux}(\h_N,\hAux)}
  & \le
  O (\epdV_N^3) ,
\\
\lbeq{ElastKbd}
    \|\Ex_{\Cnewlast{}}\theta K_N(B)\|_{T_{\varphi,\aux}(\h_N,\hAux)}
    & \le  O (\epdV_N^3 +\hAux_K(\h))
  .
\end{align}
The first three inequalities follow as in the proof of
Lemma~\ref{lem:S0} in Section~\ref{sec:S0pf} (we also use $\LT=\Tay_4$
and Lemma~\ref{lem:TayTphi} for \refeq{Varlast}), and we omit the
details.  The inequality \refeq{ElastKbd} follows as in
Lemma~\ref{lem:grand-comparison} (though here the product over blocks
has only one block), together with the norm estimates on $K$ given by
\refeq{K*norm-ell}--\refeq{Kstar-h-9} and our assumption that $K_N$
lies in the domain.

Finally, to prove the bound \eqref{e:Klastbd} on $\Knewlast'$ we first
use the chain rule to obtain
\begin{equation}
   \Knewlast'(V_N,K_N) = D_V\Knewlast(V_N,K_N)V_N' + D_K\Knewlast(V_N,K_N)K_N'.
\end{equation}
By Lemma~\ref{lem:normderiv} with $\hAux_V=g_N$ and $\hAux_K=g_N^{9/4}$, this gives
\begin{align}
    \|\Knewlast'\|_{T_\varphi(\h_N)}
    & \le
    \left(
    \hAux_V^{-1}\|V_N'\|_{T_{0}(\ell_N)}
    +
    \hAux_K^{-1} \|K_N'\|_{\Wcal_N}
    \right)
    \|\Knewlast\|_{\Wcal_N}
    \nnb & \le
    \left(
    g_N^{-1}\|V_N'\|_{T_{0}(\ell_N)}
    +
    g_N^{-9/4} \|K_N'\|_{\Wcal_N}
    \right)
    O(\vartheta_N^3 g_N^3).
\end{align}
The norms of the derivatives on the right-hand side are respectively bounded
in \refeq{nuNp-5} and \refeq{Kprime}, and using these bounds we obtain
\begin{align}
    \|\Knewlast'\|_{T_\varphi(\h_N)}
    & \le
    \left(
    g_N^{-1}\mu_N'
    +
    g_N^{-9/4} \mu_N'g_N^2
    \right)
    O(\vartheta_N^3 g_N^3) = O(\vartheta_N^3 g_N^2\mu_N').
\end{align}
This completes the proof.
\end{proof}

\begin{rk}
We emphasise that the first inequality of \refeq{Klastbd} proves a
stronger result than is required for \eqref{e:VKN-Kj-10}.  Indeed, for
the standard norm it implies that, for $(\varphi,\h)$ equal to either
$(\infty,h_N)$ or $(0,\ell_N)$,
  \begin{equation}
    \|\Knewlast(B)\|_{T_{\varphi}(\h_N)} \leq O(\epdV_N^3).
  \end{equation}
In particular, \eqref{e:VKN-Kj-10} is a consequence of this
$T_0(\ell_N)$ estimate.
\end{rk}


%% file: BS.pspdftex
\begin{picture}(0,0)%
\includegraphics{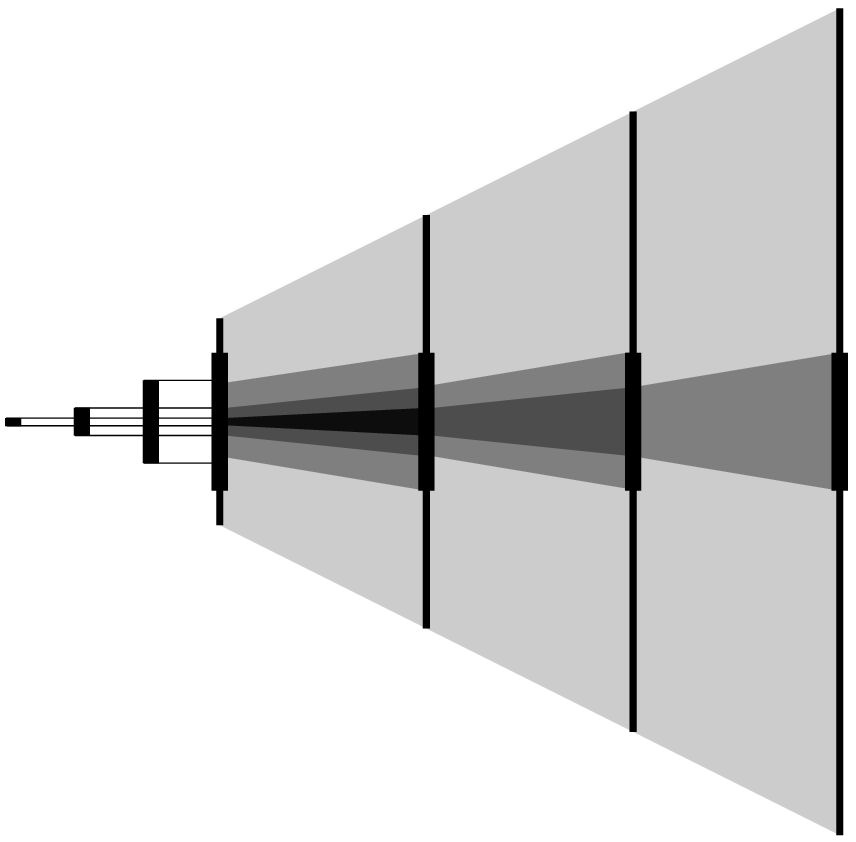}%
\end{picture}%
\setlength{\unitlength}{2901sp}%
\begingroup\makeatletter\ifx\SetFigFont\undefined%
\gdef\SetFigFont#1#2#3#4#5{%
  \reset@font\fontsize{#1}{#2pt}%
  \fontfamily{#3}\fontseries{#4}\fontshape{#5}%
  \selectfont}%
\fi\endgroup%
\begin{picture}(5636,5488)(590,-8205)
\put(3511,-6001){\makebox(0,0)[lb]{\smash{{\SetFigFont{8}{9.6}{\familydefault}{\mddefault}{\updefault}{\color[rgb]{0,0,0}$J_1$}%
}}}}
\put(4861,-6001){\makebox(0,0)[lb]{\smash{{\SetFigFont{8}{9.6}{\familydefault}{\mddefault}{\updefault}{\color[rgb]{0,0,0}$J_2$}%
}}}}
\put(6211,-6091){\makebox(0,0)[lb]{\smash{{\SetFigFont{8}{9.6}{\familydefault}{\mddefault}{\updefault}{\color[rgb]{0,0,0}$J_3$}%
}}}}
\put(1576,-6091){\makebox(0,0)[b]{\smash{{\SetFigFont{8}{9.6}{\familydefault}{\mddefault}{\updefault}{\color[rgb]{0,0,0}$I_1$}%
}}}}
\put(1126,-6091){\makebox(0,0)[b]{\smash{{\SetFigFont{8}{9.6}{\familydefault}{\mddefault}{\updefault}{\color[rgb]{0,0,0}$I_2$}%
}}}}
\put(676,-6091){\makebox(0,0)[b]{\smash{{\SetFigFont{8}{9.6}{\familydefault}{\mddefault}{\updefault}{\color[rgb]{0,0,0}$I_3$}%
}}}}
\end{picture}%

%% file: fermions.tex
\chapter{Self-avoiding walk and supersymmetry}
\label{ch:saw-int-rep}

A strength of the renormalisation group method presented in this book
is that it applies with little modification to models which
incorporate fermion fields.  This allows, in particular, for a
rigorous analysis of a version of the continuous-time weakly
self-avoiding walk (also known as the \emph{lattice Edwards model}).
The continuous-time weakly self-avoiding walk is predicted to lie in
the same universality class as the standard self-avoiding walk.  In
this chapter, whose results are not used elsewhere in the book, we
give an introduction to the continuous-time weakly self-avoiding walk
and its representation as a supersymmetric spin system.

We begin in Section~\ref{sec:saw-cb} with a brief discussion of the
critical behaviour of the standard self-avoiding walk model, and then
introduce the continuous-time weakly self-avoiding walk.  Spin systems
have been studied for many decades via their random walk
representations, and in Section~\ref{sec:rwrep} we prove the
BFS--Dynkin isomorphism theorem that implements this representation.
In Section~\ref{sec:susy}, we prove that a certain supersymmetric spin
system has a representation in terms of the continuous-time weakly
self-avoiding walk.  In contrast to the usual application of the
results of Section~\ref{sec:rwrep}, in which the random walk
representation is used to study the spin system, the results of
Section~\ref{sec:susy} have been used in reverse.  Namely, starting
with the continuous-time weakly self-avoiding walk, we use the
supersymmetric representation to convert the walk problem to a spin
problem.  Then the renormalisation group method in this book can be
applied to analyse the spin system and thereby yield results about the
weakly self-avoiding walk.  Finally, in Section~\ref{sec:susy2} we
expand on the concept of supersymmetry.

\section{Critical behaviour of self-avoiding walk}
\label{sec:saw-cb}

Our study of the continuous-time weakly self-avoiding walk is motivated
by the standard model of self-avoiding walk, which is a model of
discrete-time strictly self-avoiding walk.  In Section~\ref{sec:saw}, we
provide some background on the self-avoiding walk.
In Section~\ref{sec:ctsrw}, we discuss continuous-time random walk on $\Zd$, and
then in Section~\ref{sec:ctwsaw} we define the continuous-time weakly self-avoiding walk
and give examples of results that have been obtained using the renormalisation group method
discussed in this book.

\subsection{Self-avoiding walk}
\label{sec:saw}

The self-avoiding walk on $\Zd$ is a well-known and notoriously difficult mathematical model of
linear polymer molecules.  Further background and details can be found in \cite{MS93}.

\index{Self-avoiding walk}
\begin{defn}
\label{def:saw}
An $n$-step \emph{self-avoiding walk} is a sequence $\omega :
\{0,1,\ldots, n\} \to \Zd$ with:
 $\omega(0)=0$,
$\omega(n)=x$,
 $|\omega(i+1)-\omega(i)|=1$, and
 $\omega(i) \neq \omega(j)$ for all $i\neq j$.
We write ${\cal S}_n(x)$ for the set of $n$-step self-avoiding walks on $\Zd$
from $0$ to $x$, and write ${\cal S}_n=
\cup_{x\in\Zd} {\cal S}_n(x)$ for the set of $n$-step self-avoiding walk starting
from the origin.  We denote the cardinalities of these sets by
$c_n(x) =|{\cal S}_n(x)|$ and $c_n = |{\cal S}_n| =\sum_{x\in\Zd} c_n(x)$.
\end{defn}

\begin{figure}[h]
\begin{center}
\includegraphics[scale=0.4]{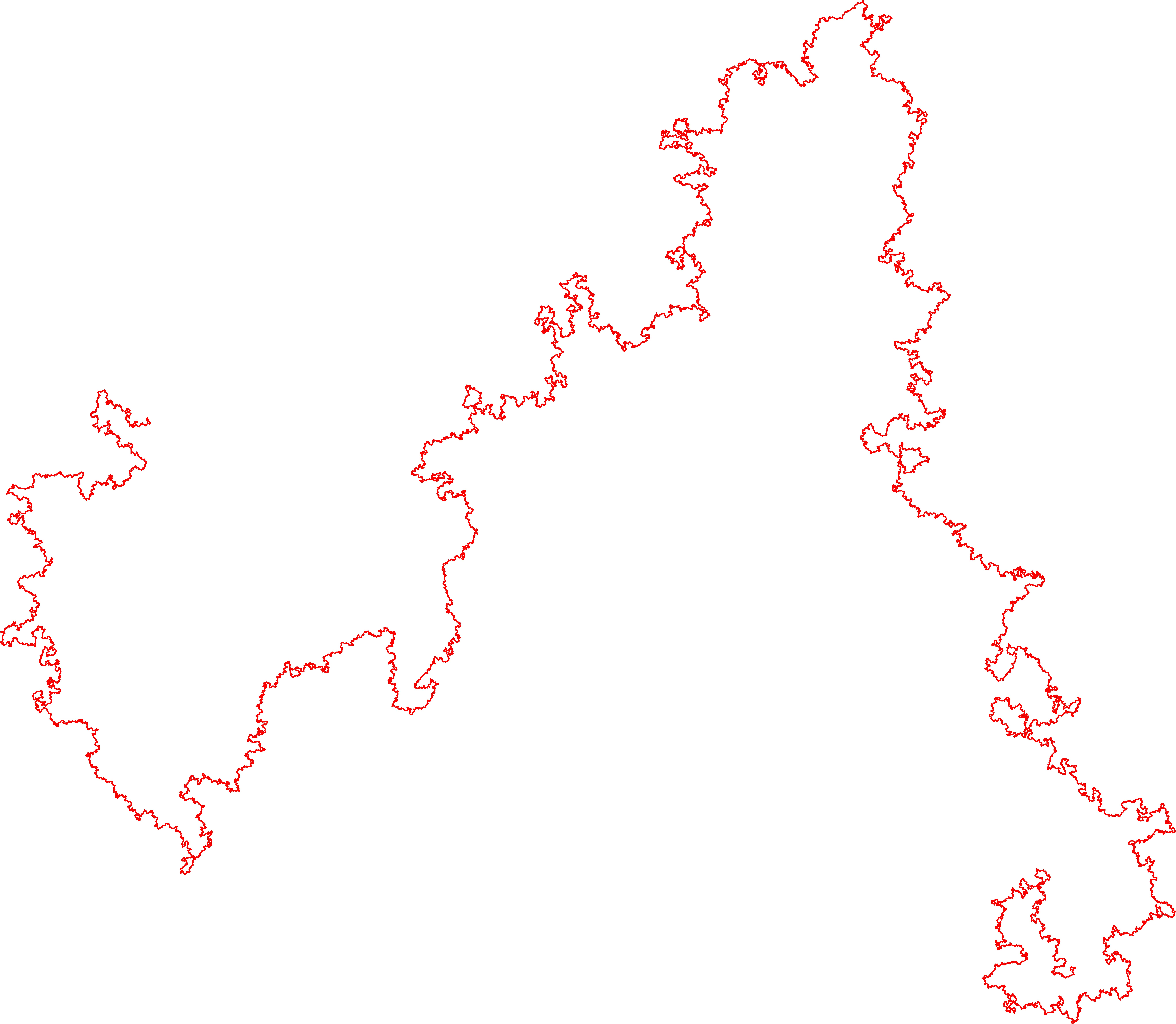}
\end{center}
\caption{
A random $10^8$-step self-avoiding walk on $\Z^2$.
Figure by Nathan Clisby (used with permission).}
\label{fig:SAW}
\end{figure}

We define a probability measure on $\Scal_n$ by declaring
all walks in ${\cal S}_n$ to be equally likely, and we write
 $E_n$ for expectation with respect to uniform measure on $\Scal_n$.  Then
each walk has probability $c_n^{-1}$.  Figure~\ref{fig:SAW} shows a random example.

It is easy to see that $c_{m+n}\le c_mc_n$.  From this, it follows from
Fekete's subadditivity lemma
(see, e.g., \cite[Lemma~1.2.2]{MS93}) that
\begin{equation}
\lbeq{connconst}
    \mu=\mu(d) = \lim_{n\to\infty}c_n^{1/n} = \inf_{n \ge 1}c_n^{1/n}.
\end{equation}
In particular, the limit exists, and $c_n \ge \mu^n$ for all $n \ge 1$.
Thus, roughly speaking, $c_n$ grows exponentially with growth rate $\mu$.
Crude bounds on the \emph{connective constant} $\mu$ are given by the
following exercise.

\begin{exercise} \label{ex:connective-constant-bds}
For $d \ge 1$, show that $\mu \in [d,2d-1]$.
\solref{connective-constant-bds}
\end{exercise}

\index{Two-point function}
The \emph{two-point function} is defined by $G_{0x}(z) = \sum_{n=0}^\infty
c_n(x) z^n$.  Its radius of convergence is $z_c=\mu^{-1}$ for all $x$
\cite[Corollary~3.2.6]{MS93}, and $z_c$ plays the role of a critical point for a spin system.
\index{Critical exponent}
It is predicted that there are universal critical exponents $\gamma,\nu,\eta$ such that
\begin{equation}
    c_n \sim A \mu^n n^{\gamma -1}, \quad
    E_n |\omega(n)|^2 \sim Dn^{2\nu},
    \quad
    G_{0x}(z_c) \sim C|x|^{-(d-2+\eta)},
\end{equation}
with $\gamma,\nu,\eta$ related by Fisher's relation  $\gamma = (2-\eta)\nu$.
Since $|\omega(n)|\le n$ by definition, it is always the case that
$\Ebold_n |\omega(n)|^2 \le n^2$, so $\nu \le 1$.  Also, \refeq{connconst} implies
that $c_n \ge \mu^n$ for all $n$, so $\gamma \ge 1$.
For simple random walk, without the self-avoidance constraint,
the number of $n$-step walks is $(2d)^n$, the mean-square displacement is equal to $n$,
and the critical two-point function is the lattice Green function which has decay
$|x|^{-(d-2)}$ (for $d >2$).  Thus the exponents for simple random walk are
$\gamma=1$, $\nu=\frac 12$, and $\eta=0$.

\index{Susceptibility}
\index{Correlation length}
The \emph{susceptibility} and \emph{correlation length} are defined by
\begin{equation}
    \chi(z) = \sum_{n=0}^\infty c_n z^n = \sum_{x \in \Zd} G_{0x}(z),
\quad\quad
    \frac{1}{\xi(z)} = - \lim_{n \to \infty}\frac{1}{n}\log G_{0,ne_1}(z),
\end{equation}
and it is predicted that
\begin{equation}
    \chi(z) \sim \frac{A' }{(z_c-z)^\gamma},
    \quad\quad
    \xi(z) \sim \frac{D' }{(z_c-z)^{\nu}}
    \quad\quad \text{as $z \uparrow z_c$}.
\end{equation}
The fact that $\gamma$ appears both for $c_n$ and its generating function $\chi(z)$
is an (in general conjectural)
Abelian/Tauberian relation.  The fact that the same exponent $\nu$
appears both for the mean-square displacement and the correlation length is an example
of the general belief that a single critical exponent governs all natural critical length scales.
For dimension $d=4$, logarithmic corrections to simple random walk scaling are
predicted \cite{Dupl86,Clis17-4dsaw} (but not for the critical two-point function):
\index{Logarithmic corrections}%
\begin{align}
\lbeq{sawchi4}
    &c_n \sim A \mu^n (\log n)^{1/4},
    &\chi(z) \sim \frac{A' |\log (z_c-z)|^{1/4}}{z_c-z},
    \\
\lbeq{sawG0x4}
    &
    \Ebold_n |\omega(n)|^2 \sim Dn (\log n)^{1/4},
    &G_{0x}(z_c) \sim C|x|^{-2}.
\end{align}

For $d \le 4$, very little has been proved.  For the end-to-end distance,
the best results are the following.

\begin{theorem}
\label{thm:msdbds}
[\cite{Madr14} (lower bound), \cite{D-CH13} (upper bound)]
For all $d \ge 2$,
\begin{equation}
    \frac{1}{6}n^{4/3d}  \le \Ebold_n |\omega(n)|^2 \le o(n^2).
\end{equation}
\end{theorem}

Theorem~\ref{thm:msdbds} can be paraphrased as $\frac{2}{3d} \le \nu \le 1^-$.
In remains an open problem in dimensions $2,3,4$ even to prove
that $\Ebold_n |\omega(n)|^2 \ge cn$ (i.e., that $\nu \ge \frac 12$),
or that $\Ebold_n |\omega(n)|^2  \le O(n^{2-\epsilon})$ for some $\epsilon>0$
(i.e., that $\nu < 1$).  This lack of proof is in spite of the fact that it seems
obvious that self-avoiding walk must move
away from the origin at least as rapidly as simple random walk, yet should not
move away from the origin with constant speed.

For dimensions $d \ge 5$, the lace expansion has been used to provide a thorough
understanding of the critical behaviour.  Some principal results
are summarised in the following theorem.
\index{Lace expansion}

\begin{theorem}
\label{thm:saw5}
\cite{HS92a,Hara08}.
For $d \ge 5$, there are positive constants $A,D,C$ (depending on $d$) such that
\[
    c_n \sim A \mu^n , \quad
    \Ebold_n |\omega(n)|^2 \sim Dn ,
    \quad
    G_{0x}(z_c) \sim C|x|^{-(d-2)},
\]
and $(\frac{1}{\sqrt{Dn}}\omega(\lfloor nt \rfloor))_{t\ge 0}$ converges
in distribution to Brownian motion $(B_t)_{t \ge 0}$.
\end{theorem}

The above theorem shows that self-avoiding walk behaves like simple random walk
when the dimension is above $4$, in the sense that both models have $\gamma=1$, $\nu=\frac 12$
and $\eta=0$, and in both cases the scaling limit is Brownian motion.
Some indication of the special role of $d=4$ is provided by
Exercise~\ref{ex:bubble1}, which shows that the expected number of intersections
of two independent simple random walks is finite if and only if $d>4$.  This suggests
that elimination of self intersections may not play a big role in the global behaviour
when $d>4$.  The proof of Theorem~\ref{thm:saw5} relies heavily on the fact that the bubble diagram
(see Section~\ref{sec:bubble})
is finite in dimensions $d \ge 5$, and indeed that it is not very large for $d=5$.

\index{SLE}
For
$d=2$, there is a complete set of predictions:
$\gamma = \frac{43}{32}$, $\nu = \frac 34$, $\eta = \frac{5}{24}$,
and that the
scaling limit is the Schramm--Loewner Evolution
${\rm SLE}_{8/3}$ \cite{Nien82,LSW04}, but none of this has been rigorously proved.
For $d=3$ there are good numerical results, e.g., $\nu = 0.58759700(40)$ \cite{CD16}.

\subsection{Continuous-time random walk}
\label{sec:ctsrw}

The definition of the continuous-time weakly self-avoiding walk is
based on a continuous-time random walk.  We provide background on the
latter here.  For simplicity, we first consider the case of random
walk on a finite set $\Lambda$, which may be but need not be a subset
of $\Zd$.

\index{Infinitesimal generator}
\index{$Q$-matrix}
\index{Simple random walk}
A continuous-time random walk $X$ on $\Lambda$ can be defined via
specification of an \emph{infinitesimal generator}, also called a
$Q$-matrix \cite{Norr97}, namely a $\Lambda\times\Lambda$ matrix
$(Q_{xy})$ with the properties that $Q_{xx}<0$, $Q_{xy}\ge 0$ for
$x\neq y$, and $\sum_{y\in\Lambda} Q_{xy}=0$.  Such a random walk
takes independent steps from $x$ at rate $-Q_{xx}$, and jumps to $y$
with probability $-\frac{Q_{xy}}{Q_{xx}}$.  The statement that steps
from $x$ occur at rate $-Q_{xx}$ means that when the random walk is in
state $x$, it waits a random time $\sigma$ before taking its next
step, where $\sigma$ has exponential distribution of rate $-Q_{xx}$
(i.e., with mean $-\frac{1}{Q_{xx}}$).  The waiting times for each
visit to a state are independent of each other and are also
independent of all steps taken.  The transition probabilities are
given in terms of the infinitesimal generator by
\begin{equation}
\lbeq{Pxy}
    P_x(X(t)=y) = E_x(\1_{X(t)=y}) = (e^{tQ})_{xy} \qquad (t \ge 0).
\end{equation}
Here $P$ denotes the probability measure associated with $X$, and $E$
is the corresponding expectation.  The subscripts on $P_x$ and $E_x$
specify that the initial state of the random walk is $X(0)=x$.

Let $\beta=(\beta_{xy})$ be a $\Lambda \times\Lambda$ symmetric matrix
with non-negative entries.  As in \refeq{Lapbeta}, we define the
Laplacian matrix $\Delta_\beta$ by
\begin{equation}
\lbeq{Lapbeta-11}
  (\Delta_\beta f)_x = \sum_{y\in\Lambda} \beta_{xy} (f_y-f_x)
  .
\end{equation}
Equivalently,
\begin{equation}
    \Delta_{\beta; xy} =
    \begin{cases}
    \beta_{xy} & ( y \neq x)
    \\
    -\sum_{z\in\Lambda: z \neq x}\beta_{xz} & (y =x).
    \end{cases}
\end{equation}
Thus $\Delta_\beta$ is a $Q$-matrix.  We fix $\beta$ and consider the
random walk $X$ with generator $\Delta_\beta$.

For example, if $\Delta$ is the nearest-neighbour Laplace operator on
a finite discrete $d$-dimensional torus $\Lambda$ approximating $\Zd$,
defined by $\beta_{xy} = \1_{x\sim y}$, then $X$ is the
continuous-time stochastic process $X$ on $\Lambda$ which takes steps
uniformly to a nearest-neighbour of its current position, at the times
of the events of a rate-$2d$ Poisson process.  This follows from the
fact that the events of a rate-$\lambda$ Poisson process are separated
by independent exponential random variables with mean
$\frac{1}{\lambda}$.  In fact, for this choice of $\beta$ the above
definition of the continuous time random walk applies directly also to
the case where the state space of the walk is $\Zd$ rather than a
finite torus: at the times of a rate-$2d$ Poisson process the walk
steps to a uniformly chosen one of the $2d$ neighbours.  We will use
this infinite-volume random walk in Section~\ref{sec:ctwsaw}.

For the continuous-time weakly self-avoiding walk, we need two random variables.
The first is
\index{Local time}
the \emph{local time of $X$ at $u\in\Lambda$ up to time $T$}, defined by
\begin{equation}
\lbeq{LTudef}
    L_{T,u} = \int_0^T \1_{X(s)=u} ds.
\end{equation}
\index{Self-intersection local time}
The second is the \emph{self-intersection local time of $X$ up to time $T$}, defined by
\begin{equation}
    I(T)
    = \sum_{u\in\Lambda} L_{T,u}^2
    = \int_0^T \int_0^T \1_{X(s)=X(t)}ds \, dt
    .
\end{equation}
As its name suggests, $I(T)$ increases with the amount of time that
 the random walk path spends intersecting itself.

\subsection{Continuous-time weakly self-avoiding walk}
\label{sec:ctwsaw}

The continuous-time weakly self-avoiding walk is a modification of the
\index{Weakly self-avoiding walk}
self-avoiding walk of Section~\ref{sec:saw} in two respects.  Firstly,
an additional source of randomness is introduced by basing the model
on the continuous-time simple random walk on $\Zd$ whose infinitesimal
generator is the standard Laplacian $\Delta$ on $\Zd$, rather than on
a discrete-time walk.  Secondly, walks with self intersections are not
eliminated, but instead receive lower probability.  Thus,
\index{Two-point function} given $g > 0$ and $\nu \in \R$, we define
the \emph{two-point function}
\begin{equation}
\lbeq{WSAW2pt}
    G_{0x}(g,\nu)
    = \int_0^\infty
    E_0 \left( e^{-gI(T)}\; \1_{X(T)=x} \right) e^{-\nu T}
    dT.
\end{equation}
In comparison with the two-point function $\sum_{n=0}^\infty
c_n(x)z^n$ for the self-avoiding walk, now the integral over $T$ plays
the role of the sum over $n$, the variable $z$ is replaced by
$e^{-\nu}$, and $c_n(x)$ is replaced by $E_0 ( e^{-gI(T)}\;
\1_{X(T)=x} )$.  This expectation gives positive weight to all walks
$X$, but the factor $e^{-gI(T)}$ assigns reduced weight for self intersections.

\index{Susceptibility}
The \emph{susceptibility} is defined by
\begin{equation}
    \chi(g,\nu) = \sum_{x\in\Zd} G_{0x}(g,\nu).
\end{equation}
A subadditivity argument \cite[Lemma~A.1]{BBS-saw4-log}
\index{Subadditivity}
shows that there exists $\nu_c(g)\in (-\infty,0]$, depending on $d$,
such that
\begin{align}
&\chi(g,\nu)
<\infty \quad \text{if and only if $\nu> \nu_c(g)$}.
\end{align}
In particular, $\chi(g,\nu_c)=\infty$.

The continuous-time weakly self-avoiding walk is predicted to be in
the same universality class as the strictly self-avoiding walk, for
all $g>0$.  In particular, critical exponents and scaling limits are
predicted to be the same for both models, including the powers of
logarithmic corrections for $d=4$.  The following theorem is an
example of this for small $g>0$.

\begin{theorem}
\label{thm:wsaw4}
\cite{BBS-saw4-log,BBS-saw4}.
Let $d = 4$, and consider the weakly self-avoiding walk on $\Z^4$
defined by the nearest-neighbour Laplacian.  For small $g>0$ and for
$\nu=\nu_c + \epsilon$, as $\epsilon \downarrow 0$,
\begin{equation}
\lbeq{chiwsaw}
    \chi(g,\nu)
    \sim A_g \frac{1}{\epsilon} (\log \epsilon^{-1})^{1/4}
    .
\end{equation}
As $|x|\to\infty$,
\begin{equation}
    G_{0x}(g,\nu_c)
    = \frac{c_g}{|x|^{2}}\left( 1 + O\left(\frac{1}{\log |x|} \right)\right)
    .
\end{equation}
As $g \downarrow 0$, the amplitude $A_g$ and critical value obey $A_g
\sim (g/2\pi^2)^{1/4}$ and $\nu_c(g) \sim - 2 N_4 g $ (with $N_4 =
(-\Delta)^{-1}_{00}$).
\end{theorem}

The logarithmic factor for the susceptibility, and the absence of a
logarithmic correction for the critical two-point function, are
consistent with the predictions for self-avoiding walk in
\refeq{sawchi4}--\refeq{sawG0x4}.  Since the strictly self-avoiding
walk corresponds to $g=\infty$ \cite{BDS12}, Theorem~\ref{thm:wsaw4}
shows that the weakly self-avoiding walk demonstrates behaviour like
the $g=\infty$ case, not the $g=0$ case.

Theorem~\ref{thm:wsaw4} is quantitatively similar to results for the
$4$-dimensional $n$-component $|\varphi|^4$ model in
Theorems~\ref{thm:phi4} and \ref{thm:ST-phi4}.  Indeed,
\refeq{chiwsaw} corresponds exactly to \eqref{e:thmphi4-chi} with $n$
replaced by $n=0$, and the situation is similar for the asymptotic
formulas for $A_{g,n}$ and $\nu_c(g,n)$ in Theorems~\ref{thm:phi4}:
with $n=0$ they give the corresponding results for the continuous-time
weakly self-avoiding walk in Theorem~\ref{thm:wsaw4}.  This is an
instance of the observation of de Gennes \cite{Genn72} that spins with
``$n = 0$'' components correspond to self-avoiding walk, which we
discuss in more detail in Section~\ref{sec:susy}.  The ``$n=0$''
connection is an important element of the proof of
Theorem~\ref{thm:wsaw4}.

Several extensions of Theorem~\ref{thm:wsaw4} have been proved.  These
include the critical behaviour of the correlation length of order $p$
in dimension $4$ \cite{BSTW-clp}, the lack of effect of a small
contact self-attraction in dimension $4$ \cite{BSW-saw-sa},
the construction of the tricritical \emph{theta point} for polymer collapse
in dimension 3 \cite{BLS19},
and the
computation of non-Gaussian critical exponents for a long-range model
below the upper critical dimension \cite{Slad17,LSW17}.  In
particular, versions of Theorems~\ref{thm:phi4}--\ref{thm:long-range}
have all been proved for the continuous-time weakly self-avoiding
walk.

\index{Hierarchical model} Related and stronger results have been
proved for a $4$-dimensional hierarchical version of the
continuous-time weakly self-avoiding walk \cite{BEI92,BI03c,BI03d},
including the predicted behaviour $T^{1/2}|\log T|^{1/8}$ for the mean
end-to-end distance.  This continuous-time weakly self-avoiding walk
is defined in terms of the hierarchical random walk of
Exercise~\ref{ex:hier-rw} via a penalisation of self intersections
using the self-interaction local time as in \refeq{WSAW2pt}.

A model related to the 4-dimensional weakly self-avoiding walk is
studied in \cite{IM94} via a different renormalisation group approach.

\section{Random walk representation of spin systems}
\label{sec:rwrep}

Random walk representations of integrals arising in mathematical
physics have been used for about half a century.  Early references
include the
work of Symanzik \cite{Syma69} in quantum field theory and the work of
Fisher on statistical mechanics \cite{Fish67}.  Random walk
representations have been used extensively in classical statistical
mechanics, e.g., in \cite{ACF83,BFS83II,BFS82,Dynk83,FFS92}.  In this
section, we present an important example: the BFS--Dynkin isomorphism
\cite{BFS82,Dynk83}.  The BFS--Dynkin isomorphism is the foundation
upon which a supersymmetric version can be built.  The supersymmetric
version and its relation to the weakly self-avoiding walk are the
topic of Section~\ref{sec:susy}.

\subsection{Continuous-time random walk and the Laplacian}

This section is devoted to a special case of the BFS--Dynkin
isomorphism, in Lemma~\ref{lem:CTSRW}.  This special case is also a
version of the \emph{Feynman--Kac formula}.

Lemma~\ref{lem:srw} indicates that the Laplacian and simple random
walk are closely related.  The next exercise extends
Lemma~\ref{lem:srw} to more general random walks on a finite set
$\Lambda$.

\begin{exercise}
\label{ex:srwbeta} Let $\Lambda$ be a finite set.  Let $\beta =
(\beta_{xy})$ be a symmetric $\Lambda \times \Lambda$ matrix with
non-negative entries.  Let $V$ be a complex diagonal matrix $V$ with
${\rm Re} \, v_x \ge c>0$ for all $x\in \Lambda$.  Let
$\bar\beta_x=\sum_{y\in\Lambda: y \neq x}\beta_{xy}$, and assume that
$\bar\beta_x>0$ for all $x$.  Then
\begin{equation}
\lbeq{srwbeta}
    (-\Delta_\beta + V)^{-1}_{xy}
    =
    \sum_{Y \in \Wcal^*(x,y)}
    \prod_{i=1}^{|\omega|} \beta_{Y_{i-1}Y_i}
    \prod_{j=0}^{|\omega|} \frac{1}{\bar\beta_{Y_j} + v_{Y_j}},
\end{equation}
where $\Wcal^{*}(x,y)$ consists of the union, over non-negative
integers $n$, of $n$-step walks $Y=(Y_0,Y_1,\ldots, Y_n)$ with
$Y_0=x$, $Y_n=y$, and $Y_{j+1}\neq Y_j$ for each $j$.  For the special
case where $\Lambda$ is a discrete $d$-dimensional torus and
$\Delta_{\beta;xy} = \1_{x\sim y}$, the right-hand side of
\refeq{srwbeta} gives a finite-volume version of \refeq{Csrwrep}.
\solref{srwbeta}
\end{exercise}

The following lemma provides a version of the relationship expressed
by Exercise~\ref{ex:srwbeta}, but now in terms of the
\emph{continuous-time} random walk $X$ with generator $\Delta_\beta$.
We denote expectation for $X$ with $X(0)=x$ by $E_x$.
\index{Local time}
Recall that the local time of $X$ at $u \in
\Lambda$ up to time $T \geq 0$ is the random variable $L_{T,u}$ given
by \refeq{LTudef}.  Since $\sum_{u \in \Lambda}L_{T,u}=T$, a special
case of \refeq{CTSRWLapz} is
\begin{equation}
\lbeq{CTSRWLap}
    (-\Delta_{\beta} + m^2)^{-1}_{xy}
    = \int_0^\infty
    E_x \left(  \1_{X(T)=y} \right) e^{-m^2 T}
    dT .
\end{equation}

\index{Feynman--Kac formula}
\begin{lemma}
\label{lem:CTSRW} Let $\Lambda$ be a finite set, and let $V$ be a
complex diagonal matrix with rows and columns indexed by $\Lambda$,
whose elements obey ${\rm Re} \, v_x \ge c >0$ for some positive $c$.
Then
\begin{equation}
\lbeq{CTSRWLapz}
    (-\Delta_{\beta} + V)^{-1}_{xy}
    = \int_0^\infty
    E_x \left( e^{-\sum_{u} v_u  L_{T,u}}\; \1_{X(T)=y} \right)
    dT .
\end{equation}
\end{lemma}

\begin{proof}
Let $\bar\beta_x = \sum_{y \neq x} \beta_{xy}$.  We can and do regard
$X$ as a discrete-time random walk $Y$ whose steps have transition
probabilities $p_{xy}= \beta_{xy}/\bar\beta_x$ (for $x \neq y$), which
are taken at rate $\bar\beta_x$, as discussed in
Section~\ref{sec:ctsrw}.  Thus, at each visit to $x$, the time
$\sigma_{x}$ spent at $x$ until the next step is an independent
Exponential random variable with mean $1/\bar\beta_x$.  Given an
$n$-step walk $Y$ and $j \le n$, we set $\gamma_{j} = \sum_{i = 0}^{j}
\sigma_{Y_i}$.  We also write
\begin{equation}
    p(Y) = \prod_{i=1}^n p_{Y_{i-1}Y_i}.
\end{equation}
Then the right-hand side of \refeq{CTSRWLapz} is equal to
\begin{align}
& \sum_{n=0}^\infty \sum_{Y \in \mathcal{W}_{n}(x,y)} p(Y)
\int_0^\infty
E_0 \left[e^{-\sum_{j=0}^{n-1} v_{Y_j} \sigma_{Y_j}}
\left.
e^{-v_{Y_n} (t - \gamma_{n-1})}
\1_{\gamma_{n-1} < t < \gamma_{n}}
\; \right| \; Y \right] dt .
\lbeq{CTSRWE}
\end{align}

We use Fubini's theorem to interchange the expectation and integral.
Since the holding times $\sigma_x$ are independent of $Y$, given $Y$
the integral in \refeq{CTSRWE} is equal to
\begin{align}
&
E \left[e^{-\sum_{j=0}^{n-1} v_{Y_j} \sigma_{Y_j}} \int_{\gamma_{n-1}}^{\gamma_n}
e^{-v_{Y_n} (t - \gamma_{n-1}) } dt \right]
\nnb
&\quad =
E \left[\left(e^{-\sum_{j=0}^{n-1} v_{Y_j} \sigma_{Y_j}}\right)
\left( -\frac{1}{v_{Y_n}}\right) \left(e^{-v_{Y_n} \sigma_{n}} - 1\right)\right].
\end{align}
Since the random variables $\sigma_x$ are independent, the expectation
factors to become
\begin{align}
&
\left(
\prod_{j=0}^{n-1}E\left[e^{- v_{Y_j} \sigma_{Y_j}}\right] \right)
\left(\frac{1}{v_{Y_n}}\right)
E\left[1-e^{-v_{Y_n} \sigma_{Y_n}}\right]
\nnb
&\quad =
\left(\prod_{j=0}^{n-1}\frac{\bar\beta_{Y_j}}{\bar\beta_{Y_j} + v_{Y_j}}\right)
\left(\frac{1}{v_{Y_n}}\right)
\left(1-\frac{\bar\beta_{Y_n}}{\bar\beta_{Y_n} + v_{Y_n}} \right)
\nnb
&\quad =
\left(\prod_{j=0}^{n-1}\frac{\bar\beta_{Y_j}}{\bar\beta_{Y_j} + v_{Y_j}}\right)
\left(\frac{1}{\bar\beta_{Y_n} + v_{Y_n}}  \right) .
\end{align}
When we substitute this into \refeq{CTSRWE}, and use the definition of
$p(Y)$, we find that
\begin{align}
\int_0^\infty
    E_x \left( e^{-\sum_{u} v_u  L_{T,u}}\; \1_{X(T)=y} \right)   dT
&= \sum_{n=0}^\infty \sum_{Y \in \mathcal{W}_{n}(x,y)}
\prod_{i=1}^{|\omega|} \beta_{\omega_{i-1}\omega_i}
    \prod_{j=0}^{|\omega|} \frac{1}{\bar\beta_{\omega_j} + v_{\omega_j}}.
\lbeq{CTSRWab}
\end{align}
The above right-hand side is $(-\Delta_\beta + V)^{-1}_{xy}$, by
Exercise~\ref{ex:srwbeta}, and the proof is
complete.
\end{proof}

\subsection{BFS--Dynkin isomorphism}
\label{sec:BFS--Dynkin-1}

We now prove the \emph{BFS--Dynkin isomorphism} \cite{BFS82,Dynk83},
which relates the local time of the continuous-time random walk
$X=(X(t))$ with generator $\Delta_\beta$ to the $n$-component GFF
specified in terms of the same coupling constants $\beta$ (see
Section~\ref{sec:gff}).

For an $n$-component
field $(\varphi_x)_{x\in\Lambda}$, we define $(\tau_x)_{x\in\Lambda}$
by
\begin{equation} \label{e:bfstaudef}
  \tau_x = \frac12 |\varphi|^2_x = \frac12( \varphi_x^1\varphi_x^1 + \cdots + \varphi_x^n\varphi_x^n ).
\end{equation}
We again write $E_x$ for the expectation when the initial condition is
$X(0)=x$, and write $L_{T} = (L_{T,u})_{u\in\Lambda}$ for the local
time field.

The term ``isomorphism'' is commonly used as an expression of
\refeq{BFS-Dynkin} as the statement that
\begin{equation}
\lbeq{tausigned}
    \tau \;\; \text{under the signed measure $\varphi_x^1\varphi_y^1 \mathbb{P}_{\rm GFF}$}
\end{equation}
and
\begin{equation}
\lbeq{taupositive}
    L+ \tau \;\; \text{under the positive measure $ P_{xy}\mathbb{P}_{\rm GFF}$}
\end{equation}
have the same distribution, where $P_{xy}$ is the random walk measure
(integrated over $T$) and $\mathbb{P}_{\rm GFF}$ is the GFF measure.
For a systematic development of the isomorphism theorem and its
applications, see \cite{Szni12}.

\begin{theorem}
\label{thm:BFS-Dynkin} Let $n \geq 1$.  Let $F: \R_{+}^{\Lambda} \to
\R$ be such that there exists an $\epsilon >0$ such that $e^{\epsilon
\sum_{z\in \Lambda}t_{z}} F (t)$ is a bounded Borel function.  Then
\begin{multline}\label{e:BFS-Dynkin}
  \int_{\R^{n \Lambda}} e^{-\frac12 (\varphi,-\Delta_\beta\varphi)}
  F(\tau) \varphi_x^1\varphi_y^1
  \, d\varphi
  \\
  =
  \int_{\R^{n \Lambda}} e^{-\frac12 (\varphi,-\Delta_\beta\varphi)}
  \int_0^\infty
  E_x\big(F(\tau+L_T) 1_{X(T)=y}\big) \, dT
  \, d\varphi
  .
\end{multline}
\end{theorem}

\begin{proof}
We first consider the special case $F(t) = e^{-(v,t)}$ with $\text{Re}
\, v_z > 0$ for all $z\in\Lambda$.  In this case, by
Lemma~\ref{lem:CTSRW}, the right-hand side of \eqref{e:BFS-Dynkin} is
\begin{equation}
\lbeq{BFSpf1}
  (-\Delta_\beta+V)^{-1}_{xy}
  \int_{\R^{n\Lambda}} e^{-\frac12 (\varphi,(-\Delta_\beta+V)\varphi)}
  \, d\varphi
  .
\end{equation}
On the other hand, except for a missing normalisation, the
left-hand
side of \eqref{e:BFS-Dynkin} is a Gaussian correlation.  By
\eqref{e:Gauss-moments124}, it is also equal to \refeq{BFSpf1}.  This
proves \eqref{e:BFS-Dynkin} for the special case $F(t) = e^{-(v,t)}$.
In the rest of the proof, we reduce the general case to this special
case by writing $F$ as a superposition of exponentials using the
Fourier inversion theorem.

By hypothesis, $|F(t)| \le C e^{-\epsilon \sum_{z\in \Lambda}t_{z}}$
for some constant $C$.  Therefore the integrands in the left- and
right-hand sides of \eqref{e:BFS-Dynkin} are integrable by the
previous paragraph. Integrability is all that the rest of the proof
requires.

By considering the positive and negative parts of $F$ it suffices to
consider $F \ge 0$, and by replacing $F$ by its product with a
compactly supported characteristic function and using the monotone
convergence theorem, we may assume that $F$ has compact support in the
quadrant $\R_{+}^{\Lambda}$. By extending $F$ by zero outside the
quadrant we regard it as a function of compact support in
$\R^{\Lambda}$. By convolving $F$ by a smooth approximate identity of
compact support and using dominated convergence, we can further assume
that $F$ is smooth and compactly supported in $\R^{\Lambda}$.
Finally, we define a smooth compactly supported function $G$ by $G(t)
= F (t) e^{\sum_{z}t_{z}}$.

Since $G \in \Ccal_{0}^{\infty}$, it is a Schwartz function.
Let $\hat G(r)$ denote its Fourier transform.
By
applying the Fourier inversion theorem to $G$,
\begin{equation}
    F (t)
    =
    G (t) e^{- \sum_{z} t_{z}}
    =
    (2\pi)^{-|\Lambda|}
    \int_{\R^{\Lambda}} e^{-\sum_{z\in\Lambda} (1-ir_{z}) t_z} \hat G(r) dr ,
\end{equation}
and the integral converges absolutely because $\hat G$ is a Schwartz
function. By inserting this formula into the left- and right-hand sides
of \eqref{e:BFS-Dynkin} and bringing the integral over $r$ outside all
other integrals, \eqref{e:BFS-Dynkin} is reduced to the exponential
case established in the first paragraph. The proof is complete.
\end{proof}

Let $g>0$ and $\nu \in \R$.  With the choice $F(t) =
e^{-\sum_{x\in\Lambda} (gt_x^2+\nu t_x)}$, the left-hand side of
\eqref{e:BFS-Dynkin} becomes the (unnormalised) two-point function of
the $|\varphi|^4$ model.  Thus the right-hand side provides a random
walk representation for the $|\varphi|^4$ two-point function.
This random walk representation can be a point of departure for the
analysis of the $|\varphi|^4$ model and is used, e.g., in
\cite{Froh82,BFS83,BHH18}.

\section{Supersymmetric representation}
\label{sec:susy}

In this section, we derive a supersymmetric integral representation of
the two-point function \refeq{WSAW2pt} for the continuous-time weakly
self-avoiding walk in finite volume.  The representation is given in
\refeq{intrepC}.  It involves the introduction of an anti-commuting
fermion field, which we present as the differential of the boson
field.  We provide here a self-contained introduction to the fermion
field.

\subsection{The case \texorpdfstring{$n=0$}{n=0}}
\label{sec:n0}

In 1972, de Gennes \cite{Genn72} argued that the self-avoiding
walk corresponds to the case ``$n=0$'' of an $n$-component spin model.
De Gennes's observation has been very productive in physics, and leads
to predictions for critical exponents for self-avoiding walk by
setting $n=0$ in the $n$-dependent formulas for the critical exponents
of the $n$-component $|\varphi|^4$.  However, it has been much less
productive in mathematics, where the notion of a zero-component field
raises obvious concerns, and a rigorous link between the critical
behaviour of the self-avoiding walk and $n$-component spins has been
elusive.  An exception is Theorem~\ref{thm:wsaw4} and its related
results, where the $n=0$ connection plays a central role.

As noted already above, the results of Theorem~\ref{thm:wsaw4} agree
with the result of setting $n=0$ in Theorem~\ref{thm:phi4}, consistent
with de Gennes's prediction.  In fact, the renormalisation group
method used to prove Theorems~\ref{thm:wsaw4} and \ref{thm:phi4} is
mainly the same and the proofs are largely simultaneous.  Here the
correspondence between the self-avoiding walk and $n=0$ arises from
another mechanism.  Roughly speaking, this mechanism is based on the
observation that an $n$-component boson field contributes a factor $n$
for every loop, but an $n$-component fermion (anti-commuting) field
contributes $-n$.  Combined, all loops cancel.  This observation was
first made in the physics literature \cite{PS80,McKa80,Lutt83}, and
mathematically rigorous versions are developed in
\cite{LeJa87,BM91,BEI92,BIS09}.  Applications in this spirit can be found in
\cite{MS08,BBS-saw4-log,BBS-phi4-log,ST-phi4}.

Supersymmetric representations have had wider application than just to
self-avoiding walks.
Linearly reinforced walks are related to spin systems with hyperbolic
symmetry.  In particular, a relation between supersymmetric hyperbolic
sigma models and reinforced walks was found in \cite{ST15} and a
hyperbolic analogue of the BFS--Dynkin isomorphism theorem in
\cite{BHS18}.  Supersymmetric hyperbolic sigma models have been
studied in particular in \cite{DSZ10,DS10,DMR14}.  For further
references, see \cite{BHS18}.

In the remainder of this chapter, we provide an introduction to
supersymmetry and demonstrate the $n=0$ correspondence, by obtaining a
functional integral representation for the continuous-time weakly
self-avoiding walk that is a supersymmetric version of the 2-component
$|\varphi|^4$ model.  The supersymmetric representation places the
weakly self-avoiding walk within a similar framework as the
$|\varphi|^4$ model, with the important new ingredient that a fermion
(anti-commuting) field appears.  It is via this framework that we are
able to treat the self-avoiding walk as the $n=0$ version of the
$|\varphi|^4$ model.

\subsection{Integration of differential forms}
\label{sec:forms}

For our treatment of the fermion field, we require some minimal
background on the integration of differential forms, which we discuss
now.  An elementary introduction to differential forms can be found in
\cite{Rudi76}.

\index{Grassmann algebra}
\index{Differential forms}
Let $\Lambda$ be a finite set. For $x\in\Lambda$,
let $(u_x,v_x)$ be real coordinates.  The 1-forms
$du_x,dv_x$, for $x\in\Lambda$,
generate the Grassmann algebra of differential
forms on $\R^{2 \Lambda}$, with multiplication given by the anti-commuting
wedge product.  In particular,
\begin{equation}
    du_x \wedge du_y = - du_y \wedge du_x,
    \quad
    du_x \wedge dv_y = - dv_y \wedge du_x,
    \quad
    dv_x \wedge dv_y = -dv_y \wedge dv_x.
\end{equation}
It follows that, e.g., $du_x \wedge du_x =0$.

For $p \ge 0$, a $p$-\emph{form} is a function of $u,v$ times a
product of $p$ differentials, or any sum of such terms.  A \emph{form}
$K$ is a sum of $p$-forms with terms possibly having different values
of $p$.  The largest such $p$ is called the \emph{degree} of $K$, and
the $p$-form contribution to $K$ is called its \emph{degree}-$p$ part.
A form which is a sum of $p$-forms for even $p$ only is called
\emph{even}.  The wedge product of any form with itself is zero, by
anti-commutativity of the product.  The \emph{standard volume form} on
$\R^{2\Lambda}$ is
\begin{equation}
    du_1 \wedge dv_1 \wedge \cdots du_{|\Lambda|} \wedge dv_{|\Lambda|},
\end{equation}
where $1,\dots, |\Lambda|$ is any fixed enumeration of $\Lambda$.
Any  $2|\Lambda|$-form $K$ can be written as
\begin{equation}
    K = f(u,v) du_1 \wedge dv_1 \wedge \cdots du_{|\Lambda|} \wedge dv_{|\Lambda|}.
\end{equation}
There is no non-zero form of degree greater than $2|\Lambda|$, so degree $2|\Lambda|$
is naturally referred to as \emph{top degree}.

We define the \emph{integral} as a linear map from forms to $\R$, with
\begin{equation}
\label{e:Kintdef}
    \int K =
    \begin{cases}
    0 & (\deg K < 2|\Lambda|)
    \\
    \int_{{\mathbb R}^{2\Lambda}} f(u,v) du_1 dv_1 \cdots du_{|\Lambda|} dv_{|\Lambda|}
    & (K \; \text{is a $2|\Lambda|$-form}),
    \end{cases}
\end{equation}
where the integral on the right-hand side is the Lebesgue integral of
$f$ over $\R^{2\Lambda}$.  It is natural to define the integral to be zero
when $\deg K < 2|\Lambda|$, just as we do not give a significance to
$\int_{\R^2} f(u_1,v_1)du_1$, but rather require $\int_{\R^2}
f(u_1,v_1)du_1dv_1$ instead.  We define a form $f(u,v) \, du_{x_1}
\wedge \cdots \wedge du_{x_k} \wedge dv_{y_1} \wedge \cdots dv_{y_l}$
to be \emph{integrable} if $f$ is Lebesgue integrable on
$\R^{2\Lambda}$. Any form $K$ is a sum of such forms and we define $K$ to be
integrable if all terms in this sum are integrable. In particular, the
integral \refeq{Kintdef} exists.

The above formalism leads to attractive formulas when translated
into complex variables.  For this, we define
\begin{equation}
\label{e:cbf}
    \phi_x = u_x+iv_x, \quad \bar\phi_x = u_x-iv_x, \quad
    d \phi_x = du_x+idv_x, \quad d\bar\phi_x = du_x-idv_x.
\end{equation}
We call $(\phi,\bar\phi)$ the complex \emph{boson field}.
By definition,
\begin{equation}
    d\bar\phi_x \wedge d\phi_x = 2i du_x \wedge dv_x.
\end{equation}
The product
\begin{equation}
\lbeq{phiLeb}
    \bigwedge_{x\in\Lambda} (d\bar\phi_x \wedge d\phi_x)
    = (2i)^{|\Lambda|} du_1\wedge dv_1\wedge \cdots \wedge du_{|\Lambda|}\wedge dv_{|\Lambda|}
\end{equation}
defines a top degree form, which we abbreviate as $d\bar\phi
d\phi$. Thus $d\bar\phi d\phi$ is $(2i)^{|\Lambda|}$ times the standard volume
form, which becomes the Lebesgue measure under an integral over
$\R^{2M}$.  The order of the product on the left-hand side of
\refeq{phiLeb} is unimportant, since each factor is an even form.
However a change in the order of the product on the right-hand side
may introduce a sign change.

We write (with any fixed choice of the square root)
\begin{equation}
\lbeq{psipsibardef}
    \psi_x = \frac{1}{\sqrt{2\pi i}} d\phi_x, \quad
    \bar\psi_x = \frac{1}{\sqrt{2\pi i}} d\bar\phi_x,
\end{equation}
and call $(\psi,\bar\psi)$  the \emph{fermion field}.
Then
\begin{equation}
\lbeq{psibarpsi}
    \bar\psi_x\wedge\psi_x = \frac{1}{2\pi i}d\bar\phi_x \wedge d\phi_x
    = \frac{1}{\pi} du_x \wedge dv_x.
\end{equation}
Given a $\Lambda \times \Lambda$ complex matrix $A$, we define
\begin{equation}
\label{e:SAdef}
    S_A = \phi A\bar\phi + \psi  A \bar\psi
    = \sum_{x,y\in\Lambda} \left( \phi_x A_{xy}\bar\phi_y + \psi_x \wedge A_{xy}\bar\psi_y \right).
\end{equation}

For $J \in \N$, consider a $C^\infty$ function $F : \R^{J} \to\mathbb
C$.  Let $K=(K_j)_{j \le J}$ be a collection of even forms, and assume
that the degree-zero part $K_j^0$ of each $K_j$ is real.  We define a
form denoted $F(K)$ by Taylor series about the degree-zero part of
$K$, i.e.,
\begin{equation}
\label{e:Fdef}
    F(K) = \sum_{\alpha} \frac{1}{\alpha !}
    F^{(\alpha)}(K^{0})
    (K - K^{0})^{\alpha}.
\end{equation}
Here $\alpha = (\alpha)_{j \le J}$ is a multi-index, with $\alpha ! =
\prod_{j=1}^{ J}\alpha_j !$ and $(K - K^{0})^{\alpha}
=\bigwedge_{j=1}^{ J} (K_{j} - K_{j}^{0})^{\alpha_{j}}$.  The order of
the product does not matter since each $K_j-K_j^0$ is even by
assumption.  Also, the summation terminates as soon as $\sum_{j=1}^{
J}\alpha_j=M$ since each non-zero $K_j-K_j^0$ has degree at least $2$,
so $(K-K^0)^\alpha$ is a sum of $p$-forms with $p>2M$ when
$\sum_{j=1}^{ J}\alpha_j>M$, and forms beyond top degree vanish.  Thus
all forms $F(K)$ defined in this way are polynomials in the fermion
field.  For example,
\begin{align}
    e^{-(\phi_x\bar\phi_x +\psi_x\wedge\bar\psi_x)}
    &= e^{-\phi_x\bar\phi_x} e^{-\psi_x\wedge\bar\psi_x}
    \nnb &
    = e^{-\phi_x\bar\phi_x} \sum_{n=0}^\infty \frac{(-1)^n}{n!}(\psi_x\wedge\bar\psi_x)^{n}
    = e^{-\phi_x\bar\phi_x} (1-\psi_x\wedge\bar\psi_x),
\\
\lbeq{eSAexp}
    e^{-S_A} &= e^{-\phi A\bar\phi} \sum_{n=0}^M \frac{(-1)^n}{n!}(\psi A\bar\psi)^n.
\end{align}

\begin{example} \label{example:detA-1d}
Let $|\Lambda|=1$.  By definition of the integral and \refeq{psibarpsi},
\begin{align}
    \int e^{-a\phi\bar\phi - a\psi\wedge \bar\psi}
    & =
    \int e^{-a\phi\bar\phi} a(-\psi \wedge \bar\psi)
    \nnb & =
    \int e^{-a(u^2+v^2)} \frac{a}{\pi} du  dv
    =
    \left(\frac{1}{\sqrt{\pi}} \int e^{-t^2} dt\right)^2 = 1.
\lbeq{selfnorm1}
\end{align}
The factors $\frac{1}{\sqrt{2\pi i}}$ are included in
\refeq{psipsibardef} precisely in order to normalise the above
integral.
\end{example}

The scaling of the constant $a$ in \refeq{selfnorm1} could also have
been done earlier, by scaling $\phi$ and $\psi$ simultaneously:
\begin{align}
    \int e^{-a\phi\bar\phi - a\psi\wedge \bar\psi}
    & =
    \int e^{-\phi\bar\phi - \psi\wedge \bar\psi}.
\end{align}
This properly accounts for the change of variables in the Lebesgue
integral, since $\psi$ is proportional to $d\phi$.  This principle
generalises to higher dimensional integrals and is used in the proof
of the next lemma.  Its hypothesis that $A$ is an $M\times M$ matrix
with positive definite Hermitian part means that $\frac 12 (A+A^*)$ is
a strictly positive definite matrix, or, more explicitly, that
$\sum_{x,y\in\Lambda} \phi_x(A_{xy}+\bar{A}_{yx})\bar\phi_y >0$ for all
nonzero $\phi$.

\begin{lemma} \label{lem:detA-multi}
Let $\phi$ have components $\phi_x$ for $x\in\Lambda$, and let $A$ be a $\Lambda \times \Lambda$
matrix with positive definite Hermitian part.  Then
\begin{equation}
\lbeq{self-norm}
    \int e^{-S_A}
    =1.
\end{equation}
\end{lemma}

\begin{proof}
Consider first the case where $A$ is Hermitian, so there is a unitary
matrix $U$ and a diagonal matrix $D$ such that $A=U^{-1}DU$, so $\phi
A \phib = w D \bar{w}$ with $w = \bar{U}\phi$.  Then the change of
variables which replaces $\bar{U}\phi$ by $\phi$ and $\bar{U}\psi$ by
$\psi$ leads to
\begin{equation}
\lbeq{eSApf1}
    \int e^{-S_A}
    = \int e^{-\phi D\phib - \psi D \psib}.
\end{equation}
The integral on the right-hand side factors into a product of $|\Lambda|$
1-dimensional integrals which are all equal to $1$ by
Example~\ref{example:detA-1d}.  This proves the result in the
Hermitian case.

For the general case, we write $A (z) =G+iz H$ with $G= \frac 12
(A+A^*)$, $H=\frac{1}{2i}(A-A^*)$ and $z=1$.  Since $\phi (iH) \phib$
is imaginary, when $G$ is positive definite the integral of $e^{-S_A}$
converges and defines an analytic function of $z$ in a neighborhood of
the real axis.  Furthermore, for $z$ small and purely imaginary, $A
(z)$ is Hermitian and positive definite, and hence \eqref{e:eSApf1}
holds in this case.  Therefore \refeq{eSApf1} hold for all real $z$
(in particular for $z=1$) by uniqueness of analytic extension.
\end{proof}

The following exercise gives an instructive alternate proof of
Lemma~\ref{lem:detA-multi}, via an argument involving cancellation of
determinants.

\begin{exercise} \label{ex:detcancellation}
Let $A$ be an $\Lambda\times\Lambda$ matrix with positive definite Hermitian
part, and let $\phi$ be a complex field indexed by $\Lambda$.

\smallskip\noindent
(i)
Show that
\begin{equation}
\lbeq{Aint}
    \int_{\R^{2\Lambda}}
    e^{-\phi A\bar\phi} \;d\bar\phi d\phi =
    \frac{(2\pi i)^{|\Lambda|}}{\det A}.
\end{equation}
Since $d\bar\phi d\phi$ is a multiple of the standard volume form, the
integral in \refeq{Aint} is the Lebesgue integral of a complex
function.

\smallskip \noindent
(ii)
Show that the degree-$2|\Lambda|$ part of $e^{-\psi A\bar\psi}$ is $(\det A)
\bar\psi_1\wedge \psi_1 \wedge \cdots \wedge \bar\psi_{|\Lambda|}\wedge \psi_{|\Lambda|}$,
and use this with \refeq{eSAexp} to show that the degree-$2|\Lambda|$ part of
$e^{-S_A}$ is
\begin{equation}
\lbeq{topdeg}
  (\det A) e^{-\phi A \phib} \; \frac{d\phib d\phi}{(2\pi i)^{|\Lambda|}},
\end{equation}
and hence that
\begin{equation}
  \lbeq{self-norm-exercise}
  \int e^{-S_A}
  =1
  .
\end{equation}
\solref{detcancellation}
\end{exercise}

The following exercise makes a connection between integration of
$0$-forms and Gaussian integration as discussed in
Chapter~\ref{ch:gauss}.

\begin{exercise} \label{ex:fermionsGauss}
Let $C$ be a positive definite real symmetric $\Lambda\times\Lambda$ matrix, and
let $A=C^{-1}$.  Let $\phi=u+iv$ be a complex field indexed by $\Lambda$.
We write $u_x=\frac{1}{\sqrt{2}}\varphi_x^1$ and $v_x =
\frac{1}{\sqrt{2}}\varphi_x^2$, and regard
$\varphi=(\varphi^1,\varphi^2)$ as a 2-component real field.  Let
$\Ex_C$ be the Gaussian measure with respect to which $\varphi$ is a
2-component Gaussian field with covariance $C$ as in
Example~\ref{example:gauss-vect}.  Show that if $f$ is a $0$-form
(function) then
\begin{equation}
\lbeq{SAGaussian}
    \int e^{-S_A} f = \Ex_C f.
\end{equation}
In particular,
\begin{equation} \label{e:SAGauss}
  \int e^{-S_A} \phi_x\phib_y = C_{xy}
  .
\end{equation}
\solref{fermionsGauss}
\end{exercise}

Theorem~\ref{thm:BFS-Dynkin} is a representation for the two-point
function of an $n$-component boson field.  The next proposition
extends this representation to include the fermion field.
Let
\begin{equation}
    \tau_{x}
    =
    \phi_{x}\bar\phi_{x} + \psi_{x}\wedge \bar \psi_{x} .
\end{equation}

\begin{prop}
\label{prop:BFS-Dynkin} Let $A = -\Delta_{\beta}$.
Let $F:\R^{\Lambda}\to \R$ be
such that $e^{\epsilon \sum_{z\in \Lambda}t_{z}} F (t)$ is a Schwartz
function for some $\epsilon >0$.  Then
\begin{equation}\label{e:BFS-Dynkin-2}
  \int
  e^{-S_{A}}
  F(\tau)  \phi_x\bar\phi_y
  =
  \int
  e^{-S_{A}}
  \int_0^\infty
  E_x\big(F(\tau+L_T) 1_{X(T)=y}\big) \, dT
  .
\end{equation}
\end{prop}

\begin{proof}
We follow the same strategy as in the proof of
Theorem~\ref{thm:BFS-Dynkin}, and first consider the special case
$F(t) = e^{-(v,t)}$ with $\text{Re} \, v_z > 0$ for all $z\in\Lambda$.
In this case, the right-hand side of \eqref{e:BFS-Dynkin-2} is
\begin{align}
  \int
  e^{-S_{A}}
  \int_0^\infty
  E_x\big(F(\tau+L_T) 1_{X(T)=y}\big) \, dT
  & =
  \int e^{-S_{A+V}}
  E_x\big(e^{-\sum_u v_u L_{T,u}} 1_{X(T)=y}\big) \, dT
  \nnb & =
  (A+V)^{-1}_{xy} \int e^{-S_{A+V}}
  \nnb & =
  (A+V)^{-1}_{xy}
  ,
\end{align}
where we used Lemma~\ref{lem:CTSRW} and then
Lemma~\ref{lem:detA-multi} for the last two equalities.  On the other
hand, by \eqref{e:SAGauss} the left-hand side of
\eqref{e:BFS-Dynkin-2} is now
\begin{align}
  \int
  e^{-S_{A+V}}
  \phi_x\bar\phi_y = (A+V)^{-1}_{xy}.
\end{align}
This proves \refeq{BFS-Dynkin-2} for $F(t) = e^{-(v,t)}$.

For the general case, let $G$ be the Schwartz function given by $F (t)
= G (t) e^{- \epsilon \sum_{z} t_{z}}$.  As in the proof of
Theorem~\ref{thm:BFS-Dynkin}, we again write $F$ in terms of the
Fourier transform of $G$, as
\begin{equation}
    F (t)
    =
    G (t) e^{- \sum_{z} t_{z}}
    =
    (2\pi)^{-|\Lambda|}
    \int_{\R^{\Lambda}} e^{-\sum_{z\in\Lambda} (1-ir_{z}) t_z} \hat G(r) dr . 
\end{equation}
This equation remains valid with $\tau$ in place of $t$ because
equality for all $t$ implies both sides have the same Taylor
expansions about $t$, and we can again interchange the order of
integration to conclude the general case from the special case already
verified.
\end{proof}

Since the left-hand side of \eqref{e:BFS-Dynkin-2} depends only on the
restriction of $F$ to the quadrant $\R_{+}^{\Lambda}$, it is more
natural to formulate Proposition~\ref{prop:BFS-Dynkin} for a smooth
function $F$ defined on $\R_{+}^{\Lambda}$; such a formulation can be
found in \cite[Proposition~4.4]{BIS09}.

Proposition~\ref{prop:BFS-Dynkin} generalises to models with
$n$-component boson fields and $m$-component fermion fields with $m$
even, but we do not make this claim precise because we are in the
special situation where the fermions have been identified with
differential forms.  The more general concept of Grassmann integration
is needed for the case where $m \not = n$. In the present case
Proposition~\ref{prop:BFS-Dynkin} has a surprising simplification (see
Corollary~\ref{cor:BFS-Dynkin}), via the \emph{localisation theorem}
discussed below.

\subsection{Localisation theorem and weakly self-avoiding walk}
\label{sec:BFS--Dynkin-2}

The following theorem is the localisation theorem.  A more general
theorem is proved in Section~\ref{sec:susy2}, with a different and
revealing proof.

\index{Localisation theorem}
\begin{theorem}
\label{thm:F0}
For $F$ as in Proposition~\ref{prop:BFS-Dynkin} and any $A$ with
non-negative real part,
\begin{equation}
\label{e:F0}
    \int e^{-S_A} F(\tau) = F(0).
\end{equation}
\end{theorem}

\begin{proof}
The proof again follows by checking the case $F(t) = e^{-(v,t)}$, as
in the proof of Proposition~\ref{prop:BFS-Dynkin}.  Instead of
Lemma~\ref{lem:CTSRW}, the special case is handled using
\refeq{self-norm}.
\end{proof}

\begin{cor}\label{cor:BFS-Dynkin}
With $A=-\Delta_{\beta}$ and $F$ as in
Proposition~\ref{prop:BFS-Dynkin},
\begin{equation}
\label{e:FDynkin}
    \int e^{-S_A} F(\tau) \bar\phi_x \phi_y
    =
    \int_0^\infty
    E_x\big(F(L_T) 1_{X(T)=y}\big) \, dT
  .
\end{equation}
\end{cor}

\begin{proof}
The left-hand side of \eqref{e:FDynkin} is equal to the right-hand side of
\refeq{BFS-Dynkin-2}, and the latter is equal to
$\int_0^\infty  E_x\big(F(0+L_T) 1_{X(T)=y}\big) \, dT$ by
Theorem~\ref{thm:F0}.
\end{proof}

The right-hand side of \refeq{FDynkin} arises precisely because the
outer integral on the right-hand side of \refeq{BFS-Dynkin-2} simply
evaluates the inner integral at $\tau=0$.  This is a rigorous
implementation of the idea that the case of ``$n=0$'' components
corresponds to self-avoiding walk.  As with
\refeq{tausigned}--\refeq{taupositive}, the identity \refeq{FDynkin}
is commonly expressed as an ``isomorphism,'' in the sense that $\tau$
on the left-hand side can be loosely interpreted as having the same
distribution under $e^{-S_A}$ as does $L_T$ under $E_x$ on the
right-hand side.

Corollary~\ref{cor:BFS-Dynkin} provides the supersymmetric
representation for the weakly self-avoiding walk
two-point function.  The supersymmetric representation is actually a
representation for a finite-volume version of the two-point function.
The convergence of the finite-volume two-point function to its
infinite-volume counterpart is not difficult, and can be found in
\cite{BBS-saw4-log}.  Here we restrict attention to the finite-volume
two-point function.

We fix $N$, let $\Lambda=\Lambda_N ={\mathbb Z}^d/L^N {\mathbb Z}^d$,
and let $E_0^N$ denote the expectation for the continuous-time simple
random walk on the discrete torus $\Lambda_N$.  For $g>0$ and $\nu \in \R$, as in
\refeq{WSAW2pt}, the \emph{finite-volume} two-point function is
defined by
\begin{equation}
\lbeq{GN}
    G_{0x}^N(g,\nu)
    = \int_0^\infty
    E_0^N
    \left( e^{-gI(T)} \1_{X(T)=x} \right) e^{-\nu T}
    dT.
\end{equation}

\begin{exercise} \label{ex:fin-vol-G0x}
Show that $G_{0x}^N(g,\nu)$ is finite for all $\nu \in \R$, provided
$g>0$.  This is clearly not the case when $g=0$.  Hint: use the
Cauchy--Schwarz for $\sum_{u\in \Lambda} L_{T,u}$.
\solref{fin-vol-G0x}
\end{exercise}

Our analysis of the $|\varphi|^4$ model relies on its formulation as a
perturbation of a Gaussian free field, and on the convolution property
of the corresponding Gaussian expectation.  The extension to the
supersymmetric setting relies on analogous properties of the Gaussian
super-expectation.

\begin{defn}
\label{def:super-expectation}
Let $C$ be a real symmetric positive definite $\Lambda \times \Lambda$ matrix.
Let $A=C^{-1}$.  The Gaussian \emph{super-expectation} with covariance
$C$, of a form $K$, is defined by
\begin{equation}
   {\sf E}_C K = \int K e^{-S_A},
\end{equation}
where the integral on the right-hand side is defined by
\refeq{Kintdef}.
\end{defn}

According to Exercise~\ref{ex:fermionsGauss}, the super-expectation of
a $0$-form $f$ is equal to the usual Gaussian expectation, i.e.,
\begin{equation}
\lbeq{sfECf}
    {\sf E}_C f
    =
    \Ex_C f.
\end{equation}
However, the super-expectation can also be applied to an arbitrary
differential form.  Particular cases of \refeq{sfECf}, seen already in
Lemma~\ref{lem:detA-multi}, are the self-normalising property ${\sf
E}_C 1 = 1$ and the identity ${\sf E}_C \bar\phi_x\phi_y = C_{xy}$.
In many ways, the properties of the Gaussian super-expectation
parallel those of the ordinary Gaussian expectation. In particular, it
satisfies a version of the convolution property. A systematic
introduction is provided in \cite{BS-rg-norm}.

Given any $m^2 >0$, let
\begin{equation}
  \nu_0=\nu-m^2, \qquad C=(-\Delta +m^2)^{-1}.
\end{equation}
Then Corollary~\ref{cor:BFS-Dynkin} can be restated in terms of the
super-expectation as
\begin{align} \label{e:intrepC}
    G_{0x}^N(g,\nu)
    =
    {\sf E}_C \left( e^{-\sum_{y\in\Lambda} ( g\tau_y^2 + \nu_0\tau_y  )}
    \bar{\phi}_0 \phi_x  \right).
\end{align}
This gives a supersymmetric representation for $G_{g,\nu}^{N}(x)$.
Its origins include \cite{PS80,McKa80,Lutt83,LeJa87} and, in the form
presented here, \cite{BI03d,BIS09}.

Note that there is no dependence on $m^2$ in \refeq{intrepC}, and its
introduction is simply to regularise the Laplacian so that $C$ is
well-defined.  The right-hand side of \refeq{intrepC} is the two-point
function of a supersymmetric field theory with boson field $(\phi,
\bar\phi)$ and fermion field $(\psi,\bar\psi)$.  The supersymmetric
representation allows a unified treatment of both weakly self-avoiding
walk and $n$-component $|\varphi|^4$, with the former behaving as the
$n=0$ version of the latter.  It bears a strong resemblance to the
corresponding identity \refeq{Fexpectation} for $|\varphi|^4$, with
the simplification that the denominator (partition function) in
\refeq{Fexpectation} is replaced here by $1$.

\subsection{Localisation theorem and strictly self-avoiding walk}

Other models of self-avoiding walk also have integral representations.
In this section, we use the localisation theorem to obtain integral
representations for the strictly self-avoiding walk, from \cite{BIS09}.
We also present a representation for the edge self-avoiding walks known
as self-avoiding trails, from \cite{RS01}.
The proofs use the integration by parts formula given in the following
exercise.

\begin{exercise} \label{ex:SUSY-ibp}
Let $\Lambda$ be a finite set.  Extend Exercise~\ref{ex:ibp} to the
super-expectation defined in Definition~\ref{def:super-expectation},
i.e., verify the Gaussian integration by parts formula
\begin{equation}
\lbeq{complexibp}
    {\sf E}_C(\phib_x K) = \sum_{y \in \Lambda} C_{xy} {\sf E}_C\left (\ddp{K}{\phi_y}\right)
\end{equation}
for any form $K$ such that both sides converge absolutely.
\solref{SUSY-ibp}
\end{exercise}

We define the two-point function of weighted strictly self-avoiding
walk on an arbitrary finite set $\Lambda$, as follows.  For $n \ge 1$,
let $\Scal_n(x,y)$ denote the set of sequences $\omega=(\omega(0),
\omega(1),\ldots, \omega(n))$ with $\omega(i)\in\Lambda$ for all $i$,
$\omega(0)=x$, $\omega(n)=y$, and with $\omega(i) \neq \omega(j)$ for
all $i \neq j$.  Let $\Scal_0(x,y)$ be empty if $x \neq y$, and let
$\Scal_0(x,x)$ consist of the zero-step walk $\omega(0)=x$.  Let
$\Scal(x,y)=\cup_{n=0}^\infty \Scal_n(x,y)$.
Given a symmetric $\Lambda \times \Lambda$ matrix $W$ of edge weights,
for $\omega \in \Scal_n(x,y)$ we set $W^\omega = \prod_{i=1}^n W_{\omega(i-1),\omega(i)}$.
As usual, the empty product equals $1$ when $n=0$.
We define the \emph{weighted two-point function}
to be
\begin{equation}
\lbeq{Wsaw}
    \sum_{\omega \in \Scal(x,y)}W^\omega.
\end{equation}

The following proposition gives a representation for the two-point
function \refeq{Wsaw} with weights given by
\emph{positive definite} matrix $W=C$.

\begin{prop}
\label{prop:strictSAW}
  Let $A$ be positive definite and $C=A^{-1}$.
  Then for $x\neq y$
  \begin{equation}
  \lbeq{sawrep-strict}
    \sum_{\omega \in \Scal(x,y)}C^\omega =
    \int \phib_x\phi_y
    \prod_{z \in \Lambda\setminus\{x,y\}} (1+\tau_z) e^{-S_A} .
  \end{equation}
\end{prop}

\begin{proof}
The right-hand side of \eqref{e:sawrep-strict} is equal to
${\sf E}_C \bar{\phi}_x F$
with $F$ given by
$F = \phi_y \prod_{z \neq x,y}(1+\tau_z)$.  Computation of the derivative gives
\begin{equation}
\label{e:SUSY-Fder}
    \frac{\partial F}{\partial \phi_v}
    =
    \delta_{vy} \prod_{z \neq x,y}(1+\tau_z)
    +
    \1_{v \ne x,y} \phi_y \bar{\phi}_v \prod_{z \neq x,y,v}(1+\tau_z).
\end{equation}
Substitution of \eqref{e:SUSY-Fder} into the integration by parts
formula \eqref{e:complexibp}, followed by application of
the localisation theorem \eqref{e:F0},
gives
\begin{equation}
    {\sf E}_C \bar{\phi}_x F
    =
    C_{xy} + \sum_{v\ne x,y} C_{xv} {\sf E}_C
    \bar{\phi}_v \phi_y \prod_{z \neq x,y,v}(1+\tau_z).
\end{equation}
After iteration, the right-hand side gives the left-hand side of \refeq{sawrep-strict}.
\end{proof}

The strictly self-avoiding walk requires that no vertex be visited
more than once.  An alternate model allows vertices to be revisited but
prohibits edges from being visited more than once.  These edge self-avoiding
walks are commonly called \emph{self-avoiding trails} \cite{Hugh95}.
A precise definition is as follows.

Let $\Lambda$ be a finite set and let $E$ denote the set of all unordered
pairs of distinct points in $\Lambda$.  Then $(\Lambda,E)$ is the complete
graph on $|\Lambda|$ vertices.
For $n \geq 1$, let $\Tcal_n(x,y)$ denote the set of sequences $\omega=(\omega(0),
\omega(1),\ldots, \omega(n))$ with $\omega(i)\in\Lambda$, $\omega(0)=x$, $\omega(n)=y$, and
with the undirected edges $\{\omega(i),\omega(i+1)\}$ distinct for all $i < n$.  Let
$\Tcal_0(x,y)$ be empty if $x \neq y$, and let $\Tcal_0(x,x)$ consist
of the zero-step walk.
Let $\Tcal(x,y)=\cup_{n=0}^\infty \Tcal_n(x,y)$.
Given a $\Lambda \times \Lambda$ matrix $W$,
the \emph{weighted two-point function} of self-avoiding trails is defined by
\begin{equation} \label{e:edgesaw-defn}
  \sum_{\omega\in \Tcal(x,y)} W^{\omega}.
\end{equation}

The following exercise
provides an integral representation for this weighted two-point function
when the weights are given by a \emph{symmetric} matrix $W=\beta$
(not necessarily positive definite).
The representation is analogous to Proposition~\ref{prop:strictSAW},
and essentially appears in \cite{RS01}.
A similar formula holds for walks which do not revisit \emph{directed} edges.

The representation is stated in terms of the form
  \begin{equation} \label{e:tauxy-0}
    \tau_{xy} = \frac12  \left(
    \phi_x\bar\phi_y + \psi_x\wedge\bar\psi_y
    + \phi_y\bar\phi_x + \psi_y\wedge\bar\psi_x \right).
  \end{equation}
The solution to the exercise uses the extension of the localisation theorem
given in Theorem~\ref{thm:F0-bis} below, which implies in particular that
the identity \refeq{F0}, i.e., $\int e^{-S_A}F(\tau)=F(0)$, holds also when
$F$ is a function of $(\tau_{xy})$ rather than just a function of $(\tau_x)$ as
in Theorem~\ref{thm:F0}.

\begin{exercise} \label{ex:edgesaw}
Let $\beta$ be a symmetric matrix.  Show that
\begin{equation} \label{e:edgesaw-repr}
  \sum_{\omega\in \Tcal(x,y)} \beta^{\omega}
  =
  \int
  \phib_x\phi_y
  \prod_{\{u,v\}\in E} (1+2\beta_{uv} \tau_{uv})
  \prod_{w\in \Lambda} e^{-\tau_w},
\end{equation}
where $E$ is the set of edges in the complete graph $(\Lambda,E)$.
\solref{edgesaw}
\end{exercise}

\section{Supersymmetry and the localisation theorem}
\label{sec:susy2}

Integrals such as $\int e^{-S_A} F(\tau)$ are unchanged if we formally
interchange the pairs $\phi,\phib$ and $\psi,\psib$.
By \eqref{e:F0}, it is also true that $\int e^{-S_A}
F(\tau)\phib_a\phi_b = \int e^{-S_A} F(\tau)\psib_a\psi_b$ (the
difference is $\int e^{-S_A}\tau F(\tau) = 0$).  This is a
manifestation of a symmetry between bosons and fermions, called
\emph{supersymmetry}.  In this section, we use methods of
supersymmetry to provide an alternate proof of the localisation
theorem, Theorem~\ref{thm:F0}.  Ideas of this nature are discussed in
much more generality in \cite[Section~2]{Witt92}.  The localisation
theorem is related to the Duistermaat--Heckman formula and equivariant
cohomology; see, e.g, \cite{AB84,DH82,SZ97,Witt92}.

\subsection{The localisation theorem}

\index{Anti-derivation} We start with some definitions.  An
\emph{anti-derivation} $\Gamma$ is a linear map from the space of
forms to itself which obeys
\begin{equation}
    \Gamma(K_1\wedge K_2)
    = (\Gamma K_1)\wedge K_2 + (-1)^{p_1} K_1 \wedge (\Gamma K_2)
\end{equation}
when $K_1$ is a $p_1$-form.

For $x \in \Lambda$, we define
\begin{align}
\lbeq{ddphi}
    \frac{\partial}{\partial \phi_x}
    &=
    \frac{1}{2}\left(\frac{\partial}
    {\partial u_x} - i \frac{\partial}{\partial v_x} \right),
\quad
    \frac{\partial}{\partial \bar{\phi}_x}
    =
    \frac{1}{2}\left(\frac{\partial}
    {\partial u_x} + i \frac{\partial}{\partial v_x} \right).
\end{align}
\index{Grassmann calculus}%
The following definition provides a notion of differentiation of a
form with respect to the fermion field.  This is a standard notion in
Grassmann calculus (see, e.g., \cite{Bere66,FKT02,Salm99}).
For $x\in\Lambda$,
the derivatives $\frac{\partial}{\partial \psi_{x}}$
and $\frac{\partial}{\partial \psib_{x}}$ are defined as the
anti-derivations which obey the conditions:
\begin{align}
    &\frac{\partial \psi_{y}}{\partial \psi_{x}} = \frac{\partial \psib_{y}}{\partial \psib_{x}}
    = \delta_{xy},
    \quad
    \frac{\partial \psib_{y}}{\partial \psi_{x}} = \frac{\partial \psi_{y}}{\partial \psib_{x}}=0,
    \quad
    \frac{\partial f}{\partial \psi_{x}} = \frac{\partial f}{\partial \psib_{x}}=0,
\end{align}
for any $0$-form $f$.  It follows from the definition that the
derivatives anti-commute, i.e.,
\begin{equation}
    \frac{\partial}{\partial \psi_{x}}\frac{\partial}{\partial \psi_{y}}
    =
    -\frac{\partial}{\partial \psi_{y}}\frac{\partial}{\partial \psi_{x}},
    \quad
    \frac{\partial}{\partial \psib_{x}}\frac{\partial}{\partial \psib_{y}}
    =
    -\frac{\partial}{\partial \psib_{y}}\frac{\partial}{\partial \psib_{x}},
        \quad
    \frac{\partial}{\partial \psi_{x}}\frac{\partial}{\partial \psib_{y}}
    =
    -\frac{\partial}{\partial \psib_{y}}\frac{\partial}{\partial \psi_{x}}.
\end{equation}

\begin{example} \label{ex:SUSY-intderiv}
  For notational simplicity, let
  $\partial_{\psi_x} = \ddp{}{\psi_x}$ and $\partial_{\psib_x} = \ddp{}{\psib_x}$,
  and let $\partial_\psi \partial_{\psib}$ denote the product
    $\prod_{x\in\Lambda} \partial_{\psi_x}\partial_{\psib_x}$.
    For any $x\in\Lambda$,
  \begin{equation}
    K = \psi_x \partial_{\psi_x} K = \psib_x\psi_x \partial_{\psi_x}\partial_{\psib_x} K.
  \end{equation}
  In particular,
  \begin{equation}
    \int K =
    \frac{1}{\pi^{|\Lambda|}} \int_{\R^{2\Lambda}} \partial_\psi \partial_{\psib}K
    \frac{d\phib d\phi}{(2 i)^{|\Lambda|}},
  \end{equation}
  so $\int K$ is the Lebesgue integral over $\R^{2\Lambda}$ of the function ($0$-form)
  $\pi^{-|\Lambda|} \partial_\psi \partial_{\psib}K$.
\end{example}

\index{Supersymmetry generator}
\index{Supersymmetric}
\index{$Q$-closed}
\index{$Q$-exact}
The \emph{supersymmetry generator} $Q$ is the anti-derivation defined by
\begin{equation}
\label{e:Qhat}
    Q
    =
    \sum_{x\in \Lambda}
    \left(
    \psi_{x} \frac{\partial}{\partial \phi_{x}} +
    \bar\psi_{x} \frac{\partial}{\partial \bar\phi_{x}}-
    \phi_{x} \frac{\partial}{\partial \psi_{x}}  +
    \bar\phi_{x} \frac{\partial}{\partial \bar\psi_{x}}
    \right).
\end{equation}
In particular,
\begin{align}
\label{e:Qaction}
   &
   Q\phi_{x} = \psi_{x},
   \quad\quad
   Q\bar\phi_{x} = \psib_{x},
   \quad \quad
   Q\psi_{x} = - \phi_{x},
   \quad \quad
   Q\psib_{x} = \bar\phi_{x}
   .
\end{align}
An form $K$ is said to be \emph{supersymmetric} or $Q$-\emph{closed}
if $QK=0$.  A form $K$ that is in the image of $Q$ is called
$Q$-\emph{exact}.  Note that the integral of any $Q$-\emph{exact} form
is zero (assuming that the form decays appropriately at infinity),
since integration acts only on forms of top degree $2N$ and the
$\psi$-derivatives in \refeq{Qhat} reduce the degree to at most $2N -
1$, while the integral of the $\phi$-derivatives is zero by the
Fundamental Theorem of Calculus (recall \refeq{ddphi} and the fact
that integrals ultimately are evaluated as Lebesgue integrals).

\begin{example} \label{example:tauexact}
The form
  \begin{equation} \label{e:tauxy}
    \tau_{xy} = \frac12  \left(
    \phi_x\bar\phi_y + \psi_x\wedge\bar\psi_y
    + \phi_y\bar\phi_x + \psi_y\wedge\bar\psi_x \right)
  \end{equation}
  is both $Q$-exact and $Q$-closed.
  (Note that the degree-zero part of $\tau_{xy}$ is real.)
  In fact, it follows from \refeq{Qaction} that
  $Q(\phi_x\bar\psi_y) =  \phi_x\bar\phi_y + \psi_x\wedge\bar\psi_y$ and hence
  \begin{equation}
      \tau_{xy} = Q \lambda_{xy}, \quad
      \text{where} \quad
      \lambda_{xy} = \frac12( \phi_x\bar\psi_y +     \phi_y\bar\psi_x).
  \end{equation}
Similarly, it is easily verified that $Q\tau_{xy}=0$.
\end{example}

\begin{exercise}\label{ex:Qchain}
Prove that $Q$ obeys the chain rule for even forms, in the sense that
if $K = (K_j)_{j \le J}$ is a finite collection of even forms, and if
$F : \R^J \to \C$ is $C^\infty$, then
\begin{equation}
 \label{e:Qcr}
     Q(F(K)) = \sum_{i=1}^J F_i(K)
     QK_i,
\end{equation}
where $F_i$ denotes the partial derivative.
\solref{Qchain}
\end{exercise}

\begin{example}\label{example:FtauQ}
  Let $F : \R^{\Lambda \times \Lambda} \to \C$
  be a smooth function, and let $\tau=(\tau_{xy})$.
  Then $Q(F(\tau))=0$.
  In particular, $Qe^{-S_{A}}=0$ for any symmetric $\Lambda\times\Lambda$ matrix $A$.
\end{example}

\begin{proof}
  This follows from Exercise~\ref{ex:Qchain} and $Q\tau_{xy}=0$.
  For $e^{-S_A}$, it can alternatively be seen by expanding $e^{-S_{A}}$
  and applying $Q$ term by term using $Q\tau_{xy}=0$.
\end{proof}

The following version of the localisation theorem generalises
Theorem~\ref{thm:F0} and provides an alternative proof.

\index{Localisation theorem}
\index{Duistermaat--Heckman theorem}
\begin{theorem}
\label{thm:F0-bis}
Let $K$ be a smooth integrable $Q$-closed form, so $QK=0$. Then
\begin{equation}
\label{e:F0-bis}
    \int K = K^{0}(0) ,
\end{equation}
where $K^{0}(0)$ is the evaluation of the degree-zero part $K^0$ of
$K$ at $\varphi=0$.
\end{theorem}

\begin{proof}
Any integrable form $K$ can be written as $K=\sum_{\alpha}K^{\alpha
}\psi^{\alpha}$, where $\psi^{\alpha}$ is a monomial in
$\psi_{x},\psib_{x}, \, x \in \Lambda$, and $K^{\alpha}$ is an
integrable function of $\phi ,\phib$.  To emphasise this, we write
$K=K(\phi, \phib, \psi, \psib)$.  Let $S= \sum_{x\in\Lambda} (\phi_x\phib_x
+ \psi_x\wedge\psib_x)$.  Thus $S=S_A$ with $A=\Id$.

\smallskip\noindent
\emph{Step 1.} We prove the following version of Laplace's Principle:
\begin{equation} \label{e:localisation-Laplace}
  \lim_{t\to\infty} \int e^{-t S}K = K^0(0) .
\end{equation}
Let $t>0$.
We make the change of variables
$\phi_{x} = \frac{1}{\sqrt{t}}\phi'_{x}$ and $\psi_{x} =
\frac{1}{\sqrt{t}}\psi'_{x}$; since $\psi_{x}$ is proportional
to $d\phi_{x}$ this correctly implements the change of variables.
Let $\omega = -\sum_{x\in\Lambda} \psi_{x} \wedge \psib_{x}$.
After dropping the primes, we obtain
\begin{equation}
    \int e^{-t S}K
    =
    \int e^{-\sum_{x}\phi_{x} \phib_{x} + \omega}
    K(
    \tfrac{1}{\sqrt{t}}\phi,
    \tfrac{1}{\sqrt{t}}\phib,
    \tfrac{1}{\sqrt{t}}\psi,
    \tfrac{1}{\sqrt{t}}\psib) .
\end{equation}
To evaluate the right-hand side,
we expand $e^{\omega}$ and
and obtain
\begin{align}
\lbeq{etSK}
    \int e^{-t S}K
    &=
    \sum_{n=0}^{|\Lambda|}
    \int e^{-\sum_{x}\phi_{x} \phib_{x}} \frac{1}{n!}\omega^{n}
    K(
    \tfrac{1}{\sqrt{t}}\phi,
    \tfrac{1}{\sqrt{t}}\phib,
    \tfrac{1}{\sqrt{t}}\psi,
    \tfrac{1}{\sqrt{t}}\psib)
    .
\end{align}
We write $K=K^0+G$, where $G=K-K^0$ contains no degree-zero part.  The
contribution of $K^0$ to to \refeq{etSK} involves only the $n=|\Lambda|$ term
and equals
\begin{equation} \label{e:etSK0}
  \int e^{-t S}K^0 =
  \int e^{-\sum_{x}\phi_{x} \phib_{x}} \frac{1}{|\Lambda|!} \omega^{|\Lambda|}
    K^0(
    \tfrac{1}{\sqrt{t}}\phi,
    \tfrac{1}{\sqrt{t}}\phib)
    ,
\end{equation}
so by the continuity of $K^0$,
\begin{align}
    \lim_{t\to\infty}
    \int e^{-t S}K^0
    &=
    K^{0}(0)
    \int e^{-\sum_{x}\phi_{x} \phib_{x}} \frac{1}{|\Lambda|!}\omega^{|\Lambda|}
    =
    K^{0}(0)
    \int e^{- S}
    .
\end{align}
By  Lemma~\ref{lem:detA-multi} (with $A=\Id$), this proves that
\begin{align}
    \lim_{t\to\infty}
    \int e^{-t S}K^0
    &=
    K^{0}(0)
    .
\end{align}

To complete the proof of \eqref{e:localisation-Laplace}, it remains to
show that $\lim_{t\to\infty}\int e^{-t S}G = 0$.  As above,
\begin{equation}
    \int e^{-t S}G
    =
    \sum_{n=0}^{|\Lambda|} \int e^{-\sum_{x}\phi_{x} \phib_{x}} \frac{1}{n!}
    \omega^{n}\,
    G \left(
    \tfrac{1}{\sqrt{t}}\phi,
    \tfrac{1}{\sqrt{t}}\phib,
    \tfrac{1}{\sqrt{t}}\psi,
    \tfrac{1}{\sqrt{t}}\psib
    \right).
\end{equation}
Since $G$ has no degree-zero part, the term with $n=|\Lambda|$ is zero.  Terms
with smaller $n$ require factors $\psi\psib$ from $G$, which carry
inverse powers of $t$.  They therefore vanish in the limit, and the
proof of \eqref{e:localisation-Laplace} is complete.

\smallskip
\noindent
\emph{Step 2.}
The Laplace approximation is exact:
\begin{equation} \label{e:Laplace-exact}
  \int e^{-t S}K \;\; \text{ is independent of $t \ge 0$.}
\end{equation}
To prove this, recall from Example~\ref{example:tauexact} that $\tau_x
= Q\lambda_{x}$ where $\lambda_x = \lambda_{xx}$.  Let $\lambda =
\sum_{x\in\Lambda} \lambda_{x}$.  Then
\begin{equation}
    S
    =
    \sum_{x\in\Lambda}\tau_{x}
    =
    \sum_{x\in\Lambda} Q\lambda_{x}
    =
    Q \lambda .
\end{equation}
Also, $Q e^{-S}=0$ by Example~\ref{example:FtauQ}, and $QK=0$ by
assumption.  Therefore,
\begin{equation}
     \label{e:dlam}
    \frac{d}{dt} \int e^{-t S}K
    =
    -\int e^{-t S} S  K
    =
    -\int e^{-t S} (Q \lambda)  K
    =
    -\int Q\left(e^{-t S} \lambda K \right)
    =
    0 ,
\end{equation}
since the integral of any $Q$-exact form is zero.

\smallskip\noindent \emph{Step 3.} Finally, we combine Laplace's
Principle \eqref{e:localisation-Laplace} and the exactness of the
Laplace approximation \eqref{e:Laplace-exact}, to obtain the desired
result
\begin{equation}
    \int K
    =
    \lim_{t\rightarrow \infty}
    \int e^{-t S}K
    =
    K^{0} (0).
\end{equation}
This completes the proof.
\end{proof}

\begin{proof}[Alternate proof of Theorem~\ref{thm:F0}]
  By Example~\ref{example:FtauQ}, $Q(F(\tau))=0$ and $Qe^{-S_A}=0$.
  Since $Q$ is an anti-derivation, this gives
  $Q(e^{-S_A}F) = (Q e^{-S_{A}})F + e^{-S_A}QF =0$.
  Also, $e^{-S_A}F$ is integrable by the decay assumption in Theorem~\ref{thm:F0}.
  The claim therefore follows from Theorem~\ref{thm:F0-bis}.
\end{proof}

\subsection{Supersymmetry and exterior calculus}

As a final observation, we indicate how the supersymmetry generator
$Q$ can be expressed in terms of standard operations in differential
geometry, namely the exterior derivative, the interior product, and
the Lie derivative.

\index{Exterior derivative}
The {\em exterior derivative} $d$ is the  anti-derivation that maps
a form of degree $p$ to a form of degree $p+1$, defined by $d^2=0$ and,
for a zero form $f$,
\begin{equation}
    df = \sum_{x\in \Lambda} \Big(\frac{\partial f}{\partial \phi_x} d\phi_x
    + \frac{\partial f}{\partial \phib_x} d\phib_x \Big).
\end{equation}

Consider the flow acting on $\C^{\Lambda}$ defined by $\phi_x \mapsto
e^{ -2 \pi i \theta }\phi_x$.  This flow is generated by the vector
field $X$ defined by $X(\phi_x) = -2 \pi i \phi_x$, and $X(\phib_x) =
2 \pi i \phib_x$.  The action by pullback of the flow on forms is
 \begin{equation}
 d\phi_x
 \mapsto d (e^{-2 \pi i \theta}\phi_x)
  = e^{- 2 \pi i \theta} \, d\phi_x,  \quad \quad
 d\phib_x \mapsto e^{ 2 \pi i \theta}\, d\phib_x.
 \end{equation}
 The {\em interior product} ${\ci} = \ci_X$ with the vector field $X$
 is the  anti-derivation that maps forms of degree $p$ to forms of
 degree $p-1$ (and maps forms of degree zero to zero), given by
 \begin{equation}
 {\ci} d\phi_x = -2 \pi i \phi_x,  \quad\quad {\ci} d\phib_x = 2 \pi
 i \phib_x.
 \end{equation}
 The interior product obeys $\ci^2=0$.

\index{Lie derivative}
  The \emph{Lie derivative} $\cL = \cL_X$ is the infinitesimal flow
 obtained by differentiating with respect to the flow at $\theta = 0$.
 Thus, for example,
 \begin{equation}
  {\cL} \,d\phi_x = \frac{d}{d\theta} e^{-2\pi i \theta} d\phi_x
   \big|_{\theta =0} = - 2 \pi i \,d\phi_x .
 \end{equation}
 A form $K$ is defined to be \emph{invariant under the flow of} $X$ if $\cL K = 0$.
 For example, the form
 \begin{equation}
 \label{e:uxydef}
     u_{xy}=\phi_x d\phib_y
 \end{equation}
 is invariant since it is constant under the flow of $X$.

\begin{prop}
The supersymmetry generator is given by $Q= \frac{1}{\sqrt{2\pi
i}}(d+\ci)$, and $Q^2= \frac{1}{2\pi i}\cL$.  In particular, a
supersymmetric form is invariant under the flow of $X$.
\end{prop}

\begin{proof}
  By the definitions of $d$ and $\ci$, and of $\psi,\psib$,
  \begin{align}
    d &= \sqrt{2\pi i }
    \sum_{x=1}^M \left(
      \psi_{x} \frac{\partial}{\partial \phi_{x}} +
      \bar\psi_{x} \frac{\partial}{\partial \bar\phi_{x}}\right),
    \\
    \ci &=
    \sqrt{2\pi i }
    \sum_{x=1}^M \left(
      -\phi_{x} \frac{\partial}{\partial \psi_{x}} +
          \bar\phi_{x} \frac{\partial}{\partial \bar\psi_{x}}\right)
          .
  \end{align}
  The identity $Q= \frac{1}{\sqrt{2\pi i}}(d+\ci)$
  then follows immediately from the definition of $Q$ in \eqref{e:Qhat}.

  Cartan's formula asserts that $\cL = d \,\ci + \ci \,d$ (see, e.g., \cite[Prop.~2.25]{Warn83}
  or \cite[p.~146]{GHV72}).
  Since $d^{2}=0$ and $\ci^2=0$, it follows that
  $\cL = 2\pi iQ^2$.
\end{proof}


%% file: euclidean-notes.tex
\chapter{Extension to Euclidean models}
\label{app:Euclidean}

In this book, we have applied the renormalisation group method to
analyse the 4-dimensional hierarchical model.  We now briefly describe
some of the modifications needed to extend the method from the
hierarchical to the Euclidean setting, and also point out where in the
literature these extensions are carried out in detail.

The Euclidean setting refers to models defined on $\Zd$.  As usual, we
work first with finite volume, followed by an infinite volume limit.
(The renormalisation group map can however be defined directly in
infinite volume, as explained in \cite[Section~1.8.3]{BS-rg-step}.)
To preserve translation invariance in finite volume, we use the
$d$-dimensional discrete torus $\Lambda$ of period $L^N$.  The
decomposition \eqref{e:hierarchical-decomp} of the covariance in the
hierarchical setting is supplied directly by the definition of the
hierarchical Laplacian.  In the Euclidean setting, we instead use the
the finite-range decomposition of the Laplacian on the Euclidean torus
\eqref{e:torus-frd}.  As discussed in
Section~\ref{sec:hierarchical-fields}, the finite-range decomposition
bears similarities to the decomposition of the hierarchical Laplacian,
but it is less simple and details differ.

The main issues discussed below are:
\begin{itemize}
\item
  Unlike the hierarchical expectation at a given scale,
  the finite-range expectation does not factorise
  over blocks.
  A more general version of local coordinates is needed.
\item
  Unlike the hierarchical fields at a given scale,
  the finite-range fields are not constant within
  blocks, but only approximately so. This requires careful control
  not only of large fields but also of large gradients.
\item
  Unlike the hierarchical fields at a given scale, the covariances
  do not have the zero-sum property \refeq{hierarchical-C}.
  This necessitates an analogue of
  the term $W$ as in \eqref{e:ExW5} not only in the last step,
  but throughout all renormalisation group steps.
\end{itemize}
In particular, the above points require the following generalisations:
\begin{itemize}
\item
  The generalisations of the $T_\varphi$-seminorms must
  account for the fact that fields are only approximately constant.
\item
  The generalisation of the $\Wcal$-norm must
  account for the fact that the analogue of $K$ does not factorise over blocks.
\item
  In addition to the
  coupling constants $(g,\nu)$,
  the marginal monomial $\varphi (\Delta \varphi)$
  with coupling
  constant denoted by $z$
  must be tracked carefully.
  This coupling constant $z$ is associated to the field strength or stiffness of the field.
\end{itemize}

\section{Perturbative renormalisation group coordinate}

For the hierarchical model, the interaction $I : \Bcal \to \Ncal$ is
defined in Section~\ref{sec:local-coords} as
\begin{equation}
  I(b) = e^{-V(b)}, \quad V(b) = \sum_{x\in b} \left(\frac14 g|\varphi|^4 + \frac12 \nu |\varphi|^2\right).
\end{equation}
This requires two generalisations: field-strength renormalisation
and the second-order irrelevant contribution to the interaction.

\subsubsection*{Field-strength renormalisation}

\index{Field-strength renormalisation}
For the Euclidean model, the $n$-component field $\varphi=(\varphi_x)_{x\in \Lambda}$
is an element of $(\R^n)^\Lambda$.
Since fields are not constant on blocks,
the polynomial $V$ acquires a marginal monomial
$\varphi\cdot (-\Delta\varphi)$,
and a corresponding coupling constant $z$
called the \emph{field-strength renormalisation}.
There are three   coupling constants $(g,\nu,z)$, and
$V$ has the form
\begin{equation}
\label{e:Veuc}
  V(b) = \sum_{x\in b} \Big(\frac14 g|\varphi_x|^4 + \frac12 \nu |\varphi_x|^2 + \frac12 z \varphi_x \cdot(-\Delta \varphi_x)\Big).
\end{equation}

Now there are  two marginal monomials: $|\varphi|^4$ and
$\varphi\cdot(-\Delta \varphi)$.
As a consequence, the clever Bleher--Sinai argument applied
in Section~\ref{sec:BS1} to construct the critical point
is replaced by a more robust dynamical
systems argument \cite{BBS-rg-flow}.

\subsubsection*{Second-order irrelevant contribution}

For the hierarchical model, the second-order irrelevant term $W$
arises from the expectation in \refeq{ExW5}, but it vanishes at all
scales except the last scale due to the zero-sum condition
\refeq{hierarchical-C} on the
covariance.  For the finite-range decomposition, the zero-sum
condition does not hold, and the term $W$ occurs at all nonzero scales
and must be incorporated into the interaction $I$.
Thus we set
\begin{equation}
  I_j(V,b) = e^{-V_j(b)}(1+W_j(V,b)).
\end{equation}
The term $W$ is an explicit quadratic
polynomial in $V$ and is defined by
\begin{equation}
  W_j(V,b) = \frac12 (1-\Loc_b) F_{w_j}(V(b),V(\Lambda))
\end{equation}
where $F_C$ is given by \eqref{e:FCex} and
\begin{equation}
  w_j = C_1 + \dots + C_j
\end{equation}
is the total covariance that has been integrated so far.
The above formula for $W$ involves the
Euclidean localisation operator $\Loc_X$, which is defined for
arbitrary subsets $X \subset
\Lambda$ on smooth functions of the field $\varphi$.
Its output is a local polynomial summed over $X$, of the form
$\sum_{x\in X} V(\varphi_x)$,  with $V$ in a class of
local polynomials which includes the $\varphi\cdot (-\Delta\varphi)$ term.
See \cite[Section~2.4]{BBS-phi4-log} for more details, and \cite{BS-rg-loc} for
the general theory.

The definition \eqref{e:Vplus-hier-def} of the renormalised polynomial
$\Vpt$ must be generalised.  For the Euclidean model, it is
given by
\begin{equation}
    \label{e:beta-function2-H}
    \Vpt = \Ex_{+}\theta V - P,
\end{equation}
where $P$ is the local polynomial defined for $B \in \Bcal_+$ by
\begin{equation}
    \label{e:PdefF-tri}
    P (B)
    =
    \LT_B\,\left(
    \Ex_{+}\theta W(V,B)
    + \frac 12
    \Ex_{+}\big(\theta V(B);\theta V (\Lambda)\big)
    \right).
\end{equation}
The formulas \eqref{e:beta-function2-H}--\eqref{e:PdefF-tri}
reduce to \eqref{e:Vplus-hier-def} when the field is hierarchical
because there $W=0$ except for the last step and also $V(\Lambda\setminus B)$
is independent of $V(B)$ and hence drops out of the covariance term.
The definition \eqref{e:beta-function2-H} of $\Vpt$ is discussed at
length in \cite{BBS-rg-pt} (see also
\cite[Section~3.2]{BBS-phi4-log}).
In particular, the equation for $P_x$ in \cite[(2.12)]{BBS-rg-pt}
is equivalent to \refeq{PdefF-tri} because of the relation between
$F_{C_+}$ and $\Ex_+$ given in Exercise~\ref{ex:Fexpand}, and because
of the fact that $\sum_{x\in B}\LT_x$ can be replaced here by $\LT_B$ due
to \cite[Proposition~1.8]{BS-rg-loc}.

The main achievement of \eqref{e:beta-function2-H} and
\eqref{e:PdefF-tri} is the following lemma, which is an extension of
Lemma~\ref{lem:K0PT}.  It differs from \eqref{e:ExW5} in the sense
that the left-hand side of \eqref{e:ExW5} is of the form $e^{-U(B)}$
rather than $e^{-U(B)}(1+W(B))$.  In the statement of the lemma,
$O(U^p)$ denotes an error of $p^{\rm th}$ order in the coupling constants
$U$, which need not be uniform in the field $\varphi$ or the volume
$\Lambda$.  A proof of the lemma (in the supersymmetric case) is given
in \cite[Proposition~2.1]{BBS-rg-pt}.

\begin{lemma}
\label{lem:K0PTwithW} For any polynomial $V$ as in \refeq{Veuc} such
that the expectation exists, and for $B \in \Bcal_+$,
\begin{align}
\lbeq{ExW5withW}
    \Ex_{+} \theta I(\Lambda)
    &=
    I_+(\Vpt, \Lambda)
    + O(V^3)
    .
\end{align}
\end{lemma}

\section{Approximate factorisation}

The conceptually most significant generalisation that is required is
that of only approximate factorisation.

\subsection{Factorisation of expectation}

\index{Block}
From Definition~\ref{def:Bcal},
recall that $\Bcal$ denotes the set of \emph{blocks} at scale $j$.

\index{Polymer}\index{Connected component}
\begin{defn}
A \emph{polymer} is a union of blocks from $\Bcal$.  We define $\Pcal$
to be the set of polymers.  We write $\Bcal(X)$ and $\Pcal(X)$ for the
sets of blocks and polymers contained in the polymer $X$.  We say that
$X$ and $Y$ are \emph{disjoint} if $X\cap Y = \varnothing$ and that
$X$ and $Y$ are \emph{disconnected} if there is no pair of
blocks $B \in X$ and $B' \in Y$ that touch (as in
Definition~\ref{def:Bcal}).  A polymer is \emph{connected} if it is
not the union of two disconnected polymers; this gives a partition
${\rm Comp}(X)$ of a polymer $X$ into its connected components.
\end{defn}

\index{Disjoint}\index{Disconnected}\index{Factorisation}
Let $\Ncal$ denote the algebra of sufficiently smooth
functions of the field $\varphi = (\varphi_x)_{x\in \Lambda}$,
i.e., maps from $(\R^n)^\Lambda$ to $\R$.
We say that $F: \Pcal \to \Ncal$ is \emph{strictly local} if $F(X)$ depends only on $\varphi|_{X}$.
Let $F,G$ be strictly local. The hierarchical expectation has the factorisation property:
\begin{equation} \label{e:eucl-factor-hier}
  \Ex_+
  (F(X)G(Y)) = \Ex_{+}(F(X))\Ex_{+}(G(Y))
  \quad
  \text{if $X,Y\in \Pcal_+$ are \emph{disjoint}.}
\end{equation}
The finite-range expectation has the weaker factorisation property:
\begin{equation} \label{e:eucl-factor-fr}
  \Ex_{+}(F(X)G(Y)) = \Ex_{+}(F(X))\Ex_{+}(G(Y))
  \quad
  \text{if $X,Y \in \Pcal_+$ are \emph{disconnected}.}
\end{equation}
In fact, for \eqref{e:eucl-factor-fr}, the condition that $F$ and $G$ be strictly local
can be weakened and at times needs to be weakened.

\subsection{Circle product}

The hierarchical model
is written in the \emph{factorised} form (see \eqref{e:Z0prod}, \eqref{e:FjIjKj})
\begin{equation} \label{e:eucl-hierfact}
  \prod_{b\in\Bcal} e^{-u|b|}(I(b)+K(b))
  =
  e^{-u|\Lambda|} \prod_{b\in\Bcal} (I(b)+K(b)),
\end{equation}
and this form is preserved by the hierarchical expectation due to \eqref{e:eucl-factor-hier} (see \eqref{e:ZZ+}).
The expectation with \emph{finite-range} covariance
does not preserve this strong factorisation and a generalisation
is required.

\index{Circle product}
\begin{defn}
For $F, G: \Pcal \to \Ncal$
we define the \emph{circle product} $F \circ G: \Pcal \to \Ncal$ by
\begin{equation}
  (F \circ G)(X) = \sum_{Y \in \Pcal(X)} F(Y)G(X\setminus Y)
  \quad\quad (X \in \Pcal).
\end{equation}
\end{defn}

\index{Factorisation}
Let $F: \Pcal \to \Ncal$. We say that:
\begin{itemize}
\item $F$ \emph{factorises over blocks} if $F(X) = \prod_{b\in \Bcal(X)} F(b)$ holds for any $X \in \Pcal$;
\item $F$ \emph{factorises over connected components} if $F(X) = \prod_{Y \in {\rm Comp}(X)} F(Y)$.
\end{itemize}
When $F$ factorises over blocks we write $F^X=F(X)$.
The circle product has
the following properties which we use below
(see \cite{BS-rg-step}), namely:
\begin{itemize}
\item Commutativity:  $ F\circ G = G \circ F$.
\item
  Associativity: $(F \circ G) \circ H = F \circ (G\circ H)$.
\item
  Suppose that $F$ and $G$ factorise over blocks. Then
  \begin{equation} \label{e:eucl-FG-factB}
    (F \circ G)(X) = \sum_{Y \in \Pcal(X)} F^YG^{X\setminus Y} = \prod_{b\in\Bcal(X)} (F(b)+G(b))
    =(F+G)^X.
  \end{equation}
\end{itemize}

By \eqref{e:eucl-FG-factB},
the representation \eqref{e:eucl-hierfact} equals $e^{-u|\Lambda|} (I \circ K)(\Lambda)$
when  $I$ and $K$ both factorise over blocks, as they do in the hierarchical setting.
In the Euclidean setting,
$I$ does factorise over blocks, but $K$ only factorises over connected components.
Hence we work
with $(I \circ K)(\Lambda)$,
and we must maintain this form after taking the expectation.
More precisely, in the Euclidean setting
the hierarchical formula \refeq{ZZ+} becomes
\begin{equation}
\lbeq{ZZ+-euc}
    Z_+ = e^{-u_+|\Lambda|}(I_+\circ K_+)(\Lambda)
    =
    e^{-u|\Lambda|}\Ex_+ \theta (I\circ K)(\Lambda) = \Ex_+\theta Z
    .
\end{equation}
The circle product is scale dependent: in \eqref{e:ZZ+-euc}
$I \circ K$ is a scale-$j$
product whereas  $I_+\circ K_+$ is at scale-$(j+1)$.

\section{Change of coordinates}

The hierarchical representation $Z=\prod_b (I(b)+K(b))$ is not unique because it permits division of $I(b)+K(b)$ into terms $I$ and $K$ in different ways.
The circle product representation $Z=(I\circ K)(\Lambda)$ of a given $Z \in \Ncal$ in terms
of the coordinates $I$ and $K$ is further from being unique
under the constraints that $I$ factors over blocks and $K$ over connected components,
as it allows parts of $K(X)$ to be redistributed over different polymers.

The essential difficulty is to obtain a representation $Z=(I\circ K)(\Lambda)$
with the property that $K$ does not grow with the scale.  This requires
the transfer of dangerous parts of $K$ into $I$, via exploitation of the nonuniqueness of the
circle product representation.  This is done via two mechanisms of change of coordinates,
which we now demonstrate.

\subsection{Block cancellation}

A change of coordinates for the hierarchical model is performed in
\refeq{eVKhats}, where, given $V,K,\Vhat$, we find $\Khat$ such that
\begin{equation}
    e^{-V(B)}+K(B) = e^{-\Vhat(B)}+\Khat(B).
\end{equation}
Indeed $\Khat(B)$ is simply given in \refeq{KofKhat}
as the solution to this equation.
For the Euclidean model, the corresponding step would be easy to perform if
we only wished to alter $K$ on blocks and not on larger polymers.
Indeed, by the associative property of the circle product and
the identity \refeq{eucl-FG-factB}, given any $I,\tilde I, K$, we can set
$\delta I = \tilde I-I$ and $\tilde K= \delta I \circ K$ and obtain
\begin{equation}
\lbeq{chvar1}
    I\circ K = (\tilde I+\delta I)\circ K = (\tilde I \circ \delta I) \circ K
    =
    \tilde I \circ \tilde K.
\end{equation}
In particular,
\begin{equation}
    \tilde K(B) = \delta I(B) + K(B).
\end{equation}
By choosing $\delta I$ appropriately, we can cancel the relevant
and marginal parts from $K(B)$ by transferring them into $\tilde I$.
This is what we did in the hierarchical setting in
\refeq{eVKhats}, and it was sufficient.

\subsection{Small set cancellation}
\label{sec:smallsetcancellation}

The procedure used in \refeq{chvar1}
does not cancel the relevant parts from $K(X)$
when $X$ is not a single block.
It turns out to
be necessary to cancel the relevant parts from $K(X)$
only for the restricted class of \emph{small sets} $X\in \Scal$,
where  $\Scal$ is the set of connected polymers which consist of
at most $2^d$ blocks.
\index{Small set}
\index{$\Scal$}
Indeed, $K(X)$ contracts for geometric reasons when the polymer
$X$ is not a small set, making
it unnecessary to extract relevant parts
(see Lemma~\ref{lem:largesetcontract} below).

The small set cancellation lies at the heart of the non-hierarchical problem,
and is achieved by a different mechanism than \refeq{chvar1}.
Instead, given any $I,K$, we produce $K'$ so that
\begin{equation}
\lbeq{chvar2}
    (I \circ K)(\Lambda) = (I \circ K')(\Lambda).
\end{equation}
Note that the \emph{same} $I$ appears on both sides of \refeq{chvar2}.
(Unlike \refeq{chvar1}, we do \emph{not} have equality of
$(I \circ K)(X)$ and $(I \circ K')(X)$ for every polymer $X$.)
The new coordinate $K'$ will effectively move the unwanted part of $K(X)$
  when $X$ is a small set that is not a block into $K(B)$;
  in particular $K(B)$ will not undergo a cancellation.
However, we can subsequently apply a version of \refeq{chvar1} to deal
with $K(B)$.  A precursor of \refeq{chvar2} appears in
\cite[Theorem~A]{BY90}.

For simplicity, we illustrate \refeq{chvar2} for the case $I=1$.
For a small set $X$ that is not a block, let $\bar J(X)$ be the portion of $K(X)$
that we wish to cancel.  A key example is to have $\bar J(X)$
equal to $\Loc_XK(X)$.  This is a local polynomial
in the field, summed over the polymer $X$.  It can therefore be written
as $\bar J(X) = \sum_{B\in \Bcal(X)} J(X,B)$ where $J(X,B)$ is the restriction
of $\bar J(X)$ to summation over the block $B$.  Now we \emph{define} $J(B,B)$
by
\begin{equation}
    J(B,B) = - \sum_{X \supset B: X \neq B}J(X,B).
\end{equation}
Thus we assume that we are given $\bar J(X) = \sum_{B \in \Bcal(X)}J(X,B)$ with
\begin{equation}
\lbeq{Jprops}
    J(X,B) = 0 \quad\text{if $X \not\in \Scal$ or $B \not\subset X$}
    ,\qquad
    \sum_{X \supset B} J(X,B) = 0.
\end{equation}

A change of coordinates in this situation is given by
\cite[Proposition~D.1]{BS-rg-step}, whose conclusion is that, given
\refeq{Jprops}, the identity \refeq{chvar2} holds with $K'$ obeying
component factorisation, good estimates, and the desirable property
\begin{align}
\lbeq{Kcancel}
    K'(X) & = K(X) - \bar J(X) + \text{remainder}
    \qquad (X \in \Scal).
\end{align}
Thus, for $X \in \Scal\setminus \Bcal$, $K'(X)$ is approximately equal to
$K(X)$ with its relevant and marginal parts subtracted.  The price to
be paid for this is that
\begin{equation}
    K'(B) = K(B)-\bar J(B)+ \text{remainder} = K(B) + \sum_{X \supset B: X \neq B}J(X,B)+ \text{remainder}.
\end{equation}
Thus $K'(B)$ not only fails to make a cancellation in $K(B)$, but it
also receives the dangerous parts of $K(X)$ from small sets $X$
that contain but do not equal $B$.

To fix this defect in $K'(B)$, as mentioned already above, we can use \refeq{chvar1}.
Moreover,
that repair does not do harm to $K'(X)$ for polymers
that are not a single block.

\subsection{Application}

In Sections~\ref{sec:Ex-euc}--\ref{sec:lc}, we will apply
each of the changes of coordinates \refeq{chvar1}--\refeq{chvar2} twice,
as follows.

\subsubsection*{Perturbation theory}

In Section~\ref{sec:Ex-euc}, we choose $\tilde I$ suggested by
perturbation theory, apply \refeq{chvar1} and take the expectation
to obtain \refeq{ZZ+-euc} in the form $\Ex_+\theta (I \circ K)(\Lambda)
= (\tilde I \circ \tilde K)(\Lambda)$.  The resulting
$\tilde K$ is unsatisfactory, as it contains
second-order contributions.  When $I=1$ these bad contributions are
called $h(U)$.  Because of our choice of $\Vpt$,
to second order we can find $h(U,B)$ such that $h(U)$ equals
$\bar h(U)=\sum_{B \in \Bcal(U)}h(U,B)$ with $\sum_{U \supset B}h(U,B)=0$.
From \refeq{Kcancel} with $\bar J = \bar h$, we obtain now
\begin{align}
\lbeq{hcancel}
    h'(X) & = h(X) - \bar h(X) + \text{remainder}  = \text{remainder}
    \qquad (X \in \Scal).
\end{align}
(This holds also for $X=B$.)
Now $h'$ is third order.  This is carried out in Section~\ref{sec:lcp}.
There we indicate why \refeq{chvar2} holds with \refeq{hcancel}
for this easier special case
which has $h \approx \bar h$.

\subsubsection*{Relevant and marginal parts}

After the above has been carried out, we have a third-order $K$, but it contains
relevant and marginal parts which would grow uncontrollably as the scale
advances.
For $X$ a small set that is not a block,
we transfer these parts from $K(X)$ to $I$ using the mechanism
 described in Section~\ref{sec:smallsetcancellation}
with $\bar J = \Loc K$.
The details are given in Section~\ref{sec:lcnp}, where we indicate
why \refeq{chvar2} holds with \refeq{Kcancel} in the general case.
Finally, the unwanted parts of $K(B)$ for blocks $B$ are removed by an application
of \refeq{chvar1}.

\section{Expectation, change of scale, and reblocking}
\label{sec:Ex-euc}

For the hierarchical model (see \eqref{e:FplusB}),
we showed that for any choice of $U_+$ (in fact any choice of $I_+$)
we could choose $K_+$ as in \eqref{e:K+B-bis} to obtain the representation
\begin{equation}
  e^{-u |\Lambda|} \Ex_{+}\prod_{b\in\Bcal} (I(b)+K(b))
  = e^{-u_+|\Lambda|} \prod_{B\in\Bcal_+} (I_+(B)+K_+(B))
\end{equation}
where $\Ex_+$ is the hierarchical expectation.
A Euclidean version of this is given in the following proposition.
The proposition shows
that given any choice of $\tilde I_+$ we can find an appropriate $\Ktilde_+$,
in the
more general setting of the circle product.  The scale of the circle
product becomes increased in this operation, and this requires a reblocking step.
Proposition~\ref{prop:EIK}
provides the defining element of \emph{Map 3}
in \cite[Section~5.1]{BS-rg-step}.
For its statement and proof, we need the following definition.

\begin{defn}
\label{def:closure}
The \emph{closure} of a polymer $X\in \Pcal$ is the smallest polymer
$\overline X \in \Pcal_+$ such that $X \subset \overline X$.
\end{defn}

\begin{prop}
\label{prop:EIK}
Let $I,\tilde I_+$ factorise over blocks $b \in \Bcal_j$
and let  $\delta I (b)=\theta I (b) -\tilde I_+(b)$.
Let $K$ factorise over connected components at scale $j$.  Then
\begin{equation}
\label{e:EIK1}
    \Ex_{+}\theta (I \circ K)(\Lambda)
    = (\tilde I_+ \circ \Ktilde_+)(\Lambda)
\end{equation}
with
\begin{equation}
\label{e:EIK2}
    \Ktilde_+ (U)
=
    \sum_{X \in\Pcal(U)}
    \tilde I_+^{U\setminus X}
    \Ex_{+}
    (\delta I
    \circ
    \theta K
    )(X)
    \1_{\overline X =U}
    \quad\quad(U \in \Pcal_{+}),
\end{equation}
and $\Ktilde_+$ factorises over connected components at scale $j+1$.
\end{prop}

\begin{proof}
Let $P = \delta I \circ \theta K$.  By \refeq{chvar1} at scale $j$,
\begin{equation}
    (\theta I)\circ (\theta K) = \tilde I \circ P.
\end{equation}
Since $\tilde I_+$ does not depend on the fluctuation field,
\begin{align}
    \Ex_{+}\theta
    (
    I \circ K  )(\Lambda)
    &   =
       (
    \tilde I_+   \circ \Ex_{+} P )(\Lambda)
    =
    \sum_{X \in \Pcal}
    \tilde I_+^{\Lambda \setminus X}
    \Ex_{+}\big(P(X) \big)
    \nnb
    & =
    \sum_{U \in \Pcal_{+}}
    \tilde I_+^{\Lambda \setminus U}
    \sum_{X \in \Pcal }
    \tilde I_+^{U \setminus X}
    \Ex_{+}\big(P(X) \big)
    \1_{\overline X =U}
    .
\end{align}
The right-hand side is \refeq{EIK1} with $\Ktilde_+$ given by \refeq{EIK2}, and
the proof of the identity is complete.

It is not difficult to verify that $\Ktilde_+$ factorises over connected components.
The geometry of the identity \refeq{EIK2} defining $\Ktilde_+(U)$ is illustrated in Figure~\reffg{reblock-euc}, which is helpful for the verification of factorisation.
\end{proof}

\begin{figure}
\begin{center}
\includegraphics[scale = 0.2]{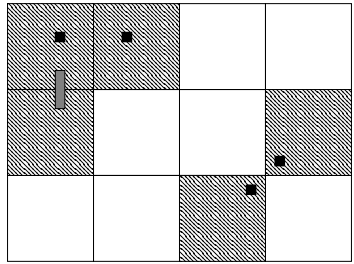}
\end{center}
\caption{\lbfg{reblock-euc}
The five large shaded blocks represent $U$, which is the closure of the
union of the four small dark blocks (the support of $\delta I$) and the
small shaded polymer (the support of $K$).}
\end{figure}

For $X\in \Pcal$, let $|\Bcal(X)|$ denote the number of scale-$j$ blocks in $X$.
Similarly, we write $|\Bcal_+(U)|$ for the number of scale-$(j+1)$ blocks in
$U\in \Pcal_+$.
Let ${\cal C} \subset {\cal P}$
denote the set of connected polymers.  
In the formula \refeq{EIK2} for $\tilde K_+(U)$, it is helpful if $|\Bcal(X)|$
is large, as this brings small factors from $(\delta I \circ K)(X)$.
The following lemma shows that for large connected sets $X \in \Ccal\setminus\Scal$,
the constraint $\overline{X}=U$
in \refeq{EIK2} forces $|\Bcal(X)|$ to be
strictly larger than $|\Bcal_+(U)|$;
for small sets $|\Bcal(X)|=|\Bcal_+(U)|$ is possible
and the choice of $2^d$ in the definition of $\Scal$
is precisely due to this possibility.
It is this geometric fact---the excess of $|\Bcal(X)|$ over $|\Bcal_+(U)|$ for large
connected sets---that
allows the main focus to be placed on the control of small sets.
Large sets are \emph{irrelevant}.
A proof of Lemma~\ref{lem:largesetcontract} is given in
\cite[Lemma~C.3]{BS-rg-step}, and an earlier statement is \cite[Lemma~2]{DH92}.
Its application in the Euclidean
setting occurs in
\cite[Lemma~5.6]{BS-rg-step}.

\begin{lemma} \label{lem:largesetcontract}
  Let $d \ge 1$.  There is an $\eta = \eta(d) >1$ such that
  for all $L \geq 2^d+1$ and for all $X \in \Ccal\setminus \Scal$,
  \begin{equation}
    |\Bcal(X)| \geq \eta |\Bcal_+(\overline{X})|.
  \end{equation}
\end{lemma}

The following example indicates a mechanism in which
Lemma~\ref{lem:largesetcontract} is applied.  It illustrates why the
focus can be restricted to small sets.

\begin{example}
  Let $A>1$ and define a norm on $F: \Ccal \to \Ncal$  by
  \begin{equation}
    \|F\| = \sup_{X\in \Ccal} A^{|\Bcal(X)|} |F(X)|.
  \end{equation}
  We extend $F: \Ccal \to \Ncal$ to $F: \Pcal \to \Ncal$ by component factorisation,
  and define $\overline{F}: \Ccal_+ \to \Ncal$ by
  \begin{equation}
    F(U) = \sum_{\overline{X}=U} F(X).
  \end{equation}
  The map $F\mapsto \overline{F}$ is a prototype for the map $K \mapsto K_+$
  that captures the reblocking aspect.
  Suppose that $F(X) = 0$ if $X \in \Scal$. We claim that
  \begin{equation} \label{e:Fbarcontract}
    A^{|\Bcal_+(U)|}|F(U)| \leq  \left( A^{|\Bcal_+(U)|} \sum_{\overline{X}=U} A^{-|\Bcal(X)|} \1_{X \not\in\Scal} \right) \|F\|
    \leq o(1) \|F\|,
  \end{equation}
  with the second inequality valid
  as $A \to\infty$. Therefore, with $A$ sufficiently large, there
  exists $\kappa<1$ such that
  \begin{equation}
  \label{e:Fkappa}
    \|\overline{F}\|_+ \leq \kappa \|F\|
  \end{equation}
  for all $F$ with $F(X)=0$ for $X \in \Scal$.
  The inequality \refeq{Fkappa} shows that large sets are
  not important for the simple prototype
  $F \mapsto \overline{F}$ for the map $K \mapsto K_+$.

It remains to prove \eqref{e:Fbarcontract}.  The first inequality holds
by definition of the norm.  For the second,
we bound the number of terms $X$ in the sum
by $2^{|\Bcal(U)|} = 2^{L^{d} |\Bcal_+(U)|}$,
and apply Lemma~\ref{lem:largesetcontract}
to obtain $A^{-|\Bcal(X)|} \1_{X \not\in\Scal} \leq A^{-\eta|\Bcal_+(U)|}$.
This gives
  \begin{align}
    A^{|\Bcal_+(U)|} \sum_{\overline{X}=U} A^{-|\Bcal(X)|} \1_{X \not\in\Scal}
    &\leq
      ( A 2^{L^d} A^{-\eta})^{|\Bcal_+(U)|}
    \leq
      2^{L^d} A^{1-\eta}
    ,
  \end{align}
with the last inequality valid assuming $A^{\eta-1} \ge 2^{L^d}$ (which does hold for
large $A$ since $\eta>1$).
The right-hand side becomes arbitrarily small for $A$ sufficiently large.
\end{example}

\section{Cancellation via change of coordinates}
\label{sec:lc}

\index{Coordinates}

\subsection{Local cancellation: perturbative}
\label{sec:lcp}

The formula for $\Ktilde_+$ in \refeq{EIK2} is not adequate
even when $\tilde I_+$ is well chosen as $\tilde I_+(\Vpt)$,
due to the presence of
perturbative contributions to $\Ktilde_+$ that are
manifestly second order in $V$.
In this section, we sketch an argument to explain how the
change of coordinates \refeq{chvar2} can be
used to correct this problem.
We also sketch a proof of \refeq{chvar2} in this special case.
This discussion reveals what lies at the heart of
\emph{Map~4} in \cite[Section~5.3]{BS-rg-step}.

\subsubsection*{Second-order contribution to $\Ktilde_+$}

Let $U$ be a connected polymer in $\Ccal_+$. There is a contribution
to the right-hand side of \refeq{EIK2} of the form
\begin{equation}
\lbeq{HUdef}
    H(U) =  \tilde I_+^U h(U),
    \quad\quad
    h(U)=
    \sum_{X \in\Pcal(U) : |X|=1,2}
    \tilde I_+^{-X} \Ex_+ \delta \tilde I^X \1_{\overline X =U}
    ,
\end{equation}
with the closure $\overline X$ of $X$ defined in Definition~\ref{def:closure}.
For the terms in \refeq{HUdef}
where $X$ consists of a single block, $h(U)=0$ unless
$U$ is a single block, and when $X$ consists of two blocks then $h(U)=0$
unless $U$ consists of one or two blocks.
We extend the definition of $H$ and $h$ to disconnected polymers by imposing
component factorisation.
The contribution of $H$ to $(\tilde I_+ \circ \Ktilde_+)(\Lambda)$ is
\begin{equation}
    (\tilde I_+ \circ H)(\Lambda) = \tilde I_+^\Lambda (1 \circ h)(\Lambda).
\end{equation}

The terms in the formula for $\Ktilde_+$ that
are first or second order in $\delta I$
are isolated in $h$. Naively, we expect each factor of $\delta I(b)$ to
provide a factor $O(V)$, so that three or more
factors of $\delta I(b)$ will ensure an estimate
$O(V^3)$.  In $h(U)$ there are only one or two such factors when
$U \in \Ccal_+$.
The apparently first-order terms with $|X|=1$ are in fact second order in $V$,
because in $\Ex_+\delta I(b) = \Ex_+\theta I - I_+$ there is
cancellation of the first-order term in \refeq{beta-function2-H} due to
our use of $\Vpt$ to define $\tilde I_+$. Thus $h$ is $O(V^2)$.

As we will argue at the end of Section~\ref{sec:lcp},
the second-order part of $h(U)$ has the form
\begin{equation}
\label{e:hUB2}
    h(U) \equiv \sum_{B \in \Bcal_+(U)}h(U,B) ,
\end{equation}
with $h(U,B)$ second order and obeying the local cancellation
\begin{equation}
\label{e:hzerosum}
    \sum_{U \in \Ccal_+ : U \supset B}  h(U,B) \equiv 0,
\end{equation}
where $F\equiv G$ denotes that $F=G +O(V^3)$.

\subsubsection*{Local cancellation in $\Ktilde_+$}

We now
apply \refeq{hUB2}--\refeq{hzerosum} and exploit the non-uniqueness of the
circle product representation, to show that it is possible to reapportion the
second-order contributions to $h$ in such a way that there is a third-order
$h_+$ such that
\begin{equation}
\lbeq{IHhbar}
    (1 \circ  h)(\Lambda)
    =
    (1\circ  h_+)(\Lambda),
\end{equation}
with $h_+(U)\equiv 0$ when $U \in \Ccal_+$.
This gives a version of \refeq{chvar2} at scale $j+1$,
with $J$ and $K$ both given by $h$ on small sets and with $I=1$.

We use the component factorisation property of $h$ and \refeq{hUB2} to obtain
\begin{align}
    (1 \circ  h)(\Lambda)
    & =
    \sum_{Y \in \Pcal_+}  h(Y)
    =
    \sum_{Y \in \Pcal_+} \prod_{Y_i \in {\rm Comp}(Y)}  h(Y_i)
    \nnb &
    =
    \sum_{Y \in \Pcal_+}
    \prod_{Y_i \in {\rm Comp}(Y)}
    \sum_{B_i \in \Bcal_+(Y_i)}
    h(Y_i,B_i).
\end{align}
Given a block $B$, let $B^{(2)}$ denote the polymer which is the union
of $B$ and all two-block connected polymers that contain
$B$. For example, when the dimension is $d=2$ then $B^{(2)}$ is the union of $B$
with the eight blocks that touch $B$. We partition
the summation on the right-hand side according to the polymer $\cup_i
B_i^{(2)}$, to obtain
\begin{align}
    (1 \circ  h)(\Lambda)
    & =
    \sum_{U\in \Pcal_+} \sum_{Y \in \Pcal_+}
    \prod_{Y_i \in {\rm Comp}(Y)}
    \sum_{B_i \in \Bcal(Y_i)}  h(Y_i,B_i) \1_{\cup_i B_i^{(2)}=U}
    \nnb & =
    \sum_{U\in \Pcal_+}  h_+(U)
    \nnb &
    = (1 \circ h_+)(\Lambda),
\end{align}
where the second equality defines the terms $ h_+(U)$.
It can be checked that $h_+$ has the component factorisation property.

A second-order contribution to $ h_+(U)$ can occur only when
$U$ is connected (otherwise $h_{+} (U)$ factors into
contributions from each connected component of $U$,
each of which is second order).
Therefore the only possible
second-order contribution to $ h_+(U)$ is
\begin{align}
    \sum_{Y \in \Pcal_+}
    \sum_{B \in \Bcal_+(Y)}  h(Y,B) \1_{B^{(2)} = U}.
\end{align}
Given $U$, the condition $B^{(2)} =U$ uniquely determines $B$ (or there is no
such $B$).  With that particular $B=B(U)$, the above is equal to
\begin{align}
    \sum_{Y \in \Pcal}  h(Y,B),
\end{align}
which vanishes by \refeq{hzerosum}.  Thus we have achieved the goal
\refeq{IHhbar} with third-order $h_+$.  The calculations here illustrate
part of what occurs in the proof of \cite[Proposition~D.1]{BS-rg-step},
in a simplified setting.

\subsubsection*{Verification of \refeq{hUB2}--\refeq{hzerosum}}

We now verify \refeq{hUB2}--\refeq{hzerosum}.
That is, we will identify second-order quantities
$h(U,B)$, for $U$ a two-block polymer
containing $B$, with the properties that
\begin{equation}
\lbeq{hUB}
    h(U) \equiv \sum_{B \in \Bcal_+(U)}h(U,B) ,
\end{equation}
and that, due to our choice of $\Vpt$, there is the \emph{local cancellation}
\begin{equation}
\lbeq{hzerosum1}
    \sum_{U \in \Ccal_+ : U \supset B}  h(U,B) \equiv 0
\end{equation}
(a version of \refeq{hzerosum1} with \emph{equality} appears in
 \cite[(2.22)]{BS-rg-IE}).

To keep the focus on the main ideas, let us simplify the problem
and assume that
$I_+(U) = e^{-\Vpt(U)}$.  With $\delta V = \theta V - \Vpt$, we can
then rewrite $h$ as
\begin{equation}
    h(U) =  \sum_{X \in\Pcal(U) : |X|=1,2}
    \big( \prod_{b \in \Bcal(X)}e^{-\delta V(b)} - 1 \Big) \1_{\overline X =U}.
\end{equation}
To uncover the lower-order terms in $h(U)$, we expand
the exponential in a Taylor series and obtain \refeq{hUB}
with
\begin{align}
     h(B,B) &=
    \sum_{b \in \Bcal(B)}
    \Ex_+
    \Big( \delta V(b) + \frac 12 (\delta V(b))^2 \Big)
    +
    \frac 12
    \sum_{b,b' \in \Bcal(B): b \neq b' }
    \Ex_+
    \Big( \delta V(b) \delta V(b') \Big),
\\
     h(U,B) &=
    \frac 12
    \sum_{b \in \Bcal(B)}\sum_{b' \in \Bcal(B')}
    \Ex_+
    \Big( \delta V(b) \delta V(b') \Big).
\end{align}
The $\delta V(b)$ term in $h(B,B)$ is actually second order, not first order,
because $\Vpt$ is equal to $\Ex_+\theta V$ minus quadratic terms in $V$,
and the linear term in $V$ therefore cancels in $\Ex_+\delta V(b)$.
Thus all terms in $h(B,B)$ and $h(U,B)$ are second order.

To derive \refeq{hzerosum1},
we continue to neglect $W$, and recast Lemma~\ref{lem:K0PTwithW} as
$\Ex_+\theta e^{-V(\Lambda)} \equiv e^{-\Vpt(\Lambda)}$, i.e.,
\begin{align}
    e^{-\Vpt(\Lambda)} \Ex_+ \Big( \prod_{b \in \Bcal(\Lambda)}
    e^{-\delta V(b)} - 1 \Big)
    & \equiv 0.
\end{align}
Again only the case where the product is over one or two small blocks can
lead to a second-order contribution, and these small blocks must
either lie in the same large block or in adjacent large blocks, because
otherwise the finite-range property of the expectation produces a product
of two second-order factors and hence is fourth order.  The same Taylor
expansion used above then leads to
\begin{align}
    \Ex_+ \Big( \prod_{b \in \Bcal(\Lambda)}
    e^{-\delta V(b)} - 1 \Big)
    & \equiv
    \sum_{U \in \Ccal_+(\Lambda)} h(U)
    \nnb & =
    \sum_{U \in \Ccal_+(\Lambda)} \sum_{B \in \Bcal_+(U)}  h(U,B)
    \nnb & =
    \sum_{B \in \Bcal_+(\Lambda)} \sum_{U \in \Ccal_+ : U \supset B}  h(U,B).
\end{align}
It is natural that the right-hand side would be third order because
each term in the sum over $B$ is,
and this indeed turns out to be the case and gives \refeq{hzerosum1}.

\subsection{Local cancellation: nonperturbative}
\label{sec:lcnp}

For the hierarchical model, the marginal and relevant directions in
$K$ are absorbed into $U_+$ via the term $\Loc (e^V K)$ in
\eqref{e:U+def}.  In the Euclidean setting, the analogous manoeuvre is
more delicate because $I$ has only one degree of freedom for each
block $B$ (it factorises over blocks), while now $K$ is a function of
arbitrary polymers $X$.  Two steps are used: (i) we apply the change
of coordinates \refeq{chvar2} to move the contributions from small
sets into blocks, and (ii) we use the simpler change of coordinates
\refeq{chvar1} for single blocks, as done in \refeq{eVKhats} in the
hierarchical setting.

To explain how the cancellation on small sets is arranged, we first
write $\Loc_UF$ as $\Loc_U F = \sum_{x \in U}P_x$ and use $P$ to
define $\Loc_{U,B}F=\sum_{x \in B}P_x$ for $U\supset B$.
In particular, for $U \in \Pcal$ and
$B\in \Bcal$,
\begin{equation}
\lbeq{LocXBsum}
    \sum_{B \in \Bcal(U)}\Loc_{U,B}F = \Loc_UF.
\end{equation}
Then
we define $J(U,B)=0$ if $U$ is not a small set containing $B$, and otherwise
\begin{align}
\label{e:Map1JXB}
    J(U,B) & = \LT_{U,B}I^{-U}K(U) \quad \text{for
    $U \in \Scal$ with $U \supsetneqq B$},
\\
\label{e:Map1JBB}
    J (B,B)
    & =
    -
    \sum_{U \in \Scal :U\supsetneqq B}
    J(U,B)
    .
\end{align}
By construction,
\begin{equation}
\lbeq{Jzerosum}
    \sum_{U : U \supset B}J(U,B)=0.
\end{equation}
The local cancellation in \refeq{Jzerosum}, which holds by definition
of $J$, is as in \refeq{Jprops}.

For $U \in \Scal$, let
\begin{equation}
    \bar J(U) = \sum_{B \in \Bcal_+(U)}I^UJ(U,B),
    \quad \quad
    M(U)=K(U)-\bar J(U).
\end{equation}
The new feature compared to our analysis of $h$ in Section~\ref{sec:lcp}
is that here for small sets $U$ the
role of $h(U)$ is played by $I^{-U}K(U)=I^{-U}M(U)+ I^{-U}\bar J(U)$; the analysis for $h$ corresponds
to $M=0$ which we no longer have.  This requires more sophisticated combinatorics.

\subsubsection*{Cancellation on small sets other than blocks}

To illustrate the main idea we make the following simplifications:
\begin{itemize}
\item
We assume that $I=1$.
\item
We assume that among
connected polymers $K$ is supported on small sets only.
\end{itemize}

Then, with $V$ the union of the components of $\hat V$,
\begin{align}
    (1\circ K)(\Lambda)
    & =
    \sum_{U \in \Pcal} \prod_{U_i \in {\rm Comp}(U)}(M(U_i)+\bar J (U_i))
    \nnb & =
    \sum_{U \in \Pcal}
    \sum_{\hat V \subset {\rm Comp}(U)}M^{U\setminus V} \prod_{U\in \hat V}
    \sum_{B \in \Bcal(U)}J(U,B).
\end{align}
Given $X \in \Pcal$, let $B_1,\ldots, B_n$ be a list of
the blocks in $\Bcal(X)$, and let
\begin{align}\label{}
    \Ucal(X)
    =
    &\{
    \{(U_{B_1},B_1), \ldots, (U_{B_n},B_n)\} :
    \nnb
    &
    U_{B_i} \in \Scal, \; U_{B_i} \supset B_i,\;
    \text{
    $U_{B_i}$ does not touch $U_{B_j}$ for $i \neq j$
    }
    \}
.
\end{align}
Given an element of $\Ucal(X)$, we write $Y_J = \cup_{B
\in \Bcal(X)}U_B$, and write $\Pcal^-(Y_J)$ for the
set of polymers that do not touch $Y_J$.
The \emph{small-set neighbourhood} $X^\Box$ of a polymer $X$ is
the union of all small sets that contain a block in $X$.
By interchanging the
sums over blocks $B$ and polymers $U_B$, we
obtain
\begin{align}
    (1\circ K)(\Lambda)
    & =
    \sum_{X \in \Pcal}  \sum_{\{(U_B,B)\}\in \Ucal(X)}
    \Big( \prod_{B \in \Bcal(X)}J(U_B,B) \Big)
    \sum_{Y \in \Pcal^-(Y_J)}M^Y
    \nnb
    & =
    \sum_{W \in \Pcal}
    \sum_{X \in \Pcal}  \sum_{\{(U_B,B)\}\in \Ucal(X)}
    \Big( \prod_{B \in \Bcal(X)}J(U_B,B) \Big)
    \sum_{Y \in \Pcal^-(Y_J)}M^Y
    \1_{X^\Box \cup Y=W},
\end{align}
where the last equality is just a conditioning of the sums over $X$
and $Y$ according to the constraint  $X^\Box \cup Y=W$.
Then we define $K'(W)$ to be the summand in the sum over $W$.
It can be verified that $K'$ has the component factorisation property,
and it is proved in \cite[Proposition~D.1]{BS-rg-step} that $K'$ obeys
good estimates.

We examine two special cases:
\begin{itemize}
\item
If $W$ is a small set $S$ then we must have $X =\varnothing$ (because
otherwise $X^\Box$ cannot be contained in $S$, as $X^\Box$ is not a small set
even if $X$ is a single block) and also $Y=S$, so
\begin{equation}
    K'(S)=M(S).
\end{equation}
Therefore, for $S=U \not\in \Bcal_+$ or $S=B \in \Bcal_+$,
\begin{align}
     K'(U) & = K(U) - \sum_{B \in \Bcal_+(U)} J(U,B)
     =  K(U) - \LT_U K(U)
     & (U \not\in \Bcal_+),
\lbeq{KUnotB}
\end{align}
\begin{align}
     K'(B) & = K(B) + \sum_{U\supsetneqq B} J(U,B)
     = K(B) + \sum_{U\supsetneqq B}\LT_{U,B} K(U)
     & (B \in \Bcal_+).
\end{align}
In \refeq{KUnotB}, the subtracted term is simply
$\LT_U K(U)$ by \refeq{LocXBsum}.
Thus the relevant and marginal parts of $K(U)$ are subtracted on small sets
that are not a single block.  The price to pay is that those subtractions
have been transferred into
$K'(B)$, which additionally fails to have the relevant and marginal
parts of $K(B)$ subtracted.
\item
If $X=B$ and $Y =\varnothing$ then we must have $W=X^\Box$ and $X$ is uniquely
determined by $W$, and the contribution from this case to $K(W)=K(B^\Box)$
is
\begin{equation}
  \sum_{U \supset B}J(U,B)=0.
\end{equation}
This cancellation has the good consequence that there is no
contribution to $K'(W)$, for any polymer $W$ that is not a single
block, that consists solely of $J$ terms.  The net effect of this is
that there is no connected polymer $W$ such that $K'(W)$ is a linear
function of $J$ without any compensating $K$ factors.
\end{itemize}

The details of the above analysis can be found in the proof of
\cite[Proposition~D.1]{BS-rg-IE}.
It leads to a representation
\begin{equation}
    (I \circ K)(\Lambda) = (I \circ K')(\Lambda)
\end{equation}
where in $K'(X)$ the relevant and marginal parts of $K(X)$ have been removed
from all small sets $X$ except single blocks.
This is carried out in detail in \emph{Map~1} of \cite[Section~4.2]{BS-rg-step}.

\subsubsection*{Cancellation on blocks}

It remains to remove the relevant and marginal parts of $K'(B)$ (which
incorporate the relevant and marginal parts of $K(X)$ for all small
sets $X$), and to transfer them into $V_+$.
This is achieved by replacing $\Vpt(V)$ by $\Vpt(\Vhat)$, where
\begin{equation}\label{e:VhatEuc}
  \Vhat = V - \sum_{U \in \Scal : U \supset B} \LT_{U,B}I^{-U}K(U).
\end{equation}
Let $\Ihat = I(\Vhat)$, $\delta \Ihat = I - \Ihat$,
$\Khat = \delta \Ihat \circ K'$.
By \refeq{chvar1},
\begin{equation}
  (I \circ K')(\Lambda)
  = (\Ihat \circ \Khat)(\Lambda).
\end{equation}
The relevant and marginal parts of $K'$ are thereby transferred to
$\Vhat$ and removed from $\Khat$.  The details of this operation
are outlined in \emph{Map~2} of \cite[Section~4.3]{BS-rg-step}.
The corresponding step for the hierarchical model is performed at \refeq{hatdef}.

\section{Norms}

The norms applied in this book for the hierarchical model require
modification and extension in the Euclidean setting.  We discuss some
aspects of this here.  Full details can be found in \cite{BS-rg-step},
and a general development of properties of the norms is presented in
\cite{BS-rg-norm}.

\subsection{$T_{\varphi}$-seminorms}

For the Euclidean model, a counterpart of the hierarchical
$T_\auxx$-seminorm of \refeq{Tznormdef} is defined in
\cite{BS-rg-norm}. For simplicity, we consider the $1$-component
Euclidean $\varphi^{4}$ model, and do not include an auxiliary space
$\Aux$.  The field $\varphi = (\varphi_{x})_{x\in \Lambda }$ is a
point in $\R^{\Lambda}$, and we will define the Euclidean
$T_{\varphi}$-seminorm on the space of functions $F:\R^{\Lambda}\to \R$.  An
example of such an $F$ is the nonperturbative coordinate $K(X)$
evaluated on a polymer $X$; in this case the dependence is only on
$\varphi_x$ for $x$ in or near $X$.

Given a function $F:\R^{\Lambda}\to \R$, the derivative $F^{(p)}
(\varphi)$ is a $p$-linear function on the space $\R^{\Lambda}$ of
directions.  Let $\Phi$ be a normed vector subspace of
$\R^{\Lambda}$. We denote a direction in $\Phi$ by $\dot{\varphi}$ and
a $p$-tuple of directions by $\dot{\varphi}^{p}$.  Let $\Phi (1)$ be
the unit ball in $\Phi$.  Then $\|F^{(p)} (\varphi)\| =
\sup_{\dot{\varphi}^{p} \in \Phi(1)^{p}}|F^{(p)}
(\varphi;\dot{\varphi}^{p})|$.  The Euclidean $T_\varphi$-seminorm is
defined by
\begin{equation}
    \|F\|_{T_{\varphi}}
    = \sum_{p=0}^{p_\Ncal} \tfrac{1}{p!}\|F^{(p)}(\varphi)\|,
\end{equation}
where $p_\Ncal \in [0,\infty]$ is a parameter at our disposal.  The
example $\Phi = \R_\h^{\Lambda}$ with norm $\|\dot{\varphi} \| =
\frac{1}{\h} \max\{|\dot{\varphi}_{x}|:x\in \Lambda \}$ gives a
$T_{\varphi}$-seminorm with the product property $\|FG\|_{T_\varphi}
\le \|F\|_{T_\varphi}\|G\|_{T_\varphi}$ of
\eqref{e:T-product-prop}. Restrictions on the spaces of directions
that are consistent with the product property are discussed in
\cite{BS-rg-norm}.

The freedom to choose allows us to take into account the properties
that are imposed on typical fields by their probability
distribution. For example, hierarchical fields are constant on
blocks. Suppose that $F (\varphi)$ depends only on fields in a block
$B$. Let $\Phi$ be the subspace of directions in $\R_\h^{\Lambda}$
such that $\dot{\varphi}$ is constant on $B$, i.e.,
$\dot{\varphi}_{x}=\dot{\varphi}_{y}$ for all $x,y$ in $B$.  For
$\varphi$ constant on $B$ the Euclidean $T_{\varphi}$-seminorm with
this choice of $\Phi$ equals the $T_{\varphi}$-seminorm of
Definition~\ref{def:Tnorm}. This is true by virtue of chain rule
formulas like
\begin{equation}
    \sum_{x \in \Lambda } \tfrac{\partial}{\partial \varphi_{x}} F (\varphi)
    =
    \sum_{x \in B} \tfrac{\partial}{\partial \varphi_{x}} F (\varphi)
    =
    \tfrac{\partial}{\partial u} f (u)
\end{equation}
which is valid when the left-hand side is evaluated at $\varphi$ such
that $\varphi_{x}=u$ for all $x\in B$ and, by definition, $f (u) = F
(\varphi)$.

In the Euclidean setting, we use $\Phi$ which takes into account the
spatial variation of fields.  After $j$ renormalisation group steps,
the remaining field to be integrated is $\varphi = \zeta_{j+1}+\dots
+\zeta_{N}$, with increments as in
Corollary~\ref{cor:varphi-decomp}. The scaling estimates
\eqref{e:C-est} indicate that the variance of $\nabla^\alpha \varphi$
typically scales down with $j$ like $L^{-j (d-2)}L^{-2j|\alpha
|_{1}}$.  Fix a positive integer $p_{\Phi}$.  Let the norm on
$\Phi$ be the lattice $\Ccal^{p_{\Phi}}$-norm
\begin{equation}
\label{e:Phinormdef}
    \|\dot \varphi\|_{\Phi(\h)}
    =
    \max\{\h_j^{-1} L^{j|\alpha|_{1}}|\nabla^{\alpha}\dot{\varphi}_{x}|  :
    x\in \Lambda, |\alpha |_{1} \le p_\Phi \}.
\end{equation}
With the choice $\h=\ell_j=\ell_0L^{-j(d-2)/2}$, as in
\refeq{ell-def}, the $T_{\varphi}$-seminorm of $F$ tests the response
of $F$ to typical fluctuations of the field, in particular
fluctuations around being constant on blocks.  The choice
$\h=h_j=k_0L^{-jd/4}\ggen_j^{-1/4}$, as in \refeq{h-def}, is used to
test the response of $F$ to typical large fields.  This is all as it
is for the hierarchical model, apart from the fact that now spatial
gradients of the field are taken into account.

Both parameters $\ell$ and $h$ are combined in the
hierarchical $\Wcal$-norm defined
in \refeq{Wcal}.  A Euclidean counterpart of the $\Wcal$-norm is
defined in \cite[(1.45)]{BS-rg-step}.  The latter also involves \emph{regulators},
which we discuss next.

\subsection{Regulators}

For the hierarchical model, the crucial
Proposition~\ref{prop:crucial-0} asserts that the renormalisation
group map is contractive in $T_{\infty}$-norm. The proof uses the
following fact: if the $+$ scale field $\varphi$ is large in a block
$b$, then it is large on $B\setminus b$ where $B$ is the $+$ scale
block that contains $b$, because $\varphi$ is constant on $B$. This is
used in \eqref{e:large-field-suppression} where the factor
$P_{h_{+}}^{6}(\varphi)$ arises from the growth of $K (b)-\LT K (b)$
as $\varphi$ becomes large in $b$. If $\varphi$ is large in $b$ then
the exponential $\exp[-V (B\setminus b)]$ is small and more than
compensates for the growth of $P_{h_{+}}^{6}(\varphi)$.

For the Euclidean model the $+$ field can be large in $b$ without
being large in $B\setminus b$. We have to prove that typical fields do
not do this.  Let $\varphi_{B} = |B|^{-1}\sum_{x\in B}\varphi_{x}$ be
the average of $\varphi$ over the block $B$ and let
$\delta_{B}=\delta_{B} (\varphi)$ be the supremum over $x,y \in B$ of
$|\varphi_x-\varphi_{y}|/h_{+}$. Then
$|\varphi_{x}-\varphi_{B}|/h_{+}\le \delta_{B}$. In other words
$\varphi_{x}/h_+$ is constant to within $\delta_{B}$.  We will show that
for typical $\varphi$, $\delta_{B} = O_{L} (g^{1/4})$. Thus typical
fields are very close to constants in this sense and it should be
plausible that the hierarchical bound
\eqref{e:large-field-suppression} continues to hold for fields with
$\delta_{B} = O_{L} (\ggen^{1/4})$.

What does it mean for a field to be typical?  For intuition, recall
from \eqref{e:C-est} that the standard deviation of
$\nabla\varphi_{x}$ is $O_{L} (L^{-j}\ell_{+})$, where $\ell_+$ is
defined in \eqref{e:ell-def}.  We say a field $\varphi$ is
\emph{typical} if the maximum over $B$ of $|\nabla \varphi|$ is $O_{L}
(L^{-j}\ell_{+})$. Since $|\varphi_x-\varphi_{y}|$ is bounded by the
length $L^{j+1}$ of a path joining $x$ to $y$ times the maximum
gradient, we find, using \eqref{e:h-def}, that $\delta_{B}$ is bounded by
$O_{L} (L \ell_{+}/h_{+}) = O_{L} (\ggen^{1/4})$ as claimed above.

Although \eqref{e:large-field-suppression} does not hold for all
Euclidean $\varphi$, the inequality obtained by including an extra
factor $\exp [-\delta_{B} ]$ might hold for all $\varphi$ because it
holds for typical $\varphi$ by the arguments above and the decay of
$\exp [-\delta_{B} ]$ might compensate for atypical fields with large
$\delta_{B}$. This example leads to the idea that the
$T_{\infty}$-norm for the hierarchical model should be replaced by a weighted
$T_{\infty}$-norm where the weight will allow the hierarchical proofs
that work for $\delta =0$ to extend to the Euclidean model. The
$T_{\infty}$-norm of a function $F (\varphi)$ tests $F (\varphi)$ on
all possible $\varphi$ but a weighted $T_{\infty}$-norm focuses on the
fields that $F (\varphi)$ actually encounters when taking its
expectation.

Given $w(X,\varphi) >0$, a general weighted $T_{\infty}$-norm is
defined by
\begin{equation}
\lbeq{wnorm}
    \|K(X)\|_w
    =
    \sup_{\varphi \in \R^\Lambda} \frac{\|K(X)\|_{T_\varphi}}{w(X,\varphi)}.
\end{equation}
For the Euclidean $|\varphi|^4$ model, we use two choices of weight
function, or \emph{regulators}, corresponding to the two choices
$\h=\ell$ and $\h=h$ of the parameter $\h$ in the definition of the
$T_\varphi$-seminorm.  The specific choices are discussed in
\cite[Section~1.1.6]{BS-rg-IE}.
The systematic use of such regulators originated in
\cite{BY90}; in \cite[p.216, (1)]{GK83} the breakdown of estimates for
fields with large gradients is instead put into inductive assumptions.

We discuss here the most important case: the \emph{large-field
regulator} $w=\tilde G$.  Desirable properties of the regulator are:
\begin{enumerate}
\item[(i)]
$\tilde G(X\cup Y,\varphi)=\tilde G(X,\varphi)\tilde G(Y,\varphi)$ if $X,Y$ are disjoint.
\item[(ii)]
$\tilde G(X,\varphi+ \varphi') \le \tilde G^2(X,\varphi)\tilde G^2(X, \varphi')$.
\item[(iii)]
$\Ex_+ \tilde G^t(X) \le 2^{|\Bcal(X)|}$ for bounded powers $t$.
\item[(iv)]
$\tilde G^t \le \tilde G_+$ for bounded powers $t$.
\end{enumerate}
Property~(i) extends the product property of the $T_\varphi$-seminorm
to the norm \refeq{wnorm} when $X,Y$ are disjoint.  Properties
(ii)-(iv) allow estimates to be advanced from one scale to the next,
as in the following lemma.  The proof of the lemma uses the general
inequality
\begin{equation}
\lbeq{Exwnorm}
    \|\Ex_+ \theta F\|_{T_\varphi(h_+)}
    \le
    \Ex_+ \|F\|_{T_{\varphi + \zeta}(h)},
\end{equation}
which follows from \cite[Proposition~3.19]{BS-rg-norm} (see also
\cite[(7.2)--(7.3)]{BS-rg-IE}).  The inequality \refeq{Exwnorm} is
reminiscent of Proposition~\ref{prop:ExCthetaTphi} for the hierarchial
case.

\begin{lemma}
Suppose that $\tilde G$ obeys (ii), (iii) and (iv) above.
Then
\begin{equation}\label{e:Kwnorm}
  \|\Ex_+\theta K(X)\|_{\tilde G_+}
  \le
  2^{|\Bcal(X)|} \|K(X)\|_{\tilde G}
  .
\end{equation}
\end{lemma}

\begin{proof}
We first apply \refeq{Exwnorm}, and then the definition of the weighted
norm, to obtain
\begin{align}
      \|\Ex_+\theta K(X)\|_{T_\varphi(h_+)}
      & \le  \Ex_+ \|K(X)\|_{T_{\varphi + \zeta}(h)}
      \nnb
      & \le
      \|K(X)\|_{\tilde G} \;
      \Ex_+ \tilde G(X,\varphi + \zeta) .
\end{align}
Using properties (ii), (iii) and (iv), we find that
\begin{align}
      \|\Ex_+\theta K(X)\|_{T_\varphi(h_+)}
      & \le
       \|K(X)\|_{\tilde G} \; \tilde G^{2}(X,\varphi)
       \Ex_+ \tilde G^{2} (X,\zeta)
    \nnb & \le
    \|K(X)\|_{\tilde G} \tilde G^{2}(X,\varphi)   2^{|\Bcal(X)|}
     \nnb & \le
    \|K(X)\|_{\tilde G} \tilde G_+(X,\varphi)   2^{|\Bcal(X)|}
    .
\end{align}
This proves \refeq{Kwnorm}.
\end{proof}

To implement the above, we require a regulator $\tilde G$ which obeys
properties (i)-(iv).  A trivial choice is of course given by $\tilde G
= 1$.  However, for the weight to be helpful, $\tilde G$ should be as
large as possible.
Other authors have used
regulators based on lattice Sobolev norms, e.g.,
\cite[(47)]{DH92}.
Our choice is the regulator $\tilde G$ given in \cite[(1.41)]{BS-rg-IE}.
We conclude by presenting its definition.
Further details, including a discussion of the
\emph{fluctuation-field regulator}, can be found in
\cite[Section~1.1.6]{BS-rg-IE}.

First, for $X\subset\Lambda$ with diameter less than the period of the
torus, we define
\begin{equation}
\label{e:Phitilnorm}
    \| \varphi \|_{\tilde{\Phi} (X)}
=
    \inf \{ \| \varphi -f\|_{\Phi}  : \text{$f$ restricted to $X$
    is a linear polynomial }
    \}.
\end{equation}
The restriction on the diameter of $X$ is present so that it makes sense
to consider $f$ as a linear polynomial in \eqref{e:Phitilnorm}.
The \emph{large-field regulator} is then given by
\begin{align}
\label{e:Gtildef}
    \tilde G  (X,\varphi)
    =
    \prod_{x \in X}
    \exp \left(
    L^{-dj}
    \|\varphi\|_{\tilde\Phi (b_{x}^\Box)}^2
    \right)
    ,
\end{align}
where $b_x\in \Bcal$ is the unique block which contains the point $x$.

The above construction of $\tilde{G}$ factors out linear
polynomials.  This is a way to examine the size of
$|\nabla^{2}\varphi|$, and in that sense is related to a Sobolev
norm. Thus the regulator can bound $\nabla^{2} \varphi$, but not
$\varphi$.
In our motivation of the weighted
$T_{\infty}$-norm, we estimated how close $\varphi$ is to being constant
in a block $B$. However, the regulator \refeq{Gtildef} only enables us to
estimate how close $\varphi$ is to being a linear function. Of course
linear functions include constants and in fact we expect
that fields are
close to being constants, but it is easier to prove the weaker
statement that they are close to linear. Also it is sufficient: if
$\varphi$ is linear on $B$ and it is large on $b$ then it is large on
roughly half of $B$ so the factor $\|e^{-V (B\setminus
b)}\|_{T_{\varphi}}$ in \eqref{e:large-field-suppression} is still
exponentially small and bounds the polynomial that depends on the
field in $b$.

An advantage of the regulator \refeq{Gtildef} is that its weighted norm
leads to a complete Banach space after an additional weighted supremum
over polymers $X$ is taken in \refeq{wnorm}.  This is discussed in detail in
\cite[Appendix~A]{BS-rg-step}.  The Sobolev regulator was erroneously
claimed to produce a complete space, e.g., in \cite{BMS03}; this error was
pointed out and corrected in \cite{Abde07} in a manner than maintained
the Sobolev regulator.

Finally, we note that properties (i)-(iv) hold for \refeq{Gtildef}.
Property~(i) holds by definition, and property~(ii) is a
consequence of the elementary inequality $(a+b)^2 \le 2a^2 + 2b^2$.
Property~(iii) is a consequence of
\cite[Proposition~3.20]{BS-rg-norm} together with the fact that the
large-field regulator is less than or equal to the fluctuation-field
regulator.
According to \cite[Lemma~1.2]{BS-rg-IE}, property (iv) holds if if $L$
is large enough.


%% file: solns.tex
\chapter{Solutions to exercises}
\label{app:solutions}

\section{Chapter~\ref{ch:intro} exercises}

\begin{solution}{ex:laplace-principle}\label{soln:laplace-principle}
By replacing $V(\varphi)$ by $V(\varphi)-V(\varphi_0)$, $g(\varphi)-g(\varphi_0)$ by $g(\varphi)$,
and $\varphi$ by $\varphi+\varphi_0$,
we can assume that $\varphi_0=0$, $V(\varphi_0)=0$ and $g(\varphi_0)=0$.
For $E \subset \R^n$, let
\begin{equation}
    I_N(E,f) = \int_E f(\varphi) e^{-NV(\varphi)}d\varphi.
\end{equation}
With the above assumptions, we must prove that
\begin{equation}
    \lim_{N\to\infty} \frac{I_N(\R^n,g)}{I_N(\R^n,1)} =0.
\end{equation}

Given $t\in (0,1)$,
let
\begin{equation}
  M_{t}=\{\varphi: V(\varphi) \leq t \}.
\end{equation}
By assumption, $M_t$ is compact, and it clearly contains the set
$\{\varphi: V(\varphi) < t \}$ which is open by continuity of $V$.
This set is not empty because it contains $0$. Therefore the integral
$I_1(M_t,1)=\int_{M_t}e^{-V} \, d\varphi$ is nonzero.
On $M_t^c$, we have $e^{-NV} = e^{-(N-1)V}e^{-V} \le e^{-(N-1)t}e^{-V}$.
With this and a similar but reversed inequality on $M_{t/2}$, we obtain
\begin{align}
    |I_N(M_t^c,g)| & \le \|g\|_\infty  e^{-(N-1)t} I_1(\R^n,1),
    \\
    I_N(\R^n,1) & \ge I_N(M_{t/2},1) \ge e^{-(N-1)t/2} I_1(M_{t/2},1).
\end{align}
Thus, for a $t$-dependent constant $c_t$,
\begin{equation}
    \frac{|I_N(M_t^c,g)|}{I_N(\R^n,1)} \le c_t e^{-Nt/2}.
\end{equation}

Given $\epsilon >0$, choose $\delta>0$ small enough that
$|g(\varphi)|<\epsilon$ if $|\varphi|<\delta$.  Since $\cap_{t>0}M_t =
\{0\}$ we have $\cap_{t>0}M_{t}\cap \{|\varphi|\ge \delta \} =
\varnothing$. By $M_t \subset M_{t'}$ for $t>t'$ and the finite
intersection property for compact sets, there exists $t_{\delta}$ such
that $M_{t_{\delta}}\cap \{|\varphi|\ge \delta \} =
\varnothing$. Therefore $M_{t_\delta} \subset \{|\varphi|<\delta\}$.
Then, with $t=t_\delta$,
\begin{equation}
    |I_N(M_t,g)| \le \epsilon I_N(M_t,1) \le \epsilon I_N(\R^n,1),
\end{equation}
and hence
\begin{equation}
    \frac{|I_N(\R^n,g)|}{I_N(\R^n,1)}
    \le
    \frac{|I_N(M_t,g)|}{I_N(\R^n,1)} + c_t e^{-Nt/2}
    \le
    \epsilon + c_t e^{-Nt/2}.
\end{equation}
Consequently the $\limsup_{N\to \infty}$
of the left-hand side is at most $\epsilon$.  Since $\epsilon$
is arbitrary, the limit must exist and equal zero.
\end{solution}

\begin{solution}{ex:V-properties3} \label{soln:V-properties3}
  By definition and since $|\sigma|=1$,
  \begin{align}
    V(\varphi)
    &= -\log \int_{S^2} e^{-\frac{\beta}{2}|\varphi-\sigma|^2 + h\cdot\sigma} \, \mu(d\sigma)
    \nnb
    &= \frac{\beta}{2}|\varphi|^2+\frac{\beta}{2}-\log \int_{S^2} e^{(\beta \varphi+h)\cdot\sigma} \, \mu(d\sigma).
  \end{align}
  In spherical coordinates,
  \begin{align}
    \int_{S^2} e^{(\beta \varphi+h)\cdot\sigma} \, \mu(d\sigma)
    &=
    \frac12 \int_0^\pi e^{|\beta\varphi+h|\cos\theta} \sin\theta \, d\theta
    \nnb
    &=
    \frac12 \int_{-1}^1 e^{|\beta\varphi+h|u} du
    = \frac{\sinh(|\beta\varphi+h|)}{|\beta\varphi+h|}.
  \end{align}
\end{solution}

\begin{solution}{ex:V-properties} \label{soln:V-properties}
  Let $\varphi \in\R^n$,  $h \in \R^n$, $e \in \bbS^{n-1}$.
  Denote by $\mu_\varphi^h$ the measure from \eqref{e:muphidef} with
  external field $h\in\R^n$.
  From \eqref{e:Vr-def} it follows that
  \begin{equation}
    e \cdot \He V(\varphi)e = \beta - \beta^2 \Var_{\mu_\varphi^h}(e \cdot \sigma)
    .
  \end{equation}

  \smallskip\noindent
  (i)
  By \cite[Theorem~D.2]{DLS78},
  \begin{equation}
    \Var_{\mu_\varphi^h}(e\cdot\sigma) \leq \Var_{\mu_0^0}(e\cdot\sigma) = \frac{1}{n}.
  \end{equation}
  Therefore, $e\cdot \He V(\varphi) e \geq \beta -\beta^2/n$.  When $\beta \le n$,
  the right-hand side is indeed non-negative.

  \smallskip\noindent
  (ii)
    Note that
  \begin{equation}
    \frac{\beta}{2} (\varphi-\sigma,\varphi-\sigma) - (h,\sigma)
    = \frac{\beta}{2} (\varphi,\varphi) - (\beta\varphi+ h, \sigma)
    + \text{constant}.
  \end{equation}
  Hence, for $\beta\varphi+h=0$, the measure $\mu_\varphi^h$ on $\bbS^{n-1}$ is uniform and thus
  \begin{equation}
    \Var_{\mu_\varphi^h}(e\cdot \sigma) = \frac{1}{n}.
  \end{equation}
For $\varphi=-h/\beta$ and any $e \in \bbS^{n-1}$,
this implies that $e\cdot \He V(\varphi)e = \beta-\beta^2/n$.  The right-hand side is
negative if $\beta>n$, so $V$ is non-convex.
\end{solution}

\begin{solution}{ex:Greenf}
\label{soln:Greenf}
For $f \in \R^{\Lambda}$ define $(f,g) = \sum_{x\in \Lambda}
f_x g_x$ and $(\nabla f)_{xy} = f_x-f_y$.  By \eqref{e:Lapbeta}
\begin{align}
    (-\Delta_{\beta} f, f)
    &=
    - \sum_{x \in \Lambda} \Big(\sum_{y \in \Lambda} \beta_{xy}(\nabla f)_{yx} \Big) f_{x}
    =
    - \frac 12 \sum_{x,y \in \Lambda}  \beta_{xy}(\nabla f)_{yx} (f_{x}-f_{y})\nonumber \\
    &=
    \frac 12 \sum_{x,y \in \Lambda}  \beta_{xy}(\nabla f)^{2}_{yx} \ge 0 .
\end{align}
The second equality is obtained by interchanging
$x,y$, $(\nabla f)_{yx} = - (\nabla f)_{xy}$ and recalling that
$\beta_{xy}=\beta_{yx}$.
Also, $(-\Delta_\beta \1)_x = -\sum_{y\in\Lambda} \beta_{xy}(\nabla
\1)_{yx}=0$, and \refeq{freechim2} follows by applying the inverse
operator to $\1= m^{-2}(-\Delta_\beta + m^2)\1$.
\end{solution}

\index{Bubble diagram}
\begin{solution}{ex:bubble1z}
Since $\FDel(k) \sim |k|^2$ as $k \to 0$, we see from \refeq{freebubble}
that $B_0$ is finite if and only if $d>4$.  So it remains to prove that
\label{soln:bubble1z}
  \begin{equation}
  \lbeq{bubasy}
    B_{m^2} \sim
    b_d \times
    \begin{cases}
     m^{-(4-d)} & (d < 4)
    \\
    \log m^{-2} & (d=4),
    \end{cases}
  \end{equation}
with $b_1= \frac 18$, $b_2 = \frac{1}{4\pi}$, $b_3 = \frac{1}{8\pi}$, $b_4 = \frac{1}{16\pi^2}$.

Let $d \le 4$.  By \refeq{freebubble},
  \begin{equation}
    \label{e:freebubble-app}
    B_{m^2}
    =
    \int_{[-\pi,\pi]^d}
    \left|\frac{1}{4 \sum_{j=1}^{d} \sin^2 (\frac{k_j}{2}) +m^2}\right|^2
    \frac{dk}{(2\pi)^d}
    .
  \end{equation}
  Let $U_1$ be the ball of radius $1$ in $\R^d$.
  Then, uniformly as $m^2 \downarrow 0$,
  \begin{equation}
    \int_{[-\pi,\pi]^d \setminus U_1}
    \left|\frac{1}{4 \sum_{j=1}^{d} \sin^2 (\frac{k_j}{2}) +m^2}\right|^2
    \frac{dk}{(2\pi)^d} = O(1),
  \end{equation}
  and, uniformly in $k \in U_1$,
  \begin{equation}
    \frac{1}{4 \sum_{j=1}^{d} \sin^2 (\frac{k_j}{2}) +m^2}
    = \frac{1}{|k|^2+m^2} + O(1).
  \end{equation}
  Therefore,
  \begin{equation} \label{e:bubblemain}
    B_{m^2} \sim \int_{U_1} \left(\frac{1}{|k|^2 + m^2}\right)^2 \; \frac{dk}{(2\pi)^d}
    .
  \end{equation}
  We use polar coordinates to obtain
  \begin{equation}
    \int_{U_1} \left(\frac{1}{|k|^2 + m^2}\right)^2 \; \frac{dk}{(2\pi)^d}
    = \frac{\omega_{d-1}}{(2\pi)^d} \int_0^1 \left(\frac{1}{r^2 + m^2}\right)^2 r^{d-1} \; dr
  \end{equation}
  where $\omega_0 =1$, $\omega_1= 2\pi$, $\omega_2=4\pi$,
   $\omega_3 = 2\pi^2$ arise from the area of the unit $(d-1)$-sphere.
  With the change of variables $r=sm$, this gives
  \begin{equation}
    B_{m^2} \sim
    \frac{\omega_{d-1}}{(2\pi)^d} m^{d-4} \int_0^{m^{-1}} \left(\frac{1}{s^2 + 1}\right)^2 s^{d-1} \; ds.
  \end{equation}
If $d<4$ then the integral converges to a finite limit as $m \downarrow 0$,
and if $d=4$ then it is asymptotic to $\log m^{-1}$.
For $d<4$, the value of the integral is  given by \cite[3.241]{GR65} as
\begin{equation}
    \int_0^{\infty} \left(\frac{1}{s^2 + 1}\right)^2 s^{d-1} \; ds
    =
    \frac{1}{2} \frac{\Gamma(d/2) \Gamma(2-d/2)}{\Gamma(2)}.
\end{equation}
This leads to
\begin{equation}
    \int_0^{\infty} \left(\frac{1}{s^2 + 1}\right)^2 s^{d-1} \; ds
    =
    \begin{cases}
    \frac{\pi}{4} & (d=1)
    \\
    \frac 12  & (d=2)
    \\
    \frac{\pi}{4} & (d=3).
    \end{cases}
\end{equation}
From this we obtain the constants $b_d$ reported below \refeq{bubasy}.
\end{solution}

\begin{solution}{ex:transience}
\label{soln:transience}
  (i) Let $T_0 = 0$ and $T_k = \inf \{ n > T_{k-1}: S_n = 0 \}$.
  Then $u = P(T_1 < \infty)$, and by induction
  and the strong Markov property,
  $P(T_k < \infty) = u^k$.
  Therefore,
  \begin{equation}
    m = EN = \sum_{k \geq 0} P(T_k< \infty) = (1-u)^{-1} .
  \end{equation}

\smallskip\noindent
  (ii)
  Let $S_n$ denote simple random walk.  The equality $EN = \sum_{n \ge 0} p_n(0)$
  follows from the identity $N = \sum_{n=0}^\infty \1_{S_n=0}$.
  Let $D(x) = \frac{1}{2d}\1_{|x|_1=1}$.  Then $\hat D(k) = \frac{1}{d}\sum_{j=1}^d \cos k_j$
  and $1-\hat D(k) = \frac{1}{2d} \FDel(k)$.
  Also,
  \begin{equation}
    P(S_n = 0) = \int_{[-\pi,\pi]^d} \hat D(k)^n \frac{d^dk}{(2\pi)^d}.
  \end{equation}
  Some care is required to perform the sum over $n$ since the best uniform
  bound on $\hat D^n$ is $1$
  which is not summable.
  By monotone convergence, and then by the dominated
  convergence theorem,
  \begin{equation}
    m = \lim_{t\nearrow 1} \sum_{n \geq 0} P(S_n=0)t^n
    = \lim_{t\nearrow 1} \int_{[-\pi,\pi]^d} \frac{1}{1- t\hat D(k)}\; \frac{d^dk}{(2\pi)^d}.
  \end{equation}
  The function $\hat D$ is real valued, and
  \begin{equation}
    \frac{1}{1- t\hat D(k)}
    \leq \frac{2}{1- \hat D(k)} \quad \text{for $t\in[1/2,1]$.}
  \end{equation}
  If $(1-\hat D)^{-1} \in L^1$,
  the claim then follows by dominated convergence.
  If  $(1-\hat D)^{-1} \not\in L^1$,
  the claim follows from Fatou's lemma.

  \smallskip\noindent
  (iii)  This follows from the fact that $\FDel(k) \asymp |k|^2$ as $k \to 0$,
  and thus $1/\FDel(k)$ is integrable if and only if $d>2$.
\end{solution}

\index{Bubble diagram}
\begin{solution}{ex:bubble1}
\label{soln:bubble1}
  By definition,
  \begin{equation}
    I = \sum_{x \in \Z^d} \Big(\sum_{m= 0}^\infty
    \1_{S^1_m = x}\Big)\Big( \sum_{n= 0}^\infty \1_{S^2_n = x}\Big).
  \end{equation}
  By Lemma~\ref{lem:srw} (with monotone convergence to take the limit $m^2 \downarrow 0$),
  \begin{equation}
    E \sum_{i\geq 0} \1_{X^1_i = x} =  2d C_{0x}(0).
  \end{equation}
  Therefore, by using the independence, we have
  \begin{equation}
    E I = (2d)^2 \sum_{x \in \Z^d} (C_{0x}(0))^2 = (2d)^2 B_0.
  \end{equation}
  The sum is infinite if and only if both sides are infinite.
  The latter happens if and only if $d \le 4$, by Exercise~\ref{ex:bubble1z}.
\end{solution}

\section{Chapter~\ref{ch:gauss} exercises}

\begin{solution}{ex:ibp}
\label{soln:ibp}
For notational convenience, we consider the case where $C$ is strictly positive definite.
The semi-definite case can be handled by replacing $C$ by $C'$ as in \refeq{Gm2}.
Let $A=C^{-1}$, so that $P_C(d\varphi)$ is proportional to
$e^{-\frac 12 (\varphi,A\varphi)}d\varphi$.  Then standard integration by parts
and the symmetry of the matrix $C$ give
\begin{equation}
    \int \ddp{F}{\varphi_y} e^{-\frac 12 (\varphi,A\varphi)}d\varphi
    =
    \int (A\varphi)_y F e^{-\frac 12 (\varphi,A\varphi)}d\varphi.
\end{equation}
Now we multiply by $C_{xy}$, sum over $y$, and use $CA=I$.  This gives
\begin{equation}
    \sum_y C_{xy} \int \ddp{F}{\varphi_y} e^{-\frac 12 (\varphi,A\varphi)}d\varphi
    =
    \int F \varphi_x e^{-\frac 12 (\varphi,A\varphi)}d\varphi,
\end{equation}
as required.
\end{solution}

\begin{solution}{ex:wickpp}
\label{soln:wickpp}
By definition,
\begin{equation}
    \Cov_C (\varphi_{x}^{p},\varphi_{x'}^{p'})
    = \Ex_C(\varphi_{x}^{p}\varphi_{x'}^{p'}) - (\Ex_C\varphi_{x}^{p})(\Ex_C\varphi_{x'}^{p'}).
\end{equation}
The estimate is obtained by bounding each expectation on the right-hand side using
\refeq{ECpoly}, without any attention to cancellation between the two terms.
The operator $e^{\frac 12 \Delta_C}$ is defined by power series expansion, and
in using \refeq{ECpoly} nonzero contributions can arise only when all fields are
differentiated.  For the term $\Ex_C(\varphi_{x}^{p}\varphi_{x'}^{p'})$, this
differentiation leads to $(p+p')/2$ factors of the covariance, which can be factors
$C_{xx}$, $C_{xx'}$, or $C_{x'x'}$.  The covariance is maximal on the diagonal since
it is positive definite, so these factors are all bounded by $\|C\|$ and hence the term
$\Ex_C(\varphi_{x}^{p}\varphi_{x'}^{p'})$ obeys the desired estimate.  The subtracted term
$(\Ex_C\varphi_{x}^{p})(\Ex_C\varphi_{x'}^{p'})$ is similar.
\end{solution}

\begin{solution}{ex:Gauss-Laplace-Z0}
\label{soln:Gauss-Laplace-Z0}
This follows from \refeq{Gauss-chvar}, using (with $A=C^{-1}$)
\begin{align}
    \int e^{(f,\varphi)} Z_0(\varphi) e^{-\frac 12(\varphi,A\varphi)} d\varphi
    & =
    e^{\frac 12 (f,Cf)}
    \int  Z_0(\varphi) e^{-\frac 12(\varphi-Cf,A(\varphi-Cf))} d\varphi
    \nnb
    & =
    e^{\frac 12 (f,Cf)}
    \int  Z_0(\varphi+Cf) e^{-\frac 12(\varphi,A\varphi)} d\varphi.
\end{align}
\end{solution}

\begin{solution}{ex:gauss-On}
\label{soln:gauss-On} (i)
Let $X = \{(x,i): x\in \Lambda, i=1, \dots, n\}$ and
$\hat{C}_{(x,i),(y,j)} =\delta_{ij}C_{xy}$.
According to
Example~\ref{example:gauss-vect} and
Proposition~\ref{prop:Gauss-Laplace}, the $n$-comp\-on\-ent Gaussian
field field $\varphi = (\varphi_x^i)_{x\in\Lambda,i=1, \dots,n}$ with
covariance $C$ is characterised by
\begin{equation}
    \Ex_C(e^{(\hat{f},\varphi)})
    =
    e^{\frac12 (\hat{f},\hat{C}\hat{f})}
    \quad
    \text{for $\hat{f} = (f^{i}_{x}) \in \R^{X}$.}
\end{equation}
The form of $\hat{C}$
implies this is the same as
\begin{equation}
    \Ex_C(e^{(\hat{f},\varphi)})
    =
    \prod_{i} e^{\frac12 (f^{i},Cf^{i})}
    \quad
    \text{for $f^{i} = (f^{i}_{x})_{x\in \Lambda} \in \R^{\Lambda}$ and $i=1,\dots ,n$.}
\end{equation}
The factorisation on the right-hand side implies that the components
$\varphi^{i}$ are independent and by
Proposition~\ref{prop:Gauss-Laplace} applied to each component the
components are identically distributed Gaussian fields on $\Lambda$
with covariance~$C$, as desired.
\\
(ii) The set of functions $F$ for which
\begin{equation}
    \label{e:sol:gauss-On-1}
    \Ex_C(F(\varphi)) = \Ex_C(F(T\varphi))
\end{equation}
holds is a vector space closed under bounded convergence and under
monotone convergence. Exponential functions generate the Bore1
$\sigma$-algebra in $\R^{X}$ and form a class closed under
multiplication.  Hence if \eqref{e:sol:gauss-On-1} holds for
exponential functions then it holds for all bounded Borel functions
$F$. For exponential functions we evaluate and compare both sides of
\eqref{e:sol:gauss-On-1} using Proposition~\ref{prop:Gauss-Laplace}.
\begin{equation}
    \Ex_C(e^{(f,T\varphi)})
    =
    \Ex_C(e^{(T^{t}f,\varphi)})
    =
    e^{\frac12 (T^{t}\hat{f},\hat{C}T^{t}\hat{f})}
    =
    e^{\frac12 (\hat{f},\hat{C}\hat{f})}
    =
    \Ex_C(e^{(f,\varphi)}) .
\end{equation}
The formula $\Ex_C \theta \circ T = T \circ \Ex_C\theta$ is obtained
from \eqref{e:sol:gauss-On-1} by renaming the random variable
$\varphi$ to $\zeta$ followed by replacing $F (\zeta)$ by $F (\varphi
+ \zeta)$ where $\varphi$ is a fixed element of $\R^{X}$.
\end{solution}

\begin{solution}{ex:Gauss-infdim}
\label{soln:Gauss-infdim}
We apply (e.g.) \cite[Theorem~10.18]{Foll99} with $A=\Zd$.
By Corollary~\ref{cor:Gauss-restrict},
the finite-dimensional distributions are consistent, and the
permutation hypothesis of \cite[Theorem 10.18]{Foll99} follows from
the definition of the Gaussian measure with covariance $C_{X\times X}$
for finite $X$.
\end{solution}

\begin{solution}{ex:trunc-corr-existence}
\label{soln:trunc-corr-existence}
Let $I$ be a finite nonempty subset of natural numbers.
A partition $\pi$ of $I$ is a collection of disjoint nonempty
subsets of $I$ whose union is $I$.  In particular, $\{I \}$ is a
partition of $I$. Let $\Pi (I)$ be the set of all partitions of
$I$. Given a natural number $n$ and coefficients $\mu_{I}$ for all $I$ of
cardinality $|I| \le n$, define coefficients $\kappa_{J}$ for all
finite subsets $J$ with $|J| \le n$ to be the unique solution of the
system of equations
\begin{equation}\label{e:cumulants-0}
    \mu_{I}
    =
    \sum_{\pi \in \Pi (I)}\prod_{J \in \pi}  
    \kappa_{J},
\end{equation}
where there is one equation for each $I$ with $|I|\le n$.  To show
that this system has a unique solution, rewrite it as
\begin{equation}
    \label{e:cumulant-recursion}
    \kappa_{I}
    =
    \mu_{I} -
    \sum_{\pi \in \Pi (I)\setminus\{I\}}\prod_{J \in \pi}  
    \kappa_{J} .
\end{equation}
For any finite $I$, this defines $\kappa_{I}$ in terms of $\mu_{I}$ and
recursively in terms of $\kappa_{J}$, where $J$ runs over proper
subsets of $I$. Thus we obtain a formula for $\kappa_{I}$ in terms of
$\mu_{J}$ by inserting the recursion into itself. Since $J$ is a
proper subset of $I$, the recursion terminates after a finite number of
steps determined by the cardinality $|I|$. For $I=\{i\}$ the recursion
reduces to $\kappa_{\{i\}}=\mu_{\{i\}}$ because $\{I \}$ has no proper
subsets and empty sums are by definition zero. By induction on
$|I|$, the coefficient $\kappa_{I}$ is a finite sum of finite products
of $\mu_{J}$ with $|J|\le |I|$. Conversely, given $\kappa_{J}$ for
$|J| \le n$ the formula \eqref{e:cumulants-0} constructs $\mu_{I}$ for
$|I| \le n$.

We assume the existence of exponential moments as required by the
definition of cumulants in \eqref{e:cumulants}, and set $\mu_{I}=\Ex
(A_{i_{1}} \cdots A_{i_{n}})$ for all $I=\{i_{1},\dots ,i_{n} \}$.  We
claim that $\kappa_{i_{1},\dots ,i_{n}} = \Ex (A_{i_{1}};\cdots
;A_{i_{n}})$. This claim proves the desired result.
In particular, the cumulant of order $n$ exists
precisely when expectations up to order $n$ exist.

To prove the claim, for arbitrary $I=\{i_{1},\dots ,i_{n} \}$ let
\[
    \partial_{I}
    =
    \frac{\partial^{n}}{\partial t_{i_{1}} \cdots \partial t_{i_{n}}} ,
\]
and define $f_{I}$ by $f_{I} (t_{j_1},\dots ,t_{i_{n}}) = \log
\Ex (e^{t_{i_{1}}A_{i_{1}} + \cdots + t_{i_{m}} A_{i_{n}}})$.  By the
chain rule and induction on $|I|$,
\[
    \partial_{I} e^{f_{I}}
    =
    \left(
    \sum_{\pi \in \Pi (I)}\prod_{J \in \pi}
    \partial_{J}f_{J}
    \right) e^{f_{I}} .
\]
Set $t_{i_{1}},\dots ,t_{i_{n}}=0$. By the definition of $f_{I}$, the
left-hand side is $\mu_I=\Ex (A_{i_{1}} \cdots A_{i_{n}})$.
By comparing the above equation with
\eqref{e:cumulants-0}, and noting that $e^{f_{I}}=1$ at $t_{i_{1}},\dots
,t_{i_{n}}=0$, we have $\kappa_{I}=\partial_{I}f_{I}$ for all $I$. By
the definition of the truncated expectation, $\partial_{I}f_{I}= \Ex
(A_{i_{1}};\cdots ;A_{i_{n}})$, so we have proved the claim that $\Ex
(A_{i_{1}};\cdots ;A_{i_{n}}) = \kappa_{I}$, as desired.
\end{solution}

\begin{solution}{ex:Gauss-cum}
\label{soln:Gauss-cum}
Suppose first that $\varphi$ is Gaussian with covariance $C$.
By Proposition~\ref{prop:Gauss-Laplace},
\begin{equation}
    \Ex_C(e^{\sum_{i=1}^p t_{i} \varphi_{x_i}}) = e^{\frac12 \sum_{i,j=1}^p t_{i}t_{j}C_{x_ix_j}} .
\end{equation}
By \eqref{e:cumulants}, the cumulants are derivatives of
the right-hand side, and therefore \eqref{e:ex:Gauss-cum} holds,
as desired.

Suppose next that for all $p \in \N$ and $x_1, \dots, x_p \in X$,
\begin{equation}
  \label{e:ex:Gauss-cum-soln}
  \Ex(\varphi_{x_1}; \cdots; \varphi_{x_p})
  =
  \begin{cases} C_{x_1x_2} & (p=2)
      \\
      0 & (p\neq 2).
  \end{cases}
\end{equation}
By Exercise~\ref{ex:trunc-corr-existence}, the truncated expectations up
to order $n$ determine the expectations up to order $n$.  Therefore
all moments are the same as those of a Gaussian with covariance
$C$.  This implies that
\begin{equation}
    \sum_{n=0}^\infty \frac{1}{n!} \Ex ((f,\varphi)^n)
\end{equation}
is equal to the sum over even $n$ (because odd Gaussian moments are zero) and
therefore converges by monotone convergence to $\Ex (e^{(f,\varphi)})$.
However, since the moments are Gaussian, the above sum is equal to $e^{\frac 12 (f,Cf)}$.
It follows that  $\Ex (e^{(f,\varphi)})=e^{\frac 12 (f,Cf)}$.
By Proposition~\ref{prop:Gauss-Laplace}, this
proves that the field is Gaussian with covariance $C$, and the proof is complete.
\end{solution}

\begin{solution}{ex:Fexpand}
\label{soln:Fexpand}
Let $A,B$ be polynomials in $\varphi$ degree at most $p$.
By definition,
\begin{equation} \label{e:FC}
F_C(A,B) = e^{\frac12 \Delta_C}((e^{-\frac12 \Delta_C}A)(e^{-\frac12 \Delta_C}B)) - AB.
\end{equation}
We must show that
\begin{equation} \label{e:FC_derivatives_formula-bis}
F_C(A,B) = \sum_{n=1}^p \frac{1}{n!} \sum_{x_1,y_1} \cdots \sum_{x_n,y_n} C_{x_1,y_1} \cdots C_{x_n,y_n}
\ddp{^nA}{\varphi_{x_1}\cdots\varphi_{x_n}}
\ddp{^nB}{\varphi_{y_1}\cdots\varphi_{y_n}}.
\end{equation}
Define
\begin{align}
&\Lcal_C = \frac 12 \Delta_{C}
= \frac12 \sum_{u,v \in \Lambda} C_{u,v} \partial_{\varphi_{u}} \partial_{\varphi_{v}}
& &\Lcallr_C = \sum_{u,v \in \Lambda} C_{u,v} \partial_{\varphi_{u}'} \partial_{\varphi_{v}''} \\
&\Lcal_C' = \frac12 \sum_{u,v \in \Lambda} C_{u,v} \partial_{\varphi_{u}'} \partial_{\varphi_{v}'}
& &\Lcal_C'' = \frac12 \sum_{u,v \in \Lambda} C_{u,v} \partial_{\varphi_{u}''} \partial_{\varphi_{v}''}
.
\end{align}
Then \eqref{e:FC} becomes
\begin{equation}
\begin{aligned}
F_C(A,B)
&= e^{\Lcal_C' + \Lcal_C'' + \Lcallr_C}
((e^{-\Lcal_C'}A(\varphi'))(e^{- \Lcal_C''}B(\varphi'')))\big|_{\varphi'=\varphi''=\varphi} -AB \\
&= e^{\Lcallr_C}\big(A(\varphi')B(\varphi'')\big) \Big|_{\varphi'=\varphi''=\varphi} - AB,
\end{aligned}
\end{equation}
and \eqref{e:FC_derivatives_formula-bis} follows by
expanding the exponential.
\end{solution}

\section{Chapter~\ref{ch:decomp} exercises}

\begin{solution}{ex:uncorr-then-indep-Gauss}
\label{soln:uncorr-then-indep-Gauss}
Two random variables $X$ and $Y$ are independent if their distribution is a product measure.
Provided that both random variables
have exponential moments, this is equivalent to the factorisation
of the Laplace transform:
\begin{equation}
  \Ex(e^{tX+sY}) = \Ex(e^{tX})\Ex(e^{sY}),
\end{equation}
since the distribution of $(X,Y)$ is characterised by the Laplace transform
and the Laplace transform of independent random variables factorises.

Consider now the special case $X=\varphi_x$ and $Y=\varphi_y$, and let
$C_{xy} = \Ex(\varphi_x\varphi_y)$.  By assumption, $C_{xy}=0$ for $x\neq y$.
The above factorisation now follows from
\eqref{e:Gauss-Laplace}, which implies that
\begin{equation}
  \Ex(e^{t\varphi_x + s\varphi_y})
  = e^{\frac12 (t^2 C_{xx} + s^2 C_{yy} + 2st C_{xy})}
  = e^{\frac12 (t^2 C_{xx} + s^2 C_{yy})}
  =   \Ex(e^{t\varphi_x}) \Ex(e^{s\varphi_y}).
\end{equation}
This completes the proof.
\end{solution}

\begin{solution}{ex:posdef}
\label{soln:posdef}
If $h$ is even then $h*h$ is even since
\begin{align}
    h*h(x) & = \int_{\R^d} h(x-y)h(y) \,dy = \int_{\R^d} h(-x+y)h(y) \, dy
    \nnb &
    =\int_{\R^d} h(-x-y)h(-y) \, dy = \int_{\R^d} h(-x-y)h(y) \, dy= h*h(-x).
\end{align}
Since $\widehat{h*h}=\hat{h}^2$, and since $\hat{h}$ is real because $h$ is even,
we see that $\widehat{h*h} \ge 0$.
Thus the positive definiteness of $h*h$ follows from the more general statement about $f$.

To prove the more general statement, suppose $f$ has non-negative Fourier transform.
Then for $v \in \R^n$ we have
\begin{align}
    \sum_{l,m} v_l f(x_l-x_m)v_m & =
    \sum_{l,m} v_l v_m \int_{\R^d}\hat f(k) e^{ik\cdot (x_l-x_m)} \frac{dk}{(2\pi)^d}
    \nnb  & =
    \int_{\R^d}
    \hat f(k)
    \sum_l |v_le^{ik\cdot x_l}|^2 \frac{dk}{(2\pi)^d} \ge 0.
\end{align}
\end{solution}

\begin{solution}{ex:paley-wiener}
\label{soln:paley-wiener}
The Schwartz--Paley--Wiener Theorem states
that a Schwartz distribution $g$ on $\R^d$ has support in a ball of
radius $R$ if its Fourier transform $\hat g$ is entire on $\C^d$ and
satisfies the growth estimate
\begin{equation}
\lbeq{PWbd}
  |\hat g(k)| \leq C(1+|k|)^N e^{R |\text{Im}(k)|} \quad (k \in \C^d)
\end{equation}
for some constants $C$ and $N$.
Thus it suffices to prove that the function
$f(|k|)= \frac{1}{2\pi} \int_{-1}^1 \hat f(s)\, \cos(|k|s) \, ds$ is entire in $k$ and obeys an estimate of the
form \refeq{PWbd}.  Let $\hat g(k) = f(|k|)$.
The
function $\cos(|k|s) = \sum_{m=0}^\infty \frac{(-1)^m}{(2m)!} (s^2k^2)^m$ is a convergent
series in powers of the components $k= (k_{1},\dots ,k_{d})$ and
therefore is entire in $k$. By Morera's theorem, with interchange of integrals over $k$ and
$s$,
$\hat g$ is indeed entire.  We will prove below that
\begin{equation}
  \label{e:ex:paley-wiener-cos}
  |\cos (|k|)| \leq e^{|\text{Im}(k)|}.
\end{equation}
Given this, it follows, as desired, that
\begin{align}
  |\hat g(k)|
  &= \frac{1}{2\pi} |\int_{-1}^1 \hat f(s) \cos(|k|\, s) \, ds|
  \nnb &
  \leq \frac{1}{2\pi} \int_{-1}^1 |\hat f(s)| e^{|\text{Im}(k)|s} \, ds
  \leq C e^{|\text{Im} k|}.
\end{align}

It remains only to prove \refeq{ex:paley-wiener-cos}.
We use the branch of the square root with branch cut $(-\infty,0)$ and with positive real
numbers having positive square root.  This branch of the square root is analytic on
the cut plane $\C \setminus (-\infty,0)$.  It suffices to prove that for $k$ in the cut plane,
\begin{equation}
  \label{e:ex:paley-wiener-1}
  |\cos(\sqrt{k_1^2+\dots +k_d^2})| \leq e^{|\text{Im}(k)|}.
\end{equation}
Let
$k_1^2+\dots +k_d^2=A+Bi$ with $A,B \in \R$.
For $j=1,\dots ,d$, let $k_{j}=u_j+iv_j$ with $u_j,v_j\in \R$,
and let $u=(u_1,\ldots, u_d)$, $v = (v_1,\ldots,v_d)$.
Since $\cos
\sqrt{A+iB} = \frac{1}{2}\big(e^{i\sqrt{A+iB}}+e^{-i\sqrt{A+iB}}\big)$, it suffices to prove that
\begin{equation}
    \label{e:ex:paley-wiener-2}
    \big| \text{Im} \sqrt{A+iB} \big| \le |v| .
\end{equation}

\begin{figure}[h]
\begin{center}
  \input{paley-wiener}
\end{center}
\caption{Illustation of \eqref{e:PWABbd}. \label{fig:PWABbd}}
\end{figure}

We use polar coordinates to write
$A+iB=Re^{i\theta}$ with $\theta \in (-\pi ,\pi )$ and
$R=\sqrt{A^{2}+B^{2}}$.
From Figure~\ref{fig:PWABbd}, we see that
\begin{equation} \label{e:PWABbd}
    \big| \text{Im}\sqrt{A+iB} \big|
    =
    \sqrt{R}\, |\sin (\theta/2)|
    =
    \frac{\sqrt{R}|B|}{\sqrt{(R+A)^{2}+B^{2}}}
    =
    \frac{1}{\sqrt{2}}\,
    \frac{|B|}{\sqrt{R+A}} .
\end{equation}
It therefore suffices to prove that
\begin{equation}
    \label{e:ex:paley-wiener-3}
    \frac{B^2}{R+A}
    \le
    2 v\cdot v .
\end{equation}
By construction,
\begin{equation}
    \label{e:ex:paley-wiener-4}
    A = u\cdot u - v\cdot v ,
    \quad
    B=2(u\cdot v) ,
    \quad
    R^{2} = A^{2}+B^{2} .
\end{equation}
Thus \eqref{e:ex:paley-wiener-3} is equivalent to $B^{2} - 2
v^{2}A \le 2 v^{2} R$, which is implied by $(B^{2} - 2v^{2}A)^{2} \le 4 (v^{2})^{2} R^{2}$.
The latter is equivalent to $B^{2} - 4 v^{2} A \le 4 (v^{2})^{2}$, which in
turn is equivalent to $4 (u\cdot v)^{2} - 4 v^{2}
(u^{2}-v^{2}) \le 4 (v^{2})^{2}$.  This last inequality follows from
the Cauchy--Schwarz inequality $|u\cdot v| \le |u||v|$.
This proves \eqref{e:ex:paley-wiener-1} and completes the proof.
\end{solution}

\begin{solution}{ex:poisson-summation}\label{soln:poisson-summation}
For $p \in \Z$,
  \begin{align*}
      \int_0^{2\pi} f^*_t(x) \cos(px) \, dx
      &=
      \sum_{n \in \Z} \int_0^{2\pi} f(xt-2\pi nt) e^{ipx}\,dx
      \\
      &=
      \int_{0}^{\infty} f(xt) e^{ipx}\,dx
      =
      \frac{1}{t}\hat{f} (p/t) ,
  \end{align*}
and then \refeq{varphi*-defn2} follows by Fourier series inversion for
the even periodic function $f_t^*$.
\end{solution}

\section{Chapter~\ref{ch:hier} exercises}

\begin{solution}{ex:hier-field-tree-repr}
\label{soln:hier-field-tree-repr}
We assign a \emph{generation} to each vertex in the tree as follows: a vertex at distance
$k$ from the root has generation $N-k$.  Thus each leaf is at generation 0,
a vertex adjacent to a leaf has generation 1, and the root has generation $N$.
Any edge in the tree joins two vertices at subsequent generations $j-1$ and $j$ (say),
and we say this edge has generation $j$.  Given $j \in \{1,\ldots,N\}$, we assign to
each edge at generation $j$ an independent Gaussian random variable $\zeta_j$
with a covariance $C_j$.
Random variables from different generations are independent.
A leaf corresponds to a point $x \in \Lambda_N$.  Then we set $\varphi_x=\zeta_1+\cdots+\zeta_N$.
\end{solution}

\begin{solution}{ex:PQproj}
\label{soln:PQproj}
We first show that the range of $Q_j$ is $X_j$.  Let $\varphi \in \ell^2$ and let $x \in B$ for
some $j$-block $B$.  Then
\begin{align}
    (Q_j\varphi)_x & = \sum_{y \in \Lambda} Q_{j;xy}\varphi_y = \sum_{y\in B} L^{-dj} \varphi_y,
\end{align}
and since the right-hand side is the same for every $x\in B$, we see that $Q_j\varphi \in X_j$.
Also, if $\varphi \in X_j$, so that for all $x \in B$ we have $\varphi_x=c_B$ for some constant
$c_B$, the above calculation gives $(Q_j\varphi)_x = c_B$, so $Q_j\varphi = \varphi$.
This proves that the range of $Q_j$ is $X_j$.

Next we show that the range of $P_j$ is orthogonal to $X_j$.  Let $\psi\in X_j$, so there
are constants $c_B$ such that $\psi_x = c_{B_x}$.  Then
\begin{align}
    (\psi,P_j\varphi) & = (\psi,Q_{j-1}\varphi) - (\psi,Q_j\varphi)
    \nnb & =
    \sum_{B\in \Bcal_j} \sum_{b \in \Bcal_{j-1}(B)} \sum_{x\in b} c_B \sum_{y\in b}L^{-d(j-1)}\varphi_y
    -
    \sum_{B\in \Bcal_j} \sum_{x\in B} c_B \sum_{y\in B}L^{-dj}\varphi_y
    \nnb & =
    \sum_{B\in \Bcal_j}  c_B \sum_{y\in B}\varphi_y
    -
    \sum_{B\in \Bcal_j} c_B \sum_{y\in B}\varphi_y
    = 0.
\end{align}

Finally, we prove that the range of $P_j$ is $X_{j-1} \cap X_j^\perp$.
Clearly the range of $P_j$ is contained in $X_{j-1}$, since $X_{j-1}$ is the range of $Q_{j-1}$
and the range of $Q_j$ is $X_j \subset X_{j-1}$.
We have the direct sum decomposition
\begin{equation}
    X_{j-1} = (X_{j-1}\cap X_j) \oplus (X_{j-1} \cap X_j^\perp)
    =
    X_j \oplus (X_{j-1} \cap X_j^\perp),
\end{equation}
so $\varphi \in X_{j-1}$ can be written uniquely as $\varphi = \psi + \eta$ with
$\psi \in X_j$ and $\eta \in X_{j-1} \cap X_j^\perp$.
Then $Q_{j-1}\psi = \psi$ since $\psi \in X_{j} \subset X_{j-1}$,
$Q_{j}\psi = \psi$ since $\psi \in X_{j}$,
$Q_{j-1}\eta = \eta$ since $\eta \in  X_{j-1}$,
and $Q_j\eta=0$ since $\eta \in X_j^\perp$.  Therefore,
\begin{align}
    P_j\varphi & = Q_{j-1}\psi - Q_j\psi + Q_{j-1}\eta - Q_j\eta
    = \psi - \psi + \eta - 0 = \eta.
\end{align}
This completes the proof.
\end{solution}

\begin{solution}{ex:hier-rw}
\label{soln:hier-rw}
By \eqref{e:hierarchical-laplacian}, \eqref{e:Pj-def}, \eqref{e:Qj-def},
\begin{align}
    \Delta_{H;0,0} & = -\sum_{j=1}^{N} L^{-2(j-1)}(L^{-dj}-L^{-d(j-1)})
    \nnb &    =
    -(1-L^{-d}) \sum_{j=1}^{N} L^{-(d+2)(j-1)}
    =
    -\frac{1-L^{-d}}{1-L^{-(d+2)}} (1-L^{-(d+2)N}).
\end{align}
Also, for $x \neq 0$,
\begin{align}
    \Delta_{H;0,x} & =
    -\sum_{j=1}^{N} L^{-2(j-1)}(L^{-d(j-1)}\1_{j_x \le j-1} -L^{-dj}\1_{j_x\le j})
    \nnb & =
    -\sum_{j=j_x+1}^{N} L^{-2(j-1)}L^{-d(j-1)}
    +
    \sum_{j=j_x}^{N} L^{-2(j-1)} L^{-dj}
    \nnb & =
    -(1-L^{-d}) \sum_{j=j_x+1}^{N} L^{-(d+2)(j-1)} + L^2 L^{-(d+2)j_x}
    \nnb & =
    -\frac{1-L^{-d}}{1-L^{-(d+2)}}(L^{-(d+2)j_x} - L^{-(d+2)N}) + L^2 L^{-(d+2)j_x}
    \nnb & =
    \frac{L^2-1}{1-L^{-(d+2)}} L^{-(d+2)j_x} +
    \frac{1-L^{-d}}{1-L^{-(d+2)}} L^{-(d+2)N}.
\end{align}
For $x \neq 0$, let $n_k$ be the cardinality of $\{x : j_{x}=k\}$,
namely $n_k=L^{dk}-L^{d(k-1)}=L^{dk}(1-L^{-d})$.  Then
\begin{align}
    \sum_{x \neq 0} \Delta_{H;0x}
    & =
    \frac{L^2-1}{1-L^{-(d+2)}}\sum_{k=1}^N L^{dk}(1-L^{-d})L^{-(d+2)k}
    \nnb & \quad +
    (L^{dN}-1)\frac{1-L^{-d}}{1-L^{-(d+2)}} L^{-(d+2)N}
    \nnb & =
    \frac{1-L^{-d}}{1-L^{-(d+2)}}(1-L^{-2N})
    + (L^{dN}-1)\frac{1-L^{-d}}{1-L^{-(d+2)}} L^{-(d+2)N}
    \nnb & =
    \frac{1-L^{-d}}{1-L^{-(d+2)}}
    -\frac{1-L^{-d}}{1-L^{-(d+2)}} L^{-(d+2)N}
    \nnb & =
    -\Delta_{H;00}
    .
\end{align}
A random walk with infinitesimal generator $Q$ takes steps from a site $x$
at rate $-Q_{x,x}$, and when the step is taken it is a step to $y$ with probability
$-Q_{x,y}/Q_{x,x}$.  Here $Q=\Delta_{H,N}$ as in \refeq{DelH0x}.  The random walk can
make a step to any site, and the probability to step from $x$ to $y$ decays with a
factor $L^{-(d+2)j_{x-y}}$ where $j_{x-y}$ is the smallest scale such that $x$ and $y$
are in the same block at that scale.
\end{solution}

\begin{solution}{ex:hier-freechi}
\label{soln:hier-freechi}
By \refeq{hierarchical-decomp}, $C=\sum_{j=1}^NC_j + \Cnewlast{}$.
By \refeq{Cjhier} and \refeq{Psum0} we have $\sum_x C_{j;0x}=0$, and
by \refeq{Cjhier} and \refeq{Qj-def} we have $\sum_x \Cnewlast{;0x}=m^{-2}$.
\end{solution}

\begin{solution}{ex:cjns}
\label{soln:cjns}
Let $M_j = (1+m^2L^{2j})^{-1}$.  Then
\begin{align}
    c_{j}^{(2)}
    & = (L^{-(d-2)j}M_j)^2
    \left(L^{dj}(1-L^{-d})^2 + (L^{d(j+1)}-L^{dj}) (-L^{-d})^2 \right)
    \nnb
    & =
    L^{-(d-4)j} M_j^2 (1-L^{-d}).
\end{align}
\begin{align}
    c_{j}^{(3)}
    & = (L^{-(d-2)j}M_j)^3
    \left(L^{dj}(1-L^{-d})^3 + (L^{d(j+1)}-L^{dj}) (-L^{-d})^3 \right)
    \nnb
    & =
    L^{-(2d-6)j} M_j^3 (1-3L^{-d}+2L^{-2d}).
\end{align}
\begin{align}
    c_{j}^{(4)}
    & = (L^{-(d-2)j}M_j)^4
    \left(L^{dj}(1-L^{-d})^4 + (L^{d(j+1)}-L^{dj}) (-L^{-d})^4 \right)
    \nnb
    & =
    L^{-(3d-8)j} M_j^4 \big(1 - 4L^{-d} + 6L^{-2d} - 4 L^{-3d} + L^{-4d} +  (L^{d}-1) L^{-4d}\big)
    \nnb
    & =
    L^{-(3d-8)j} M_j^4 \big(1 - 4L^{-d} + 6L^{-2d} - 3 L^{-3d}\big) .
\end{align}
\end{solution}

\begin{solution}{ex:bubble}
\label{soln:bubble}
The infinite-volume hierarchical bubble diagram is given by
\begin{align}
    B_{m^2}^H & = \sum_x \left( \sum_{j=0}^\infty C_{j+1;0x}(m^2) \right)^2
    \nnb & =
    2 \sum_{0 \le j<k}\sum_x C_{j+1;0x}C_{k+1;0x} + \sum_{j=0}^\infty \sum_x C_{j+1;0x}^2,
\end{align}
with the sum over all $x \in \Z^d$.
The sum over $x$ in the first sum on the right-hand side is zero by \refeq{CC0}.
The second sum is $\sum_{j=0}^\infty c_j^{(2)}$, as required.

By Exercise~\ref{ex:cjns},
\begin{align}
   \sum_{j=0}^\infty c_j^{(2)}
    & =
    (1-L^{-d})\sum_{j=0}^\infty \frac{L^{(4-d)j}}{(1+L^{2j}m^2)^2},
\end{align}
which converges for $d>4$.
For $d\le 4$, the above is asymptotically $(1-L^{-d})$
times (use change of variables $y=L^{2x}$ followed
by $z=ym^2$)
\begin{align}
    \int_0^\infty \frac{L^{(4-d)x}}{(1+L^{2x}m^2)^2} dx
    & =
    \frac{1}{\log L} \int_1^\infty \frac{y^{(4-d)/2}}{(1+ym^2)^2} \frac{dy}{y}
    \nnb & =
    m^{d-4} \frac{1}{\log L} \int_{m^2}^\infty \frac{z^{(4-d)/2}}{(1+z)^2} \frac{dz}{z}.
\end{align}
The desired asymptotic behaviour then follows from the fact that the integral converges
with lower limit zero if $d<4$, whereas it diverges logarithmically if $d=4$.
\end{solution}

\begin{solution}{ex:hier-cov-asym}
\label{soln:hier-cov-asym}
(i)
By definition,
\begin{align}
\lbeq{CQ}
    C(m^2) & = \sum_{j=1}^N \gamma_j P_j + m^{-2}Q_N
    = \sum_{j=1}^N \gamma_j (Q_{j-1}-Q_{j}) + m^{-2}Q_N
    \nnb
    & = \gamma_1Q_0 + \sum_{j=1}^{N-1} (\gamma_{j+1}-\gamma_j)Q_j + (m^{-2}-\gamma_N) Q_N.
\end{align}

\smallskip \noindent (ii)
The coalescence scale $j_x$ is the smallest $j$ such that $B_x=B_0$.  Since $0$ is
at the corner of $B_0$ by Definition~\ref{def:Bcal}, $L^{j_x-1} < |x|_\infty \le L^{j_x}$.
In particular, $L^{j_x} \asymp |x|$, and also
$\log_L|x|_\infty \leq j_x \leq \log_L|x|_\infty +1$
so $j_x = \log_L |x| + O(1)$.
We use the fact that
if $j \ge j_x$, then $Q_{j;0x}=L^{-dj}$  and otherwise is it zero.

For $d>2$, we are interested in the limit as $N\to\infty$ and then $m^2\downarrow 0$ of
\refeq{CQ}.
We take the limit of the right-hand side and obtain
\begin{align}
    \lim_{m^2 \downarrow 0} \lim_{N \to \infty} C_{0x}(m^2)
    &  =
    \lim_{m^2 \downarrow 0} \sum_{j=j_x}^\infty (\gamma_{j+1}-\gamma_j)L^{-dj}
    \nnb
    & = \sum_{j=j_x}^\infty (L^{2j}-L^{2(j-1)})L^{-dj}
    = (1-L^{-2}) \sum_{j=j_x}^\infty L^{-(d-2)j}
    \nnb &
    \asymp L^{-(d-2)j_x} \asymp |x|^{-(d-2)}.
\end{align}
For $d \le 2$, we have instead
\begin{align}
    \lim_{m^2 \downarrow 0} \lim_{N \to \infty} ( C_{0x}(m^2) - C_{00}(m^2) )
    &  =
    -\lim_{m^2 \downarrow 0} \sum_{j=1}^{j_x-1}(\gamma_{j+1}-\gamma_j)L^{-dj}
    \nnb
    & = -\sum_{j=j_x-1}^\infty (L^{2j}-L^{2(j-1)})L^{-dj}
    \nnb &
    = -(1-L^{-2}) \sum_{j=1}^{j_x-1}  L^{(2-d)j}.
\end{align}
For $d=1$, the right-hand side is $\asymp -L^{j_x} \asymp -|x|$, whereas for $d=2$
it is equal to $-(1-L^{-2})(j_x -1) = -(1-L^{-2})\log_L|x| + O(1)$.
\end{solution}

\begin{solution}{ex:susceptZN-hier}
\label{soln:susceptZN-hier}
By evaluation of the derivative, we see that
\begin{equation}
    D^2 \Sigma_N(0;\1,\1) = \sum_{x,y\in\Lambda} \Ex_C (\varphi_x^1 \varphi_y^1 Z_0(\varphi))
    = |\Lambda| \sum_{x\in\Lambda} \Ex_C (\varphi_0^1 \varphi_x^1 Z_0(\varphi)),
\end{equation}
which proves the first equality of \refeq{susceptZN-hier}.
We may also compute the above derivative using the identity \refeq{Sigma1},
which states that
\begin{equation}
  \label{e:Sigma1-bis}
  \Sigma_N(f)
  = e^{\frac12 (f,Cf)} (\Ex_C\theta Z_0)(Cf)
  .
\end{equation}
In this way, since $C\1= (m^{-2},0,\ldots,0)\1$ by Exercise~\ref{ex:hier-freechi}, we obtain
\begin{equation}
    D^2 \Sigma_N(0;\1,\1)
    = \frac{1}{m^2} |\Lambda| Z_N(0) + \frac{1}{m^4} D^2 Z_N(0;\1,\1).
\end{equation}
For example, the factor $m^{-2}|\Lambda|$ in the first term on the right-hand side
arises from
\begin{equation}
    \frac{d^2}{dsdt}\Big|_{s=t=0} e^{\frac 12  ((s+t)\1,C(s+t)\1)}
    =
    \frac{d^2}{dsdt}\Big|_{s=t=0} e^{\frac 12 (s+t)^2 m^{-2}|\Lambda|}.
\end{equation}
This proves the second equality of \refeq{susceptZN-hier}.
\end{solution}

\begin{solution}{ex:gren}
(i)
By definition,
\begin{equation}
    \sum_{x,y,z\in \Lambda}
    \langle \varphi_0\varphi_x\varphi_y\varphi_z \rangle_N
    =
    \frac{1}{|\Lambda|} \frac{D^4 \Sigma_N(0;\1,\1,\1,\1)}{\Sigma_N(0)}.
\end{equation}
We write $\lambda = |\Lambda|m^{-2}$, $t_j=s_j  +\cdots s_4$,
and $u_j=m^{-2}t_j$ (for $j=1,2,3,4$).
We compute the numerator on the right-hand side
using \refeq{Sigma1} with $f=t_1\1$ and $Cf=u_1\1$.
This gives (with derivatives $D^j$ having all $j$ directions equal to $\1$)
\begin{align}
    & D^4 \Sigma_N(0;\1,\1,\1,\1)
    \nnb
    & \quad =
    \left. \frac{d^4}{ds_1ds_2ds_3ds_4} \right|_{0}
    e^{\frac 12 t_1^2 \lambda} \Znewlast(u_1\1)
    \nnb
    & \quad =
    \left. \frac{d^3}{ds_2ds_3ds_4} \right|_{0}
    e^{\frac 12 t_2^2 \lambda}
    \left( \lambda t_2 \Znewlast(u_2\1) + m^{-2} D^1\Znewlast(u_2\1) \right)
    \nnb
    & \quad =
    \left. \frac{d^2}{ds_3ds_4} \right|_{0}
    e^{\frac 12 t_3^2 \lambda}
    \left( (\lambda^2 t_3^2 +\lambda) \Znewlast(u_3\1)
    + 2\lambda t_3m^{-2} D^1\Znewlast(u_3\1)
    + m^{-4}D^2\Znewlast(u_3\1) \right)
    \nnb
    & \quad =
    \left. \frac{d}{ds_4} \right|_{0}
    e^{\frac 12 t_4^2 \lambda}
    \Big( (\lambda^3t_4^3+3\lambda^2t_4) \Znewlast(u_4\1)
    + 3(\lambda^2t_4^2 + \lambda)
    m^{-2}
    D^1\Znewlast(u_3\1)
    \nnb & \hspace{40mm}
    + 3\lambda t_4m^{-4}D^2\Znewlast(u_4\1) +m^{-6} D^3\Znewlast(u_4\1) \Big)
    \nnb & \quad =
    3\lambda^2 \Znewlast(0) + 6\lambda m^{-4}D^2\Znewlast(0;\1,\1)
    + m^{-8}D^4\Znewlast(0;\1,\1,\1,\1).
\end{align}
Therefore,
\begin{align}
    \sum_{x,y,z\in \Lambda}
    \langle \varphi_0\varphi_x\varphi_y\varphi_z \rangle_N
    & =
    \frac{3|\Lambda|}{m^4} + \frac{6}{m^6} \frac{D^2\Znewlast(0;\1,\1)}{\Znewlast(0)}
    + \frac{1}{m^8|\Lambda|} \frac{D^4\Znewlast(0;\1,\1,\1,\1)}{\Znewlast(0)}.
\end{align}
Using Exercise~\ref{ex:susceptZN-hier}, we subtract from this the quantity
\begin{align}
    3|\Lambda|\chi_N^2 & =
    3|\Lambda| \left(
    \frac{1}{m^4} + \frac{2}{m^6|\Lambda|} \frac{D^2\Znewlast(0;\1,\1)}{\Znewlast(0)}
    + \frac {1}{m^8|\Lambda|^2} \left(\frac{D^2\Znewlast(0;\1,\1)}{\Znewlast(0)} \right)^2
    \right).
\end{align}
This gives the desired formula for $\bar{u}_4$.

\smallskip \noindent
(ii)
Direct calculation gives
\begin{align}
    &\frac{D^4 e^{-V_N}(0;\1,\1,\1,\1)}{e^{-V_N} (0)}
    -
    3
    \left(\frac{D^2e^{-V_N}(0;\1,\1)}{e^{-V_N}(0)} \right)^2
    \nnb & \quad =
    [-6g_N|\Lambda| + 3(\nu_N|\Lambda|)^2]
    -
    3\left(-\nu_N|\Lambda| \right)^2
    =
    -6g_N|\Lambda|
    .
\end{align}
Therefore,
\begin{equation}
    \tilde{g}_{{\rm ren},N} = g_N  ,
\end{equation}
as desired.
The factor $\frac 16$ accounts for the fact that the natural prefactor of $\varphi^4$ in
this context
is $\frac{1}{4!}g$ rather than our convention $\frac 14 g$.
\end{solution}

\section{Chapter~\ref{ch:hierpt} exercises}

\begin{solution}{ex:ExUcal-bis}
\label{soln:ExUcal-bis}
By Proposition~\ref{prop:wick},
\begin{equation}
    \Ex_C \theta U
    =
    U
    +
    \tfrac{1}{2} \Delta_{C} (\tfrac{1}{4}g |\varphi|^4 + \tfrac{1}{2}\nu |\varphi|^2 )
    +
    \tfrac{1}{8} \Delta_{C}^2 \tfrac{1}{4}g |\varphi|^4 .
\end{equation}
The terms involving $\Delta_{C}  |\varphi|^2 $
and $\Delta_{C}^2 |\varphi|^4$ produce constants.  The remaining term involves
\begin{align}
     \Delta_{C }  |\varphi|^4
     & = C_{xx} \sum_{i=1}^n \frac{\partial^2}{\partial (\varphi^i)^2}
     \left( |\varphi|^2 \right)^2
     =
     C_{xx} \sum_{i=1}^n
     \left(
     8(\varphi^i)^2 + 4|\varphi|^2
     \right),
\end{align}
which produces a $|\varphi|^2$ term.
A complete calculation of $\Ex_C \theta U$ is given in the proof of Lemma~\ref{lem:ECV}.
\end{solution}

\begin{solution}{ex:phi4integral}
As in \eqref{e:Gdet}, we have
\begin{equation}
  \frac{1}{\sqrt{2\pi}}
  \int_{-\infty}^\infty e^{-\frac12 x^2} dx = 1.
\end{equation}
Since $e^{-gx^4} \leq 1$ for $g\geq 0$ the first bound follows.
On the other hand,
\begin{equation}
\lbeq{plusgint}
  \frac{1}{\sqrt{2\pi}}
  \int_{-\infty}^\infty e^{+gx^4} e^{-\frac12 x^2} dx =\infty.
\end{equation}
If the series \eqref{e:x4series} were to converge absolutely for
some $g \neq 0$, then by dominated convergence this also would
imply the convergence of \refeq{plusgint}. Since \refeq{plusgint}
does not converge, we conclude that neither does \eqref{e:x4series}.
\end{solution}

\begin{solution}{ex:Greeks}
\label{soln:Greeks}
Since the field is constant, we drop subscripts $x,y$, and for notational convenience
use subscripts (rather than superscripts) for component indices.
To begin, we observe that
\begin{align}
    \ddp{}{\varphi_i}|\varphi|^2 & = 2\varphi_i,
\\
    \label{e:monomial-phi-derivs2}
    \ddp{^2 |\varphi|^2}{\varphi_{i}  \partial \varphi_{j}}
    & = 2 \delta_{ij},
\\
    \ddp{}{\varphi_i}|\varphi|^4 & = 4|\varphi|^2 \varphi_i,
\\
    \ddp{^2 |\varphi|^4}{\varphi_{i}  \partial \varphi_{j}}
    & = 8\varphi_i \varphi_j + 4 |\varphi|^2 \delta_{ij},
\\
    \ddp{^3 |\varphi|^4}{\varphi_{i}  \partial \varphi_{j} \partial \varphi_{k}}
    & = 8(\varphi_i \delta_{jk} + \varphi_j \delta_{ik} + \varphi_k \delta_{ij}),
\\
    \ddp{^4 |\varphi|^4}{\varphi_{i}  \partial \varphi_{j} \partial \varphi_{k}\partial \varphi_{l}}
    & = 8(\delta_{il} \delta_{jk} + \delta_{jl} \delta_{ik} + \delta_{kl} \delta_{ij}).
\end{align}

The two terms that were not computed in the proof of Lemma~\ref{lem:ECV}
are $\kappa_{\nu}'= \frac 14 \Delta_{C}  |\varphi|^2 $
and
$\kappa_{g}'= \tfrac{1}{32} \Delta_{C}^2  |\varphi|^4 $.
From \eqref{e:monomial-phi-derivs2}, we have
\begin{align}
    \kappa_\nu' & = \tfrac{1}{4} c
    \sum_i \ddp{^2 |\varphi|^2}{\varphi_{i}^2}
    = \tfrac{1}{4} c 2n = \tfrac{1}{2}nc.
\end{align}
Similarly,
\begin{align}
    \ddp{^4 |\varphi|^4}{\varphi_{i}^2  \partial \varphi_{j}^2}
    & = 16 \delta_{ij} + 8,
\end{align}
and hence
\begin{align}
    \kappa_{g}' & = \tfrac{1}{32} \Delta_{C}^2  |\varphi|^4
    = \tfrac{1}{32} c^2 \sum_{i,j} (16 \delta_{ij} + 8)
    \nnb
    & =
    \tfrac{1}{32} c^2 (16n + 8n^2) = \tfrac{1}{4} n (n+2) c^2.
\end{align}

Now we turn to the more difficult quadratic term.
We first compute the sum over $y \in B$ of \refeq{EFid2}, which is
\begin{align}
\lbeq{EFid2-pf}
    &
    \tfrac{1}{16}g^2
     F_C \big(|\varphi|^4;|\varphi|^4\big)
    \nnb
    &
    +
    \tfrac{1}{4}\Big(g^2 (\eta')^2 + 2g\nu \eta' +  \nu^2 \Big)
    F_C \big(|\varphi|^2;|\varphi|^2\big)
    \\ \nonumber
    &
    +
    \tfrac{1}{4}
    \Big(g^2 \eta' +  g  \nu
    \Big)
    F_C \big(|\varphi|^2;|\varphi|^4\big)
     ;
\end{align}
here subscripts $x,y$ and $\sum_{y \in B}$ are implicit in the notation.
From \refeq{Fder} we obtain
\begin{align}
    F_C(|\varphi|^2;|\varphi|^2) & =
    \frac{1}{1!} c^{(1)} \sum_i 2\varphi_i 2 \varphi_i
    + \frac{1}{2!} c^{(2)} \sum_{i,j} 2\delta_{ij}2\delta_{ij}
    \nnb & =
    4c^{(1)}|\varphi|^2 + 2c^{(2)}n,
\\
    F_C(|\varphi|^2;|\varphi|^4) & =
    \frac{1}{1!} c^{(1)} \sum_i 2\varphi_i 4|\varphi|^2 \varphi_i
    + \frac{1}{2!} c^{(2)} \sum_{i,j} 2\delta_{ij}(8\varphi_i \varphi_j + 4 |\varphi|^2 \delta_{ij})
    \nnb &
    = 8c^{(1)}|\varphi|^4  +  c^{(2)}(4n+8)|\varphi|^2,
\\
    F_C(|\varphi|^4;|\varphi|^4) & =
    \frac{1}{1!} c^{(1)} \sum_i 4|\varphi|^2 \varphi_i 4|\varphi|^2 \varphi_i
    + \frac{1}{2!} c^{(2)} \sum_{i,j} (8\varphi_i \varphi_j + 4 |\varphi|^2 \delta_{ij})^2
    \nnb & \;\;\;
    + \frac{1}{3!} c^{(3)} \sum_{i,j,k} 64(\varphi_i \delta_{jk} + \varphi_j \delta_{ik} + \varphi_k \delta_{ij})^2
    \nnb & \;\;\;
    + \frac{1}{4!} c^{(4)} \sum_{i,j,k,l} 64(\delta_{il} \delta_{jk} + \delta_{jl} \delta_{ik} + \delta_{kl} \delta_{ij})^2
    \nnb &
    = 16c^{(1)}|\varphi|^6  +  8c^{(2)}(n+8)|\varphi|^4
    + 32c^{(3)}(n+2)|\varphi|^2
    + 8 c^{(4)} n(n+2).
\end{align}
Substitution into \refeq{EFid2-pf} gives
\begin{align}
\lbeq{EFid2-pf2}
    &
    \tfrac{1}{16}g^2
    \left(
    16c^{(1)}|\varphi|^6  +  8c^{(2)}(n+8)|\varphi|^4
    + 32c^{(3)}(n+2)|\varphi|^2
    + 8 c^{(4)} n(n+2)
    \right)
    \nnb
    & \quad
    +
    \tfrac{1}{4}\Big(g^2 (\eta')^2 + 2g\nu \eta' +  \nu^2 \Big)
    \left(
    4c^{(1)}|\varphi|^2 + 2c^{(2)}n
    \right)
    \nnb
    & \quad
    +
    \tfrac{1}{4}
    \Big(g^2 \eta' +  g  \nu     \Big)
    \left(
    8c^{(1)}|\varphi|^4  +  c^{(2)}(4n+8)|\varphi|^2
    \right)
    \nnb &
    =
    g^2  c^{(1)}|\varphi|^6
    \nnb & \quad
    +
    \left(
    \tfrac{1}{2}g^2c^{(2)}(n+8)
    +
    2  \Big(g^2 \eta' +  g  \nu     \Big) c^{(1)}
    \right)
    |\varphi|^4
    \nnb &
    \quad
    +
    \left(
    2 g^2 c^{(3)}(n+2)
    +
    \tfrac{1}{4}     \Big(g^2 \eta' +  g  \nu     \Big) c^{(2)}(4n+8)
    +
    \Big(g^2 (\eta')^2 + 2g\nu \eta' +  \nu^2 \Big) c^{(1)}
    \right)
    |\varphi|^2
    \nnb & \quad
    +
    \left(
    \tfrac{1}{2}g^2  c^{(4)} n(n+2)
    +
    \tfrac{1}{2}\Big(g^2 (\eta')^2 + 2g\nu \eta' +  \nu^2 \Big) c^{(2)}n
    \right)
    \nnb &
    =
    8 g^2  c^{(1)}
    \tau^3
    \nnb & \quad
    +
    4 \left(
    \tfrac{1}{2}g^2c^{(2)}(n+8)
    +
    2  \Big(g^2 \eta' +  g  \nu     \Big) c^{(1)}
    \right)
    \tau^2
    \nnb &
    \quad
    +
    2
    \left(
    2 g^2 c^{(3)}(n+2)
    +
    \   \Big(g^2 \eta' +  g  \nu     \Big) c^{(2)}(n+2)
    +
    \Big(g^2 (\eta')^2 + 2g\nu \eta' +  \nu^2 \Big) c^{(1)}
    \right)
    \tau
    \nnb & \quad
    +
    \left(
    \tfrac{1}{2}g^2  c^{(4)} n(n+2)
    +
    \tfrac{1}{2}\Big(g^2 (\eta')^2 + 2g\nu \eta' +  \nu^2 \Big) c^{(2)}n
    \right)
    .
\end{align}
The variance term enters $\Upt$ with factor $-\frac 12$, and with this factor the
above becomes
\begin{align}
\lbeq{EFid2-pf3}
    &
    -4 g^2  c^{(1)}
    \tau^3
    \nnb &
    -
    \left(
    g^2c^{(2)}(n+8)
    +
    4  \Big(g^2 \eta' +  g  \nu     \Big) c^{(1)}
    \right)
    \tau^2
    \nnb &
    -
    \left(
    2 g^2 c^{(3)}(n+2)
    +
    \   \Big(g^2 \eta' +  g  \nu     \Big) c^{(2)}(n+2)
    +
    \Big(g^2 (\eta')^2 + 2g\nu \eta' +  \nu^2 \Big) c^{(1)}
    \right)
    \tau
    \nnb &
    -
    \tfrac{1}{4}
    \left(
    g^2  c^{(4)} n(n+2)
    +
    \Big(g^2 (\eta')^2 + 2g\nu \eta' +  \nu^2 \Big) c^{(2)}n
    \right)
    .
\end{align}
According to our definitions \eqref{e:tree-coeffs2}--\eqref{e:tree-coeffs1} and \eqref{e:Greekdef}--\eqref{e:kappadef} of the coefficients,
and by the identity $c^{(2)}(n+2) = \gamma\beta'$, the above is equal to
\begin{align}
\lbeq{EFid2-pf4}
    &
    -4 g^2  c^{(1)}
    \tau^3
    \nnb &
    -
    \left(
    \beta' g^2
    +
    s'_{\tau^2}
    \right)
    \tau^2
    \nnb &
    -
    \left(
    \xi' g^2
    +
    \gamma\beta' g  \nu
    +
    s'_{\tau}
    \right)
    \tau
    \nnb &
    - \kappa_{gg}'g^2- \kappa_{g\nu}'g\nu - \kappa_{\nu\nu}'\nu^2
    .
\end{align}
The $\tau^3$ term is the one term that is not in the range of $\LT$, and hence
it is equal to $W_+$.
\end{solution}

\section{Chapter~\ref{ch:pt} exercises}

\begin{solution}{ex:gsequence}
\label{soln:gsequence} Let $\epsilon = \min\{1, a/(2M), 1/(2A+2M)\}$.
Let $0 < g_0 < \epsilon$.  We assume by induction that $0 < g_j <
\epsilon$.  Then
\begin{align}
    g_j - g_{j+1}
    & =
    a_j g_{j}^{2} - e_j \ge ag_j^2(1 - \tfrac{M}{a}g_j)
    >
    \tfrac{1}{2} a g_j^2,
\end{align}
so $g_{j+1} < g_j$.  Also, $g_{j+1} \ge g_j(1-Ag_j-Mg_j^2)$ and the
second factor on the right-hand side is greater than $\frac 12$, so
$g_{j+1} > \frac 12 g_j$.  The induction is complete, the strict
monotonicity follows, as does the inequality $0< \frac 12 g_j<
g_{j+1}<g_j$.

Let $g_\infty = \lim_{n\to \infty}g_j$, which is nonnegative.
We take the limit $j \to \infty$ in the
inequality
\begin{equation}
    ag_j^2 \le a_j g_j^2 = g_j - g_{j+1}+e_j \le g_j - g_{j+1} + M g_j^3
\end{equation}
to obtain $ag_\infty^2 \le Mg_\infty^3$.  One solution is
$g_\infty=0$.  A positive solution requires $g_{\infty} \ge
a/M$ which is not possible because it exceeds $g_0$.
\end{solution}

\begin{solution}{ex:Asequence}
\label{soln:Asequence}
Let $m^2>0$.
For $j \le j_m$,
\begin{align}
    \sum_{i=0}^{j-1} \frac{1}{(1+L^{2i}m^2)^2}
    & =
    \sum_{i=0}^{j-1} 1
    +
    \sum_{i=0}^{j-1} \left( \frac{1}{(1+L^{2i}m^2)^2} - 1 \right)
    \nnb & =
    j
    -
    \sum_{i=0}^{j-1} \frac{2m^2L^{2i}+m^4L^{4i}}{(1+L^{2i}m^2)^2}
    \nnb & =
    j + O(L^{-2(j_m-j)_+})
    .
\end{align}
Therefore, for $j > j_m$,
\begin{align}
    \sum_{i=0}^{j-1} \frac{1}{(1+L^{2i}m^2)^2}
    & =
        \sum_{i=0}^{j_m} \frac{1}{(1+L^{2i}m^2)^2}
    +
    \sum_{i=j_m}^{j-1} \frac{1}{(1+L^{2i}m^2)^2}
    \nnb & =
    (j_m+O(1)) +O(1)
    .
\end{align}
This proves the result for $A_j$.

Secondly, for large $j$,
\begin{align}
    t_j & = \frac{g_0}{1+g_0\beta_0^0 (j \wedge j_m) + O(1)}
    \nnb
    & =
    \frac{g_0}{1+g_0\beta_0^0 (j \wedge j_m) }
    \left(
    1 + \frac{O(1)}{1+g_0\beta_0^0 (j \wedge j_m)}
    \right)
    \nnb &
    \asymp
    \frac{g_0}{1+g_0\beta_0^0 (j \wedge j_m) }.
\end{align}

Finally,
for the last inequality it suffices (by the previous result) to prove it
for $s_i = t_{j\wedge j_m}(0) = \frac{g_0}{1+g_0\beta_0^0 (j \wedge j_m)}$.  If $j < j_m$ then
\begin{align}
    \sum_{i=0}^{j} \vartheta_i s_i
    & =
    \sum_{i=0}^{j} s_i
    =
    O(|\log s_j|),
\end{align}
while if $j \ge j_m$ then
\begin{align}
    \sum_{i=0}^{j} \vartheta_i s_i
    & =
    \sum_{i=0}^{j_m}  s_i + \sum_{i=j_m+1}^{j} \vartheta_i s_i
    =
    O(|\log s_{j_m}|) + O(s_{j_m}) = O(|\log s_j|).
\end{align}
\end{solution}

\begin{solution}{ex:tsequence}
\label{soln:tsequence}
By Proposition~\ref{prop:gjtj} it suffices to verify the claims for the sequence $t_j$.
By definition, $t_j - t_{j+1} = t_jt_{j+1}\beta_{j+1}$,
so
\begin{equation}
    1 - \frac{t_{j+1}}{t_j} = t_{j+1}\beta_{j+1} = O(g_0),
\end{equation}
which proves that $t_{j+1} = t_j(1+O(g_0))$.

Since $A_j(m^2)$ decreases as $m^2$ increases, $t_j(m^2) \le t_j(0)$. This proves
the first inequality when $j \le j_m$.  For $j \ge j_m$, we note instead that
\begin{equation}
    \vartheta_j(m^2) t_j(m^2) \le 2^{-(j-j_m)} t_{j_m}(m^2) \le g_0 O((1+g_0\beta_0^0 j)^{-1}).
\end{equation}

For the remaining inequality, by \refeq{tm0} it suffices
(as in the solution to Exercise~\ref{ex:Asequence}) to verify the inequality
with $t_j(m^2)$ replaced by $t_{j\wedge j_m}(0)$.
Let $p>1$. If $j < j_m$ then by comparison of the
sum with an integral,
\begin{align}
    \sum_{i=j}^{\infty} \vartheta_i s_{i }^p
    & =
    \sum_{i=j}^{j_m} s_i^p
    +
    \sum_{i=j_m+1}^{\infty} \vartheta_i s_i^p
    \le O(s_j^{p-1} + s_{j_m}^p) \le O(s_j^{p-1})
    =
    O(\vartheta_j s_j^{p-1}),
\end{align}
while if $j \ge j_m$ then
\begin{align}
    \sum_{i=j}^{\infty} \vartheta_i s_i^p
    &
    \le O(\vartheta_j s_{j}^p) .
\end{align}

\smallskip \noindent {\bf Alternate solution to Exercise~\ref{ex:tsequence}.}  The following
alternative solution is adapted from \cite[Lemma~2.1]{BBS-rg-flow}. The
identity \eqref{e:gsumbis} follows directly from the
recursion \eqref{e:gae}.  The desired bounds, including the
logarithmic bound for $p=1$ are corollaries.
The useful identity \eqref{e:gsumbis} gives an
alternative way to analyse the recursion \eqref{e:gae}.

We first show that if $\psi: \R_+ \to \R$ is absolutely continuous and
the coefficients $a_{l}$ in \eqref{e:gae} are uniformly bounded,
$|a_{l}| \le A$, then
\begin{equation}
  \label{e:gsumbis}
  \sum_{l=j}^{k} (a_l g_l^2 - e_{l})\psi(g_l)
  = \int_{g_{k+1}}^{g_{j}} \psi(t) \; dt
  + O\left(\int_{g_{k+1}}^{g_{j}} t^2 |\psi'(t)| \; dt \right)
  .
\end{equation}
To prove \eqref{e:gsumbis}, we apply \eqref{e:gae} to obtain
\begin{equation}
  \label{e:intsum}
  \sum_{l=j}^{k} (a_l g_l^2 - e_{l})\psi(g_l)
  =
  \sum_{l=j}^{k} \psi(g_{l}) (g_{l} - g_{l+1})
  .
\end{equation}
We wish to replace the Riemann sum on the right-hand side by the
corresponding integral. For this we use
\begin{equation}
  \label{e:psiint}
  \psi(g_{l}) (g_{l} - g_{l+1})
  =
  \int_{g_{l+1}}^{g_{l}} \psi(t) \,dt
  +
  \int_{g_{l+1}}^{g_{l}} \int_t^{g_l} \psi'(s) \, ds \, dt
  ,
\end{equation}
which follows by applying the fundamental theorem of calculus to the
last term.  After inserting this into \eqref{e:intsum} we have
\begin{equation}
  \label{e:intsum2}
  \sum_{l=j}^{k} (a_l g_l^2 - e_{l})\psi(g_l)
  =
  \int_{g_{k+1}}^{g_{j}} \psi(t) \; dt
  +
  \sum_{l=j}^{k}
  \int_{g_{l+1}}^{g_{l}} \int_{g_{l+1}}^s \psi'(s) \, dt \, ds
  ,
\end{equation}
where we have inverted the order of integration.  Upon evaluating the
$t$ integral we obtain a factor $s-g_{l+1}$ so \eqref{e:gsumbis} holds
if $|s-g_{l+1}|=O (s^{2})$.  This is proved as follows: by
\eqref{e:gbar9} and $g_{l}\le g_{0}$, for $s$ in the domain of
integration we have
\begin{equation}
  \label{e:s-g}
  |s-g_{l+1}|
  \leq
  |g_l - g_{l+1}|
  =
  |a_l| \bar g_l^2
  \leq
  (1+O(\bar g_0)) |a_l| \bar g_{l+1}^2
  \leq O(s^2)
  ,
\end{equation}
where we used the hypothesis $|a_{l}| \le A$.  This concludes the
proof of \eqref{e:gsumbis}.

Direct evaluation of the integrals in \eqref{e:gsumbis} with
$\psi(t) = t^{p-2}$ and $a_{l}=\beta_{l}$ gives
\begin{equation}
  \lbeq{betagbd}
  \sum_{l=j}^k (\beta_l g_l^{p} - e_{l}g_l^{p-2})
  \beta_l  g_l^p
  \leq C_{p}
  \begin{cases}
    |\log  g_{k}|   & p = 1\\
     g_j^{p-1}       & p > 1
    .
  \end{cases}
\end{equation}
We only deduce \eqref{e:gbarsum}, as the proof of \eqref{e:gbarsum1} is
similar.
Suppose first that $j \le j_m$. Then $1 = O(\beta_j)$ and $|e_l| \le
M_l g_l^3$, therefore
\begin{align}
  \sum_{l=j}^k \vartheta_l  g_l^p
  & \leq
  \sum_{l=j}^{j_m} O(\beta_l g_l^{p} - e_{l}g_l^{p-2})
  +
  \sum_{l=j_m+1}^{k} 2^{-(l-j_m)_+}  g_l^p
  .
\end{align}
By \refeq{betagbd}, the first term is bounded by $O( g_j^{p-1} ) $.
The second term (which is only present when $j_m<\infty$) obeys the
same bound using monotonicity of $\gbar_j$ in $j$.  This proves
\refeq{gbarsum} for the case $j \le j_m$.  On the other hand, if
$j>j_m$, then again using the exponential decay of $\vartheta_l$ and
$\gbar_{l+1} \leq \gbar_l$, we obtain
\begin{equation}
  \sum_{l=j}^k \vartheta_l  g_l^p
  \leq
  C\vartheta_j  g_{j}^p
  \leq
  C\gbar_0\vartheta_j  g_{j}^{p-1}
  .
\end{equation}
This completes the proof of \eqref{e:gbarsum}.
\begin{rk}
By choosing $\psi (t)=t^{-2}$, $j=0$ and replacing $k$ by $k-1$ in
\eqref{e:gsumbis} we obtain
\begin{equation}
  \sum_{l=0}^{k-1} (a_l  - e_{l}g_l^{-2})
  =  \frac{1}{g_{k}} - \frac{1}{g_{0}}
  +  O (|\log g_{k}|)
  .
\end{equation}
Let $a_{l}=\beta_{l}$. Recall from \eqref{e:Aj-def} that $A_k =
\sum_{l=0}^{k-1} \beta_l$ and insert $|e_l| \le M_l g_l^3$ to obtain
\begin{equation}
  \label{e:gsum3}
  A_{k}  - O\left(\sum_{l=0}^{k-1} g_l \right)
  =  \frac{1}{g_{k}} - \frac{1}{g_{0}}
  +  O (|\log g_{k}|) .
\end{equation}
Solving for $g_{k}$ leads to \eqref{e:gtasy}.
\end{rk}
\end{solution}

\begin{solution}{ex:gbarcomp}
\label{soln:gbarcomp} By Proposition~\ref{prop:gjtj} it suffices to
verify the claim for the sequence $t_j$.  Let $m^2>0$.  Let
$t_j=t_j(m^2)$ and $\tilde{t}_j = t_j(\mgen^2)$, and similarly for
$A_j$, $\tilde{A}_j$.  By the definition \eqref{e:Aj-def},
\begin{align}
    t_j - \tilde{t}_j & =  t_j  \tilde{t}_j (\tilde{A}_j - A_j)
    =
    t_j  \tilde{t}_j
    \beta_0^0 \sum_{i=0}^{j-1}
    \frac{2(m^2-\mgen^2) L^{2i}\ +(m^4-\mgen^4) L^{4i}}
    {(1+\mgen^2 L^{2i})^{2}(1+m^2 L^{2i})^{2}} .
\end{align}
For case $\mgen^{2}=0$, the condition $m^{2} \in
\Iint_j(0)$ implies that $m^{2}L^{2j}\le 1$ so the sum is roughly
geometric, dominated by its largest term, and therefore of order
one. Similarly, for case $\mgen^{2}>0$ the condition $m^{2} \in
\Iint_j(\mgen^2)$ implies that $m^{2}=O (\mgen^{2})$ and now the terms
in the sum such that $L^{2i}\mgen^{2}>1$ are negligible and again the
sum is of order one.  Therefore in both cases $t_j - \tilde{t}_j = O (
t_{j}\tilde{t}_j ) = O ( \tilde{t}_j^{2})$ as desired.
\end{solution}

\begin{solution}{ex:chi-asy}
\label{soln:chi-asy}
The desired conclusion is obtained by applying the following lemma, which
is \cite[Lemma~4.3]{BBS-saw4-log}, to $u(t)= (B\chi(\nu_c +t))^{-1}$.
In fact, \refeq{uprime} is the hypothesis \refeq{chipdr}, and \refeq{ut} is the
conclusion \refeq{thmphi4-hier-chi-pf}.

\smallskip \noindent
\emph{Lemma.}
  Let $\gamma \in \R$ and $\delta > 0$.
  Suppose that $u: [0,\delta) \to [0,\infty)$ is continuous,
  differentiable on $(0,\delta)$, that $u(0)=0$ and $u(t)>0$ for $t>0$, and that
  \begin{equation}
  \lbeq{uprime}
    u'(t) = (-\log u(t))^{-\gamma} ( 1+ o(1)) \quad \text{(as $t\downarrow 0$)} .
  \end{equation}
  Then
  \begin{equation}
  \lbeq{ut}
    u(t) = t(-\log t)^{-\gamma} (1+o(1))
    \quad \text{(as $t\downarrow 0$)}
    .
  \end{equation}

\begin{proof}
By hypothesis,
\begin{equation}
    \int_0^t u'(t) (-\log u(t))^{\gamma} \; dt
    =
    \int_0^t (1+o(1)) \; dt
    =
    t(1+o(1))
    .
\end{equation}
  Since $u(t) > 0$ implies that $u'(t) > 0$ for small $t$, we see that $u$ is monotone.
  By a change of variables, followed by integration by parts,
  \begin{equation}
  \begin{aligned}
    \int_0^t u'(t) (-\log u(t))^{\gamma} \; dt
    &
    =
    \int_0^{u(t)} (-\log v)^{\gamma} \; dv
    \\&
    =  u(t) (-\log u(t))^\gamma
    (1+ O((-\log u(t))^{-1}))
    .
  \end{aligned}
  \end{equation}
  Since the above two right-hand sides are equal,
  \begin{align}
    u(t) (-\log u(t))^\gamma = t(1+o(1)).
  \end{align}
  Let $f(x) = x(-\log x)^\gamma$ and $g(y) = y(-\log y)^{-\gamma}$. Then
  $f$ and $g$ are approximate inverses in the sense that
  $f(g(y)) = y(1+o(1))$.
  Thus
    $u(t) = t(-\log t)^{-\gamma}(1+o(1))$.
  This completes the proof.
\end{proof}
\end{solution}

\section{Chapter \ref{ch:Tphi} exercises}

\begin{solution}{ex:Tphi}\label{soln:Tphi}
There is no dependence on $\Aux$ so we work with the
$T_{\varphi}$-seminorm.  
By Example~\ref{example:Tphi}, for $p \le k$,
\begin{equation}
    \|F^{(p)} (\varphi)\|
    \le
    \frac{k!}{(k-p)!} \|M \|_{\Auxx_1} \left( \frac{|\varphi|}{\h} \right)^{k-p},
\end{equation}
and $F^{(p)}$ is zero if $p>k$.  We insert this bound into
Definition~\ref{def:Tnorm} and obtain
\begin{equation}
\lbeq{FMk}
    \|F\|_{T_{\varphi} (\h)}
    \le
    \|M \|_{\Auxx_1}
    \sum_{p \le k}
    \binom{k}{p}  \left( \frac{|\varphi|}{\h} \right)^{k-p}
    =
    \|M\|_{\Auxx_1} \left(1+\frac{|\varphi|}{\h} \right)^{k} .
\end{equation}
This proves \eqref{e:prodTphi}.

To prove $\| (\varphi\cdot\varphi)^p \|_{T_{\varphi}} \le (|\varphi|+\h)^{2p}$,
by the product property Lemma~\ref{lem:Tphi-prod} it suffices to consider the case $p=1$.
For this we apply \eqref{e:FMk} with $F = \varphi\cdot\varphi$ and
$M(\varphi,\psi) = \varphi\cdot\psi$.  By the definition
\eqref{e:M-norm} and the Cauchy-Schwarz inequality $|\varphi\cdot\psi|
\le |\varphi||\psi|$, we have $\|M\|_{\Auxx_1}\le \h^2$.
Therefore, by \eqref{e:FMk},
$\|\varphi\cdot\varphi\|_{T_{\varphi}} \le (|\varphi|+\h)^2$
as desired.

For the last part, which is
$\|(\zeta \cdot\varphi) (\varphi\cdot\varphi)^p \|_{T_{\varphi}} \le |\zeta| (|\varphi|+\h)^{2p+1}$,
the product property and the previous estimate
reduce the desired bound to
$\|(\zeta \cdot\varphi) \|_{T_{\varphi}} \le |\zeta| (|\varphi|+\h)$.
This follows easily from \eqref{e:FMk} and the Cauchy-Schwarz inequality.
\end{solution}

\begin{solution}{ex:Tphi-poly}
This is an immediate corollary of Lemma~\ref{lem:TayTphi}, since
$F = \Tay_{k}F$ and hence
\begin{equation}
    \|F\|_{T_{\varphi,\aux}}
    =
    \|\Tay_k F\|_{T_{\varphi,\aux}}
    \le
    \|F\|_{T_{0,\aux}}
    P_{\h}^{k}(\varphi)  .
\end{equation}
\end{solution}

\begin{solution}{ex:EthetaV-6}
By Taylor's Theorem, $U(\varphi+\zeta_x) = \sum_{|\alpha| \le 4} \frac{1}{\alpha!}
U^{(\alpha)}(\varphi) \zeta_x^\alpha$ and hence, since $U^{(\alpha)}=V^{(\alpha)}$
for $|\alpha| \ge 1$,
\begin{equation}
        \Ex_{C_{+}}\big(\theta U (B) - U (B)\big)
        =
        \sum_{x\in B}
        \sum_{1 \le |\alpha| \le 4} \frac{1}{\alpha!}V^{(\alpha)}(\varphi)
        \Ex_{C_{+}}
        \zeta_x^\alpha
        .
\end{equation}
Therefore, by \refeq{Ualphabd} and Exercise~\ref{ex:wickpp}, and since $\h \ge \mathfrak{c}_+$,
\begin{align}
    \|
    \Ex_{C_{+}}\big(\theta U (B) - U (B)\big)
    \|_{T_{\varphi,\aux} (\h,\hAux)}
    & \le
    O(1) 
    \sum_{|\alpha| \in \{2, 4\}}
    \h^{-|\alpha|}
    \|V (B)\|_{T_{0,\aux}}
    P_{\h}^{4-|\alpha|}(\varphi) \mathfrak{c}_+^{|\alpha|}
    \nnb & \le
    O(1)
    \
    \left( \frac{\mathfrak{c}_+}{\h} \right)^2
    \|V(B)\|_{T_{0,\aux}}
    P_{\h}^{2}(\varphi)
    ,
\end{align}
as required.
\end{solution}

\section{Chapter~\ref{ch:pfsus} exercises}

\begin{solution}{ex:stability}
\label{sol:stability}
By the definition of $U$ and completing the square,
\begin{align}
    U (\varphi)
    &=
    \tfrac{1}{8} g |\varphi|^{4}
    +
    \tfrac{1}{8} g |\varphi|^{4}
    +
    \nu \tfrac{1}{2}
    |\varphi|^{2}
    +
    u
    \nonumber\\
    &=
    \tfrac{1}{8} g |\varphi|^{4}
    +
    \tfrac{1}{8} g
    \left(
    |\varphi|^{2}
    +
    \tfrac{2\nu}{g}
    \right)^{2}
    -
    \tfrac{1}{8} g
    \left(
    \tfrac{2\nu}{g}
    \right)^{2}
    +
    u
    \nonumber\\
    &\ge
    \tfrac{1}{8} g |\varphi|^{4}
    -
    \tfrac{1}{8} g
    \left(
    \tfrac{2\nu}{g}
    \right)^{2}
    +
    u
    =
    \tfrac{1}{8} g |\varphi|^{4}
    -
    \tfrac{1}{2}
    \tfrac{\nu^{2}}{g}
    +
    u .
    \label{e:stability-exercise-1}
\end{align}
The bounds on coupling constants due to $U \in \DVstab$ (defined in
\eqref{e:DVstab}) imply that
\begin{equation}
    -
    \tfrac{1}{2}
    \tfrac{\nu^{2}}{g}
    +
    u
    \ge
    -
    \tfrac{1}{2}
    k_0^3 L^{-dj}
    -
    k_0^4  L^{-dj}
    \ge
    - \tfrac{3}{2} k_0^3 L^{-dj},
\end{equation}
since $k_0 \le 1$.  Combining this with the previous bound and using
$g \ge k_0 \ggen$ gives the desired bound,
\begin{equation}
    U (\varphi)
    \ge
    \tfrac{1}{8} k_0  \ggen |\varphi|^{4}
    -
    \tfrac{3}{2}k_0^3 L^{- \drb j} ,
    \label{e:stability-exercise-2}
\end{equation}
which, by the definition \eqref{e:h-def} of $h$ is equivalent to the
first inequality in \eqref{e:stability}.
\end{solution}

\begin{solution}{ex:Ucal-ident}
\label{sol:Ucal-ident}
Let
\begin{equation}
    U (\varphi_{x})
    =
    \tfrac{1}{4}g (\varphi_{x}\cdot \varphi_{x})^{2} +
    \tfrac{1}{2} \nu \varphi_{x} \cdot \varphi_{x} + u .
\end{equation}
We write $\varphi = \varphi_{x}$, because $\varphi_{x}$ is constant in
the block that contains $x$.  By Definition~\ref{def:Tnorm}, the
$T_{0}$-seminorm equals the sum of the seminorms of the monomials in $F$,
because they have different degrees. Applying the definition of the
norm to the monomial $\tfrac{1}{4}g (\varphi\cdot
\varphi)^{2}$, we have
\begin{equation}
    \|\tfrac{1}{4}g (\varphi\cdot \varphi)^{2}\|_{T_{0} (\h)}
    =
    \sup
    \tfrac{1}{4}g
    \tfrac{1}{4!}\sum_{i_{1},\dots ,i_{4}}
    |\dot{\varphi}_{i_{1}}\cdot\dot{\varphi}_{i_{2}}|
    |\dot{\varphi}_{i_{3}}\cdot\dot{\varphi}_{i_{4}}| ,
\end{equation}
where the supremum is over unit norm directions
$\dot{\varphi}_{1},\dot{\varphi}_{2},\dot{\varphi}_{3},\dot{\varphi}_{4}$,
and $i_{1},\dots ,i_{4}$ is summed over permutations of
$1,2,3,4$. Note that the sum is normalised by $\frac{1}{4!}$.  By
\eqref{e:Rn-norm}, unit norm in the space $X$ means that the Euclidean
norm is $|\dot{\varphi}_{i}|=\h$. Therefore
\begin{equation}
    \|\tfrac{1}{4}g (\varphi\cdot \varphi)^{2}\|_{T_{0} (\h)}
    \le
    \tfrac{1}{4}g \h^{4}.
\end{equation}
This upper bound is actually equality because the right-hand side is
also a lower bound on the supremum by testing the case where
$\dot{\varphi}_{1},\dot{\varphi}_{2},\dot{\varphi}_{3},\dot{\varphi}_{4}$
all equal. By a similar easier argument we find that the norm of
$\frac 12 \nu \varphi \cdot \varphi$ is $\frac 12 |\nu| \h^2$ and obtain
\begin{equation}
    \|U_x\|_{T_{0} (\h)}
    =
    \tfrac{1}{4} |g | \h^{4} +
    \tfrac{1}{2} |\nu | \h^{2} +
    |u| ,
\end{equation}
as desired.  Since derivatives are taken in directions
$\dot{\varphi}$ that are constant on the block $b$ the norm of $U (b)$
is $|b| = L^{\drb j}$ times as large and \eqref{e:T0-equivalent}
immediately follows from \eqref{e:ell-def}.
\end{solution}

\section{Chapter~\ref{ch:R+U} exercises}

\begin{solution}{ex:R+U}
We consider some examples; higher-order and mixed derivatives can be handled similarly.

Let $\dot Q(b) = \LT (e^{V(b)}\dot K)$, and note that
$D_K Q(K;\dot K) =  \dot Q$ since $\LT$ is linear.
Recall from \refeq{RUrewrite} that
\begin{equation}
\lbeq{RUrewrite-soln}
    R_+^U(B)
    =
    - \Ex_{C_{+}}\theta Q (B) + \Cov_+ (\theta (V(B)-\half   Q(B)),\theta Q(B))  .
\end{equation}
Since $\Cov_+$ is bilinear, differentiation gives
\begin{align}
    D_K R_+^U(B; \dot K)
    & =
    -\Ex_+\theta \dot Q  + \Cov_+ (\theta  (V-Q) ,\theta \dot Q(B)) ,
    \\
    D_K^2 R_+^U(B; \dot K, \ddot K)
    & =
    -\Cov_+ ( \theta \dot Q ,\theta \ddot Q(B)),
\end{align}
and higher-order $K$-derivatives are zero.  Similarly,
\begin{align}
\lbeq{DVRsoln}
    D_V R_+^U(B; \dot V)
    & =
    -\Ex_+\theta \dot V Q(B) + \Cov_+(\theta \dot V(1-\half Q(B)), \theta Q(B))
    \nnb & \quad
    + \Cov_+ (\theta  (V(B) -\half Q(B) ),\theta \dot V(B) Q(B)) .
\end{align}

It is now straightforward to estimate the derivatives.  For example,
\begin{align}
    \| D_K R_+^U(B; \dot K)\|_{+}
    & \le
    \|\dot Q(B)\|_{+}
    + O(1) \|V(B)-Q(B)\|_{+} \|\dot Q(B)\|_{+}
    \nnb & \le O(\|\dot K\|) ,
\end{align}
which gives an $O(1)$ bound for $D_K R_+^U$.  Similarly, we obtain an $O(\|K\|)$
bound for $D_V R_+^U$ from the fact that each term on the right-hand side of
\refeq{DVRsoln} contains a factor $Q$.
\end{solution}

\section{Chapter~\ref{ch:saw-int-rep} exercises}

\begin{solution}{ex:connective-constant-bds}
\label{soln:connective-constant-bds}
There are $d^n$ $n$-step walks that take steps only in positive coordinate directions,
and such walks are self-avoiding.  Also, every walk that avoids reversing its previous
step is self-avoiding, and there are $(2d)(2d-1)^{n-1}$ such $n$-step walks.
Therefore $d^n \le c_n \le (2d)(2d-1)^{n-1}$, and the result follows.
\end{solution}

\begin{solution}{ex:srwbeta}
\label{soln:srwbeta}
We denote the right-hand side of \refeq{srwbeta} by $W_{xy}$,
and write $r_x= \frac{1}{\bar\beta_{x} + v_{x}}$.
The condition on $V$ guarantees that $W_{xy}$ converges, since
\begin{align}
    |W_{xy}|
    &=
    \left|
    \sum_{\omega \in \Wcal^*(x,y)}
    r_{Y_{|\omega|}}
    \prod_{i=0}^{|\omega|-1} r_{Y_i}\beta_{Y_{i}Y_{i+1}}
    \right|
    \le
    \sum_{n=0}^\infty \frac{1}{c} (\max_x r_x\bar\beta_x)^n
    ,
\end{align}
and the sum on the right-hand side converges because
$|r_x\bar\beta_x| \le \bar\beta_x/(\bar\beta_x+c) <1$, and hence the maximum is strictly less
than $1$ since there are finitely many points $x$ in $\Lambda$.

We extract the term with $|\omega|=0$, and condition on the first step for the remaining terms,
to get
\begin{equation}
    W_{xy} = r_x \delta_{xy} + \sum_{u\neq x} r_x \beta_{xu} W_{uy}.
\end{equation}
This can be rearranged to give
\begin{equation}
    (-\Delta_\beta W)_{xy} + \left(\frac{1}{r_x} - \sum_{u\neq x} \beta_{xu} \right) W_{xy} = \delta_{xy},
\end{equation}
which is the same as
\begin{equation}
    ((-\Delta_\beta+V) W)_{xy}  = \delta_{xy}.
\end{equation}
Therefore $W = (-\Delta_\beta +V)^{-1}$.  The special case follows from
the fact that the first product in \refeq{srwbeta} then selects the nearest-neighbour walks,
and $\bar\beta_x = 2d$ for all $x$.
\end{solution}

\begin{solution}{ex:detcancellation}
\label{soln:detcancellation}
(i) This is taken from \cite[Lemma~2.1]{BIS09}, and involves steps
used in the proof of Lemma~\ref{lem:detA-multi}.
Let $M=|\Lambda|$.
Consider first the case where $A$ is Hermitian.
Then there is a unitary matrix $U$ and a diagonal
matrix $D$ such that $A=U^{-1}DU$, so
$\phi A \phib = w D \bar{w}$ with $w = \bar{U} \phi$, and
\begin{equation}
    \frac{1}{(2\pi i)^M} Z_C
    =
    \prod_{x=1}^M
    \left(
    \frac{1}{\pi}
    \int_{-\infty}^\infty e^{-d_x (u_x^2+v_x^2)}
    du_xdv_x  \right)
    = \prod_{x=1}^M\frac{1}{d_x} = \frac{1}{\det A}.
\end{equation}
For the general case, we write $A (z) =G+iz H$ with
$G= \frac 12 (A+A^*)$, $H=\frac{1}{2i}(A-A^*)$ and $z=1$.
Since $\phi (iH) \phib$ is imaginary, when $G$ is positive definite the
integral in \eqref{e:Aint} converges and defines an analytic
function of $z$ in a neighborhood of the real axis.  For
$z$ small and purely imaginary, $A (z)$ is Hermitian and
positive definite, and hence \eqref{e:Aint} holds in this case.
Since $(\det A(z))^{-1}$ is a meromorphic function of $z$,
\eqref{e:Aint} follows from the uniqueness of analytic extension.

\smallskip \noindent
(ii)
We expand the exponential and obtain
\begin{align}
e^{-\psi A\bar\psi}
&= \sum_{n=0}^M \frac{(-1)^n}{n!} \left(\psi A\bar\psi\right)^n
= \frac{(-1)^M}{M!} \left(\psi A\bar\psi\right)^M + (\text{forms of degree} < 2M).
\end{align}
Only the forms of top degree ($2M$) contribute to the integral.
In the following, for simplicity we drop the symbol $\wedge$ for the wedge product.
By definition,
$\psi A\bar\psi = \sum_{x} \psi_x (A\bar\psi)_x = \sum_{x,y} A_{xy} \psi_x\bar\psi_y$,
and hence
\begin{align}
\left(\psi A \bar\psi\right)^M
&= \sum_{x_1,y_1} \cdots \sum_{x_M,y_M} A_{x_1 y_1} \cdots A_{x_M y_M} \psi_{x_1}\bar\psi_{y_1} \cdots \psi_{x_M} \bar\psi_{y_M}   \nnb
&= \sum_{\eta \in S_M} \sum_{\sigma \in S_M} A_{\eta(1) \sigma(1)} \cdots A_{\eta(M) \sigma(M)} \psi_{\eta(1)}\bar\psi_{\sigma(1)} \cdots \psi_{\eta(M)} \bar\psi_{\sigma(M)}  \nnb
&= M! \sum_{\sigma \in S_M} A_{1 \sigma(1)} \cdots A_{M \sigma(M)} \psi_{1}\bar\psi_{\sigma(1)} \cdots \psi_{M} \bar\psi_{\sigma(M)}  \nnb
&= M! \sum_{\sigma\in S_M} \sgn (\sigma) A_{1 \sigma(1)}  \cdots A_{M \sigma(M)} \psi_{1}\bar\psi_{1} \cdots \psi_{M} \bar\psi_{M}
\end{align}
In the above the second equality follows from the fact that any product with two identical
$\psi$ factors vanishes, the third follows by rearranging $\psi\bar\psi$ pairs (which does
not introduce signs), and the fourth follows by reordering the $\psi$ factors.
This proves that
$\left(\psi A \bar\psi\right)^M = (-1)^M M!\ \left(\det A\right) \bar\psi_{1}\psi_{1} \cdots \bar\psi_{M} \psi_{M}$
and the top degree part of $e^{-\psi A\bar\psi}$ is equal to
$\left(\det A\right) \bar\psi_{1}\psi_{1} \cdots \bar\psi_{M} \psi_{M}$.
Finally,
\begin{equation}
    \int e^{-S_A}
    =
    (\det A) \int e^{-\phi A \bar\phi} \bar\psi_{1}\psi_{1} \cdots \bar\psi_{M} \psi_{M},
\end{equation}
and the right-hand side is $1$ by \refeq{Aint}.
\end{solution}

\begin{solution}{ex:fermionsGauss}
\label{soln:fermionsGauss}
By Exercise~\ref{ex:detcancellation},
\begin{equation}
    \int e^{-S_A} f  =
    \frac{\int_{\R^{2M}} f\, e^{-\phi A\bar\phi}  \;d\bar\phi d\phi}
    {\int_{\R^{2M}}  e^{-\phi A\bar\phi} \;d\bar\phi d\phi}
    .
\end{equation}
By the symmetry of $A$,
\begin{equation}
    \phi A \phib = \frac 12 (u,Au) + \frac 12 (v,Av),
\end{equation}
so
\begin{equation}
    \int e^{-S_A} f  =
    \frac{\int_{\R^{2M}}f\, e^{-\frac 12 (u,Au) + \frac 12 (v,Av)}
    \;du\, dv}
    {\int_{\R^{2M}}  e^{-\frac 12 (u,Au) + \frac 12 (v,Av)} \;du\, dv}
    =\Ex_C f.
\end{equation}
This proves \refeq{SAGaussian}, and \refeq{SAGauss} then follows from
\begin{align}
    \int e^{-S_A} \phi_x\phib_y
    & =
    \Ex_C
    \frac 12 \left(u_xv_y + v_xv_y
    - iu_xv_y + iv_xu_y   \right)
    \nnb & =
    \frac 12 \left(C_{xy} + C_{xy}
    - 0 + 0 \right) = C_{xy}.
\end{align}
\end{solution}

\begin{solution}{ex:fin-vol-G0x}
\label{soln:fin-vol-G0x}
By the Cauchy--Schwarz inequality,
$T= \sum_{x\in \Lambda} L_{T,x} \le (|\Lambda|I(T))^{1/2}$, and hence
\begin{equation}
  G_{0x}^N(g,\nu)
  \leq
  \int_0^\infty e^{-gT^2/|\Lambda_N|} e^{-\nu T} \; dT < \infty
  \quad
  \text{for all $\nu \in \R$.}
\end{equation}
\end{solution}

\begin{solution}{ex:SUSY-ibp} \label{sol:SUSY-ibp}
By linearity of both sides, we may assume that $K$ is a $p$-form.
It follows from the definition of the super-expectation that both sides vanish
unless $K$ contains
the same number of factors of $\psi$ and $\psib$.  We can therefore assume
that $K =f(\phi,\phib) \psib_{x_1}\psi_{y_1} \cdots \psib_{x_p}\psi_{y_p}$.
Then
\begin{equation}
    \phib_x K e^{-S_A}
    = e^{-\phi A\phib} \phib_x f
    \sum_{N=0}^{|\Lambda|}
    \frac{(-1)^N}{N!}
    (\psib A\psi)^N \psib_{x_1}\psi_{y_1} \cdots \psib_{x_p}\psi_{y_p} .
\end{equation}
Since only the top-degree part of this form contributes to its integral,
\begin{equation}
  \int \phib_x K e^{-S_A} = T_{A,x_1,y_1,\dots,x_p,y_p} \Ex_C \phib_x f
\end{equation}
for some constants $T_{A,x_1,y_1,\dots,x_p,y_p}$ not depending on the function $f$.
Therefore, by standard Gaussian integration by parts (Exercise~\ref{ex:ibp}),
\begin{equation}
\lbeq{cibp1}
  \int \phib_x K e^{-S_A}
  = T_{A,x_1,y_1,\dots,x_p,y_p} \sum_{y \in\Lambda} C_{xy} \Ex_C\Big(\ddp{f}{\phi_y}\Big).
\end{equation}
Since the constants $T_{A,x_1,y_1,\dots,x_p,y_p}$ do not depend on $f$, it is also the case that
\begin{equation}
\lbeq{cibp2}
  \int \ddp{K}{\phi_y} e^{-S_A} = T_{A,x_1,y_1,\dots,x_p,y_p} \Ex_C\Big(\ddp{f}{\phi_y}\Big)
  .
\end{equation}
Insertion of \refeq{cibp2} into \refeq{cibp1} gives
\begin{equation}
  \int \phib_x K e^{-S_A}
  = \sum_{y \in\Lambda} C_{xy} \int \ddp{K}{\phi_y} e^{-S_A},
\end{equation}
as claimed.
\end{solution}

\begin{solution}{ex:edgesaw}\label{sol:edgesaw}
  Let $C$ be the identity matrix and set
  \begin{equation}
    G_{xy}(\beta)
    = {\sf E}_C\Big(\phib_x\phi_y \prod_{\{u,v\}\in E} (1+2\beta_{uv} \tau_{uv})\Big).
  \end{equation}
  Then $G_{xy}(\beta)$ is the right-hand side of \eqref{e:edgesaw-repr}.
  By Gaussian integration by parts \eqref{e:complexibp},
  \begin{equation}
     G_{xy}(\beta) =
      \delta_{xy}
      {\sf E}_C\Big(\prod_{\{u,v\}} (1+2\beta_{uv}\tau_{uv})\Big)
      +
      {\sf E}_C\Big(
      \phi_y
      \ddp{}{\phi_x}
      \prod_{\{u,v\}} (1+2\beta_{uv}\tau_{uv})\Big).
  \end{equation}
  The first term involves the expectation of a function of $(\tau_{uv})$
  which evaluates to $1$ by the localisation theorem \eqref{e:F0-bis}.
  Given $x,w\in\Lambda$,
  let $\beta^{(xw)}_{uv} = \beta_{uv}$ for
  $\{u,v\} \neq \{x,w\}$
  and
  $\beta^{(xw)}_{xw}=\beta^{(xw)}_{wx}=0$.
  The second expectation is
  \begin{align}
    &{\sf E}_C\Big(\phi_y
    \ddp{}{\phi_x}
    \prod_{\{u,v\}} (1+2\beta_{uv}\tau_{uv})\Big)
    \nnb
    &
      = \sum_{w\in \Lambda} \beta_{xw}
    {\sf E}_C\Big(\phib_w\phi_y
      \prod_{\{u,v\}}(1+2\beta_{uv}^{(xw)}\tau_{uv})\Big)
      = \sum_{w\in \Lambda} \beta_{xw} G_{wy}(\beta^{(xw)})
      .
  \end{align}
  Thus we have shown that
  \begin{align}
    G_{xy}(\beta) = \delta_{xy} + \sum_{w\in\Lambda} \beta_{xw} G_{wy}(\beta^{(xw)}).
  \end{align}
  This recursion characterises the weighted two-point function
  $\sum_{\omega\in \Tcal(x,y)}\beta^\omega$ appearing on the
  left-hand side of \refeq{edgesaw-repr}.
\end{solution}

\begin{solution}{ex:Qchain}
\label{soln:Qchain}
 Suppose first that $K$ is a collection of zero forms.  Then
 \begin{align}
     QF(K) & = \sum_{x\in\Lambda}
     \left[\psi_x \frac{\partial F(K)}{\partial \phi_x}  +
     \psib_x \frac{\partial F(K)}{\partial \bar\phi_x}  \right]
      =
     \sum_{j=1}^J F_j(K)\sum_{x\in\Lambda}
     \left[\psi_x \frac{\partial K_j}{\partial \phi_x}  +
     \psib_x \frac{\partial K_j}{\partial \bar\phi_x}  \right]
     ,
 \end{align}
 where the second equality follows from the
 chain rule for zero-forms. The right-hand side is $\sum_{j}
 F_{j}(K) QK_{j}$, so this proves \eqref{e:Qcr} for zero-forms and we may
 assume now that $K$ is higher degree.

 Let $\epsilon_j$ be the multi-index that has $j^{\rm th}$ component $1$ and
 all other components $0$.  Let $K^{0} = (K^0_j)_{j \in J}$ denote the  zero-degree part of $K$.
 By \eqref{e:Fdef}, the fact that $Q$ is an anti-derivation, and the chain
 rule applied to zero-forms,
 \begin{align}
     QF(K) & =
     \sum_{\alpha} \frac{1}{\alpha !}
     [QF^{(\alpha)}(K^{0})]
     (K - K^{0})^{\alpha}
     +
     \sum_{\alpha} \frac{1}{\alpha !}
     F^{(\alpha)}(K^{0})
     Q[(K - K^{0})^{\alpha}]
     \nnb
 \label{e:Qcr1}
     & = \sum_{\alpha} \frac{1}{\alpha !}
     \sum_{j=1}^J
     F^{(\alpha + \epsilon_j)}(K^{0})[ QK^{0}_i]
     (K - K^{0})^{\alpha} \nnb
     &\quad\quad+
     \sum_{\alpha} \frac{1}{\alpha !}
     F^{(\alpha)}(K^{0})
     Q[(K - K^{0})^{\alpha}].
 \end{align}
 Since $Q$ is an anti-derivation,
 \begin{equation}
 \label{e:Qcr2}
     Q(K - K^{0})^\alpha
     =
     \sum_{j=1}^J
     \alpha_{j} (K - K^{0})^{\alpha -\epsilon_j}[QK_j - QK^{0}_j].
 \end{equation}
 The first term on the right-hand side of \eqref{e:Qcr1} is cancelled by
 the contribution to the second term of \eqref{e:Qcr1} due to the
 second term of \eqref{e:Qcr2}.  The remaining contribution to the second term
 of \eqref{e:Qcr1} due to the first term of \eqref{e:Qcr2} then gives
\begin{equation}
    QF(K)=
    \sum_{j=1}^J \left( \sum_\alpha \frac{1}{\alpha!} F^{(\alpha)}(K^0)
    \alpha_j (K - K^{0})^{\alpha -\epsilon_j}  \right) QK_j
    =
    \sum_{j=1}^J  F_j(K) QK_{j}
\end{equation}
as required.
\end{solution}


%% file: paley-wiener.tex
\begin{tikzpicture}[scale = 1]
\coordinate (c) at (0:0) {};
\path (.3,.2) node  {\small $\theta$};
\draw (.6,0) arc (0:60:.6);
\draw[thick] (c) -- (60:3) coordinate (ab);
\path (ab) ++ (.3,.2) node {\small $(A,B)$};
\path (180:3) coordinate (left);
\path (0:3) coordinate (right);
\draw (ab) -- (left);
\path(left) node at + (1,.2) {\small $\theta/2$};
\path (left) ++ (1.5,0) coordinate (arcstart);
\draw (arcstart) arc (0:30:1.5);
\draw[thick] (left) -- (right);
\draw[thick] (right) arc (0:180:3);
\path (1.5,0) coordinate (perp);
\draw[thick] (ab) -- (perp);
\draw[dashed, thick,<-] (-3,-.2) -- (-1.5,-.2) node (r) {\small $\;\; R$};
\draw[dashed, thick,->] (r) -- (-.1,-.2);

\path (.725,-.2) node (a) {\small $A$};
\draw[dashed, thick,->] (a) -- (.1,-.2);
\draw[dashed, thick,->] (a) -- (1.5,-.2);
\path (-1.0,1.8) node {\small $\sqrt{(R+A)^{2}+B^{2}}$};
\end{tikzpicture}